\documentclass[12pt,a4paper]{article}

\usepackage[english]{babel}
\usepackage[autostyle,english=british]{csquotes} 

\usepackage{latexsym}
\usepackage{amssymb,amsmath}
\usepackage{amsfonts}
\usepackage{mathtools}
\usepackage{verbatim}
\usepackage{amsthm}
\usepackage{amsbsy}
\usepackage{graphicx,color}
\usepackage{epsfig}
\usepackage{graphicx}
\usepackage{float}
\usepackage{braket}
\usepackage[mathscr]{eucal}
\usepackage{mathcomp}
\usepackage{subcaption}
\usepackage{mathdots}
\usepackage{textcomp}
\usepackage{simpler-wick}
\usepackage{url}	
\usepackage{physics}
\usepackage{cancel}
\usepackage{enumitem}
\usepackage{bm}
\usepackage{bbm}
\usepackage{slashed}
\usepackage{mathrsfs}

\usepackage[utf8]{inputenc}
\usepackage{epigraph}
\makeatletter
\newenvironment{equations}[1][]{\subequations\ifx\relax#1\relax\else\label{#1}\fi\align\ignorespaces}{\endalign\ignorespacesafterend\endsubequations}
\def\@spliteq#1{\begin{equation}\begin{split}#1\end{split}\end{equation}}
\def\splitequation{\collect@body\@spliteq}

\makeatother

\usepackage{etoolbox}
\usepackage{dynkin-diagrams}
\def\row#1/#2!{#1_{\IfStrEq{#2}{}{n}{#2}} & \dynkin{#1}{#2}\\}
\newcommand{\tble}[1]{
   \renewcommand*\do[1]{\row##1!}
   \[
      \begin{array}{ll}\docsvlist{#1}\end{array}
   \]
}

\newcommand{\diff}{\mathrm{d}}
\newcommand{\de}{\partial}
\newcommand{\vepsilon}{\varepsilon}
\newcommand{\vtheta}{\vartheta}
\newcommand{\vphi}{\varphi}
\newcommand{\vrho}{\varrho}
\newcommand{\s}{\slashed}

\newcommand{\bb}[1]{\mathbf{#1}}

\newcommand{\0}[1]{\mathring{#1}}
\newcommand{\C}{\mathbb{C}}
\newcommand{\R}{\mathbb{R}}

\newcommand{\nn}{\nonumber}

\newcommand{\Ga}{\Gamma}
\newcommand{\Ll}{[\![}
\newcommand{\Lr}{]\!]}

\usepackage[backend=bibtex, citestyle=numeric-comp, bibstyle=ieee]{biblatex}
\usepackage{enumitem}
\usepackage[a4paper, total={6in, 9in}]{geometry}
\usepackage[hidelinks]{hyperref}
\definecolor{jade}{RGB}{0, 168, 107}
\hypersetup{
 colorlinks,
 linkcolor={blue},
 citecolor={blue},
 urlcolor={blue}
}

\renewcommand{\theequation}{\thesubsection.\arabic{equation}} \csname
@addtoreset\endcsname{equation}{subsection}

\makeatletter
\newcommand{\firstsectioneqnums}{%
  \renewcommand{\theequation}{\thesection.\arabic{equation}}%
  \@addtoreset{equation}{section}%
}
\newcommand{\restoreeqnums}{%
  \renewcommand{\theequation}{\thesubsection.\arabic{equation}}%
  \@addtoreset{equation}{subsection}%
}
\makeatother

\begin{document}

\begin{titlepage}

\begin{center}

\includegraphics[width=10.5cm,height=5cm]{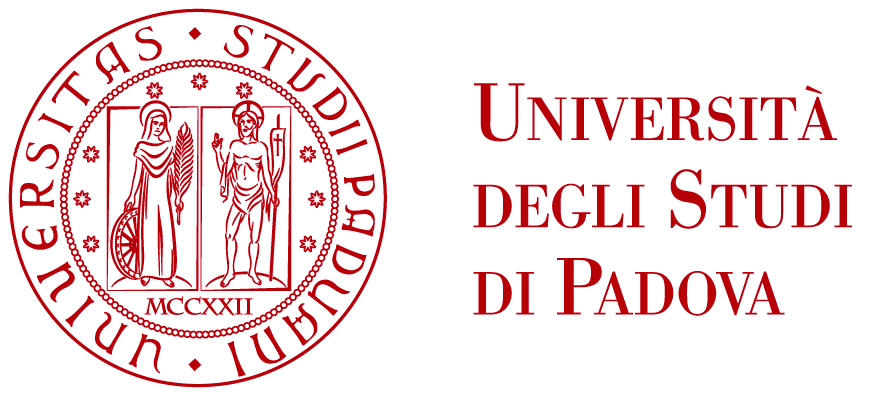}

\vspace{10mm}

{\large{\bf Università degli Studi di Padova}} 

\vspace{5mm}

{\large{\bf Dipartimento di Fisica e Astronomia ``Galileo Galilei''}} 

\vspace{5mm}

{\Large{\bf \emph{Philosophiae Doctoral Course in Physics}}}

\vspace{15mm}

{\Huge{\bf The Use of Torsion}}\\
\vspace{3mm}
{\Huge{\bf in  Supergravity Uplifts}}\\
\vspace{3mm}
{\Huge{\bf and Covariant Fractons}}

\end{center}

\vspace{10mm}

{\Large{ Coordinator: Prof. Giulio Monaco}}

\vspace{3mm}

{\Large{Supervisor: Prof. Gianluca Inverso}}

\vspace{3mm}

{\Large{Co-supervisor: Prof. Gianguido Dall'Agata}}

\hfill

\begin{flushright}
{\Large{Candidate: Davide Rovere}}
\end{flushright}

\vspace{30mm}

\rule[0.5cm]{15.8cm}{0.6mm}

\begin{center}
{\large{\bf Anno Accademico 2024/2025}}
\end{center}

\end{titlepage}

\newpage
\quad
\setcounter{page}{2}
\thispagestyle{empty}

\newpage
\quad\\
\quad\\
\quad\\
\quad\\
\quad\\
\quad\\
\quad\\
\quad\\
\begin{flushright}
\emph{a Koffi Romeo}
\end{flushright}
\thispagestyle{empty}

\newpage
\quad
\thispagestyle{empty}

\newpage
\quad\\
\quad\\
\quad\\
\quad\\
\quad\\
\quad\\
\quad\\

\epigraph{Take nothing for granted if you can check it. Even though that seem waste-work, and has nothing to do with the essentials of things, it encourages the Daemon. There are always men who by trade or calling know the fact or the inference that you put forth. If you are wrong by a hair in this, they argue ‘False in one thing, false in all.’ Having sinned, I know. Likewise, never play down to your public — not because some of them do not deserve it, but because it is bad for your hand. All your material is drawn from the lives of men. Remember, then, what David did with the water brought to him in the heat of battle.}{R. Kipling, \emph{Something of Myself}}
\thispagestyle{empty}

\newpage
\quad
\thispagestyle{empty}

\newpage
\begin{center}
\textbf{Abstract}
\end{center}

The aim of this Thesis is twofold. On the one hand, we find the necessary and sufficient conditions for a maximally supersymmetric supergravity theory in three dimensions to be a solution of eleven-dimensional supergravity  (but the result is general and also holds for ten-dimensional supergravities), with eight dimensions compactified into a coset space. The used method is based on the formalism of \emph{generalised geometry}, useful for the study of dualities in string theory and supergravity. The analysis extends the known results to the case in which the duality group of the reduced theory is $E_{8(8)}$, whose generalised geometry is still little understood.

On the other hand, we study properties of the so-called \emph{covariant fracton gauge theory}, computing the \textsc{brst} cohomology and consistent anomalies, and showing that its solutions describe a specific subsector of an extension of General Relativity, called \emph{M{\o}ller-Hayashi-Shirafuji theory}. Covariant fracton theory is the gauge theory of a symmetric rank-two tensor, invariant under gauge transformations depending on the second derivative of a scalar parameter, and is the Lorentz-covariant extension of the continuous limit of spin-chain theories admitting excitations with reduced mobility (called ``fractons"), due to the conservation of the dipole moment.

In both cases, the \emph{Weitzenb\"ock torsion} plays a crucial r\^ole. In the usual formulation of General Relativity, the spacetime curvature is responsible for the gravitational interaction. However, an alternative formulation exists, in which spacetime is flat and gravitation is an effect of the Weitzenb\"ock torsion. In this context, the M{\o}ller-Hayashi-Shirafuji theory is the extension of General Relativity, in parallelisable spacetime, which spoils local Lorentz invariance. At the linearised level, it corresponds to add an antisymmetric rank-two tensor to the symmetric rank-two tensor describing the perturbation of the metric.

Consistent reductions, studied by Scherk and Schwarz as an extension of the Kaluza-Klein compactifications, are reductions of ($d$+$n$)--dimensional (super)gravity theories to $d$--dimensional theories, the remaining $n$ dimensions being compactified into an internal space, in such a way that all solutions of the reduced theory are also solutions of the original one. Consistency requires the internal  geometry to be parallelisable with constant Weitzenb\"ock torsion. The internal space is the gauge group of the reduced theory, and the torsion corresponds to the structure constants of the group algebra. Generalised geometry allows to consider cases where the internal space is not a group, by extending the notions of parallelisability and torsion. The torsion is identified with the so-called ``embedding tensor", which captures the gauge couplings of the reduced theory.

The Thesis includes a self-contained review of the main notions needed to understand the original results, comprising, in addition to the already mentioned topics, duality in supergravity theories, the generalised Lie derivative, formulation of eleven-dimensional supergravity suitable to reductions with duality groups given by exceptional Lie groups, salient aspects of three-dimensional gravity, and the \textsc{brst} formalism for computing anomalies in field theories.

\thispagestyle{empty}

\newpage
\quad 
\thispagestyle{empty}

\newpage
\begin{center}
\textbf{Sommario}
\end{center}

Lo scopo della presente tesi è duplice. Da un lato, si trovano le condizioni necessarie e sufficienti affinché una teoria di supergravità massimamente supersimmetrica in tre dimensioni possa essere soluzione della teoria di supergravità in undici dimensioni (ma il risultato è generale ed è valido anche per supergravità in dieci dimensioni), con otto dimensioni compattificate in uno spazio di coset. Il metodo usato è basato sulla \emph{geometria generalizzata}, utile allo studio di dualità in teoria di stringa e supergravità. L'analisi estende i risultati noti al caso in cui il gruppo di dualità della teoria ridotta è $E_{8(8)}$, la cui geometria generalizzata presenta peculiarità ancora poco comprese. 

Dall'altro lato, si studiano le proprietà della \emph{teoria di gauge frattonica covariante}, calcolando coomologia \textsc{brst} e anomalie consistenti, e mostrando che le sue soluzioni descrivono un sottosettore specifico di un'estensione della relatività generale, detta \emph{teoria di M{\o}ller-Hayashi-Shirafuji}. La teoria frattonica covariante è la teoria di gauge di un campo simmetrico di rango due, invariante sotto trasformazioni di gauge dipendenti dalla derivata seconda di un parametro scalare, ed è l'estensione covariante del limite continuo di teorie di catene di spin che ammettono eccitazioni con mobilità ridotta (dette appunto ``frattoni"), a causa della conservazione del momento di dipolo. 

In entrambi i casi la \emph{torsione di Weitzenb\"ock}  svolge un ruolo essenziale. Nella formulazione usuale della relatività generale, la curvatura dello spaziotempo è responsabile dell'interazione gravitazionale. Tuttavia, è ammessa una formulazione alternativa, in cui lo spaziotempo è piatto e la gravitazione è effetto della torsione di Weitzenb\"ock. In questo contesto, la teoria di M{\o}ller-Hayashi-Shirafuji è l'estensione della relatività generale, in caso di spaziotempo parallelizzabile, che rinuncia all'invarianza di Lorentz locale. A livello linearizzato, corrisponde ad affiancare un tensore antisimmetrico di rango due al tensore simmetrico di rango due che descrive la perturbazione della metrica. 

Le riduzioni consistenti, studiate da Scherk e Schwarz come estensione della compattificazione di Kaluza-Klein, sono riduzioni di teorie di (super)gravità in $d$+$n$ dimensioni in teorie in $d$ dimensioni, le restanti $n$ dimensioni essendo compattificate in uno spazio interno, in modo che tutte le soluzioni della teoria ridotta siano anche soluzioni della teoria di partenza. La consistenza richiede che la geometria interna sia parallelizzabile con torsione di Weitzenb\"ock costante. Lo spazio interno è il gruppo di gauge della teoria ridotta, e la torsione corrisponde alle costanti di struttura dell'algebra del gruppo. La geometria generalizzata consente di considerare anche casi in cui lo spazio interno non è un gruppo, mediante l'estensione della nozione di parallelizzablità e di torsione fornita dalla geometria generalizzata.  La torsione è ora identificata con il cosiddetto ``tensore di immersione", che descrive gli accoppiamenti di gauge della teoria ridotta.

La tesi contiene un'ampia rassegna delle principali nozioni necessarie per comprendere i risultati originali, tra cui, oltre agli argomenti già citati, dualità in teorie di supergravità, derivata di Lie generalizzata, formulazione della supergravità in undici dimensioni funzionale a riduzioni con gruppi di dualità dati dai gruppi di Lie eccezionali, aspetti salienti della gravità in tre dimensioni e il formalismo \textsc{brst} per calcolare anomalie in teorie di campo.
  
\thispagestyle{empty}

\newpage
\quad
\thispagestyle{empty}

\newpage
\tableofcontents

\newpage
\quad
\thispagestyle{empty}

\newpage
\begin{center}
\textbf{Declaration}
\end{center}

The results of the original part of the present Thesis, contained in Sections \ref{4} and \ref{5}, have already appeared in the following papers:
\begin{itemize}
\item[--]  with G. Inverso, \emph{How to Uplitf 3d Maximal Supergravities}, \href{https://link.springer.com/article/10.1007/JHEP02(2025)130}{\textcolor{blue}{JHEP 02 (2025) 130}}, \href{https://arxiv.org/abs/2410.14520}{\textcolor{blue}{2410.14520}},
\item[--] \emph{Anomalies in Covariant Fracton Theories}, \href{https://link.aps.org/doi/10.1103/PhysRevD.110.085012}{\textcolor{blue}{Phys. Rev. D 110, 085012 (2024)}}, \href{https://arxiv.org/abs/2406.06686}{\textcolor{blue}{2406.06686}},
\item[--] \emph{Covariant fractons and Weitzenb\"ock torsion}, \href{https://journals.aps.org/prd/abstract/10.1103/y426-6lkj}{\textcolor{blue}{Phys. Rev. D 112, 105002 (2025)}}, \href{https://arxiv.org/pdf/2505.21022}{\textcolor{blue}{2505.21022}}.
\end{itemize}
Some marginally covered aspects are discussed in more details in 
\begin{itemize}
\item[--] with C. Imbimbo, and A. Warman, \emph{Superconformal Anomalies from Superconformal Chern-Simons Polynomials}, \href{https://link.springer.com/article/10.1007/JHEP05(2024)277}{\textcolor{blue}{JHEP 05 (2024) 277}}, \href{https://web3.arxiv.org/abs/2311.05684}{\textcolor{blue}{2311.05684}},
\item[--] \emph{Kodaira-Spencer Anomalies with Stora-Zumino Method}, \href{https://link.springer.com/article/10.1007/JHEP01(2025)073}{\textcolor{blue}{JHEP 01 (2025) 073}}, \href{https://arxiv.org/abs/2403.17071}{\textcolor{blue}{2403.17071}},
\item[--] with F. Fecit, \emph{Wordline formulations of Covariant Fracton Theories}, \href{https://arxiv.org/abs/2508.14591}{\textcolor{blue}{2508.14591}}.
\end{itemize}
The following work, which extends some aspects explored in the Thesis, appeared after the latter was completed
\begin{itemize}
\item[--] with C. Sterckx, \emph{How to uplift non-maximal gauged supergravities}, \href{https://arxiv.org/abs/2510.24850}{\textcolor{blue}{2510.24850}}.
\end{itemize}
During the PhD, the author worked also on topics concerning conformal gravity. Namely, the following work appeared after the thesis was completed
\begin{itemize}
\item[--] with N. Boulanger, \emph{8D conformal gravity with Einstein sector, and its relation to the Q-curvature}, \href{https://arxiv.org/abs/2510.24850}{\textcolor{blue}{2511.01368}},
\end{itemize}
and the following ones are currently in preparation
\begin{itemize}
\item[--] with G. Anastasiou, A. Araya, N. Boulanger, and R. Olea, \emph{Conformal renormalisation in eight dimensions}, in preparation,
\item[--] with N. Boulanger, \emph{Classification of ten-dimensional conformal anomalies}, in preparation.
\end{itemize}

\newpage
\quad
\thispagestyle{empty}

\newpage

\section{Introduction}

\textbf{\emph{Dynamics and geometry.}} A crucial challenge in the modern perspective of Theoretical Physics is the interplay between \emph{dynamics} and \emph{geometry} in field theories. The ``dynamics" of a field theory takes place in a background, which is spacetime, governed by ``geometry", but the developments that have been made have progressively led to the dismantling of this rigid distinction, particularly when gravity is considered. 

The three interactions described by the Standard Model are thought of as forces in a fixed, flat spacetime, whereas gravitational interaction is described by Einstein's theory as a geometric effect, because it amounts to spacetime curvature. Namely, the novelty of gravity compared to other interactions lies in its lack of a fixed background: reversing the perspective, saying that gravitational interaction is described in geometric terms means that the background is also part of the dynamics. Mathematically, this is a consequence of the independence of the theory from general changes of coordinates (\emph{general covariance}), that is, from the independence of the observer's point of view. Nevertheless, as Klein-Gordon theory describes the propagation of a helicity-zero mode, or Maxwell electromagnetism describes a helicity-one mode, so solutions of General Relativity having the form of small perturbations of the Minkowski vacuum describe the propagation of an helicity-two mode. 

We know of a quantisation procedure for field theories with a fixed background: classical fields are promoted to operators acting on a Hilbert space of states, and a perturbation theory is established around a classical solution with a fixed background, suitably renormalising the coupling constants when a further order in perturbation theory is taken into account.\footnote{Gauge theories present an additional difficulty due to gauge redundancy, which physics must not depend on. The \textsc{brst} procedure, based on the discovery that the quantum theory exhibits a fermionic symmetry  (i.e., a symmetry that changes bosonic/fermionic statistics of fields, without changing the spin, unlike supersymmetry) even when gauge invariance is broken by a gauge choice, has allowed to successfully quantise gauge theories as well.}
But, the mass dimension of the gravitational constant makes the theory of gravity non-renormalisable, meaning that the perturbation series around the vacuum is non-convergent and the integrals corresponding to the scattering amplitudes diverge at high energies. This problem stems from our ignorance of the high-energy physics of phenomena involving gravity, and can be understood by adopting the perspective of effective theories: General Relativity is an effective theory in the sense that it describes physics up to a certain energy scale, freezing the degrees of freedom at higher energy scales. But beyond this, another more conceptual problem affects the quantisation of gravity: it is the fact that we do not know how to quantise a field theory in a background independent way \cite{Witten:1988ze, Witten:1988xj, Labastida:1988zb}. 

\textbf{\emph{String theory.}} A conceptual paradigm more than a single theory is proposed by ``string theory". In short, the paradigm is this: it is possible to treat spacetime coordinates as fields in a field theory. Just as the coordinate describing a particle's worldline is parametrised by one parameter, the proper time, so the propagation of a string describes a two-dimensional \emph{worldsheet}, parametrised by two parameters. Worldsheet replaces spacetime as a background. The description of the worldsheet must not depend on the particular parametrisation – that is, there must be two-dimensional general covariance. The low-energy excitations arising from the first quantisation of the string coordinates always include a spin-two mode. These excitations can be made to correspond to those of a standard field theory on a fixed, flat spacetime. Since the quantised string coordinates describe a tower of other, higher-energy states, such a field theory is an effective theory that freezes the higher-energy propagating degrees of freedom. 


Geometrically, the formal requirement for the absence of anomalies – that is, that the symmetries of classical theory are also symmetries of quantum theory – constrains the number of fields in string theory, i.e., the dimension of spacetime. In presence of fermions (\emph{superstring theory}), this number is ten. The effective field theories that describe the low-energy limit of string theory are a small class of \emph{supergravity theories}. Among them, a peculiar r\^ole is played by \emph{eleven-dimensional supergravity}, discovered by Cremmer, Julia, and Scherk in 1978 \cite{Cremmer:1978km}. It is defined in one dimension more than string theory, so that its high-energy completion (\emph{M-theory}) is not completely understood yet, but it has a striking simple field content, including only the metric tensor, a three-form gauge field, and the gravitino. 

\textbf{\emph{(Super)gravity consistent reductions.}} Following this line of reasoning, our universe should be found as a solution in the string theory framework. The extra dimensions should form an additional structure in any spacetime point. Phenomenologically, if the ``internal space" is compact with characteristic length of the same order of magnitude of the Planck length, spacetime and spacetime endowed with such additional structure are experimentally indistinguishable \cite{Witten:1986}.

Now, the question to ask is: when and how is it possible to truncate a higher-dimensional theory to a lower-dimensional one, in the sense that all the solutions of the reduced theory are also solutions of the higher one? When this happens, the truncation of the higher theory is said to be \emph{consistent}. In this perspective, the first attempt was addressed by Kaluza and Klein \cite{Kaluza:1921tu, Klein:1926tv}. They were motivated to consider the truncation of five-dimensional Einstein gravity, because they realised that the five-dimensional metric can be thought as containing the components of a four-dimensional metric and of the electromagnetic gauge field. Moreover, the symmetries of five-dimensional gravity contains the symmetries of four-dimensional Einstein and Maxwell theories. The truncation was performed by imagining to make the fifth dimension periodic as a small circle, and assuming that the physics should not depend on the coordinate on such a circle as long as it is small enough. Therefore, Kaluza and Klein thought that increasing by one the number of spacetime dimensions was a promising way to study gravitation and electromagnetism in a unified way. Nevertheless, the idea was soon discharged because a scalar field is also needed to be added in order to exhaust the degrees of freedom encoded in five-dimensional metric, and such a scalar field lacked of a physical interpretation. 

Subsequently, Kaluza and Klein's idea drew renewed interest both in the contest of conformal gravity, where the additional scalar field was interpreted as a dilaton field \cite{Jordan:1955, Brans:1961sx}, and, above all, in the context of string theories, which are thought to be consistent only in ten or eleven dimensions, because of formal arguments (namely, the absence of anomalies \cite{Green:1984sg}). An extension of the compactification mechanism conceived by Kaluza and Klein would have allowed to think the additional spacetime dimensions besides the four observed ones as forming a small compact internal space in each external spacetime point. In particular, Scherk and Schwarz understood how to extend Kaluza-Klein compactification to internal spaces other then the simple circle (or hypertori in higher dimensions), providing a consistent truncation procedure when the internal space is a group \cite{Scherk:1978ta, Scherk:1979zr}.

\textbf{\emph{Uplifting a field theory.}} Nevertheless, it is quite natural to also consider the opposite perspective than the reduction of the higher-dimensional theory to get our universe as a solution (or, more in general, to find solutions of the higher-dimensional one). Namely, one may start from a class of lower-dimensional theories and asks which of them can be \emph{uplifted} to a specific higher-dimensional theory, meaning that the lower-dimensional theory can be treated as a consistent truncation of the higher-dimensional one \cite{Inverso:2017lrz, Inverso:2024xok}. 

In this Thesis we focus on the bosonic sector of eleven-dimensional supergravity as higher-dimensional theory, but the results cover also uplifts to the other type of low-energy limit of string theory, namely type IIB and type IIA supergravities, the last one also with mass deformation \cite{Ciceri:2016dmd}, and the fermionic sector could be taken into account as well. As lower-dimensional theories \emph{gauged supergravities} will be considered \cite{Samtleben:2008pe, Trigiante:2016mnt}, which are supergravity theories in which part of the big group of global symmetries of their equations of motion (\emph{duality group}) is promoted to a local gauge symmetry.

\textbf{\emph{Parallelisable manifolds and the r\^ole of the torsion.}} The uplift problem needs to deeply investigate the geometric aspect of compactifications, going beyond even the Scherk-Schwarz paradigm. By the way, consistent reductions on spheres which are not groups (so that they are not encompassed in Scherk-Schwarz compactification) were soon discovered. The most famous example is the consistent truncation of eleven-dimensional supergravity on the seven-dimensional sphere $S^7$, which brings to the maximal four-dimensional supergravity, with eight supercharges \cite{deWit:1982bul, deWit:1986oxb}. In \cite{Nastase:1999cb, Nastase:1999kf, Cvetic:2000nc, Cvetic:2000ah} there are also truncations on $S^4$ or $S^5$ of eleven-dimensional, type IIB and IIA supergravities.

$S^7$ is peculiar because, although it is not a group manifold, is a \emph{parallelisable manifold}. A manifold is said ``parallelisable" if it is possible to flatten it, as when one draws a map. For example, we can open a circle to get a straight segment: this means that the circle or $S^1$ is parallelisable. Similarly, we can ideally parallelise a doughnut, which geometrically is a two-dimensional torus or $T^2$, since it can be thought as a rectangle with opposite sides glued together. Instead, cartographers soon faced the problem of how to represent the globe on a flat map, and realised that there was no way to do so except through projections: for example, the famous Mercator map preserves the angles, that is, it is a conformal projection, but increases the area as one moves away from the equator, creating the well-known effect whereby Greenland appears to have an extension comparable to that of Africa, while in reality it is much smaller. In other words, the two-dimensional spheres or $S^2$ is not parallelisable. Mathematicians had long wondered which spheres could be parallelised. The answer came in 1958, when Bott, Milnor, and Kervaire proved that the unique parallelisable spheres are $S^1$, $S^3$, and $S^7$ \cite{Bott1958-vt, Milnor1958, Kervaire1958}.

More precisely, in a parallelisable manifold it is always possible to find a unique basis for the tangent spaces in all the points, usually called \emph{absolute frame}. This is a set of as many independent globally-defined vectors $e_a = e^\mu{}_a\,\de_\mu$ as the dimension of the manifold. The commutator between elements of the absolute frame $[e_a,e_b]$ (which is the same as the Lie derivative $\mathcal{L}_{e_a}\,e_b$) is written as a linear combination of the absolute frame, whose coefficients $-T_a{}^c{}_b$, are minus of the so-called \emph{intrinsic torsion} or \emph{Weitzenb\"ock torsion} of the manifold:
\begin{equation}
[e_a,e_b] = \mathcal{L}_{e_a}\,e_b = -T_a{}^c{}_b\,e_c.
\end{equation}
Now we can understand the reason why the Scherk-Schwarz compactification works for group manifold. Scherk and Schwarz showed that the truncation is consistent if the internal space admits an absolute frame with \emph{constant} torsion, the absolute frame giving the factorised dependence on the internal coordinates in the reduced fields. This is the case of group manifolds, because a Lie group is always a parallelisable manifold, with absolute frame given by the set of generators of the Lie algebra. Therefore, the intrinsic torsion is given by minus the structure constants $f_{ab}{}^c$, and so it is constant:
\begin{equation}
[e_a,e_b] = \mathcal{L}_{e_a}\,e_b = f_{ab}{}^c\,e_c.
\end{equation}
The Kaluza-Klein case is encompassed in this context, since the internal geometry is a torus, which is parallelisable with vanishing torsion.

$S^7$ is parallelisable, but it has no constant intrinsic torsion, since indeed it is not a group manifold. Even worse, $S^4$ or $S^5$ are not even parallelisable. The natural question is at this point if it is possible to extend the Scherk-Schwarz procedure, in order to understand why these compactifications are also consistent. 

\textbf{\emph{Exceptional field theory.}} The solution was offered by the inspection on the geometry of dualities in supergravity. By ``duality" one means a symmetry of the classical equations of motion of two different theories. The duality transformation brings one theory into another, while preserving the equations of motion. The structure of dualities appeared particularly rich in supergravity theories \cite{Julia:1980gr, Cremmer:1997ct}. Namely, when eleven-dimensional supergravity is compactified up to five, four or three dimensions, the resulting lower-dimensional theories enjoy a much wider duality group than expected: their equations of motion are invariant under the so-called \emph{exceptional Lie groups} $E_{n(n)}$, where $n=6,7,8$ is the dimension of the internal space.\footnote{$E_{n(n)}$ means that the exceptional groups are considered in their maximal non-compact form.}

The emergence of these exotic groups looked mysterious and it was quite difficult to deal with them, because the invariance under such  duality groups is not manifest after the compactification. Nevertheless, a reformulation of higher-dimensional supergravities has been developed, suited to the dimension of the reduced theory, which allows to repackage its fields in a covariant way with respect to the exceptional groups \emph{before} the reduction. This formulation is called \emph{exceptional field theory} \cite{Hohm:2013vpa, Hohm:2013uia, Hohm:2014fxa}. When the reduction is performed starting with this formulation of the higher-dimensional theory, the duality group is manifest. Moreover, exceptional field theory allows to put all the known consistent truncations in a unified setting. 

\textbf{\emph{Generalised geometry.}} The formalism which makes exceptional field theory possible is an extension of \emph{generalised geometry}, which we will refer to as ``exceptional geometry" \cite{Hull:2007zu, Grana:2008yw, PiresPacheco:2008qik, Coimbra:2011ky, Coimbra:2011nw, Coimbra:2012af, Strickland-Constable:2013xta}. ``Generalised geometry" was introduced by Hitchin, Gualtieri, and Cavalcanti for pure mathematical reasons \cite{Hitchin:2003cxu,Gualtieri:2003dx, Hitchin:2005in, Gualtieri:2007ng, Hitchin:2010qz}. It provides a formalism in which symplectic and complex geometry are unified, extending the notion of Dirac structure, which unifies Poisson and symplectic geometry and dates back to Courant, Weinstein and Dorfman's work \cite{Courant1988, Dorfman1987, Courant1990}. But generalised geometry proved very useful also in Theoretical Physics, because it captures the geometry underlying T-duality in string theory \cite{Buscher:1987sk, Buscher:1987qj}: this application is known as \emph{double field theory} \cite{Hull:2006va, Hull:2009mi, Berman:2010is, Aldazabal:2013sca, Hohm:2013bwa}. 

The idea of generalised geometry is to combine vector fields and one-forms into single objects, treated as they were vector fields. One can extend the notion of Lie derivative on the space of such generalised vectors, defining a \emph{generalised Lie derivative} $L$, which is a deformation of the usual Lie derivative $\mathcal{L}$. One realises that generalised diffeomorphisms generated by generalised vector fields encode the usual diffeomorphisms, generated by the vector components of generalised vectors, and one-form gauge transformations, generated by the one-form components. The latter ones are the gauge transformations of a two-form gauge field, which can be identified with the Ramond-Kalb field in the context of string theory \cite{Kalb:1974yc}.

There are many natural and tantalising extensions of this paradigma. On a side, one can define a generalised geometry which puts together vector fields and ($n$--2)--differential forms in $n$ dimensions. In this way, one can show that, although $S^n$ is not parallelisable unless $n=1,3$, or $7$, $S^n$ is always parallelisable in a ``generalised sense" \cite{Lee:2014mla}. This means that it is possible to produce a set of $n$ generalised vectors $E_A$ in this geometry, which serves as absolute frame for $S^n$. On the other, since the field content of eleven-dimensional supergravity encompasses a three-form gauge field, one could consider generalised vectors with vector fields and two-forms as components, the latter being the gauge parameter of the three-form gauge transformations \cite{Hull:2007zu,PiresPacheco:2008qik}. Actually, the structure of generalised diffeomorphisms is more complicated when the internal dimension is $n\geqslant 7$ \cite{Coimbra:2011ky, Coimbra:2011nw, Coimbra:2012af}, but this is the starting point in defining exceptional geometry.

\textbf{\emph{Generalised torsion.}} The notion of \emph{generalied parallelisability} is the key ingredient in extending Scherk-Schwarz consistent reduction in order to cover all the known consistent reductions of higher-dimensional supergravities. 
One naturally defines the \emph{generalised intrinsic torsion} by replacing the frame and the Lie derivative with their generalised cousins
\begin{equation}
L_{E_A}\,E_B = -T_A{}^C{}_B\,E_C.
\end{equation}
Then, the Scherk-Schwarz reduction condition is extended by requiring that the internal manifold in the compatification is parallelisable in generalised sense with constant generalised intrinsic torsion. As in the usual case the constant intrinsic torsion is equal to minus the structure constants of the group describing the internal manifold, so in the generalised case the constant generalised intrinsic torsion has to be equal to the \emph{embedding tensor} $X_{AB}{}^C$ of the lower-dimensional gauged supergravity
\begin{equation}
L_{E_A}\,E_B = -X_{AB}{}^C\,E_C.
\end{equation}
The embedding tensor completely characterised the lower-dimensional supergravity theory, describing its couplings \cite{Samtleben:2008pe, Trigiante:2016mnt}.

\textbf{\emph{$E_{8(8)}$ exceptional geometry and uplifts.}} In this Thesis, we will focus in particular on the $n=8$ case, which is less known, compared to $n\leqslant 7$, because the corresponding exceptional geometry is not completely understood \cite{Strickland-Constable:2013xta}. Basically, the problem is that the smallest non-trivial representation of the algebra of $E_{8(8)}$, which is the adjoint representation, is larger than the number of eleven-dimensional supergravity degrees of freedom. The geometric structure of supergravity dualities seems to know something about the high-energy degrees of freedom of M-theory (of which eleven-dimensional supergravity is the low-energy limit), which remains elusive \cite{Berman:2020tqn}. Our main novel contribution is to prove that the method for the uplift problem in supergravity, introduced in \cite{Inverso:2017lrz} up to $n\leqslant 7$, works also in the $n=8$ case. Despite the complications of the not completely clear $n=8$ exceptional geometry, it is possible to extend all the procedure in a quite natural way \cite{Inverso:2024xok}. The result is completely general, but the steps towards the complete proof are many and laborious: the uplift procedure will be discussed in details in the Thesis, in a self-contained way, reviewing in summary all the necessary basic notions, analysing the $n=7$ case, and then comparing with similarities and differences in the $n=8$ case.

\textbf{\emph{Gravity as gauge theory.}} The discussion of the uplift problem led us to explore the geometric aspects of gauge theories, but the possibility of treating gravity as a gauge theory, along the lines of other fundamental interactions, has been extensively investigated too. It is an old idea to consider Einstein gravity as a sort of Yang-Mills theory: this analogy can be carried on in the Cartan formalism, where the metric and the Levi-Civita connection are replaced by the vielbein and by the spin connection \cite{Cartan1922, Cartan1923, Cartan1924, Cartan1925}. They are treated as independent variables. The torsion and the Riemann tensor are the curvatures of these ``gauge fields". The equations of motion of the spin connection set the torsion to zero, so that the spin connection can be solved algebraically in terms of the vielbein, in the same way as the Levi-Civita connection is written in terms of the metric tensor. This requires the vielbein to be invertible in each point. Otherwise, singularities appear. The equations of motion of the vielbein give the Einstein equations. 

Nevertheless, the complete analogy is spoiled both at the level of symmetries and at the level of the action. In Cartan formalism, Einstein gravity is locally Lorentz invariant and generally covariant, although in the ``gauge approach" the vielbein should be the gauge field of some internal translations. They should be replaced by general change of coordinates (\emph{diffeomorphisms}), but these ones cannot equivalent to abelian internal translations, since they are generated by the Lie derivative. At the level of the action, a Yang-Mills-type action should be quadratic in the curvature, whereas Hilbert-Einstein action is proportional to the Ricci scalar.\footnote{Remarkably, the analogy between gravity and gauge theories is sounder in three-dimension, because, as proved for the first time by Witten, three-dimensional gravity is equivalent to a Chern-Simons theory \cite{Witten:1988hc, Witten:1989sx}.}

\textbf{\emph{General relativity in torsion formulation.}} There is at least another possibility, which dates back to a hundred years ago, since it had already been considered by Einstein himself \cite{Einstein1928}, based on Weitzenb\"ock's studies \cite{Weitzenbock1928}. In the usual formulation of General Relativity the gravitational degrees of freedom are encoded in the Riemann tensor $R^\mu{}_{\nu\vrho\sigma}$, meaning that the gravitational effects are explained in terms of the spacetime curvature. The Riemann tensor is the curvature tensor of the Levi-Civita connection $\Gamma_\mu{}^\vrho{}_\nu$, the unique connection compatible with the metric and with vanishing torsion $T_\mu{}^\vrho{}_\nu$. In terms of the connection, the torsion is the antisymmetric part.\footnote{This is not longer true in generalised geometry.}  

Nevertheless, one can equivalently describe General Relativity by encoding the gravity degrees of freedom in the intrinsic torsion. This formulation is usually called \emph{teleparallel gravity} \cite{deAndrade:1997gka}. Since a parallelisable spacetime is flat, this formulation is a sort of complement of the usual one: in the latter the torsion is zero and gravitation is a curvature effect, whereas in the former the curvature is zero and gravitation is a torsion effect. Since the Riemann tensor is a combination of second derivatives of the metric, whereas the torsion depends on the first derivative of the frame, if Einstein equations can be derived also in the torsion formulation, then the action must be quadratic in the torsion, unlike Hilbert-Einstein action, which is linear in the Ricci scalar. Therefore, the torsion formulation of General Relativity is of Yang-Mills type. The torsion is a rank-three tensor, with two antisymmetrised indices. This implies that there are only three independent quadratic scalar contractions. They can be chosen to be $T^{\mu\nu\vrho}\,T_{\mu\nu\vrho}$, $T^{\mu\nu\vrho}\,T_{\mu\vrho\nu}$, and $T^{\mu\nu}{}_\nu\,T_{\mu\vrho}{}^\vrho$. Each of them is invariant with respect to diffeomorphisms, since the torsion is a well-defined covariant tensor. But General Relativity is also local Lorentz-invariant. Local Lorentz-invariance is not manifest in torsion formulation. Studying the transformation of the torsion under Lorentz transformations, one discovers that there is a unique invariant combination of the three scalar contractions. Up to a total derivative term, this combination is proportional to the Ricci scalar. The result is
\begin{equation}
\int\diff^d x\,e\,R = \int\diff^d x\,e\,\bigg{[}-\frac{1}{4}\,T^{\mu\nu\vrho}\,T_{\mu\nu\vrho} - \frac{1}{2}\,T^{\mu\nu\vrho}\,T_{\mu\vrho\nu} + T_{\mu\nu}{}^\nu\,T^{\mu\vrho}{}_\vrho\bigg{]}.
\end{equation}
This formula, which by the way is the best way to study the reduction of higher-dimensional gravity theory in the compactification procedure \cite{Scherk:1979zr, Julia:1980gr}, and it is the starting point for a dual theory of gravity \cite{West:2001as, Hohm:2018qhd}, lends itself to a natural generalisation: when spacetime is parallelisable, one may investigate the theory of gravity defined by the torsion action with arbitrary free parameters in place of the above Lorentz-invariant combination. This possibility was first considered by M{\o}ller \cite{Moller1961a, Moller1961b, Moller1978}, and independently by Hayashi and Shirafuji \cite{Hayashi:1967se, Hayashi:1977jd, Hayashi:1979qx}. The new theory, which was called \emph{New General Relativity} by Hayashi and Shirafuji, attracted interest since the request that the theory admits a Schwartzschild-like solution left one parameter free. But there is a completely different context in which M{\o}ller-Hayashi-Shirafuji theory makes its renewed appearance. As in an uplift problem, all the solution of the so-called \emph{covariant fracton gauge theory} are also solutions of the linearisation of M{\o}ller-Hayashi-Shirafuji theory \cite{Rovere:2025nfj}.

\textbf{\emph{Fractons.}} In recent years, exotic excitations with restricted mobility, called ``fractons", have been extensively studied, originally in the context of Condensed Matter Physics \cite{Pretko:2017xar, Burnell:2021reh}. Fracton excitations appear in lattice spin models \cite{Haah:2011drr, Vijay:2015mka}, whose low-energy continuous limit \cite{Affleck:1986} can be captured by effective field theories \cite{Seiberg:2020bhn}. These models are characterized by single charged particles and dipoles with not only conserved charge, but also conserved dipole moment. The dipole-moment conservation implies for a single charged particle to be fixed in space, whereas dipoles are free to move. 

These features can be described by a pair of gauge fields, with scalar gauge transformations. In particular, the second one involves a double derivative. The minimal coupling with some fractonic matter current leads to a higher-derivative continuity equation, which implies that both the charge and the dipole moment are conserved. These models break manifestly the Lorentz invariance. Nevertheless, they can be recovered in the covariant framework of a family of gauge theories of a rank-two symmetric tensor $h_{\mu\nu}$ with gauge symmetry given by the double derivative of a scalar parameter
\begin{equation}
\delta\,h_{\mu\nu} = \de_\mu\,\de_\nu\,\lambda,
\end{equation}
which we will refer to as \emph{covariant fracton gauge theory}, introduced for the first time in \cite{Blasi:2022mbl}.\footnote{Although the label ``fracton" for the covariant theory is a bit misleading, we maintain it, in accordance with existing literature on the topic (\cite{Blasi:2022mbl} and the subsequent works).}

Covariant fractons cover a wider scope than the reasons why they were originally introduced. An interesting perspective has been recently suggested in \cite{Hinterbichler:2025ost}. Exploiting the similarities of the fracton gauge transformation with that of a partially massless spin-two field in de Sitter spaces, relativistic fractonic matter is coupled to the gauge field, and the Higgs mechanism for the partially massless field is studied. As a result, a superconducting phase is induced, characterized by a condensation of fractonic matter.

On the one hand, covariant fracton gauge theory can be seen as a higher rank electrodynamics \cite{Bertolini:2022ijb}, because of the scalar gauge invariance. On the other hand, covariant fracton gauge theory can be thought of as a family of generalizations of linearised gravity \cite{Blasi:2022mbl, Bertolini:2023juh}, where the invariance under linearised diffeomorphisms in the latter theory is restricted to longitudinal diffeomorphisms in the former. Longitudinal diffeomorphism means that the vector parameter generating the symmetry transformation is taken to be a derivative of a scalar parameter, so that the gauge transformation is a double derivative of such a scalar parameter. While the gauge invariance of linearised gravity reduces the propagating degrees of freedom to spin-two polarisations only, in covariant fracton gauge theory propagating spin-one and spin-zero excitations remain \cite{Afxonidis:2023pdq}, which lead to instabilities \cite{Afxonidis:2024tph} to be possibly cured by adding suitable interactions -- a first attempt is presented in \cite{Bertolini:2024apg}. 

In \cite{Rovere:2025nfj} another possibility is suggested. It turns out that, apart from a particular case, the space of solutions of covariant fracton theory, modulo the gauge invariance, is isomorphic to a precise subsector of the space of solutions of linearised M{\o}ller-Hayashi-Shirafuji theory. In other words, a theory invariant under a small group of symmetry can be seen as part of a theory defined on a larger field space, but with a larger symmetry group too. Namely, the small group of linearised longitudinal diffeomorphisms is enlarged to the whole group of linearised diffeomorphisms, and the condition which identifies the subsector of the space of solutions of linearised M{\o}ller-Hayashi-Shirafuji theory breaks the large symmetry group to the smallest one, as in spontaneous symmetry breaking. Now, since the larger theory is well-established also at a non-linear level, this could  give an insight in the direction of how to add interactions in covariant fracton theory. 

In the Thesis we will discuss in details both M{\o}ller-Hayashi-Shirafuji theory and covariant fracton gauge theory, studying the embedding of the latter in the former. It will be also taken the opportunity to elaborate on little-known aspects of the theory. In particular, the structure of possible consistent \emph{anomalies} of fractonic matter theories covariantly coupled to fractonic gauge theory will be discussed \cite{Rovere:2024nwc}. 

\textbf{\emph{BRST cohomology and anomalies.}} An ``anomaly" is a breakdown of symmetry of the classical theory at quantum level. After many diagrammatic computations \cite{Bell:1969ts}, it was realised that anomalies depend neither on the regularisation schemes one may choose in performing diagrammatic computations, nor on the dynamics of the theory, but only on the field content of the background gauge theory to which the matter is coupled, and on symmetry transformations \cite{Bardeen:1984pm}. The best way to make this manifest is to exploit the \textsc{brst} approach \cite{Becchi:1974md, Becchi:1974xu, Becchi:1975nq, Tyutin:1975qk}. In the functional approach, one considers the classical fields without assuming them to obey the classical equations of motion, and extract information on the quantum theory by summing over all the possible configurations of the classical fields. \textsc{brst} formalism provides a framework which allows to extend the functional approach in field theory to gauge theories. The gauge parameters are promoted to fields, called \emph{ghosts}, with the opposite statistics as that dictates by the spin-statistics theorem. For example, the scalar gauge parameter in Maxwell gauge theory is promoted to an \emph{anticommuting scalar field}, whereas the spinor generating a supersymmetry transformation is promoted to a \emph{commuting spinor field}. The differential \textsc{brst} operator $s$ is defined on the enlarged space of fields and ghosts, generating all the gauge transformations of the theory. Its basic property, which motivates the introduction of ghosts, is that $s$ is nilpotent: $s^2 = 0$ on the space of fields and ghosts. This allows to define \textsc{brst} cohomology $\text{ker}\,s/\text{im}\,s$, which plays a fundamental r\^ole: the physical observables of the gauge theory, as well as anomalies, must belong to it. So, the problem of computing possible anomalies in a gauge theories is reduced to the problem of computing the \textsc{brst} cohomology of the gauge theory.
In the Thesis we will study the \textsc{brst} cohomology of covariant fracton gauge theory, and we will employ the so-called Stora-Zumino method \cite{Stora:1976kd, Stora:1976LM, Stora:1984, Zumino:1983ew, Zumino:1983rz, Manes:1985df} for studying the possible consistent anomalies, comparing similarities and differences with the other known cases.

\bigskip

The structure of the Thesis is the following. In Section \ref{0} the basic tools used in the subsequent part -- namely the properties of the Lie derivative, the definition of the Weitzenb\"ock connection and torsion, the main aspects of generalised geometry and the \textsc{brst} formalism -- are described in a self-contained way. This material is preparatory for both supergravity uplifts and fracton topics. Sections \ref{1}--\ref{6} discuss the uplift problem in supergravity. Sections \ref{5}--\ref{6bis}, which are independent of the previous ones, are devoted to the investigation on covariant fractons. 

Precisely, Section \ref{1} is devoted to introduce the topic of supergravity compactifications, from Kaluza-Klein reduction to Scherk-Scharz reduction and its extension due to exceptional field theory and exceptional geometry: attention will be paid to the case of eleven-dimensional supergravity and its compactification down to four dimensions. In Section \ref{2} the uplift procedure in supergravity using exceptional geometry is discussed in details. Section \ref{3} is a detour devoting to explore some peculiarities of gravity in three dimensions. Section \ref{4} explores $E_{8(8)}$ generalised geometry and the uplift problem with eight internal dimensions. In Section \ref{6} there are some comments on the extension to non-maximal uplifts.

In Section \ref{5} M{\o}ller-Hayashi-Shirafuji theory and covariant fracton gauge theory are described, and the embedding of the latter in the former, and the \textsc{brst} cohomology and anomalies of the latter are investigated. In the concluding Section \ref{6bis} there are comments on possible future directions. 

Finally, in Appendices \ref{Conventions}--\ref{A6} conventions, details, and marginal aspects are collected.

\newpage

\section{Preliminaries}\label{0}

\subsection{Equivalent definitions of the Lie derivative}

Introducing the \emph{push-forward} and the \emph{pull-back}, one can extend in differential geometry the usual notion of derivative as incremental ratio, defining the so-called \emph{Lie derivative}. Consider two manifolds $\mathscr{M}$ and $\mathscr{N}$ and a $C^\infty$ map $\vphi$ between them: $\vphi : \mathscr{M} \rightarrow \mathscr{N}$. 
The \emph{push-forward} $\vphi_*$ in a point $x\in\mathscr{M}$ is a map between the tangent space of $\mathscr{M}$ and $\mathscr{N}$ in $x$ and $\vphi(x)$ respectively: $v_x \mapsto {\vphi_*\,v}_{\vphi(x)}: T_x\mathscr{M} \rightarrow T_{\vphi(x)}\mathscr{N}$, defined by
\begin{equation}
{\vphi_*\,v}_{\vphi(x)}[f] = v_x[f\circ \vphi].
\end{equation}
As a result, if $v$ is evaluated in a point $x$, then $\vphi_*\,v$ is evaluated in the point $\vphi(x)$. In this sense, $\vphi_*$ pushes $v$ in $x$ forward $\vphi(x)$. Similarly, the \emph{pull-back} $\vphi^*$ is a map between the cotangent spaces: $\omega_{\vphi(x)} \mapsto {\vphi^*\,\omega}_x : T^*_{\vphi(x)}\,\mathscr{N} \rightarrow T^*_x\,\mathscr{M}$, defined by the dual action of $\vphi_*$
\begin{equation}
\langle{\vphi^*\,\omega}_x,v_x \rangle =
\langle\omega_{\vphi(x)},{\vphi_*\,v}_{\vphi(x)}\rangle.
\end{equation}
So, if $\omega$ is evaluated in $\vphi(x)$, then $\vphi^*\,\omega$ is evaluated in $x$, and in this sense $\vphi^*$ pulls $\omega$ from $\vphi(x)$ back to $x$. One could ask why we did not define directly $\vphi^*\,\omega$ starting from $\omega$ evaluated in $x$. The reason is that the pairing $\langle\cdot,\cdot\rangle$ requires the two entries to be evaluated in the same point.

Consider now a path $\gamma$ in the manifold $\mathscr{M}$, $\gamma : I \subseteq \R \rightarrow \mathscr{M}$, parametrised by the a “time” coordinate $t$. $\gamma$ is the \emph{integral curve} of a vector field $\xi$, if it is the path whose tangent vector in each point is $\xi$ itself. Because of the unicity of solutions of a Cauchy initial problem, the integral curve passing through a point $x \in \mathscr{M}$ of $\xi$ is unique. We will denote it by $\gamma_x$. The evaluation of $\gamma_x$ at a time $t$ is $\gamma_x(t)$. We assume that the curve passes through $x$ at time $t=0$, that is $\gamma_x(0)=x$. If we evaluate $\gamma_x(t)$ at a fixed time $t$ as $x$ varies, we get the \emph{flow} of $\xi$. We will denote the flow at $t$ by $\sigma_t$, and the evaluation in a point $x$ by $\sigma_t(x) = \gamma_x(t)$. Notice that $\sigma_0$ is the identity map, because $\sigma_0(x)=\gamma_x(0)=x$, for each point $x$. Furthermore, $\sigma_t \circ \sigma_s = \sigma_{t+s}$. Therefore the set of $\lbrace\sigma_t\rbrace_{t \in I}$ forms a one-parameter group with the composition $\circ$ between maps. In particular, the inverse flow is obtained by changing the sign of the time: $\sigma_t^{-1} = \sigma_{-t}$. If $\sigma_t$ is a $C^\infty$ map, then the group is a group of diffeomorphisms. By definition, $\sigma_t$ takes $x$ to another point in the manifold $\sigma_t(x)$, viewed as the point at time $t$ on the integral curve of the vector field $\xi$. So, the push-forward and the pull-back induced by $\sigma_t$ can be used to move vector fields and one-forms from a point to another in the manifold. ($\sigma_t$ plays the r\^ole of $\vphi$ in previous definitions, and we are in the case in which $\mathscr{M}$ and $\mathscr{N}$ are the same manifold.)

If $\omega_{\sigma_t(x)}$ is a one-form $\omega$ evaluated in $\sigma_t(x)$, then $\sigma_t^*\,\omega_{\sigma_t(x)}$ is evaluated in $x$. So, the difference between the latter and $\omega_x$ makes sense and we can define an incremental ratio:
\begin{equation}
{\mathcal{L}_\xi\,\omega}_x = \lim_{t\rightarrow 0} \frac{\sigma_t^*\,\omega_{\sigma_t(x)}-\omega_x}{t} = 
{\frac{\diff}{\diff t}}_{|_{t=0}}\sigma^*_t\,\omega_{\sigma_t(x)},
\end{equation}
known as the Lie derivative of the one-form $\omega$ generated by the vector field $\xi$ evaluated in $x$. Notice that the pull-back of the identity map is the identity map too. Therefore, we can write $\omega_x = \sigma_0^*\,\omega_{\sigma_0(x)}$. Therefore the previous definition can be seen as an incremental ratio. We can similarly define the Lie derivative on a vector field. If $v_{\sigma_t(x)}$ is a vector field $v$ evaluated in $\sigma_t(x)$, then $(\sigma_{-t})_*\,v_{\sigma_t(x)}$ is evaluated in $x$. (Here we choose $\vphi = \sigma_t^{-1} = \sigma_{-t}$.) Then, we define the Lie derivative of $v$ along $\xi$ in $x$ by
\begin{equation}
{\mathcal{L}_\xi\,v}_x = \lim_{t\rightarrow 0} \frac{(\sigma_{-t})_*\,v_{\sigma_t(x)}-v_x}{t} = 
{\frac{\diff}{\diff t}}_{|_{t=0}}(\sigma_{-t})_*\,v_{\sigma_t(x)}.
\end{equation}
The Lie derivative on a scalar $\vphi$ is simply given by the action of the generating vector field on the scalar itself:
\begin{equation}
\mathcal{L}_\xi\,\vphi = \xi^\mu\,\de_\mu\,\vphi.
\end{equation}
Using the Leibniz identity, one extend the previous definitions to arbitrary rank tensors.\footnote{The Lie derivative is also known as “fisherman's derivative”, because “The flow carries all possible differential geometric objects past the fisherman and the fisherman sits there and differentiate them” (p. 198 of \cite{Arnold:1989who}).}

Equivalently, one can introduce the Lie derivative in a coordinate-dependent way, as the variation under  infinitesimal change of coordinates
\begin{equation}\label{TransX}
x'^\mu = x^\mu - \xi^\mu,
\end{equation}
where $\xi^\mu$ are the coordinates of a vector field $\xi = \xi^\mu\,\de_\mu$. The finite transformation of an arbitrary tensor field $T(x)$ under a change of coordinates $x^\mu \mapsto x'^\alpha(x^\mu)$ is
\begin{equation}\label{Tprime}
T^{\mu_1\dots\mu_p}_{\nu_1\dots\nu_q}(x) = \tfrac{\de\,x^{\mu_1}}{\de\,x'^{\alpha_1}}\dots\tfrac{\de\,x^{\mu_p}}{\de\,x'^{\alpha_p}}\,T'^{\alpha_1\dots\alpha_p}_{\beta_1\dots\beta_q}(x')\,\tfrac{\de\,x'^{\beta_1}}{\de\,x^{\nu_1}}\dots\tfrac{\de\,x'^{\beta_q}}{\de\,x^{\nu_p}}.
\end{equation}
In order to get the first order variation, differentiate and invert \eqref{TransX}, getting
\begin{equation}\label{derTransX}
\tfrac{\de\,x'^\mu}{\de\,x^\alpha} = \delta_\alpha{}^\mu - \de_\alpha\,\xi^\mu + \mathcal{O}(\xi^2), \quad \tfrac{\de\,x^\beta}{\de\,x'^\nu} = \delta_\nu{}^\beta + \de_\nu\,\xi^\beta + \mathcal{O}(\xi^2).
\end{equation}
Therefore, using the Taylor expansion at first order, and \eqref{derTransX}, \eqref{Tprime} becomes,\footnote{The Taylor expansion is $T'(x') = T'(x-\xi)=T'(x) - \xi^\mu\,\de_\mu\,T'(x) + \mathcal{O}(\xi^2)$, taking into account contribution in the variation due to the spacetime dependence in $T$.}
\begin{align}
T^{\mu_1\dots\mu_p}_{\nu_1\dots\nu_q}(x) &= (\delta_{\alpha_1}{}^{\mu_1}+\de_{\alpha_1}\,\xi^{\mu_1})\dots(\delta_{\alpha_1}{}^{\mu_p}+\de_{\alpha_p}\,\xi^{\mu_p})\,(T'^{\alpha_1\dots\alpha_p}_{\beta_1\dots\beta_q}(x)-\xi^\lambda\,\de_\lambda\,T'^{\alpha_p\dots\alpha_p}_{\beta_1\dots\beta_q}(x))\,\times\nn\\
&\times (\delta_{\nu_1}{}^{\beta_1}-\de_{\nu_1}\,\xi^{\beta_1})\dots(\delta_{\nu_1}{}^{\beta_p}+\de_{\nu_p}\,\xi^{\beta_p}) + \mathcal{O}(\xi^2) = \nn\\
&= T'^{\mu_1\dots\alpha_p}_{\beta_1\dots\beta_q}(x) - 
\xi^\lambda\,\de_\lambda\,T'^{\mu_1\dots\mu_p}_{\nu_1\dots\nu_q}(x) - \sum_{j=1}^q T'^{\mu_1\dots\mu_p}_{\nu_1\dots\nu_{j-1}\lambda\nu_{j+1}\nu_q}(x)\,\de_{\nu_j}\,\xi^\lambda \,+\nn\\
& + \sum_{i=1}^p \de_\lambda\,\xi^{\mu_i}\,T'^{\mu_1\dots\mu_{i-1}\lambda\mu_{i+1}\dots\mu_p}_{\nu_1\dots\nu_q}(x) + \mathcal{O}(\xi^2) = \nn\\
& = T'^{\mu_1\dots\alpha_p}_{\beta_1\dots\beta_q}(x) - 
\xi^\lambda\,\de_\lambda\,T^{\mu_1\dots\mu_p}_{\nu_1\dots\nu_q}(x) - \sum_{j=1}^q T^{\mu_1\dots\mu_p}_{\nu_1\dots\nu_{j-1}\lambda\nu_{j+1}\nu_q}(x)\,\de_{\nu_j}\,\xi^\lambda \,+\nn\\
& + \sum_{i=1}^p \de_\lambda\,\xi^{\mu_i}\,T^{\mu_1\dots\mu_{i-1}\lambda\mu_{i+1}\dots\mu_p}_{\nu_1\dots\nu_q}(x) + \mathcal{O}(\xi^2).
\end{align}
Thus, defining the Lie derivative $\mathcal{L}_\xi\,T^{\mu_1\dots\mu_p}_{\nu_1\dots\nu_q}(x)$ as the infinitesimal variation in the same spacetime point, one obtains
\begin{align}
\mathcal{L}_\xi\,T^{\mu_1\dots\mu_p}_{\nu_1\dots\nu_q}(x)  &=
T'^{\mu_1\dots\alpha_p}_{\beta_1\dots\beta_q}(x) 
-T^{\mu_1\dots\alpha_p}_{\beta_1\dots\beta_q}(x) = \nn\\
&= \xi^\lambda\,\de_\lambda\,T^{\mu_1\dots\mu_p}_{\nu_1\dots\nu_q}(x) + \sum_{j=1}^q T^{\mu_1\dots\mu_p}_{\nu_1\dots\nu_{j-1}\lambda\nu_{j+1}\nu_q}(x)\,\de_{\nu_j}\,\xi^\lambda \,+\nn\\
& - \sum_{i=1}^p \de_\lambda\,\xi^{\mu_i}\,T^{\mu_1\dots\mu_{i-1}\lambda\mu_{i+1}\dots\mu_p}_{\nu_1\dots\nu_q}(x).\label{LieComponents}
\end{align}
$\mathcal{L}_\xi\,T^{\mu_1\dots\mu_p}_{\nu_1\dots\nu_q}(x)$ encodes a \emph{translational} part, due to the spacetime point dependence of $T$, which is the action of $\xi = \xi^\mu\,\de_\mu$, and a \emph{rotational} part, due to the action of the general linear group $\text{GL}(d,\mathbb{R})$, in $d$ dimensions, on the indices of $T$. 

Defining the following operator:
\begin{equations}
&\Delta_\alpha{}^\beta\,T^\mu = - \delta_\alpha^\mu\,T^\beta,\quad
\Delta_\alpha{}^\beta\,T_\mu = \delta_\mu^\beta\,T_\alpha,\\
&\Delta_\alpha{}^\beta\,T_{\mu\nu} = \delta_\mu^\beta\,T_{\alpha\nu} + \delta_\nu^\beta\,T_{\mu\alpha},\\
& \Delta_\alpha{}^\beta\,T^\mu{}_\nu = -\delta_\alpha^\mu\,T^\beta{}_\nu + \delta_\nu{}^\beta\,T^\mu{}_\alpha,\\
& \Delta_\alpha{}^\beta\,T^{\mu\nu} = -\delta^\mu_\alpha\,T^{\beta\nu} - \delta_\alpha^\nu\,T^{\mu\beta},
\end{equations}
and so on, then the Lie derivative is
\begin{equation}\label{CompactFormulaLieDer}
\mathcal{L}_\xi = \xi^\alpha\,\de_\alpha + M^\alpha{}_\beta\,\Delta_\alpha{}^\beta, \;\;\text{where}\;\; M^\alpha{}_\beta = \de_\beta\,\xi^\alpha.
\end{equation}
The first term is the translational part, and the second one is the $\text{GL}(d,\mathbb{R})$ rotational part, with rotation matrix $M^\alpha{}_\beta = \de_\beta\,\xi^\alpha$. If the lowered indices are totally antisymmetrised, so that $T^{\mu_1\dots\mu_p}_{\nu_1\dots\nu_q}$ are the components of a rank--$p$ tensor-valued $q$--form 
\begin{equation}
T^{(q)\,\mu_1\dots\mu_p}=\tfrac{1}{q!}\,T_{\nu_1\dots\nu_q}^{\mu_1\dots\mu_p}\,\diff x^{\nu_1}\dots\diff x^{\nu_q},
\end{equation}
then one can see by explicit evaluation that the first and the second piece in \eqref{LieComponents} can be packed together in this way
\begin{equation}
\mathcal{L}_\xi\,T^{(q)\,\mu_1\dots\mu_p} = [\iota_\xi,\diff]\,T^{(q)\,\mu_1\dots\mu_p} - \sum_{i=1}^p \de_\lambda\,\xi^{\mu_i}\,T^{(q)\,\mu_1\dots\mu_{i-1}\lambda\mu_{i+1}\dots\mu_p},
\end{equation}
where $\iota_\xi$ is the contraction operator or inner derivative, defined by
\begin{equation}
\iota_\xi\,\omega^{(0)} = 0, \quad 
\iota_\xi\,\diff x^\mu = \xi^\mu,
\end{equation}
satisfying the following graded Leibniz rule
\begin{equation}
\iota_\xi\,(\omega^{(p)}\,\chi^{(q)}) = \iota_\xi\,\omega^{(p)}\,\chi^{(q)} + (-)^{p(|\xi|+1)}\,\omega^{(p)}\,\iota_\xi\,\chi^{(q)}.
\end{equation}
In other words, $\iota_\xi$ is a derivative whose degree is the opposite of the degree of $\xi$: if $\xi$ is even/odd, $\iota_\xi$ is odd/even. The commutator with the de Rham differential is the graded commutator:\footnote{In general, the graded commutator is $[A,B] = -(-)^{|A||B|}\,[B,A]$.}
\begin{equation}
[\iota_\xi,\diff] = \iota_\xi\,\diff + (-)^{|\xi|} \diff\,\iota_\xi.
\end{equation}
So, the Lie derivative on a $p$--form $\omega^{(p)}$ is 
\begin{equation}
\mathcal{L}_\xi\,\omega^{(p)} = [\iota_\xi,\diff]\,\omega^{(p)}.
\end{equation}
This is the so-called \emph{Cartan formula}.

The operator $\Delta$ satisfies the following properties
\begin{equation}
\Delta_\alpha{}^\beta\,\delta_\gamma{}^\delta = 0,\;\;
[\Delta_\alpha{}^\beta,\xi] = 0,\;\;
[\Delta_\alpha{}^\beta,\Delta_\gamma{}^\delta] = (-\delta_\alpha{}^\delta\,\delta_\gamma{}^\vepsilon\,\delta_\eta{}^\beta + \delta_\gamma{}^\beta\,\delta_\alpha{}^\vepsilon\,\delta_\eta{}^\delta)\,\Delta_\vepsilon{}^\eta,
\end{equation}
which imply
\begin{align}
[\xi,\de_\beta\,\eta^\alpha\,\Delta_\alpha{}^\beta] &= \xi^\alpha\,\de_\alpha\,\de_\delta\,\xi^\gamma\,\Delta_\gamma{}^\delta,\\
[\de_\beta\,\xi^\alpha\,\Delta_\alpha{}^\beta,\de_\delta\,\xi^\gamma\,\Delta_\gamma{}^\delta] &= (\de_\delta\,\xi^\alpha\,\de_\alpha\,\eta^\gamma - (-)^{|\xi||\eta|}\,\de_\delta\,\xi^\alpha\,\de_\alpha\,\xi^\gamma)\,\Delta_\gamma{}^\delta.
\end{align}
Moreover, the Lie derivative on vector fields is the commutator between them\footnote{\emph{Proof}. $\mathcal{L}_\xi\,\eta^\mu = \xi^\alpha\,\de_\alpha\,\xi^\mu + \de_\alpha\,\xi^\mu\,\eta^\alpha = \xi^\alpha\,\de_\alpha\,\eta^\mu - (-)^{|\xi||\eta|}\,\eta^\alpha\,\de_\alpha\,\xi^\mu = \xi\,\eta^\mu - (-)^{|\xi||\eta|}\,\eta\,\xi^\mu = [\xi,\eta]^\mu$.}
\begin{equation}
\mathcal{L}_\xi\,\eta^\mu = [\xi,\eta]^\mu,
\end{equation}
and it satisfies a graded symmetry property
\begin{equation}\label{GradedSymLieDer}
\mathcal{L}_\xi\,\eta = -(-)^{|\xi||\eta|}\,\mathcal{L}_\eta\,\xi.
\end{equation}
Using these properties, one can show that\footnote{\emph{Proof}. 
\begin{align}
[\mathcal{L}_\xi,\mathcal{L}_\eta] &= [\xi + \de_\beta\,\xi^\beta\,\Delta_\alpha{}^\beta, \eta + \de_\delta\,\eta^\gamma\,\Delta_\gamma{}^\delta] = \nn\\
&= [\xi,\eta] + [\de_\beta\,\xi^\beta\,\Delta_\alpha{}^\beta,\de_\delta\,\eta^\gamma\,\Delta_\gamma{}^\delta] + [\xi,\de_\delta\,\eta^\gamma\,\Delta_\gamma{}^\delta] + [\de_\beta\,\xi^\beta\,\Delta_\alpha{}^\beta,\eta] = \nn\\
&= (\mathcal{L}_\xi\,\eta)^\mu\,\de_\mu + (\de_\delta\,\xi^\alpha\,\de_\alpha\,\eta^\gamma - (-)^{|\xi||\eta|}\,\de_\delta\,\xi^\alpha\,\de_\alpha\,\xi^\gamma)\,\Delta_\gamma{}^\delta \,+\nn\\
& + (\xi^\alpha\,\de_\alpha\,\delta_\delta\,\xi^\gamma - (-)^{|\xi|\eta|}\,\eta^\alpha\,\de_\alpha\,\de_\delta\,\xi^\gamma)\,\Delta_\gamma{}^\delta = \nn\\
&= (\mathcal{L}_\xi\,\eta)^\mu\,\de_\mu + \de_\beta\,(\mathcal{L}_\xi\,\eta)^\alpha\,\Delta_\alpha{}^\beta = \mathcal{L}_{\mathcal{L}_\xi\,\eta}.
\end{align}}
\begin{equation}
[\mathcal{L}_\xi,\mathcal{L}_\eta] = \mathcal{L}_{\mathcal{L}_\xi\,\eta}.
\end{equation}
This property is nothing but the Leibniz identity:
\begin{equation}
\mathcal{L}_\xi\,(\mathcal{L}_\eta\,\cdot) = 
\mathcal{L}_{\mathcal{L}_\xi\,\eta}\,\cdot + (-)^{|\xi||\eta|}\,\mathcal{L}_\eta\,(\mathcal{L}_\xi\,\cdot).
\end{equation}
It is equivalent to the Jacobi identity on vectors, thanks to \eqref{GradedSymLieDer}:\footnote{\emph{Proof}. 
\begin{align}
& \mathcal{L}_\xi\,(\mathcal{L}_\eta\,\zeta) = 
\mathcal{L}_{\mathcal{L}_\xi\,\eta}\,\zeta + (-)^{|\xi||\eta|}\,\mathcal{L}_\eta\,(\mathcal{L}_\xi\,\zeta),\nn\\
\Leftrightarrow \,& [\xi,[\eta,\zeta]] = [[\xi,\eta],\zeta] + (-)^{|\xi||\eta|}\,[\eta,[\xi,\zeta]],\nn\\
\Leftrightarrow \,& [\xi,[\eta,\zeta]] = -(-)^{|\zeta|(|\xi|+|\eta|)}\,[\zeta,[\xi,\eta]] - (-)^{|\xi||\eta|}\,(-)^{|\xi||\zeta|}\,[\eta,[\zeta,\xi]],\nn\\
\Leftrightarrow \,& (-)^{|\xi||\zeta|}\,[\xi,[\eta,\zeta]] + (-)^{|\xi||\zeta|}\,\,(-)^{|\zeta|(|\xi|+|\eta|)}\,[\zeta,[\xi,\eta]] + (-)^{|\xi||\zeta|}\,(-)^{|\xi||\eta|}\,(-)^{|\xi||\zeta|}\,[\eta,[\zeta,\xi]] = 0,\nn\\
\Leftrightarrow \,& (-)^{|\xi||\zeta|}\,[\xi,[\eta,\zeta]] + (-)^{|\zeta||\eta|)}\,[\zeta,[\xi,\eta]] + (-)^{|\xi||\eta|}\,[\eta,[\zeta,\xi]] = 0.
\end{align}}
\begin{equation}
(-)^{|\xi||\zeta|}\,[\xi,[\eta,\zeta]] + (-)^{|\zeta||\eta|}\,[\zeta,[\xi,\eta]] + (-)^{|\xi||\eta|}\,[\eta,[\zeta,\xi]] = 0.
\end{equation}
This means that, on the space of vector fields,
\begin{align}
\text{graded symmetry} \,\&\,
\text{Jacobi} \Leftrightarrow 
\text{graded symmetry} \,\&\,
\text{Leibniz}.
\end{align}

Upside down, one could also assume both the Leibniz rule and the graded symmetry property on vectors as axioms, 
\begin{equations}
\mathcal{L}_\xi\,\eta &= [\xi,\eta],\\
\mathcal{L}_\xi\,(A\,B) &=\mathcal{L}_\xi\,A\,B + (-)^{|A||B|}\,A\,\mathcal{L}_\xi\,B, \\
[\mathcal{L}_\xi,\mathcal{L}_\eta] &= \mathcal{L}_{\mathcal{L}_\xi\,\eta},
\end{equations}
and derive all the results in an algebraic manner without an explicit coordinate-dependent expression of the Lie derivative. 

\subsection{Lie derivative of Levi-Civita connection}

The Levi-Civita connection
\begin{equation}
\Gamma_\mu{}^\lambda{}_\nu = \tfrac{1}{2}\,g^{\lambda\alpha}\,(\de_\mu\,g_{\alpha\nu}+\de_\nu\,g_{\alpha\mu}-\de_\alpha\,g_{\mu\nu})
\end{equation}
is not a covariant tensor. Nevertheless, we can define its Lie derivative by using the coordinate approach. Using the Taylor expansion,
\begin{equation}
\de'_\vrho\,g_{\mu\nu}(x') = \de'_\vrho\,g_{\mu\nu}(x-\xi) = \de'_\vrho\,g'_{\mu\nu}(x) - (\xi\cdot\de)\,\de'_{\vrho}\,g'_{\mu\nu}(x) + \mathcal{O}(\xi^2).
\end{equation}
On the other hand, the derivative of the metric transforms under a finite coordinate change as
\begin{equation}
\de_\gamma\,g_{\alpha\beta}(x) = \frac{\de x'^\vrho}{\de x^\gamma}\,\de'_\vrho\,\left[\frac{\de x'^\mu}{\de x^\alpha}\,\frac{\de x'^\nu}{\de x^\beta}\,g'_{\mu\nu}(x')\right],
\end{equation}
showing that $\de_\gamma\,g_{\alpha\beta}$ is not a tensor, since the derivative $\de'_\vrho$ acts also on the Jacobian matrices in the brackets. Replacing all up to first order in $\xi$, we can compute
\begin{align}
\de_\gamma\,g_{\alpha\beta}(x) &= -\delta^\vrho{}_{\gamma}\,\de'_\vrho\,\de_\alpha\,\xi^\mu\,\delta^\nu{}_\beta\,g'_{\mu\nu}(x) - 
\delta^\vrho{}_\gamma\,\delta^\mu{}_\alpha\,\de'_\vrho\,\de_\beta\,\xi^\nu\,g_{\mu\nu}(x) \,+\nonumber \\
& + (\delta^\vrho{}_\gamma - \de_\gamma\,\xi^\vrho)(\delta^\mu{}_\alpha - \de_\alpha\,\xi^\mu)(\delta^\nu{}_\beta - \de_\beta\,\xi^\nu)(\de'_\vrho\,g'_{\mu\nu}(x) - (\xi\cdot\de)\,\de'_{\vrho}\,g'_{\mu\nu}(x)) + \mathcal{O}(\xi^2) = \nonumber \\
& = \de'_\gamma\,g'_{\alpha\beta}(x) - (\xi\cdot\de)\,\de'_\gamma\,g'_{\alpha\beta}(x) - \de_\beta\,\xi^\nu\,\de'_\gamma\,g'_{\alpha\nu}(x) - \de_\alpha\,\xi^\mu\,\de'_\gamma\,\gamma'_{\mu\beta}(x) \,+\nonumber \\
& - \de_\gamma\,\xi^\vrho\,\de'_\vrho\,g'_{\alpha\beta}(x) - \de_\gamma\,\de_\alpha\,\xi^\mu\,g_{\mu\beta}(x) - \de_\gamma\,\de_\beta\,g_{\alpha\nu}(x) + \mathcal{O}(\xi^2),
\end{align}
so that
\begin{align}
\de'_\gamma\,g'_{\alpha\beta}(x) - \de_\gamma\,g_{\alpha\beta}(x) &= \xi^\mu\,\de_\mu\,\de'_\gamma\,g'_{\alpha\beta}(x) + \de_\beta\,\xi^\mu\,\de'_\gamma\,g'_{\alpha\mu}(x) + \de_\alpha\,\xi^\mu\,\de'_\gamma\,\gamma'_{\mu\beta}(x) \,+\nonumber \\
& + \de_\gamma\,\xi^\vrho\,\de'_\vrho\,g'_{\alpha\beta}(x) + \de_\gamma\,\de_\alpha\,\xi^\mu\,g_{\mu\beta}(x) + \de_\gamma\,\de_\beta\,\xi^\nu\,g_{\alpha\nu}(x) + \mathcal{O}(\xi^2).
\end{align}
We can treat the left-hand side as the definition of $\mathcal{L}_\xi\,\de_\gamma\,g_{\alpha\beta}$. The first four terms in right-hand side form what we would call the Lie derivative of $\de_\gamma\,g_{\alpha\beta}$ if it were a tensor. As a shorthand, denote them by $\pounds_\xi\,\de_\gamma\,g_{\alpha\beta}$. Of course, $\pounds_\xi\,T = \mathcal{L}_\xi\,T$, if $T$ is a covariant tensor. Therefore, we can write
\begin{equation}
\mathcal{L}_\xi\,\de_\gamma\,g_{\alpha\beta} = \pounds_\xi\,\de_\gamma\,g_{\alpha\beta} + \de_\gamma\,\de_\alpha\,\xi^\vrho\,g_{\vrho\beta} + \de_\gamma\,\de_\beta\,\xi^\vrho\,g_{\vrho\alpha}.
\end{equation}
Now, using this result, we can compute
\begin{align}
\mathcal{L}_\xi\,\Gamma_\mu{}^\lambda{}_\nu &= \tfrac{1}{2}\,\pounds_\xi\,g^{\lambda\alpha}\,(\de_\mu\,g_{\alpha\nu}+\de_\nu\,g_{\alpha\mu}-\de_\alpha\,g_{\mu\nu}) \,+\nonumber \\
& + \tfrac{1}{2}\,g^{\lambda\alpha}\,(
\pounds_\xi\,\de_\mu\,g_{\alpha\nu} + \de_\mu\,\de_\alpha\,\xi^\vrho\,g_{\vrho\nu} + \de_\mu\,\de_\nu\,\xi^\vrho\,g_{\vrho\alpha} \,+\nonumber \\
& + \pounds_\xi\,\de_\nu\,g_{\alpha\mu} + \de_\nu\,\de_\alpha\,\xi^\vrho\,g_{\vrho\mu} + \de_\nu\,\de_\mu\,\xi^\vrho\,g_{\vrho\alpha} \,+\nonumber \\
& - \pounds_\xi\,\de_\alpha\,g_{\mu\nu} - \de_\alpha\,\de_\mu\,\xi^\vrho\,g_{\vrho\nu} - \de_\alpha\,\de_\nu\,\xi^\vrho\,g_{\vrho\mu}) = \nonumber \\
& = \pounds_\xi\,\Gamma_\mu{}^\lambda{}_\nu + 
\tfrac{1}{2}\,g^{\lambda\alpha}\,2\,\de_\mu\,\de_\nu\,\xi^\vrho\,g_{\vrho\alpha} = \nonumber \\
& = \pounds_\xi\,\Gamma_\mu{}^\lambda{}_\nu + 
\de_\mu\,\de_\nu\,\xi^\lambda.
\end{align}
$\Gamma_\mu{}^\lambda{}_\nu$ can be conveniently treat as a matrix-valued one-form:
\begin{equation}
\Gamma^\lambda{}_\nu = \Gamma_\mu{}^\lambda{}_\nu\,\diff x^\mu.
\end{equation}
Finally, unpack $\pounds_\xi$, recognising the action of $[\iota_\xi,\diff]$ on the form index, introduce the covariant derivative with respect to $\Gamma$, given by $\text{D} = \diff + [\Gamma,\cdot]$, and define $M^\alpha{}_\beta = \de_\beta\,\xi^\alpha$: 
\begin{align}
\mathcal{L}_\xi\,\Gamma_\mu{}^\lambda{}_\nu &= [\iota_\xi,\diff]\,\Gamma_\mu{}^\lambda{}_\nu + \Gamma_\mu{}^\lambda{}_\vrho\,\de_\nu\,\xi^\vrho - \Gamma_\mu{}^\lambda{}_\nu\,\de_\vrho\,\xi^\lambda + \de_\mu\,\de_\nu\,\xi^\lambda = \nn\\
& = [\iota_\xi,\diff]\,\Gamma_\mu{}^\lambda{}_\nu + \de_\mu\,M^\lambda{}_\nu + \Gamma_\mu{}^\lambda{}_\vrho\,M^\vrho{}_\nu - M^\lambda{}_\vrho\,\Gamma_\mu{}^\vrho{}_\nu = \nonumber \\
& = ([\iota_\xi,\diff]\,\Gamma_\mu + \de_\mu\,M + [\Gamma_\mu,M])^\lambda{}_\nu = \nonumber \\
& = ([\iota_\xi,\diff]\,\Gamma_\mu + D_\mu\,M)^\lambda{}_\nu.
\end{align}
The last term is a non-abelian gauge transformation with gauge group $\text{GL}(d,\mathbb{R})$.

\subsection{Lie derivative of a density}

If $\nabla_\mu$ is a covariant derivative compatible with the metric $g_{\mu\nu}$, that is, $\nabla_\vrho\,g_{\mu\nu} = 0$, then the Lie derivative of the metric tensor is
\begin{equation}\label{LieDerMetric}
\mathcal{L}_\xi\,g_{\mu\nu} = \nabla_\mu\,\xi_\nu + \nabla_\nu\,\xi_\mu,
\end{equation}
where $\xi_\mu = g_{\mu\nu}\,\xi^\nu$.

In order to compute the Lie derivative of a tensor density it is sufficient to compute the Lie derivative of the determinant of the metric $|g|^{1/2} = e$, where $e$ is the determinant of the vielbein $e_\mu{}^a$, such that
\begin{equation}
g_{\mu\nu} = e_\mu{}^a\,e_\nu{}^b\,\eta_{ab}. 
\end{equation} 
Using the Lie derivative of the metric \eqref{LieDerMetric} and the variation $\delta\,e = \tfrac{1}{2}\,e\,g^{\mu\nu}\,\delta\,g_{\mu\nu}$,  
\begin{align}
\mathcal{L}_\xi\,e &= \tfrac{\delta\,e}{\delta\,g_{\mu\nu}}\,\mathcal{L}_\xi\,g_{\mu\nu} = \tfrac{1}{2}\,e\,g^{\mu\nu}\,2\,\nabla_{\mu}\,\xi_{\nu} = \nabla_\mu\,(e\,\xi^\mu),
\end{align}
which, using the formula of the covariant divergence
\begin{equation}
\nabla_\mu\,V^\mu = \de_\mu\,V^\mu + \Gamma_\mu{}^\mu{}_\nu\,V^\nu = \de_\mu\,V^\mu + e^{-1}\,V^\mu\,\de_\mu\,e,
\end{equation}
becomes a total derivative:
\begin{align}
\mathcal{L}_\xi\,e &= \de_\mu\,(e\,\xi^\mu).
\end{align}
This is the reason why the Hilbert-Einstein action is diffeomorphism invariant:
\begin{align}
\mathcal{L}_\xi\,(e\,R) &= \de_\mu\,(e\,\xi^\mu\,R),
\end{align}
so that its integral is invariant. Notice that
\begin{equation}
\mathcal{L}_\xi\,e = \xi^\mu\,\de_\mu\,e + \de_\mu\,\xi^\mu\,e,
\end{equation}
is the usual Lie derivative of a scalar plus a term proportional to the divergence of $\xi^\mu$. In general, the Lie derivative of a tensor density $T$ with density weight $\lambda$ is the usual Lie derivative plus $\lambda\,\de_\mu\,\xi^\mu\,T$. So, the formula \eqref{CompactFormulaLieDer} is enhanced as
\begin{equation}
\mathcal{L}_\xi = \xi^\alpha\,\de_\alpha + M^\alpha{}_\beta\,\Delta_\alpha{}^\beta + \lambda\,\de_\alpha\,\xi^\alpha, 
\;\;\text{where}\;\; M^\alpha{}_\beta = \de_\beta\,\xi^\alpha.
\end{equation}

\subsection{Weitzenb\"ock connection and teleparallel gravity}\label{GenRelTorsion}

The first attempts of a torsion formulation of General Relativity, usually known as \emph{teleparallel gravity}, date back to Einstein \cite{Einstein1928}, who unsuccessfully tried to unificate gravitation and electromagnetism in four dimensions, using the sixteen degrees of freedom of the vierbein, according to the works of Cartan \cite{Cartan1922, Cartan1923, Cartan1924, Cartan1925} and Weitzenb\"ock \cite{Weitzenbock1928}. For the history and prehistory of the teleparallel gravity and its extension see for example \cite{Shirafuji:1995xc}. 

In the usual formulation of General Relativity, one chooses a connection $\Gamma_\mu{}^\vrho{}_\nu$ compatible with the metric $g_{\mu\nu}$, such that the torsion vanishes and the curvature encodes the dynamical variables:
\begin{equations}
\nabla_\vrho(\Gamma)\,g_{\mu\nu} &= \de_\vrho\,g_{\mu\nu} - \Gamma_\vrho{}^\sigma{}_\mu\,g_{\sigma\nu} - \Gamma_\vrho{}^\sigma{}_\nu\,g_{\mu\sigma} = 0, \\
T_\mu{}^\vrho{}_\nu(\Gamma) &= \Gamma_\mu{}^\vrho{}_\nu - \Gamma_\nu{}^\vrho{}_\mu = 0,\\
R_{\mu\nu}{}^\vrho{}_\sigma(\Gamma) &= \de_\mu\,\Gamma_\nu{}^\vrho{}_\sigma - \de_\nu\,\Gamma_\mu{}^\vrho{}_\sigma + \Gamma_\mu{}^\vrho{}_\tau\,\Gamma_\nu{}^\tau{}_\sigma - \Gamma_\nu{}^\vrho{}_\tau\,\Gamma_\mu{}^\tau{}_\sigma.
\end{equations}
As known, there is a unique choice for such a connection, the Levi-Civita connection: 
\begin{equation}
\Gamma_\mu{}^\vrho{}_\nu = \tfrac{1}{2}\,g^{\vrho\sigma}\,(\de_\mu\,g_{\nu\sigma} + \de_\nu\,g_{\mu\sigma} - \de_\sigma\,g_{\mu\nu}).
\end{equation}
In the following we will denote with $\nabla_\mu$ the covariant derivative with respect to the Levi-Civita connection $\Gamma_\mu{}^\vrho{}_\nu$, and with $R^\vrho{}_{\sigma\mu\nu}$ the curvature of the Levi-Civita connection (Riemann tensor).

There is an alternative but equivalent formulation of General Relativity (at least classically and without boundary terms), known as \emph{teleparallel gravity}, in which one chooses a connection $W_\mu{}^\vrho{}_\nu$, compatible with the metric, such that the dynamical variables are encoded in the torsion and the curvature vanishes \cite{Einstein1928, Weitzenbock1928}:
\begin{equations}
\nabla_\vrho(W)\,g_{\mu\nu} &= \de_\vrho\,g_{\mu\nu} - W_\vrho{}^\sigma{}_\mu\,g_{\sigma\nu} - W_\vrho{}^\sigma{}_\nu\,g_{\mu\sigma} = 0, \label{WMetricCompatible}\\
T_\mu{}^\vrho{}_\nu(W) &= W_\mu{}^\vrho{}_\nu - W_\nu{}^\vrho{}_\mu,\\
R_{\mu\nu}{}^\vrho{}_\sigma(W) &= \de_\mu\,W_\nu{}^\vrho{}_\sigma - \de_\nu\,W_\mu{}^\vrho{}_\sigma + W_\mu{}^\vrho{}_\tau\,W_\nu{}^\tau{}_\sigma - W_\nu{}^\vrho{}_\tau\,W_\mu{}^\tau{}_\sigma = 0.
\end{equations}
While in the usual formulation of General Relativity the gravitational interaction is due to the curvature of spacetime, in teleparallel gravity the gravitational interaction is mediated by a Lorentz-like force in a flat spacetime, as in Yang-Mills theories \cite{deAndrade:1997gka}. Indeed, while in the former case the action is linear in the curvature, in the latter one the action is quadratic in the torsion, as one can argue by dimensional arguments and as we will see in details in the following.

Since the spacetime manifold is supposed to be flat, then it is possible to choose a system of locally inertial coordinates valid \emph{everywhere}. In other words, the frame or inverse vielbein $e^\mu{}_a$ is a set of $d$ vectors in $d$ dimensions in the tangent space of the manifold such that, when it is evaluated in a point, it forms a basis for the tangent space in that point, for all the points of the manifold. In such a case the manifold is said \emph{parallelisable}, and it is possible to choose a connection such that the vielbein is parallel or covariantly constant 
\begin{equation}\label{parallel}
\de_\mu\,e_\nu{}^a - W_\mu{}^\vrho{}_\nu\,e_\vrho{}^a = 0,
\end{equation}
The unique connection solving the condition is the \emph{Weitzenb\"ock connection} \cite{Weitzenbock1928}
\begin{equation}\label{WeitzenbockDef}
W_\mu{}^\vrho{}_\nu = e^\vrho{}_a\,\de_\mu\,e_\nu{}^a,
\end{equation}
which consistently has vanishing curvature, since it is the Cartan-Maurer form (see Section \ref{MaurerCartan}) out of the vielbein.\footnote{The spin connection corresponding to the Weitzenb\"ock connection is zero.} Moreover, the condition \eqref{parallel} implies the compatibility condition \eqref{WMetricCompatible}, since
\begin{equation}
g_{\mu\nu} = e_\mu{}^a\,e_\nu{}^b\,\eta_{ab}.
\end{equation}
In the following we will denote with $D_\mu$ the covariant derivative with respect to the Weitzenb\"ock connection $W_\mu{}^\vrho{}_\nu$, and with $T_\mu{}^\vrho{}_\nu$ the \emph{Weitzenb\"ock torsion},\footnote{Here and in the rest of the Thesis, the round (square) brackets among indices denote (anti)symmetrisation \emph{without} any numerical coefficient. For example, $A_{(\mu\nu)}=A_{\mu\nu}+A_{\nu\mu}$, and $A_{[\mu\nu]}=A_{\mu\nu}-A_{\nu\mu}$.}
\begin{equation}\label{TorsionFromW}
T_\mu{}^\vrho{}_\nu = e^\vrho{}_a\,\de_{[\mu}\,e_{\nu]}{}^a,
\end{equation}
antisymmetric in the first and in the last index, which is also known as \emph{intrinsic torsion} (see Section \ref{IntrinsicTorsion}). It satisfies the following Bianchi identity
\begin{equation}\label{BianchiWeitzenbock}
D_{[\mu}\,T_{\nu]}{}^\sigma{}_\vrho + D_\vrho\,T_\mu{}^\sigma{}_\nu -T_{[\mu}{}^{\sigma\tau}\,T_{\nu]\tau\vrho} -T_{\mu\tau\nu}\,T_\vrho{}^{\sigma\tau} = 0,
\end{equation}
whose trace reads
\begin{equation}\label{TraceBianchiWeitzenbock}
D_{[\mu}\,T_{\nu]\vrho}{}^\vrho + D_\vrho\,T_\mu{}^\vrho{}_\nu 
+ T_{\mu\vrho\nu}\,T^{\vrho\sigma}{}_\sigma = 0.
\end{equation}
Using the derivative of the vielbein determinant $\de_\mu\,e = e\,W_\mu{}^\vrho{}_\vrho$, one can show that the following identity holds for the Weitzenb\"ock covariant divergence, for any vector $V^\mu$,
\begin{equation}\label{CovariantWeitzenbockDivergence}
e\,D_\mu\,V^\mu = \de_\mu\,(e\,V^\mu) + e\,T_\mu{}^\mu{}_\nu\,V^\nu.
\end{equation}

The relation between the Levi-Civita connection and the Weitzenb\"ock one is
\begin{equation}\label{Levi-Civita-Weitzenbock}
\Gamma_\mu{}^\vrho{}_\nu = W_\mu{}^\vrho{}_\nu + K_\mu{}^\vrho{}_\nu = W_\mu{}^\vrho{}_\nu + \tfrac{1}{2}\,(T_{\mu\nu}{}^\vrho + T_{\nu\mu}{}^\vrho - T_\mu{}^\vrho{}_\nu),
\end{equation}
as one can check by explicit evaluation, where the tensor $-K_\mu{}^\vrho{}_\nu$ is called \emph{contorsion}. Using the vanishing of the curvature of the Weitzenb\"ock connection and the relation \eqref{Levi-Civita-Weitzenbock}, one can write the Riemann tensor in terms of the contorsion:
\begin{equation}
R^\mu{}_{\nu\vrho\sigma} = D_\vrho\,K_\sigma{}^\mu{}_\nu - D_\sigma\,K_\rho{}^\mu{}_\nu + T_\vrho{}^\tau{}_\sigma\,K_\tau{}^\mu{}_\nu + K_\vrho{}^\mu{}_\tau\,K_\sigma{}^\tau{}_\nu - K_\sigma{}^\mu{}_\tau\,K_\vrho{}^\tau{}_\nu.
\end{equation}
Therefore, using the trace of the Bianchi identity \eqref{TraceBianchiWeitzenbock}, the Ricci tensor is
\begin{align}
R_{\mu\nu} = R^\vrho{}_{\mu\vrho\nu} &= \tfrac{1}{2}\,D_\vrho\,T_{(\mu\nu)}{}^\vrho - \tfrac{1}{2}\,D_{(\mu}\,T_{\nu)\vrho}{}^\vrho + \tfrac{1}{2}\,T_{(\mu\nu)\vrho}\,T^{\vrho\sigma}{}_\sigma \,+\nn\\
& + \tfrac{1}{4}\,T_{\vrho\mu\sigma}\,T^\vrho{}_\nu{}^\sigma - \tfrac{1}{2}\,T_{\mu\vrho\sigma}\,T_{\nu}{}^{\vrho\sigma} - \tfrac{1}{2}\,T_{\mu\vrho\sigma}\,T_{\nu}{}^{\sigma\vrho},\label{RicciWeitzenbock}
\end{align}
and so the Ricci scalar is 
\begin{equation}\label{RicciScalarWeitzenbock}
R = R^\mu{}_\mu = -2\,D_\mu\,T^{\mu\nu}{}_\nu  - \tfrac{1}{4}\,T^{\mu\nu\vrho}\,T_{\mu\nu\vrho} - \tfrac{1}{2}\,T^{\mu\nu\vrho}\,T_{\mu\vrho\nu} - T_{\mu\nu}{}^\nu\,T^{\mu\vrho}{}_\vrho.
\end{equation}
Finally, using the Weitzenb\"ock covariant divergence \eqref{CovariantWeitzenbockDivergence}, one gets
\begin{equation}
e\,R = -\de_\mu\,(2\,e\,T^{\mu\nu}{}_\nu) - e\,(\tfrac{1}{4}\,T^{\mu\nu\vrho}\,T_{\mu\nu\vrho} + \tfrac{1}{2}\,T^{\mu\nu\vrho}\,T_{\mu\vrho\nu} - T_{\mu\nu}{}^\nu\,T^{\mu\vrho}{}_\vrho),
\end{equation}
which shows that, up to a boundary term, the Einstein-Hilbert action admits an equivalent formulation in terms of a linear combination of the quadratic scalar contractions of the Weitzenb\"ock torsion \cite{Moller1961a, Moller1961b, Moller1978, Hayashi:1967se, Hayashi:1979qx, deAndrade:1997gka}:
\begin{equation}\label{HE}
S_{\textsc{eh}} = \frac{1}{16\,\pi\,G}\int\diff^d x\,e\,R = \frac{1}{16\,\pi\,G}\int\diff^d x\,e\,\bigg{[}{-\frac{1}{4}}\,T^{\mu\nu\vrho}\,T_{\mu\nu\vrho} - \frac{1}{2}\,T^{\mu\nu\vrho}\,T_{\mu\vrho\nu} + T_{\mu\nu}{}^\nu\,T^{\mu\vrho}{}_\vrho\bigg{]}.
\end{equation}

\subsection{Lie derivative and intrinsic torsion}\label{IntrinsicTorsion}

In \eqref{TorsionFromW} the intrinsic torsion is defined as the antisymmetric part of the Weitzenb\"ock connection. This does not make manifest the fact that the intrinsic torsion is a well-defined covariant tensor. Let us see why. Using the expressions for Lie derivative on a vector $V^\mu$ along a $\xi^\mu$ 
\begin{equation}
\mathcal{L}_\xi\,V^\mu = \xi^\nu\,\de_\nu\,V^\mu - V^\nu\,\de_\nu\,\xi^\mu, 
\end{equation}
and the formula of the covariant derivative with respect to a connection $\gamma_\mu{}^\lambda{}_\nu$,
\begin{equation}
\nabla_\nu(\gamma)\,V^\mu = \de_\nu\,V^\mu + \gamma_\nu{}^\mu{}_{\lambda}\,V^\lambda, 
\end{equation}
one can check the following identity
\begin{equation}\label{IdentityLieDer}
\mathcal{L}_\xi\,V^\mu - (\xi^\nu\,\nabla_\nu(\gamma)\,V^\mu - V^\nu\,\nabla_\nu(\gamma)\,\xi^\mu) = - (\gamma_{\nu}{}^\mu{}_{\lambda}\,\xi^\nu\,V^{\lambda} - 
\gamma_{\lambda}{}^\mu{}_{\nu}\,\xi^\nu\,V^{\lambda}).
\end{equation}
Introducing the torsion of the connection $\gamma_\mu{}^\lambda{}_\nu$ as its antisymmetric part
\begin{equation}
T_\mu{}^\lambda{}_\nu(\gamma) = \gamma_\mu{}^\lambda{}_\nu - \gamma_\nu{}^\lambda{}_\mu,
\end{equation}
and notice that the last two terms in left-hand side of \eqref{IdentityLieDer} are the Lie derivative of $V^\mu$ along $\xi^\mu$, with the partial derivative replaced by the covariant derivative -- denote it as $\mathcal{L}_\xi^{\nabla(\gamma)}$ -- the previous identity writes
\begin{equation}\label{TorsionOnVector}
(\mathcal{L}_\xi-\mathcal{L}_\xi^{\nabla(\gamma)})\,V^\mu = - T_{\nu}{}^\mu{}_{\lambda}(\gamma)\,\xi^\nu\,V^\lambda.
\end{equation}
This relation shows that the torsion is a tensor, since the Lie derivative of a tensor is a tensor, and the covariant derivative of a tensor is a tensor. In particular, if we set $\xi^\mu = V^\mu = e^\mu{}_a$, where $e^\mu{}_a$ are the components of the frame, noticing that $\mathcal{L}_{e_a}^{\nabla(\gamma)}\,e_b$ vanishes if the covariant derivative is compatible with the metric, the identity becomes
\begin{equation}\label{ExprTorsionLieDer}
\mathcal{L}_{e_a}\,e^\mu{}_b = -T_\nu{}^\mu{}_\lambda(\gamma)\,e^\nu{}_a\,e^\lambda{}_b, \;\;\text{if}\;\;\nabla_\mu(\gamma)\,e^\nu{}_a = 0.
\end{equation}
This formula can be assumed as the \emph{definition} of the torsion. Notice that it does not depend on the connection, so we can simply write $T_\nu{}^\mu{}_\lambda$. For this reason, the Weitzenb\"ock torsion is also called ``intrinsic torsion". Making the left-hand side explicit and inverting, one finds that $T_\mu{}^\lambda{}_\nu$ is the Weitzenb\"ock torsion as it was defined in \eqref{TorsionFromW}:
\begin{equation}
T_\mu{}^\lambda{}_\nu = e^\lambda{}_a\,\de_{[\mu}\,e_{\nu]}{}^a = W_{[\mu}{}^\lambda{}_{\nu]}.
\end{equation}
Introducing the flattened components of the intrinsic torsion
\begin{equation}
T_a{}^c{}_b = T_\mu{}^\vrho{}_\nu\,e^\mu{}_a\,e^\nu{}_b\,e_\vrho{}^c,
\end{equation}
the formula \eqref{ExprTorsionLieDer} can be written as
\begin{equation}\label{ExprTorsionLieDer2}
\mathcal{L}_{e_a}\,e_b = [e_a,e_b] = -T_a{}^c{}_b\,e_c.
\end{equation}
This formula, and its extensions in generalised geometry, will be crucial.

The analogous expression of \eqref{TorsionOnVector} acting on a one-form $\omega_\mu$ is
\begin{equation}
(\mathcal{L}_\xi-\mathcal{L}_\xi^{\nabla(\gamma)})\,\omega_\mu = - T_{\nu}{}^\lambda{}_{\mu}(\gamma)\,\xi^\nu\,\omega_\lambda.\label{LieDerTorsion}
\end{equation}
If the connection is the Levi-Civita connection $\gamma_\mu{}^\lambda{}_\nu = \Gamma_\mu{}^\lambda{}_\nu$, so that the torsion is vanishing, and one applies the previous formula on the components of the metric, which is compatible with the connection, one finds the formula \eqref{LieDerMetric}.

\subsection{Gauge symmetries and Lie derivative}\label{GaugeLie}

A gauge theory is geometrically described as a principal fibre bundle with spacetime as base and an internal Lie group manifold $\mathscr{M}$, with dimension $n$, as fibre. A Lie group is always a parallelisable manifold, whose global frame is given by the generators for its algebra and the flattened intrinsic torsion is constant and  equal to minus the structure constants $f_{ab}{}^c$, according to the formula \eqref{ExprTorsionLieDer2}:
\begin{equation}\label{TorsionStructureConstant}
T_a{}^c{}_b = -f_a{}^c{}_b.
\end{equation}
The frame $e_a = e^\alpha{}_a\,\de_\alpha$ is a collection of $n$ internal vector fields, where $a = 1, \dots, n$ is the index in the adjoint of the group, $\alpha = 1,\dots, n$ is the internal index in $\mathscr{M}$, $n$ is the dimension of $\mathscr{M}$, and $\de_\alpha = \frac{\de}{\de y^\alpha}$ is the derivative in the internal coordinates $y^\alpha$. 

Consider two objects $\vphi^\alpha(x,y)$ and $A_\mu{}^\alpha(x,y)$ in the principal bundle, $x^\mu$ being the spacetime coordinates, which are both vectors with respect to the internal manifold, and respectively a scalar and a one-form with respect to spacetime. Suppose that the dependence on the internal coordinates is factorised, amounting to the frame:
\begin{equation}
\vphi^\alpha(x,y) = \vphi^a(x)\,e^\alpha{}_a(y), \quad
A_\mu{}^\alpha(x,y) = A_\mu{}^a(x)\,e^\alpha{}_a(y).
\end{equation}
Observe that
\begin{align}
A_\mu{}^\alpha(x)\,\de_\alpha\,\vphi^\beta(x,y) &= A_\mu{}^a(x)\,e^\alpha{}_a(y)\,\de_\alpha\,e^\beta{}_b(y)\,\vphi^b(x) = \nn\\
&= -A_\mu{}^a(x)\,e^\alpha{}_a(y)\,\vphi^b(x)\,e^\gamma{}_b(y)\,W_\alpha{}^\beta{}_\gamma(y).
\end{align}
Similarly, 
\begin{equation}
\vphi^\alpha(x,y)\,\de_\alpha\,A_\mu{}^\beta(x,y) = -A_\mu{}^a(x)\,e^\alpha{}_a(y)\,\vphi^b(x)\,e^\gamma{}_b(y)\,W_\gamma{}^\beta{}_\alpha(y).
\end{equation}
Therefore, using \eqref{TorsionStructureConstant},
\begin{equation}\label{ParMan3}
\mathcal{L}_{A_\mu}\,\vphi^\beta = A_\mu{}^a\,\vphi^b\,f_a{}^c{}_b\,e^\beta{}_c = [A_\mu,\vphi]^c\,e^\beta_c.
\end{equation}
Thus, if $A_\mu{}^a$ is a gauge connection, and $\vphi^a$ covariantly transforms with respect to the gauge group, then the covariant gauge derivative:
\begin{equation}
D_\mu\,\vphi^c = \de_\mu\,\vphi^c + [A_\mu,\vphi]^c 
\end{equation}
can be also written in terms of the Lie derivative along $A_\mu{}^a$:
\begin{equation}
D_\mu\,\vphi^c = \de_\mu\,\vphi^c + e_\beta{}^c\,\mathcal{L}_{A_\mu}\,\vphi^\beta,
\end{equation}
which means that \cite{Berman:2020tqn}
\begin{equation}\label{CovariantDerivativeGaugeTrans}
D_\mu = \de_\mu + e\circ\mathcal{L}_{A_\mu}\circ e^{-1},
\end{equation}

\subsection{Generalised double geometry}\label{DFT}

The construction of exceptional geometry and exceptional field theory, which will be studied in Section \ref{ExceptionalGeometry} and \ref{ExFT}, was inspired by the work on generalised geometry, which we will refer here as ``double geometry", by Hitchin, Gualtieri \cite{Hitchin:2003cxu, Gualtieri:2003dx, Hitchin:2005in, Bursztyn:2005vwa, Gualtieri:2007ng, Hitchin:2010qz}, inspired by the previous work by Courant, Weinstein, and Dorfman \cite{Courant1988, Courant1990, Dorfman1987}. It could be useful to summarise the main features of double geometry and compare with the exceptional case, since the double geometry setting grasps the main ideas as in exceptional geometry in simpler and neater setting. Nevertheless, the physical interpretation of the two geometries are very different, and it could be quite misleading to take a parallel of them. While $n$-dimensional exceptional geometry brings to a reformulation of eleven-dimensional supergravity, making manifest the duality symmetry of the theory when it is compactified in an $n$-dimensional internal manifold, double geometry, unifying diffeomorphisms and one-form gauge transformations of a metric tensor and of a two-form gauge field, allows to deal with T-duality in string theory. Therefore, the encompassed symmetries in the double Lie derivative are, at the level of the whole spacetime, formally doubled in order to take into account also the winding modes, whereas exceptional Lie derivative regards only the internal gauge transformations, restricted to the internal manifold, excluding the external spacetime diffeomorphisms. 

The starting point in double geometry is the definition of a double vector bundle $E$, locally identified as the direct sum of the tangent bundle of a $d$-dimensional manifold $\mathscr{M}$ and the corresponding cotangent bundle:
\begin{equation}
E \simeq T\mathscr{M} \oplus T^*\mathscr{M}.
\end{equation}
The $E$ sections are ``double vectors" with two components, a vector field and a one-form field. Double vectors are denoted with uppercase Latin letters, the vector components with the corresponding lowercase letters, and the one-form components with Greek letters. The two components can be presented in a formal sum, or in matrix notation as a column vector. For example,
\begin{equation}
X = x + \xi \equiv \begin{pmatrix} x \\ \xi \end{pmatrix}, \quad
X \in \Gamma(E), \; x \in \Gamma(T\mathscr{M}), \; \xi \in \Gamma(T^*\mathscr{M})
\end{equation}
where $\Gamma(\cdot)$ is the standard notation for the sections of a vector bundle. The \emph{pairing} between double vectors is defined by
\begin{equation}\label{DoublePairing}
\langle X,Y\rangle = \tfrac{1}{2}\,(\iota_x\,\eta + \iota_y\,\xi), \quad X = x + \xi,\; Y = y + \eta.
\end{equation}
The \emph{Dorfman derivative} of the double vector $Y=y+\eta$, generated by the double vector $X=x+\xi$, is the double vector $L_X\,Y$, defined by\footnote{Other standard notations in the literature are $[X,Y]_D$ or $X\circ Y$.}
\begin{equation}\label{DoubleDorfman}
L_X\,Y = \mathcal{L}_x\,y + (\mathcal{L}_x\,\eta - \iota_y\,\diff\xi).
\end{equation}
If the right multiplication of a differential form $B$ by a vector field $x$ is formally identified with the inner contraction of the differential form with the vector field
\begin{equation}\label{DoubleIdentification}
B\,x \equiv \iota_x\,B,
\end{equation}
then there is a matrix representation of the Dorfman derivative, given by
\begin{equation}
L_X\,Y = \begin{pmatrix}
\mathcal{L}_x & 0 \\ -\diff\,\xi & \mathcal{L}_x
\end{pmatrix}\,\begin{pmatrix}
y \\ \eta
\end{pmatrix}.
\end{equation}
The Dorfman derivative enjoys the following properties:
\begin{itemize}
\item[--] It satisfies the Leibniz rule, as a derivative should do,
\begin{equation}\label{DoubleLeibniz}
L_X\,(L_Y\,Z) = L_{L_X Y}\,Z + L_Y\,(L_X\,Z), \quad \forall\;X,Y,Z \in \Gamma(E).
\end{equation}
\item[--] It is \emph{not} antisymmetric
\begin{equation}\label{DoubleNoAnti}
L_X\,Y \neq -L_Y\,X.
\end{equation}
\item[--] It has non-trivial kernel:
\begin{equation}\label{DoubleKernel}
\exists\; 0\neq\tilde{X}\in\Gamma(E),\;\text{such that}\;L_{\tilde{X}}\,Y = 0, \;\;\forall\;Y\in\Gamma(E).
\end{equation}
\item[--] The symmetric part is always in the kernel
\begin{equation}\label{DoubleSymTriv}
L_{\frac{1}{2}(L_X Y + L_Y X)}\,Z = 0, \quad\forall\;X,Y,Z\in\Gamma(E).
\end{equation}
\item[--] The elements in the kernel are closed one-forms.
\end{itemize}
First of all, let us check the Leibniz rule \eqref{DoubleLeibniz}, which can be also written as
\begin{equation}\label{DoubleLeibnizBis}
[L_X,L_Y] = L_{L_X Y}.
\end{equation}
It follows by explicit evaluation, starting from the definition \eqref{DoubleDorfman}. If $X=x+\xi$, $Y=y+\eta$, $Z=z+\zeta$, then
\begin{equations}
L_X\,L_Y\,Z &= \mathcal{L}_x\,\mathcal{L}_y\,z + \mathcal{L}_x\,(\mathcal{L}_y\,\zeta - \iota_z\,\diff\eta) - \iota_{\mathcal{L}_x z}\,\diff\xi, \\
L_{L_X Y}\,Z &= \mathcal{L}_{\mathcal{L}_x\,y}\,z + \mathcal{L}_{\mathcal{L}_x\,y}\,\zeta - \iota_z\,\diff(\mathcal{L}_x\,\eta - \iota_y\,\diff\xi),
\end{equations}
so that
\begin{align}
L_X\,L_Y\,Z - L_{L_X Y}\,Z - L_Y\,L_X\,Z &= 
(\mathcal{L}_x\,\mathcal{L}_y - \mathcal{L}_{\mathcal{L}_x\,y}-\mathcal{L}_y\,\mathcal{L}_x)\,z \,+\nn\\
& + (\mathcal{L}_x\,\mathcal{L}_y - \mathcal{L}_{\mathcal{L}_x y} - \mathcal{L}_y\,\mathcal{L}_x)\,\zeta \,+\nn\\
& + (-\mathcal{L}_x\,\iota_z\,\diff \eta - \iota_\zeta\,\diff\,\mathcal{L}_x\,\eta + \iota_{\mathcal{L}_x\,z}\,\diff \eta) \,+\nn\\
& + (-\iota_{\mathcal{L}_y\,z}\,\diff \xi - \iota_z\,\diff\,\iota_y\,\diff\xi + \mathcal{L}_y\,\diff_z\,\diff\xi) = \nn\\
&= ([\mathcal{L}_x,\mathcal{L}_y] -\mathcal{L}_{\mathcal{L}_x y})\,(z+\zeta) \,+\nn\\
& + (-\mathcal{L}_x\,\iota_z - \iota_\zeta\,\mathcal{L}_x + \iota_{\mathcal{L}_x z})\,\diff\eta \,+\nn\\
& + ([\mathcal{L}_x,\iota_z] - \iota_{\mathcal{L}_x z})\,\diff\eta \,+\nn\\
& + ([\mathcal{L}_y,\iota_z]-\iota_{\mathcal{L}_y z})\,\diff \xi = 0,
\end{align}
as we wanted to show.

The Dorfman derivative is manifestly non-antisymmetric, by definition \eqref{DoubleDorfman}. As a consequence, the Leibniz rule is not equivalent to a Jacobi identity, which so is not fulfilled by the Dorfman derivative in general. The Leibniz rule written in the form \eqref{DoubleLeibnizBis} shows that the symmetric part is an element in the kernel, as in \eqref{DoubleSymTriv}, since the left-hand side of \eqref{DoubleLeibnizBis} is antisymmetric in $X \leftrightarrow Y$, so also the right-hand side must be. Looking at the definition \eqref{DoubleDorfman}, one can notice that the one-form component $\xi$ of the generator double vector $X$ appears only through its external differential $\diff\xi$. Therefore, if we take $\tilde{X} = 0 + \diff\vphi$, we get always a trivial parameter. The symmetric part is always an exact one-form too. Indeed, if $X=x+\xi$ and $Y=y+\eta$,
\begin{equations}
\tfrac{1}{2}\,(L_X\,Y + L_Y\,X) &= \tfrac{1}{2}\,(\mathcal{L}_x\,y + \mathcal{L}_y\,x) + \tfrac{1}{2}\,(\mathcal{L}_x\,\eta - \iota_y\,\diff\xi + \iota_y\,\diff\xi + \mathcal{L}_y\,\xi - \iota_x\,\diff \eta) = \nn\\
&= \tfrac{1}{2}\,(\iota_x\,\diff\eta + \diff\,\iota_x\,\eta - \iota_y\,\diff\xi + \iota_y\,\diff\xi + \diff\,\iota_y\,\xi - \iota_x\diff \eta) =\nn\\
& = \tfrac{1}{2}\,\diff\,(\iota_x\,\eta + \iota_y\,\xi) = \diff\langle X,Y \rangle.\label{DoubleSym}
\end{equations}
The antisymmetric part of the Dorfman derivative is called \emph{Courant bracket}:\footnote{The standard notation is $[\cdot,\cdot]_C$ or simply $[\cdot,\cdot]$.}
\begin{equation}
C_X\,Y = \tfrac{1}{2}\,(L_X\,Y - L_Y\,X) = \mathcal{L}_x\,y + [\mathcal{L}_x\,\eta - \mathcal{L}_y\,\xi - \tfrac{1}{2}\,\diff\,(\iota_x\,\eta - \iota_y\,\xi)],
\end{equation}
so that, using also \eqref{DoubleSym}, we can write
\begin{equation}
L_X\,Y = C_X\,Y + \diff\langle X,Y \rangle.
\end{equation}
Moreover, combining \eqref{DoubleSymTriv} and the previous decomposition,
\begin{equation}\label{DoubleDoubleDec}
L_{L_X Y}\,Z = C_{C_X\,Y}\,Z + \diff\langle C_X\,Y,Z \rangle.
\end{equation}
Although the Courant bracket is antisymmetric by definition, it does not even satisfy the Jacobi identity. Nevertheless, the failure of the Jacobi identity is always an element in the kernel, that is, an exact one-form. To show it, let define the \emph{Jacobiator}, which would vanish if the Jacobi identity were satisfied,
\begin{equation}
\text{Jac}(X,Y,Z) = C_{C_X Y}\,Z + C_{C_Y Z}\,X + C_{C_Z X}\,Y.
\end{equation}
Define moreover the so-called \emph{Nijenhuis operator}
\begin{equation}
\text{Nij}(X,Y,Z) = \langle C_X\,Y,Z \rangle + \langle C_Y\,Z,X \rangle + \langle C_Z\,X, Y \rangle.
\end{equation}
Since
\begin{align}
C_{C_X Y}\,Z &= \tfrac{1}{2}\,(L_{C_X Y}\,Z - L_Z\,C_X\,Y) =
\tfrac{1}{2}\,(L_{L_X Y}\,Z - \tfrac{1}{2}\,L_Z\,(L_X\,Y-L_Y\,X)) = \nn\\
&= \tfrac{1}{2}\,(L_X\,L_Y\,Z - L_Y\,L_X\,Z - \tfrac{1}{2}\,L_Z\,L_X\,Y + \tfrac{1}{2}\,L_Z\,L_Y\,X),
\end{align}
then, using also \eqref{DoubleDoubleDec}, 
\begin{align}
\text{Jac}(X,Y,Z) &= \tfrac{1}{2}\,(L_X\,L_Y\,Z + L_Y\,L_Z\,X + L_Z\,L_X\,Y \,+\nn\\
& - L_Y\,L_X\,Z - L_X\,L_Z\,Y - L_Z\,L_Y\,X \,+\nn\\
& -\tfrac{1}{2}\,L_Z\,L_X\,Y - \tfrac{1}{2}\,L_X\,L_Y\,Z - \tfrac{1}{2}\,L_Y\,L_Z\,X \,+\nn\\
& +\tfrac{1}{2}\,L_Z\,L_Y\,X + \tfrac{1}{2}\,L_Y\,L_X\,Z + \tfrac{1}{2}\,L_X\,L_Z\,Y) = \nn\\
&= \tfrac{1}{4}\,(L_X\,L_Y\,Z + L_Y\,L_Z\,X + L_Z\,L_X\,Y \,+\nn\\
& -L_Y\,L_X\,Z - L_X\,L_Z\,Y - L_Z\,L_Y\,X) = \nn\\
&= \tfrac{1}{4}\,(L_{L_X Y}\,Z + L_{L_Y Z}\,X + L_{L_Z X}\,Y) = \nn\\
&= \tfrac{1}{2}\,(C_{C_X Y}\,Z + C_{C_Y Z}\,X + C_{C_Z}\,Y) \,+\nn\\
& + \tfrac{1}{4}\,\diff\,(\langle C_X\,Y,Z\rangle + \langle C_Y\,Z,X\rangle + \langle C_Z\,X,Y\rangle) = \nn\\
&= \tfrac{1}{4}\,\text{Jac}(X,Y,Z) + \tfrac{3}{4}\,\diff\,\text{Nij}(X,Y,Z).
\end{align}
Thus, we conclude that
\begin{equation}
\text{Jac}(X,Y,Z) =  \diff\,\text{Nij}(X,Y,Z).
\end{equation}

Consider now the  following $2\,d\times 2\,d$ matrix, where $B$ is a two-form,
\begin{equation}
\begin{pmatrix}
I_d & 0 \\ B & I_d
\end{pmatrix}, \;\;\text{such that}\;\;\begin{pmatrix}
I_d & 0 \\ B & I_d
\end{pmatrix}^n = \begin{pmatrix}
I_d & 0 \\ n\,B & I_d.
\end{pmatrix}
\end{equation}
Therefore it can be identified  with an exponential $e^B$. Its action on $X = x + \xi$ is 
\begin{equation}
e^B\,X = \begin{pmatrix}
I_d & 0 \\ B & I_d
\end{pmatrix}\,\begin{pmatrix}
x \\ \xi
\end{pmatrix} = x + B\,x + \xi = x + \xi + \iota_x\,B,
\end{equation}
where we used the identification \eqref{DoubleIdentification}
in the last step. The moltiplication by $e^B$ of a double vector is called \emph{twisting}: $e^B\,X$ is the twisted $X$.\footnote{Another possible name is \emph{dressing}.} Notice that the pairing is twisting-invariant. Indeed, using $[\iota_x\,\iota_y] = 0$:
\begin{align}
\langle e^B\,X,e^B\,Y\rangle &= \tfrac{1}{2}\,(\iota_x\,(\eta + \iota_y\,B) + \iota_y\,(\xi + \iota_x\,B)) = \nn\\
&= \tfrac{1}{2}\,(\iota_x\,\eta + \iota_y\,\xi) + \tfrac{1}{2}\,(\iota_\xi\,\iota_y + \iota_y\,\iota_x)\,B = \langle X,Y \rangle.
\end{align}
Let us study the behaviour of the Dorfman derivative under the twisting of its entries:
\begin{align}
L_{e^B X}\,e^B\,Y &= \mathcal{L}_x\,y + \mathcal{L}_x\,(\eta + \iota_y\,B) - \iota_y\,\diff(\xi + \iota_x\,B) = \nn\\
&= \mathcal{L}_x\,y + \mathcal{L}_x\,\eta + \mathcal{L}_x\,\iota_y\,B - \iota_y\,\diff\xi - \iota_y\,\diff\,\iota_x\,B = \nn\\
&= L_X\,Y + \mathcal{L}_x\,\iota_y\,B - \iota_y\,\mathcal{L}_x\,B + \iota_y\,\iota_x\,\diff B = \nn\\
&= L_X\,Y + \iota_{\mathcal{L}_x y}\,B + \iota_y\,\iota_x\,\diff B = \nn\\
&= e^B\,L_X\,Y + \iota_y\,\iota_x\,\diff B. \label{DoubleTwistingDorfman}
\end{align}
Defining the \emph{twisted Dorfman derivative} as
\begin{equation}
L_X^{(H)}\,Y = L_X\,Y + \iota_y\,\iota_x\,H,\;\;\text{where}\;H\;\text{is a three-form},
\end{equation}
then, the result computed in \eqref{DoubleTwistingDorfman} writes as
\begin{equation}
L_{e^B X}\,e^B\,Y = e^B\,L_X^{(\diff B)}\,Y.
\end{equation}
In other words, whenever the deformation $H$ in the twisted Dorfman derivative is closed, so that it is locally integrable as an exact form $H = \diff B$, for some two-form $B$, then the deformation in the twisted derivative is absorbed in the twisting of the entries of the untwisted derivative by $e^B$. On the other hand, one can check that, without any assumption on the three-form $H$, the Leibniz rule for the twisted derivative is spoiled by a term proportional to $\diff H$: 
\begin{equation}
[L^{(H)}_X,L^{(H)}_Y]\,Z = L^{(H)}_{L^{(H)}_X Y}\,Z + \iota_z\,\iota_y\,\iota_x\,\diff H.
\end{equation}
Therefore the twisted derivative satisfies the Leibniz rule if an only if $H$ is closed, which means locally exact:
\begin{equation}
[L^{(H)}_X,L^{(H)}_Y] = L^{(H)}_{L^{(H)}_X Y} \Leftrightarrow H = \diff B\;\text{for some two-form}\;B.
\end{equation}
And this is the same case in which the deformation can be absorbed in the twisting of the entries. So, the consistency of the twisted derivative is equivalent to the possibility to absorb the $H$ deformation, which is equivalent to the $H$ integrability:
\begin{equation}
L^{(H)}\;\text{is consistent} \Leftrightarrow
\text{the}\;H\;\text{deformation can be absorbed} 
\Leftrightarrow H\;\text{is integrable}.
\end{equation}

Looking at the first line of \eqref{DoubleTwistingDorfman}, one can see that the action of $L_{e^B X}$ consists in a Lie derivative generated by the vector component $x$ of $e^B\,X$, and by a one-form gauge transformation generated by the one-form components $\xi + \iota_x\,B =\xi'$ of $e^B\,X$. This is clearer in matrix notation, in which it is simple to extract the action of $L_{e^B X}\circ e^B$:
\begin{equation}
L_{e^B X} \circ e^B = \begin{pmatrix}
\mathcal{L}_x & 0 \\
-\diff(\xi + \iota_x\,B) & \mathcal{L}_x 
\end{pmatrix}\begin{pmatrix}
I_d & 0 \\ B & I_d 
\end{pmatrix} = \begin{pmatrix}
\mathcal{L}_x & 0 \\
\mathcal{L}_x\,B -\diff\xi' & \mathcal{L}_x
\end{pmatrix}.
\end{equation}

We can see this adopting to another point of view \cite{Berman:2020tqn}. The pairing introduced in \eqref{DoublePairing}, writes in matrix notation as
\begin{equation}
\tfrac{1}{2}\,X^m\,\eta_{mn}\,Y^n = \langle X,Y \rangle = \tfrac{1}{2}\begin{pmatrix}
x^\mu & \xi_\mu \end{pmatrix}
\begin{pmatrix}
0 & \delta^\nu_\mu \\ \delta^\mu_\nu & 0 
\end{pmatrix}\begin{pmatrix}
y^\nu \\ \eta_\nu \end{pmatrix},
\end{equation}
\begin{equation}
X = (X^m)_{m=1\dots,2d} = (x^\mu,\xi_\mu)_{\mu = 1,\dots d}.
\end{equation}
$\eta_{MN}$ are the components of the symmetric matrix associated to the pairing:
\begin{equation}
\eta_{MN} = \begin{pmatrix}
0 & \delta^\nu_\mu \\ \delta^\mu_\nu & 0 
\end{pmatrix}.
\end{equation}
Now, let us define the \emph{double Lie derivative} as a deformation of the usual Lie derivative, acting on and generated by double vectors
\begin{equation}\label{DoubleLieDer}
L_X\,Y^m = X^n\,\de_n\,Y^m - Y^n\,\de_n\,X^m + \eta^{mn}\,\eta_{pq}\,\de_n\,X^p\,Y^q.
\end{equation}
The first two terms are the usual derivative between vector fields. $\de_m$ splits into spacetime partial derivative $\de_\mu$ and we assume that the second components ``$\tilde{\de}^\mu$" are vanishing:
\begin{equation}
\de_m = (\de_\mu, 0).
\end{equation}
As the notation suggests, this is nothing but the Dorfman derivative, written in covariant way. Indeed, splitting the dummy indices in \eqref{DoubleLieDer},
\begin{align}
(L_X\,Y)^\mu &= (x^\nu\,\de_\nu\,y^\mu - y^\nu\,\de_\nu\,x^\mu, x^\nu\,\de_\nu\,\eta_\mu - y^\nu\,\de_\nu\,\xi_\mu + \delta_\mu^\nu\,\delta_\vrho^\sigma \de_\nu\,x^\vrho\,\eta_\sigma + \eta^\mu_\nu\,\delta^\vrho_\sigma\,\de_\nu\,\xi_\vrho\,y^\sigma) = \nn\\
&= (\mathcal{L}_x\,y^\mu, x^\nu\,\de_\nu\,\eta_\mu - y^\nu\,\de_\nu\,\xi_\mu + \eta_\nu\,\de_\mu\,x^\nu + y^\nu\,\de_\mu\,\xi_\nu) = \nn\\
&= (\mathcal{L}_x\,y^\mu, \mathcal{L}_x\,\eta_\mu - (\iota_y\,\diff \xi)_\mu) \rightarrow \mathcal{L}_x\,y + \mathcal{L}_x\,\eta - \iota_y\,\diff \xi.
\end{align}
But one can derive the main properties of the derivative without assuming the expression in components. Namely, the double Lie derivative is not antisymmetric because of the deformation term. In order to be a well-defined derivative, it must satisfy the Leibniz rule
\begin{equation}\label{DoubleLeibnizL}
[L_X,L_Y] = L_{\frac{1}{2}(L_X Y - L_Y X)},
\end{equation}
which implies the symmetric part to be in the kernel
\begin{equation}
L_{\tfrac{1}{2}\,(L_X Y + L_Y X)} = 0.
\end{equation}
By explicit evaluation, one can see that the identity \eqref{DoubleLeibnizL} requires a \emph{section constraint} to be fulfilled
\begin{equation}\label{DoubleSectionConstraint}
\eta^{mn}\,\de_m\,\de_n = 0.
\end{equation}
This is indeed the case, because the action of $\eta$ on a double vector swaps the component and the second component of $\de_m$ is vanishing. Nevertheless, more general solutions of the section constraints are possible, leaving the opportunity not to set ``$\tilde{\de}^\mu$" equal to zero.

In general, the double vectors in the kernel are of the form 
\begin{equation}\label{DoubleTrivialL}
\tilde{X}^m = \eta^{mn}\,\de_n\,\vphi,\;\text{for any scalar}\;\vphi.
\end{equation}
It follows by explicit evaluation:
\begin{equation}
L_{\tilde{X}}\,Y^m = \eta^{pq}\,\de_q\,\vphi\,\de_p\,Y^m - Y^n\,\eta^{mq}\,\de_n\,\de_q\,\vphi + \eta^{mn}\,\eta_{pq}\,\de_n\,\eta^{pr}\,\de_r\,\vphi\,Y^q = 0, 
\end{equation}
since the last two sums to zero, and the first one drops out by means of \eqref{DoubleSectionConstraint}. Also the symmetric part of the double derivative takes the form in \eqref{DoubleTrivialL}:
\begin{align}
\tfrac{1}{2}\,(L_X\,Y + L_Y\,X)^m &= \tfrac{1}{2}\,(\eta^{mn}\,\eta_{pq}\,\de_n\,X^p\,Y^q + \eta^{mn}\,\eta_{pq}\,\de_n\,Y^p\,X^q) = \nn\\
&= \tfrac{1}{2}\,\eta^{mn}\,\eta_{pq}\,\de_n(X^p\,Y^q) = \eta^{mn}\,\de_n\,(\tfrac{1}{2}\,\eta_{pq}\,X^p\,Y^q) =\nn\\
&= \eta^{mn}\,\de_n\,\langle X,Y\rangle.
\end{align}

One can see that the transformations encoded in the double Lie derivative are diffeomorphisms and gauge transformations of a one-form by considering a ``double metric" $H_{mn}$, encoding the components of a metric tensor $g_{\mu\nu}$ and a two-form $B_{\mu\nu}$:
\begin{equation}
H_{mn} = \begin{pmatrix}
g_{\mu\nu} - B_{\mu\vrho}\,g^{\vrho\sigma}\,B_{\sigma\nu} & B_{\mu\vrho}\,g^{\vrho\nu} \\
-g^{\mu\vrho}\,B_{\vrho\nu} & g^{\mu\nu}
\end{pmatrix}.
\end{equation}
$H_{mn}$ is supposed to transform according to the double Lie derivative, generated by a double vector $X^m = (x^\mu,\xi_\mu)$:
\begin{equation}
\delta_X\,H_{mn} = L_X\,H_{mn} = X^p\,\de_p\,H_{mn} + H_{p(m}\,\de_{n)}\,X^p - \eta^{pq}\,\eta_{r(m}\,H_{n)q}\de_p\,X^r.
\end{equation} 
One can see by explicit evaluation that the double Lie derivative of the double metric is equivalent to the diffeomorphisms generated by $x^\mu$ on $g_{\mu\nu}$ and $B_{\mu\nu}$, and to the one-form gauge transformation generated by $\xi_\mu$ on $B_{\mu\nu}$
\begin{equation}
\delta_X\,H_{mn} = L_X\,H_{mn} \Leftrightarrow
\begin{cases}
\delta_{x,\xi}\,g_{\mu\nu} = \mathcal{L}_x\,g_{\mu\nu},\\ 
\delta_{x,\xi}\,B_{\mu\nu} = \mathcal{L}_x\,B_{\mu\nu} - \de_{[\mu}\,\xi_{\nu]}.
\end{cases}
\end{equation}

We conclude this overview by noticing that it is possible to extend the notions parallelisability, Weitzenb\"ock connection and torsion in double geometry in a natural way. A global frame in double geometry is a set of $d$ independent, globally defined, non-vanishing double vectors $E_a$, $a=1,\dots,d$. If such a frame exists, one says that the manifold is parallelisable in the sense of double geometry. One can use the frame to define the analogue of the Weitzenb\"ock connection, according to the definition \eqref{WeitzenbockDef}, in which the simple frame is replaced by the double one
\begin{equation}
W_m{}^r{}_n(E) = E^r{}_a\,\de_m\,E_n{}^a.
\end{equation}
We have to maintain the defining property which identify the Weitzenb\"ock torsion as a tensor for a correct generalisation in double geometry. Namely, the Weitzenb\"ock torsion in double geometry is not simply the antisymmetric part of the double Weitzenb\"ock torsion (compare with \eqref{TorsionFromW}), since it would not be possible to express it in terms of the double Lie derivative, according to \eqref{ExprTorsionLieDer2}. Instead, if we take \eqref{ExprTorsionLieDer2} as the definition of the torsion, extending it to double geometry by replacing the simple Lie derivative with the double one, we obtain a well-behaved extension of the Weitzenb\"ock torsion in double geometry
\begin{equation}\label{ParallDFT}
L_{E_a}\,E_b = -T_a{}^c{}_b(E)\,E_c.
\end{equation}
Since the double Lie derivative is not antisymmetric, the Weitzenb\"ock torsion acquires a non-antisymmetric part in double geometry. Explicitly, one can compute
\begin{align}
L_{E_a}\,E^m{}_b &= E^n{}_a\,\de_n\,E^m{}_b - E^n{}_b\,\de_n\,E^n{}_a + \eta^{mn}\,\eta_{pq}\,\de_n\,E^p{}_a\,E^q{}_b = \nn\\
&= -(E^n{}_a\,E^s{}_b\,\de_n\,E_s{}^c - E^n{}_b\,E^s{}_a\,\de_n\,E_s{}^c + \eta^{ce}\,\eta_{bd}\,E^n{}_e\,E^s{}_a\,\de_n\,E_s{}^d)\,E^m{}_c.
\end{align}
Recognising in the last line, the flattened version of the Weitzenb\"ock connection 
\begin{equation}
W_a{}^c{}_b(E) = E^m{}_a\,E^n{}_b\,E_r{}^c\,W_m{}^r{}_c(E) = E^m{}_b\,E^n{}_a\,\de_m\,E_n{}^c,
\end{equation}
we find the explicit expression for the double torsion, in which the last term breaks the antisymmetry in the lowered indices,
\begin{equation}\label{TorsionDFT}
T_a{}^c{}_b(E) = W_a{}^c{}_b(E)-W_b{}^c{}_a(E) + \eta^{ce}\,\eta_{bd}\,W_e{}^d{}_a(E).
\end{equation}

\subsection{Parallelisability of spheres}

In 1958 Bott, Milnor, and Kervaire proved that the $n$-dimensional hypersphere $S^n$ is parallelisable if and only if $n=1,3,7$ \cite{Bott1958-vt, Milnor1958, Kervaire1958}. 
As an instructive example, let us see in details the simplest case. $S^1$ is the unitary circle. It can be described as a submanifold of $\mathbb{R}^2$ with constrained coordinates. If $(x,y)$ is any point in $\mathbb{R}^2$, with $x,y\in\mathbb{R}$, then $(x,y)\in S^1$ if $f(x,y)=x^2+y^2-1=0$:
\begin{equation}
S^1 \simeq \{ (x,y) \in \mathbb{R}^2\,:\,x^2+y^2=1\}.
\end{equation}
The tangent bundle of a manifold is the set of all the vector fields of the manifold, that is, the vector space generated by the partial derivatives of the coordinates. Since in the case of $S^1$ we use constrained embedding coordinates, also the partial derivatives are constrained. In particular, taking the external differential of $f$, 
\begin{equation}
\diff f = (\diff x\,\de_x + \diff y\,\de_y)\,(x^2+y^2-1) = 2\,x\,\diff x + 2\,y\,\diff y.
\end{equation}
When this one-form is evaluated on an arbitrary $\mathbb{R}^2$ vector field $u\,\de_x + v\,\de_y$, $u,v\in\mathbb{R}$, it gives
\begin{equation}
\diff f(u,v) = \langle 2\,x\,\diff x + 2\,y\,\diff y,u\,\de_x + v\,\de_y \rangle = 2\,x\,u + 2\,y\,v.
\end{equation}
Since $f(x,y)=0$ on $S^1$ for all $(x,y)\in\mathbb{R}^2$, we requires $\diff\,f(u,v) = 0$, for all $(u,v)\in\mathbb{R}^2$, and this condition identifies the tangent space $T S^1$ of $S^1$. Therefore, $T S^1$ can be described as the submanifold of $\mathbb{R}^2 \times \mathbb{R}^2$, such that the coordinates $(x,y)$ in the first copy of $\mathbb{R}^2$ are constrained as the coordinates of $\mathbb{R}^2$ describing $S^1$, and the coordinates $(u,v)$ of the second copy of $\mathbb{R}^2$ are constrained by $\frac{1}{2}\,\diff f(u,v) = u\,x + v\,y = 0$:
\begin{equation}
T S^1 = \{((x,y),(u,v)) \in \mathbb{R}^2 \times \mathbb{R}^2 \,:\, x^2 + y^2 = 1, \; u\,x + v\,y = 0 \}.
\end{equation}
Now, we can see that $T S^1$ admits a global, non-vanishing frame, so that $S^1$ is parallelisable. It is sufficient to consider a solution of the condition $u\,x + v\,y = 0$, such as $(-y,x)$, and choose as a global frame the vector field defined by
\begin{equation}
(x,y) \mapsto ((x,y),(-y,x))\,:\,S^1 \rightarrow T S^1,
\end{equation}
which is non-vanishing everywhere, since $((x,y),(-y,x))$ is vanishing if and only if $x=y=0$, but $(x,y)=(0,0)$ does not belong to $S^1$.

\subsection{Generalised parallelisability of spheres}\label{ParallSn}

Although only $S^1$, $S^3$, and $S^7$ are parallelisable, it was shown in \cite{Lee:2014mla} that a generalised geometry can be defined, such that $S^n$ is parallelisable  for all $n$ in a generalised sense. By parallelisation in generalised sense one means the existence of a global, non-vanishing frame in the generalised bundle. Basically, the result is that, although $S^n$ is not parallelisable in general, 
\begin{equation}
T S^n \oplus \wedge^{n-2}\,T^* S^n
\end{equation}
is always parallelisable. Notice that it is simpler to find a global, non-vanishing frame in the generalised bundle, since it is sufficient that the two components of the generalised frame do not simultaneously vanish for the non-vanishing condition to be fulfilled, rather than they do not vanish at all. Let us see the details, following faithfully \cite{Lee:2014mla}.

In double geometry one introduces a generalised bundle whose sections are the sum of a vector and of a one-from. Here the sum $V=v+\lambda$ of a vector $v$ and of a $(n-2)$--form $\lambda$ is considered in $n$ dimensions, which are sections of the following generalised bundle:
\begin{equation}
E \simeq T\mathscr{M} \oplus \wedge^{n-2}\,T^*\mathscr{M}.
\end{equation}
In components, $V$ is
\begin{equation}
V^M = \begin{pmatrix}
v^m \\ \lambda_{m_1\dots m_{n-2}}
\end{pmatrix}.
\end{equation}
This is relevant for $S^n$ since one can show that an $n$-dimensional sphere with radius $R$ is a solution for the following equations
\begin{equation}
R_{mn} = \tfrac{1}{n!(n-1)}\,F^{m_1\dots m_n}\,F_{m_1\dots m_n}\,g_{mn}, \quad F = \tfrac{n-1}{R}\,\text{vol}_g,
\end{equation}
where $R_{mn}$ is the Ricci tensor associated to the metric tensor $g_{mn}$, $F = \diff A$ is the field strength of the $(n-1)$--form potential $A$, and the volume form can be written as
\begin{equation}
\text{vol}_g = \tfrac{R^n}{n!}\,\vepsilon_{i_1\dots i_{n+1}}\,y^{i_1}\dot\diff y^{i_2}\dots \diff y^{n+1},
\end{equation}
where $y^i$, $i=1,\dots,n+1$, are the Cartesian embedding coordinates of the sphere, constrained by $\delta_{ij}\,y^i\,y^j = 1$, such that the metric of the sphere $S^n$ can be written as
\begin{equation}
\diff s^2(S^n) = \delta_{ij}\,\diff y^i\otimes\diff y^j.
\end{equation}
The components of a generalised vector $X$ in the bundle $E$ encompasses the vector field generating the $n$-dimensional diffeomorphisms and the $(n-2)$--form generating the gauge transformation of $A$, similarly to the case of double geometry, which can be used as the geometric arena for describing a theory invariant under diffeomorphism and gauge transformations of the two-form.

As in double geometry, one can define a Dorfman derivative  and a twist operation, the last one depending on the potential $A$. The Dorfman derivative is 
\begin{equation}
L_V\,W = \mathcal{L}_v\,w + \mathcal{L}_v\,\mu - \iota_w\,\diff\lambda,
\end{equation}
where $V=v+\lambda$, and $W=w+\mu$. Notice that the form component $\lambda$ of $V$ appears only with its differential, so that $\tilde{V} = 0+ \diff\alpha$, $\alpha$ being an arbitrary $(n-3)$--form, is a trivial parameter. This is the effect of the reducibility of the gauge transformation. The twist operation is defined by
\begin{equation}
V = v + \lambda \mapsto e^A\,V = v + \lambda + \iota_v\,A.
\end{equation}
The following isomorphism
\begin{equation}
\wedge^{n-2}\,T^*\mathscr{M} \simeq \text{det}\,T^*\,\mathscr{M} \otimes \wedge^2\,T\mathscr{M}
\end{equation}
says that it is possible to replace the form component $\lambda$ of a generalised vector with a dual antisymmetric rank-two tensor density, according to 
\begin{equation}
\lambda^{mn} = \tfrac{1}{(n-2)!}\,\vepsilon^{mnp_1\dots p_{n-2}}\,\lambda_{p_1\dots p_{n-2}}.
\end{equation}
This suggests that a generalised vector index can be replaced by an antisymmetric rank-two index in $\text{GL}(n+1)$. When the indices take value from 1 to $n$, one gets the dual form component, and when one of the two indices take the last value, one gets the vector component:
\begin{equation}
V^M = V^{ij} = \begin{cases}
V^{m,n+1} = v^m, \\
V^{mn} = \lambda^{mn},
\end{cases}
\end{equation}
where $i,j=1,\dots,n+1$, $m,n=1,\dots,n$, and $V^{ij}=-V^{ji}$. In other words, one introduces a new generalised bundle
\begin{align}
W &\simeq (\text{det}\,T^*\mathscr{M})^{1/2}\otimes (T\mathscr{M} \oplus \wedge^n\,T\mathscr{M}) \simeq\nn\\
& \simeq (\text{det}\,T^*\mathscr{M})^{1/2}\otimes T\mathscr{M} \oplus (\text{det}\,T^*\mathscr{M})^{-1/2},
\end{align}
with elements $K = k + t$, $k$ being a vector density with weight $\frac{1}{2}$ and components $k^i$, and $t$ being a scalar density with weight $-\frac{1}{2}$, and then one takes the antisymmetrisation
\begin{align}
\wedge^2\,W &\simeq T\mathscr{M}  \oplus \text{det}\,T^*\mathscr{M} \otimes \wedge^2\,T\mathscr{M} \simeq \nn\\
& \simeq T\mathscr{M} \oplus \wedge^{n-2}\,T^*\mathscr{M} \simeq E.
\end{align}

In order to prove that $S^n$ is parallelisable in generalised sense it is sufficient to exhibit a generalised global, non-vanishing frame. In \cite{Lee:2014mla} it is proposed the following twisted frame:
\begin{equation}
\hat{E}_{ij} = e^A\,E_{ij} = u_{ij} + \sigma_{ij} + \iota_{v_{ij}}\,A,
\end{equation}
where $ij$ are antisymmetrised $\text{GL}(n+1)$ indices, $u_{ij}$ is the following vector field 
\begin{equation}
u_{ij} = R^{-1}\,(y_{i}\,\de_{j} -y_{j}\,\de_{i}), 
\end{equation}
and $\sigma_{ij}$ is the following $(n-2)$--form
\begin{equation}
\sigma_{ij} = {\star(R^2\,\diff y_i\,\diff y_j)} = \tfrac{R^{n-2}}{(n-2)!}\,\vepsilon_{ijk_1\dots k_{n-2}}\,y^{k_1}\,\diff y^{k_2}\dots \diff y^{k_{n-2}}.
\end{equation}
$\hat{E}_{ij}$ is globally defined. Moreover, $u_{ij}$ vanishes if and only if $y_a = y_b = 0$, whereas $\diff y_a\,\diff y_b$, and so $\sigma_{ab}$, vanish if and only if $y_a^2 + y_b^2 = 1$. Since the two conditions are not compatible, $u_{ab} + \sigma_{ab}$ never vanishes. Therefore, $S^n$ is parallelisable in generalised sense, that is, $T\,S^n \oplus \wedge^{n-2}\,T^*\,S^n$ is parallelisable.

Finally, as regards the torsion of the global frame, one can compute
\begin{align}
L_{\hat{E}_{ij}}\,\hat{E}_{kl} &= R^{-1}\,(\delta_{ik}\,\hat{E}_{lj} - \delta_{il}\,\hat{E}_{kj} - \delta_{jk}\,\hat{E}_{li} + \delta_{jl}\,\hat{E}_{ki}) = \nn\\
&= R^{-1}\,(\delta_{ik}\,\delta_{l}^{g}\,\delta_{j}^{h} - \delta_{il}\,\delta_{k}^{g}\,\delta_{j}^{h} - \delta_{jk}\,\delta_{l}^{g}\,\delta_{i}^{h} + \delta_{jl}\,\delta_{k}^{g}\,\delta_{i}^{h})\,\hat{E}_{gh} = -T_{ij}{}^{hg}{}_{kl}\,\hat{E}_{gh},
\end{align}
which says that the generalised torsion is constant and proportional to the structure constant of $\text{Lie}\,\text{SO}(n+1)$. Indeed, $S^n$ is equivalent to a coset space whose gauge group is $\text{SO}(n+1)$:
\begin{equation}
S^n \simeq \text{SO}(n+1)/\text{SO}(n).
\end{equation}

\subsection{BRST formalism for gauge theories} 

In \textsc{brst} formalism for gauge theories \cite{Becchi:1974md, Becchi:1974xu, Becchi:1975nq, Tyutin:1975qk}, the infinitesimal parameters of the gauge transformations are promoted to fields with opposite statistics with respect to the corresponding gauge symmetry: If the gauge symmetry is bosonic/fermionic, the ghost is an anticommuting/commuting field. An odd differential operator $s$ is introduced, generating all the gauge transformations of the theory, considered as global rotations in the space of fields and ghosts, evaluated in a given spacetime point:
\begin{equation}
s = \text{all infinitesimal gauge symmetries enjoyed by a theory},
\end{equation}
\begin{equation}
s\,\text{field} = \text{ghost}\,\times\,\text{field}.
\end{equation}
$s$ is required to be \emph{nilpotent}:
\begin{equation}
s^2 = 0.
\end{equation}
The nilpotency on the fields fixes the transformations of the ghosts:
\begin{equation}
s^2\,\text{field} = 0 \Rightarrow 
s\,\text{ghost} = \text{ghost}\,\times\,\text{ghost}.
\end{equation}
The fact that the ghost transformations are quadratic in the ghosts corresponds to the fact that the gauge transformations form an algebra, that is, the commutator of two transformations is a transformation too:
\begin{equation}
\text{The gauge transformations form an algebra} \Leftrightarrow s\,\text{ghost} = \text{ghost}\,\times\,\text{ghost}.
\end{equation}

The nilpotency of $s$ on the ghosts corresponds to the Jacobi identity of the algebra of gauge transformations.
\begin{equation}
\text{Jacobi identity} \Leftrightarrow s^2\,\text{ghost} = 0.
\end{equation}
Assuming the nilpotency of $s$ and the algebra of transformations, one can also deduce the transformations on the ghosts without knowing those of fields. Nevertheless, the nilpotency requirement is more general than the closure into an algebra: one can consider nilpotent \textsc{brst} transformations which do not form an algebra.\footnote{One can also weaken the nilpotency condition: this generalisation is studied by the BV formalism.}

The grading induced by $s$ is called \emph{ghost number}, since it counts the number of ghosts. $s$ increases the ghost number by one, in the same way in which the external differential increases by one the form degree. In \textsc{brst} formulation, the observables of a field theory are classified according to the $s$-cohomology $H_g(s)$ at a given ghost number $g$:
\begin{equation}
H_g(s) = \tfrac{\text{ker}_g\,s}{\text{im}_g\,s} 
\end{equation}
where $g(\psi)$ denotes the ghost number of $\psi$. This means that $H_g(s)$ is defined as the set of elements $\psi$ with ghost number $g(\psi)=g$ in the $s$ kernel $\text{ker}_g\,s$ ($s\,\psi = 0$) identifying elements which differ by an element in the $s$ image $\text{im}_g\,s$ (given $\psi,\psi'\in \text{ker}_g\,s$, $\psi \sim \psi' \Leftrightarrow \psi - \psi' = s\,\chi$, for some $\chi$).

We will denote with $H(s)$ the union of the cohomologies at a given ghost number over all the possible ghost numbers:
\begin{equation}
H(s) = \cup_g\,H_g(s).
\end{equation}
The cohomology is \emph{local} if any of its elements is an expression of the fields and of the ghosts of the theory and their derivatives, evaluated in the same spacetime point. 

Some definitions:
\begin{itemize}
\item[--] An \emph{action} $\Gamma$ is an element in the integrated local cohomology on fields without dependence on the ghosts (ghost number zero);
\item[--] The \emph{physical states} in the Fock space of the quantum field theory are the states in \textsc{brst} cohomology on the Fourier modes;
\item[--] The \emph{gauge-fixing}, required in a gauge theory to make the kinetic operator invertible, in order to compute propagators and in order to define the path integral, consists in adding to the action $s$-exact terms, breaking the gauge invariance but preserving the \textsc{brst} symmetry, in such a way that the gauge-fixed action sits in the same class of cohomology of the initial one:
\begin{equation}
\Gamma \rightarrow \Gamma + s\,\psi.
\end{equation}
\item[--] The action of a \emph{topological field theory} is $s$-exact, that is, cohomologically zero.
\end{itemize}

\subsection{Yang-Mills theory and BRST polyforms}\label{YMpolyform}

As an example, consider the Yang-Mills theory \cite{Yang:1954ek}. In Yang-Mills theory one considers the one-form connection and the ghost 
\begin{equation}
A = A_\mu^a\,\diff x^\mu\,T_a,  \quad c=c^a\,T_a,
\end{equation}
where $T_a$ are the generators of the gauge group. The field strength is defined by 
\begin{equation}
F=\diff\,A + A^2.
\end{equation}
$A$ and $F$ can be also seen as independent fields, and the previous expression can be thought as the definition of how $\diff$ algebraically acts on $A$: 
\begin{equation}
\diff\, A = F-A^2.
\end{equation}
The \textsc{brst} transformations for Yang-Mills theory are  \cite{Stora:1976kd, Stora:1976LM, Stora:1984} 
\begin{equation}
s\,c=-\tfrac{1}{2}\,[c,c] = -c^2, \quad 
s\,A = -\,\diff c - [A,c]= -\,\text{D}\,c,
\end{equation}
where $\text{D}$ is the covariant derivative and the commutator is in the gauge algebra. One can assume $s\,A$ as a definition, compute $s^2\,A = \text{D}\,(s\,c+\tfrac{1}{2}\,[c,c])$ and finding $s\,c$ by imposing $s^2\,A=0$. Then, one can check that $s^2\,c=0$ is equivalent to the Jacobi identity. The \textsc{brst} variation of $F$ is
\begin{equation}
s\,F=-[c,F],
\end{equation}
as one can check by using the definition of $F$ and assuming $\diff$ and $s$ to anticommute, since they act on local fields and they are both odd differentials
\begin{equation}
[s,\diff] = s\,\diff + \diff\,s = 0.
\end{equation}
Equivalently, if $A$ and $F$ are treated as independent, $s\,F$ is a definition, and one can algebraically deduce that $s$ and $\diff$ on the space $\{F,A,c,\diff c\}$ are both nilpotent and anticommuting. Moreover, using this approach, one can check the $\diff$-cohomology on $\{F,A\}$ to be trivial: this result, known as \emph{algebraic Poincaré lemma}, can be proved using the \emph{filtering theorem} \cite{Piguet:1995er}. 

The action is
\begin{equation}
S = -\int\diff^dx\,\tfrac{1}{4}\,F^{\mu\nu}\,F_{\mu\nu}.
\end{equation}
A covariant gauge-fixing is obtained by introducing a trivial doublet $(\overline{c},b)$
\begin{equation}
s\,\overline{c} = b, \quad s\,b = 0,
\end{equation}
taking value in the gauge algebra $b = b^a\,T_a$, $\overline{c} = \overline{c}^a\,T_a$, with $g(b)=0$, and so $g(\overline{c})=-1$. $b$ is called \emph{Nakanishi-Lautrup field}, and $\overline{c}$ is called \emph{antighost}. Consider the following gauge-fixed action
\begin{align}
S_{\text{gf}} &= -\tfrac{1}{4}\,F^{\mu\nu}\,F_{\mu\nu} + s\,(\overline{c}^a\,(\de_\mu\,A^\mu_a + \tfrac{1}{2}\,\xi\,b_a)) = \nn\\
&= -\tfrac{1}{4}\,F^{\mu\nu}\,F_{\mu\nu} + b^a\,\de_\mu\,A^\mu_a + \tfrac{1}{2}\,\xi\,b^a\,b_a) - \overline{c}^a\,\de_\mu\,D^\mu\,c^a,
\end{align}
where $\xi$ is an arbitrary constant.

The \textsc{brst} rules for Yang-Mills theory can be elegantly recasted in terms of the so-called \emph{polyforms} \cite{Stora:1976kd, Stora:1976LM, Stora:1984, Zumino:1983ew, Zumino:1983rz, Manes:1985df}. In particular, they are useful in computing anomalies, as it will be shown in Section \ref{BRSAnomalies}. One defines 
\begin{equation}\label{YMpoly}
\bm{A} = A + c, \quad
\delta = \diff + s, \quad
\end{equation}
We will call $\bm{A}$ Yang-Mills \emph{polyconnection}. The corresponding \emph{polycurvature} $\bm{F}$ is obtained by using $\bm{A}$ and $\delta$ in place of $A$ and $\diff$:
\begin{equation}
\bm{F} = \delta\, \bm{A} + \bm{A}^2.
\end{equation}
Since the algebraic relations satisfied by $\bm{A}$ and $\delta$ are formally the same as those satisfied by $A$ and $\diff$, we can immediately argue that $\bm{F}$ satisfies the analogue of the Bianchi identity $\diff\,F + [A,F]$, that is
\begin{equation}
\delta\,\bm{F} + [\bm{A},\bm{F}] = 0.
\end{equation}

Crucially, by expanding the polyform and using the \textsc{brst} rules, 
\begin{equation}\label{HorizontalityYM}
\bm{F} = F,
\end{equation}
and the opposite direction is true too. When a polyform has no nonvanishing components with nonvanishing ghost number, one says the polyform to be \emph{horizontal}. Therefore, horizontality condition on the Yang-Mills polycurvature \eqref{HorizontalityYM} is equivalent to the Yang-Mill \textsc{brst} rules:
\begin{equation}
s\,c = -c^2,\;\; s\,A = -\diff c - [A,c], \;\; F = \diff A + A^2 \Leftrightarrow \bm{F} = F.
\end{equation}

\subsection{Symmetries in gravity and supergravity}

A gravity theory is a theory enjoying diffeomorphism invariance. Diffeomorphisms are generated by the Lie derivative
\begin{equation}
s = \mathcal{L}_\xi,
\end{equation}
where $\xi = \xi^\mu\,\de_\mu$ is an anticommuting ghost. Given any covariant field $\vphi$,
\begin{equation}
s\,\vphi = \mathcal{L}_\xi\,\vphi.
\end{equation}
The requirement of the nilpotency of the previous transformation 
\begin{equation}
0 = s^2\,\vphi = s\,\mathcal{L}_\xi\,\vphi = 
\mathcal{L}_{s\,\xi}\,\vphi - \mathcal{L}_\xi^2\,\vphi =
\mathcal{L}_{s\,\xi - \frac{1}{2}\,\mathcal{L}_\xi\,\xi}\,\vphi
\end{equation}
fixes the \textsc{brst} transformation of the ghost:
\begin{equation}
s\,\xi^\mu = \tfrac{1}{2}\,\mathcal{L}_{\xi}\,\xi^\mu = \xi^\nu\,\de_\nu\,\xi^\mu,
\end{equation}
which is uniquely fixed since the Lie derivative has trivial kernel. This transformation is nilpotent, as one can check by explicit evaluation:
\begin{align}
s^2\,\xi^\mu &= s\,(\xi^\nu\,\de_\nu\,\xi^\mu) =
\xi^\vrho\,\de_\vrho\,\xi^\nu\,\de_\nu\,\xi^\mu - \xi^\nu\,\de_\nu\,(\xi^\vrho\,\de_\vrho\,\xi^\mu) = \nn\\
& = \xi^\nu\,\de_\nu\,\xi^\vrho\,\de_\vrho\,\xi^\mu 
- \xi^\nu\,\de_\nu\,\xi^\vrho\,\de_\vrho\,\xi^\mu 
- \xi^\nu\,\xi^\vrho\,\de_\nu\,\de_\vrho\,\xi^\mu = 0.
\end{align}

Let us deform the \textsc{brst} ghost transformation of a gravity theory by an arbitrary shift $\gamma^\mu$ \cite{
Becchi:1997jg, Imbimbo:2009dy}
\begin{equation}\label{sxideformed}
s\,\xi^\mu = \xi^\nu\,\de_\nu\,\xi^\mu + \gamma^\mu.
\end{equation}
The requirement of nilpotency fixes $\gamma^\mu$'s \textsc{brst} transformation:
\begin{align}
0 = &\, s^2\,\xi^\mu = s\,(\xi^\nu\,\de_\nu\,\xi^\mu + \gamma^\mu) = \nn\\
& = (\xi^\vrho\,\de_\vrho\,\xi^\nu + \gamma^\nu)\,\de_\nu\,\xi^\mu - \xi^\nu\,\de_\nu\,(\xi^\vrho\,\de_\vrho\,\xi^\mu + \gamma^\mu) + s\,\gamma^\mu = \nn\\
& = \gamma^\nu\,\de_\nu\,\xi^\mu - \xi^\nu\de_\nu\,\gamma^\mu + s\,\gamma^\mu = -\mathcal{L}_\xi\,\gamma^\mu + s\,\gamma^\mu \Leftrightarrow s\,\gamma^\mu = \mathcal{L}_\xi\,\gamma^\mu.
\end{align}

The deformation in \eqref{sxideformed} is useful in supergravity. A supergravity theory is a gravity theory with local supersymmetry. The supersymmetry ghost is a \emph{commuting} Majorana spinor $\zeta$, because supersymmetry is a fermionic symmetry (its parameter is a spinor and spinor are Grassmannian variables). We can decompose the \textsc{brst} operator of a supergravity theory in three pieces \cite{Imbimbo:2018duh, Bae:2015eoa}:
\begin{equation}
s = \mathcal{L}_\xi - \delta_c + S,
\end{equation}
where $\delta_c$ generates possible bosonic gauge transformations, with anticommuting ghost $c$, and $S$ generates supersymmetry:
\begin{equation*}
\delta_c = \text{bosonic gauge transformations, with anticommuting ghost}\;c,
\end{equation*}
\begin{equation*}
S = \text{supersymmetry, with commuting spinor ghost}\;\zeta.
\end{equation*} 
In general, $S$ is not required to be nilpotent. This reflects the fact that supersymmetry does not off-shell close into an algebra, but it closes either taking into account the equations of motion, or, if it is possible, by adding auxiliary fields. This last chance seems to be possible only in simple supergravity, where there is a single supersymmetry charge. In the case of extended supersymmetry, one has $\mathcal{N}$ supersymmetry generators and there is a ghost $\zeta_i$, $i=1,\dots,\mathcal{N}$, for each of them. There is no known example of extended supergravity in which the supersymmetry transformations off-shell close by means of a set of appropriate auxiliary fields.

If $e^a = e_\mu{}^a\,\diff x^\mu$ is the vielbein, and $\psi = \psi_\mu\,\diff x^\mu$ is the gravitino, then each supergravity theory must satisfy
\begin{equation}
S\,e^a = \overline{\zeta}\,\Gamma^a\,\psi,
\end{equation}
where $\Gamma^a$ are the Clifford matrices. Supersymmetry enters in the \textsc{brst} transformations as a topological deformation with a peculiar $\gamma^\mu$, as one can argue by looking at the commutator of two supersymmetry charges:
\begin{equation}
\{Q,Q\}\sim P \Leftrightarrow
\gamma^\mu = \tfrac{1}{2}\,\overline{\zeta}\,\Gamma^a\,\zeta\,e^\mu{}_a,
\end{equation}
where $Q$ and $P$ are the generators of supersymmetry and of translations. The \textsc{brst} transformation of $\zeta$ is fixed by knowning the transformations of $\gamma^\mu$ and of the vielbein:
\begin{equation}
s\,\gamma^\mu = \mathcal{L}_\xi\,\gamma^\mu \Leftrightarrow
S\,\gamma^\mu = 0 \Leftrightarrow
S\,\zeta = \iota_\gamma\,\psi,
\end{equation}
where $\gamma^\mu$ is assumed to be gauge invariant. Notice that $S\,\zeta$ depends on the gravitino and it is not quadratic in the ghosts. This is the first example we meet of a ghost transformation which depends on the fields. This reflects the fact that in general there is no underlying supersymmetry algebra.

Observe that \cite{Imbimbo:2018duh}
\begin{align}
S^2 &= (s-\mathcal{L}_\xi + \delta_c)^2 = s^2 + \mathcal{L}_\xi^2 + \delta_c^2 - [s,\mathcal{L}_\xi] + [s,\delta_c] - [\mathcal{L}_\xi,\delta_c] = \nn\\
& = \mathcal{L}_{\frac{1}{2}\,\mathcal{L}_{\xi}\xi} + \delta_{c^2} - \mathcal{L}_{s\,\xi} + \delta_{s\,c} - \delta_{\mathcal{L}_\xi\,c} = 
\mathcal{L}_\gamma + \delta_{s\,c+c^2-\mathcal{L}_\xi\,c},
\end{align}
which becomes 
\begin{equation}
S^2 = \mathcal{L}_{\gamma}\;\text{on gauge invariants expressions}.
\end{equation}

\subsection{Kalkman identity}

Consider the \textsc{brst} formulation of a (super)gravity theory, with anticommuting diffeomorphism ghost $\xi$, and \textsc{brst} transformation
\begin{equation}
s\,\xi = \tfrac{1}{2}\,\mathcal{L}_\xi\,\xi + \gamma.
\end{equation} 
Then, on the space of differential forms, \cite{Kalkman:1993zp}
\begin{equation}
e^{-\iota_\xi}\,(\diff + s)\,e^{\iota_\xi} = \diff + s -\mathcal{L}_\xi + \iota_\gamma,
\end{equation}
which we will called \emph{Kalkman identity}.

\bigskip

\emph{Proof}. Consider that  
\begin{equation}
\mathcal{L}_\xi = [\iota_\xi,\diff], \quad
[s,\iota_\xi] = \iota_{s\,\xi},\quad
[\mathcal{L}_\xi, \iota_\xi] = \iota_{\mathcal{L}_\xi\,\xi}, \quad
[\iota_\xi,\iota_\zeta]=0,
\end{equation}
for any anticommuting vectors $\xi,\zeta$. Then, one can compute
\begin{align}
& [\iota_\xi,\diff+s]= [\iota_\xi,\diff] - [s,\iota_\xi] = \mathcal{L}_\xi-\iota_{s\,\xi},\\
& [\iota_\xi,[\iota_\xi,\diff+s]]=[\iota_\xi,\mathcal{L}_\xi-\iota_{s\,\xi}]= -[\mathcal{L}_\xi,\iota_\xi] = -\iota_{\mathcal{L}_\xi\,\xi},\\
& [\iota_\xi,[\iota_\xi,[\iota_\xi,\diff+s]]]=-[\iota_\xi,\iota_{\mathcal{L}_\xi\xi}]=0,\\
& [\iota_\xi,\cdot]^{n>3}(\diff+s)=0.
\end{align}
Therefore, 
\begin{align}
e^{-\iota_\xi}\,(\diff + s)\,e^{\iota_\xi} &= e^{-[\iota_\xi,\cdot]}\,(\diff + s) = \nn\\
&= (\mathbbm{1} - [\iota_\xi,\cdot] + \tfrac{1}{2!}\,[\iota_\xi,[\iota_\xi,\cdot]] - \tfrac{1}{3!}\,[\iota_\xi,[\iota_\xi,[\iota_\xi,\cdot]]] + \dots)\,(\diff + s) = \nn\\
&= \diff + s - \mathcal{L}_\xi + \iota_{s\,\iota_\xi} - \tfrac{1}{2}\,\iota_{\mathcal{L}_\xi\,\xi} = \nn\\
&= \diff + s - \mathcal{L}_\xi + \iota_{s\,\xi - \frac{1}{2}\,\mathcal{L}_\xi\,\xi} = \diff + s - \mathcal{L}_\xi,
\end{align}
as we wanted to show. 

The Kalkman identity says that the ($\diff + s$)-cohomology on the space of ghosts and fields -- let it be $\mathscr{F}$ -- and the ($\diff + s - \mathcal{L}_\xi$)-cohomology on $\mathscr{F}\setminus\{\xi\}$ (undifferentiated $\xi$ is pulled out of $\mathscr{F}$) are isomorphic:
\begin{equation}
H_{\mathscr{F}}(\diff + s) \simeq
H_{\mathscr{F}\setminus\{\xi\}}(\diff + s - \mathcal{L}_\xi).
\end{equation} 
Indeed, if $Q \in H_{\mathscr{F}\setminus\{\xi\}}(\diff + s - \mathcal{L}_\xi)$, then $e^{\iota_\xi}\,Q \in  H_{\mathscr{F}}(\diff + s)$:
\begin{equation}
0 = e^{\iota_\xi}\,(\diff + s - \mathcal{L}_\xi)\,Q = (\diff + s)\,e^{\iota_\xi}\,Q.
\end{equation}
The multiplication by $e^{\iota_\xi}$ defines an isomorphism, since $e^{\iota_\xi}$ is invertible.

\newpage
\quad
\thispagestyle{empty}

\newpage

\section{Supergravity Compactifications}\label{1}

\subsection{Dualities} 

A transformation is a \emph{symmetry} for a field theory if it leaves its action $S$ invariant. If the transformation is a continuous one, and the infinitesimal variation is denoted with $\delta$, then 
\begin{equation}
\delta\,S = 0.
\end{equation}
Instead, a \emph{duality} is a correspondence between two field theories which preserves the equations of motion. Since the equations of motion are obtained from the first variation of the action, the invariance of the equations of motion under an infinitesimal continuous transformation makes the second variation of the action vanishing:
\begin{equation}
\delta^2\,S = 0.
\end{equation}
The simplest example of duality is the \emph{electromagnetic duality}: the Maxwell equations at vacuum
\begin{equation}
\diff F = 0, \quad \diff\,{\star F} = 0
\end{equation}
do not change if we rotate the field strength $F$ and its Hodge dual ${\star F}$ according to a constant $\text{U}(1)$ rotation 
\begin{equation}
F' + i\,{\star F'} = e^{i\,\vtheta}\,(F + i\,{\star F}).
\end{equation}
If $\vtheta = \frac{\pi}{2}$, $F$ becomes $-{\star F}$, and ${\star F}$ becomes $F$. This corresponds to reverse the electric and the magnetic field. 

There are two possible generalisation of the eletromagnetic duality, the \emph{Cremmer-Julia duality} and the \emph{Gaillard-Zumino duality}. The former is the extension of the electromagnetic duality to $p$-form fields $A^{(p)}$ \cite{Cremmer:1979up}. If $F^{(p+1)} = \diff A^{(p)}$ is the field strength, the equations of motion which generalise the Maxwell theory are
\begin{equation}
\diff\,F^{(p+1)} = 0, \quad \diff\,{\star F^{(p+1)}}= 0,
\end{equation}
The field strength and its Hodge dual have the same form degree in $d$ dimensions only in the case in which $p = \frac{d-2}{2}$, as in 4d Maxwell theory, where $p=1$. But we can nonetheless consider the discrete transformation swapping the field strength and its Hodge dual, which obviously preserves the equations of motion. In particular, if we set $ G^{(d-p-1)} = \star F^{(p+1)}$, they become
\begin{equation}
\diff\,{\star G^{(d-p-1)}}= 0, \quad \diff\,G^{(d-p-1)} = 0,
\end{equation}
which are the equations of motion of $(d-p-2)$--form $B^{(d-p-2)}$, since the second equation implies locally that $G^{(d-p-1)}$ is the field strength of a $(d-p-2)$--form $B^{(d-p-2)}$:
\begin{equation}
G^{(d-p-1)} = \diff B^{(d-p-2)}.
\end{equation}
This means that the $p$--form field theory in $d$ dimensions is dual to the $(d-p-2)$--form field theory in $d$ dimensions. Cremmer-Julia duality is important in supergravity, since the bosonic sector of maximal supergravities in dimensions lower than eleven, which are obtained by compactifying eleven-dimensional supergravity on hypertori, enjoys a global duality under $E_{11-d(11-d)}$ when the fields are expressed according the lowest form degree in the dualisation. For example, in four dimensions all the two-forms coming from the reduction of the bosonic fields of eleven-dimensional supergravity should be dualised in scalar fields (since two-form in four dimensions are dual to $4-2-2 = 0$ forms) in order to make the $E_{7(7)}$ duality manifest (see for example \cite{Cremmer:1997ct}).

Gaillard and Zumino considered in \cite{Gaillard:1981rj} (for a review see also \cite{Zumino:1981pt, Gaillard:1997rt}) a four-dimensional field theory whose field content is given by $n$ gauge potentials $A_\mu^a$, $a = 1,\dots, n$, and some additional fields $\chi^i$. Suppose that the Lagrangian density $\mathcal{L}$ depends on $A_\mu^a$ only through the corresponding field strength $F_{\mu\nu}^a = \de_\mu\,A_\nu^a - \de_\nu\,A_\mu^a$, and on $\chi^i$ only up to the first derivative $\chi_\mu^i := \de_\mu\,\chi^i$ (modulo integration by parts), in such a way that the equations of motion of $\chi^i$ depends only on $\chi_\mu^i$. Denote the components of the Hodge dual of the field strength with $\tilde{F}_{\mu\nu}^a$ and the dual of twice the derivative of the Lagrangian density with respect to $F_{\mu\nu}^a$ with $\tilde{G}_{\mu\nu}^a$:
\begin{equation}\label{Gdefinition}
\tilde{F}_{\mu\nu}^a = \tfrac{1}{2}\,\vepsilon_{\mu\nu\vrho\sigma}\,F^{\vrho\sigma\,a}, \quad
\tilde{G}^a_{\mu\nu} = \tfrac{1}{2}\,\vepsilon_{\mu\nu\vrho\sigma}\,G^{\vrho\sigma\,a} = 2\,\frac{\de\,\mathcal{L}}{\de\,F_{\mu\nu}^a}.
\end{equation}
The Bianchi identity and the equations of motion reads
\begin{equation}\label{GZEom}
\de^\mu\tilde{F}_{\mu\nu}^a = 0, \quad 
\de^\mu\tilde{G}_{\mu\nu}^a = 0.
\end{equation}
One can show that, if the fields $\chi^i$ are assumed to infinitesimally transform according to $\delta\,\chi^i =\xi^i(\chi)$, where $\xi^i$ are arbitrary functions of $\chi^i$ but not of $\chi_\mu^i$, then the most general rotation of the $F^a$'s and of the $G^a$'s preserving the equations \eqref{GZEom} is an element of the symplectic algebra $\text{Lie}\,\text{Sp}(2n,\mathbb{R})$, parametrised by 
\begin{equation}\label{ZGrotation}
\delta\begin{pmatrix} F \\ G \end{pmatrix} = \begin{pmatrix} A & B \\ C & -A^t \end{pmatrix} \begin{pmatrix} F \\ G \end{pmatrix},
\end{equation} 
where $A$ is an arbitrary $n\times n$ matrix, and $B$ and $C$ are $n \times n$ symmetric matrices.\footnote{The sympletic group is $\text{Sp}(2n,\mathbb{R}) = \lbrace
M \in \text{GL}(2n,\mathbb{R})\;|\; M^t\,\Omega_{2n}\,M = \Omega_{2n} \rbrace$, where $\Omega_{2n} = \begin{pmatrix} 0 & \mathbbm{1}_n \\ -\mathbbm{1}_n & 0 \end{pmatrix}$. Its algebra is $\text{Lie}\,\text{Sp}(2n,\mathbb{R}) = \{X\in\text{Mat}_{2n}\,|\,(\Omega_{2n}\,X)^t = \Omega_{2n}\,X\}$, so that $X$ is parametrised as the matrix above.} Moreover, one can show that, even if the Lagrangian density is not invariant in general, its derivative with respect to an invariant parameter or background field is invariant, so that the stress-energy tensor, which is given by the variation of  the action with respect to the background metric, is invariant under duality rotation. Finally, the scalar fields $\chi^i$ take values in the coset space $\text{Sp}(2n,\mathbb{R})/\text{U}(n)$, where $\text{U}(n)$ is the maximal compact subgroup of $\text{Sp}(2n,\mathbb{R})$. 

\subsection{Proof of Gaillard-Zumino duality}

The equations of motion of $A_\mu^a$ are given by the second equation in \eqref{GZEom} since, using the hypothesis on the dependence on $A_\mu^a$ in the Lagrangian density and the definition in \eqref{Gdefinition},
\begin{equations}
\frac{\delta\,S}{\delta\,A_\nu^a} &= \frac{\de\,\mathcal{L}}{\de\,A_\nu^a} - \de_\nu\,\frac{\de\,\mathcal{L}}{\de\,\de_\mu\,A_\nu^a} = - \de_\mu\,\frac{\de\,\mathcal{L}}{\de\,\de_\mu\,A_\nu^a} = -\de_\mu\,\big{(}\frac{\de\,\mathcal{L}}{\de\,F_{\vrho\sigma}^b}\,\frac{\de\,F_{\vrho\sigma}^b}{\de\,\de_\mu\,A_\nu^a}\big{)} = \nn\\
&= -\de_\mu\,\big{(}\frac{\de\,\mathcal{L}}{\de\,F_{\vrho\sigma}^b}\,\delta^\mu_{[\vrho}\,\delta^\nu_{\sigma]}\,\delta^a_b\big{)} = -2\,\de_\mu\,\big{(}\frac{\de\,\mathcal{L}}{\de\,F_{\mu\nu}^a}\big{)} = -\de_\mu\tilde{G}^{a\mu\nu}.
\end{equations}
In order to find the most general rotation which preserves the equations of motion, we have to compute the \emph{second} variation of the action, imposing it to vanish, because in general the action could not be invariant. The first variation of the action is
\begin{align}
\delta\,S &= \frac{\delta S}{\delta \chi^i}\,\delta\,\chi^i + \frac{\de S}{\de F^a}\,\delta\,F^a = \frac{\de\,\mathcal{L}}{\de \chi^i}\,\delta\,\chi^i + \frac{\de\,\mathcal{L}}{\de \chi_\mu^i}\,\delta\,\chi^i_\mu + \frac{\de\,\mathcal{L}}{\de F^a}\,\delta\,F^a = \nn\\
& = \left[\xi^i\,\frac{\de}{\de \chi^i} + \chi^j_\mu\,\frac{\de \xi^i}{\de \chi^j}\frac{\de}{\de \chi_\mu^i} + (A^b{}_c\,F^c + B^b{}_c\,G^c)\,\frac{\de}{\de F^b}\right]\mathcal{L}.
\end{align}
The second variation is
\begin{equation} \label{delta2L}
\delta^2\,S = \delta\,(\delta\,S) = 
\frac{\de\,\delta\,S}{\de F^a}\,\delta\,F^a + \frac{\delta\,\delta\,S}{\delta \chi^i}\,\delta\,\chi^i,
\end{equation}
where the first and the second piece can be computed to be equal to
\begin{align}
2\,\frac{\de\,\delta\,S}{\de F^a} &= 
2\,(D_{ab} + A_{ba})\,\frac{\de\,\mathcal{L}}{\de\,F^b} + \frac{1}{2}\,\frac{\de\,G^c}{\de\,F^a}\,(B_{bc}-B_{cb})\,\tilde{G}^b \,+\nn\\
& + \frac{1}{2}\,(C_{ab}-C_{ba})\,\tilde{F}^b + \frac{1}{2}\,\frac{\de}{\de F^a}\,(F\,C\,\tilde{F} + G\,B^t\,\tilde{G}),\label{FirstTerm}\\
\frac{\delta\,\delta\,S}{\delta \chi^i} &= \frac{\de\,\xi^j}{\de\,\chi^i}\,E_j + \delta\,E_i + \frac{1}{2}\,\frac{\de \,G^c}{\de\,\chi^i}\,B^{bc}\,\tilde{G}^b -\frac{1}{2}\,\de_\mu\left(\frac{\de\,G^c}{\de\,\chi_\mu^i}\,B^{bc}\,\tilde{G}^b\right), \label{SecondTerm}
\end{align}
where $E_i := \frac{\delta S}{\delta \chi^i}$, and $F\,C\,\tilde{F}$ is a shorthand for $F_{\mu\nu}^a\,C_a{}^b\,\tilde{F}^{\mu\nu}_b$. Since the left-hand side of \eqref{FirstTerm} is a derivative, the right-hand side must be a derivative too, so that we have to impose
\begin{equation}\label{ZGCond1}
B = B^t, \quad
C = C^t, \quad 
D^{ba} + A^{ba} = \tfrac{1}{4}\,\vepsilon\,\delta^{ab},
\end{equation}
for some constant $\vepsilon$ (the factor $\frac{1}{4}$ is for convenience). $E_i$ can be treated as a one-form in the $\chi^i$ coordinates. The covariance of $E_i$ means that under the infinitesimal diffeomorphism $\delta\,\chi^i = \xi^i$, it must transform according to minus the Lie derivative generated by the vector field $\xi^i$:
\begin{equation}\label{ZGCond2}
\delta\,E_i = -\mathcal{L}_\xi\,E_i = -\xi^j\,\frac{\de}{\de \vphi^j}\,E_i - \frac{\de \xi^j}{\de \chi^i}\,E_j = - \frac{\de \xi^j}{\de \chi^i}\,E_j,
\end{equation}
where the first term vanishes since the equation of motion does not depend on $\chi^i$, but only on its derivatives. Therefore, using \eqref{ZGCond1} and \eqref{ZGCond2} in \eqref{FirstTerm}--\eqref{SecondTerm}, one gets
\begin{align}
4\,\frac{\de\,\delta\,S}{\de\,F^a} &= \frac{\de}{\de\,F^a}\,(F\,C\,\tilde{F} + G\,B\,\tilde{G} + \vepsilon\,\mathcal{L}),\\
4\,\frac{\delta\,\delta\,S}{\delta\,\chi^i} &= 
\frac{\delta}{\delta\,\chi^i}\,(F\,C\,\tilde{F} + G\,B\,\tilde{G}).
\end{align}
In the second piece it was possible to include the term $\frac{1}{4}\,F\,C\,\tilde{F}$, since it is constant with respect to $\chi^i$ and $\chi^i_\mu$. Replacing in \eqref{delta2L},
\begin{align}
4\,\delta\,(\delta\,\mathcal{L}) &= \delta\,(F\,C\,\tilde{F} + G\,B\,\tilde{G}) + \vepsilon\,\delta\,F^a\,\frac{\de\,\mathcal{L}}{\de F^a},
\end{align}
so that $\vepsilon = 0$. Finally, back again in  \eqref{ZGCond1},
\begin{equation}
B=B^t, \quad C=C^t, \quad D = - A^t.
\end{equation} 
Therefore, the rotation in \eqref{ZGrotation} is an element in the algebra of the symplectic group.

\subsection{Eleven-dimensional supergravity} 

The eleven-dimensional supergravity, discovered by Cremmer, Julia and Scherk in 1978 \cite{Cremmer:1978km,}, has a very simple field content:
\begin{equation}
g_{\hat{\mu}\hat{\nu}}, \quad A^{(3)}, \quad \psi_{\hat{\mu}},
\end{equation}
where $\hat{\mu}, \hat{\nu} = 0, \dots, 10$. $g_{\hat{\mu}\hat{\nu}}$ is the metric tensor, $A^{(3)}$ is a three-form field, and $\psi_{\hat{\mu}}$ is vector-valued Majorana spinor (gravitino). The number of on-shell degrees of freedom are \footnote{In arbitrary $d$ dimensions, the number of on-shell degrees of freedom for each field is
\begin{equation}
\#(g_{\hat{\mu}\hat{\nu}}) = \frac{d\,(d-3)}{2},\quad
\#(A^{(3)}) = {{d-2}\choose{p}},\quad
\#(\psi_{\hat{\mu}}) = 2^{\lfloor d/2 \rfloor -1 -\alpha}\,(d-3),
\end{equation}
where $\alpha = 0, 1$ if the gravitino is Majorana spinor, or a Majorana-Weyl spinor. See next section for details.} 
\begin{equation}
\#(g_{\hat{\mu}\hat{\nu}}) = 44, \quad
\#(A^{(3)}) = 84, \quad
\#(\psi_{\hat{\mu}}) = 128,
\end{equation}
so that the number of bosonic and fermionic degrees of freedom match ($44+84=128$), as required in any supersymmetric theory. The action reads
\begin{align}
S &= \frac{1}{2\kappa^2}\int\diff^{11} x\,\hat{e}\bigg{[}
\hat{\mathscr{R}} - \tfrac{1}{24}\,F^{\hat{\mu}\hat{\nu}\hat{\vrho}\hat{\sigma}}\,F_{\hat{\mu}\hat{\nu}\hat{\vrho}\hat{\sigma}} 
- \bar{\psi}_{\hat{\mu}}\,\Gamma^{\hat{\mu}\hat{\nu}\hat{\vrho}}\,D_{\hat{\nu}}\,\psi_{\hat{\vrho}} \,+\nonumber \\
& - \frac{\sqrt{2}}{192}\,\bar{\psi}_{\hat{\mu}}\,\left(\Gamma^{\hat{\alpha}\hat{\beta}\hat{\gamma}\hat{\delta}\hat{\mu}\hat{\nu}}+12\,\Gamma^{\hat{\alpha}\hat{\beta}}\,g^{\hat{\gamma}\hat{\mu}}\,g^{\hat{\delta}\hat{\nu}}\right)\,\psi_{\hat{\nu}}\,\left(F_{\hat{\alpha}\hat{\beta}\,\hat{\gamma}\hat{\delta}} + \tilde{F}_{\hat{\alpha}\hat{\beta}\hat{\gamma}\hat{\delta}}\right)\,+\nonumber \\
& - \frac{\sqrt{2}}{(144)^2}\,\vepsilon^{\hat{\alpha}_1\dots\hat{\alpha}_4\hat{\beta}_1\dots\hat{\beta}_4\hat{\mu}\hat{\nu}\hat{\vrho}}\,F_{\hat{\alpha}_1\dots\hat{\alpha}_4}\,F_{\hat{\beta}_1\dots\hat{\beta}_4}\,A_{\hat{\mu}\hat{\nu}\,\hat{\vrho}}
\bigg{]},\label{11dSugraAction}
\end{align}
where $\kappa$ is the gravitational coupling constant, which in eleven dimensions has mass dimension $[\kappa] = -9/2$; $F_{\hat{\mu}\hat{\nu}\hat{\vrho}\hat{\sigma}}$ are the components of the four-form field strength of three-form gauge potential $F^{(4)} = \diff A^{(3)}$; the covariant derivative $D_{\hat{\mu}}$ is computed using $(\omega + \tilde{\omega})/2$ as spin connection, $\omega$ being the spin connection with vanishing supersymmetrised torsion; the tilde denotes supercovariantization (one adds fermionic terms in such a way that the supersymmetric variation does not depend on the derivative of the supersymmetry parameter). The action \eqref{11dSugraAction} is invariant under diffeomorphisms, local Lorentz transformations, local supersymmetry, and three-form gauge transformations, the last one amounting
\begin{equation}\label{GaugeThreeForm}
\delta\,A^{(3)} = \diff\Lambda^{(2)}.
\end{equation} 

The action was obtained by using the so-called ``Noether method". One starts with the combination of the Einstein-Hilbert action, the Rarita-Schwinger action, and Maxwell action for $A^{(3)}$. The supersymmetry variation of the first two terms vanishes, as in simple supergravity. To compensate the variation of the last term, one adds a term of the type $\psi\,X\,\psi\,F$, for some $X$, and one deforms the  supersymmetry variation on $\delta\,\psi_{\hat{\mu}} = (D_{\hat{\mu}} + (Z\,F)_{\hat{\mu}})\,\zeta$, for some $Z$. Then, one requires the gravitino equations of motion to be  superconvariant: $\Gamma^
{\hat{\mu}\hat{\nu}\hat{\vrho}}\,(D_{\hat{\mu}} + (Z\,F)_{\hat{\mu}})\,\psi_{\hat{\vrho}}=0$. In this way, $X$ and $Z$ are fixed, modulo a coefficient. This coefficient can be fixed in such a way that all of the terms in the variation of the action drop out, except for a term with nine $\Gamma$-matrices. To cancel out this term one has to include a Chern-Simons-like term, whose variation on this term suggests to set the following supersymmetry variation of $A^{(3)}$: $\delta\,A_{\hat{\mu}\hat{\nu}\hat{\vrho}} = b\,\bar{\zeta}\,\Gamma_{[\hat{\mu}\hat{\nu}}\psi_{\hat{\vrho}]}$.  The coefficient $b$ is fixed in such a way that $\bar{\zeta}\,\de\,\psi\,F$ and $\bar{\zeta}\,\psi\,\de\,F$ cancel out. At this stage, it remains to cancel out trilinear and quadrilinear terms in the gravitino. To do this, one replaces $F$ and $\omega$ in $\delta\,\psi$ by the supercovariant versions $\tilde{F}$ and $\tilde{\omega}$.

Using differential forms, the bosonic part of the action \eqref{11dSugraAction} can be written as
\begin{equation}
S = \frac{1}{2\,\kappa^2}\,\int \diff^{11}x\,\hat{e}\,\hat{\mathscr{R}} - \frac{1}{4\,\kappa^2}\int F^{(4)}\star F^{(4)} - \frac{1}{12\,\kappa^2}\int A^{(3)}\,F^{(4)}\,F^{(4)}.
\end{equation}
Because of the presence of the last term -- which is known as Chern-Simons term, since it is background independent -- the equations of motion  $A^{(3)}$ are
\begin{equation}\label{Aeoms}
\diff\,F^{(4)} = 0, \quad 
\diff\,{\star F^{(4)}} - \tfrac{1}{2}\,(F^{(4)})^2 = 0.
\end{equation}
The first one is a Bianchi identity, which locally implies the definition of the four-form field strength as external differential of the three-form gauge potential. Denoting the Hodge dual of the field strength with $G^{(7)} = {\star F^{(4)}}$, the equations \eqref{Aeoms} becomes
\begin{equation}
\diff\,{\star G^{(7)}} = 0, \quad
\diff\,G^{(7)} = \tfrac{1}{2}\,(F^{(4)})^2.
\end{equation}
According to Cremmer-Julia duality, the Bianchi identity and the equations of motion swap r\^oles upon dualising. Namely, the second equation can be viewed as a Bianchi identity, since it is solved by
\begin{equation}
G^{(7)} = \diff\,\tilde{A}^{(6)} - \tfrac{1}{2}\,A^{(3)}\,F^{(4)},
\end{equation}
where $\tilde{A}^{(6)}$ is the dual six-form potential of the three-form one. The gauge transformation on the dual potential, which leaves field strength invariant, is
\begin{equation}
\delta\,\tilde{A}^{(6)} = \diff\,\Lambda^{(5)} + \tfrac{1}{2}\,\Lambda^{(2)}\,F^{(4)},
\end{equation}
where $\Lambda^{(2)}$ is the same two-form parameter appearing in \eqref{GaugeThreeForm}.

\subsection{Counting the degrees of freedom of fields}\label{Counting}

In this section we summarise the counting of the number of  degrees of freedom of various fields in arbitrary spacetime dimension. 

As degrees of freedom one means the transverse modes of the linearised equations of motion in a field theory \cite{Julia:1980gr}. When one considers a gauge field, the number of degrees of freedom is given by the number of components of the corresponding irreducible tensor in $d-2$ dimensions (the number of transverse directions in $d$ dimension: one discharges time and longitudinal direction to the wave propagation). For example, a one-form gauge field $A_\mu$, with gauge redundancy $\delta\,A_\mu = \de_\mu\,\alpha$, has $d$ components in $d$ dimensions, so the number of degrees of freedom is $d-2$. This result can be extended to arbitrary $p$-form gauge fields $A^{(p)}$. Since a $p$--form is an totally rank-$p$ antisymmetric tensor, it has $\binom{d}{p}$ components, so that the number of degrees of freedom is $\binom{d-2}{p}$. A symmetric transverse, rank two tensor $h_{\mu\nu}$ describes a massless spin-two particle, if $d$ components are fixed, due to the gauge redundancy $\delta\,h_{\mu\nu} = \de_{(\mu}\,\xi_{\nu)}$. So, the number of degrees of freedom is $\frac{d(d+1)}{2}-d-d=\frac{d(d-3)}{2}$, which indeed is the number of components of a \emph{traceless}, symmetric, rank-two tensor in $d-2$ dimensions, since $\frac{d(d+1)}{2}-1\,|_{d \rightarrow d-2} = \frac{d(d-3)}{3}$.

Consider spinor fields $d=2n$ dimensions. A Dirac spinor $\lambda$ is an element in the spinor representation of $\text{SO}(2n)$ (or $\text{SO}(2n-1,1)$ in Minkowski signature), with dimension $2^n$. Its charge conjugate spinor is defined by $\lambda^c := B^{-1}\,\lambda^*$, where $B$ is the matrix which implements the automorphism $-(\Gamma^\mu)^*$ of the Clifford algebra. If 
\begin{equation}
\lambda^c = \lambda,
\end{equation}
which is the equivalent of the reality condition, $\lambda$ is called \emph{Majorana spinor}. A Majorana spinor is the natural supersymmetric partner of a real bosonic field. $B$ is such that $B^*\,B=\vepsilon\,\mathbbm{1}$, where $\vepsilon = \pm 1$. Majorana spinors exist if and only $\vepsilon = 1$. Futhermore, $\vepsilon = 1$ if and only if a rapresentation of Clifford algebra exists such that all of the $\Gamma^\mu$'s are imaginary. Majorana found such a representation in $d=4$ \cite{Majorana:1932ga}. In general, Gliozzi, Olive, and Scherk proved that $\vepsilon$ depends on the dimension of spacetime \cite{Gliozzi:1976qd}:
\begin{equation}
\vepsilon = \begin{cases} +1 \quad \text{if}\;d=2,4\;\text{mod}\,8,\\
-1\quad\text{if}\;d=6,8\;\text{mod}\,8.
\end{cases}
\end{equation}
The chirality matrix $\Gamma_{d+1}$ is a matrix which anticommutes with all of the $\Gamma^\mu$'s and whose square is $\mathbbm{1}$. So, its eigenvalues are $\pm 1$. It is the matrix one has to add in order to get the Clifford algebra in $d+1$ dimensions. So, a chirality matrix makes sense only in even dimensions. If a Dirac spinor is an eigenvector of $\Gamma_{d+1}$ 
\begin{equation}
\Gamma_{d+1}\,\lambda = \pm\,\lambda,
\end{equation}
we say that $\lambda$ is a \emph{Weyl spinor} (left-handed if the eigenvalue is negative, right-handed if it is positive). One can prove that Majorana-Weyl spinors exist if and only if
\begin{equation}
d = 2\;\text{mod}\,8 = 2, 10, 18, \dots
\end{equation} 
A Dirac spinor has $2^{d/2+1}$ real components. Majorana condition halves it, so a Majorana spinor has $2^{d/2}$ real components. Weyl condition further halves the components, so a Majorana-Weyl spinor has $2^{d/2-1}$ real components. $\lambda$ satisfies on-shell the Dirac equation
\begin{equation}
\Gamma^\mu\de_\mu\,\lambda = 0.
\end{equation}
In Weyl representation, the Dirac equation allows to write half of the components of $\lambda$ in terms of the other half. These components satisfy the wave equation. Indeed, if we square Dirac operator and use Clifford algebra, we obtain the wave operator $\Ga^\mu\Ga^\nu\de_\mu\de_\nu = \de^2$. Therefore, the number of degrees of freedom is 
\begin{equation}
\#(\lambda) = \begin{cases}
2^{d/2-1}\quad\text{if}\;\lambda\;\text{Majorana or Weyl spinor},\\
2^{d/2-2}\quad\text{id}\;\lambda\;\text{Majorana-Weyl spinor}.
\end{cases}
\end{equation}
A gravitino is a Majorana spinor (or Majorana-Weyl spinor, if possible) with a vector index $\psi_\mu$. The gauge redundancy $\delta\,\psi_\mu = \de_\mu\,\zeta$ leaves invariant the kinetic term of Rarita--Schwinger Lagrangian $\psi_\mu\,\Ga^{\mu\nu\vrho}\,\de_\nu\,\psi_\vrho$ \cite{Rarita:1941mf}, where $\zeta$ is a Majorana(--Weyl) spinor.  We may expect that the number of degrees of freedom is $2^{d/2-2}(d-2)$ ($\#$ Majorana--Weyl spinor $\times$ $\#$ gauge vector field). But the representation of $\psi_\mu$ is too big. For example, if four dimensions, it is $\frac{1}{2}\otimes 1$, which decomposes according to $ \frac{3}{2} \oplus \frac{1}{2}$, and only the $\frac{3}{2}$ component describes the gravitino. So, we have to introduce a gauge-fixing condition in order to eliminate spin $\frac{1}{2}$ component. We could use
\begin{equation}
\Ga^\mu\psi_\mu = 0.
\end{equation}
Indeed, again in four dimensions for example, this constraint imposes $2^{4/2-1}=2$ conditions, the same as the dimension of Dirac $\frac{1}{2}$ representation. In general, the gauge-fixing condition imposes $2^{d/2-2}$ (if Majorana-Weyl spinor is considered). In conclusion, we have 
\begin{equation}
\#(\psi_\mu) = \begin{cases}
2^{d/2-1}(d-3)\quad\text{if}\,\psi_\mu\,\text{Majorana gravitino},\\
2^{d/2-2}(d-3)\quad\text{if}\,\psi_\mu\,\text{Majorana-Weyl gravitino}.
\end{cases}
\end{equation}

In summary ($\alpha = 0, 1$ if the spinor is Majorana or Majorana--Weyl):
\begin{table}[H]
\begin{tabular}{lll}
$\qquad$ & Field & $\#$ \\
$\qquad$ & $\quad$ & $\quad$ \\
$\qquad$ & $A^{(p)}$ & $\binom{d-2}{p}$ \\
$\qquad$ & $\quad$ & $\quad$ \\
$\qquad$ & $h_{\mu\nu}$ & $\frac{d(d-3)}{2}$ \\
$\qquad$ & $\quad$ & $\quad$ \\
$\qquad$ & $\lambda$ & $2^{d/2-1-\alpha}$ \\
$\qquad$ & $\quad$ & $\quad$ \\
$\qquad$ & $\psi_\mu$ & $2^{d/2-1-\alpha}(d-3)$ 
\end{tabular}
\end{table}

\subsection{Trombone duality}

Another example of duality is the invariance of Einstein equations under a global Weyl scaling of the metric:
\begin{equation}\label{WeylScaling}
g_{\mu\nu} \rightarrow e^{2\,\sigma}\,g_{\mu\nu},
\end{equation}
Correspondingly, the Hilbert-Einstein action scales homogeneously, according to 
\begin{equation}
e\,R \rightarrow e^{(d-2)\,\sigma}\,e\,R.
\end{equation}
This duality explains why the solutions of Einstein equations depend on free parameters \cite{Cremmer:1997xj}. For example, the mass is left free in the Schwarzschild solution. Indeed, if we rescale the Schwarzschild solution with mass $M$, according to \eqref{WeylScaling}, we obtain again a Schwarzschild solution, but with a rescaled mass $e^{\sigma}\,M$, so a global Weyl scaling of the Schwarzschild solution is the same as changing the mass.   
\begin{align}
e^{2\,\sigma}\,&\,[-(1-\tfrac{2\,G\,M}{r})\,\diff\,t^2 + (1-\tfrac{2\,G\,M}{r})^{-1}\,\diff\,r^2 + r^2\,\diff\,\Omega_2^2] = \nn\\
&=-(1-\tfrac{2\,G\,(e^\sigma\,M)}{e^\sigma\,r})\,\diff\,(e^\sigma\,t)^2 + (1-\tfrac{2\,G\,(e^\sigma\,M)}{e^\sigma\,r})^{-1}\,\diff\,(e^\sigma\,r)^2 + (e^\sigma\,r)^2\,\diff\,\Omega_2^2.
\end{align}
The same phenomenon occurs in supergravity, if the gravitini and the $p$--form fields are suitably rescaled. In particular, in the case of eleven-dimensional supergravity, if 
\begin{equation}
g_{\hat{\mu}\hat{\nu}} \rightarrow  e^{2\,\sigma}\,g_{\hat{\mu}\hat{\nu}}, \quad
A_{\hat{\mu}\hat{\nu}\hat{\vrho}} \rightarrow e^{3\,\sigma}\,A_{\hat{\mu}\hat{\nu}\hat{\vrho}}, \quad
\psi_{\hat{\mu}} \rightarrow  e^{\sigma/2}\,\psi_{\hat{\mu}},
\end{equation}
then the action \eqref{11dSugraAction} is homogeneously rescaled 
\begin{equation}\label{GlobScal11dSugra}
S \rightarrow e^{(11-2)\,\sigma}\,S,
\end{equation}
and, as a consequence, the equations of motion are left invariant. This duality of eleven-dimensional supergravity is called \emph{trombone duality}.\footnote{“Because they allow one to scale magnitudes in and out, we shall call this scaling symmetries ‘trombone’ symmetries” \cite{Cremmer:1997xj}.} More in details, the scaling \eqref{GlobScal11dSugra} implies that
\begin{equations}
& \omega \rightarrow \omega, \\
& \hat{e} \rightarrow e^{11\,\sigma}\,\hat{e}, \\
& \hat{R} \rightarrow e^{-2\,\sigma}\,\hat{R},\\
& F^{(4)} \rightarrow e^{3\,\sigma}\,F^{(4)}, \\
& {\star F^{(4)}} \rightarrow e^{(11+3-2\cdot 4)\,\sigma}\,{\star F^{(4)}},\\
& \Gamma_{\hat{\mu_1}\dots\hat{\mu}_r} \rightarrow e^{r\,\sigma}\,\Gamma_{\hat{\mu_1}\dots\hat{\mu}_r}, \\
& \Gamma^{\hat{\mu_1}\dots\hat{\mu}_r} \rightarrow e^{-r\,\sigma}\,\Gamma^{\hat{\mu_1}\dots\hat{\mu}_r}.
\end{equations}
The spin connection is invariant since $\omega \sim e^{-1}\de\,e$; the $\Gamma$-matrices with curved indices are rescaled because they contain the metric $\Gamma_\mu = e_\mu{}^a\,\Gamma_a$, $\Gamma^\mu = e^\mu{}_a\,\Gamma^a$; the scaling of the gravitino ensures that the fermionic correction in $\tilde{\omega}$ and in $\tilde{F}$ are consistent with the scaling of $\omega$ and $F$ respectively. Therefore, the variation of each of the terms in the action \eqref{11dSugraAction} are:
\begin{equations}
& \hat{e}\,\hat{R} \rightarrow e^{(11-2)\,\sigma}\,\hat{e}\,\hat{R},\\
& F^{(4)}{\star{F^{(4)}}} \rightarrow e^{(3+6)\,\sigma}\,F{\star{F}},\\
& F^{(4)}\,F^{(4)}\,A^{(3)}  \rightarrow e^{(3+3+3)\,\sigma}\,F^{(4)}\,F^{(4)}\,A^{(3)}, \\
& \hat{e}\,\overline{\psi}\,\Gamma^{(3)}\,D\,\psi \rightarrow 
e^{(11+1/2-3+1/2)\sigma}\,\hat{e}\,\overline{\psi}\,\Gamma^{(3)}\,D\,\psi,\\
& \hat{e}\,\overline{\psi}\,\Gamma^{(6)}\,\psi\,F \rightarrow
e^{(11+1/2-6+1/2+3)\sigma}\,\hat{e}\,\overline{\psi}\,\Gamma^{(6)}\,\psi\,F,\\
& \hat{e}\,\overline{\psi}\,\Gamma^{(2)}\,g^{-1}\,g^{-1}\,\psi\,F \rightarrow e^{(11+1/2-2-2\cdot 2 + 1/2 + 3)\sigma}\,\hat{e}\,\overline{\psi}\,\Gamma^{(2)}\,g^{-1}\,g^{-1}\,\psi\,F,
\end{equations} 
that is, all the terms scale with the weight is $e^{9\,\sigma}$, justifying \eqref{GlobScal11dSugra}. 

\subsection{Kaluza-Klein compactification of scalar theory}

Consider a $\hat{d}$-dimensional theory. Compactify $n$ dimensions in a hypertorus $T^n$. Denote $x^{\hat{\mu}} = (x^\mu, y^m)$, with $\hat{\mu} = 0, \dots, \hat{d}-1$, $\mu = 0, \dots, d-1$, $m=1,\dots, n$. The periodicity in the internal coordinates $y^m$ dictates how the fields depend on such coordinates: the dependence on the internal coordinates is given by the Fourier modes. There is a field tower in the $d$-dimensional theory (\emph{Kaluza-Klein tower}) for each field in the $\hat{d}$-dimensional theory. Selecting the massless modes is the same as requiring that there is no dependence on the internal coordinates in the fields of the $d$-dimensional theory
\begin{equation}\label{KKcondition}
\de_m = 0.
\end{equation} 
(\emph{Kaluza-Klein truncation}). As a simple example of Kaluza-Klein compactification, consider a theory of a massless real scalar field $\vphi$ in $\hat{d}$-dimensions, whose action is the Klein-Gordon action:
\begin{equation}\label{ActionExKK}
S = -\frac{1}{2}\int \diff^{\hat{d}}x\,\de_{\hat{\mu}}\,\vphi\,\de^{\hat{\mu}}\,\vphi.
\end{equation}
The mass dimension of the scalar field is $[\vphi] = \frac{\hat{d}-2}{2}$, in such a way that the action is dimensionless. Let us compactify the last dimension in a circle with length $\ell$. The coordinate on the circle is periodic with period $\ell$: 
\begin{equation}
y \sim y+ n\,\ell.
\end{equation}
This implies that the dependence on $y$ in the scalar field is given by
\begin{equation}
\vphi(x,y) = \ell^{-1/2}\sum_{n\in \mathbb{Z}} \vphi_n(x)\,e^{2\pi i n y / \ell},
\end{equation}
where the factor $\ell^{-1/2}$ ensures the modes $\vphi_n$ to have mass dimension $\frac{d-2}{2}$. Since
\begin{equation}
\vphi^*(x,y) = \ell^{-1/2}\sum_{n \in \mathbb{Z}}\vphi_{-n}^*(x)\,e^{2 \pi i n y /\ell},
\end{equation}
The reality condition of the scalar field implies 
\begin{equation}\label{RealityConditionPhi}
\vphi_{-n} = \vphi_{n}^*.
\end{equation}
Moreover, we can compute
\begin{equations}
& \de_\mu\,\vphi(x,y) = \ell^{-1/2}\sum_{n\in\mathbb{Z}} \de_\mu\,\vphi_n(x)\,e^{2 \pi i n y /\ell}, \\
& \de_y\,\vphi(x,y) = \ell^{-1/2}\,\frac{2\,\pi\,i}{\ell}\sum_{n\in\mathbb{Z}} n\,\vphi_n(x)\,e^{2 \pi i n y /\ell}.
\end{equations}
Therefore, if the convention on the signature of the metric is $(-,+,\dots,+)$, such that $\eta^{yy} = 1$, one has
\begin{equation}
\de_{\hat{\mu}}\,\vphi\,\de^{\hat{\mu}}\,\vphi = 
\ell^{-1}\sum_{n,m\in\mathbb{Z}} \left(\de_\mu\,\vphi_n\,\de^\mu\,\vphi_{-m} + \tfrac{4\,\pi^2}{\ell^2}\,n\,m\,\vphi_n\,\vphi_{-m}\right)\,e^{2 \pi i (n-m) y / \ell}.
\end{equation}
Using the orthogonality condition $\int \diff y\,e^{2 \pi i (n-m) y / \ell} = \ell\,\delta_{nm}$, and the reality condition \eqref{RealityConditionPhi} in the action \eqref{ActionExKK}, one gets
\begin{equation}
S = \sum_{n\in\mathbb{Z}}\int \diff^d x\,(-\tfrac{1}{2}\,\de_\mu\,\vphi_n\,\de^\mu\,\vphi^*_n + 2\,\pi^2\,n^2\,\ell^{-2}\,\vphi_n\,\vphi_n^*),
\end{equation}
which is the action for a set of infinite many complex scalar fields $\vphi_n$, with mass $\sqrt{2}\,\pi\,|n|\,\ell^{-1}$. Notice that the null mode $\vphi_0$ is real and massless, and the Kaluza-Klein truncation condition \eqref{KKcondition} selects the massless mode only.

\subsection{Eleven-dimensional supergravity reduction}\label{11reduction}

In this paragraph we study how Kaluza-Klein compactification works in a theory involving gravity, focusing in particular on the case of eleven-dimensional supergravity. 

We start with a $\hat{d}$-dimensional (super)gravity. We will denote the curved $\hat{d}$-dimensional index with $\hat{\mu},\hat{\nu},\dots$, and the flattened $\hat{d}$-dimensional index with $\hat{a},\hat{b},\dots$. Let us reduce the theory from $\hat{d}$ to $d$ dimensions, compactifying $\hat{d}-d = n$ dimensions in circles. Thus, the internal manifold is an $n$-torus $\mathscr{T}^n$. The splitting of the indices is denoted in the following way: $\hat{\mu}=(\mu,\mu')$ and $\hat{a} = (a,a')$, with $\mu$ and $a$ in $d$ dimensions, and $\mu', a'$ in $n$ dimensions. The $\hat{d}$--bein $e_{\hat{\mu}}{}^{\hat{a}}$ decomposes according to 
\begin{equation}
e_{\hat{\mu}}{}^{\hat{a}} = \begin{pmatrix}
\hat{e}_\mu{}^a & \hat{e}_\mu{}^{a'} \\
\hat{e}_{\mu'}{}^a & \hat{e}_{\mu'}{}^{a'}
\end{pmatrix}.
\end{equation}
The following identity
\begin{equation}\label{IdentityCountingKK}
\hat{d}^2 - \hat{d} - \tfrac{\hat{d}(\hat{d}-1)}{2} = d^2 - d - \tfrac{d(d-1)}{2} +
n\,(d-1) + n^2 - \tfrac{n(n-1)}{2},\;\text{if}\;\hat{d}=d+n,
\end{equation}
shows that $\hat{e}_\mu{}^{a}$, $\hat{e}_\mu{}^{a'}$ and $\hat{e}_{\mu'}{}^{a'}$ have the same number of (off-shell) degrees of freedom as those of a $d$--bein, $n$ abelian one-forms, and $\frac{n(n+1)}{2}$ scalar fields, parametrising the coset space $\text{GL}(n,\mathbb{R})/\text{SO}(n)$. Then, one can set $\hat{e}_{\mu'}{}^a = 0$, without loss of generality. Namely, we consider the following parametrisation, known as \emph{Kaluza-Klein ansatz}, 
\begin{equation}\label{ParametrVielbein}
e_{\hat{\mu}}{}^{\hat{a}} = 
\begin{pmatrix}
\hat{e}_\mu{}^a & \hat{e}_\mu{}^{a'} \\
\hat{e}_{\mu'}{}^a & \hat{e}_{\mu'}{}^{a'}
\end{pmatrix} = \begin{pmatrix}
\vphi^{k}\,e_\mu{}^a & A_\mu{}^{\mu'}\,\vphi_{\mu'}{}^{a'} \\
0 & \vphi_{\mu'}{}^{a'}
\end{pmatrix},
\end{equation}
where $e_\mu{}^a$ is the $d$-bein of the external spacetime, $\vphi_{\mu'}{}^{a'}$ is symmetric matrix, which encodes the $\frac{n(n+1)}{2}$ scalar fields, $\vphi$ is the determinant of $\vphi_{\mu'}{}^{a'}$, $k$ is constant, and $A_\mu{}^{\mu'}$ are called \emph{Kaluza-Klein vectors}. $A_\mu{}^{\mu'}$ and $\vphi_{\mu'}{}^{a'}$ are dimensionless -- in order to recover the canonical mass dimension they should be rescaled using the $d$-dimensional gravitational constant $\kappa$. Moreover, all the components are fields in the external spacetime coordinates $x^\mu$, but they are supposed to be independent of the internal coordinates $y^{\mu'}$, according to the Kaluza-Klein truncation condition 
\begin{equation}\label{KaluzaKleinTruncation}
\de_{\mu'} = \frac{\de}{\de\,y^{\mu'}} = 0.
\end{equation}
One can compute the inverse $\hat{d}$-bein 
\begin{equation}\label{ParametrVielbeinInv}
(\hat{e}^{\hat{\mu}}{}_{\hat{a}}) = 
\begin{pmatrix}
\hat{e}^\mu{}_a & \hat{e}^{\mu'}{}_a \\
\hat{e}^\mu{}_{a'} & \hat{e}^{\mu'}{}_{a'} 
\end{pmatrix} = \begin{pmatrix}
\vphi^{-k}\,e^\mu{}_a & -\vphi^{-k}\,e^\nu{}_a\,A_\nu{}^{\mu'} \\
0 & \vphi^{\mu'}{}_{a'},
\end{pmatrix},
\end{equation}
where $e^\mu{}_a$ is the inverse $d$-bein, and $\vphi^{\mu'}{}_{a'}$ is the inverse of $\vphi_{\mu'}{}^{a'}$. The metric tensor is
\begin{equation}
\hat{g}_{\hat{\mu}\hat{\nu}} = \hat{e}_{\hat{\mu}}{}^{\hat{a}}\,\hat{e}_{\hat{\nu}}{}^{\hat{b}}\,\hat{\eta}_{\hat{a}\hat{b}} = \begin{pmatrix}
\hat{g}_{\mu\nu} & \hat{g}_{\mu\nu'} \\
\hat{g}_{\mu'\nu} & \hat{g}_{\mu'\nu'}
\end{pmatrix} = \begin{pmatrix}
\vphi^{2k}\,g_{\mu\nu} + A_\mu{}^{\mu'}\,A_{\nu}{}^{\nu'}\,h_{\mu'\nu'} & A_{\mu}{}^{\mu'}\,h_{\mu'\nu} \\
h_{\mu'\nu'}\,A_{\nu}{}^{\nu'} & h_{\mu'\nu'}
\end{pmatrix},
\end{equation}
where $h_{\mu'\nu'} = \vphi_{\mu'}{}^{a'}\,\vphi_{\nu'}{}^{b'}\,\delta_{a'b'}$. The inverse metric is
\begin{equation}
\hat{g}^{\hat\mu\hat\nu} = \begin{pmatrix}
\hat{g}^{\mu\nu} & \hat{g}^{\mu\nu'} \\
\hat{g}^{\mu'\nu} & \hat{g}^{\mu'\nu'} 
\end{pmatrix} = \begin{pmatrix}
\vphi^{-2k}\,g^{\mu\nu} & -\vphi^{-2k}\,A^{\mu\nu'} \\
-\vphi^{-2k}\,A^{\nu\mu'} & \vphi^{-2k}\,g^{\mu\nu}\,A_\mu{}^{\mu'}\,A_{\nu}{}^{\nu'} + h^{\mu'\nu'}
\end{pmatrix},
\end{equation}
where $g^{\mu\nu}$ and $h^{\mu'\nu'}$ are the inverse of $g_{\mu\nu}$ and $h_{\mu'\nu'}$, and $A^{\mu\nu'} = g^{\mu\nu}\,A_\nu{}^{\nu'}$. Finally, the line element is
\begin{equation}\label{LineElement}
\hat{g}_{\hat\mu\hat\nu}\,\diff x^{\hat\mu}\otimes\diff x^{\hat\nu} = \vphi^{2k}\,g_{\mu\nu}\,\diff x^\mu\otimes\diff x^{\nu} + h_{\mu'\nu'}\,(A^{\mu'} + \diff y^{\mu'})\otimes(A^{\nu'} + \diff y^{\nu'}),
\end{equation}
where $A^{\mu'} = A_\mu{}^{\mu'}\,\diff x^\mu$. 

As regards symmetries, the $\hat{d}$-dimensional theory is locally invariant under $\hat{d}$-dimensional diffeomorphisms with parameter $\hat{\xi}^{\hat\mu} = \hat{\xi}^{\hat\mu}(x,y)$, and under $\hat{d}$-dimensional Lorentz transformation with parameter $\hat{\Omega}_{ab} = \hat{\Omega}_{ab}(x,y) = -\hat{\Omega}_{ba}(x,y)$. Moreover, we can also consider the global scaling symmetry of the equations of motion (trombone), parametrised by the constant $\sigma$. These transformations act on the $\hat{d}$-bein as
\begin{equation}
\delta\,\hat{e}_{\hat\mu}{}^{\hat{a}} = \mathcal{L}_{\hat\xi}\,\hat{e}_{\hat\mu}{}^{\hat{a}} + \sigma\,e_{\hat\mu}{}^{\hat a} = \hat{\xi}^{\hat\vrho}\,\de_{\hat\vrho}\,\hat{e}_{\hat\mu}{}^{\hat{a}} + e_{\hat\vrho}{}^{\hat{a}}\,\de_{\hat\mu}\,\hat{\xi}^{\hat\vrho} + \sigma\,e_{\hat\mu}{}^{\hat a}.
\end{equation}
We want to find which of these transformations preserve the Kaluza-Klein ansatz \eqref{ParametrVielbein}. The counting in the identity \eqref{IdentityCountingKK} suggests that, besides possible global transformations, $e_\mu{}^a$, $A_\mu{}^{\mu'}$, and $\vphi_{\mu'}{}^{a'}$ transform covariantly under $d$-dimensional diffeomorphisms, and that $e_\mu{}^a$ transforms also as a vector with respect to local Lorentz transformations, $A_\mu{}^{\mu'}$ enjoys also a local one-parameter symmetry, and the scalars $\vphi_{\mu'}{}^{a'}$ are rotated by local Lorentz transformations. The ansatz requires that the variation of the fields in the reduced theory does not depend on the internal coordinates and that $\delta\,\hat{e}_{\mu'}{}^a = 0$, in such a way that the transformed $\hat{d}$-bein remains upper triangular. Computing $\delta\,\hat{e}_{\mu'}{}^a $ and setting it equal to zero, one finds
\begin{equation}
0 = \vphi^k\,e_\nu{}^a\,\de_{\mu'}\,\hat{\xi}^\nu + \hat{\Omega}^a{}_{b'}\,\vphi_{\mu'}{}^{b'}
\Leftrightarrow
\hat{\xi}^\mu(x,y) = \xi^\mu(x), \;
\hat{\Omega}_{a'b} = \hat{\Omega}_{ab'} = 0.
\end{equation}
Computing $\delta\,\hat{e}_{\mu'}{}^{a'}$ and using the previous results, one finds
\begin{equation}
\delta\,\vphi_{\mu'}{}^{a'} = \xi^\vrho\,\de_\vrho\,\vphi_{\mu'}{}^{a'} + \hat{\Omega}^{a'}{}_{b'}\,\vphi_{\mu'}{}^{b'} + \vphi_{\nu'}{}^{a'}\,\de_{\mu'}\,\hat{\xi}^{\nu'},
\end{equation}
where the first term is the $d$-dimensional Lie derivative on a scalar, and the second term is the local Lorentz rotation. There is no dependence on the internal coordinates if and only if  
\begin{equation}
\hat{\xi}^{\nu'}(x,y) = \Lambda^{\nu'}{}_{\mu'}\,y^{\mu'} + c\,y^{\nu'} + \alpha^{\nu'}(x),\quad
\hat{\Omega}_{a'b'}(x,y) = \omega_{a'b'}(x),
\end{equation}
where $c$ is a constant, and $\Lambda$ is traceless constant matrix. Now, we can compute $\delta\,\hat{e}_{\mu}{}^{a'}$ and $\delta\,\hat{e}_{\mu}{}^a$, which fix the transformation of $A_\mu{}^{\mu'}$ and $e_\mu{}^a$, finding as unique new restriction the requirement that $\hat{\Omega}_{ab}$ does not depend on the internal coordinates
\begin{equation}
\hat{\Omega}_{ab}(x,y) = \Omega_{ab}(x).
\end{equation}
The resulting transformations of the $d$-dimensional fields are
\begin{equations}
\delta\,e_\mu{}^a &= \mathcal{L}_\xi\,e_\mu{}^a + \Omega^a{}_b\,e_\mu{}^b - (c + (d-1)\,\sigma)\,e_\mu{}^a,\\
\delta\,A_{\mu'}{}^\mu &= \mathcal{L}_\xi\,A_\mu{}^{\mu'} + \de_{\mu}\,\alpha^{\mu'} - \Lambda^{\mu'}{}_{\nu'}\,A_\mu{}^{\nu'} - c\,A_\mu{}^{\mu'},\\
\delta\,\vphi_{\mu'}{}^{a'} &= \mathcal{L}_\xi\,\vphi_{\mu'}{}^{a'} + \omega^{a'}{}_{b'}\,\vphi_{\mu'}{}^{b'} + \Lambda^{\nu'}{}_{\mu'}\,\vphi_{\nu'}{}^{a'} + (\sigma + c)\,\vphi_{\mu'}{}^{a'}.
\end{equations}
where the Lie derivative acts only on the external spacetime index, so that $e_\mu{}^a$ and $A_\mu{}^{\mu'}$ are covariant one-forms, and $\vphi_{\mu'}{}^{a'}$ is a covariant scalar.
Therefore, we conclude that the dependence of the internal coordinate in the symmetry parameter is at most linear, and that the local $\hat{d}$-dimensional diffeomorphisms are reduced to local $d$-dimensional diffeomorphisms; $n$ independent local $\text{U}(1)$ gauge transformations, generated by $\alpha^{\mu'}$, $\mu' = 1,\dots, n$, with gauge field $A_\mu{}^{\mu'}$; a global rotation of $A_\mu{}^{\mu'}$ and $\vphi_{\mu'}{}^{a'}$, parametrised by $\Lambda$; and a constant shift, parametrised by $c$. Importantly, the emergence of the $U(1)^n$ symmetry was already manifest in the line element \eqref{LineElement}, where the compactified dimensions $y^{\mu'}$ are cyclic coordinates\footnote{Cyclic coordinate means that the metric does not depend on it: as in Kaluza-Klein compactification, a coordinate is cyclic when it can be thought as the result of a compactification on a circle.}, according to \eqref{KaluzaKleinTruncation}, so that $\de_{\mu'}$ is a Killing vector, generating a $U(1)$ isometry. In other words, the emergent $U(1)^n$ symmetry is the isometry group of internal hypertorus $T^n$.

The factor $\vphi^k$ in the first block in \eqref{ParametrVielbein} ensures that the dimensional reduction of the $\hat{d}$-dimensional Hilbert-Einstein action gives the $d$-dimensional Hilbert-Einstein action with the same coefficient, for a suitable choice of $k$ (this choice is called ``Einstein frame"). Indeed, since the determinant of the $\hat{d}$-bein is
\begin{equation}
\hat{e} = \text{det}\,\hat{e}_{\hat{\mu}}{}^{\hat{a}}
= \text{det}\,(\vphi^k\,e_\mu{}^a)\,\text{det}\,\vphi_{\mu'}{}^{a'} = \vphi^{kd+1}\,e,
\end{equation}
then
\begin{equation}
\hat{e}\,\hat{R} = \hat{e}\,g^{\hat{\mu}\hat{\nu}}\,\hat{R}_{\hat{\mu}\hat{\nu}} = (\vphi^{kd+1}\,e)\,\left[(\vphi^{-2k}\,g^{\mu\nu})\,R_{\mu\nu} + \dots\right] = \vphi^{(d-2)k+1}\,e\,R + \dots,
\end{equation}
so that,
\begin{equation}
\hat{e}\,\hat{R} = e\,R + \dots, \;\;
\text{if}\;\;k = \tfrac{1}{2-d}.
\end{equation}
If we require the reduction of the measure of integration to be given by
\begin{equation}
\frac{1}{2\,\hat{\kappa}^2}\int\diff^{\hat{d}}x = 
\frac{1}{2\,\hat{\kappa}^2}\int\diff^dx\int_{\mathscr{T}^n}\diff^ny = \frac{1}{2\,\kappa^2}\int\diff^d x,
\end{equation}
where the $\hat{d}$- and $d$-dimensional gravitational constants are related by
\begin{equation}\label{KKGravCost}
\kappa = \frac{\hat{\kappa}}{\sqrt{\text{Vol}(\mathscr{T}^n)}},
\end{equation}
consistently with mass dimension $[\kappa^{(d)}] = (2-d)/2$. The integral over the internal manifold factors out and it gives the internal volume, since the integrand does not depend on the internal coordinates $y^{\mu'}$, thanks to \eqref{KaluzaKleinTruncation}.

The full dimensional reduction of the $\hat{d}$-dimensional Hilbert-Einstein action is remarkably simplified if the torsion formulation studied in Section \ref{GenRelTorsion} is exploited. In the present setting, the equation \eqref{HE} becomes
\begin{equation}
\int\diff^{\hat{d}}x\,\hat{e}\,\hat{R} = \int\diff^{\hat{d}}x\,\hat{e}\,(-\tfrac{1}{4}\,\hat{T}^{\hat\mu\hat\nu\hat\vrho}\,\hat{T}_{\hat\mu\hat\nu\hat\vrho} - \tfrac{1}{2}\,\hat{T}^{\hat\mu\hat\nu\hat\vrho}\,\hat{T}_{\hat\mu\hat\vrho\hat\nu} + \hat{T}_{\hat\mu\hat\nu}{}^{\hat\nu}\,\hat{T}^{\hat\mu\hat\vrho}{}_{\hat\vrho}).
\end{equation}
Indeed, it is simple to decompose the torsion $\hat{T}_{\hat\mu}{}^{\hat\vrho}{}_{\hat\nu} = e^{\hat\vrho}{}_{\hat{a}}\,\de_{[\hat\mu}\,e_{\hat\nu]}{}^{\hat{a}}$ (compare with \eqref{TorsionFromW}) using the parametrisation of the \eqref{ParametrVielbein} and the inverse \eqref{ParametrVielbeinInv}:
\begin{equations}
\hat{T}_\mu{}^\vrho{}_\nu &= T_\mu{}^\vrho{}_\nu - k\,\delta_{[\mu}^\vrho\,\de_{\nu]}\,\text{log}\,\vphi,\\
\hat{T}_{\mu}{}^{\vrho'}{}_\nu &= -A_{\vrho}{}^{\vrho'}\,T_{\mu}{}^\vrho{}_\nu + k\,A_{[\mu}{}^{\vrho'}\,\de_{\nu]}\,\text{log}\,\vphi + h^{\vrho'}{}_{\nu'}\,F_{\mu\nu}{}^{\nu'} - \vphi^{\vrho'}{}_{a'}\,\de_{[\mu}{}^{\nu'}\,\de_{\nu]}\,\vphi_{\nu'}{}^{a'},\\
\hat{T}_\mu{}^{\vrho'}{}_{\nu'} &= \vphi^{\vrho'}{}_{a'}\,\de_\mu\,\vphi_{\nu'}{}^{a'}, \quad
\hat{T}_{\mu'}{}^\vrho{}_\nu = -\vphi^{\vrho'}{}_{a'}\,\de_\nu\,\vphi_{\mu'}{}^{a'},\\
\hat{T}_{\mu'}{}^\vrho{}_\nu &= \hat{T}_{\mu'}{}^\vrho{}_{\nu'} = \hat{T}_{\mu'}{}^{\vrho'}{}_{\nu'} = 0.
\end{equations}
The result is
\begin{align}
S &= \int\diff^d x\,e\,[-\tfrac{1}{2\,\kappa^2}\,R - \tfrac{1}{4}\,\vphi^{2/(d-2)}\,h_{\vrho'\sigma'}\,F^{\mu\nu\vrho'}\,F_{\mu\nu}{}^{\sigma'} \,+\nonumber\\
& \qquad\qquad\quad - \tfrac{1}{8\,\kappa^2}\,\de_\mu\,h_{\vrho'\sigma'}\,\de^\mu\,h^{\vrho'\sigma'} - \tfrac{1}{2\,\kappa^2\,(d-2)}\,\de_\mu\,\text{log}\,\vphi\,\de^\mu\,\text{log}\,\vphi],\label{ReductionActionKK}
\end{align}
where $F_{\mu\nu}{}^{\vrho'}$ is the field strength of the Kaluza-Klein vectors:
\begin{equation}
F_{\mu\nu}{}^{\vrho'} = \de_{\mu}\,A_{\nu}{}^{\vrho'} - \de_{\nu}\,A_{\mu}{}^{\vrho'}.
\end{equation}

Consider now the reduction of a $\hat{d}$-dimensional $p$--form $\hat{A}^{(p)}$. Its components $\hat{A}_{\hat{\mu}_1\dots \hat{\mu}_p}$ are decomposed into ${{n}\choose{p-q}}$ $q$--forms $A^{(q)}_{\mu'_1\dots\mu'_{p-q}}$, with components
\begin{equation}
\hat{A}_{\mu_1\dots \mu_q\mu'_1\dots\mu'_{p-q}}(x,y) =
A_{\mu_1\dots \mu_q\mu'_1\dots\mu'_{p-q}} (x), 
\end{equation} 
for $0\leqslant q \leqslant p$, with antisymmetrised internal indices $\mu'_1, \dots, \mu'_{p-q}$:
\begin{align}
\hat{A}^{(p)} &= \sum_{q=0}^{p} \frac{1}{q!}\,\diff x^{\mu_1}\dots\diff x^{\mu_q}\,A_{\mu_1\dots\mu_q\mu'_1\dots\mu'_{p-q}}\,\diff y^{\mu'_1}\dots\diff y^{\mu'_{p-q}}\,\frac{1}{(p-q)!} = \nn\\
& = \sum_{q=0}^{p} A^{(q)}_{\mu'_1\dots\mu'_{p-q}}\,\diff y^{\mu'_1}\dots\diff y^{\mu'_{p-q}}\,\frac{1}{(p-q)!} = \sum_{q=0}^{p} A^{(q,p-q)},
\end{align}
where we introduced as a shorthand the notation $A^{(q,p-q)}$ for $q$--form with $p-q$ internal indices, which indeed is a $(p-q)$--form according to the internal space point of view. The gauge transformation is 
\begin{equation}
\delta\,\hat{A}^{(p)} = \hat{d}\,\hat\Lambda^{(p-1)} = 
\diff\,\hat\Lambda^{(p-1)} + \diff y^{\mu'}\,\de_{\mu'}\,\hat\Lambda^{(p-1)}.
\end{equation}
In order to obtain a transformation for the components of the reduction of $\hat{A}^{(p)}$ which does not depend on the internal coordinates, the dependence on the internal coordinate in the gauge parameter is at most linear:
\begin{align}
\hat{\Lambda}^{(p-1)} &= \sum_{q=0}^{p-1} \Lambda^{(q,p-1-q)} + \lambda^{(q,p-1-q)}_{\mu'}\,y^{\mu'},
\end{align}
where the components of $\Lambda^{(q,p-1-q)} $ depends only on the external coordinates, whereas those of $\lambda_{\mu'}^{(q,p-1-q)}$ are constant
\begin{equations}
\Lambda^{(q,p-1-q)} &= \tfrac{1}{q!}\,\diff x^{\mu_1}\dots\diff x^{\mu_q}\,\Lambda_{\mu_1\dots\mu_q\mu'_{1}\dots\mu_{p-1-q}}(x)\,\diff y^{\mu'_1}\dots\diff y^{\mu'_{p-1-q}}\,\tfrac{1}{(p-1-q)!},\\
\lambda^{(q,p-1-q)}_{\mu'}\,y^{\mu'} &= \tfrac{1}{q!}\,\diff x^{\mu_1}\dots\diff x^{\mu_q}\,\lambda_{\mu',\mu_1\dots\mu_q\mu'_1\dots\mu'_{p-1-q}}\,\diff y^{\mu'_1}\dots\diff y^{\mu'_{p-1-q}}\,\tfrac{1}{(p-1-q)!}.
\end{equations}
In this way the gauge transformations are reduced to 
\begin{equation}
\delta\,A^{(q,p-q)} = \diff\,\Lambda^{(q-1,p-q)} + \lambda^{(q,p-q)},
\end{equation}
where $\lambda^{(q-1,p+1-q)} = \diff y^{\mu'}\,\lambda^{(q-1,p-q)}_{\mu'}$. The first term is a gauge transformation for a $q$--form; the second one is a constant shift (\emph{$p$--form shift}). In particular, in the case $p=3$, which is the case of eleven-dimensional supergravity 
\begin{equation}
\delta\,\hat{A}^{(3)} = \hat{\diff}\,\Lambda^{(2)} \Rightarrow 
\begin{cases}
\delta\,\hat{A}^{(3,0)} = \diff\,\Lambda^{(2,0)},\\
\delta\,\hat{A}^{(2,1)} = \diff\,\Lambda^{(1,1)} + \lambda^{(2,1)},\\
\delta\,\hat{A}^{(1,2)} = \diff\,\Lambda^{(0,2)}  + \lambda^{(1,2)},\\
\delta\,\hat{A}^{(0,3)} = \lambda^{(0,3)}.
\end{cases}
\end{equation}

Finally, the gravitino $\psi_{\hat{\mu}\hat{\iota}}$, where $\hat{\iota}=1,\dots,2^{\lfloor\hat{d}/2\rfloor}$ is the spinor index in $\hat{d}$ dimensions, is decomposed into $\psi_{\mu i, \iota}$ and into $\psi_{mi\iota} = \chi_{i,m\iota}$, where $i=1,\dots,2^{\lfloor d/2 \rfloor}$ is the spinor index in $d$ dimensions and $\iota = 1,\dots, 2^{\lfloor n/2 \rfloor}$ is the spinor index in $n$ dimensions. $\psi_{\mu i, \iota}$ corresponds to $2^{\lfloor n/2 \rfloor}$ gravitini, indexed by $\iota$; $\chi_{i,m\iota}$ corresponds to $n \times 2^{\lfloor n/2 \rfloor}$ spin--$\frac{1}{2}$ fields, indexed by $m\iota$. Notice that the number of degrees of freedom matches, since
\begin{equation}
2^{\lfloor n/2 \rfloor} \times 2^{\lfloor d/2 \rfloor -1}(d-3) + n\,2^{\lfloor n/2 \rfloor} \times 2^{\lfloor d/2 \rfloor -1} = 2^{\lfloor \hat{d}/2 \rfloor -1}\,(\hat{d}-3).
\end{equation}

The coset $\text{GL}(n,\mathbb{R})/\text{SO}(n)$ does not take into account all the scalars of the reduced theory, since one has also to consider the scalars coming from the reduction of the $p$--form, whose number is ${{n}\choose{p}}$, if $0\leqslant p \leqslant n$. Moreover, one has to consider the dual scalar fields, according to the Cremmer-Julia duality. The scalars are dual to the $(d-2)$--forms coming from the reduction of the $p$--form, whose number is ${{n}\choose {p-d+2}}$, if $0\leqslant p-d+2 \leqslant n$. 

In conclusion, the coset space describing the scalar fields coming from the reduction in $d$-dimensions of a $\hat{d}$-dimensional (super)gravity theory, whose bosonic field content is made up of graviton and a $p$--form, is
\begin{equation}\label{coset1}
\frac{\text{GL}(n,\R) \times \R) \ltimes \R^{\#}}{\text{SO}(n)\times \R},
\end{equation}
where $\#$ is the number of scalars coming from the $p$--form, and $\R^{\#}$ indicates the constant shift acting on these scalars. We included also a $\R$ term, describing the trombone symmetry of the reduced $d$-dimensional theory. In this way, the numerator is the \emph{manifest} duality group of the theory. In the denominator we quotiented an $\R$ factor, because the trombone does not correspond to any scalar field. 

If the Kaluza-Klein compactification of the $\hat{d}$-dimensional supergravity theory maintains all the supersymmetry charges in the $d$-dimensional theory, the latter is called \emph{maximal supergravity}. The duality group of maximal supergravities is larger than the manifest duality group found above (for this reason we called it ``manifest"). Let us denote the whole duality group with $G_D\times \R$, where we factorised the trombone symmetry. Then, the coset space describing the scalars is
\begin{equation}\label{coset2}
\frac{G_D \times\R}{K(G_D)\times\R},
\end{equation}
where $K(G_D)$ denotes the maximal compact subgroup of $G_D$. Since the number of scalars is the same as before, the dimensions of the two coset spaces \eqref{coset1} and \eqref{coset2} must be the same. Thus, we have a constraint on the dimension of $G_D$:
\begin{equation}
\dim{G_D} - \dim{K(G_D)} = n^2 + \# - \tfrac{n(n-1)}{2}.
\end{equation}

Let us focus now on eleven-dimensional supergravity ($\hat{d}=11$), reduced to $d=4$. Since $n=7$, the scalars coming from the graviton are $\frac{7(7+1)}{2}=28$; those coming from the three-form are ${{7}\choose{3}}=35$. The one-forms coming from the graviton are $7$; those coming from the three-form are ${{7}\choose{2}}=21$. The two-forms from the three-form are ${{7}\choose{1}}=\textcolor{red}{7}$. The scalars, the one-forms and the two-forms are the unique propagating fields in four dimensions. Moreover, in four dimensions the two-forms are equivalent to the scalars, according to the Cremmer-Julia duality. So, all the bosonic degrees of freedom are described by scalars and one-forms. The total number of scalars is $28+35+\textcolor{red}{7}=70$; the total number of vectors is $7+21=28$. Notice that the number of scalars $70 = \dim{E_{7(7)}/\text{SL}(8,\mathbb{R})}$. $E_{7(7)}$ is a subgroup of $\text{Sp}(14,\mathbb{R})$, consistently with the Zumino-Gaillard duality. Finally, there are the two degrees of freedom of the four-dimensional metric. Thus, the total number of bosonic degrees of freedom is $70 + 2\times 28 + 2 = 128$, as expected.

Similarly, consider the reduction from $\hat{d}=11$ to $d=3$. Since $n=8$, the scalars coming from the graviton are $\frac{8(8+1)}{2}=36$; those coming from the three-form are ${{8}\choose{3}}=56$. The one-forms coming from the graviton are $\textcolor{teal}{8}$; those coming from the three-form are ${{8}\choose{2}}=\textcolor{blue}{28}$. The scalars and the one-forms are the unique propagating fields in three dimensions. Moreover, in three dimensions the one-forms are equivalent to the scalars, according to the Cremmer-Julia duality. So, all the bosonic degrees of freedom are described by scalars: 128 scalars are expected. Indeed, $36+56+\textcolor{teal}{8}+\textcolor{blue}{28}=128$. Notice that $128 = \dim{E_{8(8)}/\text{SO}(16)}$. $E_{8(8)}$ is a subgroup of $\text{Sp}(16,\mathbb{R})$, consistently with the Zumino-Gaillard duality. 

In general, $E_{n(n)}$ is the duality group of the  $11-n$ dimensional theory, obtained by compactifying the eleven-dimensional supergravity on a hypertorus $T^n$. This compactification preserves all the supersymmetry: for this reason we have taken into account the bosonic sector only. Moreover, the duality group does not mix bosonic and fermionic degrees of freedom. Notice that the reduction of the eleven-dimensional gravitino brings to the maximal number of supercharges, such that the supersymmetry multiplets contain a single graviton and they do not contain higher-spin particles This number is $\mathcal{N}=8$ in $d=4$, or $\mathcal{N}=16$ in $d=3$. 

An alternative but equivalent way to compute the bosonic degrees of freedom without using the Cremmer-Julia duality in the reduced theory is to consider in eleven dimensions the dual of $A^{(3)}$, which is a six-form $\tilde{A}^{(6)}$, and the dual of $g_{\hat{\mu}\hat{\nu}}$ (\emph{dual graviton}), and counting the number of vectors or scalars (if the reduced theory is in four or three dimensions) coming from them. The dual graviton can defined in the following way. Consider the linearised graviton $\eta_{\hat\mu\hat\nu} + h_{\hat\mu\hat\nu}$; dualise one of the two symmetric indices $h_{\hat\mu\hat\nu}$, obtaining $\tilde{h}_{\hat{\mu}_1\dots\hat{\mu}_8,\hat\nu}$ (in eleven dimensions, with $p=1$, the dual is $d-p-2=8$); consider this object as an eight-form with a spectator index, such that totally antisymmetric part is vanishing $\tilde{h}_{[\hat{\mu}_1\dots\hat{\mu}_8,\hat\nu]}$ (since $h_{\mu\nu}$ is symmetric); its gauge symmetry is parametrised by a seven-form $\tilde{\lambda}_{\hat{\mu}_1\dots\hat{\mu}_7,\hat{\nu}}$ with a spectator index. This gauge symmetry is the dual of the linearised diffeomorphism symmetry $\delta\,h_{\hat\mu\hat\nu} = \de_{(\hat\mu}\,\lambda_{\hat\nu)}$.\footnote{See Appendix \ref{DualGraviton} for more details.}

The scalars coming from $\tilde{A}^{(6)}$ are ${{n}\choose{6}}$, which is \textcolor{red}{7} if $n=7$ and \textcolor{blue}{28} if $n=8$; the vectors coming from $\tilde{A}^{(6)}$ are ${{n}\choose{5}}$, which is 21 if $n=7$, and 56 if $n=8$. The scalars coming from $\tilde{h}_{\hat{\mu}_1\dots\hat{\mu}_8,\hat\nu}$ are $n\,{{n}\choose{8}}$, which is 0 if $n=7$, and \textcolor{teal}{8} if $n=8$; the vectors coming from $\tilde{h}_{\hat{\mu}_1\dots\hat{\mu}_8,\hat\nu}$ are $n\,{{n}\choose{7}}$, which is 7 if $n=7$, and 64 if $n=8$.

\subsection{Scherk-Schwarz compactifications}

If a more general dependence on the internal coordinates $y^{\mu'}$ is allowed in the fields, weakening the Kaluza-Klein truncation condition \eqref{KaluzaKleinTruncation}, then geometries other than the hypertorus $T^n$ are possible for the internal manifold. As a result, the gauge symmetry resulting in the reduction of the $\hat{d}$-dimensional diffeomorphisms becomes in general a non-abelian one. But the dependence on the internal coordinates should be simple enough to obtain the Kaluza-Klein compactification  as a limit case, and the internal coordinates should cancel out in reducing the field symmetry laws and the action. 

The problem of finding the suitable dependence on internal coordinates was first addressed by Scherk and Schwarz, in \cite{Scherk:1979zr}. They noticed that a generalisation of Kaluza-Klein compactification is always possible when the internal manifold is parallelisable with constant Weitzenb\"ock torsion. This is the case of group manifolds, where the Weitzenb\"ock torsion is the structure constants of the group algebra. The dependence on the internal coordinates is given by the global frame of the internal manifold $E^{\mu'}{}_{a'}$ -- as in the previous section, Greek/Latin primed indices denote the curved/flattened internal indices. One replaces the components in Kaluza-Klein ansatz and the parameters in the transformation laws according to
\begin{equations}
\xi^\mu(x) &\rightarrow \xi^\mu(x,y) = \xi^\mu(x),\\
\alpha^{\mu'}(x) &\rightarrow \alpha^{\mu'}(x,y) = E^{\mu'}{}_{a'}(y)\,\alpha^{a'}(x), \\
e_\mu{}^a(x) &\rightarrow e_\mu{}^a(x,y) = e_\mu{}^a(x),\\
\vphi(x) &\rightarrow \vphi(x,y) = \vphi(x), \\
\vphi_{\mu'}{}^{a'}(x) &\rightarrow \vphi_{\mu'}{}^{a'}(x,y) = E_{\mu'}{}^{b'}(y)\,\vphi_{b'}{}^{a'}(x),\\
A_\mu{}^{\mu'}(x) &\rightarrow A_\mu{}^{\mu'}(x,y) = E^{\mu'}{}_{a'}(y)\,A_\mu{}^a(x),
\end{equations} 
where $E_{\mu'}{}^{a'}$ is the inverse internal frame ($n$--bein). The ansatz for the $\hat{d}$--bein becomes
\begin{equation}
\hat{e}_{\hat\mu}{}^{\hat{a}} = \begin{pmatrix}
\hat{e}_\mu{}^a & \hat{e}_\mu{}^{a'} \\
\hat{e}_{\mu'}{}^a & \hat{e}_{\mu'}{}^{a'} 
\end{pmatrix} = \begin{pmatrix}
\vphi^k\,e_\mu{}^a & E^{\mu'}{}_{b'}\,A_{\mu}{}^{b'}\,\vphi_{\mu'}{}^{a'} \\
0 & E_{\mu'}{}^{b'}\,\vphi_{b'}{}^{a'}
\end{pmatrix},
\end{equation}
where the only part depending on the internal coordinates is given by the frame, and the other components are fields in the external coordinates.

Consider now symmetry transformation reduction. Two vector fields $\hat\xi^{\hat\mu}$ and $\hat\eta{}^{\hat\mu}$ in the $\hat{d}$-dimensional theory are reduced to 
\begin{equation}
\hat{\xi}^{\hat\mu} = (\xi^\mu,E^{\mu'}{}_{a'}\,\alpha^{a'}), \quad
\hat{\eta}^{\hat\mu} = (\eta^\mu,E^{\mu'}{}_{a'}\,\beta^{a'}).
\end{equation}
One can compute
\begin{equation}\label{SSLieDer}
\mathcal{L}_{\hat\xi}\,\hat\eta^{\hat\mu} = (\mathcal{L}_\xi\,\eta^\mu, E^{\mu'}{}_{a'}\,\mathcal{L}_\xi\,\beta^{a'} - E^{\mu'}{}_{a'}\,\mathcal{L}_\eta\,\alpha^{a'} - \alpha^{a'}\,\beta^{b'}\,E^{\vrho'}{}_{a'}\,E^{\nu'}{}_{b'}\,T_{\vrho'}{}^{\mu'}{}_{\nu'}),
\end{equation}
where
\begin{equation}
T_{\mu'}{}^{\vrho'}{}_{\nu'} = E^{\vrho'}{}_{a'}\,\de_{[\mu'}\,E_{\nu']}{}^{a'}
\end{equation}
is the Weitzenb\"ock torsion for the internal manifold. This computation shows that the $\hat{d}$-dimensional diffeomorphisms, with parameter $\hat\xi^{\hat\mu}$, split into $d$-dimensional diffeomorphisms, with parameter $\xi^\mu$, and in a non-abelian gauge transformation whose structure constants are the flattened Weitzenb\"ock torsion
\begin{equation}
T_{a'}{}^{c'}{}_{b'} = E_{\vrho'}{}^{c'}\,E^{\mu'}{}_{a'}\,E^{\nu'}{}_{b'}\,T_{\mu'}{}^{\vrho'}{}_{\nu'}.
\end{equation}
Indeed, setting $\xi^\mu = \eta^\mu = 0$, and $\alpha^{a'} = \beta^{a'} = 1$ in \eqref{SSLieDer}, one gets
\begin{equation}
\mathcal{L}_{E_{a'}}\,E^{\mu'}{}_{b'} = - T_{a'}{}^{c'}{}_{b'}\,E^{\mu'}{}_{c'},
\end{equation}
where $E_{a'} = E^{\mu'}{}_{a'}\,\de_{\mu'}$, consistent with \eqref{ExprTorsionLieDer2}. Similarly, studying the reduction of the action of $\hat{d}$-dimensional diffeomorphisms on the components of the $\hat{d}$--bein, one finds
\begin{equations}
\delta\,e_\mu{}^a &= \mathcal{L}_\xi\,e_\mu{}^a = \xi^\vrho\,\de_\vrho\,e_\mu{}^a + e_\vrho{}^a\,\de_\mu\,\xi^\vrho,\\
\delta\,A_\mu{}^{a'} &= \mathcal{L}_\xi\,A_\mu{}^{a'} + \de_\mu\,\alpha^{a'} + T_{b'}{}^{a'}{}_{c'}\,\alpha^{b'}\,A_\mu{}^{c'},\\
\delta\,\vphi_{a'}{}^{b'} &= \mathcal{L}_\xi\,\vphi_{a'}{}^{b'} + T_{c'}{}^{b'}{}_{d'}\,\alpha^{c'}\,\vphi_{a'}{}^{d'},
\end{equations}
showing in particular that $\alpha^{a'}$ is the parameter of a non-abelian gauge transformation, whose connection is the Kaluza-Klein vectors $A_\mu{}^{a'}$. The gauge transformation corresponds to the isometry group of the internal manifold. Thus, \emph{Scherk-Schwarz compactification} provides a non-abelian extension of Kaluza-Klein compactification. One can also decompose in a similar fashion the higher form gauge fields included in the supergravity multiplets. For example, for the three-form potential $\hat{A}_{\hat\mu\hat\nu\hat\vrho}$ the Scherk-Schwarz ansatz is
\begin{equation}
\hat{A}_{\hat\mu\hat\nu\hat\vrho} = 
(A_{\mu\nu\vrho}, E_{\vrho'}{}^{a'}\,A_{\mu\nu a'}, E_{\nu'}{}^{a'}\,E_{\vrho'}{}^{b'}\,A_{\mu a'b'},E_{\mu'}{}^{a'}\,E_{\nu'}{}^{b'}\,E_{\vrho'}{}^{c'}\,A_{a'b'c'}),
\end{equation}
with a three-form, $n$ two-forms, $\frac{n(n-1)}{2}$ one-forms, and $\frac{n(n-1)(n-2)}{6}$ zero-forms.

Since the Lagrangian density in the Hilbert-Einstein action is a scalar density, it transforms as a total derivative under diffeomorphisms. This property should remain true also after the reduction, but now this is not trivial as in Kaluza-Klein compactification, since a dependence on the internal coordinates in the fields is introduced, and the integral of a total derivative in the internal space is not guaranteed to be equal to zero, since it corresponds to a boundary term. In particular, under a $\hat{d}$-dimensional diffeomorphism $\delta\,(\hat{e}\,\hat{R}) = \de_{\hat\mu}\,(\hat{\xi}^{\hat\mu}\,\hat{e}\,\hat{R})$. Then,
\begin{equation}
\delta\int\diff^{\hat{d}}x\,\hat{e}\,\hat{R} = \int\diff^d x\int\diff^n y\,\de_\mu\,(\xi^\mu\,\vphi^{kd+1}\,e\,\hat{R})\,E + \de_{\mu'}\,(E^{\mu'}{}_{a'}\,E)\,\alpha^{a'}\,\vphi^{kd+1}\,e\,\hat{R},
\end{equation}
where $E$ is the determinant of the $n$--bein, and we use the fact fact $\hat{R}$ does not depend on the internal coordinates. The first piece is the internal volume $\int\diff^ny\,E$ times the integral of a total derivative in the external spacetime, so it vanishes; in order for the second term to vanish too, one has to impose the constraint
\begin{equation}
\de_{\mu'}\,(E^{\mu'}{}_{a'}\,E) = 0,
\end{equation}
which, using the formula \eqref{CovariantWeitzenbockDivergence}, is equivalent to
\begin{equation}
T_{a'}{}^{b'}{}_{b'} = 0,
\end{equation}
which is always true in the case in which the group algebra is semi-simple.

Studying the reduction of the Hilbert-Einstein action one finds the natural non-Abelian extension of the result in \eqref{ReductionActionKK}, with the abelian Kaluza-Klein field strength replaced by the non-abelian corresponding $F_{\mu'\nu'}{}^{a'} = \de_{\mu'}\,A_{\nu'}{}^{a'} - T_{b'}{}^{a'}{}_{c'}\,A_{\mu'}{}^{b'}\,A_{\nu'}{}^{c'}$, plus a \emph {scalar potential}:
\begin{align}
S &= \int\diff^dx\,e\,[-\tfrac{1}{2\,\kappa^2}\,R -\tfrac{1}{4}\,\vphi^{2/(d-2)}\,h_{a'b'}\,F^{\mu\nu a'}\,F_{\mu\nu}{}^{b'} \,+\nonumber \\
& \qquad\qquad\;\; - \tfrac{1}{8\,\kappa^2}\,D_\mu\,h_{a'b'}\,D^{\mu}\,h^{a'b'} - \tfrac{1}{2\,\kappa^2\,(d-2)}\,\de_\mu\,\text{log}\,\vphi\,\de^\mu\,\text{log}\,\vphi \,+\nonumber \\
& \qquad\qquad\;\; - \tfrac{1}{8\,\kappa^2}\,\vphi^{-2/(d-2)}\,T_{b'}{}^{a'}{}_{c'}\,(2\,T_{a'}{}^{b'}{}_{d'}\,h^{c'd'} + f_{d'}{}^{f'}{}_{e'}\,h^{d'b'}\,h^{e'c'}\,h_{f'a'})],
\end{align}
where $D_\mu$ is the covariant derivative with respect to the Kaluza-Klein vectors $A_\mu{}^{a'}$, and, similarly to the Kaluza-Klein case \eqref{KKGravCost}, the relation between the gravitational constants $\hat\kappa$ and $\kappa$ is 
\begin{equation}
\kappa = \frac{\hat\kappa}{\sqrt{\text{Vol}\,\mathscr{M}_n}},
\end{equation}
where $\mathscr{M}_n$ is the internal manifold.

The gauge group of the vectors in the reduced theory is a subgroup of the duality group of the maximal lower-dimensional supergravity theory. Such supergravity theories are called  \emph{gauged supergravities}. A Kaluza-Klein compactification would have brought to an \emph{un}gauged supergravity theory. A gauged supergravity obtained using Scherk-Schwarz compatification can be equivalently obtained by gauging the isometry group of the internal space in Scherk-Schwarz compactification in the ungauged theory obtained using Kaluza-Klein compactification. In the next paragraphs the general features of gauged supergravities will be described in details.

\subsection{Gauged supergravities}

Consider a supergravity theory with duality group $G_D$. A gauged supergravity is a supergravity theory in which a subgroup $G$ of $G_D$ is gauged, that is, $G$ acts as a local symmetry. If we want to preserve the supersymmetry in the gauging, it is necessary to find the gauge fields inside supersymmetry multiplets. Denote the chosen vectors with $A_\mu^a$, transforming the fundamental representation of $\text{Lie}\,G_D$, defined on a $N$-dimensional vector space $V$, with index $a = 1,\dots,N$. $N$ has to be at least equal to the dimension of the algebra of $G$:
\begin{equation}
\dim{\text{Lie}\,G} \leqslant N.
\end{equation}
In order for the set of $A_\mu^a$'s to form the components of a $G$ connection, it is necessary to project the $a$ index in the \emph{adjoint} representation of $\text{Lie}\,G$ -- denote it with $\alpha = 1,\dots \dim{\text{Lie}\,G}$. This projection is realised by a mixed tensor $\vtheta_a{}^\alpha$, called \emph{embedding tensor}, in such a way that the vectors 
\begin{equation}
(A^\vtheta)_\mu{}^\alpha = A_\mu{}^a\,\vtheta_a{}^\alpha.
\end{equation}
are in $\text{Lie}\,G$. In other words, $\vtheta$ is a linear map on $V$, taking values in $\text{Lie}\,G_D$
\begin{equation}
\vtheta : v \in V \mapsto \vtheta(v) \in \text{Lie}\,G_D,
\end{equation}
with the requirement
\begin{equation} \label{ImmagineTheta}
\mathfrak{Im}\,\vtheta = \text{Lie}\,G \subset \text{Lie}\,G_D.
\end{equation}
Using $\vtheta$ in index-free notation, the gauge one-form field is \cite{Lavau:2017tvi, Lavau:2020pwa}
\begin{equation}
(A^\vtheta)_\mu\,\diff x^\mu = \vtheta(A_\mu)\,\diff x^\mu =:\vtheta(A).
\end{equation} 

In order for the gauging to preserve the supersymmetry, $\vtheta_a{}^\alpha$ must take values only on supersymmetry compatible representations. The irreducible representation content of $\vtheta_a{}^\alpha$ is obtained studying the decomposition of the tensor product between the fundamental and the adjoint of $\text{Lie}\,G_D$. For example, the duality group of maximal supergravity four dimensions is $E_{7(7)}$. Its fundamental representation is $\bb{56}$, and its adjoint representation is $\bb{133}$. They are such that
\begin{equation}
\bb{56} \otimes \bb{133} =  \bb{912} \oplus \bb{56} \oplus \bb{6480}.
\end{equation}
One can show that the compatibility with supersymmetry selects $\bb{912}$ as the representation of the embedding tensor. Similarly, in $d=3$, the duality group of maximal supergravity is $E_{8(8)}$. The fundamental representation and the adjoint one are the same $\bb{248}$, such that
\begin{equation}
\textbf{248} \otimes \textbf{248} = 
\left[\textbf{1} \oplus \textbf{3875} \oplus \textbf{27000}\right]_s \oplus \left[\textbf{248} \oplus \textbf{30380}\right]_a.
\end{equation}
The supersymmetry compatibility selects the symmetric part $\textbf{1} \oplus \textbf{3875} \oplus \textbf{27000}$. The selection required by supersymmetry to be preserved is called \emph{representation constraint} or \emph{linear constraint} on the embedding tensor.

Another constraint follows from the fact that $\mathfrak{Im}\,\vtheta$ must be a Lie algebra, as a consequence of the requirement \eqref{ImmagineTheta}. So, the following closedness condition must hold 
\begin{equation}
[\vtheta(v'),\vtheta(v)]=\vtheta(w(v',v)),
\end{equation}
for each $v,v'$ and for some $w=w(v',v)$ in $V$. A sufficient condition is requiring the embedding tensor to be invariant with respect to the action of $G_D$, 
\begin{equation} \label{ThetaInvariante}
\delta_{g'}\,\vtheta = 0, \quad \forall\;g'\in G_D.
\end{equation}
Let us prove this claim. The embedding tensor defines a bilinear on   $V \otimes (\text{Lie}\,G_D)^*$, taking values in $\mathbb{R}$, where $(\text{Lie}\,G_D)^*$ is the dual Lie algebra
\begin{equation}
(v^a,\gamma_\alpha) \in V \otimes (\text{Lie}\,G_D)^* \mapsto \langle\gamma,\vtheta(v)\rangle = \gamma_\alpha\,\vtheta_a{}^\alpha\,v^a \in \mathbb{R}.
\end{equation}
If the elements of $\text{Lie}\,G_D$ transform in the adjoint representation
\begin{equation}
\delta_{g'}\,g = \text{ad}_{g'}\,g = [g',g], \quad
g,g' \in \text{Lie}\,G_D,
\end{equation}
those in $(\text{Lie}\,G_D)^*$ transform in the dual adjoint one:
\begin{equation}
\delta_{g'}\,\gamma = \text{ad}^*_{g'}\,\gamma, \quad \gamma \in (\text{Lie}\,G_D)^*,
\end{equation} 
as a consequence of the requirement of the bilinear invariance:
\begin{equation}
0 = \delta_{g'}\,\langle\gamma,g\rangle = \langle\text{ad}_{g'}^*\,\gamma,g \rangle + \langle\gamma,\text{ad}_{g'}\,g\rangle.
\end{equation}
Now, denote with $\vrho$ the fundamental representation
\begin{equation}
\delta_{g'}\,v = \vrho_{g'}\,v, \quad v \in V, g' \in \text{Lie}\,G_D.
\end{equation}
Then, supposing the embedding tensor to be invariant, according to \eqref{ThetaInvariante}, we can compute
\begin{equation}
0 = \delta_{g'}\,\langle\gamma,\vtheta(v)\rangle = \langle\text{ad}^*_{g'}\,\gamma, \vtheta(v)\rangle +
\langle\gamma,\vtheta(\vrho_{g'}\,v)\rangle = 
-\langle\gamma,[g',\vtheta(v)]-\vtheta(\vrho_{g'}\,v)\rangle.
\end{equation}
Namely, for $g'=\vtheta(v')$, one finds 
\begin{equation}
[\vtheta(v'),\vtheta(v)]=\vtheta(\vrho_{g'}\,v),
\end{equation}
since $\gamma$ is arbitrary. The last one is the required closedness condition. In components it reads
\begin{equation}\label{QC1}
\vtheta_a{}^\alpha\,\vtheta_b{}^\beta\,f_{\alpha\beta}{}^\gamma = \vtheta_c{}^\gamma\,\vtheta_a{}^\alpha\,(T_\alpha)^c{}_b,
\end{equation}
where $(T_\alpha)^c{}_b$ is a generator of $\text{Lie}\,G_D$ in the fundamental. This condition on the embedding tensor is usually called \emph{quadratic constraint}, since it is quadratic in the embedding tensor. As a digression, let us write the quadratic constraint in a form which is common in literature \cite{Samtleben:2008pe, Trigiante:2016mnt}, and which will be useful in the following. Define
\begin{equation}\label{FalseBasis}
X_{ab}{}^c = (X_a)^c{}_b = \vtheta_a{}^\alpha\,(T_\alpha)^c{}_b.
\end{equation}
Therefore, \eqref{QC1} can be also written as\footnote{Computing the adjoint action $\delta_{X_f}$ generated by $X_f$ on the components $X_{ab}{}^c$
\begin{align}
\delta_{X_f}\,X_{ab}{}^c &= X_{fa}{}^e\,X_{eb}{}^c + X_{fb}{}^e\,X_{ae}{}^c - X_{fe}{}^c\,X_{ab}{}^e = ([X_f,X_a]+X_{fa}{}^e\,X_e)^c{}_b,
\end{align} 
we get an alternative formulation of the quadratic constraint
\begin{equation}
[X_f,X_a] = - X_{fa}{}^e\,X_e \Leftrightarrow \delta_{X_f}\,X_{ab}{}^c = 0.
\end{equation}}
\begin{equation}\label{QC2}
X_{ac}{}^e\,X_{be}{}^d - X_{bc}{}^e\,X_{ae}{}^d + X_{ab}{}^e\,X_{ec}{}^d = 0.
\end{equation}

The closedness condition does not properly define a Lie algebra, but a \emph{Leibniz algebra}. To see why, define
\begin{equation}
\Ll v,v' \Lr := \vrho_{\vtheta(v)}\,v', \quad\forall\;v,v'\in V.
\end{equation}
The closedness condition is an \emph{intertwiner} between $\Ll \cdot,\cdot \Lr$ and the Lie product  $[\cdot,\cdot]$ in $\text{Lie}\,G$:
\begin{equation}
[\vtheta(v),\vtheta(v')] = \vtheta(\Ll v,v' \Lr).
\end{equation}

To show that $\Ll \cdot, \cdot \Lr$ is a Leibniz product, and so that $(V,\Ll \cdot, \cdot \Lr)$ is a Leibniz algebra, it is sufficient to show that $\Ll \cdot,\cdot \Lr$ satisfies the Leibniz identity. Using the compatibility condition, 
\begin{equation}
[\vrho_g,\vrho_{g'}]=\vrho_{[g,g']},\quad \forall\;g,g'\in \mathfrak{g},
\end{equation}
ensuring $\vrho$ to be a representative, we can compute for each $u,v,w\in V$,
\begin{align}
\Ll u, \Ll v, w \Lr \Lr &= 
(\vrho_{\vtheta(u)} \circ \vrho_{\vtheta(v)})(w)
= \nonumber \\
& = \vrho_{[\vtheta(u),\vtheta(v)]}(w) + (\vrho_{\vtheta(v)} \circ \vrho_{\vtheta(u)})(w) = \nonumber \\
& = \vrho_{\vtheta(\Ll u,v \Lr)}(w) + \Ll v, \Ll u,w \Lr \Lr = \nonumber \\
& = \Ll \Ll u,v \Lr, w \Lr + \Ll v, \Ll u,w \Lr \Lr,
\end{align}
which is the Leibniz identity, as we wanted.

The gauge transformation of the connection $\vtheta(A)$ with respect to $\vtheta(v)$ is
\begin{align}
\delta_{\vtheta(v)}\,\vtheta(A) &= -(\diff + \delta_{\vtheta(A)})\,\vtheta(v) = \nn\\
& = -\diff \vtheta(v) - [\vtheta(A),\vtheta(v)] = \nn\\
& = -\vtheta(\diff v + \Ll A,v \Lr) =: -\vtheta(\text{D}\,v).
\end{align}
The curvature or field strength of $\vtheta(A)$ is 
\begin{equations}
F^\vtheta &= \diff\vtheta(A) + \tfrac{1}{2}\,[\vtheta(A),\vtheta(A)] = \nn\\
& = \vtheta(\diff A + \tfrac{1}{2}\,\Ll A,A \Lr) =: \vtheta(F).
\end{equations}
The gauge transformation of the curvature is\footnote{Use $\delta\,F = \delta\,(\diff A + A^2) = \diff\,\delta\,A + [A,\delta\,A] = (\diff + [A,\cdot])\,\delta\,A = \text{D}\,\delta\,A$.}
\begin{align}
\delta_{\vtheta(v)}\,F^\vtheta &= (\diff + \delta_{\vtheta(A)})\,\delta_{\vtheta(v)}\,\vtheta(A) = \nn\\
& = -(\diff + \delta_{\vtheta(A)})\,\vtheta(\diff v + \Ll A,v \Lr) = \nn\\
& = -\vtheta(\diff\,\Ll A,v \Lr) - \delta_{\vtheta(A)}\,\vtheta(\diff v + \Ll A,v \Lr) = \nn\\
& = -\vtheta(\diff\,\Ll A,v \Lr + \Ll A,\diff v + \Ll A,v \Lr) = \nn\\
& = -\vtheta(\Ll \diff A,v \Lr + \Ll A, \Ll A,v \Lr \Lr) = \nn\\
& = -\vtheta(\Ll \diff A + \tfrac{1}{2}\,\Ll A,A \Lr,v\Lr) = -\vtheta(\Ll F,v \Lr).
\end{align}
So, we obtained the usual expressions, “dressed" with the embedding tensors, with the Lie bracket replaced by the Leibniz bracket.

\subsection{Tensor hierarchy}

The set $\{X_a\}$, where $X_a$ is defined in \eqref{FalseBasis}, generates the algebra of the gauge group $G$, and we can use $A = A_\mu{}^a\,X_a\,\diff x^\mu$ as (non-abelian) connection. Nevertheless, $\{X_a\}$ is not a basis in general, because the fundamental index can assume more values than the adjoint one. So, the $X_a$'s are linearly dependent, and the components $A_\mu{}^a$ are not uniquely defined, since one can always add some $t_\mu{}^a$, such that they vanish when they are saturated with the generators $t_\mu{}^a\,X_a = 0$,
\begin{equation}
A_\mu = A_\mu{}^a\,X_a = (A_\mu{}^a + t_\mu{}^a)\,X_a.
\end{equation}
We call the $t_\mu{}^a$ terms \emph{trivial parameters} and we denote with $\overset{t}{=}$ the equal up to trivial parameters:
\begin{equation}
A_\mu{}^a\,X_a = A'_\mu{}^a\,X_a  \Leftrightarrow
A_\mu{}^a = A'_\mu{}^a + t_\mu{}^a, \;\;t_\mu{}^a\,X_a = 0 \Leftrightarrow A_\mu{}^a \overset{t}{=} A'_\mu{}^a.
\end{equation}
It is convenient to extract from the set of $X_a$ a basis $\lbrace X_{a'}\rbrace_{a'=1,\dots,\text{dim}\,\text{Lie}\,G}$ for $\text{Lie}\,G$. Denote the structure constants with $f_{a'b'}{}^{c'}$:
\begin{equation}
[X_{a'},X_{b'}] = f_{a'b'}{}^{c'}\,X_{c'}.
\end{equation}
The remaining $X_{a''}$, $a''=\dim\,\text{Lie}\,G, \dots, N$, are linear combinations of the $X_{a'}$:
\begin{equation}
X_{a''} = C_{a''}{}^{a'}\,X_{a'},
\end{equation} 
for some constant $C_{a''}{}^{a'}$. Since
\begin{align}
[X_{a''},X_{b'}] &= C_{a''}{}^{a'}\,f_{a'b'}{}^{c'}\,X_{c'},\\
[X_{a''},X_{b''}] &= C_{a''}{}^{a'}\,C_{b''}{}^{b'}\,f_{a'b'}{}^{c'}\,X_{c'},
\end{align}
we can write
\begin{equation}
[X_a,X_b] = f_{ab}{}^c\,X_c
\label{comm}
\end{equation}
defining
\begin{equation}
f_{ab}{}^{c''} = 0,\quad
f_{a''b'}{}^{c'} = C_{a''}{}^{a'}\,f_{a'b'}{}^{c'},\quad
f_{a''b''}{}^{c'} = C_{a''}{}^{a'}\,C_{b''}{}^{b'}\,f_{a'b'}{}^{c'}.
\end{equation}
So, $f_{ab}{}^c$ are not in general totally antisymmetric. Nevertheless, the expression \eqref{comm} involves the antisymmetric part only, so that 
\begin{equation}
f_{(ab)}{}^c\,X_c = 0.
\end{equation}
The $\text{Lie}\,G$ gauge transformation is
\begin{equation}\label{GaugeTrueBasis}
\delta_g\,\tilde{A}_\mu{}^{a'} = - D_\mu\,g^{a'} =  -\de_\mu\,g^{a'} - f_{b'c'}{}^{a'}\,\tilde{A}_\mu{}^{b'}\,g^{c'},
\end{equation}
where $\tilde{A}_\mu{}^{a'}$ are the unique components of $A_\mu$ with respect to the true basis $\lbrace X_{a'}\rbrace$
\begin{equation}
A_\mu = \tilde{A}_\mu{}^{a'}\,X_{a'}.
\end{equation}
On the other hand, 
\begin{equation}
A_\mu = A_\mu{}^a\,X_a = A_\mu{}^{a'}\,X_{a'} + A_\mu{}^{a''}\,X_{a''} = (A_\mu{}^{a'} + A_\mu{}^{a''}\,C_{a''}{}^{a'})\,X_{a'}.
\end{equation}
So, up to trivial parameters, we obtain the relation
\begin{equation}
\tilde{A}_\mu{}^{a'} \overset{t}{=} A_\mu{}^{a'} + X_{a''}{}^{a'}\,A_\mu{}^{a''}.
\end{equation}
Replacing the tilde components in \eqref{GaugeTrueBasis}, 
\begin{equation}
\left[\delta\,A_\mu{}^{a'} + (\de_\mu\,g^{a'} + f_{bc}{}^{a'}\,A_\mu{}^b\,g^c)\right] +
C_{a''}{}^{a'}\,\left[\delta\,A_\mu{}^{a''} + \de_\mu\,g^{a''}\right] \overset{t}{=} 0,
\end{equation}
so that, adding in the second term $f_{bc}{}^{a''}\,A_\mu{}^b\,g^c = 0$, one gets
\begin{equation}
\delta\,A_\mu{}^a \overset{t}{=} -\de_\mu\,g^a - f_{bc}{}^a\,A_\mu{}^b\,g^c.
\end{equation}
Formally, this is the same expression as that of $\delta\,\tilde{A}_\mu{}^{a'}$. But there is a crucial difference: now $f_{ab}{}^c$ is not antisymmetric.  

Consider the field strength
\begin{equation}
F_{\mu\nu} = \de_{[\mu}\,A_{\nu]} + [A_\mu,A_\nu],
\end{equation} 
whose components $\tilde{F}_{\mu\nu}{}^{a'}$ with respect to the true basis, and $F_{\mu\nu}{}^a$ with respect to the redundant basis, with a natural choice for the trivial parameters, read
\begin{equation}
\tilde{F}_{\mu\nu}{}^{a'} = \de_{[\mu}\,\tilde{A}_{\nu]}{}^{a'} + f_{b'c'}{}^{a'}\,\tilde{A}_{\mu}{}^{b'}\,\tilde{A}_{\nu}{}^{c'},\quad
F_{\mu\nu}{}^a = \de_{[\mu}\,A_{\nu]}{}^a + \tfrac{1}{2}\,f_{[bc]}{}^a\,A_{\mu}{}^{b'}\,A_{\nu}{}^{c'}.
\end{equation}

The gauge transformation of the curvature in the true basis is
\begin{equation}
\delta_g\,\tilde{F}_{\mu\nu}{}^{a'} =  f_{c'b'}{}^{a'}\,\tilde{\mathscr{F}}_{\mu\nu}{}^{b'}\,g^{c'},
\end{equation}
whereas in the redundant basis is
\begin{align}
\delta_g\,F_{\mu\nu}{}^a &= 
-f_{bc}{}^a\,F_{\mu\nu}{}^b\,g^c - \tfrac{1}{2}\,f_{(bc)}{}^a\,A_{[\mu}{}^b\,\delta_g\,A_{\nu]}{}^c = \nn\\
& = f_{cb}{}^a\,\tilde{F}_{\mu\nu}{}^b\,g^c - \tfrac{1}{2}\,f_{(bc)}{}^a\,\tilde{F}_{\mu\nu}{}^b\,g^c - \tfrac{1}{2}\,f_{(bc)}{}^a\,A_{[\mu}{}^b\,\delta_g\,A_{\nu]}{}^c-
\end{align}
Therefore, $\tilde{F}_{\mu\nu}{}^{a'}$ homogeneously transforms, whereas $F_{\mu\nu}{}^a$ does not. We want to find the components of$H_{\mu\nu}{}^a$ of a curvature, in the redundant basis, which is equivalent to $F_{\mu\nu}{}^a$ up to trivial terms, but covariant
\begin{equation}
H_{\mu\nu}{}^a \overset{t}{=} F_{\mu\nu}{}^a, \quad
\delta_g\,H_{\mu\nu}{}^{a} =  f_{cb}{}^{a}\,H_{\mu\nu}{}^{b}\,g^{c}.
\end{equation}
We can get a suitable $H_{\mu\nu}{}^a$ by deforming $F_{\mu\nu}{}^a$ with a term proportional to $f_{(bc)}{}^a$, which ensures the equivalence with $F_{\mu\nu}{}^a$, and proportional to a two-form $B_{\mu\nu}{}^{bc}$, with a proper gauge transformation to ensure the covariance:
\begin{equation}
H_{\mu\nu}{}^a = \tilde{F}_{\mu\nu}{}^a + f_{(bc)}{}^a\,B_{\mu\nu}{}^{bc},
\end{equation}
where we choose the two-form symmetric in the upper indices
\begin{equation}
B_{\mu\nu}{}^{ab} = B_{\mu\nu}{}^{ba}.
\end{equation}
The right transformation turns out to be
\begin{equation}
\delta\,B_{\mu\nu}{}^{ab} = D_{[\mu}\,h_{\nu]}{}^{ab} + H_{\mu\nu}{}^a\,g^b + A_{[\mu}{}^{(a}\,\delta\,A_{\nu]}{}^{b)}.
\end{equation}
The last two terms in $\delta\,B_{\mu\nu}{}^{bc}$  cancel out the undesidered terms in $\delta\,\tilde{F}_{\mu\nu}{}^a$. We would like to include also a gauge transformation for the two-form itself, which is generated by a one-form of components $h_\mu{}^{ab}$, symmetric in its upper indices. This term induces in turn a modification in the transformation of $A_\mu{}^a$, but this modification is a simple shift in $h_\mu{}^{ab}$, because they appear in the covariant derivative:
\begin{equation}
\delta\,A_\mu{}^a = -D_\mu\,g^a + \tfrac{1}{2}\,f_{(bc)}{}^a\,h_\mu{}^{bc}.
\end{equation}
Moreover, this shift is a trivial parameter, so that the new transformation of $A_\mu{}^a$ is equivalent to the previous one.  

The addition of a two-form induces the so-called \emph{tensor hierarchy}. Indeed, when the curvature of $B_{\mu\nu}{}^{ab}$ is introduced, one may ask if it is covariant. If it is not, one has to modify it by introducing a three-form, and so on. The procedure ends when the dimension of the spacetime is reached, because a $p$--form in $d$ dimensions can be non-vanishing only if $p\leqslant d$.

\subsection{Examples of gauged supergravities}

The first discovered example of gauged supergravity was the four-dimensional $\text{SO}(8)$ gauged supergravity, obtained as Scherk-Schwarz compactification of eleven-dimensional supergravity on $S^7$ \cite{deWit:1982bul, deWit:1986oxb}. $\text{SO}(8)$ is indeed the isometry group of $S^7 \simeq \text{SO}(8)/\text{SO}(7)$.\footnote{We computed that, compactifying up to four dimensions, one gets 28 vectors, and the dimension of $\text{SO}(8)$ is precisely $28$.}

If we want to obtain a five-dimensional gauged supergravity, we have to consider 27 vectors, but there is no simple Lie group with this dimension. The vectors transform in the \textbf{27} representation of $E_{6(6)}$, and the gauge group should be a subgroup of $E_{6(6)}$. So, we can seek the largest subgroup of $E_{6(6)}$, such that the $\textbf{27}$ representation of $E_{6(6)}$ containes the adjoint of the gauge group, as the embedding tensor procedure suggests. But, remember that $2 \equiv 1$ in five dimensions. So, we can consider 15 vectors out of the 27 ones, using $\text{SO}(6)$ as gauge group, since it is contained into $E_{6(6)}$ and its dimension is precisely 15, and transforming the remaining 12 vectors into the same number of two-form fields $A^{(2)}$. They satisfy the equation of motion $A^{(2)} = c\,{\star{F^{(3)}}}$, where $F^{(3)}$ is the corresponding curvature. So, they have no gauge invariance, and the gauged theory one obtains is consistent. The previous equation is called odd self-dual equation. Given a $p$--form $A^{(p)}$, we say it is self-dual if $A^{(p)} = {\star{A^{(p)}}}$, but this is true if $p = d-p$, that is, only in even dimension. The previous condition is similar to the last one, but it involves the dual of the curvature, instead of the $p$--form itself, and it holds in odd dimensions, because $p = d-(p+1)$ \cite{Gunaydin:1985cu}.

The same trick has to be exploited to obtain the $d=7$ gauged supergravity on $S^4$, where the two-form fields must be dualised into three-form fields with no gauge invariance, due to odd self-dual equations of motion on the three-forms \cite{Pernici:1984xx}.

\subsection{$E_{7(7)}$ generalised geometry}\label{ExceptionalGeometry}

Studying the dimensional reduction of eleven-dimensional supergravity up to, say, four dimensions, we saw that the 128 bosonic degrees of freedom of the starting higher-dimensional theory are repackaged into 70 scalars, 56 vectors (including their duals), with two degrees of freedom each, and a four-dimensional metric, with two degrees of freedom. We argue that the manifest duality group of the compactified theory is enlarged to the maximal non-compact form of the exceptional Lie group $E_{7(7)}$, since the scalar fields can be seen as the coordinates on the coset space $E_{7(7)}/\text{SU}(8)$. The vectors can be related to $E_{7(7)}$ too. Indeed, 56 is the dimension of fundamental representation of $E_{7(7)}$, which we denote with $\textbf{56}$. In other words, we expect that the vectors of the four-dimensional theory and their dual can be rearranged in forming a vector in $E_{7(7)}$. The partition $56 = 7+21+21+7$ of the vectors, coming from the metric $g_{\hat\mu\hat\nu}$, the three-form $A^{(3)}$, the dual six-form $\tilde{A}^{(6)}$, and the linearised dual graviton $\tilde{h}_{\hat{\mu}_1\dots\hat{\mu}_8,\hat\nu}$ in eleven-dimensional supergravity, corresponds to the decomposition of $\textbf{56}$ according to $\text{GL}(7,\mathbb{R})$, which is the group of linearised diffeomorphisms in the seven-dimensional compact internal manifold. 

This observation suggests that it is possible to formulate the degrees of freedom of eleven-dimensional supergravity in a manifest $E_{7(7)}$-covariant way. This is the idea underlying the so-called \emph{exceptional field theory}. In the next sections, besides the four-dimensional case, we will discuss in details also the three-dimensional one, which presents some peculiar features. But the procedure is general and can be extended to all the dimensions of the internal manifold $1 \leqslant n \leqslant 8$, in which case the duality group is $E_{n(n)}$.\footnote{The Dynkin diagram of $E_7$, whose dimension is 133, is obtained starting from that of $E_8$, whose dimension is 248, by removing a node; similarly for $E_6$, with dimension 78, starting from $E_7$. Continuing in removing nodes  successively, one could obtain $E_5$, $E_4$, $E_3$, $E_2$, and $E_1$ (the so-called exceptional series"), but the resulting Diagrams are those of $D_5$, $D_4$, $A_3$, $A_1 \oplus A_1$, and $A_1$. The maximal non-compact associated groups are $\text{SO}(5,5)$, $\text{SL}(5)$, $\text{SL}(3)\otimes\text{SL}(2)$, $\text{SL}(2)\otimes\mathbb{R}$, $\mathbb{R}$ respectively, with dimensions $45, 28, 11, 4, 1$.
\tble{E/8,E/7,E/6,D/5,D/4,A/3,A/1}}

Now we study in details the decomposition of the fundamental representation $\textbf{56}$ of $E_{7(7)}$ according to $\text{GL}(7,\mathbb{R})$. Let $E$ be the vector bundle in the $\textbf{56}$. According to $\text{SL}(8,\mathbb{R})$,
\begin{equation}
\textbf{56} \rightarrow \textbf{28} \oplus \overline{\textbf{28}}, 
\end{equation}
where $\textbf{28}$ is the rank 2 antisymmetric of $\text{SL}(8,\mathbb{R})$, with dimension $\frac{8(8-1)}{2}=28$. If $V$ is the fundamental representation of $\text{SL}(8,\mathbb{R})$, then we can write \cite{PiresPacheco:2008qik, Coimbra:2011ky}
\begin{equation}
E \simeq \wedge^2\,V \oplus \wedge^2\,V^*.
\end{equation}
We want to write $V$ in terms of the fundamental of $\text{GL}(7,\mathbb{R})$, denoted with $F$. This means that one has to write a $8\times 8$ matrix $\mathscr{M}$ with $\det = 1$ in terms of a $7\times 7$ matrix $M$ with nonvanishing determinant:
\begin{equation}
\mathscr{M} = \begin{pmatrix}
(\text{det}\,M)^\alpha\,M & 0 \\ 0 & (\text{det}\,M)^\beta
\end{pmatrix},
\end{equation}
whose determinant is $1 = \text{det}\,\mathscr{M} = (\text{det}\,M)^{7\alpha + 1 + \beta}$. Choosing $\alpha = -\frac{1}{4}$, we have $\beta = \frac{3}{4}$.

If $v \in V$, $v^M = (v^m,v^8)$, $M=1,\dots,8$, $m=1,\dots,7$: $v^m$ is a $\text{GL}(8,\mathbb{R})$ vector with weight $-\frac{1}{4}$; $v^8$ is a $\text{GL}(7,\mathbb{R})$ scalar with weight $\frac{3}{4}$. Since the Levi-Civita tensor in $\text{GL}(7,\mathbb{R})$, which is in $\wedge^7\,F^*$, has weight 1 ($\text{det}\,M=\vepsilon_{m_1\dots m_7}\,M^{m_1}{}_1\dots M^{m_7}{}_7$), one can use it to build tensor densities. Notice that the duality between $\tilde{v}_{m_1\dots m_\alpha}$ and $\vepsilon_{m_1\dots m_\alpha\,n_{\alpha+1}\dots n_7}\,V^{n_{\alpha+1}\dots n_7}$ is equivalent to write
\begin{equation}
\wedge^\alpha\,F^* \simeq (\wedge^7\,F^*) \otimes \wedge^{7-\alpha}\,F.
\end{equation}
Therefore, we find that
\begin{equation}
V \simeq (\wedge^7\,F)^{-1/4} \otimes F \oplus (\wedge^7\,F)^{3/4}.
\end{equation}
Moreover, $\vepsilon^{m_1\dots m_7}$, which sits in $\wedge^7\,F$, has opposite weight of $\vepsilon_{m_1\dots m_7}$:
\begin{equation}
(\Lambda^7\,F)^\alpha \simeq (\Lambda^7\,F^*)^{-\alpha}.
\end{equation}
In order to build $\wedge^2\,V$ and $\wedge^2\,V^*$, consider that an object $v^{MN}$ in $\wedge^2\,V$ splits according to $(v^{mn},v^{m8})$, with weight $-\frac{1}{4}-\frac{1}{4}=-\frac{1}{2}$ and $-\frac{1}{4}+\frac{3}{4}=\frac{1}{2}$. Therefore,
\begin{align}
\wedge^2\,V &\simeq (\wedge^7\,F)^{-1/2} \otimes
\wedge^2\,F \oplus (\wedge^7\,F)^{1/2}\,\otimes F \simeq \nn\\
& \simeq (\wedge^7\,F^*)^{1/2}\otimes [(\wedge^7\,F^*)^{-1} \otimes \wedge^5\,F^*]\otimes (\wedge^7\,F^*)^{-1/2} \otimes F \simeq\nn\\
& \simeq (\wedge^7\,F^*)^{-1/2} \otimes [\wedge^5\,F^* \oplus F].
\end{align}
The tensors with lowered indices have the same weight as those with raised ones, since the indices are lowered/raised using the metric tensor, which does not change the weight. Therefore,
\begin{align}
\wedge^2\,V^* &\simeq (\wedge^7\,F)^{-1/2} \otimes
\wedge^2\,F \oplus (\wedge^7\,F^*)^{1/2}\,\otimes F^* \simeq \nn\\
& \simeq (\wedge^7\,F^*)^{-1/2}\,\otimes [\wedge^2\,F^* \oplus \Lambda^7\,F^* \otimes F^*].
\end{align}
Finally, we have obtained that \cite{PiresPacheco:2008qik, Coimbra:2011ky}
\begin{equation}
E \simeq (\wedge^7\,F^*)^{-1/2} \otimes [F \oplus \wedge^2\,F^* \oplus \wedge^5\,F^* \oplus (\wedge^7\,F^* \otimes F^*)],
\end{equation}
where the overall factor $(\wedge^7\,F^*)^{-1/2}$ can be omitted by a suitable isomorphism. Summarising, if $F \rightarrow T\mathscr{M}$ is the tangent space of the internal manifold $\mathscr{M}$, 
\begin{equation}
E \simeq T\mathscr{M} \oplus \wedge^2\,T^*\mathscr{M}\oplus \wedge^5\,T^*\mathscr{M} \oplus (\wedge^7\,T^*\mathscr{M} \otimes T^*\mathscr{M}).
\end{equation}
Consistently with the previous counting, the dimensions of the various pieces are 7, 21, 21, and 7 respectively. 

\subsection{$E_{7(7)}$ generalised Lie derivative}

The $\text{GL}(7,\mathbb{R})$ decomposition of the representation $\mathbf{56}$ can be also interpreted in terms of the \emph{internal symmetries} of eleven-dimensional supergravity with seven internal dimensions. The internal symmetries are the internal diffeomorphisms, parametrised by $\xi^m$ ($\# = 7$); the gauge transformation of the three-form $A^{(3)}$, with a two-form parameter $\omega^{(2)}$ ($\# = {{7}\choose{2}}=21$ components in seven dimensions); the gauge transformation of the dual six-form $\tilde{A}^{(6)}$, with a five-form parameter $\sigma^{(5)}$ ($\# = {{7}\choose{5}}=21$ components in seven dimensions); the gauge transformation of the linearised dual graviton, with a vector-valued seven-form parameter $\tau_{m_1\dots m_7,n}$ ($\# = 7\,{{7}\choose{7}}=7$ components in seven dimensions). Thus, the previous four parameters can be arranged in a $E_{7(7)}$ vector
\begin{equation}
\xi + \omega + \sigma + \tau = X \in E.
\end{equation}

We can define a \emph{generalised Lie derivative} $L_X: E \rightarrow E$, generating both the internal diffeomorphisms and the gauge symmetries, as the Dorfman derivative associated to the generalised bundle $E$. It is required to satisfy the  Leibniz identity 
\begin{equation}\label{LeibnizIdentity}
L_X\,(L_V\,\cdot) =
L_{L_X V}\cdot + L_V\,L_X\,\cdot,
\end{equation}
which can be also written as 
\begin{equation}\label{LeibnizIdentityAlgebra}
[L_X,L_V] = L_{L_X V}.
\end{equation}
The generalised Lie derivative corresponds to the usual Lie derivative, with the rotational piece projected on the adjoint representation of (the algebra of) the duality group (in the case under discussion $E_{7(7)}$). This means that the Lie derivative generated by $X^M$ on the vector density $V^M$ with density weight $\lambda$
\begin{equation}
\mathcal{L}_X\,V^M = X^N\,\de_N\,V^M - V^N\,\de_N\,X^M + \lambda\,\de_N\,X^N\,V^M
\end{equation}
is replaced by \cite{Berman:2012vc}
\begin{equation}\label{GenLieDerP}
L_X\,V^M = X^N\,\de_N\,V^M - \alpha\,V^N\,(P^{(\text{adj)}})^M{}_N{}^P{}_Q\,\de_P\,X^Q + \lambda\,\de_N\,X^N\,V^M,
\end{equation}
where $P^{(\text{adj})}$ is the projector on the adjoint representation, and $\alpha$ is a constant. The index $M,N=1,\dots, 56$ is in the fundamental representation of the duality group. It is necessary to set to zero some representations, which sit in the symmetrised product of the derivatives $\de_M$, in order for the Leibniz identity \eqref{LeibnizIdentity} to be satisfied. Denoting these representations with $N$, and with $P^{(N)}$ the corresponding projector, one has to \emph{impose}
\begin{equation}\label{SectionConstraintP}
(P^{(N)})^M{}_N{}^P{}_Q\,\de_M\,\de_P = 0,
\end{equation}
which is called \emph{section constraint}. 

A widespread parametrisation of the generalised Lie derivative is the following:
\begin{equation}\label{GenLieDerY}
\mathcal{L}_X\,V^M = X^N\,\de_N\,V^M - V^N\,\de_ N\,X^M + (\lambda-\omega)\,\de_N\,X^N\,V^M + Y^{MP}{}_{QN}\,\de_P\,X^Q\,V^N,
\end{equation}
for some constant $\omega$. It is equivalent to \eqref{GenLieDerP} if 
\begin{equation}\label{YTensor}
Y^{MP}{}_{QN} = \delta^M_Q\,\delta^P_N + \omega\,\delta^M_N\,\delta^P_Q - \alpha\,(P^{(\text{adj})})^M{}_N{}^P{}_Q.
\end{equation}
This tensor is useful because it measures the deviation of the generalised Lie derivative from the usual one. The section constraint \eqref{SectionConstraintP} can be written as
\begin{equation}\label{SectionConstraintY}
Y^{MP}{}_{QN}\,\de_M\,\de_P = 0.
\end{equation}
The section constraint says that not all the derivatives $\de_M$, treated algebraically (consider them as the momentum in Fourier space), are independent:
\begin{equation}
\de_M = \mathcal{E}_M{}^m\,\frac{\de}{\de y^m} = \mathcal{E}_M{}^m\,\de_m,
\end{equation}
where $y^m$ are the coordinates of the internal manifold, $m=1,\dots,n$, and $\mathcal{E}_M{}^m$ is a constant rectangular matrix, known \emph{section matrix}. In this way, the section constraint \eqref{SectionConstraintY} reads
\begin{equation}\label{SectionConstraintYSectionMatrix}
Y^{MP}{}_{QN}\,\mathcal{E}_M{}^m\,\mathcal{E}_P{}^p = 0.
\end{equation}
In general, we say that an object $Z_M$ is \emph{on section} if its components are not independent. Selecting in $\{Z_M\}_M$ a subset $\{z_m\}_m$ of independent components, we can write
\begin{equation}\label{OnSectionObject}
Z_M = \mathcal{E}_M{}^m\,z_m.
\end{equation}

As in the other examples of generalised geometry we met in \ref{DFT} and \ref{ParallSn}, the generalised Lie derivative has a non-trivial kernel. This means that there are non-vanishing vectors $\tilde{X}$, called \emph{trivial parameters}, such that the generalised Lie derivative generated by them vanishes whatever the argument of the derivative is
\begin{equation}
L_{\tilde{X}}\,V^M = 0, \;\forall\;V^M.
\end{equation}
Moreover, the generalised Lie derivative is not antisymmetric:
\begin{equation}
L_X\,V \neq -L_V\,X,
\end{equation}
but the symmetric part is always a trivial parameter:
\begin{equation}
L_{\frac{1}{2}(L_X\,V +L_V\,X)}\,W^M = 0, \;\forall\;W^M.
\end{equation}
Therefore,
\begin{equation}
L_{L_X\,V}\,W^M = 
L_{\frac{1}{2}(L_X V -L_V X)}\,W^M \;\forall\;W^M,
\end{equation}
and this is consistent with the Leibniz identity \eqref{LeibnizIdentityAlgebra}, which requires the right-hand side to be antisymmetric in swapping $X \leftrightarrow V$.

Let us focus on the $E_{7(7)}$ case. We have seen in the previous Section that each element $V$ in the generalised vector bundle $E$ can be decomposed according to $\text{GL}(7,\mathbb{R})$ as
\begin{equation}
V = V^{(1,0)} + V^{(0,2)} + V^{(0,5)} + V^{(1,7)},
\end{equation}
where the component $V^{(i,j)}$ has $i=0,1$ vector indices, and $j=2,5,7$ form indices. The decomposition of $L_X\,V$ is the following:
\begin{align}
L_X\,V &= \mathcal{L}_{X^{(1,0)}}\,V^{(1,0)} + (\mathcal{L}_{X^{(1,0)}}\,V^{(0,2)} - \iota_{V^{(1,0)}}\,\diff X^{(0,2)}) \,+\nn\\
& + (\mathcal{L}_{X^{(1,0)}}\,V^{(0,5)} - \iota_{V^{(1,0)}}\,\diff X^{(0,5)} - V^{(0,2)}\,\diff X^{(0,2)}) \,+\nn\\
& + (\mathcal{L}_{X^{(1,0)}}\,V^{(0,7+1)} - j\,V^{(0,5)}\,\diff X^{(0,2)} - j\,V^{(0,2)}\,\diff X^{(0,5)}),
\end{align}
where the operator $j$ acts formally on a eight-form (which should vanish in a seven-dimensional manifold) of the form $\alpha^{(p+1)}\,\beta^{(7-p)}$ by selecting the component $\alpha_{m[n_1\dots n_p}\,\beta_{n_{p+1}\dots n_7]}$. Notice that $X^{(0,2)}$ and $X^{(0,5)}$ appear only as $\diff\,X^{(0,2)}$ and $\diff\,X^{(0,5)}$, as in gauge transformations, and $X^{(0,7+1)}$ does not appear. Therefore, $0+\diff \tilde{X}^{(0,1)} + \diff \tilde{X}^{(0,4)} + \tilde{X}^{(0,7+1)}$ is a trivial parameter, for all $\tilde{X}^{(0,1)}$, $\tilde{X}^{(0,4)}$ and $\tilde{X}^{(0,7+1)}$:
\begin{equation}
L_{0+\diff \tilde{X}^{(0,1)} + \diff \tilde{X}^{(0,4)} + \tilde{X}^{(0,7+1)}}\,V^M = 0, \quad \forall\;V^M.
\end{equation}
Thus, the non-empty kernel of the generalised Lie derivative reflects the reducibility of the gauge transformation in eleven-dimensional supergravity and the decoupling of the dual graviton degrees of freedom.

Notice that the generalised Lie derivative is closed on the generalised vectors whose unique non-vanishing component is the vector one, and in that case the generalised Lie derivative is equal to the usual Lie derivative:
\begin{equation}
L_{X^{(1,0)}+0+0+0}\,(V^{(1,0)}+0+0+0) = \mathcal{L}_{X^{(1,0)}}\,V^{(1,0)} + 0 + 0 + 0.
\end{equation}
The symmetric part is not vanishing in general, but it is always equal to a trivial parameter:
\begin{align}
L_X\,W + L_W\,X &= 0 + \diff\,(\iota_{X^{(1,0)}}\,W^{(0,2)}+\iota_{W^{(1,0)}}\,X^{(0,2)}) \,+\nn\\
& + \diff\,(\iota_{X^{(1,0)}}\,W^{(0,5)}+\iota_{W^{(1,0)}}\,X^{(0,5)}-X^{(0,2)}\,W^{(0,2)}) + (\dots),
\end{align}
where we used $\mathcal{L}_{X^{(1,0)}} = \iota_{X^{(1,0)}}\,\diff + \diff\,\iota_{X^{(1,0)}}$.

One can also find trivial parameters in $E_{7(7)}$-covariant way. Consider $\tilde{X}^M = \Omega^{MN}\,Z_N$, where $\Omega^{MN}$ are the components of the invariant symplectic form of $E_{7(7)}$ and $Z_N$ is on section, as in \eqref{OnSectionObject}. Then, $\tilde{X}^M$ is a trivial parameter for the $n=7$ Lie derivative \cite{Hohm:2013uia}. The first index in $\Omega^{MN}$ takes values only in the $\overline{\mathbf{7}}$ representation, since the index of $Z_N$ must sit in the $\overline{\mathbf{7}}$ representation. So, $\tilde{X}^M$ sits in the $\overline{\mathbf{7}}$ representation as well. 

\subsection{Exceptional field theory}\label{ExFT}

When eleven-dimensional supergravity is compactified in an $n$-dimensional internal space, by replacing the Kaluza-Klein ansatz (or, more in general, by the Scherk-Schwarz one) in the general covariant action, the duality group $E_{n(n)} \times \mathbb{R}^+$ of the reduced theory is not manifest. \emph{Exceptional field theory} provides a reformulation of eleven-dimensional supergravity, by means of generalised geometry, which allows to organise the fields of the reduced theory in multiplets of the duality group \cite{Hohm:2013vpa, Hohm:2013uia, Hohm:2014fxa}.\footnote{Type IIB and type IIA supergravity (with or without Romans mass deformation) can also been formulated in exceptional way \cite{Ciceri:2016dmd, Hohm:2013vpa}.} The idea is to partially break the local Lorentz invariance of eleven-dimensional supergravity, as in Kaluza-Klein and Scherk-Schwarz reduction ansatz, and to split the eleven-dimensional coordinates $\hat{x}^{\hat\mu}$ in $d$ ``external coordinates" $x^\mu$ and $n = 11-d$ ``internal coordinates" $y^m$, but without any restriction on the dependence of the internal coordinates in the fields. The theory is formulated in a non-covariant fashion with respect to eleven-dimensional diffeomorphisms, but what it is earned is an $E_{n(n)}$-covariant formulation of the field content, which is maintained even when the theory is reduced to lower dimension. 

Indeed, the fields are organised in a metric tensor $\hat{g}_{\mu\nu}(x,y)$, which is a symmetric rank-two tensor with respect to $d$-dimensional diffeomorphisms, and it is a scalar with respect to the duality group; a vector $\hat{A}_\mu{}^M(x,y)$, which is a one-form with respect to $d$-dimensional diffeomorphisms, and it is a vector with respect to the duality group, $M$ being an index in a representation of $E_{n(n)}$, which we denote with $\mathbf{f}_n$ -- in particular $\mathbf{f}_7 = \mathbf{56}$, and $\mathbf{f}_8 = \mathbf{248}$; a coset of scalar fields $E_{n(n)}/K(E_{n(n)})$, where $K(E_{n(n)})$ is the maximum compact subgroup of $E_{n(n)}$ -- in particular, it is $\text{SU}(8)$ if $n=7$, and $\text{Spin}(16)$ if $n=8$ -- parametrised by an $f^{(n)}$-bein $\hat{V}_M{}^A(x,y)$; and eventually propagating higher $p$-form gauge fields, whose Cremmer-Julia duals are of rank greater than $p$ (if $n=7,8$ there no such fields). 

The action of the theory is fixed by requiring the invariance under the $E_{n(n)}$ generalised Lie derivative. As seen in the previous Section, the generalised Lie derivative encodes the internal symmetry of eleven-dimensional supergravity when studied in an $n$-dimensional internal space, as one can see by decomposing the fundamental representation of $E_{n(n)}$ in irreducible representation of $\text{GL}(n,\mathbb{R})$. The gauge symmetries are geometrised, in the same way as seen in Section \ref{GaugeLie}. This means that the covariant derivative with respect to the internal symmetries is defined by \eqref{CovariantDerivativeGaugeTrans}, with the simple Lie derivative replaced by the generalised Lie derivative generated by $\hat{A}_\mu{}^M$.

Consistency of generalised Lie derivative requires the partial derivatives $\de_M$ to satisfy the section constraint. There are two possible independent solutions for a section matrix $\mathcal{E}_M{}^m$ with maximum rank, which is $n$ since $\mathcal{E}_M{}^n$ is a rectangular $\dim R_v \times n$ matrix. One of these two maximum-rank corresponds to the field content of eleven-dimensional supergravity.\footnote{The other solution corresponds to IIB supergravity. Instead, IIA supergravity can be obtained by deforming the generalised Lie derivative \cite{Ciceri:2016dmd}.} The section matrix selects in the $\text{GL}(n,\mathbb{R})$ decomposition of $\de_M$ only the components corresponding to the $n$ internal partial derivatives $\de_m = \frac{\de}{\de y^m}$. For example, in $n=7$ case, using the decomposition of the fundamental representation $\mathbf{56}$ according to $\text{SU}(8)$, and then according to $\text{GL}(7)$, 
\begin{equation}
\mathbf{56} \rightarrow \overline{\mathbf{28}} \oplus \mathbf{28}\rightarrow \overline{\mathbf{7}} \oplus \mathbf{21} \oplus \overline{\mathbf{21}} \oplus \mathbf{7},
\end{equation}
and denoting with $i,j=1,\dots,8$ the eight-dimensional index, and with $m,n=1,\dots,7$ the seven dimensional index, one gets correspondingly
\begin{equation}
\de_M \rightarrow (\de_{ij},\de^{ij}) \rightarrow (\de_{m8},\de^{mn},\de_{mn},\de^{m8}).
\end{equation}
The solution of section constraint corresponds to identify $\de_{m8}$ with the internal partial derivatives, and to set to zero the remaining components:
\begin{equation}
\de_{m8} = \de_m, \quad
\de^{mn} = \de_{mn} = \de^{m8} = 0,
\end{equation}
In terms of the section matrix $\mathcal{E}_M{}^m$, such that $\de_M = \mathcal{E}_M{}^m\,\de_m$, this corresponds to a $56 \times 7$ rectangular matrix whose first seven rows form the seven-dimensional identity matrix and the remaining part is equal to zero:
\begin{equation}
\mathcal{E}_M{}^m = \begin{pmatrix}
I_{7 \times 7} \\
O_{49\times 7}
\end{pmatrix}.
\end{equation}
Lower-rank solutions for the section matrix capture reductions of the theories corresponding to the maximal-rank solution, since in this way conditions also on the internal partial derivatives are imposed. In particular, the Kaluza-Klein truncation corresponds to set to zero all the internal derivatives, as in \eqref{KKcondition}.

\newpage

\section{Supergravity Uplifts}\label{2}

\subsection{Generalised Scherk-Schwarz compactification}

The Scherk-Schwarz reduction does not cover all the consistent truncations we know. A famous example is the reduction of eleven-dimensional supergravity on a $S^7$. The resulting reduced theory is $\text{SO}(8)$-invariant four-dimensional gauged supergravity \cite{deWit:1986oxb}. As we know, $S^7$ is parallelisable, but it is \emph{not} a group manifold ($S^7 \simeq \text{Spin}(7)/G_2$), so this reduction does not fit in the Scherk-Schwarz paradigma. Other examples of consistent truncations, which are not covered by Scherk-Schwarz reduction, are eleven-dimensional supergravity in $S^4$, type IIB supergravity in $S^5$, or type IIA supergravity in $S^4$ \cite{Nastase:1999cb, Nastase:1999kf, Cvetic:2000nc, Cvetic:2000ah}.

These consistent truncations can be understood by means of the generalised geometry. The idea is to formally use the same setting of Scherk-Schwarz reduction, with generalised geometric objects in place of the usual ones. The global frame of the internal manifold is replaced by a set of generalised vectors $\hat{E}_A$, $A=1\dots,f^{(n)}$ in the generalised bundle, $n$ being the dimension of the internal space. The parallelisation condition is 
\begin{equation}
L_{\hat{E}_A}\,\hat{E}_B = - X_{AB}{}^C\,\hat{E}_C, 
\end{equation}
where $L$ is the generalised Lie derivative, and $X_{AB}{}^C$ are the components of the embedding tensor. In other words, the generalised torsion $T_{AB}{}^C(\hat{E})$ of the generalised frame $\hat{E}_A$ should be constant and equal to $X_{AB}{}^C$. This identification is in principle possible, since one can show that the torsion sits in the same representation as the embedding tensor. The components of the embedding tensor $X_{AB}{}^C$ play the r\^ole of the structure constants of the gauge group in Scherk-Schwarz reduction. Finally, the generalised frame encodes the dependence of the internal coordinates in the fields of the reduced theory, on the same footing as in simple Scherk-Schwarz reduction.

Now, the opposite problem can be also addressed. By ``uplift problem" we mean the search of necessary and sufficient conditions to determine which $d$-dimensional maximally supersymmetric gauged supergravity theories can be obtained as consistent truncations of a higher-dimensional supergravity theory. The embedding of the lower-dimensional theory in the higher one is called ``uplift". As we know, consistent truncation means that all the solutions of the classical equations of motion of the lower-dimensional theory should also be solutions of the equations of the higher-dimensional one. Therefore, if we are able to establish if a theory admits an uplift, we can also find solutions of the higher-dimensional theory starting from the lower one -- an easier task than finding directly solutions of the higher-dimensional theory. Moreover, the distinction between theories which admit an uplift and theories which do not is a first step in classifying all the possible gaugings in supergravity.

The uplift problem can be conveniently addressed by using the $E_{n(n)}$--covariant formulation of higher-dimensional supergravity furnished by exceptional field theory and exceptional geometry. Using the generalised Scherk-Schwarz ansatz, the dependence on the coordinates of the internal manifold are factorised by a generalised frame. The latter takes values in the duality group, and it twists the fields of the theory in the reduction. It exists if the internal manifold is parallelisable in generalised sense, as the hyperspheres in Section \ref{ParallSn}. The corresponding torsion has to be identified with the components of the embedding tensor of the gauged supergravity, so that the torsion is required to be constant for the uplift to exist. This means that, in the uplift problem, one starts from the embedding tensor, looking for an internal manifold and for a generalised frame which solves the generalised parallelisability condition, with the components of the embedding tensor playing the r\^ole of the torsion.

Since only a subset of all the possible gauging of a maximal supergravity theory can be obtained from a consistent truncation, the aim of the uplift problem is to find conditions on the embedding tensor of the gauged supergravity which has to be fulfilled in order for an uplift to exist. Actually, the procedure allows to explicitly build the uplift. A key ingredient in finding these conditions is the consistent deformation of the exceptional Lie derivative. By consistent deformation we mean the addition of a flux-term to the generalised Lie derivative, in such a way that the resulting derivative still satisfies the Leibniz identity. If a frame is known such that the parallelisability condition is satisfied by its deformed generalised Lie derivative, then an uplift exists when the flux is integrable, that is, when a twist matrix exists such that its torsion is the flux. Indeed, in this case, the deformation can be absorbed in the twisting of the parameters of the derivatives, so that the twisted frame satisfies the parallelisability condition, and it can be used to define the uplift.

A procedure for checking and defining the uplift is systematically known when the dimension of the internal space is $n\leqslant 7$ \cite{Inverso:2017lrz}. The initial data are the duality group $G_D$, the gauge group $G \subset G_D$ of the gauged supergravity, the embedding tensor $\vtheta_A{}^\alpha$, which captures the features of the gauged supergravity one starts with, and a subgroup $H \subset G$, such that the coset $G/H$ describes an internal geometry. The generalised Scherk-Schwarz compactification is assumed to be the unique way to produce a consistent reduced theory out of a higher-dimensional one, or, at least, the conditions we find in order for an uplift to exist assume that the reduced theory we start with is recovered, performing a generalised Scherk-Schwarz compactification of the uplift. The generalised Scherk-Schwarz reduction is indeed the most general known procedure, and no example of reduction of supergravity theories is known, which is not encompassed in this setting.

More in details, the projection of the embedding tensor onto the coset generators is shown to provide a solution of the section constraint. So, this projection teaches how to define a $d+n$ dimensional supergravity out of the formalism of the generalised geometry. Moreover, one can built a frame only using the geometric data of the coset. Nevertheless, it fails to satisfy the parallelisability condition in general. But one can find a suitable flux, such that the frame satisfies the parallelisability condition, with the generalised Lie derivative replaced by the deformed one. Now the problem is to study which conditions the flux has to satisfy in order for the deformation not to spoil the Leibniz identity. One finds that three independent conditions have to be fulfilled. Two of them are automatically satisfied by the peculiar flux one has in this setting. Moreover, these two conditions allow to conclude that the flux is integrable, when one takes into account the solution of the section constraint given by the embedding tensor of the lower-dimensional gauged supergravity. Instead, the last condition is fulfilled if and only if the embedding tensor satisfies an algebraic constraint, which therefore distinguishes which gauged supergravities can be upliftable and which do not.

\subsection{Uplift problem with $n\leqslant 7$}

The procedure for determining which gauged supergravity admits an uplift, and for defining the uplift itself, can be described in the following way:\footnote{Here we consider uplifts of maximal theories, but exceptional field theory and exceptional geometry can
be used also when the internal manifold is a deformation of the coset space (see Section \ref{nonMax} and, for example, \cite{Cassani:2019vcl, Blair:2024ofc, Rovere:2025jks}).} \cite{Inverso:2017lrz}
\begin{itemize}
\item[1.] Consider a $d$-dimensional gauged supergravity with gauge group $G$ (which is trivial $G=\mathbbm{1}$ in the ungauged case), whose features are captured by an embedding tensor $\vtheta_A{}^\alpha$, where $A$ is an index in a representation $\mathbf{f}_n$ of the duality group $E_{n(n)}\times\mathbb{R}^+$, and $\alpha$ is in the adjoint representation of $G$. 
\item[2.] Consider an $n$-dimensional internal manifold, described by a coset space $G/H$, where $H$ is a subgroup of $G$, such that $\text{dim}\, H = \text{dim}\, G - n$. When the internal manifold is a group manifold $H$ is trivial.
\item[3.] Assume that the reduction procedure, when the $d$-dimensional theory admits an uplift, is the (generalised) Scherk-Schwarz compactification, which is the most general known procedure furnishing consistent truncations. Although we will focus on eleven-dimensional supergravity as the higher-dimensional theory to which the lower-dimensional one is uplifted, the cases in which the higher-dimensional theory is IIB, IIA, or massive IIA supergravities can also be considered within the same formalism \cite{Ciceri:2016dmd}.
\item[4.] In order for the uplift to be defined a global (generalised) frame $\hat{E}^M{}_A$ is needed, whose torsion $T(\hat{E})$ is equal to the components $X_{AB}{}^C$ of the embedding tensor $\vtheta_A{}^\alpha$
\begin{equation}\label{ParallEhat}
L_{\hat{E}_A}\,\hat{E}_B = -X_{AB}{}^X\,\hat{E}_C.
\end{equation}
This is the starting point of (generalised) Scherk-Schwarz compactification, which ensures the dependence of the internal coordinates in the fields to be factorisable in the transformation laws of the fields and in the action.
\item[5.] A solution of the section constraint $\mathcal{E}_M{}^m = \mathcal{E}_M{}^m(y)$ breaks the whole duality group $E_{n(n)}\times\mathbb{R}^+$ into $(\text{GL}(n,\mathbb{R})\times \mathbb{R}^+) \ltimes P$, where $\mathbb{R}^+$ corresponds to the trombone transformation of the uplift theory, and $P$ includes the $p$--form constant shifts. $\mathbb{R}^+$ and $P$ preserves the section choice. In particular, if $C^M{}_N$ is in $P$, 
\begin{equation}
C^N{}_M\,\mathcal{E}_N{}^m = \mathcal{E}_M{}^m,\;C \in P.
\end{equation}
Since the derivatives in the exceptional Lie derivative are defined to be on section $\de_M = \mathcal{E}_M{}^m\,\de_m$, the elements in $P$ preserve such derivatives: $C^N{}_M\,\de_N = \de_M$. Instead, an element $g^m{}_n$ in $\text{GL}(n,\mathbb{R})$ acts on the section matrix as
\begin{equation}
G^N{}_M\,\mathcal{E}_N{}^m = \mathcal{E}_M{}^n\,g^m{}_n,\;g \in \text{GL}(n,\mathbb{R}),
\end{equation}
which defines the embedding $G^N{}_M$ of $g^n{}_m$ in $E_{n(n)}$.
\item[6.] The projection of the embedding tensor on the coset generators $\vtheta_A{}^a$, where $\alpha = (a,i)$, $i$ being an index in the adjoint of $H$, can be used as section matrix, provided that it satisfies the section constraint:
\begin{equation}
Y^{AB}{}_{CD}\,\vtheta_A{}^a\,\vtheta_B{}^b = 0.
\end{equation}
For our purposes, we consider at this stage only the embedding tensors compatible with the section choice corresponding to eleven-dimensional supergravity, but this is not necessary.
\item[7.] A frame $E^M{}_A$ can be built using the geometric data of the coset, but it fails to satisfy the parallelisability condition in general.
\begin{equation}
L_{E_A}\,E_B \neq -X_{AB}{}^C\,E_C.
\end{equation}
\item[8.] Nevertheless, if we consider the following flux
\begin{equation}\label{DefFlux}
F_{MN}{}^P = E_M{}^A\,E_N{}^B\,(X_{AB}{}^C - T_A{}^C{}_B(E))\,E^P{}_C,
\end{equation}
then $E^M{}_A$ satisfies a parallelisability condition with the generalised Lie derivative replaced by a flux-deformed generalised Lie derivative
\begin{equation}\label{DeformedParall}
L^{(F)}_{E_A}\,E_B = -X_{AB}{}^C\,E_C,
\end{equation}
where the flux-deformed generalised Lie derivative is defined by
\begin{equation}
L_X^{(F)}\,V^M = L_X\,V^M + X^P\,F_{PQ}{}^M\,V^Q.
\end{equation}
Consistently, had the torsion $T(E)$ been equal to $X_{AB}{}^C$, the flux $F$ in \eqref{DefFlux} would have been vanishing.
\item[9.] The flux-deformation can be absorbed in an undeformed generalised Lie derivative whenever the flux is the torsion $T(C)$ of a twist matrix $C^M{}_N$ compatible with the section choice, because in such a case the flux-deformed generalised Lie derivative turns out to be equal to the undeformed Lie derivative with twisted parameters
\begin{equation}
(L_{X}^{(F)}\,V)' = L_{X'}\,V'
\end{equation}
whenever $C^M{}_N$ exists, preserving the section choice
\begin{equation}
C^N{}_M\,\mathcal{E}_N{}^n = \mathcal{E}_N{}^n,\;\; X'^M = C^M{}_N\,X^N, \;\; V'^M = C^M{}_N\,V^N,
\end{equation} 
such that
\begin{equation}
F = T(C).
\end{equation}
\item[10.] The flux-deformed generalised Lie derivative is a consistent derivative, satisfying the Leibniz identity, if and only if three constraints are fulfilled. They are:
\begin{itemize}
\item[--] The \emph{first section constraint}
\begin{equation}\label{FirstConstraintForFlux}
F_{MN}{}^P\,\mathcal{E}_P{}^p = 0,
\end{equation}
for any choice of the section matrix $\mathcal{E}_M{}^m$ defining the generalised geometry. 
\item[--] The \emph{second section constraint} 
\begin{equation}\label{SecondConstraintForFlux}
Y^{MP}{}_{QN}\,F_{MR}{}^R\,\mathcal{E}_P{}^p = 0,
\end{equation}
\item[--] The \emph{Bianchi constraint}  
\begin{equation}\label{ThirdConstraintForFlux}
(\de_Q + \delta_{F_Q})\,F_{MN}{}^P = T_M{}^P{}_N\{\de_M\,F_{QN}{}^P\},
\end{equation}
where $\delta_{F_M}$ is the adjoint action of $(F_M)_N{}^P = F_{MN}{}^P$, according to 
\begin{equation}
\delta_{F_M}\,V^P = - F_{MN}{}^P\,V^N, \quad
\delta_{F_M}\,U_N = F_{MN}{}^P\,U_P,
\end{equation}
and $T_{MN}{}^P\{\de_M\,F_{QN}{}^P\}$ is the torsion of the ``connection" $(W_Q)_M{}^P{}_N = \de_M\,F_{QN}{}^P$, $Q$ being a spectator index.
\end{itemize}
\item[11.] The constraint \eqref{SecondConstraintForFlux} is satisfied by the flux in \eqref{DefFlux} if and only if the embedding tensor obeys the following condition
\begin{equation}\label{ConditionOnTheta}
Y^{AB}{}_{CD}\,\vtheta_B{}^b\,(X_{AE}{}^E + \omega\,f^{(n)}\,\vtheta_E{}^a\,(T_a)_A{}^E)=0,
\end{equation}
where $f^{(n)}$ is the dimension of $\mathbf{f}_n$. 
\item[12.] Instead, the flux $F$ in \eqref{DefFlux} automatically satisfies the constraints \eqref{FirstConstraintForFlux} and \eqref{ThirdConstraintForFlux}, which moreover ensure $F$ to be integrable, on the solution of the section matrix furnished by the projected embedding tensor, in the case compatible with eleven-dimensional supergravity.
In particular, on the section choice, the constraint \eqref{FirstConstraintForFlux} implies that $F$ contains only the $\text{GL}(n)$ representations corresponding to a four-form, a seven-form, which correspond to the field strengths of the three-form and the dual six-form of eleven-dimensional supergravity respectively, and the flux associated with the trombone duality; instead, the constraint \eqref{ThirdConstraintForFlux} implies that the $F$ is integrable.
\item[13.] Since $F$ is integrable, a twist matrix $C^M{}_N$ exists, such that the twisted frame
\begin{equation}
C^M{}_N\,E^N{}_A = \hat{E}^M{}_A
\end{equation}
satisfies the undeformed parallelisability condition \eqref{ParallEhat}.
\item[14.] The twist matrix is of the form
\begin{equation}
C = \vrho\,e^{a}\,e^{\tilde{a}},
\end{equation}
where $a$ and $\tilde{a}$ are the three-form $a = a_{mnp}\,t^{mnp}$, and the dual six-form contribution $\tilde{a} = a_{m_1\dots m_6}\,t^{m_1\dots m_6}$, where $t^{mnp}$ and $t^{m_1\dots m_6}$ are the $\text{Lie}\,E_{n(n)}$ generators associated to the three-form and six-form components in $\text{Lie}\,E_{n(n)}$ when the algebra is decomposed according to $\text{GL}(n)$. Finally, $\vrho \in \mathbb{R}^+$ is the trombone contribution.
\item[15.] The twist matrix allows to define the $f^{(n)}$-bein parametrising the coset $E_{(n(n)}/K(E_{n(n)})$ of scalars, $K(\cdot)$ denoting the maximal compact subgroup,
\begin{equation}
\hat{V}_M{}^A = (\text{det}\,e)^{-f^{(n)}}\,C^M{}_N\,e_N{}^A,
\end{equation}
where the density factor ensures the generalised metric to be unimodular
\begin{equation}
\hat{M}_{MN} = \hat{V}_M{}^A\,\hat{V}_N{}^A.
\end{equation}
\item[16.] Finally, the (generalised) Scherk-Schwarz ansatz is written as follows. Since $\hat{E}^M{}_A$ has weight one with respect to the trombone transformations, it is somehow convenient to factor out the $\mathbb{R}^+$ term from the genuine $E_{n(n)}$ component:
\begin{equation}
\hat{E}^M{}_A = U^M{}_A\,r^{-1}, \quad 
U \in E_{n(n)},\; r \in \mathbb{R}^+.
\end{equation}
$U$ and $r$ are used to factor out the dependence of the internal coordinates in the fields, as in the Scherk-Schwarz ansatz
\begin{equations}
\hat{g}_{\mu\nu}(x,y) &= r^{-2}(y)\,g_{\mu\nu}(x), \\
\hat{A}_\mu{}^M(x,y) &= r^{-1}(y)\,U^M{}_A(y)\,A_\mu{}^A(x) = \hat{E}^M{}_A(y)\,A_\mu{}^A(x),\\
\hat{M}_{MN}(x,y) &= U_M{}^A(y)\,U_N{}^B(y)\,M_{AB}(x).
\end{equations}
\end{itemize}
In the next Sections we will describe the details of the full procedure, following and expanding \cite{Inverso:2017lrz,Inverso:2024xok}. In particular, in Sections \ref{MaurerCartan}, \ref{Coset}, \ref{SigmaModels} there is a self-contained review of coset spaces. In Section \ref{GenWT} the Weitzenb\"ock connection and its torsion are introduced in the context of generalised geometry. In Section \ref{TwistedGenLieDer} the twist of the generalised Lie derivative is studied, and in Section \ref{FluxGenLieDer} the deformation of the generalised Lie derivative by a flux term is introduced, and the consistency constraints \eqref{FirstConstraintForFlux}, \eqref{SecondConstraintForFlux}, and \eqref{ThirdConstraintForFlux} the flux has to satisfy are derived. In Section \ref{ThetaSection} it is shown that the embedding tensor furnishes a solution for the section choice, and an expression for the frame $E_A$, depending only on the coset data of the gauged supergravity to be uplifted, is found. Finally, in Sections \ref{FluxConstraint1} and \ref{FluxConstraint3} it is shown that the flux defined in \eqref{DefFlux} satisfies the constraints \eqref{FirstConstraintForFlux} and \eqref{ThirdConstraintForFlux}, and in Section \ref{FluxConstraint2} it is shown that the constraint \eqref{SecondConstraintForFlux} imposes the condition \eqref{ConditionOnTheta} on the embedding tensor.

\subsection{Maurer-Cartan form}\label{MaurerCartan}

We begin by proving two formulas \cite{Marcus:1983hb} which will be useful.
\bigskip

\emph{Lemma 1.} Consider a Lie algebra and a pair of its elements $\vphi,x$. Denoting with $\text{ad}_\vphi = [\vphi,\cdot]$ the adjoint action, then
\begin{equation}\label{AdjointRelation}
e^{-\vphi}\,x\,e^{\vphi} = e^{-\text{ad}_\vphi}\,x.
\end{equation}

\bigskip

\emph{Proof.} Since
\begin{equation}\label{expansionexpphi}
e^{-\vphi} = 1 - \vphi + \tfrac{1}{2!}\,\vphi^2 + \dots,\quad
e^\vphi = 1 + \vphi + \tfrac{1}{2!}\,\vphi^2 + \dots,
\end{equation}
then,
\begin{align}
e^{-\vphi}\,x\,e^{\vphi} &= 
x - (\vphi\,x-x\,\vphi) + \tfrac{1}{2!}\,(\vphi^2\,x - 2\,\vphi\,x\,\vphi + x\,\vphi^2) \,+\nonumber\\
& \quad - \tfrac{1}{3!}\,(\vphi^3\,x - 3\,\vphi^2\,x\,\vphi + 3\,\vphi\,x\,\vphi^2 - x\,\vphi^3) + \dots = \nonumber\\
& = x - [\vphi,x] + \tfrac{1}{2!}\,[\vphi,[\vphi,x]] + \tfrac{1}{3!}\,[\vphi,[\vphi,[\vphi,x]]] + \dots = \nonumber \\
&= x - \text{ad}_\vphi\,x + \tfrac{1}{2!}\,\text{ad}_\vphi^2\,x + \tfrac{1}{3!}\,\text{ad}_\vphi^3\,x + \dots = e^{-\text{ad}_\vphi}\,x.
\end{align}

\emph{Lemma 2.}
\begin{equation} \label{FundCosetFormula}
e^{-\vphi}\,\delta\,e^{\vphi} = \tfrac{1-e^{-\text{ad}_\vphi}}{\text{ad}_\vphi}\,\delta\,\vphi.
\end{equation}

\bigskip

\emph{Proof}.  Using \eqref{expansionexpphi},
\begin{equation}
\delta\,e^\vphi = \delta\,\vphi + \tfrac{1}{2!}\,(\delta\,\vphi\,\vphi + \vphi\,\delta\,\vphi) +
\tfrac{1}{3!}\,(\delta\,\vphi\,\vphi^2 + \vphi\,\delta\,\vphi\,\vphi + \vphi^2\,\delta\,\vphi) + \dots,
\end{equation}
so that
\begin{align}
e^{-\vphi}\,\delta\,e^\vphi &= 
\delta\,\vphi + \tfrac{1}{2!}\,(\delta\,\vphi - \vphi\,\delta) + \tfrac{1}{3!}\,(\delta\,\vphi\,\vphi^2 - 2\,\vphi\,\delta\,\vphi\,\vphi + \vphi^2\,\delta\,\vphi) + \dots = \nonumber \\
& = \delta\,\vphi - \tfrac{1}{2!}\,[\vphi,\delta\,\vphi] + \tfrac{1}{3!}\,[\vphi,[\vphi,\delta\,\vphi]] + \dots = \nonumber \\
& = (\mathbbm{1} - \tfrac{1}{2!}\,\text{ad}_\vphi + \tfrac{1}{3!}\,\text{ad}_\vphi^2 + \dots)\,\delta\,\vphi = \nonumber \\
& = \tfrac{1}{\text{ad}_\vphi}\,[\mathbbm{1}-(\mathbbm{1}-\text{ad}_\vphi + \tfrac{1}{2!}\,\text{ad}_\vphi^2 - \tfrac{1}{3!}\,\text{ad}_\vphi^3 + \dots)]\,\delta\,\vphi = \tfrac{\mathbbm{1}-e^{-\text{ad}_\vphi}}{\text{ad}_\vphi}\,\delta\,\vphi,
\end{align}
as we wanted to show. Alternatively, one can use an homotopy trick in the following way:
\begin{align}
e^{-\vphi}\,\delta\,e^{\vphi} &= 
\int_0^1\diff t\,\tfrac{\diff}{\diff t}\,(e^{-t\,\vphi}\,\delta\,e^{t\,\vphi}) = \int_0^1\,\diff t\, e^{-t\,\vphi}\,\delta\,\vphi\,e^{t\,\vphi} = \nn\\
&= \int_0^1\diff t\,e^{-t\,\text{ad}_\vphi}\,\delta\,\vphi = \frac{\mathbbm{1}-e^{\text{ad}_\vphi}}{\text{ad}_\vphi}\,\delta\,\vphi.
\end{align}

Consider a Lie group $G$. Denote with $\{e_a\}$ a global basis of the tangent bundle of the associated group manifold $\mathscr{M}_G$. The \emph{Maurer-Cartan form} is defined by
\begin{equation}\label{MaurerCartanForm}
\omega(g) = g^{-1}\,\diff g,
\end{equation}
for some $g \in G$. Since 
\begin{equation}
0 = \diff(g^{-1}\,g) = \diff g^{-1}\,g + g^{-1}\,\diff g,
\label{gg}
\end{equation}
it can be equivalently written as
\begin{equation}
\omega(g)  = - \diff g^{-1}\,g.
\end{equation}
The Maurer-Cartan form has two main properties
\begin{itemize}
\item[--] It is in the Lie algebra $\text{Lie}\,G$ of $G$;
\item[--] It is flat, that is, its curvature vanishes
\begin{equation}
\diff\omega(g) + \omega^2(g) = 0.
\end{equation}
\end{itemize}
The first property follows from Lemma 2. Indeed, setting $g = e^\vphi$ and $\delta = \diff$,
\begin{equation}
\omega(g) = \tfrac{\mathbbm{1}-e^{-\text{ad}_\vphi}}{\text{ad}_\vphi}\,\diff\vphi,
\end{equation}
which is manifestly in the algebra, since $\vphi$ is in the algebra by definition and the adjoint action $\text{ad}_\vphi$ sends the algebra in itself. On the other hand, the second property follows by explicit evaluation, using $\diff g^{-1} = - g^{-1}\,\diff g\,g^{-1}$:
\begin{align}
\diff \omega(g) &= \diff (g^{-1}\,\diff g) = \diff g^{-1}\,\diff g = \nn\\
&= - g^{-1}\,\diff g g^{-1}\,\diff g = -(g^{-1}\,\diff g)^2 = - \omega^2(g).
\end{align}

\subsection{Homogeneous manifolds and coset spaces}\label{Coset}

A Lie group $G$ acts \emph{transitively} on an $n$-dimensional manifold $\mathscr{M}$ if, for all pairs $x,y$ of points in $\mathscr{M}$, there is an element $g_{x,y}$ in $G$, such that
\begin{equation}
y = g_{x,y}\,x,
\end{equation}
where the left multiplication of an element in the group on a point of the manifolds formally defines the action of the group on the manifold. This is equivalent to say that, given any fixed \emph{reference point} $x_0$ in $\mathscr{M}$, any other point $x$ can be reached from $x_0$ by the action of an element $g_{x}$ of the group
\begin{equation}
x = g_x\,x_0, \quad \text{with}\;\;g\in G, \;\; \forall\;x\in \mathscr{M}.
\end{equation}
When this is the case, $\mathscr{M}$ is said \emph{homogeneous}. The transitive action is equivalent to the homogeneity condition since, on a side, if there is a transitive action, then the homogeneity property is satisfied by choosing $g_x = g_{x_0,x}$, and, on the other side, if the manifold is homogeneous, then a transitive action is defined by choosing $g_{x,y} = g_y\,g_x^{-1}$. Indeed, if $x=g_x\,x_0$ and $y = g_y\,x_0$, then $y = g_y\,(g_x^{-1}\,x)$.

Define the quotient space or \emph{coset} $G/\sim$ with respect to the equivalence relation 
\begin{equation}
g' \sim_{x_0} g, \quad g, g' \in G \Leftrightarrow g'\,x_0 = x = g\,x_0
\end{equation}
as the set of the equivalence classes of $\sim_{x_0}$ in $G$:
\begin{equation}
[g]_{x_0} = \lbrace g' \in G\;|\; g' \sim_{x_0} g \rbrace.
\end{equation}
Observe that $\sim_{x_0}$ does not depend on $x_0$. Indeed, supposing that $g' \sim_{x_0} g$ and that $g\,x_0'=x$, then,
\begin{equation}
g'\,x_0' = g'\,(g^{-1}\,x) = g'\,g^{-1}\,(g\,x_0) = g'\,x_0 = x,
\end{equation}
that is $g' \sim_{x_0'} g$. From now on we will simply denote $\sim_{x_0}$ and $[g]_{x_0}$ as $\sim$ and $[g]$.

The element in ${G/{\sim}}$ are in one-to-one correspondence with those of $\mathscr{M}$, since the elements in $G$ sending $x_0$ in the same point are identified in the coset space:
\begin{equation}
{G/{\sim}} \simeq \mathscr{M}.
\end{equation}
Consider the subset of the elements of $G$ preserving $x$ (the set of all $g\in G$ such that $x$ is a \emph{fixed point}):
\begin{equation}
H_x = \lbrace g \in G \;|\; g\,x = x\rbrace.
\end{equation}
It does not depend on $x$:
\begin{equation}
H_x \simeq H_{x'} \equiv H, \;\;\forall\; x,x' \in \mathscr{M}.
\end{equation}
Indeed, as a consequence of homogeneity, $g\in G$ exists such that $x' = g\,x$, so that, given $h_x \in H_x$ and $h_{x'} \in H_{x'}$, 
\begin{equation}
h_{x'}\,g\, x = h_{x'}\,x = x' = g\,x = g\,h_x\,x \Rightarrow h_{x'}\,g = g\,h_x \Rightarrow h_{x'} = g\,h_x\,g^{-1}.
\end{equation}
The set $H$ is called \emph{little group} or \emph{isotropy group} of $G$. Define $G/H$ as the set of the equivalence classes with respect to the following equivalence relation:\footnote{Notice that we could have defined the moltiplication for $h$ also on the left or replacing it by conjugation: $g' = h\,g$ or $g' = h\,g\,h^{-1}$. In all the cases, the equivalence classes are equivalent.}
\begin{equation}
g' \sim_H g \Leftrightarrow g' = h\,g, \quad g, g' \in G,\;\; h \in H, \;\; g\,x_0 = x.
\end{equation}
Observe that $g' \sim_H g \Rightarrow g' \sim g$. Indeed, $g' = h\,g \Rightarrow g'\,x_0 = h\,g\,x_0 = h\,x = x$. \emph{Vice versa}, $g' \sim g \Rightarrow g' \sim_H g$. Indeed, by definition of group, $g'' \in G$ exists such that $g' = g''\,g$. Therefore, $x=g'\,x_0 = g''\,g\,x_0 = g''\,x \Rightarrow g'' \in H$. This shows that ${G/{\sim}} \simeq G/H$. Thus, we can identify
\begin{equation}
\mathscr{M} \simeq G/H.
\end{equation}

Let us denote with $\lbrace T_I \rbrace_{I=1,\dots,\dim{\text{Lie}\,G}}$ a basis of generators of $\text{Lie}\,G$, such that the first $\dim{\text{Lie}\,H}$ generators form a basis $\lbrace T_i \rbrace_{i=1,\dots, \dim\text{Lie}H}$ of $\text{Lie}\,H$. The remaining generators are denoted by $T_a$, $a=1,\dots, \dim{\text{Lie}\,G}-\dim{\text{Lie}\,H}$. If the structure constants of $\text{Lie}\,G$ are $f_{IJ}{}^K$, the commutation rules $[T_I,T_J]=f_{IJ}{}^K\,T_K$ split in this way \cite{Castellani:1991et}:
\begin{equation}
[T_i,T_j]=f_{ij}{}^k\,T_k,\quad
[T_i,T_a]=f_{ia}{}^b\,T_b + f_{ia}{}^j\,T_j,\quad
[T_a,T_b]=f_{ab}{}^c\,T_c + f_{ab}{}^j\,T_j.
\end{equation}

It is of interest in supergravity to consider the case in which $G$ is non-compact and $H$ is its maximal compact subgroup. A non-compact realisation of a compact group can be obtained by replacing some generators $T$ with $i\,T$ (\emph{Weyl unitary trick}). The fact that $H$ is compact implies that the $T_i$'s do not have to be replaced, whereas the fact that $H$ is the maximal compact subgroup implies that all the $T_a$'s have to be replaced with $i\,T_a$. Then, one has $f_{ia}{}^j=f_{ab}{}^c =0$. Otherwise, it would not be possible any redefinition of the generators ensuring the closeness of the algebra with real structure constants. So, after a suitable redefinition, the commutation rules are:
\begin{equation}\label{CosetRules}
[T_i,T_j]=f_{ij}{}^k\,T_k,\quad
[T_i,T_a]=f_{ia}{}^b\,T_b,\quad
[T_a,T_b]=f_{ab}{}^j\,T_j.
\end{equation}
A coset satisfying these commutation rules is said to be \emph{reductive} (if $f_{ia}{}^j=0$) and \emph{symmetric} (if $f_{ab}{}^c=0$).

Given a class $[g](x)$ associated with $x \in \mathscr{M}$, we call \emph{coset representative} any element $L(x)$ chosen in $[g](x)$. The action of an arbitrary element $f \in G$ on $\mathscr{M}$ is $f\,x=x'$. The natural extension to the coset classes is $f\,[g](x) = [g'](x')$. In the right-hand side there is some $g'$ in place of $g$ because we are not guaranteed the action of $f$ to preserve the choice of the representative of the equivalence class. For the same reason, the extension to the coset representatives should be $f\,L(x) = L'(x')$. Observe that, since $L'(x) \sim L(x)$, then  $h(x) \in H$ exists such that $L'(x)=L(x)\,h(x)$. In conclusion,
\begin{equation}\label{Coset1}
f\,L(x) = L(x')\,h(x').
\end{equation}

Consider an infinitesimal transformation in $G$:
\begin{equation}\label{Coset2}
f = \mathbbm{1} + \vepsilon^I\,T_I + \mathcal{O}(\vepsilon^2).
\end{equation}
An infinitesimal transformation in $H$ is 
\begin{equation}\label{Coset3}
h = \mathbbm{1} + \vepsilon^i\,T_i + \mathcal{O}(\vepsilon^2) = \mathbbm{1} - \vepsilon^I\,w_I{}^i\,T_i + \mathcal{O}(\vepsilon^2).
\end{equation}
$w_I{}^i$ describes how the $\vepsilon^i$ may be embedded in $\text{Lie}\,G$ (the minus sign is for future convenience). It plays a r\^ole similar to the embedding tensor in gauged supergravity. Finally, consider
\begin{equation}\label{Coset4}
x'^m = x^m + \vepsilon^I\,K_I{}^m, 
\end{equation}
where $m$ is a curved index and the vector field $K_I = K_I{}^m\,\de_m$ is called \emph{Killing vector} associated to $T_I$. The Killing vectors (with minus sign) provide a representation of the $\text{Lie}\,G$. To see why, replace \eqref{Coset2}, \eqref{Coset3}, and \eqref{Coset4} in \eqref{Coset1}, finding
\begin{equation}\label{Coset0}
T_I\,L(x) = K_I\,L(x)- L(x)\,w_I{}^i\,T_i.
\end{equation}
Using this, we can compute\footnote{\emph{Proof}. It follows by explicit evaluation: 
\begin{align}
[T_I,T_J]\,L &=T_I\,(K_J\,L-L\,T_j\,w_J^j)-(I\leftrightarrow J) = \nn\\
&= K_J\,(K_I\,L-L\,T_i\,w_I^i)-(K_I\,L-L\,T_i\,w_I^i)\,T_j\,w_J^j -(I\leftrightarrow J) = \nn\\
&= -[K_I,K_J]\,L-(K_J\,L\,T_i\,w_I^i+L\,T_i\,K_J\,w_I^i+K_I\,L\,T_j\,w_J^j \,+\nn\\
& -K_I\,L\,T_j\,w_J^j-L\,T_i\,K_I\,w_J^i-K_J\,L\,T_i\,w_I^i) + L\,T_i\,T_j\,w_{[I}^i\,w_{J]}^j = \nn\\
&=-[K_I,K_J]\,L-L\,T_i\,K_{[I}\,w_{J]}^i + L\,[T_i,T_j]\,w_I^i\,w_J^j = \nn\\
& =-[K_I,K_J]\,L-L\,T_i\,(K_{[I}\,w_{J]}^i - \,f_{jk}{}^i\,w_I^j\,w_J^k).
\end{align}}
\begin{equation}
[T_I,T_J]\,L(x) = -[K_I,K_J]\,L(x)-L(x)\,T_i\,(K_{[I}\,w_{J]}{}^i - f_{jk}{}^i\,w_I{}^i\,w_J{}^j).
\end{equation}
On the other side,
\begin{equation}
[T_I,T_J]\,L(x)=f_{IJ}{}^K\,T_K\,L(x)= f_{IJ}{}^K\,(K_K\,L(x)-L(x)\,w_K{}^i\,T_i).
\end{equation}
Equating one gets in particular \cite{Castellani:1991et}
\begin{equation}\label{RuleKilling}
[-K_I,-K_J]=f_{IJ}{}^K\,(-K_K), 
\end{equation}
showing that $-K_I$ represents the algebra.

Consider now the Cartan-Maurer form $\omega(L(x)) \equiv \Omega(x)$ according to \eqref{MaurerCartanForm}:
\begin{equation}
\Omega(x) = L^{-1}\,\diff L(x).
\end{equation}
Since it lies in the algebra, it can be decomposed as
\begin{equation}\label{DecompositionMaurerCartan}
\Omega = e^a\,T_a + \omega^i\,T_i,
\end{equation}
where $e^a = e_\mu{}^a\,\diff x^\mu$ defines the vielbein on $G/H$, and $\omega^i=\omega_\alpha{}^i\,\diff x^\alpha$ is called $H$-connection because it transforms as a gauge connection for the group $H$. Indeed, using $L(x')=f\,L(x)\,h^{-1}(x)$,
\begin{align}
e'^a\,T_a + \omega'^i\,T_i &= \Omega' =(f\,L\,h^{-1})^{-1}\,\diff\,(f\,L\,h^{-1})=
h\,L^{-1}\,\diff L \,h^{-1}+h\,\diff h^{-1} = \nn\\
& = h\,\Omega\,h^{-1} + h\,\diff h^{-1} =
h\,e^a\,T_a\,h^{-1} + h\,\omega^i\,T_i\,h^{-1} + h\,\diff\,h^{-1},
\end{align}
where the first term is in the span of $\{T_a\}$ because of the relation \eqref{AdjointRelation}, the second term is in $\text{Lie}\,H$ for the same reason, and the last term is in $\text{Lie}\,H$ too, since it is equal to the Maurer Cartan form $\omega(h^{-1})$, which lies in the algebra. Therefore, $e^a$ transforms homogeneously under a transformation in $H$, and $\omega^i$ transforms as a gauge connection.

We conclude this section finding an explicit expressions for the Killing vectors $K_I$. It is sufficient to take \eqref{Coset0}, multiplying on the left by $L^{-1}$, recognising the Maurer-Cartan form $\Omega$, replacing the decomposition \eqref{DecompositionMaurerCartan}, and projecting on the coset generators. The result is \cite{Castellani:1991et}
\begin{equation}\label{ExplicitKilling}
K_I{}^m = (L^{-1}\,T_I\,L)^a\,e^m{}_a.
\end{equation}

\subsection{Non-linear $\sigma$-models}\label{SigmaModels}

In this Section the relevant features of non-linear $\sigma$-models for supergravity are reviewed \cite{Marcus:1983hb, Cremmer:1997ct, Samtleben:2008pe}. Consider a symmetric and reductive coset $G/H$. An arbitrary element $\mathcal{V} \in G$ can be parametrised as
\begin{equation}
\mathcal{V} = e^{\chi + \vphi},
\end{equation}
where $\chi = \chi^i\,T_i$ is in $\text{Lie}\,H$, and $\vphi = \vphi^a\,T_a$ is in the coset span. Physically, $\vphi^a$ are scalar fields parametrising the coset space and $\chi^i$ are some auxiliary fields. A theory of scalars $\vphi^a$, globally invariant  under $G$ and locally invariant under $H$, is called \emph{non-linear $\sigma$-model} on the coset $G/H$.\footnote{The name $\sigma$-model is for historical reason. The first example of non-linear $\sigma$-model was introduced by Gell-Mann and Lévy, in studying models of pions ($\sigma$ denoted a spinless pion) \cite{Gell-Mann:1960mvl}.} The adjective “non-linear” will be clear in the following of the discussion. A combined global and local transformation acts on $\mathcal{V}$ as 
\begin{equation}
\mathcal{V} \mapsto  e^{x'+y}\,g\,e^{x},
\end{equation}
where $x$ and $x'$ are in $\text{Lie}\,H$, and $y$ is in the span generated by the coset generators. Here we suppose that $G$ acts on the left and $H$ on the right (the other possibilities are equivalent), so that $x'$ and $y$ are constant. The corresponding infinitesimal transformation is
\begin{equation}
\delta\,\mathcal{V} = (x'+y)\,\mathcal{V} + \mathcal{V}\,x.
\end{equation}
We want to find which transformations preserve the gauge choice which eliminates the auxiliary fields $\chi^i$:
\begin{equation}
\mathcal{V} = e^\vphi.
\end{equation}
This means that $\delta\,g$ must sit in the span generated by the coset generators. Consider separately the case in which  $y=0$ or $x'=0$. In the first case, we have necessarily to set $x=-x'$, getting
\begin{equation}
\delta_x\,\mathcal{V} = [x',\mathcal{V}],
\end{equation}
because the adjoint action of $\text{Lie}\,H$ is in the coset span as a consequence of \eqref{CosetRules}. In the second case, 
\begin{align}
\delta\,\mathcal{V} = y\,\mathcal{V} + \mathcal{V}\,x &\Leftrightarrow
\mathcal{V}^{-1}\,\delta\,\mathcal{V} = \mathcal{V}^{-1}\,y\,\mathcal{V} + x 
\Leftrightarrow e^{-\vphi}\,\delta\,e^{\vphi} =
e^{-\vphi}\,y\,e^{\vphi} + x \Leftrightarrow \nn\\
& \Leftrightarrow e^{-\vphi}\,\delta\,e^{\vphi} = 
e^{-\text{ad}_\vphi}\,y + x
\Leftrightarrow \tfrac{\mathbbm{1}-e^{-\text{ad}_\vphi}}{\text{ad}_\vphi}\,\delta\,\vphi = e^{-\text{ad}_\vphi}\,y + x,\label{InfCosetTrans}
\end{align}
where in the last-but-one step we used the formula \eqref{AdjointRelation}, and in the last step we used the formula \eqref{FundCosetFormula}. The last equation is solved in terms of $y$ by\footnote{\emph{Proof.}
\begin{align}
\tfrac{\mathbbm{1}-e^{-\text{ad}_\vphi}}{\text{ad}_\vphi}\,\delta\,\vphi &= \tfrac{\mathbbm{1}-e^{-\text{ad}_\vphi}}{\text{ad}_\vphi}\,\tfrac{\text{ad}_\vphi}{\text{tanh}\,\text{ad}_\vphi}\,y =\tfrac{(\mathbbm{1}-e^{-\text{ad}_\vphi})(\mathbbm{1}+e^{-2\text{ad}_\vphi)}}{\mathbbm{1}-e^{-2\text{ad}_\vphi}}\,y =\nn\\
&= \tfrac{\mathbbm{1}+e^{-2\text{ad}_\vphi}}{\mathbbm{1}+e^{-\text{ad}_\vphi}}\,y =
 \tfrac{(\mathbbm{1}+e^{-\text{ad}_\vphi})\,e^{-\text{ad}_\vphi}+\mathbbm{1}-e^{-\text{ad}_\vphi}}{1+e^{-\text{ad}_\vphi}}\,y =\nn\\
&= e^{-\text{ad}_\vphi}\,y + \tfrac{\mathbbm{1}-e^{-\text{ad}_\vphi}}{\mathbbm{1}+e^{-\text{ad}_\vphi}}\,y = e^{-\text{ad}_\vphi}\,y + \text{tanh}\,\tfrac{\text{ad}_\vphi}{2}\,y = e^{-\text{ad}_\vphi}\,y + x.
\end{align}}
\begin{equation}
\delta\,\vphi = \tfrac{\text{ad}_\vphi}{\text{tanh}\,\text{ad}_\vphi}\,y, \quad
x = \text{tanh}\,\tfrac{\text{ad}_\vphi}{2}\,y.
\end{equation}
Combining the two cases we studied separately, we obtain the general expression for a transformation, preserving the gauge choice $\mathcal{V} = e^\vphi$,
\begin{equation}
\delta\,e^\vphi = [x',e^\vphi]+y\,e^\vphi + e^\vphi\,\text{tanh}\,\tfrac{\text{ad}_\vphi}{2}\,y.
\end{equation}
So, $e^\vphi$ has to transform in a highly non-linear way: this is the reason why this model of scalars is called \emph{non-linear} $\sigma$-model. To construct a kinetic Lagrangian for a theory of scalars, invariant under this transformation, one considers the Maurer-Cartan form out of $e^\vphi$, whose transformation is
\begin{equation}\label{CosetMaurerCartan}
\delta\,(e^{-\vphi}\,\de_\mu\,e^{\vphi}) = 
\de_\mu\,x'' + [e^{-\vphi}\,\de_\mu\,e^\vphi,x''],\quad\text{where}\;\;x''=\text{tanh}\,\tfrac{\text{ad}_\vphi}{2}\,y - x'.
\end{equation}
Notice that the Maurer-Cartan form sits in $\text{Lie}\,H$, as a consequence of \eqref{FundCosetFormula} and \eqref{CosetRules}. We can decompose it according to
\begin{equation}
e^{-\vphi}\,\de_\mu\,e^{\vphi} = \0{P}_\mu + \0{Q}_\mu,
\end{equation}
where 
\begin{equations}
\0{P}_\mu &= \tfrac{1}{2}\,(
e^{-\vphi}\,\de_\mu\,e^{\vphi} - e^{\vphi}\,\de_\mu\,e^{-\vphi}) = \tfrac{\text{sinh}\,\text{ad}_\vphi}{\text{ad}_\vphi}\,\de_\mu\,\vphi,\label{QP1}\\
\0{Q}_\mu &= \tfrac{1}{2}\,(
e^{-\vphi}\,\de_\mu\,e^{\vphi} + e^{\vphi}\,\de_\mu\,e^{-\vphi}) = \tfrac{\mathbbm{1}-\text{cosh}\,\text{ad}_\vphi}{\text{ad}_\vphi}\,\de_\mu\,\vphi.\label{QP2}
\end{equations}
Using \eqref{CosetMaurerCartan}, one can see that $\0{Q}_\mu$ transforms as a gauge connection for $H$, with gauge parameter $x''$, whereas $\0{P}_\mu$ transforms homogeneously:
\begin{equation}
\delta\,\0{Q}_\mu = \de_\mu\,x'' + [\0{Q}_\mu,x''], \quad
\delta\,\0{P}_\mu = [\0{P}_\mu,x''].
\end{equation}
Therefore, 
\begin{equation}
\text{Tr}\,\0{P}_\mu\,\0{P}^\mu,
\end{equation}
is invariant, and it can be used to define an action. Notice that, expanding \eqref{QP1}--\eqref{QP2},
\begin{equations}
Q_\mu = \tfrac{1}{2}\,[\vphi,\de_\mu\,\vphi] + \dots,\quad
P_\mu = \de_\mu\,\vphi + \tfrac{1}{3!}\,[\vphi,[\vphi,\de_\mu\,\vphi]] + \dots. 
\end{equations}
and replacing the first in the second, one gets
\begin{equation}
P_\mu = \de_\mu\,\vphi - \tfrac{1}{3}[Q_\mu,\vphi] + \dots 
\end{equation}
Thus, $P_\mu$ can be thought as a non-linear covariant derivative, whose connection is $-\frac{1}{3}\,Q_\mu$, up to higher-order correction. 

Non-linear $\sigma$-models of scalar appear in supergravity as a result of compacfications. In this case it is useful to define the theory of scalars in an $H$-invariant way. Consider again the transformation of $\mathcal{V}$, this time without assuming the latter to sit in the coset span,
\begin{equation}\label{deltaV}
\delta\,\mathcal{V} = \Lambda\,\mathcal{V} + \mathcal{V}\,x,
\end{equation}
where $\Lambda = x' + y$ is an arbitrary global transformations in $\text{Lie}\;G$. The analogous computation of the transformation of the Maurer-Cartan form out of $\mathcal{V}$ shows that it is $\Lambda$-invariant
\begin{equation}\label{TransCosetMaurerCartan}
\delta\,(\mathcal{V}^{-1}\,\de_\mu\,\mathcal{V}) = 
\de_\mu\,x + [\mathcal{V}^{-1}\,\de_\mu\,\mathcal{V},x].
\end{equation}
If we decompose $\mathcal{V}^{-1}\,\de_\mu\,\mathcal{V}$, which takes values in $\text{Lie}\,G$, in its part in the coset span and its part in $\text{Lie}\,H$, 
\begin{equation}
\mathcal{V}^{-1}\,\de_\mu\,\mathcal{V} = P_\mu + Q_\mu, \quad P_\mu \in \text{Span}\,\{T_a\},\; Q_\mu \in \text{Lie}\,H,
\end{equation}
then the transformation \eqref{TransCosetMaurerCartan} is decomposed as 
\begin{equation}
\delta\,P_\mu = [P_\mu,x], \quad
\delta\,Q_\mu = \de_\mu\,x + [Q_\mu,x],
\end{equation}
showing that $P_\mu$ homogeneously transforms and $Q_\mu$ transforms as gauge connection for $H$, on the same footing as the gauge-fixed analogues $\0{P}_\mu$ and $\0{Q}_\mu$ respectively. Again $\text{Tr}\,P_\mu\,P^\mu$ is invariant and can be used to define an action. One can also introduce a symmetric matrix 
\begin{equation}
\mathcal{M} = \mathcal{V}\,\Delta\,\mathcal{V}^t,
\end{equation}
where $\Delta$ is a constant, positive-definite, $H$-invariant matrix. Using the transformation \eqref{deltaV}, one can see that $\mathcal{M}$ is invariant under the local transformation:
\begin{align}
\delta\,\mathcal{M} &= \delta\,\mathcal{V}\,\Delta\,\mathcal{V}^t + \mathcal{V}\,\Delta\,\delta\,\mathcal{V}^t = (\Lambda\,\mathcal{V} + \mathcal{V}\,x)\,\Delta\,\mathcal{V}^t + \mathcal{V}\,\Delta\,(\mathcal{V}^t\,\Lambda^t + x^t\,\mathcal{V}^t) = \nn\\
&= \Lambda\,\mathcal{M} + \mathcal{M}\,\Lambda^t + \mathcal{V}\,(x\,\Delta + \Delta\,x^t)\,\mathcal{V}^t = \Lambda\,\mathcal{M} + \mathcal{M}\,\Lambda^t,
\end{align}
where in the last step the second term vanishes because of the $H$-invariance of $\Delta$. One can check by explicit evaluation that $\text{Tr}\,(\de_\mu\,\mathcal{M}\,\de^\mu\,\mathcal{M}^{-1})$ is invariant under the previous transformation, giving and alternative way to write the action of the theory. Indeed, it should be proportional to $\text{Tr}\,P_\mu\,P^\mu$. This is the case, and the precise constant of proportionality is in the following:
\begin{align}
-\tfrac{1}{4}\,\text{Tr}\,(\de_\mu\,\mathcal{M}\,\de^\mu\,\mathcal{M}^{-1}) &= \tfrac{1}{4}\,\text{Tr}\,(\mathcal{M}^{-1}\,\de_\mu\,\mathcal{M}\,\mathcal{M}^{-1}\,\de^\mu\,\mathcal{M}) = \nn\\
&= \tfrac{1}{2}\,\text{Tr}\,(\mathcal{V}^{-1}\,\de_\mu\,\mathcal{V}\,\mathcal{V}^{-1}\,\de^\mu\,\mathcal{V}) + \text{Tr}\,(\mathcal{V}^{-1}\,\de_\mu\,\mathcal{V}\,(\mathcal{V}^{-1}\,\de^\mu\,\mathcal{V})^t) = \nn\\
& = \text{Tr}\,P_\mu\,P^\mu.
\end{align}
For example, a possible parametrisation of $\mathcal{V}$ for $E_{8(8)}/\text{SO}(16)$ is \cite{Hohm:2014fxa}\footnote{In \cite{Coimbra:2011ky} the authors take a unique exponential, with the sum the single exponents. The two choices are equivalent, modulo appropriate redefinition of the exponents, because of the Baker-Campbell-Haussdorf formula and because the generators $t_{(n)}$ form an algebra, so the results of the commutators involved in the formula can always be written in terms of the generators themselves.}
\begin{equation}
\mathcal{V}^t = \exp(\vphi\,t_{(0)})\,\mathcal{V}_8\,\exp(a_{m_1m_2m_3}\,t_{(1)}^{m_1 m_2 m_3})\,\exp(\vepsilon^{m_1\dots m_8}\,\tilde{a}_{m_1 \dots m_6}\,t_{(2)m_7 m_8}),
\end{equation}
where $\mathcal{V}_8$ is a vielbein parametrising the coset of $\text{SL}(8,\mathbb{R})$, filled with the degrees of freedom of the gravitational sector, $t_{(0)}$ is the generator associated with the trombone, with parameter $\vphi$, and $t_{(1)}$ and $t_{(2)}$ are the $E_{8(8)}$ generators in the representation of weights one and two, respectively, which corresponds to the three-form and to the dual six-form. We are excluding the degrees of freedom coming from the dual graviton and the other exotic representations within $\mathbf{248}$, since they are expected to be gauged-away from the final theory.

\subsection{Generalised torsion and Weitzenb\"ock connection}\label{GenWT}

Considering a global frame $E_A = E^M{}_A\,\de_M$, we can extend the definition of Weitzenb\"ock torsion in \eqref{ExprTorsionLieDer2} to the generalised Lie derivative \eqref{GenLieDerP} in order to define a \emph{generalised Weitzenb\"ock torsion} \cite{PiresPacheco:2008qik, Coimbra:2011nw, Coimbra:2011ky, Coimbra:2012af}:
\begin{align}\label{TorsionExFT}
-T_A{}^C{}_B(E)\,E^M{}_C &= L_{E_A}\,E_B = E^N{}_{[A}\,\de_N\,E^M{}_{B]} + Y^{MP}{}_{QN}\,\de_P\,E^Q{}_A{}E^N{}_B.
\end{align}
Introducing the \emph{generalised Weitzenb\"ock connection}, $W_A{}^C{}_B(E)$ of the generalised frame $E^M{}_A$, in analogy with \eqref{WeitzenbockDef},
\begin{equation}\label{GenWconnection}
W_A{}^C{}_B(E) = -E_M{}^C\,E^N{}_A\,\de_N\,E^M{}_B = E^N{}_B{}\,E^M{}_A{}\,\de_M\,E_N{}^C,
\end{equation}
where $E_M{}^A$ are the components of the inverse frame ($f^{(n)}$-bein),
and using the invariance of the $Y$ tensor
\begin{equation}
Y^{CE}{}_{FB}\,E_N{}^F = Y^{CE}{}_{NB}, \quad
Y^{CE}{}_{FB}\,E^M{}_E = Y^{CM}{}_{FB},
\end{equation}
the generalised torsion can be written in terms of the Weitzenb\"ock connection as
\begin{equation}\label{GenTorsionW}
T_A{}^C{}_B(E) = W_A{}^C{}_B(E) - W_B{}^C{}_A(E) + Y^{CE}{}_{FB}\,W_E{}^F{}_A(E),
\end{equation}
where, as expected, the $Y$ tensor parametrises the deformation with respect to the familiar case. Compare this result with the analogous in double geometry \eqref{ParallDFT} and \eqref{TorsionDFT}.

\subsection{Twisted generalised Lie derivative}\label{TwistedGenLieDer}

Define the twisting operation of a generalised vector $V^A$ and of a generalised one-form $U_A$ by an invertible matrix $E^M{}_A$  (\emph{twisting matrix}), which could be a frame, but this is not necessary \cite{Ciceri:2016dmd, Inverso:2017lrz}:
\begin{equation}
V'^M = E^M{}_A\,V^A, \quad
U'_M = E_M{}^A\,U_A,
\end{equation}
where $E_M{}^A$ are the components of the inverse of $E^M{}_A$. A twisting matrix $C{}^M{}_N$ is said to preserve the section choice $\mathcal{E}_M{}^m$ if
\begin{equation}
C^P{}_R\,\mathcal{E}_P{}^p = \mathcal{E}_R{}^p.
\end{equation}
This implies that, taking a derivative $\de_M$  at both sides, 
\begin{equation}
\de_M\,C^P{}_R\,\mathcal{E}_P{}^p = 0,
\end{equation}
so that the Weitzenb\"ock connection $W_M{}^P{}_N(C)=-C_N{}^R\,\de_M\,C^P{}_R$ out of $C^M{}_N$ satisfies
\begin{equation}
W[C]_M{}^P{}_N\,\mathcal{E}_P{}^p = 0.
\end{equation}
Instead, taking the $\de_p$ derivative of both sides, and using $\mathcal{E}_P{}^p\,\de_p =\de_P$,
\begin{equation}
\de_P\,C^P{}_R = 0.
\end{equation}

Now, consider the generalised Lie derivative $\mathcal{L}_{X'}\,V'^A$ of a twisted vector $V'$, generated by a twisted vector $X'$. Taking $X'^M = E^M{}_A\,X^A$ and $V'^M = E^M{}_A\,V^A$, one can derive the following formula:
\begin{equation} \label{DressedLieDerivative}
L_{X'}\,V'^M = E^M{}_A\,(L_X\,V)^A - T_A{}^C{}_B(E)\,E^M{}_C\,X^A\,V^B,
\end{equation}
where $T_A{}^C{}_B(E)$ is the torsion out of the twisting matrix (but remember that the latter is not required to be a global frame in general), and in $(L_X\,V)^A$ all the derivatives are formally replaced by $E^M{}_A\,\de_M$, as it follows:
\begin{align}
(L_X\,V)^A &= X^B\,(E^M{}_B\,\de_M)\,V^A - V^B\,(E^M{}_B\,\de_M)\,X^A \,+\nn\\
& + (\lambda-\omega)\,(E^M{}_B\,\de_M)\,X^B\,V^A + Y^{AC}{}_{DB}\,(E^M{}_C\,\de_M)\,X^D\,V^B.
\end{align}
In the computation, the $Y$ tensor is assumed to be invariant under $E^M{}_A$, but the section is not assumed to be preserved by the twisting. The formula \eqref{DressedLieDerivative} shows that the (generalised) Lie derivative of twisted vectors is equal to the twisted Lie derivative, with twisted partial derivatives, and a correction, which is the adjoint action of $X^A\,T_A$ on $V$, where $T_A$ is the matrix with components $(T_A)^C{}_B = T_A{}^C{}_B$:
\begin{equation}
\delta_{T_A}\,V^C = -T_A{}^C{}_B,\,V^B, \quad
\delta_{T_A}\,U_C = T_A{}^B{}_C\,U_B.
\end{equation}
If the twisting matrix $C_M{}^N$ preserves the section choice, then \eqref{DressedLieDerivative} becomes
\begin{equation}\label{DressedDerivativeSectionPreserving}
L_{X'}\,V'^M = C^M{}_N\,L_X\,V^N - T_P{}^N{}_Q(C)\,C^M{}_N\,X^P\,V^Q,
\end{equation}
where now $X'^M = C^M{}_N\,X^N$ and $V'^M = C^M{}_N\,V^N$. The first term reproduces the generalised Lie derivative, because $C^M{}_N\,\de_M = \de_N$, and $T(C)$ is the torsion of $C^M{}_N$:
\begin{equation}
-C^P{}_Q\,T_M{}^Q{}_N(C) = L_{C_M}\,C^P{}_N =
C^Q{}_{[M}\,\de_Q\,C^P{}_{N]} + Y^{PR}{}_{ST}\,\de_R\,C^S{}_M\,C^T{}_N.
\end{equation}

\subsection{Flux-deformed generalised Lie derivative}\label{FluxGenLieDer}

Consider the following deformation of the generalised Lie derivative, which we will call \emph{flux-deformed Lie derivative} (or $F$-deformed Lie derivative, whenever we need to stress which flux is used): 
\begin{equation}
L_\xi^{(F)} = L_\xi + \xi^M\,\delta_{F_M},
\end{equation}
where the ``flux"  $(F_M)_N{}^P$ is a priori the torsion of no frame, and $\delta_{F_M}$ is its adjoint action. Explicitly,
\begin{align}
L^{(F)}_\xi\,V^M &= \xi^N\,\de_N\,V^M - V^N\,\de_N\,\xi^M + (\lambda-\omega)\,\de_N\,\xi^N\,V^M \,+\nn\\
& + Y^{MP}{}_{QN}\,\de_P\,\xi^Q\,V^N - F_{PQ}{}^M\,\xi^P\,V^Q,\\
L^{(F)}_\xi\,U_M &= \xi^N\,\de_N\,U_M + U_N\,\de_M\,\xi^N + (\lambda+\omega)\,\de_N\,\xi^N\,U_M \,+\nn\\
& - Y^{NP}{}_{QM}\,\de_P\,\xi^Q\,U_N + F_{NM}{}^P\,\xi^N\,U_P.
\end{align}
Using the flux-deformed Lie derivative, the formula \eqref{DressedDerivativeSectionPreserving} can be written as
\begin{equation}
L_{X'}\,V'^M =  C_N{}^M\,L_{X}^{(T(C))}\,V^N.
\end{equation} 
The flux-deformed Lie derivative has to satisfy the Leibniz identity to be a consistent derivative:
\begin{equation}
[L_X^{(F)},L_Y^{(F)}] = L^{(F)}_{L_X^{(F)}Y},
\end{equation}
which, as usual, splits in symmetric and antisymmetric parts:
\begin{equation}
[L_X^{(F)},L_Y^{(F)}] = L^{(F)}_{\frac{1}{2}(L_X^{(F)}Y-L_Y^{(F)}X)}, \quad
L^{(F)}_{\frac{1}{2}(L_X^{(F)}Y + L_X^{(F)}Y)} = 0.
\end{equation}
One can check that these requirements are fulfilled if and only if the flux satisfies the following constraints \cite{Ciceri:2016dmd, Inverso:2017lrz}:
\begin{itemize}
\item[1.] \emph{(First) section constraint} 
\begin{equation}\label{FirstSectionConstraint}
F_{PQ}{}^M\,\mathcal{E}_M{}^m = 0. 
\end{equation}
\item[2.] \emph{Linear constraint}\footnote{One can verify that the antisymmetric part of the Leibniz identity imposes a constraint, which is equivalent to the constraint imposed by the symmetric part. To see it, one has to use the invariance of the $Y$-tensor under the adjoint action of the flux $\delta_{F_A}\,Y=0$. Nevertheless, the constraint reported here is simply the sum of the constraint coming from the antisymmetric and the symmetric part, because it will be more convenient in this form.}
\begin{align}\label{FluxLinearConstraint}
& F_{SR}{}^M\,\mathcal{E}_N{}^p - Y^{QP}{}_{RS}\,F_{QN}{}^M\,\mathcal{E}_P{}^p \,+\nn\\
& - Y^{QP}{}_{RN}\,F_{SQ}{}^M\,\mathcal{E}_P{}^p + Y^{MP}{}_{RQ}\,F_{SN}{}^Q\,\mathcal{E}_P{}^p = \nn\\
& = - F_{RS}{}^M\,\mathcal{E}_N{}^p
+ Y^{PR}{}_{QN}\,F_{RS}{}^Q\,\mathcal{E}_P{}^p.
\end{align}
\item[3.] \emph{Bianchi constraint}
\begin{align}\label{BianchiConstraintFlux}
\de_M\,F_{NP}{}^Q  &+ F_{MN}^R\,F_{RP}{}^Q + F_{MP}{}^R\,F_{NR}{}^Q - F_{MR}{}^Q\,F_{NP}{}^R =\nn\\
& =\de_N\,F_{MP}{}^Q - \de_P\,F_{MN}{}^Q+ Y^{QR}{}_{SP}\,\de_R\,F_{MN}{}^S,
\end{align}
which can be conveniently written as 
\begin{equation}\label{FluxBianchi}
\delta_{F_M}\,F_{NP}{}^Q = -\de_M\,F_{NP}{}^Q + \de_N\,F_{MP}{}^Q - \de_P\,F_{MN}{}^Q + Y^{QR}{}_{SP}\,\de_R\,F_{MN}{}^S.
\end{equation}
\end{itemize}
Actually, one can verify that the linear constraint can be replaced by:
\begin{itemize}
\item[2.] (bis) \emph{Second section constraint}
\begin{equation}\label{SecondSectionConstraint}
Y^{MP}{}_{QN}\,F_{MR}{}^R\,\mathcal{E}_P{}^p = 0,
\end{equation}
\end{itemize}
which means that the tensor product between the trace $F_{MR}{}^R$ of the flux and an object satisfying the section constraint has to be on section.\footnote{We checked that this condition is equivalent to the more complicated linear constraint only numerically.} We call this constraint \emph{second section constraint}, to distinguish it from the (first) one in \eqref{FirstSectionConstraint}.

We know that the flux deformation can be absorbed in the twisting of the parameters, if the flux is the torsion of a frame. In such a case, the Leibniz identity is automatically satisfied. Nevertheless, it will be useful to write the Bianchi identity for the torsion $T_A{}^C{}_B(E) = T_A{}^C{}_B$ of a frame $E^M{}_A$, seen as a twisting matrix, without assuming it to preserve a given section choice $\mathcal{E}_M{}^m$. To compute the Bianchi identity, consider the Leibniz identity with parameter $X'$ and $Z'$; evaluate it on an vector $V'$; set $X'^M = E^M{}_A\,X^A$, $Z'^M = E^M{}_A\,Z^A$, and $V'^M = E^M{}_A\,V^M$; select the coefficient of the term $V\,X\,Z$ -- which is the same as setting $V$, $X$ and $Z$ to be constant. The result is
\begin{align}
& E^M{}_F\,\de_M\,T_A{}^C{}_B + T_F{}^D{}_B\,T_A{}^C{}_D - T_A{}^D{}_B\,T_F{}^C{}_D + T_F{}^D{}_A\,T_D{}^C{}_B \,=\nn\\
& = E^M{}_A\,\de_M\,T_F{}^C{}_B - E^N{}_B\,\de_M\,T_F{}^C{}_A + Y^{CD}{}_{EB}\,E^M{}_D\,\de_M\,T_F{}^E{}_A.
\end{align}
This is the same as the Bianchi constraint for an arbitrary flux \eqref{BianchiConstraintFlux}, with the flux $F$ replaced by the torsion $T$, and with the simple derivative replaced by the twisted one. This Bianchi identity can be also written as
\begin{equation}\label{BianchiT}
(E^M{}_F\,\de_M + \delta_{T_F})\,T_A{}^C{}_B = T_A{}^C{}_B\{E^M{}_A\,\de_M\,T_F{}^C{}_B\},
\end{equation}
where the adjoint action $\delta_{T_A}$ corresponds to the last three terms in the first line in the explicit expression of the constraint. In the right-hand side there is the torsion out of the Weitzenb\"ock connection in curly bracket, $A$ being a spectator index. The operator in the left-hand side is a sort of (twisted) covariant derivative. This is the reason why the previous identity is called Bianchi identity.\footnote{In the torsion formulation of General Relativity, the expression of the covariant derivative of the torsion is the analogous of the Bianchi identity for the Riemann curvature (compare with \eqref{TraceBianchiWeitzenbock}).}

Let us conclude this section with more details on the derivation of the flux constraints. One has to compute
\begin{equation}
([L_X^{(F)},L_Z^{(F)}] -L_{\frac{1}{2}(L_X^{(F)}Z-L_Z^{(F)}X)})\,V^M, \quad
L^{(F)}_{\frac{1}{2}(L_X^{(F)}Z + L_X^{(F)}Z)}\,V^M,
\end{equation}
with arbitrary vectors $X, Z, V$. Only the terms proportional to $F$ or $\de\,F$ have to be selected, since, if we set $F, \de\,F \rightarrow 0$, the usual generalised Lie derivative is recovered (which satisfies the Leibniz identity, upon using the section constraint). The unique possible independent terms are of the type $F\,\de\,V\,\xi\,\zeta$, $F^2\,V\,X\,Z$, $\de\,F\,V\,X\,Z$, and $F\,V\,(X\,\de\,Z \pm X\,\de\,Z)$ (the sign $\pm$ depending on whether the symmetric or the antisymmetric part is considered). In order to eliminate the first type of terms, one has to impose the symmetric and the antisymmetric part in $R,S$ of the section constraint. The coefficient multiplying $V\,X\,Z$, which is a combination of $F^2$ and $\de\,F$, is the symmetric or the antisymmetric part in $R,S$ of the quadratic constraint. Finally, the coefficient of $V\,(X\,\de\,Z \pm X\,\de\,Z)$ is 
\begin{equation}
\tfrac{1}{2}\,(\delta_N^P\,F_{(RS)}{}^M - F_{QN}{}^M\,Y^{QP}{}_{RS}-F_{(RS)}{}^Q\,Y^{MP}{}_{QN})\,\mathcal{E}_P{}^p = 0.
\end{equation}
for the symmetric part; it is equal to the following expression for the antisymmetric one:
\begin{align}
(\tfrac{1}{2}\,\delta_N^P\,F_{(RS)}{}^M &- \tfrac{1}{2}F_{QN}{}^M\,Y^{QP}{}_{RS}-\tfrac{1}{2}\,F_{[RS]}{}^Q\,Y^{MP}{}_{QN} \,+\nn\\
& - F_{SQ}{}^M\,Y^{QP}{}_{RN} + F_{SN}{}^Q\,Y^{MP}{}_{RQ})\,\mathcal{E}_P{}^p.
\end{align}
Although it is not completely evident at a first glance, this coefficient is equal to the symmetric one. To see why, rewrite it as follows\footnote{In this way it is clear that, if we sum the symmetric and the antisymmetric constraints, we find the linear constraint as in \eqref{FluxLinearConstraint}.}
\begin{align}
(\tfrac{1}{2}\,\delta_N^P\,F_{(RS)}{}^M &- \tfrac{1}{2}F_{QN}{}^M\,Y^{QP}{}_{RS}-\tfrac{1}{2}F_{(RS)}{}^Q\,Y^{MP}{}_{QN} \,+\nn\\
& - (F_{SQ}{}^M\,Y^{QP}{}_{RN} - F_{SN}{}^Q\,Y^{MP}{}_{RQ}-F_{SR}{}^Q\,Y^{MP}{}_{QN})\,\mathcal{E}_P{}^p
\end{align}
(the last term in the first line and the last in the second one sum into the last term in the first line in the starting expression). Notice that the first line is the linear constraint times $\tfrac{1}{2}$, so we only have to prove the second line to vanish. To this aim, consider the invariance of the $Y$ tensor under the adjoint action of the flux:
\begin{align}
\delta_{F_S}\,Y^{MP}{}_{RN} &= F_{SQ}{}^M\,Y^{QP}{}_{RN} + F_{SQ}{}^P\,Y^{MQ}{}_{RN} \,+\nn\\
& - F_{SN}{}^Q\,Y^{MP}{}_{RQ} - F_{SR}{}^Q\,Y^{MP}{}_{QN} = 0.
\end{align}
Multiplying it by $\mathcal{E}_P{}^p$ and using the section constraint of the flux (which eliminates the second term),
\begin{equation}
(F_{SQ}{}^M\,Y^{QP}{}_{RN} - F_{SN}{}^Q\,Y^{MP}{}_{RQ}- F_{SR}{}^Q\,Y^{MP}{}_{QN})\,\mathcal{E}_P{}^p = 0.
\end{equation}
But this is precisely the second line in the starting expression, as we wanted.

\subsection{Section constraint for the embedding tensor}\label{ThetaSection}

Let us show that any embedding tensor $\vtheta_A{}^I$ satisfies the section constraint, providing a candidate for the section matrix. Following the notation in Section \eqref{Coset}, let $\{ T_I \}$ the generators of the algebra of the gauge group $G \in E_{n(n)}$. Splitting the index $I$, which is the adjoint of the algebra of $G$, in $(a,i)$, where $i=1,\dots,\text{dim}\,\text{Lie}\,H$ is the adjoint index of the algebra of $H$, and $a = 1,\dots, n$ lists the coset generators, \cite{Inverso:2017lrz}
\begin{equation}\label{DecompositionEmbeddingTensor}
\vtheta_A{}^I\,T_I = \vtheta_A{}^a\,T_a + \vtheta_A{}^i\,\,T_i.
\end{equation}
The embedding tensor selects the subalgebra of the algebra of the duality group to be gauged in the gauged supergravity, choosing the gauge fields among the vectors $\hat{A}_\mu{}^M$. In the particular case of $n=7$, the embedding tensor sits in the representation $\mathbf{56}\oplus\mathbf{912}$, where the first representation is due to the trombone component.  

$e^m{}_a = e^m{}_a(y)$ denotes the frame for the $n$-dimensional internal manifold, where $y^m$, $m = 1,\dots,n$, are the coordinates of the internal manifold, and $a = 1,\dots,n$ is a flattened index. The section matrix $\mathcal{E}_M{}^m$ teaches how to embed the inverse frame $e_m{}^a$ in a $E_{n(n)}$ covariant object $e_M{}^A = e_M{}^A(y)$:
\begin{equation}\label{EmbeddingE}
\mathcal{E}_M{}^m\,e_m{}^a = e_M{}^A\,\mathcal{E}_A{}^a.
\end{equation}

Consider a generalised frame $\hat{E}^M{}_A$ satisfying the parallelisability condition with constant intrinsic torsion given by the components of the embedding tensor
\begin{equation}\label{CondPar}
L_{\hat{E}_A}\,\hat{E}_B = - X_{AB}{}^C\,\hat{E}_C.
\end{equation}
Multiplying by the section matrix $\mathcal{E}_M{}^m$, and unpacking the generalised Lie derivative,
\begin{equation}
\hat{E}^N{}_{[A}\,\de_N\,\hat{E}^M{}_{B]}\,\mathcal{E}_M{}^m + Y^{MP}{}_{QN}\,\hat{E}^N{}_A\,\de_P\,\hat{E}^Q{}_B\,\mathcal{E}_M{}^m = -X_{AB}{}^C\,\hat{E}^M{}_C\,\mathcal{E}_M{}^m.
\end{equation}
Using $\de_M = \mathcal{E}_M{}^m\,\de_m$ and the section constraint, making the $Y$-term vanishing, 
\begin{equation}
\hat{E}^N{}_{[A}\,\mathcal{E}_N{}^n\,\de_n\,(\hat{E}^M{}_{B]}\,\mathcal{E}_M{}^m) = -X_{AB}{}^C\,\hat{E}^M{}_C\,\mathcal{E}_M{}^m,
\end{equation}
where one recognises the commutator of the vector field $\hat{E}^M{}_A\,\mathcal{E}_M{}^m$, 
\begin{equation}\label{CommKilling}
[\hat{K}_A,\hat{K}_B]^m = -X_{AB}{}^C\,\hat{K}_C,\quad \text{with}\;K_A{}^m = \hat{E}^M{}_A\,\mathcal{E}_M{}^m.
\end{equation}
As a solution, one can choose the Killing vectors $K_I{}^m$, twisted by $\vtheta_A{}^I$,
\begin{equation}\label{SolKilling}
K_A{}^m = \vtheta_A{}^I\,K_I{}^m.
\end{equation}
Indeed, replacing \eqref{SolKilling} in \eqref{CommKilling}, 
\begin{equation}
\vtheta_A{}^I\,\vtheta_B{}^J\,[K_I,K_J]^m = -X_{AB}{}^C\,\vtheta_C{}^K\,K_K{}^m.
\end{equation}
Using the quadratic constraint of the embedding tensor \eqref{QC1}
\begin{equation}
\vtheta_A{}^I\,\vtheta_B{}^J\,[K_I,K_J]^m = -\vtheta_A{}^I\,\vtheta_B{}^J\,f_{IJ}{}^K\,K_K{}^m,
\end{equation}
which is an identity, as a consequence of the commutation rule \eqref{RuleKilling}.

The Killing vectors in a coset space $G/H$ are given by \eqref{ExplicitKilling}. Multiplying by $\vtheta_A{}^I$, and using the gauge invariance of $\vtheta_A{}^I$ and the decomposition of \eqref{DecompositionEmbeddingTensor},
\begin{equation}
\vtheta_A{}^I\,K_I{}^m = (L^{-1}\,X_A\,L)^a\,e^m{}_a = (L_A{}^B\,X_B)^a\,e^m{}_a = 
L_A{}^B\,\vtheta_B{}^a\,e^m{}_a.
\end{equation}
This means that the expression \eqref{SolKilling} becomes 
\begin{equation}
K_A{}^m = \vtheta_B{}^a\,L_A{}^B\,e^m{}_a.
\end{equation}
This expression can be viewed as a choice for the section matrix provided by the embedding tensor:
\begin{equation}\label{RelationEhattheta}
\hat{E}^M{}_A\,\mathcal{E}_M{}^m\,e_m{}^a = L_A{}^B\,\vtheta_B{}^a.
\end{equation}
Notice that the frame in the above expression is defined up to the twisting by a matrix $C^M{}_N$ preserving the section choice 
\begin{equation}
\hat{E}^M{}_A\,\mathcal{E}_M{}^m = C^M{}_N\,E^N{}_A\,\mathcal{E}_M{}^m = E^M{}_A\,\mathcal{E}_M{}^m.
\end{equation}
As a consistency condition, the embedding tensor has to satisfy the section constraint.
\begin{align}
& Y^{MN}{}_{PQ}\,\mathcal{E}_M{}^m\,\mathcal{E}_N{}^n = 0 \Leftrightarrow \nn\\
\Leftrightarrow &\, \hat{E}_P{}^C\,\hat{E}_Q{}^D\,Y^{AB}{}_{CD}\,(\hat{E}^M{}_A\,\mathcal{E}_M{}^m)\,(\hat{E}^N{}_B\,\mathcal{E}_N{}^n) = 0 \Leftrightarrow \nn\\
\Leftrightarrow &\,Y^{AB}{}_{CD}\,(\hat{E}^M{}_A\,\mathcal{E}_M{}^m\,e_m{}^a)\,(\hat{E}^N{}_B\,\mathcal{E}_N{}^n\,e_n{}^b) = 0 \Leftrightarrow \nn\\
\Leftrightarrow &\,Y^{AB}{}_{CD}\,(L_A{}^E\,\vtheta_E{}^a)\,(L_B{}^F\,\vtheta_F{}^b) = 0 \Leftrightarrow \nn\\
\Leftrightarrow &\, Y^{EF}{}_{CD}\,\vtheta_E{}^a\,\vtheta_F{}^b = 0.
\end{align}
In the first step we used the invariance of $Y$ under $\hat{E}$, in the last one the invariance under $L$. 

In conclusion, the embedding tensor defines a choice of the section, differing from any  fixed $\mathcal{E}_M{}^m$ only by a rotation in the duality group, which can be absorbed in the definition of $\hat{E}_A{}^M$, providing that the embedding tensor satisfies the section constraint: 
\begin{equation}\label{SectionChoiceTheta}
\mathcal{E}_A{}^a = \vtheta_A{}^a,
\end{equation}
or, using \eqref{EmbeddingE},
\begin{equation}
\mathcal{E}_M{}^m = e_M{}^A\,e^m{}_a\,\vtheta_A{}^a.
\end{equation}

The generalised frame defined by \eqref{RelationEhattheta} up to a rotation in the duality group, which we denote with $E^M{}_A$, is obtained by replacing \eqref{SectionChoiceTheta}
\begin{equation}
E^M{}_A\,e_M{}^B\,\vtheta_B{}^a = L_A{}^B\,\vtheta_B{}^a
\end{equation}
so that the natural choice is 
\begin{equation}\label{ExpressionForE}
E^M{}_A = e^M{}_B\,L_A{}^B.
\end{equation}
This frame does not satisfy the parallelisability condition, but the freedom in twisting by duality rotation can be exploited to get a frame $\hat{E}^M{}_A$, fulfilling \eqref{CondPar}.

\subsection{The flux satisfies the first section constraint}\label{FluxConstraint1}

Now, we want to show that the flux $F_{MN}{}^P$ satisfies the first section constraint \eqref{FirstSectionConstraint}. Observe that, since the dressing matrix $C^M{}_N$ preserves the section choice, the Killing vectors $K_A{}^m$ can be equivalently defined using either $\hat{E}^M{}_A$ or $E^M{}_A$:
\begin{equation}
K_A{}^m = \hat{E}^M{}_A\,\mathcal{E}_M{}^m = E^N{}_A\,C^M{}_N\,\mathcal{E}_M{}^m = E^N{}_A\,\mathcal{E}_N{}^m.
\end{equation}
Therefore, not only
\begin{equation}
L_{\hat{E}_A}\,\hat{E}^M{}_B\,\mathcal{E}_M{}^m = \mathcal{L}_{K_A}\,K_B{}^m,
\end{equation}
but also
\begin{equation}
L_{E_A}\,E^M{}_B\,\mathcal{E}_M{}^m = \mathcal{L}_{K_A}\,K_B{}^m.
\end{equation}
Using this,
\begin{align}\label{SectionFirst1}
L_{E_A}^{(F)}\,E^M{}_B\,\mathcal{E}_M{}^m &= L_{E_A}\,E^M{}_B\,\mathcal{E}_M{}^m - F_{NP}{}^M\,E^N{}_A\,E^P{}_B\,\mathcal{E}_M{}^m = \nn\\
& = \mathcal{L}_{K_A}\,K_B{}^m - (F_{NP}{}^M\,\mathcal{E}_M{}^n)\,E^N{}_A\,E^P{}_B = \nn\\
& = -X_{AB}{}^C\,K_C{}^m - (F_{NP}{}^M\,\mathcal{E}_M{}^m)\,E^N{}_A\,E^P{}_B.
\end{align}
Since the parallelisation condition for $\hat{E}^M{}_A$ can be written making $E^M{}_A$ explicit by means of the flux-deformed generalised Lie derivative in the following way
\begin{equation}\label{SectionFirst2}
L_{E_A}^{(F)}\,E^M{}_B\,\mathcal{E}_M{}^m = -X_{AB}{}^C\,E^M{}_C\,\mathcal{E}_M{}^m = -X_{AB}{}^C\,K_C{}^m,
\end{equation}
comparing the two expressions, we deduce that
\begin{equation}
F_{NP}{}^M\,\mathcal{E}_M{}^m = 0,
\end{equation}
which is the first section constraint for the flux.

\subsection{The second section constraint implies a condition for the embedding tensor}\label{FluxConstraint2}

Let us now study the implication of the second section constraint \eqref{SecondSectionConstraint}.
From the definition of the flux in \eqref{DefFlux}, 
its trace reads
\begin{equation}\label{TraceFlux}
F_{MR}{}^R = E_M{}^A\,(X_{AB}{}^B -T_A{}^B{}_B).
\end{equation}
Remember the formula \eqref{GenTorsionW} for the torsion in terms of the Weitzenb\"ock connection, one can compute its trace. Since the $Y$ tensor is given by \eqref{YTensor},
its trace reads
\begin{equation}
Y^{BD}{}_{EB} = (1+\omega\,f^{(n)}),\delta^D_E,
\end{equation}
where $f^{(n)}$ is the dimension of the fundamental representation of the duality group. 
Thus, the trace of the torsion is
\begin{equation}
T_A{}^B{}_B = W_A{}^B{}_B + \omega\,f^{(n)}\,W_B{}^B{}_A.
\end{equation}
Using the definition of the Weitzenb\"ock connection \eqref{GenWconnection}, and replacing the last expression in \eqref{TraceFlux}, one gets 
\begin{equation}\label{TraceFlux2}
F_{MR}{}^R = E_M{}^A\,X_{AB}{}^B - E^P{}_A\,\de_M\,E_P{}^A + \omega\,f^{(n)}\,E_M{}^A\,\de_N\,E^N{}_A.
\end{equation}
Using the explicit expression for the frame in\eqref{ExpressionForE}, and the decomposition of the Maurer-Cartan form in \eqref{DecompositionMaurerCartan}, the last term can be written as
\begin{align}
E_M{}^A\,\de_N\,E^N{}_A &= (e_M{}^B\,L^A{}_B)\,(\mathcal{E}_N{}^n\,\de_n)\,(e^N{}_C\,L_A{}^C) =\nn\\
&= (e\,\de_n\,e^{-1})^N{}_M\,\mathcal{E}_N{}^n + e_M{}^B\,e^N{}_C\,(L^{-1}\,\de_n\,L)^C{}_B\,\mathcal{E}_N{}^n = \nn\\
&= (e\,\de_N\,e^{-1})^N{}_M
+ e_M{}^B\,\vtheta_C{}^a\,(T_a)^C{}_B + \omega_n{}^i\,(T_i)^N{}_M\,\mathcal{E}_N{}^n,
\end{align}
where we used \eqref{SectionChoiceTheta} for writing the section matrix in terms of the embedding tensor. Back to \eqref{TraceFlux2}, 
\begin{align}
F_{MR}{}^R &= E_M{}^A\,X_{AB}{}^B +\,\omega\,f^{(n)}\,e_M{}^B\,\vtheta_C{}^a\,(T_a)^C{}_B \,+\nn\\
& - E_A{}^P\,\de_M\,E_P{}^A +\omega\,f^{(n)}\,(e\,\de_N\,e^{-1})^N{}_M 
+ \omega\,f^{(n)}\,\omega_n{}^i\,(T_i)^N{}_M\,\mathcal{E}_N{}^n .\label{TraceFlux3}
\end{align}
The first term in the second line is manifestly on section; in the second term in the second line a Maurer-Cartan form is recognised, so the $M$ index is rotated by a $\text{Lie}\,\text{GL}(n)$ rotation, preserving the section choice; the last term in the second line is a transformation in $H$, preserving the section choice. Thus, replacing the trace of the flux in the second section constraint \eqref{SecondSectionConstraint}, according to the last expression \eqref{TraceFlux3}, only the first line in the above expression survives:
\begin{align}
0= Y^{MN}{}_{PQ}\,F_{MR}{}^R\,\mathcal{E}_N{}^n = Y^{AB}{}_{PQ}\,(X_{AC}{}^C + \omega\,f^{(n)}\,\vtheta_C{}^a\,(T_a)_C{}^A)\,e_b{}^n\,\vtheta_B{}^b,
\end{align}
where we used again \eqref{SectionChoiceTheta}. Denoting $\vtheta_A{}^n = \vtheta_A{}^a\,e_a{}^n$, and flattening the last two indices of the $Y$ tensor, we obtain the condition \eqref{ConditionOnTheta}.

\subsection{The flux satisfies the Bianchi constraint}\label{FluxConstraint3}

Finally, we want to prove that the flux satisfies the Bianchi constraint. Compute the flux-deformed Lie derivative of the flux itself
\begin{align}
L_\xi^{(F)}\,F_{MN}{}^P &= \xi^Q\,\de_Q\,F_{MN}{}^P + F_{QN}{}^P\,\de_M\,\xi^Q + F_{MQ}{}^P\,\de_N\,\xi^Q - F_{MN}{}^Q\,\de_Q\,\xi^P \,+\nn\\
& -Y^{QR}{}_{SM}\,\de_R\,\xi^S\,F_{QN}{}^P - Y^{QR}{}_{SN}\,\de_R\,\xi^S\,F_{MQ}{}^P + Y^{PR}{}_{SQ}\,\de_R\,\xi^S\,F_{MN}{}^Q \,+\nn\\
& + (F_{QN}{}^R\,F_{MR}{}^P - F_{QR}{}^P\,F_{MN}{}^R + F_{QM}{}^R\,F_{RN}{}P)\,\xi^Q.
\end{align}
Notice that the first term in the first line vanishes because of the first section constraint \eqref{FirstSectionConstraint}, which holds, as we verified before. The last-but-one term in the first line and all the term in the second line can be replaced by means of the linear constraint \eqref{FluxLinearConstraint} (saturated with $\xi^S$), which holds too, being equivalent to the second section constraint \eqref{SecondSectionConstraint} we already verified (we assume that the condition \eqref{ConditionOnTheta} on the embedding tensor is fulfilled). The last line is the adjoint action of the flux on itself $\xi^Q\,\delta_{F_Q}\,F_{MN}{}^P$, by definition of flux-deformed Lie derivative. So, using the already verified constraints, we can write
\begin{align}
L_\xi^{(F)}\,F_{MN}{}^P &= \xi^Q\,\de_Q\,F_{MN}{}^P + F_{QN}{}^P\,\de_M\,\xi^Q \,+\nn\\
& - F_{QM}{}^P\,\de_N\,\xi^Q + Y^{PR}{}_{SN}\,F_{QM}{}^S\,\de_R\,\xi^Q \,+\nn\\
& + \xi^Q\,\delta_{F_Q}\,F_{MN}{}^P.
\end{align}

Remembering that the Bianchi constraint \eqref{FluxBianchi}
we want to verify fixes the adjoint action of the flux on itself, we can write
\begin{align}
\text{The Bianchi constraint holds} \Leftrightarrow
L_\xi^{(F)}\,F_{MN}{}^P &= \xi^Q\,\de_Q\,F_{MN}{}^P + F_{QN}{}^P\,\de_M\,\xi^Q \,+\nn\\
& - F_{QM}{}^P\,\de_N\,\xi^Q + Y^{PR}{}_{SN}\,F_{QM}{}^S\,\de_R\,\xi^Q \,+\nn\\
& +\xi^Q\,(- \de_Q\,F_{MN}{}^P + \de_M\,F_{QN}{}^P \,+\nn\\
& - \de_N\,F_{QM}{}^P + Y^{PR}{}_{SN}\,\de_R\,F_{QM}{}^S) = \nn\\
& = \de_M\,(\xi^Q\,F_{QN}{}^P) - \de_N\,(\xi^Q\,F_{QM}{}^P) \,+\nn\\
& + Y^{PR}{}_{SN}\,\de_R\,(\xi^Q\,F_{QM}{}^S).
\end{align}
But the right-hand-side is the torsion out of the connection $\de_M\,(\xi^Q\,F_{QN}{}^P)$. In particular, choosing for the vector field $\xi^M$ a frame $E_F{}^M$, we obtain
\begin{equation}\label{Bianchi_iff1}
\text{The Bianchi constraint holds} \Leftrightarrow 
L_{E_F}^{(F)}\,F_{MN}{}^P = T_M{}^P{}_N\{\de_M\,(E_F{}^Q\,F_{QN}{}^P)\}.
\end{equation}
Now, if $\xi^M$ and $F_{MN}{}^P$ are supposed to be twisted by the frame $E_A{}^M$,
\begin{equation}
\xi^M = E^M{}_A\,\tilde{\xi}^A, \quad
F_{MN}{}^P = E^M{}_A\,E^N{}_B\,E^P{}_C\,\tilde{F}_{AB}{}^C,
\end{equation}
then, using the twisting formula for the generalised Lie derivative, we can write, 
\begin{align}\label{LieDerFlux}
L_\xi\,^{(F)}F_{MN}{}^P &= 
E_M{}^A\,E_N{}^B\,E^P{}_C\,(\tilde{L}_{\tilde{\xi}}\,\tilde{F}_{AB}{}^C + \tilde{\xi}^F\,\delta_{T_F}\,\tilde{F}_{AB}{}^C) + \xi^Q\,\delta_{F_Q}\,F_{MN}{}^P = \nn\\
& = E_M{}^A\,E_N{}^B\,E^P{}_C\,(\tilde{L}_{\tilde{\xi}}\,\tilde{F}_{AB}{}^C + \tilde{\xi}^F\,\delta_{T_F}\,\tilde{F}_{AB}{}^C + \tilde{\xi}^F\,\delta_{\tilde{F}_F}\,\tilde{F}_{AB}{}^C).
\end{align}
where $\tilde{L}$ means that the simple derivative $\de_M$ has to be replaced by the dressed one $E^M{}_A\,\de_M$. In the second step, the flux-deformation can be put inside the twisting, because it does not involve any derivative. From the definition of the flux \eqref{DefFlux}, we see that \begin{equation}
\tilde{F}_{AB}{}^C = X_{AB}{}^C - T_A{}^C{}_B.
\end{equation}
Therefore, using the quadratic constraint of the embedding tensor as in \eqref{QC2},
\begin{equation}
\delta_{T_F}\,\tilde{F}_{AB}{}^C + \delta_{\tilde{F}_F}\,\tilde{F}_{AB}{}^C = - \delta_{X_F}\,T_A{}^C{}_B,
\end{equation}
The choice $\xi^M \rightarrow E^M{}_F$ corresponds to $\tilde{\xi}^G \rightarrow \delta^G_F$. In this case, only the transport term survives in the Lie derivative, since it does not involve derivatives of $\tilde{\xi}^G$:
\begin{equation}
\tilde{L}_{\tilde{\xi}}\,\tilde{F}_{AB}{}^C =
\tilde{\xi}^G\,E^M{}_G\,\de_M\,F_{AB}{}^C + \dots  \rightarrow E^M{}_F\,\de_M\,\tilde{F}_{AB}{}^C = - E^M{}_F\,\de_M\,T_A{}^C{}_B.
\end{equation}
Back to \eqref{LieDerFlux}, we obtain
\begin{align}\label{Bianchi_iff2}
L_{E_F}^{(F)}\,F_{MN}{}^P &= E_M{}^A\,E_N{}^B\,E^P{}_C\,(-E^M{}_F\,\de_M\,T_A{}^C{}_B - \delta_{X_F}\,T_A{}^C{}_B) = \nn\\
& = -E_M{}^A\,E_N{}^B\,E^P{}_C\,(E^M{}_F\,\de_M + \delta_{X_F})\,T_A{}^C{}_B.
\end{align}
Therefore, replacing in \eqref{Bianchi_iff1}, 
\begin{equation*}
\text{The Bianchi constraint holds} \Leftrightarrow 
\end{equation*}
\begin{equation}
\Leftrightarrow -(E^M{}_F\,\de_M + \delta_{X_F})\,T_A{}^C{}_B= E^M{}_A\,E^N{}_B\,E_P{}^C\,T_M{}^P{}_N\{\de_M\,(E^Q{}_A\,F_{QN}{}^P)\}.
\end{equation}
Consider again the right-hand side. Using \eqref{GenWconnection} and \eqref{GenTorsionW}, it can be equivalently written as
\begin{equation}\label{Bianchi_iff3}
E^M{}_A\,E^N{}_B\,E_P{}^C\,
 T_M{}^P{}_N\{\de_M\,(E^Q{}_A\,F_{QN}{}^P)\} = 
 T_A{}^C{}_B\{[W_A,\tilde{F}_F]^C{}_B + E^M{}_A\,\de_M\,\tilde{F}_{FB}{}^C\},
\end{equation}
where $(W_A)^C{}_B = W_A{}^C{}_B$ and $(\tilde{F}_A)^C{}_B = \tilde{F}_A{}^C{}_B$. Now, since 
\begin{align}
\delta_{\tilde{F}_F}\,W_A{}^C{}_B &= W_E{}^C{}_B\,\tilde{F}_{FA}{}^E + W_A{}^C{}_E\,\tilde{F}_{FB}{}^E - W_A{}^E{}_B\,\tilde{F}_{EB}{}^C = \nn\\
& = \tilde{F}_{FA}{}^E\,W_E{}^C{}_B - [W_A,\tilde{F}_F]^C{}_B = - [W_A,\tilde{F}_F]^C{}_B,
\end{align}
where the first term in the last step vanishes thanks to the first section constraint, we can replace the commutator $[W_A,\tilde{F}_F]^C{}_B$:
\begin{align}
E^M{}_A\,E^N{}_B,E_P{}^C\,
 T_M{}^P{}_N\{\de_M\,(E^Q{}_A\,F_{QN}{}^P)\} &= 
 T_A{}^C{}_B\{-\delta_{\tilde{F}_F}\,W_A{}^C{}_B+ E^M{}_A\,\de_M\,\tilde{F}_{FB}{}^C\} = \nn\\
& = -\delta_{\tilde{F}_F}\,T_A{}^C{}_B + T_A{}^C{}_B\{E^M{}_A\,\de_M\,\tilde{F}_{FB}{}^C\} = \nn\\
& = -\delta_{\tilde{F}_F}\,T_A{}^C{}_B - T_A{}^C{}_B\{E^M{}_A\,\de_M\,T_F{}^C{}_B\}.
\end{align}
Using this result in \eqref{Bianchi_iff3}, we finally arrive to the conclusion that 
\begin{align}
\text{The Bianchi constraint holds} \Leftrightarrow & \,(E^M{}_F\,\de_M + \delta_{T_F})\,T_A{}^C{}_B = T_A{}^C{}_B\{E^M{}_A\,\de_M\,T_F{}^C{}_B\}.
\end{align}
But this is precisely the Bianchi identity \eqref{BianchiT} for the torsion $T_A{}^C{}_B$ of the frame $E^M{}_A$. This shows that the flux defined in \eqref{DefFlux} satisfies the Bianchi constraint.

\newpage

\section{Detour: Three-dimensional gravity}\label{3}

The analogy between gravity and gauge theories is peculiar and intriguing in three dimensions. First of all, the equations of motion of three-dimensional  gravity in Cartan formalism set to zero not only the torsion, but also the Riemann curvature. This is due to the fact that a vanishing Ricci tensor implies a vanishing Riemann tensor in three dimensions.

The vanishing of both the curvatures suggests that the theory could be treated as a topological field theory. Indeed, if we consider the “dreibein” and the spin connection as the components of a unique gauge field $A$, the three-dimensional Einstein-Hilbert action turns out to be equivalent to the three-dimensional  Chern-Simons action. This was shown by Achucarro and Townsend \cite{Giddings:1983es, Achucarro:1986uwr, Achucarro:1989gm}, and by Witten \cite{Witten:1988hc, Witten:1989sx}.

The three-dimensional Chern-Simons theory is obtained by starting from the invariant $\text{Tr}\,F^2$ in four dimensions, where $F = \diff A + A^2$ is the curvature out of $A$. This invariant, as realised by Chern \cite{Chern:1944}, is a total derivative of a polynomial $\Omega^{(3)}$ in $A$ and $F$, not only in the abelian case -- this is quite trivial, because $F = \diff A$, and so $\diff \text{Tr}\,(A\,F) = \text{Tr}\,F^2$ -- but also in the non-abelian case. So, the four-dimensional integral of $\text{Tr}\,F^2$ over a manifold $\mathscr{M}$ can be written as the three-dimensional integral of $\Omega^{(3)}$ over the boundary of $\mathscr{M}$. 

Moreover, the analogy between three-dimensional Einstein gravity and Chern-Simons theory also holds at the level of symmetries. This is shown in Witten's papers. Witten considers only the field transformations. In the following, we will adopt the \textsc{brst} formalism and we will check that the argument holds also in the ghost transformations. 

Why can we not quantise gravity on the same footing as the other field theories? There are at least two main motivations:
\begin{itemize}
\item[1.] \emph{Tecnical problem}: Einstein gravity is not renormalisable. This should be due to the ignorance of the high energy complection of the theory. (Einstein theory should be viewed as an effective field theory.)
\item[2.] \emph{Conceptual problem}: We do not know what it means to quantise a field theory without a fixed background, that is, in a reparametrisation invariant way. But Einstein gravity is reparametrisation invariant, according to the general covariance principle. In other terms, in Einstein gravity there is no fixed background since the metric is dynamical, and the background is seen as the responsible of the gravitational interaction. 
\end{itemize}

Apart from the fact that it would be desirable to understand how to describe physics beyond the Planck scale, the first problem can be set aside, adopting the perspective of effective field theories. They describe physics up to a certain scale, neglecting the excited degrees of freedom at higher energies. Anyway, the three-dimensional gravity is free of the first problem, because it is renormalisable and perfectly finite, as the gauge correspondence suggests. This is due to the fact that the gauge theory is topological, that is, locally trivial from a dynamical point of view. In other words, in three dimensions there is no new physics to be described at higher energy, and so there is no problem of renormalisability.

The second problem relies intrinsically on gravitation, which is described in a reparametrisation invariant way. Einstein theory can be described as a classical field theory, treating the metric as dynamical field. The Hilbert-Einstein action has two derivatives of the metric, but it is highly nonlinear. The kinetic term of the metric is recognised in the linearised limit, which is obtained by expanding around a flat background:
\begin{equation}
g_{\mu\nu} = \eta_{\mu\nu} + \kappa\,h_{\mu\nu} + \mathcal{O}(\kappa^2).
\end{equation}
In this way, one can perform the usual approach to quantum field theory: One has a field $h_{\mu\nu}$ obeying some classical fields equations, and a \emph{fixed} background. If the field is free (in this case this is true only in first approximation), the classical equations of motion are wave equations. Their solution are waves, which can be expanded in Fourier modes. They are treated as ladder operators in a Fock space. In this way, particles are described as local excitation of the background. But in gravity there is no fixed background, since the background, describing the geometry of spacetime, is dynamical. Quantising gravity does \emph{not} simply mean performing a quantisation of some matter fields in a curved spacetime. It means performing a quantisation when the spacetime is not fixed. From the point of view of symmetries, we are not able to perform a quantisation in a reparametrisation invariant way. Choosing a fixed background means breaking the reparametrisation invariance of Einstein gravity. The known quantisation procedures are not compatible with the general covariance principle. 

If we do not want to give the principle of general covariance up, we can try to understand something of the quantisation starting with \emph{background-independent} field theories. Such theories do not presume a spacetime metric. Which is the minimal request to define spacetime geometry regretting a metric tensor? The answer could be the spacetime to be a \emph{topological space}. A field theory on a topological space is a field theory which does not depend on the metric. We are able to quantise it if it does not admit local degrees of freedom, because there is no particle to be described and the presence of a fixed background is not necessary. Such theory is a \emph{topological field theory}. This explains way topological field theory could be useful to get a more insight in quantising gravity. 

A topological space is a space in which the notion of \emph{closedness} is defined, although in general the notion of \emph{distance} is not defined (due to the lack of a metric). In other terms, we can say \emph{if} two points are close to each other in a topological space, but we cannot measure in general \emph{how much} they are close. A topological space is given by a set $\mathscr{X}$ and a collection $\mathscr{F}$ of subsets of $\mathscr{X}$, containing the empty set and $\mathscr{X}$ itself, stable under arbitrary union and under finite intersection of its elements. The collection of $\mathscr{F}$ is a \emph{topology} of $\mathscr{X}$. Its elements are called \emph{open sets} of $\mathscr{X}$. Two points are close if they are in a same open set.

On a side, three-dimensional Einstein gravity seems   \emph{trivial} because it has no propagating degree of freedom. On the other side, it is said to be \emph{inconsistent} at quantum level because it is supposed to be unrenormalisable by power counting.\footnote{An integral corresponding to a $n$-vertexed Feynman diagram is $\sim g^n\int\frac{\diff^d k}{k^m}$, where $g$ is the coupling constant of the theory, $n$ is a positive integer and $m$ is an integer to be fixed in order for the integral to be dimensionless. The condition is $0=n\,x+d-m$, where $x$ is the mass dimension of $g$. The integral is \textsc{uv} convergent (or at least logarithmically divergent) if $m\geqslant d$, that is, if $x\geqslant 0$. A theory satisfying this condition is called \emph{renormalisable by power counting}. Let us consider three example: 1) Scalar theory $\mathcal{L} = \frac{1}{2}\,\de_\mu\,\vphi\,\de^\mu\,\vphi  + \frac{\lambda}{4!}\,\vphi^4$. The kinetic term fixes the mass dimension of $\vphi$ to be $x_\vphi = (d-2)/2$. Then, the mass dimension of $\lambda$ is $x_\lambda = 4-d$. So, the theory is renormalisable if $d=2,3,4$. 2) Fermionic theory $\mathcal{L} = \bar{\psi}\,(\s{\de}+m)\,\psi + \lambda\,(\bar{\psi}\,\psi)^2$. The kinetic term fixes $x_\psi = (d-1)/2$. Then, $x_\lambda = 2-d$. The theory is renormalisable if $d=2$. 3) General Relativity $\mathcal{L} = \frac{1}{2\kappa^2}\,e\,R$. As we know, $x_\kappa = \frac{2-d}{2}$, so it is renormalisable only in $d=2$.} But how can a trivial theory be inconsistent?

One can carry on a covariant quantisation by means of the \textsc{brst} formalism. The space of states is defined as the cohomology of the nilpotent \textsc{brst} charge. In this way, the physical states are selected, modulo the gauge redundancy, and a Hilbert space with a positive definite inner product is defined, with a probabilistic interpretation of the amplitudes. This quantisation is well-known in Chern-Simons theory. The cohomology is trivial, so that the Hilbert space of states does not admit local particle excitations. Moreover, the theory is perfectly renormalisable, as can be algebraically proved .

Chern-Simons theory is not only renormalisable, but also finite at all order in perturbation theory. Indeed, one can perform an exact computation of the partition function of the theory. This was carried on by Witten, who provided also a physical interpretation of the result, from the gravity point of view \cite{Witten:1989sx}.

In perturbation theory, one chooses a classical stable solution, and expands around it. The solution $e=\omega=0$ does not spontaneously break the symmetries, but it has no gravitational interpretation, since the metric is set to zero. When we treat $\omega=\omega(e) \sim e^{-1}\,\de\,e$, as we should do if we want to deal with Einstein gravity, the inverse dreibein is included, so that the solution $e=0$ is not allowed. This is the inconsistency in quantising the theory. And this is why the theory is not renormalisable as a gravity theory. 

In the following we will analyse in details the main characteristic of three-dimensional gravity and of three-dimensional Chern-Simons theory, according to the classical point of view. Then, we will explain the correspondence between the two theories. Finally, we will quantise the Chern-Simons theory, according to the \textsc{brst} formalism. In this Sction we will denote with $g_{\mu\nu}$, $\mu, \nu = 0,1,2$, the three-dimensional metric tensor. $g$ is its determinant. $(-)^s$ is its signature, $s$ being the number of negative eigenvalues. We will adopt the cosmological convention $(-,+,+)$ for the signature of the metric, so that $(-)^s = -1$. Notice that the alternative convention $(+,-,-)$ gives an opposite sign. (Instead in four dimensions the two conventions give the same sign.) For the following, the contraction formulas of the Levi-Civita symbol in three dimensions will be very useful (see Appendix \ref{Conventions} for the general case):
\begin{equations}
\vepsilon^{\mu\nu\vrho}\,\vepsilon_{\alpha\beta\gamma} &= (-)^s\,\delta^{[\mu}_{\alpha}\,\delta^{\nu}_{\beta}\,\delta^{\vrho]}_{\gamma},\\
\vepsilon^{\mu\nu\vrho}\,\vepsilon_{\mu\beta\gamma} &= (-)^s\,\delta^{[\nu}_\beta\,\delta^{\vrho]}_{\gamma},\\
\vepsilon^{\mu\nu\vrho}\,\vepsilon_{\mu\nu\gamma} &= (-)^s\,2\,\delta^\vrho_\gamma,\\
\vepsilon^{\mu\nu\vrho}\,\vepsilon_{\mu\nu\vrho} &= (-)^s\,6.
\end{equations}
Using $\vepsilon^{\mu\nu\vrho}$ we can pass from an antisymmetric tensor $V_{\nu\vrho}$ to the dual vector $V^\mu$. Group-theoretically, this is because the fundamental representation and the two-indexed antisymmetric have the same dimensions. If one defines
\begin{equation}
V^\mu = \tfrac{1}{2}\,\vepsilon^{\mu\nu\vrho}\,V_{\nu\vrho},
\end{equation}
the inverse is\footnote{\emph{Proof.} $\vepsilon_{\mu\nu\vrho}\,V^\vrho = \tfrac{1}{2}\,\vepsilon_{\mu\nu\vrho}\,\vepsilon^{\vrho\alpha\beta}\,V_{\alpha\beta} = (-)^s\,\tfrac{1}{2}\,\delta^\alpha_\mu\,\delta^\beta_\nu\,2\,V^{\alpha\beta} = (-)^s\,V_{\mu\nu}.$}
\begin{equation}
V_{\nu\vrho} = (-)^s\,\vepsilon_{\nu\vrho\mu}\,V^\mu
\end{equation}

\subsection{Three-dimensional Einstein gravity}

The Hilbert-Einstein action in $d$ dimensions is
\begin{equation}
S_{\textsc{eh}} = \frac{1}{2\,\kappa^2}\int\diff^d x\,\sqrt{|g|}\,R,
\end{equation}
where $\kappa^2 = 8\,\pi\,G$, $G$ being the Newton constant, if $c=1$. The dimension of an action is $M\,$ and $[R]=L^{-2}$, because it contains two derivatives. Therefore the dimension of $\kappa^2$ is constrained by $M\,L=[\kappa^2]^{-1}\,L^{-2}\,L^d$, which is solved by\footnote{The same result can be obtained by looking at the Einstein equations $G_{\mu\nu}=\kappa^2\,T_{\mu\nu}$. The time component of the stress-energy tensor $T_{tt}$ is equal to the mass density in space, so $[T_{tt}] = M\,L^{d-1}$. On the other hand, the Einstein tensor has dimension $L^{-2}$, such as the Riemann and the Ricci tensor. Equating, the previous result is recovered.}
\begin{equation}
[\kappa^2]=M^{-1}\,L^{d-3}.
\end{equation}
Remarkably, in three dimensions it is the inverse of a scale mass.

The Hilbert-Einstein action can be written in terms of differential forms. Consider the three-dimensional case. Define the dreibein, the spin connection and its curvature:
\begin{equation}
e^a = e_\mu{}^a\,\diff x^\mu, \quad
\omega_\mu{}^{ab}\,\diff x^\mu, \quad
R = \diff\omega + \omega^2.
\end{equation}
Then we can compute
\begin{align}
\int R^{ab}\,e^c\,\vepsilon_{abc} &= \frac{1}{2}\int R_{\mu\nu}{}^{ab}\,\diff x^\mu\diff x^\nu\,e_\vrho{}^c\,\diff x^\vrho\,\vepsilon_{abc} = \nn\\
&= \frac{1}{2}\int\,\diff^3x\,\vepsilon^{\mu\nu\vrho}\,R_{\mu\nu}{}^{\alpha\beta}\,e_\alpha{}^a\,e_\beta{}^b\,e_\vrho{}^c\,\vepsilon_{abc} = \nn\\
&= \frac{1}{2}\,(-)^s\int \diff^3x\,e\,\vepsilon^{\mu\nu\vrho}\,R_{\mu\nu}{}^{\alpha\beta}\,\vepsilon_{\alpha\beta\vrho} = \nn\\
&= \frac{1}{2}\,\int\diff^3x\,e\,\delta_\alpha^\mu\,\delta^\nu_\beta\,2\,R_{\mu\nu}{}^{\alpha\beta} = \nn\\
&= \int\diff^3x\,e\,R_{\mu\nu}{}^{\mu\nu} = \int\diff^3x\,e\,R,
\end{align}
We denoted by $e$ the \emph{absolute value} of the determinant of the dreibein.

Remember that, in $d$ dimensions, the metric tensor has $\frac{d(d+1)}{2}$ components, being a two-indexed symmetric tensor. Subtracting the $d$ diffeomorphisms, we get $\frac{d(d-1)}{2}$ off-shell degrees of freedom. The number of on-shell degrees of freedom is obtained subtracting $d$ transversality conditions $\de^\mu\,h_{\mu\nu}=0$, where $h_{\mu\nu}$ is defined by $g_{\mu\nu} = \eta_{\mu\nu} + \kappa\,h_{\mu\nu}$.  These conditions select the transversal polarisations. The result is $\frac{d(d-3)}{2}$, as in Section \ref{Counting}. Therefore, \emph{in $d=3$ dimensions there is no propagating degrees of freedom for the metric tensor}. 

Another way to reach to the same conclusion is to analyse the number of components of the Riemann tensor. By definition, it is  antisymmetric the pairs of indices: \cite{Giddings:1983es}
\begin{equation} \label{SymR1}
R_{\mu\nu,\alpha\beta} = -R_{\nu\mu,\alpha\beta} = -R_{\mu\nu,\beta\alpha}
\end{equation}
The algebraic Bianchi identity
\begin{equation}
R_{\mu[\nu,\alpha\beta]} = 0
\end{equation}
is equivalent to\footnote{This is a consequence of the following identities (for example in four dimensions), which can be proved by explicit evaluation, only using the symmetries in \eqref{SymR1}:
\begin{equations}
R_{\mu\nu,\alpha\beta}-R_{\alpha\beta,\mu\nu} &= \tfrac{1}{4}\,(R_{\mu[\nu,\alpha\beta]} - R_{\nu[\alpha,\beta\mu]} - R_{\alpha[\beta,\mu\nu]} + R_{\beta[\mu,\nu\alpha]}),\\
R_{\mu[\nu,\alpha\beta]} &= \tfrac{1}{4}\,R_{[\mu\nu,\alpha\beta]} - \tfrac{1}{2}\,\vepsilon_{\gamma\nu\alpha\beta}\,\vepsilon^{\gamma\vrho\sigma\tau}\,(R_{\mu\vrho,\sigma\tau}-R_{\sigma\tau,\mu\vrho}).
\end{equations}}
\begin{equation}
R_{\mu\nu,\alpha\beta} = R_{\alpha\beta,\mu\nu}, \quad
R_{[\mu\nu,\alpha\beta]} = 0.
\end{equation}
Therefore, the number of independent components of the Riemann tensor is given by
\begin{equation}
\frac{\frac{d(d-1)}{2}(\frac{d(d-1)}{2}+1)}{2} - \frac{d(d-1)(d-2)(d-3)}{4!} = \frac{d^2(d^2-1)}{12}.
\end{equation}
Thus, in $d=3$, there are six independent components. This is the same number of components of a two-indexed symmetric tensor, such as the Ricci tensor. This suggests that the Riemann tensor and the Ricci one are equivalent. 

More precisely, consider that the irreps of the Lorentz group in three dimensions are classified by the spin $j \in \frac{n}{2}\,\mathbb{N}$, having dimension $2\,j+1$. The two-indexed antisymmetric representation has the same dimension of the $j=1$ irrep, so they are the same. This is possible because of the Levi-Civita invariant $\vepsilon^{abc}$, linking $V^a$ and $V_{bc}$, as already noticed. So, the irreps transported by the three-dimensional Riemann tensor are
\begin{equation}
\bb{1}\otimes\bb{1} = \bb{1}_a\oplus(\bb{0}\oplus\bb{2})_s
\end{equation}
where $a$ and $s$ denote the [anti]symmetric parts. The symmetry in swapping the pairs of indices set to zero the antisymmetric part.\footnote{This shows that in three dimensions, the symmetry in the pair of indices is equivalent to the algebraic Bianchi identity.} The remaining part is the representation of a two-indexed symmetric tensor, which is the Ricci tensor. In particular, the representation $\bb{0}$ corresponds to the Ricci scalar.

Let us find the precise relation between the Riemann and the Ricci tensor. We start with the following ansatz: \cite{Staruszkiewicz:1963zza}
\begin{equation}
R_{\mu\nu,\alpha\beta} =\alpha\,g_{[\mu[\alpha}\,R_{\nu]\beta]} + \beta\,g_{\alpha[\mu}\,g_{\nu]\beta}\,R,
\end{equation}
where $\alpha$ and $\beta$ are constant to be determined. Saturating the first and the third index,
\begin{equation}
R_{\mu\nu}{}^{\mu\vrho} = \alpha\,R + (\alpha+2\,\beta)\,\delta^\vrho_\nu\,R.
\end{equation}
So, we have to set $\beta = -\alpha/2$. Saturating the second and the last index,
\begin{equation}
R = R_{\mu\nu}{}^{\mu\nu} = \alpha\,R,
\end{equation}
so we have to set $\alpha=1$. Thus, $\beta = -1/2$. Finally,
\begin{equation}
R_{\mu\nu,\alpha\beta} = g_{[\mu[\alpha}\,R_{\nu]\beta]} - \tfrac{1}{2}\,g_{\alpha[\mu}\,g_{\nu]\beta}\,R.
\end{equation}
This identity can be written in an alternative way \cite{Jackiw:1988sd}. Consider
\begin{align}
(-)^s\,\vepsilon_{\mu\nu\vrho}\,\vepsilon^{\alpha\beta\sigma}\,G^\vrho{}_\sigma &= \delta_\mu^{[\alpha}\,\delta_\nu^{\beta}\,\delta_\vrho^{\sigma]}\,G^\vrho{}_\sigma = -\delta_{[\mu}^{[\alpha}\,G^{\beta]}{}_{\nu]} + \delta_{\mu}^{[\alpha}\,\delta_\nu^{\beta]}\,G = \nn\\
& = -\delta_{[\mu}^{[\alpha}\,R^{\beta]}{}_{\nu]} +\tfrac{1}{2}\,\delta_{[\mu}^{[\alpha}\,\delta^{\beta]}{}_{\nu]}\,R -\tfrac{1}{2}\, \delta_{\mu}^{[\alpha}\,\delta_\nu^{\beta]}\,R= \nn\\
& = -(\delta_{[\mu}^{[\alpha}\,R_{\nu]}{}^{\beta]}-\tfrac{1}{2}\,\delta_\mu^{[\alpha}\,\delta_\nu{}^{\beta]}\,R).
\end{align}
Therefore,
\begin{equation} \label{RG}
R_{\mu\nu,\alpha\beta} = (-)^{s+1}\,\vepsilon_{\mu\nu\vrho}\,\vepsilon_{\alpha\beta\sigma}\,G^{\vrho\sigma}.
\end{equation}
This formula clearly shows that the relation between the Riemann tensor and the Ricci tensor (or the Einstein tensor) is intrinsically three-dimensional, since $\vepsilon^{\alpha\beta\gamma}$ is involved.

The fact that the Riemann and the Ricci tensor (or the Einstein tensor) are equivalent means that the Weyl tensor is intrinsically vanishing in three dimensions. Indeed, the Weyl tensor controls gravitational waves, and in three dimensions there are no gravitational waves because there are no propagating degrees of freedom. Moreover, by means of the Einstein equation, the Ricci tensor (or the Einstein one) are completely determined by the stress-energy tensor. So, in three dimensions, the whole Riemann tensor is fixed by the mass sources, and the Einstein equations in the vacuum fix the Riemann tensor to zero. So, each solution is \emph{locally} flat (isometrically equivalent to the Minkowski flat spacetime). Nevertheless, as we will see, global defects are possible.

The fact that there are no gravitational waves or gravitons means that there is no gravitational interaction. A test particle does not detect any acceleration. From the classical point of view, there is no attraction force between massive particles. A dust of massive particles cannot gravitate, forming clusters or stars: the gravitational collapse is not possible. So, there exist static configurations of particle dust. Another consequence is that the Newton theory in three dimensions cannot be obtained as a limit of the Einstein theory in three dimensions.

As a nontrivial example of solution in three-dimensional gravity, let us consider a point-like massive particle, with mass $m$, in the origin of space \cite{Deser:1983tn, Gott:1982qg, Ouvry:1988mm}. The Einstein equations are $R_{\mu\nu,\alpha\beta} = 0$ everywhere except in a timelike worldline, corresponding to the spacetime trajectory of the source. We look for a Schwarzschild-like solution, that is, we assume: the spherical symmetry; the solution to be static (no mixed terms involving the time coordinate and no dependence on time); the solution to be asymptotically flat, so that the signature is $(-,+,+)$ in our convention. Using polar coordinates,
\begin{equation}
\diff s^2 = - e^{2\alpha(r)}\,\diff t^2 + e^{2\beta(r)}\,\diff r^2 + r^2\,\diff\vphi^2,
\end{equation}
where $\alpha$ and $\beta$ are arbitrary functions of $f$. We can set $\alpha = 0$ without loss of generality, because the dependence on $\alpha$ is eliminated by the reparametrisation $t \mapsto e^{\alpha}\,t$:\footnote{Indeed, if $t'=e^\alpha\,t$, using the transformation rule of the metric tensor, $g_{t't'} = (\frac{\de t}{\de t'})^2\,g_{tt} = -e^{-2\alpha}\,e^{2\alpha} = -1$. Trivially, $g_{t'\alpha}=0$. The other components do not change.}
\begin{equation}
\diff s^2 = - \diff t^2 + e^{2\beta(r)}\,\diff r^2 + r^2\,\diff\vphi^2.
\end{equation}
We can compute the Christoffel symbols, finding that the unique non-vanishing ones are
\begin{equation}
\Gamma^r_{rr} = \beta', \quad
\Gamma^\vphi_{\vphi r} = \tfrac{1}{r}, \quad
\Gamma^r_{\vphi\vphi} = -r\,e^{-2\beta},
\end{equation}
where $\beta' = \de_r\,\beta$. Then, we can find the components of the Riemann tensor, finding that the unique non-vanishing ones are
\begin{equation}
R^r{}_{\vphi r \vphi} = r\,\beta'\,e^{-2\beta}, \quad
R^\vphi{}_{r\vphi r} = \beta'/r.
\end{equation}
They correspond to
\begin{equation}
R_{r\vphi r\vphi} = r\,\beta'.
\end{equation}
Therefore, everywhere except in the worldline of the source, the Einstein equations imply $\beta$ to be constant. Before determining it in terms of the mass of the source, consider the following reparametrisation
\begin{equation}
R = e^\beta\,r, \quad \Phi = e^{-\beta}\,\vphi.
\end{equation}
The metric becomes
\begin{equation}
\diff s^2 = - \diff t^2 + \diff R^2 + R^2\,\diff\Phi^2,
\end{equation}
which is the Minkowski metric in polar coordinates. But, observe that, if $0 \leqslant \phi < 2\,\pi$, then $0\leqslant \Phi < 2\,\pi\,e^{-\beta}$. So, if $\beta > 0$, then the solution at each $t$ is the flat two-dimensional plane, with the source in the origin and without a wedge of angular width equal to $\Delta = 2\,\pi\,(1-e^{-\beta})$. This manifold is isometrically equivalent to a \emph{single-pitched cone} whose apex is the origin of the plane. This singularity is called \emph{conical singularity}. A manifold admitting conical singularities is called \emph{conifold}.

We expect that $\beta$ depends on the mass of the source, which appears in the stress-energy tensor:\footnote{$T_{tt}$ has the dimension of a mass density, since $[\delta^{(2)}(\vec{x})]=L^{-2}$.}
\begin{equation}
T_{tt} = m\,\delta^{(2)}(\vec{x}), \quad
T_{\mu\nu} = 0 \;\;\text{otherwise}.
\end{equation}
Nevertheless, Einstein equations do not determined a well-defined problem for Green functions, because the equations are not linear. In order to regularise the problem, we can round off the tip of the cone with a spherical cap. The metric of a spherical space of radius $r_0$ is
\begin{equation}
\diff s^2 = - \diff t^2 + r_0^2\,(\diff\vtheta^2 + \sin^2\,\vtheta\,\diff\vphi^2).
\end{equation}
By explicit evaluation, one sees that the unique non-vanishing components of the Einstein tensor out of this metric is
\begin{equation}
G_{tt} = \frac{1}{r_0^2}
\end{equation}
So, the stress-energy tensor must be
\begin{equation}
T_{tt} = \frac{1}{\kappa\,r_0^2},
\end{equation}
which can be viewed as the mass density of a mass $4\,\pi\,\kappa^{-1}$ uniformly distributed on the sphere surface, whose area is $4\,\pi\,r_0^2$.

Now, suppose the spherical cap to have an angular width with $\vtheta_M$ and radius $r_0$. Then, the boundary of the truncated cone has radius $r_0\,\sin{\vtheta_M}$. The cone is described by the coordinates $r, \vphi$. We can describe the spherical cap by a couple of angles $\vtheta, \vphi$. Let us make the change of coordinates from $r$ to $\vtheta$ at the boundary, by using $r = r_0\,\sin{\vtheta}$. One has $r_0^2 = g_{\vtheta\vtheta}={\left(\frac{\de r}{\de \vtheta}\right)^2 g_{rr}} = {r_0\,\cos^2{\vtheta}\,e^{2\beta}}$. So, if $0\leqslant \vtheta_M \leqslant \pi/2$,
\begin{equation}
\cos{\vtheta_M} = e^{-\beta}.
\end{equation}
The area of the spherical cap is
\begin{equation}
r_0^2\,2\,\pi\int_0^{\vtheta_M}\diff \vtheta \sin{\vtheta} = r_0^2\,2\,\pi\,(1-\cos{\vtheta_M}).
\end{equation}
The mass $m$ of the particle is now uniformly distributed on the spherical cap (remember that the mass density is $1/\kappa\,r_0^2$):
\begin{equation}
m = \frac{2\,\pi}{\kappa}\,(1-e^{-\beta}).
\end{equation}
One can recognise the conical defect $\Delta$ in the right-hand-side, so that the mass of the particle and the conical defect are proportional:
\begin{equation}
\Delta = \kappa\,m.
\end{equation}
Solving for $\beta$ and using $\kappa = 8\,\pi\,G$, 
\begin{equation}
e^{\beta} = \frac{1}{1-4\,G\,m},
\end{equation}
which imposes a constraint on the possible masses, ensuring the right-hand-side of the previous expression to be positive:
\begin{equation}
m < 1/4\,G.
\end{equation}
Finally, the solution is 
\begin{equation}
\diff s^2 = - \diff t^2 + \frac{1}{(1-4\,G\,m)^2}\,\diff r^2 + r^2\,\diff\vphi^2.
\end{equation}

In these notes we always assume vanishing cosmological constant. Nevertheless, we mention that the correspondence between Chern-Simons and three-dimensional gravity it was long studied in the Anti de Sitter case. Indeed, if a negative cosmological constant is included, there is an exact 2+1 dimensional solution describing a black hole. This solution was discovered by Bañados, Teitelboim and Zanelli (\textsc{btz} \emph{black hole}) in 1986 \cite{Banados:1992wn}, and it came as a surprise, because three-dimensional gravity was thought as trivial in its dynamics up to that moment. The solution is
\begin{equation}
\diff s^2 = - f(r)^2\,\diff t^2 + f(r)^{-2}\,\diff r^2 + r^2\,\diff \vphi^2,
\end{equation}
where $f(r) = (r^2/\ell^2 - M)^{1/2}$, $\ell$ being the AdS radius and $M$ the mass of the black hole. This metric can be verified to satisfy the AdS three-dimensional Einstein equations $R_{\mu\nu}+\frac{2}{\ell^2}\,g_{\mu\nu}=0$.
This is a Schwarzschild-like solution. It can be generalised to a Kerr-like one.

\subsection{BRST formulation of Chern-Simons theory}

Consider the four-dimensional topological Yang-Mills theory, with action
\begin{equation}
S_{\text{inv}} = \frac{1}{2}\int_{\mathscr{M}} \text{Tr}\,F^2,
\end{equation}
on a manifold $\mathscr{M}$. It differs from the usual Yang-Mills theory because the Hodge dual, and so a metric, is not involved. Given an arbitrary infinitesimal variation  $\delta$, one can compute
\begin{equations}
& \delta\,A^2 = (-)^{|\delta|}\,[A,\delta\,A],\\
& [A,\delta\,A] = A\,\delta\,A + (-)^{|\delta|}\,\delta\,A\,A,
\end{equations}
which give
\begin{equation}
\delta\,F = (-)^{|\delta|}\,\text{D}\,(\delta\,A).
\end{equation}
Therefore, by using the Bianchi identity,
\begin{equation}
\delta\,\text{Tr}\,F^2 = 2\,(-)^\delta\,\diff\,\text{Tr}\,(F\,\delta\,A),
\end{equation}
that is, the variation is $\diff$-exact. This implies that:
\begin{itemize}
\item[1.] The classical equations of motion are trivial, if there is no boundary:
\begin{equation}
\delta\,S_{\text{inv}} = \int_{\mathscr{M}} \diff\,\text{Tr}\,(F\,\delta\,A) = 0.
\end{equation}
\item[2.] The action is \textsc{brst} invariant:
\begin{equation}
s\int\text{Tr}\,F^2 = -2\int\diff\,\text{Tr}\,(F\,s\,A) = 0,
\end{equation}
whatsoever be $s\,A$. In general, we can write $s\,A = \lambda$.
\item[3.] If there is a boundary, the three-dimensional \emph{Chern-Simons theory} is obtained:
\begin{equation}
S_{\text{cs}} = \frac{1}{2}\int_{\de\,\mathscr{M}}\text{Tr}\,(A\,F -\tfrac{1}{3}\,A^3) =
\frac{1}{2}\int_{\de\,\mathscr{M}}\text{Tr}\,(A\,\diff A +\tfrac{2}{3}\,A^3).
\end{equation}
\end{itemize} 

Let us compute the variation of the Chern-Simons polynomial
\begin{equations}
\delta\,(A\,\diff A) + \tfrac{2}{3}\,\delta A^3 &=  \delta A\,\diff A + A\,\diff\,\delta A + \tfrac{2}{3}\,\delta A\,A^2 + \tfrac{2}{3}\,(-)^\delta A\,\delta A^2 =\nn\\
& = \delta A\,\diff A - \diff\,(A\,\delta A) + \diff A\,\delta A + \tfrac{2}{3}\,\delta A\,A^2 + \tfrac{2}{3}\,A\,[A,\delta A] = \nn\\
& = 2\,\delta A\,\diff A - \diff\,(A\,\delta A) + \tfrac{2}{3}\,\delta A\,A^2 + \tfrac{2}{3}\,A\,(A\,\delta A + (-)^\delta\,\delta A\,A) = \nn\\
& = 2\,\delta A\,\diff A - \diff\,(A\,\delta A) + 2\,\delta A\,A^2 = \nn\\
& = 2\,\delta A\,F - \diff\,(A\,\delta A). 
\end{equations}
Therefore,
\begin{equation}
\delta\,S_{\textsc{cs}} = \int_{\de\,\mathscr{M}}\text{Tr}\,(\delta A\,F).
\end{equation}
The equations of motion set to zero the curvature
\begin{equation}
F = 0
\end{equation}
and, if $s\,A = -\text{D}\,c$, as in Yang-Mills theory, then
\begin{equation}
s\,S_{\textsc{cs}} = -\int_{\de\,\mathscr{M}}\text{Tr}\,(\text{D}\,c\,F) = \int_{\de\,\mathscr{M}}\diff\,\text{Tr}\,(c\,F) = 0,
\end{equation}
where we used the Bianchi identity. This means that the action is gauge invariant. Actually, this proves the theory to be invariant under \emph{infinitesimal} gauge transformations. Instead, if we consider a \emph{finite} transformation
\begin{equation}
A' = g^{-1}\,(\diff + A)\,g,
\end{equation}
then one can show that
\begin{equation}
\Omega^{(3)}(A') = \Omega^{(3)}(A) + \Omega^{(3)}(g^{-1}\,\diff g) + \diff\,\text{Tr}\,(\diff g\,g^{-1}\,A).
\label{CS_inv}
\end{equation}
Nevertheless, the quantum theory is invariant. Indeed,  the action appears in the exponential in the path integral. One can see that the integral $\int\Omega^{(2)}(g^{-1}\,\diff g)$ is quantised. Up to a suitable constant, it is equal to $2\,\pi\,N$, with $N \in \mathbb{Z}$. Thus, it does not contribute to the path integral.

\subsection{Chern-Simons polynomial}

Consider 
\begin{equation}
P^{(4)} := \text{Tr}\,F^2.
\end{equation}
It is $\diff$-closed, because of the Bianchi identity $\text{D}\,F=0$. By means of the algebraic Poincaré lemma, the $\diff$-cohomology is locally trivial. So $P^{(4)}$ must be locally $\diff$-exact:
\begin{equation}
P^{(4)} = \diff\,\Omega^{(3)},
\end{equation}
where $\Omega^{(3)}$ is a three-form in $A$ and $F$. We want to determine $\Omega^{(3)}$. 

We can start from an ansatz for $\Omega^{(3)}$:
\begin{equation}
\Omega^{(3)} = \text{Tr}\,(\alpha\,A\,F + \beta\,A^3),
\end{equation}
whose form is dictated by the fact that it must be a three-form an we can only use $A$, which is a one-form, and $F$, which is a two-form. Then, we can compute 
\begin{equation}
\diff \Omega^{(3)} = \text{Tr}\,(\alpha\,F^2 + (\alpha + 3\,\beta)\,A^2\,F), 
\end{equation}
where we used the cyclicity of the trace and the fact that $\text{Tr}\,A^4 = 0$, because $A$ is odd. Finally, the coefficients are fixed in such a way that $\text{Tr}\,F^2$ is recovered:
\begin{equation}
\alpha = 1, \quad \beta = -\tfrac{1}{3}.
\end{equation}
So, the result is
\begin{equation}
\Omega^{(3)} = \text{Tr}\,(A\,F -\tfrac{1}{3}\,A^3),
\end{equation}
which is the \emph{three-dimensional Chern-Simons polynomial}.

The previous result can be produced in a constructive way. The definition of $F$ and the Bianchi identity as the action of $\diff$ on $A$ and $F$:
\begin{equations}
\diff \,A &= F - A^2, \\
\diff \,F &= -[A,F].
\end{equations}
Define 
\begin{equations}
& \diff_0 \,A = F, \quad \diff_1 \,A = - A^2,\\
& \diff_0 \,F = 0, \quad \diff_1 \,F = -[A,F],
\end{equations}
in such a way that
\begin{equation}
\diff = \diff_0 + \diff_1.
\end{equation}
Notice that $\diff_0$ is trivially nilpotent: 
\begin{equation}
\diff_0^2 = 0.
\end{equation}
One can compute
\begin{align}
\diff_1 &= -A^2\,\tfrac{\delta}{\delta A} - [A,F]\,\tfrac{\delta}{\delta F} = \nn\\
&= -\tfrac{1}{2}[A,A]\,\tfrac{\delta}{\delta A} - [A,F]\,\tfrac{\delta}{\delta F} = \nn\\
& =-[A,A]\,\tfrac{\delta}{\delta A} - [A,F]\,\tfrac{\delta}{\delta F} + \tfrac{1}{2}\,[A,A]\,\tfrac{\delta}{\delta A} = \nn\\
&= [A,\cdot] + A^2\,\tfrac{\delta}{\delta A}.
\end{align}
Therefore $\diff_1$ acts as $A^2\,\tfrac{\delta}{\delta A}$ on invariant polynomials in the adjoint representation. 

Define
\begin{equation}
\iota_0 := A\,\tfrac{\delta}{\delta F},
\end{equation}
Notice that $\iota_0$ is nilpotent 
\begin{equation}
\iota_0^2 = 0.
\end{equation}
Furthermore,
\begin{equation}
\lbrace \diff_0,\iota_0 \rbrace = A\,\tfrac{\delta}{\delta A} + F\,\tfrac{\delta}{\delta F} =: \ell_0.
\end{equation}
where $\ell_0$ counts the number of fields in a monomial. Thus, $P^{(4)} =: P^{(4)}_2$ is an eigenfunction of $\ell_0$ with eigenvalue $2$:
\begin{equation}
\ell_0\,P^{(4)}_2 = 2\,P^{(4)}_2.
\end{equation}

The equation $\diff P^{(4)}_2 = 0$ corresponds to
\begin{equations} 
\diff_0\,P^{(4)}_2 &= 0,\\
\diff_1\,P^{(4)}_2 &= 0.
\end{equations}
Similarly, we can split $\diff \Omega^{(3)} = P^{(4)}_2$ as
\begin{equation}
(\diff_0 + \diff_1)\,(\Omega^{(3)}_2 + \Omega^{(3)}_3) = P^{(4)},
\end{equation}
where $\Omega^{(3)}_n$ is the contribution with $n$ fields in $\Omega^{(3)}$. It corresponds to a descent
\begin{equations}
\diff_0\,\Omega^{(3)}_2 &= P^{(4)}_2,\\
\diff_0\,\Omega^{(3)}_3 &= -\diff_1\,\Omega^{(3)}_2,\\
\diff_1\,\Omega^{(3)}_3 &= 0.
\end{equations}
Consider that
\begin{equation}
P^{(4)}_2 = \tfrac{1}{2}\,\ell_0\,P^{(4)}_2 = \tfrac{1}{2}\,(\diff_0\,\iota_0 + \iota_0\,\diff_0)\,P^{(4)} = \tfrac{1}{2}\,\diff_0\,(\iota_0\,P^{(4)}_2).
\end{equation}
Therefore, up to $\diff_0$-exact terms,
\begin{equation} \label{Omega2}
\Omega_2 = \tfrac{1}{2}\,\iota_0\,P^{(4)} = \text{Tr}\,(A\,F).
\end{equation}
Now define 
\begin{equation}
P_3 := -\diff_1\,\Omega_2.
\end{equation}
By means of the second equation of the descent, it is $\diff_0$-closed
\begin{equation}
\diff_0\,P_3 = 0.
\end{equation}
Thus,
\begin{equation}
P_3 = \tfrac{1}{3}\,\ell_0\,P_3 = \tfrac{1}{3}\,\diff_0\,(\iota_0\,P_3),
\end{equation}
which implies that, up to $\diff_0$-exact terms,
\begin{equation}
\Omega_3 = \tfrac{1}{3}\,\iota_0\,P_3.
\end{equation}
But
\begin{equation}
\iota_0\,P_3 = -\iota_0\,\diff_1\,\Omega_2 = \diff_1\,\iota_0\,\Omega_2 - \ell_1\,\Omega_2 = - \ell_1\,\Omega_2,
\end{equation}
where we introduced the anticommutator
\begin{equation}
\ell_1 := \lbrace \diff_1, \iota_0 \rbrace,
\end{equation}
and in the last step we used $\iota_0^2=0$ and the form of $\Omega_2$ in \eqref{Omega2}. Notice that
\begin{equations}
\ell_1\,A &= \iota_0\,\diff_1\,A = \iota_0\,A^2 = 0, \\
\ell_1\,F &= \delta_1\,\iota_0\,F = \diff_1\,A = A^2.
\end{equations}
Therefore
\begin{equation}
\ell_1 = A^2\,\tfrac{\delta}{\delta F}.
\end{equation}
Finally
\begin{equation}
\Omega_3 = -\tfrac{1}{3}\,A^2\,\tfrac{\delta}{\delta F}\,\Omega_2 = -\tfrac{1}{3}\text{Tr}\,A^3,
\end{equation}
and so
\begin{equation}
\Omega^{(3)} = \text{Tr}\,(A\,F -\tfrac{1}{3}\,A^3),
\end{equation}
which is the expected result.

\subsection{Gauge transformation of Chern-Simons polynomial}

The finite gauge transformation is 
\begin{equation}
A' = g^{-1}\,(\diff + A)\,g = g^{-1}\,A\,g + g^{-1}\,\diff g =: g^{-1}\,A\,g + G.
\end{equation}
By explicit computation, one can prove the following identities:
\begin{equations}
& \diff g^{-1} = - g^{-1}\,\diff g\,g^{-1},\label{p1}\\
& g\,\diff\,g^{-1} = -\diff g\,g^{-1},\label{p2}\\
& g\,G\,g^{-1} = \diff g\,g^{-1},\label{p3}\\
& g\,G^2\,g^{-1} = (\diff g\,g^{-1})^2,\label{p4}\\
& \diff\,(\diff g\,g^{-1}\,A) = -(\diff g\,g^{-1})^2\,A + (\diff g\,g^{-1})\,\diff A,\label{p5}\\
& \diff G + G^2 = 0\label{p6}.
\end{equations}
Using these identities, one can compute
\begin{align}
\text{Tr}\,(A'\,\diff A') &= \text{Tr}\,[(g^{-1}\,A\,g + G)\,\diff\,(g^{-1}\,A\,g + G)] = \nn\\
& = \text{Tr}\,[(g^{-1}\,A\,g)\,(\diff g^{-1}\,A\,g + g^{-1}\,\diff A\,g - g^{-1}\,A\,\diff g) + (g^{-1}\,A\,g)\,\diff G)] \,+\nn\\
& + \text{Tr}\,[G\,(\diff g^{-1}\,A\,g + g^{-1}\,\diff A\,g - g^{-1}\,A\,\diff g) + G\,\diff G] = \nn\\
& = \text{Tr}\,[-A\,\diff g\,g^{-1}\,A + A\,\diff A - A^2\,\diff\,g^{-1} - A\,(\diff g\,g^{-1})^2] \,+\nn\\
& - \text{Tr}\,[(\diff g\,g^{-1})^2\,A + (\diff g\,g^{-1})\,\diff A - (\diff g\,g^{-1})^2\,A + G\,\diff G] = \nn\\
& = \text{Tr}\,(A\,\diff A) + \text{Tr}\,(G\,\diff G) \,+\nn\\
& + \text{Tr}\,[-2\,(\diff g\,g^{-1})\,A^2 + (\diff g\,g^{-1})\,\diff A - 3\,(\diff g\,g^{-1})^2\,A],
\end{align}
\begin{align}
\text{Tr}\,A'^3 &= \text{Tr}\,[(g^{-1}\,A\,g + G)\,(g^{-1}\,A^2\,g + g^{-1}\,A\,g\,G + G\,g^{-1}\,A\,g + G^2] = \nn\\
& = \text{Tr}\,[g^{-1}\,A^3\,g + g^{-1}\,A^2\,g\,G + G\,g^{-1}\,A^2\,g + g^{-1}\,A\,g\,G^2] \,+\nn\\
& + \text{Tr}\,[G\,g^{-1}\,A^2\,g + g^{-1}\,A\,g\,G^2 + G^2\,g^{-1}\,A\,g + G^3] = \nn\\
& = \text{Tr}\,A^3 + \text{Tr}\,G^3 + 3\,\text{Tr}\,[A^2\,\diff g\,g^{-1} + A\,(\diff g\,g^{-1})^2].
\end{align}
Therefore,
\begin{align}
\Omega^{(3)}(A') &= \Omega^{(3)}(A) + \Omega^{(3)}(G) + \text{Tr}\,[(\diff g\,g^{-1})\,\diff A - (\diff g\,g^{-1})^2\,A] = \nn\\
& = \Omega^{(3)}(A) + \Omega^{(3)}(G) + \diff\,\text{Tr}\,(\diff g\,g^{-1}\,A),
\end{align}
which is the relation \eqref{CS_inv}.

\subsection{Three-dimensional gravity as a Chern-Simons theory}

The Poincaré group in three dimensions with Minkowski signature is $\text{ISO}(2,1)$. Its algebra is defined by the following commutators:
\begin{equations}
& [J_i,J_j] = \vepsilon_{ijk}\,J^k,\\
& [J_i,P_a] = \vepsilon_{iab}\,P^b,\\
& [P_a,P_b] = 0,
\end{equations}
where $J_a$ generates the Lorentz transformations and $P_a$ the translations, $a=0,1,2$. $J_a$ is a peculiarity of the three dimensions: In general the generators are described by an antisymmetric tensor $J^{ab}$, but by means of $\vepsilon_{abc}$ one can define the dual as $J_a = \frac{1}{2}\,\vepsilon_{abc}\,J^{bc}$. 

A gauge theory of this algebra is defined by a connection $A$ and a ghost $c$, which decompose as follows \cite{Witten:1988hc}
\begin{equations}
& A = e^a\,P_a + \omega^i\,J_i,\\
& c = \vrho^a\,P_a + \tau^i\,J_i.
\end{equations}
$e^a$ and $\omega^i$ will play the r\^ole of the dreibein and of the spin connection. The curvature $F=\diff A + A^2$ decomposes as
\begin{equation}
F = T^a\,P_a + R^i\,J_i,
\end{equation}
where $T^a$ and $R^a$ will play the r\^ole of the torsion and of the Riemann curvature, being the curvatures of $e^a$ and $\omega^i$ respectively.

The \textsc{brst} transformations are
\begin{equations}
s\,c &= -\,c^2 = -\,\tfrac{1}{2}\,[c,c],\\
s\,A &= -\,\text{D}\,c = -\,\diff c - [A,c].
\end{equations}
In order to compute the component of $T^a$ and $R^a$ and the transformation on the fields $e^a$, $\omega^a$ and on the ghosts $\vrho^a$, $\tau^a$, it is very useful to introduce the polyform formalism. We define 
\begin{equations}
& \delta = \diff + s,\\
& \bm{A} = A + c,\\
& \bm{F} = \delta\,\bm{A} + \bm{A}^2.
\end{equations}
The previous transformations are equivalent to the horizontality condition
\begin{equation}
\bm{F} = F.
\end{equation}
In three dimensions we can define
\begin{equations}
(a \wedge b)^i &= \vepsilon^{ijk}\,a_j\,b_k,\\
a^2 &= \tfrac{1}{2}\,a\wedge a,
\end{equations}
so that, using 
\begin{equation}
\bm{A} = \bm{E} + \bm{O} = (e + \vrho) + (\omega + \tau),
\end{equation}
we can write
\begin{equation}
\bm{F} = \delta\,\bm{E} + \delta\,\bm{O} + \bm{O} \wedge \bm{E} + \bm{O}^2.
\end{equation}
Using the horizontality condition and splitting in traslational and rotational parts,
\begin{equations}
\delta\,\bm{E} &= T - \bm{O} \wedge \bm{E},\\
\delta\,\bm{O} &= R - \bm{O}^2.
\end{equations}
Defining the covariant derivative
\begin{equation}
\text{D} = \diff + \omega\wedge\cdot
\end{equation}
and filtering according to the ghost number, we get
\begin{equations}
& T = \text{D}\,e, \quad 
R = \diff\omega + \omega^2,\\
& s\,e = -\,\text{D}\,\vrho - \tau\wedge e, \quad
s\,\omega = -\,\text{D}\,\tau,\\
& s\,\vrho = -\,\tau\wedge\vrho, \quad
s\,\tau = -\,\tau^2,
\end{equations}
at ghost number 0 (curvatures), 1 (field transformations) and 2 (ghost transformations) respectively. In order to recover the \textsc{brst} transformations, we set
\begin{equation}
\vrho^a = -\iota_\xi\,e^a, \quad
\tau^a = \Omega - \iota_\xi\,\omega^a,
\end{equation}
where $\xi = \xi^\mu\,\de_\mu$ is a vector field and $\Omega$ is in the fundamental representation.

Replacing in $s\,e$,
\begin{align}
s\,e &= \text{D}\,\iota_\xi\,e - \Omega\wedge e + \iota_\xi\,\omega \wedge e = \nn\\
& = \iota_\xi\,\text{D}\,e - \mathcal{L}_\xi\,e - \iota_\xi\,\omega \wedge e - \Omega \wedge e + \iota_\xi\,\omega \wedge e = \nn\\
& = \iota_\xi\,T - \mathcal{L}_\xi\,e - \Omega \wedge e,
\end{align}
where we used
\begin{equation}
[\iota_\xi,\text{D}]= \mathcal{L}_\xi + \iota_\xi\,\omega\wedge\cdot\;.
\end{equation}

Replacing in $s\,\omega$, 
\begin{align}
s\,\omega &= - \text{D}\,\Omega + \text{D}\,\iota_\xi\,\omega = \nn\\
& = - \text{D}\,\Omega + \iota_\xi\,\text{D}\,\omega - \mathcal{L}_\xi\,\omega - \iota_\xi\,\omega\wedge \omega = \nn\\
& = - \text{D}\,\Omega + \iota_\xi\,\diff\omega + \iota_\xi(\omega \wedge \omega) - \mathcal{L}_\xi\,\omega - \iota_\xi\,\omega\wedge\omega = \nn\\
& =  \iota_\xi\,R - \mathcal{L}_\xi\,\omega - \text{D}\,\Omega,
\end{align}
where we used
\begin{equation}
\iota_\xi\,\omega^2 = \iota_\xi\,\omega \wedge \omega = \omega \wedge \iota_\xi\,\omega.
\end{equation}

Replacing in $s\,\vrho$, on a side
\begin{align}
s\,\vrho &= -(\Omega - \iota_\xi\,\omega) \wedge (-\iota_\xi\,e) = \nn\\
& = \Omega \wedge \iota_\xi\,e - \iota_\xi\,\omega \wedge \iota_\xi\,e.
\end{align}
On the other,
\begin{align}
s\,\vrho &= -s\,(\iota_\xi\,e) = -\iota_{s\xi}\,e - \iota_\xi\,s\,e = \nn\\
& =  - \iota_{s\,\xi}\,e - \iota_\xi^2\,T + \iota_\xi\,\mathcal{L}_\xi\,e + \iota_\xi\,(\Omega \wedge e) = \nn\\
& = -\iota_{s\,\xi}\,e - \iota_\xi^2\,\diff e - \iota_\xi^2\,(\omega\wedge e) + \iota_\xi^2\,\diff e - \iota_\xi\,\diff\,\iota_\xi\,e + \Omega \wedge \iota_\xi\, e = \nn\\
& = -\iota_{s\,\xi}\,e - 2\,\iota_\xi\,\omega\wedge\iota_\xi\,e - \iota_\xi\,\diff\,\iota_\xi e + \Omega \wedge \iota_\xi \,e.
\end{align}
Therefore,
\begin{equation}
\iota_{s\,\xi}\,e = -\,\iota_\xi\,\diff\,\iota_\xi\,e - \iota_\xi\,\omega\wedge\iota_\xi\,e.
\end{equation}
In components
\begin{align}
(s\,\xi)^\lambda\,e_\lambda{}^a &= -\,(\xi^\nu\,\de_\nu\,\xi^\lambda)\,e_\lambda{}^a - \xi^\mu\,\xi^\nu\,(\de_{\mu}\,e_{\nu}{}^a + (\omega_\mu\wedge e_\nu )^a) = \nn\\
& = -\tfrac{1}{2}\,\mathcal{L}_\xi\,\xi^\lambda\,e_\lambda{}^a - \tfrac{1}{2}\,\xi^\mu\,\xi^\nu\,T_{\mu\nu}{}^a = \nn\\
& = -\tfrac{1}{2}\,\mathcal{L}_\xi\,\xi^\lambda\,e_\lambda{}^a - \tfrac{1}{2}\,\iota_\xi^2\,T^a.
\end{align}
Multiplying by $e_\mu{}^a$.
\begin{equation}
s\,\xi^\mu = -\tfrac{1}{2}\,\mathcal{L}_\xi\,\xi^\mu - \tfrac{1}{2}\,\iota_\xi^2\,T^\mu.
\end{equation}

Finally, replacing in $s\,\tau$, on a side
\begin{align}
s\,\tau &= -(\Omega - \iota_\xi\,\omega)^2 = \nn\\
& = -\Omega^2 + \iota_\xi\,\omega\wedge \Omega - \tfrac{1}{2}\,\iota_\xi\,\omega \wedge \iota_\xi\,\omega = \nn\\
& = -\Omega^2 + \iota_\xi\,\omega\wedge \Omega  -\tfrac{1}{2}\,\iota_\xi^2\,\omega^2.
\end{align}
On the other,
\begin{align}
s\,\tau &= s\,\Omega - s\,(\iota_\xi\,\omega) = \nn\\
& = s\,\Omega - \iota_{s\,\xi}\,\omega -  \iota_\xi\,(\iota_\xi\,R - \mathcal{L}_\xi\,\omega - \diff\Omega - \omega \wedge \Omega) = \nn\\
& = s\,\Omega + \tfrac{1}{2}\,\mathcal{L}_\xi\,\iota_\xi\,\omega - \tfrac{1}{2}\,\iota_\xi\,\mathcal{L}_\xi\,\omega + \tfrac{1}{2}\,\iota_{\iota_\xi^2\,T}\,\omega - \iota_\xi^2\,\diff\omega - \iota_\xi^2\,\omega \,+\nn\\
& \quad + \iota_\xi\,\mathcal{L}_\xi\,\omega + \mathcal{L}_\xi\,\Omega + \iota_\xi\,\omega\wedge\Omega = \nn\\
& = s\,\Omega + \tfrac{1}{2}\,\iota_\xi\,\diff\,\iota_\xi\,\omega + \tfrac{1}{2}\,\iota_\xi^2\,\diff \omega - \tfrac{1}{2}\,\iota_\xi\,\diff\,\iota_\xi\,\omega - \iota_\xi^2\,\diff \omega \,+\nn\\
& \quad + \tfrac{1}{2}\,\iota_{\iota_\xi^2\,T}\,\omega - \iota_\xi^2\,\omega^2 + \mathcal{L}_\xi\,\Omega  + \iota_\xi\,\omega\wedge \Omega = \nn\\
& = s\,\Omega - \tfrac{1}{2}\,\iota_\xi^2\,\diff \omega - \iota_\xi^2\,\omega^2 + \tfrac{1}{2}\,\iota_{\iota_\xi^2\,T}\,\omega + \mathcal{L}_\xi\,\Omega  + \iota_\xi\,\omega\wedge \Omega,
\end{align}
where we used
\begin{align}
\iota_{s\,\xi}\,\omega &= -\tfrac{1}{2}\,\iota_{\mathcal{L}_\xi\,\xi}\,\omega - \tfrac{1}{2}\,\iota_{\iota_\xi^2\,T}\,\omega = \nn\\
& = -\tfrac{1}{2}\,[\mathcal{L}_\xi,\iota_\xi]\,\omega - \tfrac{1}{2}\,\iota_{\iota_\xi^2\,T}\,\omega.
\end{align}
Therefore,
\begin{align}
s\,\Omega &= -\mathcal{L}_\xi\,\Omega - \Omega^2 + \tfrac{1}{2}\,\iota_\xi^2\,(\diff\omega + \omega^2) - \tfrac{1}{2}\,\iota_{\iota_\xi^2\,T}\,\omega = \nn\\
& =  -\mathcal{L}_\xi\,\Omega - \Omega^2 + \tfrac{1}{2}\,\iota_\xi^2\,R - \tfrac{1}{2}\,\iota_{\iota_\xi^2\,T}\,\omega.
\end{align}

Summarising, we obtained
\begin{equations}
s\,\xi^\mu &= -\tfrac{1}{2}\,\mathcal{L}_\xi\,\xi^\mu - \tfrac{1}{2}\,\iota_\xi^2\,T^\mu,\\
s\,\Omega &= -\mathcal{L}_\xi\,\Omega - \Omega^2 + \tfrac{1}{2}\,\iota_\xi^2\,R - \tfrac{1}{2}\,\iota_{\iota_\xi^2\,T}\,\omega,\\
s\,e &= \iota_\xi\,T - \mathcal{L}_\xi\,e - \Omega \wedge e,\\
s\,\omega &= \iota_\xi\,R - \mathcal{L}_\xi\,\omega - \text{D}\,\Omega,
\end{equations}
If we use the equations of motion of the 3d Einstein gravity
\begin{equation}
T^a = 0, \quad R^i = 0,
\end{equation}
the transformations become
\begin{equations}
s\,\xi^\mu &= -\tfrac{1}{2}\,\mathcal{L}_\xi\,\xi^\mu,\\
s\,\Omega &= -\mathcal{L}_\xi\,\Omega - \Omega^2,\\
s\,e &= - \mathcal{L}_\xi\,e - \Omega \wedge e,\\
s\,\omega &= - \mathcal{L}_\xi\,\omega - \text{D}\,\Omega,
\end{equations}
which are the correct transformations for a gravity theory in the Lorentz frame formalism.

Let us now consider the action. Define the following bilinear
\begin{equations}
& \text{Tr}\,(J_i\,J_j) = 0,\\
& \text{Tr}\,(P_a\,P_b) = 0,\\
& \text{Tr}\,(J_i\,P_a) = \delta_{ia}.
\end{equations}
Then,
\begin{equation}
\tfrac{1}{2}\text{Tr}\,F^2 = \tfrac{1}{2}\text{Tr}\,(R+T)\,(R+T) = \text{Tr}\,(R\,T) = R_a\,T^a.
\end{equation}
Moreover,
\begin{align}
\diff\left(\tfrac{1}{2}\,\vepsilon_{ija}\,R^{ij}\,e^a\right) &= \text{D}\left(\tfrac{1}{2}\,\vepsilon_{ija}\,R^{ij}\,e^a\right) = \nn\\
& = \tfrac{1}{2}\,\vepsilon_{ija}\,\text{D}\,R^{ij}\,e^a + \tfrac{1}{2}\,\vepsilon_ {ija}\,R^{ij}\,\text{D}\,e^a = \nn\\
& = \tfrac{1}{2}\,\vepsilon_ {ija}\,R^{ij}\,T^a  
= R_a\,T^a.
\end{align}
Therefore
\begin{equation}
\frac{1}{2}\int_{\mathscr{M}_4} \text{Tr}\,F^2 =
\int_{\de\mathscr{M}_4} \vepsilon_{ija}\,R^{ij}\,e^a.
\end{equation}
But the right-hand-side is the three-dimensional Hilbert-Einstein action. In this sense, it can be thought as a Chern-Simons action.

\subsection{BRST quantisation of Chern-Simons theory}

In quantising the Chern-Simons theory, we can follow the same procedure as in Yang-Mills theory. So, we need to add to the action a gauge-fixing term, according to the Fadeev-Popov prescription. We choose a Landau-type gauge-fixing, imposing the condition \cite{Piguet:1995er}
\begin{equation}
\text{div}\,A={}\star\diff\star A = 0,
\end{equation}
by means of a Lagrange multiplier $b$. We have also to add a ghost term in the ghost $c$ and in the antighost $\bar{c}$:
\begin{equation}
S_{\textsc{cs}} + S_{\textsc{gf}} = \frac{1}{2}\int \text{Tr}\,(A\,\diff A + \tfrac{2}{3}\,A^3) + \int\text{Tr}\,(b\,\star\diff\star A - \bar{c}\,\star\diff\star \text{D}\,c).
\end{equation}
This action is \textsc{brst} invariant with respect to
\begin{equations}
s\,c &= -c^2,\\
s\,A &= - \text{D}\,c,\\
s\,\bar{c} &= b,\\
s\,b &= 0,
\end{equations}
because the gauge fixing term is $s$-exact
\begin{equation}
s\,(\bar{c}\,\star\diff\star A) = b\,\star\diff\star A - \bar{c}\,\star\diff\star\text{D}\,c,
\end{equation}
and the \textsc{brst} operator is nilpotent:
\begin{equation}
s^2 = 0.
\end{equation}
Notice that the gauge-fixing term introduces a metric dependence, because of the Hodge dual. The action is the integral of a three-form of ghost number zero. So, denoting the form degree and the ghost number by $(p,g)$, $A$ is a $(1,1)$--form. From its \textsc{brst} transformation, we see that $c$ is a $(0,1)$--form. Remember that the Hodge dual sends a $p$--form into a $(d-p)$--form in $d$ dimensions. So, ${}\star\diff\star\text{D}\,c$ is a $(0,1)$--form. Therefore, $\bar{c}$ is a $(3,-1)$--form. From its \textsc{brst} transformation, $b$ is a $(3,0)$--form, which indeed is compatible with the action, because ${}\star\diff\star A$ is a $(0,0)$--form.

The complete action is obtained by adding the source terms for $s\,A$ and $s\,c$, the so-called \emph{antifields}. Let us denote them with $A^*$ and $c^*$ (not to be confused with the complex conjugation):
\begin{equation}
\Sigma = S_{\textsc{cs}} + S_{\textsc{gf}} + \int (s\,A)\,A^* + (s\,c)\,c^*.
\end{equation}
Therefore, $A^*$ must be a $(2,-1)$--form and $c^*$ must be a $(3,-2)$--form. We have to fix the transformations of the antifields, in such a way the action to remain \textsc{brst} invariant. Besides the trivial choice ($A^*$ and $c^*$ are invariant), there is the following possibility
\begin{equations}
s\,A^*  &= - F - [A^*,c],\\
s\,c^* &= -\text{D}\,A^* - [c^*,c].
\end{equations}
Indeed, the variation of the source term is $\diff$-exact:
\begin{align}
s\,\text{Tr}\,((s\,A)\,A^* + (s\,c)\,c^*) &= \text{Tr}\,(s\,A\,s\,A^* + s\,c\,s\,c^*) = \nn\\
& = \text{Tr}\,(\text{D}\,c\,(F + [A^*,c])+c^2\,(\text{D}\,A + [c^*,c])) = \nn\\
& = \text{Tr}\,(\text{D}\,(c\,F) + \text{D}\,c\,A^*\,c + \text{D}\,c\,c\,A^* + c^2\,\text{D}\,A^* + c^2\,c^*\,c+ c^3\,c^*) = \nn\\
& = \text{Tr}\,(\text{D}\,(c\,F)+\text{D}\,(c\,A^*\,c) - c^3\,c^* + c^3\,c^*) = \nn\\
& = \diff\,\text{Tr}\,(c\,F + c\,A^*\,c),
\end{align}
where we used the cyclicity of the trace. Moreover, the proposed transformations are nilpotent:
\begin{align}
s^2\,A^* &= s\,(-F-[A^*,c]) = \nn\\
& = -[F,c]-[s\,A^*,c] + [A^*,s\,c] = \nn\\
& = -[F,c]+[F,c]+[[A^*,c],c]-[A^*,c^2] = 0,\\
s^2\,c^* &= s\,(-\text{D}\,A^* -[c^*,c]) = \nn\\
& = \diff\,s\,A^* - [s\,A,A^*] + [A,s\,A^*] -[s\,c^*,c] + [c^*,s\,c] = \nn\\
& = -\,\diff F - \diff\,[A^*,c] + [\diff c,A^*]+[[A,c],A^*] - [A,F] -[A,[A^*,c]] \,+\nn\\
&\quad + [\diff A^*,c] + [[A,A^*],c] + [[c^*,c],c] - [c^*,c^2] = 0,
\end{align}
where we used the Jacobi identity and the Bianchi identity.\footnote{The \textsc{brst} transformations of the antifield we proposed do not come out of the blue. Actually, they are strongly motivated within the context of the \textsc{bv} formalism.} 

Summarising, the \textsc{brst} transformations of the Chern-Simons theory are
\begin{equations}
& s\,c= -c^2,\\
& s\,A=-\text{D}\,c,\\
& s\,A^* = -F - [A^*,c],\\
& s\,c^* = - \text{D}\,A^* - [c^*,c],\\
& s\,\bar{c} = b,\\
& s\,b = 0.
\end{equations}
$\bar{c}$ and $b$ form a doublet, which decouples from the other transformations. The transformations on $c, A, A^*, c^*$ can be written in compact form by means of the polyform formalism. Introduce the polyform of total degree one \cite{Imbimbo:2009dy}
\begin{equation}
\bm{A} = c + A + A^* + c^*,
\end{equation}
which is well-defined, since the form degree and the ghost number of each components sum to 1, and the operator
\begin{equation}
\delta = s + \diff.
\end{equation}
Then, the transformations correspond to
\begin{equation}
\delta\,\bm{A} + \bm{A}^2 = 0,
\end{equation}
as it follows by explicit evaluation:
\begin{align}
0 &= \delta\,\bm{A} + \bm{A}^2 = \nn\\
& = (\diff + s)\,(c + A + A^* + c^*) + (c + A + A^* + c^*)^2 = \nn\\
& = s\,c + c^2 +
s\,A + \diff c + [A,c] \,+\nn\\
& \quad + s\,A^* + \diff A + A^2 + [A^*,c] \,+\nn\\
& \quad + s\,c^* + \diff A^* + [A,A^*] + [c,c^*],
\end{align}
because $(c^*)^2 = (A^*)^2 = [A,c^*] = [A^*,c^*] = 0$ in $d=3$. 

If the theory can be quantised, it should be anomaly free. This is the case, as we want to prove. From the doublet theorem, we know that $\bar{c}$ and $b$ do not belong to the cohomology, because they form a doublet. So, let us consider the space of polynomials generated by $c, A, A^*, c^*$ and their derivatives.

Define the following operator $\ell$, which sends $(p,g)$ forms into $(p+1,g-1)$ forms:
\begin{equations}
& \ell\,c = A,\\
& \ell\,A = 2\,A^*,\\
& \ell\,A^* = 3\,c^*,\\
& \ell\,c^* = 0.
\end{equations}
One can verify that it fulfills:
\begin{equation}
[\ell,\diff] = 0, \quad
[\ell,s] = \diff.
\end{equation}
Indeed,
\begin{equations}
(\ell\,s-s\,\ell)\,c &= -\ell\,c^2 - s\,A = \nn\\
& = -[c,\ell\,c] + \diff c + [A,c] = \nn\\
& = -[c,A] + \diff c + [A,c] = \diff c,\\
(\ell\,s-s\,\ell)\,A &= \ell\,(-\diff c - [A,c]) - 2\,s\,A^* = \nn\\
& = -\diff\,\ell\,c - [\ell\,A,c] - [A,\ell\,c] -2\,(-\diff A - A^2 - [A^*,c]) = \nn\\
& = -\diff A - 2\,[A^*,c] - [A,A] + 2\,\diff A + 2\,A^2 + 2\,[A^*,c] = \diff A,\\
(\ell\,s-s\,\ell)\,A^* &= \ell\,(-\diff A - A^2 -[A^*,c]) - 3\,s\,c^* = \nn\\
& = \diff\,\ell\,A - [A,\ell\,A] - [\ell\,A^*,c] - [A^*,\ell\,c] + 2\,\diff A^* + 2\,[A,A^*] + 2\,[c^*,c] = \nn\\
& = -2\,\diff A^* - 2\,[A,A^*] - 3\,[c^*,c] - [A^*,A] + 3\,\diff A^* + 3\,[A,A^*] + 3\,[c^*,c] = \nn\\
& = \diff A^*.
\end{equations}
The other checks are trivial.

It is possible to write the most general solution of the anomaly descent 
\begin{equations}
& s\,\omega^{(3)}_1 = -\,\diff \omega^{(2)}_2,\\
& s\,\omega^{(2)}_2 = -\,\diff \omega^{(1)}_3,\\
& s\,\omega^{(1)}_3 = -\,\diff \omega^{(0)}_4,\\
& s\,\omega^{(0)}_4 = 0.
\end{equations}
by means of the $\ell$ operator \cite{Piguet:1995er}. Let us denote with $\natural$ the elements in the $s$ cohomologies. The general solution of the bottom equation is 
\begin{equation}
\omega_4^{(0)} = {\omega^\natural}_4^{(0)} + s\,\tilde{\omega}_3^{(0)}.
\end{equation}
We can compute
\begin{align}
\diff \omega_4^{(0)} &= (\ell\,s-s\,\ell)\,{\omega^\natural}_4^{(0)} - s\,\diff\,\tilde{\omega}_3^{(0)} = \nn\\
& = -\,s\,(\ell\,{\omega^\natural}_4^{(0)} + \diff\tilde{\omega}_3^{(0)}).
\end{align}
Replacing in the subsequent equation,
\begin{equation}
s\,(\omega_3^{(1)} - \ell\,{\omega^\natural}_4^{(0)} - \diff \tilde{\omega}_3^{(0)}) = 0,
\end{equation}
which is solved by
\begin{equation}
\omega_3^{(1)} = {\omega^\natural}_3^{(1)} + \ell\,{\omega^\natural}_4^{(0)} + \diff \tilde{\omega}_3^{(0)} + s\,\tilde{\omega}^{(1)}_2.
\end{equation}
We can compute
\begin{align}
\diff\,\ell\,{\omega^\natural}_4^{(0)} &= (\ell\,s-s\,\ell)\,\ell\,{\omega^\natural}_4^{(0)} = \nn\\
& = \ell\,(\ell\,s-\diff)\,{\omega^\natural}_4^{(0)} - s\,\ell^2\,{\omega^\natural}_4^{(0)} = \nn\\
& = - \diff\,\ell\,{\omega^\natural}_4^{(0)} - s\,\ell^2\,{\omega^\natural}_4^{(0)},
\end{align}
so that
\begin{equation}
\diff\,\ell\,{\omega^\natural}_4^{(0)} = - \tfrac{1}{2}\,s\,\ell^2\,{\omega^\natural}_4^{(0)}.
\end{equation}
Therefore,
\begin{align}
\diff \omega_3^{(1)} &= -s\,\ell\,{\omega^\natural}_3^{(1)} + \diff\,\ell\,{\omega^\natural}_4^{(0)} - s\,\diff \tilde{\omega}_2^{(1)} = \nn\\
& = - s\,(\ell\,{\omega^\natural}_3^{(1)} + \tfrac{1}{2}\,\ell^2\,{\omega^\natural}_4^{(0)} + \diff \tilde{\omega}_2^{(1)}).
\end{align}
Replacing in the subsequent equation,
\begin{equation}
s\,(\omega_2^{(2)} - \ell\,{\omega^\natural}_3^{(1)}-\tfrac{1}{2}\,\ell^2\,{\omega^\natural}_4^{(0)} - \diff \tilde{\omega}_2^{(1)}) = 0,
\end{equation}
which is solved by
\begin{equation}
\omega_2^{(2)} = {\omega^\natural}_2^{(2)} + \ell\,{\omega^\natural}_3^{(1)} + \tfrac{1}{2}\,\ell^2\,{\omega^\natural}_4^{(0)} + \diff \tilde{\omega}_2^{(1)} + s\,\tilde{\omega}_1^{(2)}.
\end{equation}
We can compute
\begin{align}
\diff\,\ell^2\,{\omega^\natural}_4^{(0)} &= 
-\,s\,\ell^3\,{\omega^\natural}_4^{(0)} + \ell\,(\ell\,s-\diff)\,\ell\,{\omega^\natural}_4^{(0)} = \nn\\
& = -\,s\,\ell^3\,{\omega^\natural}_4^{(0)} + \ell\,(\ell^2\,s - \ell\,\diff - \diff\,\ell)\,{\omega^\natural}_4^{(0)},
\end{align}
so that
\begin{equation}
\diff\,\ell^2\,{\omega^\natural}_4^{(0)} = - \tfrac{1}{3}\,s\,\ell^3\,{\omega^\natural}_4^{(0)},
\end{equation}
Therefore
\begin{equation}
\diff \omega_2^{(2)} = -s\,(\ell\,{\omega^\natural}_2^{(2)} + \tfrac{1}{2}\,\ell^2\,{\omega^\natural}_3^{(1)} + \tfrac{1}{3!}\,\ell^3\,{\omega^\natural}_4^{(0)} + \diff\tilde{\omega}_1^{(2)}).
\end{equation}
Replacing in the top equation,
\begin{equation}
s\,(\omega_1^{(3)} - \ell\,{\omega^\natural}_2^{(2)} - \tfrac{1}{2}\,\ell^2\,{\omega^\natural}_3^{(1)} - \tfrac{1}{3!}\,\ell^3\,{\omega^\natural}_4^{(0)} - \diff\tilde{\omega}_1^{(2)}) = 0,
\end{equation}
which is solved by
\begin{equation}
\omega_1^{(3)} = {\omega^\natural}_1^{(3)} + \ell\,{\omega^\natural}_2^{(2)} + \tfrac{1}{2}\,\ell^2\,{\omega^\natural}_3^{(1)} + \tfrac{1}{3!}\,\ell^3\,{\omega^\natural}_4^{(0)} + \diff\tilde{\omega}_1^{(2)} + s\,\tilde{\omega}_0^{(3)}.
\end{equation}
Summarising,
\begin{equations}
\omega_4^{(0)} &= {\omega^\natural}_4^{(0)} + s\,\tilde{\omega}_3^{(0)},\\
\omega_3^{(1)} &= {\omega^\natural}_3^{(1)} + \ell\,{\omega^\natural}_4^{(0)} + \diff \tilde{\omega}_3^{(0)} + s\,\tilde{\omega}^{(1)}_2,\\
\omega_2^{(2)} &= {\omega^\natural}_2^{(2)} + \ell\,{\omega^\natural}_3^{(1)} + \tfrac{1}{2}\,\ell^2\,{\omega^\natural}_4^{(0)} + \diff \tilde{\omega}_2^{(1)} + s\,\tilde{\omega}_1^{(2)},\\
\omega_1^{(3)} &= {\omega^\natural}_1^{(3)} + \ell\,{\omega^\natural}_2^{(2)} + \tfrac{1}{2}\,\ell^2\,{\omega^\natural}_3^{(1)} + \tfrac{1}{3!}\,\ell^3\,{\omega^\natural}_4^{(0)} + \diff\tilde{\omega}_1^{(2)} + s\,\tilde{\omega}_0^{(3)}.
\end{equations}
These expressions can be recasted in a polyform
\begin{equation}
\omega_4 = e^\ell\,\omega_4^\natural + \delta\,\tilde{\omega}_3,
\end{equation}
where
\begin{equations}
& \delta = \diff + s,\\
& \omega_4 = \omega_1^{(3)} + \omega_2^{(2)} + \omega_3^{(1)} + \omega_4^{(0)},\\
& \omega^\natural_4 = {\omega^\natural}_1^{(3)} + {\omega^\natural}_2^{(2)} + {\omega^\natural}_3^{(1)} + {\omega^\natural}_4^{(0)},\\
& \tilde{\omega}_3 = \tilde{\omega}_0^{(3)} + \tilde{\omega}_1^{(2)} + \tilde{\omega}_2^{(1)} + \tilde{\omega}_3^{(0)}.
\end{equations}
This is a general fact. The descent in $d$ dimension, which can be recasted in the Stora-Zumino equation \ref{BRSAnomalies}
\begin{equation}
\delta\,\omega_{d+1}=0,
\end{equation}
admits as the most general solution
\begin{equation}
\omega_{d+1} = e^\ell\,\omega_{d+1}^\natural + \delta\,\tilde{\omega}_d,
\end{equation}
if an $\ell$ operator exists.

Back to our problem, we have to compute the most general $\omega_4^\natural$. Consider the following filtering
\begin{equations}
& s_0\,c^* = -\diff A^*,\\
& s_0\,A^* = -\diff A,\\
& s_0\,A = -\diff c,\\
& s_0\,c = 0
\end{equations}
and $s_1 = s-s_0$. The $s_0$ transformations form a descent. Therefore, as a consequence of the doublet theorem, the $s_0$-cohomology is generated only by $c$. Therefore, the $s_0$-cohomology on the $p$-forms, with $p>0$, is empty. Moreover, the $s_0$-cohomology on the $0$--forms of even ghost number is empty too, because $\text{Tr}\,c^{2\,n} = 0$ (use the cyclicity of the trace and the anticommutativity of $c$). By the filtering theorem, also the corresponding $s$ cohomologies are empty. In particular
\begin{equation}
H_4^{(0)} \simeq H_3^{(1)} \simeq H_2^{(2)} \simeq H_1^{(3)}= \varnothing.
\end{equation}
Therefore, $\omega_4^\natural = 0$. So, there is no \emph{anomaly in three-dimensional Chern-Simons theory}.

There is another simpler way to prove that there is no anomaly. We know that the $s$ transformations can be recasted in $\delta\,\bm{A} = -\bm{A}^2$. Anomalies are the ghost number one components of the nontrivial solutions of $\delta\,\omega_4 = 0$. So, we have to compute the $\delta$-cohomology on polyforms of total degree four. The unique candidate is $\text{Tr}\,\bm{A}^4$, which vanishes for the same reason as $\text{Tr}\,c^4 = 0$. So, the Chern-Simons theory is anomaly-free.

\subsection{BRST topological field theory}

Consider the \textsc{brst} transformation of the ghost of diffeomorphisms \cite{Imbimbo:2009dy, Ouvry:1988mm, Baulieu:1988xs}
\begin{equation}
s\,\xi = -\tfrac{1}{2}\,\mathcal{L}_\xi\,\xi,
\end{equation}
where $\xi$ is an anticommuting vector field and the Lie derivative $\mathcal{L}_\xi$ on differential forms is given by the \emph{Cartan formula}:
\begin{equation}
\mathcal{L}_\xi = [\iota_\xi, \diff] = \iota_\xi\,\diff + (-)^{|\xi|}\,\diff \iota_\xi.
\end{equation}
The contraction operator $\iota_\xi$ is odd [even] when the parameter $\xi$ is even [odd]. This transformation is nilpotent by means of the Jacobi identity. Let us deform it, by adding a commuting vector field $\gamma$ with ghost number 2:
\begin{equation}
s\,\xi = -\tfrac{1}{2}\,\mathcal{L}_\xi\,\xi + \gamma.
\end{equation}
The requirement of nilpotency fixes the $\gamma$ transformation:
\begin{equation}
s\,\gamma = - \mathcal{L}_\xi\,\gamma.
\end{equation}
The $\gamma$ deformation induces to deform also the transformation of a covariant field $\Phi$, which would transform with $-\mathcal{L}_\xi\,\Phi$, if one requires $s^2\,\Phi = 0$. For future convenience, let us introduce the \emph{equivariant operator} with respect to $\xi$:  
\begin{equation}
\hat{s} = s + \mathcal{L}_\xi,
\end{equation}
which is defined on the space of ghosts and fields which excludes $\xi$. We can think at $\hat{s}$ as a simple redefinition of $s$ for the moment, though its meaning is pretty deeper. The nilpotency of $s$ is equivalent to the following condition on $\hat{s}$:
\begin{equation}
s^2 = 0 \Leftrightarrow \hat{s}^2 = \mathcal{L}_\gamma.
\end{equation}
Suppose that
\begin{equation}
\hat{s}\,\Phi = \psi,
\end{equation}
for some $\psi$ with ghost number 1, and require that
\begin{equation}
\hat{s}^2\,\Phi = \mathcal{L}_\gamma\,\Phi,
\end{equation}
which implies that
\begin{equation}
\hat{s}\,\psi = \mathcal{L}_\gamma\,\Phi.
\end{equation}
Therefore, we obtained the following nilpotent transformations
\begin{equations}
s\,\xi &= -\tfrac{1}{2}\,\mathcal{L}_\xi\,\xi + \gamma,\\
s\,\gamma &= - \mathcal{L}_\xi\,\gamma,\\
s\,\Phi &= -\mathcal{L}_\xi\,\Phi + \psi,\\
s\,\psi &= -\mathcal{L}_\xi\,\psi + \mathcal{L}_\gamma\,\Phi.
\end{equations}
These are the \textsc{brst} transformations of the \emph{topological gravity}, because the transformation of an arbitrary field $\Phi$ is deformed by an arbitrary contribution $\psi$. In particular, for the metric tensor $g_{\mu\nu}$, we can write:
\begin{equations}
s\,g_{\mu\nu} = -\mathcal{L}_\xi\,g_{\mu\nu} + \psi_{\mu\nu}, \\
s\,\psi_{\mu\nu} = -\mathcal{L}_\xi\,\psi + \mathcal{L}_\gamma\,g_{\mu\nu},
\end{equations}
where $\psi_{\mu\nu}$ is called \emph{topological gravitino}.

Let us include a gauge transformation, generated by $\delta_c$, where $c$ is an anticommuting ghost:
\begin{equation}
\delta_c\,\Phi = [c,\Phi],
\end{equation}
for any $\Phi$ in the adjoint representation of the gauge group. The corresponding connection $A$ is such that
\begin{equation}
\delta_c\,A = (-)^{1+|c|}\,\text{D}\,c,
\end{equation}
where the covariant derivative is defined as
\begin{equation}
\text{D} = \diff + \delta_A.
\end{equation}
The transformation of the connection is fixed by the curvature
\begin{equation}
F = \diff A + A^2 
\end{equation}
to be omogeneous in the adjoint represetation
\begin{equation}
\delta_c\,F = [c,F].
\end{equation}
At the ghost level, one has in general:
\begin{equation}
\hat{s}\,c = - c^2,
\end{equation}
which is nilpotent if $\gamma = 0$. We deform it by adding a ghost number 2 contribution $\kappa$:
\begin{equation}
\hat{s}\,c =-c^2 + \kappa.
\end{equation}
The nilpotency requirement fixes the transformation of $\kappa$:
\begin{equation}
\hat{s}\,\kappa = - \delta_c\,\kappa + \iota_\gamma\,\diff c.
\end{equation}
The presence of the inhomogeneous term suggests that $\kappa$ depends on the connection. So, the $\gamma$ deformation introduces a field dependence in the ghost transformations. We can choose
\begin{equation}
\kappa = \iota_\gamma\,A + \vphi,
\end{equation}
where $\vphi$ has ghost number 2:
\begin{equation}
\delta_c\,\vphi = [c,\vphi].
\end{equation}
Indeed, by including a $\lambda$ deformation in the transformation of $A$,
\begin{equation}
\hat{s}\,A = -\text{D}\,c + \lambda, 
\end{equation}
and supposing that
\begin{equation}
\hat{s}\,\vphi = -[c,\vphi]+ \iota_\gamma\,\lambda,
\end{equation}
we get the right transformation for $\kappa$.

It is useful to introduce an equivariant operator also with respect to $c$ \cite{Imbimbo:2014pla, Imbimbo:2018duh, Bae:2015eoa}:
\begin{equation}
S = \hat{s} + \delta_c,
\end{equation}
which is such that
\begin{equation}
s^2 = 0 \Leftrightarrow S^2 = \mathcal{L}_\gamma + \delta_{\iota_\gamma A + \vphi}
\end{equation}
Therefore
\begin{equation}
S\,\gamma = 0, \quad
S\,A=\lambda, \quad
S\,\phi = \iota_\gamma\,\lambda.
\end{equation}
The transformation of a regular field $\Phi$
\begin{equation}
S\,\Phi = \psi,
\end{equation}
fixes the transformation of $\psi$
\begin{equation}
S\,\psi = \mathcal{L}_\gamma\,\Phi + \delta_{\iota_\gamma A + \vphi}\,\Phi,
\end{equation}
which is nilpotent by explicit evaluation. $S\,\lambda$ is fixed by the nilpotency on $A$:
\begin{equation}
S\,\lambda = \mathcal{L}_\gamma\,A + \delta_{\iota_\gamma A + \vphi}\,A = \iota_\gamma\,F - \text{D}\,\vphi,
\end{equation}
which is nilpotent by explicit evaluation. In this way we can check the nilpotency on $\vphi$. Finally, the transformation of the curvature is
\begin{equation}
S\,F = -\,\text{D}\,\lambda,
\end{equation}
which is nilpotent by means of the Bianchi identity
\begin{equation}
\text{D}\,F = 0.
\end{equation}
Summarising, we obtained
\begin{equations}
& s\,\xi = -\tfrac{1}{2}\,\mathcal{L}_\xi\,\xi + \gamma,\\
& s\,\gamma = -\mathcal{L}_\xi\,\gamma,\\
& s\,c = -\mathcal{L}_\xi\,c - c^2 + \iota_\gamma\,A + \vphi,\\
& s\,\vphi = -\mathcal{L}_\xi\,\vphi - [c,\vphi] + \iota_\gamma\,\lambda,\\
& s\,A = -\mathcal{L}_\xi\,A - \text{D}\,c + \lambda,\\
& s\,\lambda = -\mathcal{L}_\xi\,\lambda - [c,\lambda]+ \iota_\gamma\,F - \text{D}\,\vphi,\\
& s\,F = -\mathcal{L}_\xi\,F - [c,F] -\text{D}\,\lambda,\\
& s\,\Phi = -\mathcal{L}_\xi\,\Phi - [c,\Phi] + \psi,\\
& s\,\psi = -\mathcal{L}_\xi\,\psi - [c,\psi] + \mathcal{L}_\gamma\,\Phi + [\iota_\gamma\,A + \vphi,\Phi].
\end{equations}
In absence of gravity, they reduces to
\begin{equations}
& s\,c = - c^2 + \vphi,\\
& s\,\vphi = - [c,\vphi],\\
& s\,A = - \text{D}\,c + \lambda,\\
& s\,\lambda = - [c,\lambda] - \text{D}\,\vphi,\\
& s\,F = - [c,F] -\text{D}\,\lambda,\\
& s\,\Phi = - [c,\Phi] + \psi,\\
& s\,\psi = - [c,\psi] + [\vphi,\Phi],
\end{equations}
which are the transformations of a topological gauge field theory. 

The previous algebra can be recasted in terms of polyforms in the following way. Define
\begin{equation}
\bm{A} = A + c, \quad
\delta = \diff + s, \quad
\bm{F} = \delta\,\bm{A} + \bm{A}^2.
\end{equation}
$\delta$ can be treated as a de Rham differential on the space of connections $\bm{A}$, and $\bm{F}$ is the corresponding curvature. We can compute
\begin{equations}
\bm{F} = \delta\,\bm{A} + \bm{A}^2 &= (\diff + s)\,(A + c) + (A+c)^2  = \nn\\
& = \diff A + A^2 + s\,A + \diff c + [A,c] + s\,c + c^2 = \nn\\
& = F + \lambda + \vphi + (s\,A + \text{D}\,c -\lambda) + (s\,c + c^2 - \vphi).
\end{equations}
Therefore the algebra on $A$ and $c$ is equivalent to
\begin{equation}
\bm{F} = F + \lambda + \vphi.
\end{equation}
Moreover, the generalised Bianchi identity holds, because $\delta, \bm{A}, \bm{F}$  satisfy the same algebraic relations as $\diff, A, F$:
\begin{equation}
\delta\,\bm{F} + [\bm{A},\bm{F}] = 0.
\end{equation}
Notice that this relations provides the \textsc{brst} rules on $F,\lambda,\vphi$. Indeed,
\begin{equations}
\delta\,\bm{F} + [\bm{A},\bm{F}] &= 
(\diff + s)\,(F + \lambda + \vphi) + [A+c,F+\lambda+\vphi] = \nn\\
& = \diff F + [A,F] + s\,F + \diff \lambda + [A,\lambda] + [c,F] \,+\nn\\
& \quad + s\,\lambda + \diff\vphi + [A,\vphi] + [c,\lambda] + s\,\vphi + [c,\vphi] = \nn\\
& = \text{D}\,F + (s\,F + \text{D}\,\lambda + [c,F]) \,+\nn\\
& \quad + (s\,\lambda + \text{D}\,\vphi + [c,\lambda]) + (s\,\vphi + [c,\vphi]).
\end{equations}
By means of the polyforms the descent for observables. For example, we can consider $\text{Tr}\,\bm{F}^2$, which satisfies 
\begin{equation}
\delta\,\text{Tr}\,\bm{F}^2 = 0,
\end{equation}
by means of the generalised Bianchi identity. Filtering in the ghost number, we get the following descent
\begin{equations}
& s\,\text{Tr}\,F^2 = \diff\,\text{Tr}\,(2\,F\,\lambda),\\
& s\,\text{Tr}\,(F\,\lambda) = -\diff\,\text{Tr}\,(\tfrac{1}{2}\,\lambda^2 + F\,\vphi),\\
& s\,\text{Tr}\,(\lambda^2 + 2\,F\,\lambda) = -\diff\,\text{Tr}\,(2\,\lambda\,\vphi),\\
& s\,\text{Tr}\,(\lambda\,\vphi) = -\diff\,\text{Tr}\,(\tfrac{1}{2}\,\vphi^2),\\
& s\,\text{Tr}\,\vphi^2 = 0.
\end{equations} 

In the \textsc{brst} perspective, it is very simple to couple the Chern-Simons theory with topological gravity, that is, to study the Chern-Simons theory in a curved background \cite{Becchi:1997jg, Imbimbo:2009dy}. The $s$ operator defined previously (which from now on is denoted as $S$) is extended by including the Lie derivative and the $\gamma$ deformation
\begin{equations}
s\,g_{\mu\nu} &= -\mathcal{L}_\xi\,g_{\mu\nu} +\psi_{\mu\nu},\\
s\,\psi_{\mu\nu} &= -\mathcal{L}_\xi\,\psi_{\mu\nu} + \mathcal{L}_\gamma\,g_{\mu\nu},\\
s\,\gamma &= -\mathcal{L}_\xi\,\gamma,\\
s\,c &= -\mathcal{L}_\xi\,c -c^2 + \iota_\gamma\,A,\\
s\,A &= -\mathcal{L}_\xi\,A - \text{D}\,c + \iota_\gamma\,A^*,\\
s\,A^* &= -\mathcal{L}_\xi\,A^* -F- [A^*,c] + \iota_\gamma\,c^*,\\
s\,c^* &= -\mathcal{L}_\xi\,c^* -\text{D}\,A^* - [c^*,c],\\
s\,\bar{c} &= -\mathcal{L}_\xi\,c + b,\\
s\,b &= -\mathcal{L}_\xi\,b + \mathcal{L}_\gamma\,\bar{c}.
\end{equations}
The $\gamma$ deformation individuates an operator, which we denote by $G_\gamma$, whose action is given by 
\begin{equations}
& G_\gamma\,\psi_{\mu\nu} = \mathcal{L}_\gamma\,g_{\mu\nu}, \quad G_\gamma\,g_{\mu\nu} = 0, \quad G_\gamma\,\gamma = 0,\\
& G_\gamma\,c = \iota_\gamma\,A, \quad
G_\gamma\,A = \iota_\gamma\,A^*,\quad
G_\gamma\,A^* = \iota_\gamma\,c^*,\quad
G_\gamma\,c^* = 0,\\
& G_\gamma\,b = \mathcal{L}_\gamma\,\bar{c}, \quad G_\gamma\,\bar{c} = 0.
\end{equations}
Notice that
\begin{equation}
s + \mathcal{L}_\xi = S + G_\gamma, \quad 
S^2 = G_\gamma^2 = 0, \quad
\lbrace S, G_\gamma \rbrace = \mathcal{L}_\gamma.
\end{equation}
So, $G_\gamma$ defines a new \textsc{brst} charge. The $G_\gamma$ transformations of $c, A, A^*, c^*$ are rephrased in terms of $\bm{A}$:
\begin{equation}
G_\gamma\,\bm{A} = \iota_\gamma\,\bm{A}.
\end{equation}
Following the computation without gravity, and replacing $\delta$ with $\diff + S + \mathcal{L}_\xi$, we see that the $\gamma$-independent part of the transformations of $c, A, A^*, c^*$ is equivalent to
\begin{equation}
(\diff + S + \mathcal{L}_\xi)\,\bm{A} + \bm{A}^2 = 0.
\end{equation}
Then, summing up the last two equations, we obtain
\begin{equation}
\delta\,\bm{A} + \bm{A}^2 = 0, 
\end{equation}
where  $\delta = \diff + s + \mathcal{L}_\xi - \iota_\gamma$.

\newpage
\quad
\thispagestyle{empty}

\newpage

\section{$E_{8(8)}$ generalised geometry and uplifts}\label{4}

\subsection{$E_{8(8)}$ generalised bundle and Lie derivative}\label{E8gengeo}

In trying to define a generalised geometry for an eight-dimensional internal manifold, on the same footing as in the lower-dimensional cases, several problems arise. As previously seen, when eleven-dimensional supergravity is compactified up to three dimensions, gravity is not propagating and the resulting three-dimensional theory is completely described by a non-linear $\sigma$-model with coset space $E_{8(8)}/\text{Spin}(16)$, which therefore encompasses all the 128 bosonic degrees of freedom of the original higher-dimensional theory. In order to define the generalised geometry, one has to take into account also the dual degrees of freedom. The scalars coming from the metric and the three-form are $36+56$; the one-forms are $8+28$; the dual one-forms are $56+64$; and the dual scalars are $28+8$. Unlike the $n=7$ case, the vectors $8 + 28 + 56 + 64 = 156$ do not fill the fundamental/adjoint representation $\mathbf{248}$ of $E_{8(8)}$, which is larger. It seems that a generalised vector transforming in $\mathbf{248}$ encodes other degrees of freedom which are not captured by eleven-dimensional supergravity. The decomposition of the representation $\mathbf{248}$ according to $\text{GL}(8,\mathbb{R})$ is given by \cite{Hohm:2014fxa}:
\begin{equation}\label{Dec248}
\mathbf{248}_1 \rightarrow \mathbf{8}_{3} \oplus \overline{\mathbf{28}}_{2} \oplus \mathbf{56}_1 \oplus (\mathbf{63} \oplus \mathbf{1})_0 \oplus \overline{\mathbf{56}}_{-1} \oplus \mathbf{28}_{-2} \oplus \overline{\mathbf{8}}_{-3},
\end{equation} 
where the subscript denotes a choice for the trombone $\text{GL}(1)$ weight.\footnote{In \cite{Strickland-Constable:2013xta} another equivalent choice of the weight is adopted. The corresponding components of a generalised one-form have opposite weights.} The corresponding generalised vector bundle is
\begin{align}
& \qquad\qquad\quad T\mathscr{M} \oplus \wedge^2\,T^*\mathscr{M} \oplus
\wedge^5\,T^*\mathscr{M} \oplus (T^*\mathscr{M} \otimes \wedge^7\,T^*\mathscr{M}) \,\oplus\nn\\
& \oplus (\wedge^3\,T^*\mathscr{M} \otimes \wedge^8\,T^*\mathscr{M}) \oplus (\wedge^6\,T^*\,\mathscr{M} \otimes \wedge^8\,T^*\mathscr{M}) \oplus (T^*\mathscr{M} \otimes (\wedge^8\,T^*\mathscr{M})^2).
\end{align}
Consistently, the dimensions of each component are
\begin{equation}
8 + {{8}\choose{2}} + {{8}\choose{5}} + 8\,{{8}\choose{7}} + {{8}\choose{3}} + {{8}\choose{6}} + 8 = 248.
\end{equation}
The one-forms $A_\mu{}^n$, $A_{\mu m_1 m_2}$, $\tilde{A}_{\mu m_1 \dots m_5}$, and $\tilde{h}_{\mu m_1 \dots m_7, n}$ transform according to $\mathbf{8}_3$, $\overline{\mathbf{28}}_2$, $\mathbf{56}_{1}$, and $\mathbf{63}_0$ respectively. Thus, the representations with negative weights are those not filled by the eleven-dimensional supergravity degrees of freedom.

The non-negative-weighted representations are associated to internal diffeomorphisms, three-form gauge transformation, dual six-form gauge transformation, and dual graviton gauge transformation with parameters
\begin{equation}
\xi^m, \quad \Lambda_{m_1 m_2}, \quad \Lambda_{m_1 \dots m_5}, \quad \xi_{m_1 \dots m_7, n}, \quad \xi_{m_1 \dots m_8}.
\end{equation}
$\Lambda_{m_1 m_2}$ and $\Lambda_{m_1 \dots m_5}$ are the components of a two-form and of a five-form in eight dimensions, respectively; $\xi_{m_1 \dots m_7, n}$ are the components of a mixed-symmetry tensor, antisymmetric in the first seven indices and with vanishing totally antisymmetric part $\xi_{[m_1 \dots m_7, n]}=0$ (so, it has ${\binom{8}{7}}\,7-{\binom{8}{8}} = 63$ components); $\xi_{m_1 \dots m_8}$ is an eight-form in eight dimensions (so it has a single component).\footnote{See Appendix \ref{DualGraviton} for more details.} Instead, the negative-weighted representations have no obvious corresponding symmetry in eleven-dimensional supergravity. Nevertheless, we could expect that the latter representations and those associated with the gauge transformation of the dual graviton act as trivial parameters in the to-be-defined $n=8$ generalised Lie derivative.

Not only the structure of the $n=8$ generalised bundle is not completely clear, but also the definition of the generalised Lie derivative is afflicted by issues, if the lower-dimensional cases are faithfully followed. Indeed, using the definition \eqref{GenLieDerP} with the projector in the fundamental representation $\mathbf{248}$ given by 
\begin{equation}
(P^{(\textbf{248})})^M{}_N{}^P{}_Q = f^{RM}{}_N\,f_R{}^P{}_Q,
\end{equation}
where $f_{MNP}$ are the structure constants of the algebra, $M,N,\dots = 1,\dots, 248$, then one can check that the Leibniz identity \eqref{LeibnizIdentityAlgebra} is not satisfied. Putting in evidence a density term with weight one in the last term, which is due to the $\mathbb{R}^+$ part of the duality group $E_{8(8)} \times \mathbb{R}^+$, the $n=8$ generalised Lie derivative reads
\begin{equation}\label{GenLieDerE8}
L_X\,V^M = X^N\,\de_N\,V^M + \lambda\,\de_N\,X^N\,V^M - V^P\,f^{NM}{}_P\,f_N{}^R{}_S\,\de_R\,X^S + \de_N\,X^N\,V^M.
\end{equation}
In the following, we will always consider one-weighted density vectors, therefore taking $\lambda = 0$. The $Y$-tensor reads
\begin{equation}\label{E8Y}
Y^{MP}{}_{QN} = \delta^M_Q\,\delta^P_N + \delta^M_N\,\delta^P_Q - f^{RM}{}_N\,f_R{}^P{}_Q.
\end{equation}
The section constraint \eqref{SectionConstraintYSectionMatrix} amounts to set to zero the representation $\mathbf{1}$, $\mathbf{248}$, and $\mathbf{3875}$ in $\mathbf{248}\times\mathbf{248}$. As shown in Section \ref{AppendixE8}, this means that  
\begin{equations}
& \eta^{MP}\,\mathcal{E}_M{}^m\,\mathcal{E}_P{}^p = 0,\\
& f^{MPR}\mathcal{E}_M{}^m\,\mathcal{E}_P{}^p = 0,\label{SecConstr8_1}\\
& (\delta^{(M}_Q\,\delta^{P)}_N-\tfrac{1}{2}\,f^{(M}{}_{QR}\,f^{P)}{}_N{}^R)\,\mathcal{E}_M{}^m\,\mathcal{E}_P{}^p = 0.
\end{equations}
But these conditions are not sufficient to ensure the generalised Lie derivative to satisfy the Leibniz identity, since one can compute that
\begin{equation}\label{LeibnizSpoiling0}
([L_X,L_V] - L_{\frac{1}{2}(L_X V - L_V X)})\,W^M = -\tfrac{1}{2}\,f^P{}_{RS}\,(V^R\,\de_P\,\de_Q\,X^S - X^R\,\de_P\,\de_Q\,V^S)\,f^{QM}{}_N{}\,W^N,
\end{equation}
for any $X^M, V^M$, and $W^M$. The right-hand side is the adjoint action of the combination 
\begin{equation}
S(V,W) = -\tfrac{1}{2}\,f^P{}_{RS}\,(V^R\,\de_P\,\de_Q\,X^S-X^R\,\de_P\,\de_Q\,V^S)
\end{equation}
on $W^M$. Denoting it with $\delta_{S(V,W)}$, 
\begin{equation}
\delta_{S(V,W)}\,W^M = -S_P\,f^{PM}{}_N\,W^N,
\end{equation}
so that the equation \eqref{LeibnizSpoiling} becomes
\begin{equation}\label{LeibnizSpoiling}
[L_X,L_V] = L_{\frac{1}{2}(L_X V - L_V X)} +\delta_{S(X,V)}\,W^M.
\end{equation}
This expression says that not only $L_X$ does not satisfy the Leibniz identity, but it does not even close into an algebra. Anyway, as argued in \cite{Hohm:2014fxa}, the $\delta$-piece in \eqref{LeibnizSpoiling} can be absorbed by deforming the generalised Lie derivative in the following way:
\begin{equation}
L_{X,S} = L_X + \delta_S.
\end{equation}
This is a generalised Lie derivative depending on two parameters: besides the generalised vector $X^M$ there is also the generalised one-form $S_M$, which will be called \emph{ancillary parameter}, transforming in the dual adjoint representation. Explicitly, using \eqref{GenLieDerE8},
\begin{equation}\label{GenLieDerE8bis}
L_{X,S}\,V^M = X^N\,\de_N\,V^M + \lambda\,\de_N\,X^N\,V^M + \de_N\,X^N\,V^M - f^{NM}{}_P\,V_P\,(S_N + f_N{}^R{}_S\,\de_R\,X^S).
\end{equation}
Thus, using \eqref{LeibnizSpoiling}, we can compute
\begin{align}\label{Leibniz8}
[L_{X,S},L_{V,T}] &= [L_X,L_V] + [L_X,\delta_T]-[L_V,\delta_S] + [\delta_S,\delta_T] = \nn\\
&= L_{\frac{1}{2}(L_X V - L_V X)} + \delta_{S(X,V)} + \delta_{L_X T-L_V S} + \delta_{\delta_S\,T}.
\end{align}
Now, if we require the ancillary parameters $S,T$ to be on section
\begin{equation}
S_M = \mathcal{E}_M{}^m\,s_m, \quad T_M = \mathcal{E}_M{}^m\,t_m,
\end{equation}
then the last term vanishes, since $\delta_S\,T^M = S_N\,f^{NM}{}_P\,T^P = 0$, thanks to \eqref{SecConstr8_1}. Therefore, the relation \eqref{Leibniz8} reduces to
\begin{align}\label{Leibniz8bis}
[L_{X,S},L_{V,T}] &= L_{\frac{1}{2}(L_X V - L_V X),\,L_X T-L_V S + S(X,V)},
\end{align}
which means that the generalised Lie derivative $L_{X,S}$ closes into an algebra. Nevertheless, the above relation is not yet the Leibniz identity. In order to obtain a generalised Lie derivative satisfying the Leibniz identity it is necessary to extend the action of the generalised Lie derivative on the ancillary parameters too \cite{Hohm:2017wtr}, as we will see in Section \ref{E8Dorfman}.

We conclude this section by discussing the structure of trivial parameters of the $L_{X,S}$. As shown in \cite{Hohm:2014fxa}, the following is a trivial parameter:
\begin{equation}\label{TrivialParE8}
(\tilde{X}^M,\tilde{S}_M) = (f^{MP}{}_Q\,\Omega_P{}^Q,\de_M\,\Omega_N{}^N - \de_N\,\Omega_M{}^N), \;\;\text{with}\;\;
\Omega_M{}^N = \mathcal{E}_M{}^m\,\omega_m{}^N.
\end{equation}
Since the first index of $\Omega_M{}^N$ is on section, it takes value only in the $\mathbf{8}_{3}$ representation. Consider the first component of the trivial parameter. The index $P$ in $f^{MP}{}_Q$ takes values only in $\overline{\mathbf{8}}_{-3}$, because it is saturated with the first index of $\Omega_P{}^Q$. If $q$ is the weight of any component in the index $M$, any component of the index $Q$ has weight $-q-3$, and so the second index in $\Omega_P{}^Q$ has only components with weight $q+3$. This weight should be between $-3$ and $-3$, so $q=-3,-2,-1,0$ and correspondingly $q+3=0,1,2,3$. So, if we choose $\Omega_P{}^Q$ with the second index taking values only in the representations with weights $0,1,2$, then the $\tilde{S}_M$ component of the trivial parameter vanishes. Indeed, $\Omega_P{}^Q$ enters in $S_M$ by its trace or with its second index saturated with a derivative. Thus, since the first index of $\Omega_P{}^Q$ and any derivative are on section, the second index must sit in the representation $\mathbf{8}_3$ to have possibly a non-vanishing second components. But the $\mathbf{8}_3$ representation is excluded, as said before. Therefore, we can conclude that
\begin{itemize}
\item[--] If we choose any $\tilde{X}^M$ taking values only in the representations with weights $-3,-2,-1$, then $(\tilde{X}^M,0)$ defines a trivial parameter, because we can always choose an $\Omega_P{}^Q$ such that the $\tilde{X}^M$ we started with fits in the general form of \eqref{TrivialParE8}, with $\tilde{S}_M = 0$ (the relation between $\tilde{X}^M$ and $\Omega_P{}^Q$ via the structure constant is invertible).
\item[--] If we choose $\tilde{X}^M$ taking values only in $(\mathbf{63}\oplus\mathbf{1})_0$, then $\Omega_P{}^Q$ exists such that $\tilde{X}^M$ fits in a trivial parameter, with a priori non-vanishing second component $\tilde{S}_M$; in other words, there is an ancillary parameter generating a transformation which trivialises the action of any dual graviton gauge transformation with weight zero.
\end{itemize}
In conclusion, these reasonings show that \emph{any parameter in the negative-weighted representations in the decomposition \eqref{Dec248} defines  a trivial parameter for the $n=8$ generalised Lie derivative} \eqref{GenLieDerE8bis}.

\subsection{$E_{8(8)}$ section constraint}\label{AppendixE8}

In this section, more details on $n=8$ section constraint are provided. Denote with $\lbrace T^A \rbrace_{A=1,\dots,248}$ a basis of generators of the Lie algebra of $E_8$ and with $f^{AB}{}_C$ the structure constants \cite{Green:1987sp, Green:1987mn}:
\begin{equation}
[T^A,T^B]=f^{AB}{}_C\,T^C.
\end{equation}
The Cartan-Killing metric is defined by
\begin{equation}
f^{ABC}\,f_{ABD}=-60\,\delta^C_D.
\end{equation}
The Jacobi identity is
\begin{equation}\label{E8Jacobi}
f^{[A}{}_{CE}\,f^{B]}{}_D{}^E = f^{AB}{}_E\,f_{CD}{}^E,
\end{equation}
which allows to decompose the product of two $f$'s into symmetric and antisymmetric part as
\begin{equation}
f^A{}_{CE}\,f^B{}_D{}^E = \tfrac{1}{2}\,f^{(A}{}_{CE}\,f^{B)}{}_D{}^E + \tfrac{1}{2}\,f^{AB}{}_E\,f_{CD}{}^E.
\end{equation}
The tensor product of two adjoint representations is \cite{Koepsell:1999uj}
\begin{equation} 
\textbf{248} \otimes \textbf{248} = 
\textbf{1} \oplus \textbf{3875} \oplus \textbf{27000} \oplus \textbf{248} \oplus \textbf{30380}.
\end{equation}
The first three representations form the symmetric part of the product, and the last two the antisymmetric part. The projectors on the first irreducible representations are 
\begin{equations}
(P^{(\mathbf{1})})^{MP}{}_{QN} &= \tfrac{1}{248}\,\eta^{MP}\,\eta_{QN},\\
(P^{(\mathbf{248})})^{MP}{}_{QN} &= -\tfrac{1}{60}\,f^{MPR}\,f_{QNR},\\
(P^{(\mathbf{3875})})^{MP}{}_{QN} &= \tfrac{1}{14}\,\delta^{(M}_Q\,\delta^{P)}_N - \tfrac{1}{56}\,\eta^{MP}\,\eta_{QN} - \tfrac{1}{28}\,f^{(M}{}_{QR}\,f^{P)}{}_N{}^R.
\end{equations}
The following relation holds, as one can check by using the Jacobi identity:
\begin{align}
\delta^M_Q\,\delta_N{}^P + \delta^M_N\,\delta^P_Q &+ 60\,{P^{(\textbf{248})}}^M{}_N{}^P{}_Q = \nn\\
& (14\,P^{(\textbf{3875})} + 62\,P^{(\textbf{1})} - 30\,P^{(\textbf{248})})^{MP}{}_{QN}.
\end{align}
The left-hand side is precisely the $n=8$ $Y$-tensor \eqref{E8Y}, which therefore admits the following equivalent expression:
\begin{equation}
Y^{MP}{}_{QN} = (14\,P^{(\mathbf{3875})} + 62\,P^{(\mathbf{1})} - 30\,P^{(\mathbf{248})})^{MP}{}_{QN}.
\end{equation}
This shows that the section constraint \eqref{SectionConstraintYSectionMatrix} is equivalent to 
\begin{equations}
(P^{(\mathbf{1})})^{MP}{}_{QN}\,\mathcal{E}_M{}^m\,\mathcal{E}_P{}^p &= 0,\\
(P^{(\mathbf{248})})^{MP}{}_{QN}\,\mathcal{E}_M{}^m\,\mathcal{E}_P{}^p &= 0,\\
(P^{(\mathbf{3875})})^{MP}{}_{QN}\,\mathcal{E}_M{}^m\,\mathcal{E}_P{}^p &= 0.
\end{equations}
Replacing the projectors, 
\begin{equations}
& \eta^{MP}\,\mathcal{E}_M{}^m\,\mathcal{E}_P{}^p = 0,\label{section_constraint1}\\
& f^{MPR}\mathcal{E}_M{}^m\,\mathcal{E}_P{}^p = 0,\label{section_constraint2}\\
& (\delta^{(M}_Q\,\delta^{P)}_N-\tfrac{1}{2}\,f^{(M}{}_{QR}\,f^{P)}{}_N{}^R)\,\mathcal{E}_M{}^m\,\mathcal{E}_P{}^p = 0.\label{section_constraint3}
\end{equations}
A useful property, which is a consequence of the section constraint, is the following:
\begin{equation}
f_{RAB}\,f^M{}_N{}^A\,f^P{}_Q{}^B\,\mathcal{E}_M{}^m\,\mathcal{E}_P{}^p = (\delta^M_Q\,f_N{}^P{}_R + \delta^P_N\,f^M{}_{QR} + \tfrac{1}{2}\,\delta^{(M}_R\,f_N{}^{P)}{}_Q)\,\mathcal{E}_M{}^m\,\mathcal{E}_P{}^p.
\end{equation}
To prove it, consider the symmetric part in $M,P$ in the left-hand side:
\begin{align}
f_{RAB}\,f^{(M|}{}_N{}^A\,f^{|P)}{}_Q{}^B\,\mathcal{E}_M{}^m\,\mathcal{E}_P{}^p &= 
\tfrac{1}{2}\,f_{RAB}\,f^{(M|}{}_{[N}{}^A\,f^{|P)}{}_{Q]}{}^B\,\mathcal{E}_M{}^m\,\mathcal{E}_P{}^p = \nn\\
& = (-\tfrac{1}{2}\,f_{RA}{}^{(M|}\,f^{A|P)B}\,f_{BNQ} \,+\nn\\
& - \tfrac{1}{2}\,f_{RA[N|}\,f^{BA(M|}\,f_{B|Q]}{}^{|P)})\,\mathcal{E}_M{}^m\,\mathcal{E}_P{}^p = \nn\\
& = -(\delta^{(M}_R\,\delta^{P)}_B\,f^B{}_{NQ} + \delta^{(M}_A\,\delta^{P)}_{[Q|}\,f_R{}^A{}_{|N]})\,\mathcal{E}_M{}^m\,\mathcal{E}_P{}^p = \nn\\
& = -(\delta^{(M}_R\,\delta^{P)}_B\,f^B{}_{NQ} + \delta^{(P}_{[Q|}\,f_R{}^{M)}{}_{|N]})\,\mathcal{E}_M{}^m\,\mathcal{E}_P{}^p = \nn\\
& = -(\delta^{(M}_R\,\delta^{P)}_B\,f^B{}_{NQ} + \delta^{(M}_{[N|}\,f^{P)}{}_{R|Q]})\,\mathcal{E}_M{}^m\,\mathcal{E}_P{}^p.
\end{align}
In the first step we use the fact that this product of three $f$'s is antisymmetric in $N,Q$; the second step follows by applying the Jacobi identity twice;\footnote{Indeed,
\begin{align}
f_{RAB}\,f^M{}_{[N}{}^A\,f^P{}_{Q]}{}^B &= 
-f_{RBA}\,f^M{}_{[N}{}^A\,f^P{}_{Q]}{}^B = \nn\\
& = f_R{}^M{}_A\,f_{[N|B}{}^A\,f^P{}_{|Q]}{}^B + f_{R[N|A}\,f_B{}^{MA}\,f^P{}_{|Q]}{}^B = \nn\\
& = -f_{RA}{}^M\,f_{[N}{}^{AB}\,f_{Q]}{}^P{}_B - f_{RA[N|}\,f^{BAM}\,f_{B|Q]}{}^P = \nn\\
& = -f_{RA}{}^M\,f_{NQB}\,f^{APB} - f_{RA[N|}\,f^{BAM}\,f_{B|Q]}{}^P = \nn\\
& = -f_{RA}{}^M\,f^{APB}\,f_{BNQ} - f_{RA[N|}\,f^{BAM}\,f_{B|Q]}{}^P.
\end{align}}
in the third step the section constraint \eqref{section_constraint3} is used. Now consider the antisymmetric part. It turns out to be symmetric in $N,Q$. Then one can use the previous identity by changing $M\leftrightarrow N$ and $P\leftrightarrow Q$. Finally, one uses the section constraint \eqref{section_constraint2}:
\begin{align}
f_{RAB}\,f^{[M|}{}_N{}^A\,f^{|P]}{}_Q{}^B\,\mathcal{E}_M{}^m\,\mathcal{E}_P{}^p &= \tfrac{1}{2}\,f_{RAB}\,f^{[M|}{}_{(N}{}^A\,f^{|P]}{}_{Q)}{}^B\,\mathcal{E}_M{}^m\,\mathcal{E}_P{}^p = \nn\\
& = \tfrac{1}{2}\,f_{RAB}\,f_{(N|}{}^{[M|A}\,f_{Q)}{}^{|P]B}\,\mathcal{E}_M{}^m\,\mathcal{E}_P{}^p = \nn\\
& = -\tfrac{1}{2}\,(2\,\eta_{R(N}\,f_{Q)}{}^{MP} + 2\,\delta^{[M}_{(N}\,f_{Q)R}{}^{P]})\mathcal{E}_M{}^m\,\mathcal{E}_P{}^p = \nn\\
& = -\delta^{[M}_{(N}\,f_{Q)R}{}^{P]}\mathcal{E}_M{}^m\,\mathcal{E}_P{}^p = \nn\\
& = \delta^{[M}_{(N}\,f^{P]}{}_{R|Q)}\mathcal{E}_M{}^m\,\mathcal{E}_P{}^p.
\end{align}
It remains to sum the two parts:
\begin{align}
f_{RAB}\,f^{M}{}_N{}^A\,f^{P}{}_Q{}^B\,\mathcal{E}_M{}^m\,\mathcal{E}_P{}^p &= \tfrac{1}{2}\,
(-\delta^{(M}_R\,\delta^{P)}_B\,f^B{}_{NQ} - \delta^{(M}_{[N|}\,f^{P)}{}_{R|Q]} + \delta^{[M}_{(N}\,f^{P]}{}_{R|Q)})\,\mathcal{E}_M{}^m\,\mathcal{E}_P{}^p = \nn\\
& = (\delta^M_Q\,f_N{}^P{}_R + \delta^P_N\,f^M{}_{QR} + \tfrac{1}{2}\,\delta^{(M}_R\,f_N{}^{P)}{}_Q)\,\mathcal{E}_M{}^m\,\mathcal{E}_P{}^p,
\end{align}
as we wanted to show.

\subsection{$E_{8(8)}$ Dorfman derivative}\label{E8Dorfman}

In order to define a new $n=8$ generalised Lie derivative which satisfies the Leibniz identity, the idea is to extend the action of the generalised Lie derivative on the ancillary parameters \cite{Hohm:2017wtr}. To this aim, we introduce the notion of \emph{double vector}. A double vector $\mathbb{X}$ is a pair of a $n=8$ generalised vector $\mathbb{X}^{(1)}$ and of an on-section generalised one-form $\mathbb{X}^{(2)}$,
\begin{equation}
\mathbb{X} = ({\mathbb{X}^{(1)}}^M,{\mathbb{X}^{(2)}}_M), \;\;\text{with}\;\;{\mathbb{X}^{(2)}}_M = \mathcal{E}_M{}^m\,x_m.
\end{equation}
Then, we define a new generalised Lie derivative $D_{\mathbb{X}}$, generated by a double vector $\mathbb{X}$. We will call this new derivative $E_{8(8)}$ \emph{Dorfman derivative}, to stress that it is required to satisfy the Leibniz identity, as the Dorfman derivative in double geometry \eqref{DoubleDorfman}:  
\begin{equation}\label{LeibnizIdE8Dorfman}
D_{\mathbb{X}}\,(D_{\mathbb{Y}}\,\cdot) = 
D_{D_{\mathbb{X}}\mathbb{Y}} + D_{\mathbb{Y}}\,D_{\mathbb{X}}, \;\;\text{or equivalently}\;\;
[D_{\mathbb{X}},D_{\mathbb{Y}}] = D_{D_{\mathbb{X}}\mathbb{Y}}.
\end{equation}
$E_{8(8)}$ Dorfman derivative generated by $\mathbb{X}$ and acting on $\mathbb{V}$ is double vector defined by
\begin{equation}\label{E8DorfDerDef}
D_{\mathbb{X}}\,\mathbb{V} = (L_{X,S}\,V^M, L_{X,S}\,U_M + V^N\,f_N{}^P{}_Q\,\de_M\,S^{(X)}_P), 
\end{equation}
where
\begin{equations}
\mathbb{X} &= (X^M,S_M) = (X^M,\mathcal{E}_M{}^m\,s_m), \\
\mathbb{V} &= (V^M,U_M) = (V^M,\mathcal{E}_M{}^m\,u_m), \\
S^{(X)}_N &= S_N + f_N{}^R{}_S\,\de_R\,X^S,\\
L_{X,S}\,V^M &= X^N\,\de_N\,V^M + (\lambda + 1)\,\de_N\,X^N\,V^M - f^{NM}{}_P\,S^{(X)}_{N}\,V^P,\\
L_{X,S}\,U_M &= X^N\,\de_N\,U_M + \lambda\,\de_N\,X^N\,U_M - f^N{}_M{}^P\,S^{(X)}_N\,U_P = \nn\\
& = X^N\,\de_N\,U_M + (\lambda + 1)\,\de_N\,X^N\,U_M + U_N\,\de_M\,X^M.
\end{equations}
Notice that the first component is the same as the generalised Lie derivative with ancillary parameter, introduced in Section \ref{E8gengeo}, so that the discussion on trivial parameters remains true. The alternative form of $L_{X,S}\,U_M$ relies on the fact that $U_M$ is on section. It shows that the generalised Lie derivative of an on-section parameter is the same as the usual Lie derivative. Indeed, the $Y$-tensor term drops thanks to the section constraint \eqref{SecConstr8_1}.

As the generalised Lie derivative, the $E_{8(8)}$ Dorfman derivative is not antisymmetric
\begin{equation}
D_{\mathbb{X}}\,\mathbb{Y} \neq - D_{\mathbb{Y}}\,\mathbb{X},\quad\forall\;\mathbb{X}, \mathbb{Y},
\end{equation}
and its symmetric part is trivial:
\begin{equation}\label{TrivSym8}
D_{D_{\mathbb{X}}\mathbb{Y} + D_{\mathbb{Y}}\mathbb{X}}\,\mathbb{V} = 0,\quad\forall\;\mathbb{V}.
\end{equation}
Moreover, a trivial parameter $(\hat{X},\hat{S})$ exists, such that \cite{Hohm:2017wtr}
\begin{equations}
\tfrac{1}{2}\,(L_X\, V - L_V\, X) &= \hat{X} + \tfrac{1}{2}\,(D_{\mathbb{X}}\,\mathbb{V} - D_{\mathbb{X}}\,\mathbb{V})^{(1)}, \\
L_X U-L_V S + S(X,V) &= \hat{S} + \tfrac{1}{2}\,(D_{\mathbb{X}}\,\mathbb{V} - D_{\mathbb{X}}\,\mathbb{V})^{(2)},
\end{equations}
where the left-hand sides are the parameters in \eqref{Leibniz8bis}. Therefore, using also the trivial parameter in \eqref{TrivSym8}, we can check the Leibniz identity \eqref{LeibnizIdE8Dorfman} on the first component of the $E_{8(8)}$ Dorfman derivative
\begin{align}
[D_{\mathbb{X}},D_{\mathbb{Y}}]^{(1)} &= [L_{X,S},L_{V,U}] =
L_{\frac{1}{2}(L_X V - L_V X),L_X U - L_V S + S(X,V)} = \nn\\
&= L_{\frac{1}{2}(D_{\mathbb{X}}\mathbb{Y}-D_{\mathbb{Y}}\mathbb{X})^{(1)},\frac{1}{2}(D_{\mathbb{X}}\mathbb{Y}-D_{\mathbb{Y}}\mathbb{X})^{(2)}} = L_{(D_{\mathbb{X}}\mathbb{Y})^{(1)},D_{(\mathbb{X}}\mathbb{Y})^{(2)}} = (D_{D_{\mathbb{X}}\mathbb{Y}})^{(1)}.
\end{align}
The Leibniz identity \eqref{LeibnizIdE8Dorfman} is also true on the second components, as one can check by explicit evaluation.\footnote{We perform this check using the package \texttt{xAct} in \texttt{Mathematica}.}

\subsection{$E_{8(8)}$ exceptional field theory}

A dual formulation of linearised gravity should involve $\tilde{h}_{\mu_1\dots\mu_{d-3},\alpha}$ in place of $h_{\mu\alpha}$,\footnote{Some details on the dual graviton can be found in Appendix \ref{DualGraviton}.} in such a way that the physical content is the same as in usual linearised gravity (massless Fierz-Pauli action). This is done in \cite{West:2001as}, where the starting point is the torsion formulation of Hilbert-Einstein action \eqref{HE}.

One may ask if a non-linear extension of such a theory is possible, on the same way as General Relativity is the non-linear extension of linearised gravity. The theory should be covariant and local. Nevertheless, it was shown in \cite{Bekaert:2002uh, Bekaert:2004dz} that some no-go theorems prevent it.

Alternatively, one may consider a theory in which, upon linearising, both the graviton and the dual graviton simultaneously appear, but accomplished with some gauge symmetry, which makes the dual graviton pure-gauge. This is done in \cite{West:2002jj}, through an equivalent reinterpretation of the action found in \cite{West:2001as}. Such a formulation is what is required by exceptional field theory in formulating eleven-dimensional supergravity in $E_{8(8)}$-covariant way before the compactification up to three-dimensions, as explained in \cite{Hohm:2018qhd}. 

Indeed, as studied at the end of Section \ref{11reduction}, scalar fields coming from the dual graviton are necessary to be taken into account in order to obtain the full number of scalar fields which describe the $E_{8(8)}/\text{Spin}(16)$ coset space of the three-dimensional reduction. This means that a formulation of eleven-dimensional supergravity involving the dual graviton is necessary if one wishes to make the $E_{8(8)}$ duality manifest before the reduction. The ancillary parameters, emerging as a necessary feature of $n=8$ generalised Lie derivative, can be now viewed as responsible for the gauge symmetry which makes the dual graviton contribution non-dynamical in the reduced theory. 

More precisely, in writing the action of $n=8$ exceptional field theory formulation of eleven-dimensional supergravity, the vectors $\hat{A}_\mu{}^M$ are accompanied by one-forms $\hat{B}_{\mu\,M}$, supposed to be on section. It is quite natural to introduce such an object, since $\hat{A}_\mu{}^M$ should be accompanied by a 
partner in order to form a double vector 
\begin{equation}
\mathbb{A}_\mu = (\hat{A}_\mu{}^M, \hat{B}_{\mu\,M}),
\end{equation}
as necessary in extending the formula \eqref{CovariantDerivativeGaugeTrans} with the simple Lie derivative replaced by the $n=8$ generalised Lie derivative, as discussed in general in the end of in Section \ref{ExFT}. Upon the section choice $\de_M = (\de_m,0,\dots,0)$, $\hat{B}_{\mu\,M}$ is reduced to eight external vectors $\hat{B}_{\mu m}$. They act as gauge fields, whose gauge parameter is given by the ancillary parameters in $L_{X,S}$, according to 
\begin{equation}\label{GaugeDualGravExFT}
\delta_S\,B_{\mu m} = \de_\mu\,S_m.
\end{equation}
The generalised vielbein parametrising the coset space $\hat{V}_M{}^A$ encodes the scalar fields coming from the reduction of the bosonic fields the eleven-dimensional supergravity and their duals. As seen in Section \ref{11reduction}, 
they are 36 scalars from the metric, 56 from the three-form, 28 from the dual six-form, and 8 from the dual graviton. The latter, given by the 
components $\tilde{h}_{m_1\dots m_8,n}$ can be replaced by eight dual scalars $\vphi_n$, by means of the eight-dimensional Levi-Civita symbol
\begin{equation}
\tilde{h}_{m_1\dots m_8,n} = \vepsilon_{m_1\dots m_8}\,\vphi_n.
\end{equation}
Now, the gauge transformation \eqref{GaugeDualGravExFT} acts on $\vphi_m$ as a shift symmetry
\begin{equation}
\delta\,\vphi_m = S_m,
\end{equation}
so that $\vphi_m$ can be gauge-fixed to zero, resulting that the dual graviton contribution, although necessary for the $E_{8(8)}$-duality, is not propagating.

\subsection{Introduction to $E_{8(8)}$ uplift problem}

In extending the uplift procedure in the $n=8$ case, we have to learn how to modify the twisting operation and the flux-deformed generalised Lie derivative in order to take into account the peculiar features of $E_{8(8)}$ generalised geometry, namely the presence of ancillary parameters. The highlights of $n=8$ uplift problem are summarised in the following, while all details are reported in the next Sections \cite{Inverso:2024xok}:
\begin{itemize}
\item[--] A frame is a double vector of the form
\begin{equation}\label{FrameE8}
\mathbb{E}_A = (E^M{}_A,h_{MA}(E,r)),
\end{equation}
whose first component $E^M{}_A$ is an element in $E_{8(8)}\times \mathbb{R}^+$, parametrised as
\begin{equation}
E^M{}_A = r^{-1}\,U^M{}_A,\;\; U \in E_{8(8)},\;\; r^{-1}\in \mathbb{R}^+.
\end{equation} 
\item[--] The generalised Lie derivative with ancillary parameters of the first component of the frame defines the intrinsic torsion, as in \eqref{ExprTorsionLieDer2} in the simplest case, \eqref{ParallDFT} in double geometry, and \eqref{TorsionExFT} in exceptional generalised geometry with $n\leqslant 7$:
\begin{equation}\label{E8TorsionImplicit}
L_{E_A,h_A(E,r)}\,E^M{}_B = -T_A{}^C{}_B\,(E,r)\,E^M{}_C.
\end{equation}
Explicitly, the torsion is the same as in the $n\leqslant 7$ case up to an additional term depending on the second component $h_{MA}(E,r)$ of the frame (see Section \ref{E8Torsion} for details). 
\item[--] The second component of the frame is defined by
\begin{equation}
h_{MA}(E,r) = \tfrac{1}{60}\,r^{-1}\,f_{AD}{}^C\,E_N{}^D\,\de_M\,E^N{}_C.
\end{equation}
\item[--] This definition ensures that 
\begin{equation}
D_{\mathbb{E}_A}\,\mathbb{E}_B = -T_A{}^C{}_B\,\mathbb{E}_C - (0,\tfrac{1}{60}\,r^{-1}\,f_{BD}{}^C\,\de_M\,T_A{}^D{}_C),
\end{equation}
where the first term is the right-hand side is fixed by \eqref{E8TorsionImplicit}, and the second term is a double vector whose first component is zero, and whose second component is vanishing \emph{when the torsion is constant}. 
\item[--] This means that $n=8$ generalised parallelisability condition formally assumes the same form as in \eqref{ParallEhat}, with the generalised Lie derivative replaced by the $E_{8(8)}$ Dorfman derivative:
\begin{equation}\label{E8ParallCond}
D_{\hat{\mathbb{E}}_A}\,\hat{\mathbb{E}}_B = -X_{AB}{}^C\,\hat{\mathbb{E}}_C,
\end{equation}
where $\hat{\mathbb{E}}_A$ is a suitable frame of the form in \eqref{FrameE8}, such that the corresponding intrinsic torsion $T_A{}^C{}_B(\hat{\mathbb{E}})$ is constant and equal to the components $X_{AB}{}^C$ of the embedding tensor of the gauged supergravity theory to be uplifted.
\item[--] We can define the \emph{twist of a double vector}, by requiring the first component of a twisted double vector to be the analogue of the $n\leqslant 7$ case. A twist matrix is a double vector $\mathbb{C}_N $ of the form  
\begin{equation}
\mathbb{C}_N = (C^M{}_N,h_{MN}(C,\vrho)),
\end{equation}
The twisted frame $\hat{\mathbb{E}}_A$ is 
\begin{equation}
\hat{\mathbb{E}}_A = (C^M{}_N\,E^N{}_A, \vrho^{-1}\,h_{MA}(E,r)+ h_{MN}(C,\vrho)\,E^M{}_A),
\end{equation}
In analogy with the $n\leqslant 7$, we will also formally write
\begin{equation}
\hat{\mathbb{E}}^M{}_A = \mathbb{C}^M{}_N\,\mathbb{E}^N{}_A.
\end{equation}
\item[--] We can compute the twisted $E_{8(8)}$ Dorfman derivative and use it to define the \emph{flux-deformed $E_{8(8)}$ Dorfman derivative}:
\begin{equations}
D_{\mathbb{X}}^{(F)}\,\mathbb{V} &= D_{\mathbb{X}}\,\mathbb{V} - (F_{PQ}{}^M, \tfrac{1}{60}\,f_{QS}{}^R\,\de_M\,F_{PR}{}^S)\,X^P\,V^Q \,+\nn\\
& - (0, \tfrac{1}{60}\,F_{PR}{}^S\,f_{QS}{}^R\,\de_M\,(X^P\,V^Q) + \tfrac{1}{248}\,F_{PQ}{}^Q\,X^P\,U_M),
\end{equations}
where $\mathbb{X} = (X^M,S_M)$, and $\mathbb{V} = (V^M,U_M)$.
\item[--] Remarkably, the flux-deform $E_{8(8)}$ Dorfman derivative  satisfies the Leibniz identity if and only if the flux satisfies the \emph{same} constraints as in the case $n\leqslant 7$.
\item[--] If the flux $F$ is the intrinsic torsion of some twist $C$ (\emph{integrability}), then
\begin{equation}
D_{\hat{\mathbb{E}}_A}\,\hat{\mathbb{E}}^M{}_B = \mathbb{C}^M{}_N\,D^{(F)}_{\mathbb{E}_A}\,\mathbb{E}^N{}_B.
\end{equation}
\item[--] Starting from the expression of the twisted Dorfman derivative, we write the Bianchi identity satisfied by the (first component of) the intrinsic torsion of a frame, which does not necessarily preserve the section choice. 
\item[--] The second component of the flux-deformed $E_{8(8)}$ Dorfman derivative satisfies the Leibniz identity, using the three known flux constraints, without imposing any further constraint.
\end{itemize}

\subsection{$E_{8(8)}$ twisting}

Consider a double vector $\mathbb{X} = (X^M,S_M)$, and a $E^M{}_A$ in $E_{8(8)}\times \mathbb{R}^+$, taking the factorised form
\begin{equation}
E^M{}_A = U^M{}_A\,r^{-1}, \quad \text{with}\; U^M{}_A \in E_{8(8)}, \; r^{-1} \in \mathbb{R}^+.
\end{equation}
Denoting with $U_M{}^A$ and $E_M{}^A$ the components of the inverse of $U^M{}_A$ and $E^M{}_A$ respectively, we have the following relations
\begin{equations}
& U^M{}_A = E^M{}_A\,r, \\
& U_M{}^A = E_M{}^A\,r^{-1}, \quad
E_M{}^A = U_M{}^A\,r,\\
& E^M{}_A\,E_M{}^B = \delta^A_B, \quad
E^M{}_A\,E_N{}^A = \delta^M_N, \\
& U^M{}_A\,U_M{}^B = \delta^A_B, \quad
U^M{}_A\,U_N{}^A = \delta^M_N, \\
& \de_M\,r^{-1} = - r^{-2}\,\de_M\,r.
\end{equations}
The the power of the $r^{-1}$ factor defines an $\mathbb{R}^+$ weight for any tensor $T$, which we denote with $[T]$. It is the same as the density weight in the generalised Lie derivative: if
\begin{equation}
L_{X,S}\,V^M = \dots + (\lambda + 1)\,\de_N\,X^N,\quad
L_{X,S}\,U_M = \dots + \lambda\,\de_N\,X^N\,U_M,
\end{equation}
then
\begin{equation}
[V^M] = \lambda + 1, \quad
[U_M] = \lambda = [V^M] -1.
\end{equation}
If the generalised Lie derivative does not change the weight $[\mathcal{L}_\xi\,V]=[V]$, then, $[\de_M]=-1$; if also $[\delta_{MN}]=[f_{MNP}]=0$, then $[\sigma^{(\xi)}]=[\sigma]=[\de\,\xi]=[\xi]-1$. In general, for any double vector $\mathbb{X}$, the weight of the second component is the same as the weight of the first one, lowered by one:
\begin{equation}
[\mathbb{X}^{(2)}] = [\mathbb{X}^{(1)}] - 1.
\end{equation}
Taking into account the weights and starting from the double vector $\mathbb{X}$, we define the twisted double vector $\mathbb{X}'$ by $(U,r)$ as
\begin{align}
\mathbb{X}' = (X'^M,S'_M) &= (r^{-1}\,U^M{}_A\,X^A,U_M{}^A\,(S_A + h_{AB}\,X^B)) = \nn\\
& = (E^M{}_A\,X^A, E_M{}^A\,r^{-1}\,(S_A + h_{AB}\,X^B)),\label{E8Twist}
\end{align}
for some $h_{AB}$, which at this stage is arbitrary. It will be useful to define 
\begin{equation}
h'_{MA} = U_M{}^B\,h_{AB},
\end{equation}
which is zero-weighted and it is assumed to be on section on the first component 
\begin{equation}
[h'_{MA}]=0, \quad 
h'_{MA} = \mathcal{E}_M{}^m\,\tilde{h}_{mA},
\end{equation}
for some $\tilde{h}_{mA}$. In this way 
\begin{equation}
(E^M{}_A, h'_{MA}) = \mathbb{E}_A,
\end{equation}
is a well-defined double vector.  We will formally write
\begin{equation}
\mathbb{X}'^M = \mathbb{E}^M{}_A\,\mathbb{X}^A.
\end{equation}
Notice that $E_M{}^A\,\sigma_A$ is not on section in general, but it does if the twisting preserves the section: in such a case $E_M{}^A\,\sigma_A = \sigma_M$. Finally, observe that the structure constants should be invariant under the $U$ action, because they are zero-weighted:
\begin{equation}
f_M{}^{BC}\,U^M{}_A = f_A{}^{BC}, \quad
f^{MBC}\,U_M{}^A = f^{ABC}.
\end{equation}
Equivalently,
\begin{equation}
f_M{}^{BC}\,E^M{}_A = f_A{}^{BC}\,r^{-1},\quad
f^{MBC}\,E_M{}^A = f^{ABC}\,r.
\end{equation}

\subsection{$E_{8(8)}$ intrinsic torsion}\label{E8Torsion}

The $n=8$ intrinsic torsion is the double vector defined by
\begin{equation}\label{n8IntrinsicTorsion}
D_{\mathbb{E}_A}\,\mathbb{E}_B = -(T_A{}^C{}_B\,E^M{}_A, t_{MAB}).
\end{equation}
Using the expression for the first component of $E_{8(8)}$ Dorfman derivative \eqref{E8DorfDerDef},
\begin{equation}
- T_A{}^C{}_B\,E_C{}^M = L_{E_A,h'_A}\,E^M{}_B.
\end{equation}
Setting $h'_{MA} \rightarrow 0$ in the first component, the definition of $n \leqslant 7$ generalised intrinsic torsion in \eqref{TorsionExFT} is recovered. The intrinsic torsion and the Weitzenb\"ock connection can be decomposed in the following components:
\begin{equations}
& T_A{}^C{}_B = \vtheta_A\,\delta_B^C + \tfrac{1}{2}\,f^{DC}{}_B\,f_{DA}{}^E\,\vtheta_E - f^{DC}{}_B\,\vtheta_{AD},\label{E8Tcomponents}\\
& W_A{}^C{}_B = W_{AD}\,f^D{}_B{}^C + \tfrac{1}{2}\,W_A\,\delta_B{}^C.\label{E8Wcomponents}
\end{equations}
Since
\begin{equation}
f^{ABC}\,f_{ABD} = - 60\,\delta^C_D, \quad
\delta^A_A = 248,
\end{equation}
the components can be extracted 
\begin{equations}
& W_{AB} = - \tfrac{1}{60}\,f_B{}^C{}_D\,W_A{}^D{}_C,\\
& W_A = \tfrac{1}{124}\,W_A{}^B{}_B,\\
& \vtheta_{AB} = \tfrac{1}{60}\,T_A{}^D{}_C\,f_B{}^C{}_D - \tfrac{1}{2}\,f_A{}^C{}_D\,\vtheta_C,\\
& \vtheta_A = \tfrac{1}{248}\,T_A{}^B{}_B.
\end{equations}
Since the torsion can be written in terms of the Weitzenb\"ock connection, we can write the components of the torsion in terms of those of the connection. A tedious computation shows that 
\begin{align}
-E_M{}^C\,L_{E_A}\,E^M{}_B &= W_{[A}{}^C{}_{B]} + Y^{CE}{}_{FB}\,W_E{}^F{}_A = \nn\\
& = (W_A + f_A{}^{ED}\,W_{ED})\,\delta_B^C \,+\nn\\
& + \tfrac{1}{2}\,f^{HC}{}_B\,f_{HA}{}^F\,(W_F + f_F{}^{ED}\,W_{ED}) \,+\nn\\
& - f^{HC}{}_B\,(W_{AH} - \tfrac{1}{2}\,f_{(H}{}^{EF}\,f_{A)}{}^D{}_F\,W_{ED}).
\end{align}
In the first step, consider the $n\leqslant 7$ intrinsic torsion in \eqref{TorsionExFT}; in the second step, replace the $Y$-tensor \eqref{E8Y} and expand the Weitzenb\"ock connection \eqref{E8Wcomponents}; at the very end, the Jacobi identity \eqref{E8Jacobi} has to be employed. It follows that
\begin{align}
T_A{}^C{}_B &= -E_M{}^C\,L_{E_A,h'_A}\,E^M{}_B = 
-E_M{}^C\,L_{E_A}\,E^m{}_B - E_M{}^C\,\delta_{h'_A}\,E^M{}_B = \nn\\
& = [(W_A + f_A{}^{ED}\,W_{ED})\,\delta_B^C + \tfrac{1}{2}\,f^{HC}{}_B\,f_{HA}{}^F\,(W_F + f_F{}^{ED}\,W_{ED}) \,+\nn\\
& - f^{HC}{}_B\,(W_{AH} - \tfrac{1}{2}\,f_{(H}{}^{EF}\,f_{A)}{}^D{}_F\,W_{ED})] + f^{HC}{}_B\,h_{HA}.
\end{align}
Finally, we obtain an expression for the intrinsic torsion 
\begin{align}
T_A{}^C{}_B &= (W_A + f_A{}^{ED}\,W_{ED})\,\delta_B^C + \tfrac{1}{2}\,f^{HC}{}_B\,f_{HA}{}^F\,(W_F + f_F{}^{ED}\,W_{ED}) \,+\nn\\
& -f^{HC}{}_B\,(W_{AH}-h_{HA}-\tfrac{1}{2}\,f_{(H}{}^{EF}\,f_{A)}{}^D{}_F\,W_{ED}), 
\end{align}
in which, comparing with \eqref{E8Tcomponents}, we recognise
\begin{equations}
& \vtheta_A = W_A + f_A{}^{ED}\,W_{ED},\\
& \vtheta_{AB} = W_{AB} - h_{BA} - \tfrac{1}{2}\,f_{(H}{}^{EF}\,f_{A)}{}^D{}_F\,W_{ED}.
\end{equations}
We want the intrinsic torsion to sit in the same representation as the embedding tensor $\mathbf{1} \oplus \mathbf{248} \oplus \mathbf{3875}$, where $\mathbf{248}$ is the contribution due to the trombone component. $\vtheta_A$ corresponds to the $\mathbf{248}$. In order for $\vtheta_{AB}$ to sit in $\mathbf{1} \oplus \mathbf{3875}$, it is sufficient it to be symmetric. Indeed, the projector on $\mathbf{1}$ and on the $\mathbf{3875}$ are  
\begin{equations}
(P^{(\mathbf{1})})^A{}_B{}^C{}_D &= \tfrac{1}{248}\,\delta_B{}^A\,\delta_D{}^C,\\
(P^{(\mathbf{3875})})^A{}_B{}^C{}_D &= -\tfrac{1}{56}\,\delta_B{}^A\,\delta_D{}^C + \tfrac{1}{14}\,(\delta^C_B\,\delta^A{}_D + \eta^{AC}\,\eta_{BD}) \,+\nn\\
& - \tfrac{1}{28}\,(f^A{}_{DF}\,f_B{}^{CF} + f^{AC}{}_F\,f_{BD}{}^F).
\end{equations}
Therefore,
\begin{align}
(x\,P^{(\mathbf{1})} + y\,P^{(\mathbf{3875})})_{AB}{}^{CD}\,W_{CD} &= (\tfrac{x}{248}-\tfrac{y}{56})\,\eta_{AB}\,W_C{}^C \,+\nn\\
& - \tfrac{y}{14}\,(W_{(AB)} - \tfrac{1}{2}\,f_{(A}{}^{EC}\,f_{B)}{}^F{}_C\,W_{EF}),
\end{align}
for any constant $x,y$. Choosing $x=62$ and $y=14$, the first coefficient vanishes and the second is one, so that
\begin{equation}
W_{(AB)} - \tfrac{1}{2}\,f_{(A}{}^{EC}\,f_{B)}{}^F{}_C\,W_{EF} = (62\,P^{(\mathbf{1})} + 14\,P^{(\mathbf{3875})})_{AB}{}^{CD}\,W_{CD}.
\end{equation}
To reproduce the left-hand side in $\vtheta_{AB}$ we set
\begin{equation}
h_{BA} = W_{BA} + \tilde{h}_{BA},
\end{equation}
where $\tilde{h}_{BA}$ is on section in its first index. It has to sit in $\mathbf{1} \oplus \mathbf{3875}$, so it has to be symmetric. Then, it is on section in the second index too. But the section constraint imposes the tensor product of the index in the $\mathbf{248}$ not to contain the components in $\mathbf{1} \oplus \mathbf{248} \oplus \mathbf{3875}$. So, $\tilde{h}_{BA} = 0$. Finally, the choice
\begin{equation}
h_{BA} = U^M{}_B\,h'_{MA} = E^M{}_B\,r\,h'_{MA} = - W_{BA}
\end{equation}
ensures the intrinsic torsion to stay in $\mathbf{1} \oplus \mathbf{248} \oplus \mathbf{3875}$. Summarising, we have 
\begin{equations}
\vtheta_A &= W_A + f_A{}^{BC}\,W_{BC},\\
\vtheta_{AB} &= W_{(AB)} - \tfrac{1}{2}\,f_{(A}{}^{CE}\,f_{B)}{}^D{}_E\,W_{CD},\\
h_{BA} &= \tfrac{1}{60}\,f_{AD}{}^C\,E_N{}^D\,E^M{}_B\,\de_M\,E^N{}_C,\\
h'_{MA} &= \tfrac{1}{60}\,r^{-1}\,f_{AD}{}^C\,E_N{}^D\,\de_M\,E^N{}_C.
\end{equations}

Let us conclude with some useful relations involving the components of the Weitzenb\"ock connection. Taking the trace of \eqref{E8Wcomponents},
\begin{equation}
W_A{}^A{}_B = \tfrac{1}{2}\,W_A + f_A{}^{BC}\,W_{BC}.
\end{equation}
If the twisting preserves the choice of the section, $W_A{}^C{}_B\,x_C = 0$, where $x$ is on section. Expanding in the components,
\begin{equation}
(W_{AD}\,f^D{}_B{}^C + \tfrac{1}{2}\,W_A\,\delta^C{}_B)\,x_C = 0 \Rightarrow W_{AD}\,f^{D}{}_B{}^C\,x_C = -\tfrac{1}{2}\,W_A\,x_B.
\end{equation}

Observe that
\begin{align}
W_A{}^C{}_B &= -E_N{}^C\,E^M{}_A\,\de_M\,E^N{}_B = \nn\\
& = -U_N{}^C\,r\,(E^M{}_A\,\de_M)\,(U^N{}_B\,r^{-1}) = \nn\\
& = U_N{}^C\,(E^M{}_A\,\de_M)\,U^N{}_B - \delta^C_B\,r\,(E^M{}_A\,\de_M)\,r^{-1}.
\end{align}
Therefore,
\begin{equations}
\tfrac{1}{2}\,W_A &= -r\,E^M{}_A\,\de_M\,r^{-1} = E^M{}_A\,r^{-1}\,\de_M\,r,\\
W_{AD}\,f^D{}_B{}^C &= -U_N{}^C\,E^M{}_A\,\de_M\,U^N{}_B.
\end{equations}

Given the definition of the Weitzenb\"ock connection in terms of the derivatives of the frame \eqref{GenWconnection}, one can show that the following is an identity:
\begin{equation}
E^N{}_{[A}\,\de_N\,W_{B]}{}^C{}_F + W_{[A|}{}^E{}_F\,W_{B]}{}^C{}_E - W_{[A}{}^E{}_{B]}\,W_E{}^C{}_F = 0.
\end{equation}
In matrix notation, with $(W_A)_B{}^C = W_A{}^C{}_B$,
\begin{equation}
[W_A,W_B]-W_{[A}{}^C{}_{B]}\,W_C + E_{[A}{}^N\,\de_N\,W_{B]} = 0.
\end{equation}
Writing $W_A{}^C{}_B = E_A{}^M\,W'_M{}^C{}_B$, and $(W'_A)_B{}^C = W'_A{}^C{}_B$, it assumes a simpler form
\begin{equation}\label{FlatnessW}
(\de_{[M}\,W'_{N} + [W'_M,W'_N])_B{}^A = 0,
\end{equation}
showing that the Weitzenb\"ock connection is flat. The components satisfy the following conditions:
\begin{equation}
\de_{[M}\,W'_{N]} = 0, \quad
\de_{[M}\,W'_{N]E} = f^{AB}{}_C\,W'_{MA}\, W'_{NB}.\label{tmp}
\end{equation}
To get the second relation \eqref{tmp}, one has to separate the symmetric and the antisymmetric part of \eqref{FlatnessW} in the matrix indices, the symmetric part being trivial, and to use the Jacobi identity. Finally, multiplying \eqref{tmp} by $E_A{}^M$, one obtains
\begin{equation}\label{IntegrabilityRelation}
E^M{}_A\,\de_M\,W'_{NB} = \de_N\,(E^M{}_A\,W'_{MB}) - \de_N\,E^M{}_A\,W'_{MB} + E^M{}_A\,f^{CD}{}_B\,W'_{MC}\,W'_{ND}.
\end{equation}

\subsection{Second component of the intrinsic torsion}

The second component of the intrinsic torsion in \eqref{n8IntrinsicTorsion} can be shown to be equal to
\begin{equation}
t_{MAB} = T_A{}^C{}_B\,h_{MC} + \tfrac{1}{60}\,r^{-1}\,\de_M\,T_A{}^D{}_C\,f_{BD}{}^C,
\end{equation}
so that the whole intrinsic torsion reads
\begin{equation}
D_{\mathbb{E}_A}\,\mathbb{E}_B = - T_A{}^C{}_B\,\mathbb{E}_C - (0, \tfrac{1}{60}\,r^{-1}\,\de_M\,T_A{}^D{}_C\,f_{BD}{}^C).
\end{equation}
The computation is straightforward but quite cumbersome. In the following the main steps in finding the result are summarised. $W^{(3)}, W^{(2)}, W^{(1)}$ denote here the Weitzenb\"ock connection and its components with two or one indices in \eqref{E8Wcomponents}.
\begin{itemize}
\item[--] Write $t_{MAB} = (D_{\mathbb{E}_A}\,\mathbb{E}_B)_M$ in terms of $W^{(3)}$;
\item[--] Factor out a coefficient $r^{-1}$;
\item[--] Write $W^{(3)}$ in terms of its components $W^{(2)}$ and $W^{(1)}$;
\item[--] Replace $f\,r^{-1}$ with $f\,E$ or $f\,E^{-1}$;
\item[--] Use the identity \eqref{IntegrabilityRelation};
\item[--] Use the Leibniz identity for the simple derivative;
\item[--] Replace $\de\,E$ with $W^{(3)}$, and then with $W^{(2)}$ and $W^{(1)}$;
\item[--] Use the Jacobi identity;
\item[--] Subtract $-r^{-1}T_A{}^C{}_B\,W'_{MC}$;
\item[--] Extract $\de_M\,(\vtheta_{AB}+\tfrac{1}{2}\,f_{AB}{}^C\,\vtheta_C)= \tfrac{1}{60}\,\de_M\,T_A{}^D{}_C\,f_{BD}{}^C$;
\item[--] Use the Jacobi identity to show that the remaining terms vanish;
\item[--] At the very end, one remains with
\begin{equation}
r\,[t_{MAB} - (-r^{-1}T_A{}^C{}_B\,W'_{MC})] = \tfrac{1}{60}\,\de_M\,T_A{}^D{}_C\,f_{BD}{}^C,
\end{equation}
where $-r^{-1}\,W'_{MC} = h'_{MC}$, as we wanted to show.
\end{itemize}

\subsection{Twisted $E_{8(8)}$ Dorfman derivative}

Our aim is to perform the following two computations:
\begin{itemize}
\item[1.] Consider the first component of the twisted $E_{8(8)}$ Dorfman derivative, without assuming the twisting to preserve the section, and compute the integrability condition satisfied by the intrinsic torsion. In order to do this, it is sufficient to consider the Leibniz identity for the $E_{8(8)}$ Dorfman derivative, in the case in which the first components of the double vectors are constant and the second components vanish. In order to extract the requested identity, one has to write the twisted derivative in terms of the intrinsic torsion.
\item[2.] Consider the twisted derivative in terms of the intrinsic torsion. Replace the intrinsic torsion with an arbitrary flux, assuming the twisting to preserve the section choice. Then, compute which constraints have to be satisfied by the flux in order for the flux-deformed $E_{8(8)}$ Dorfman derivative to fulfill the Leibniz identity. 
\end{itemize}

Recall that the twisting of a double vector is (see \eqref{E8Twist}):
\begin{equation}
\mathbb{X}' = (X'^M,S'_M) = (E^M{}_A\,X^A,r^{-1}\,E_M{}^A\,(S_A-W_{AB}\,S^B)),
\end{equation}
where we replace $h$ with $W^{(2)}$. Consider $D_{\mathbb{X}'}\,\mathbb{V}'$, with $\mathbb{X}'=(X',S')$ and $\mathbb{V}'=(V',U')$ and extract the intrinsic torsion as far as possible. The terms in $D_{\mathbb{X}'}\,\mathbb{V}'$ are of the form
\begin{equation}
X\,\de\,V, \quad V\,\de\,X, \quad V\,S, \quad V\,X.
\end{equation}
The result is
\begin{align}
(D_{\mathbb{X}'}\,\mathbb{V}')^{M} &= E^M{}_C\,[X^A\,(E^N{}_A\,\de_N)\,V^C + V^C\,(E^N{}_A\,\de_N)\,X^A \,+\nn\\
& - f^{AC}{}_B\,V^B\,(S_A + f_A{}^D{}_E\,(E^N{}_D\,\de_N)\,X^E) - T_A{}^C{}_B\,X^A\,V^B] = \nn\\
& = E^M{}_C\,[(D_{\mathbb{X}}\,\mathbb{V})'^{C} - T_A{}^C{}_B\,X^A\,V^B],
\end{align}
where in $(D_{\mathbb{X}}\,\mathbb{V})'$ means that the derivative are formally replaced by $E^M{}_A\,\de_M$. Let us proceed similarly for the second component. The possible terms are
\begin{equation}
X\,\de\,U, \quad U\,X, \quad U\,\de\,X,
\end{equation}
\begin{equation}
V\,\de\,S, \quad V\,\de\,S, \quad V\,S, \quad V\,\de\,\de\,X, \quad X\,\de\,V, \quad V\,\de\,X, \quad V\,X.
\end{equation}
The result is quite cumbersome:
\begin{align}
(D_{\mathbb{X}'}\,\mathbb{V}')_M &=
E_M{}^E\,r^{-1}\,[\textcolor{blue}{X^B\,(E^N{}_B\,\de_N)\,U_E + U_E\,(E^N{}_B\,\de_N)\,X^B + U_B\,(E^N{}_E\,\de_N)\,X^B \,+}\nn\\
& \textcolor{blue}{+ V^A\,(E^N{}_E\,\de_N)\,S_A + f_{BD}{}^A\,(E^N{}_E\,E^P{}_A\,\de_N\,\de_P)\,\xi^D\,V^B} \,+\nn\\
& -(E^N{}_E\,\de_N)\,(\tfrac{1}{60}\,T_B{}^F{}_G\,f_{CF}{}^G)\,X^B\,V^C - (\tfrac{1}{60}\,T_B{}^F{}_G\,f_{CF}{}^G)\,(E^N{}_E\,\de_N)\,X^B\,V^C] \,+\nn\\
& + E_M{}^E\,r^{-1}\,[\textcolor{teal}{\tfrac{1}{2}\,V^D\,f^F{}_{DB}\,(\tfrac{1}{2}\,\delta^A_F\,W_E - f_F{}^{AC}\,W_{EC})\,(E^N{}_A\,\de_N)\,X^B \,+}\nn\\
& \textcolor{teal}{+\, [f^A{}_{[B}{}^C\,W_{E]C} - \delta_E^A\,f_B{}^{CF}\,W_{CF} - \tfrac{1}{2}\,\delta_{(B}^A\,W_{E)}]\,U_A\,X^B}] \,+\nn\\
& - E_M{}^E\,r^{-1}\,W_{ED}\,[\textcolor{red}{(E^N{}_A\,\de_N)\,X^A\,V^D + X^D\,(E^N{}_B\,\de_N)\,V^D \,+}\nn\\
& \textcolor{red}{- f^{FD}{}_C\,f_F{}^A{}_B\,V^C\,(E^N{}_A\,\de_N)\,X^B- f^{BD}{}_A\,V^A\,S_B} - T_A{}^D{}_B\,X^A\,V^B],
\end{align}
The \textcolor{blue}{blue} part corresponds to $\textcolor{blue}{(D_{\mathbb{X}}\,\mathbb{V})'_E}$. Similarly, the \textcolor{red}{red} part is $\textcolor{red}{(D_{\mathbb{X}}\,\mathbb{V})'^{D}}$. Finally call the \textcolor{teal}{teal} part $\textcolor{teal}{Z_E}$:
\begin{align}
(D_{\mathbb{X}'}\,\mathbb{V}')_M &= 
(E_M{}^E\,r^{-1}\,[\textcolor{blue}{(D_{\mathbb{X}}\,\mathbb{V})'_E} - (E^N{}_E\,\de_N)\,(\tfrac{1}{60}\,T_B{}^F{}_G\,f_{CF}{}^G)\,X^B\,V^C \,+\nn\\
& - (\tfrac{1}{60}\,T_B{}^F{}_G\,f_{CF}{}^G)\,(E^N{}_C\,\de_N)\,X^B\,V^C] + (E_M{}^E\,r^{-1}\,\textcolor{teal}{Z_E} \,+\nn\\
& - E_M{}^E\,r^{-1}\,W_{ED}\,[\textcolor{red}{(D_{\mathbb{X}}\,\mathbb{V})'^{D}}- T_A{}^D{}_B\,X^A\,V^B].
\end{align}
If the twisting preserves the choice of the section, then the term $Z_E$ reduces to
\begin{equation}
\textcolor{teal}{Z_E} \rightarrow \textcolor{teal}{- U_E\,\tfrac{1}{248}\,T_B{}^C{}_C\,X^B}.
\end{equation}
In obtaining this result, one has to use the following relations, which follow because of the twist to preserve the section:
\begin{equation}
U_A\,f^C{}_E{}^A\,W_{BC} = -\tfrac{1}{2}\,U_E\,W_B,\quad
f_A{}^{BC}\,W_{BC} = -\tfrac{1}{2}\,W_A,\quad
\vtheta_A = \tfrac{1}{2}\,W_A.
\end{equation}
So, the second component of the dressed Dorfman derivative is
\begin{align}
(D_{\mathbb{X}'}\,\mathbb{V}')_M &=
E_M{}^E\,r^{-1}\,\textcolor{blue}{(D_{\mathbb{X}}\,\mathbb{V})'_E} \,+\nn\\
& - E_M{}^E\,r^{-1}\,W_{ED}\,[\textcolor{red}{(D_{\mathbb{X}}\,\mathbb{V})'^{D}}- T_A{}^D{}_B\,X^A\,V^B] \,+\nn\\
& -(E_M{}^E\,r^{-1}\,[
(E^N{}_E\,\de_N)\,(\tfrac{1}{60}\,T_B{}^F{}_G\,f_{CF}{}^G)\,X^B\,V^C +\,\nn\\
& + (\tfrac{1}{60}\,T_B{}^F{}_G\,f_{CF}{}^G)\,(E^N{}_C\,\de_N)\,X^B\,V^C]\,+\nn\\
& + E_M{}^E\,r^{-1}\,\textcolor{teal}{- U_E\,\tfrac{1}{248}\,T_B{}^C{}_C\,X^B}.
\end{align}
From the previous computation we read the definition of the flux-deformed $E_{8(8)}$ Dorfman derivative, in the same way as the flux-deformed generalised Lie derivarive is defined out of the dressed derivative in the $n\leqslant 7$ case. The first component of the flux-deformed $E_{8(8)}$ Dorfman derivative is in agreement with the $n\leqslant 7$ expression; the second component is a new result. 

The ingredients for the two computations we set out to do are:
\begin{itemize}
\item[1.] Without assuming the twisting to preserve the section choice, we consider the deformation of $E_{8(8)}$ Dorfman derivative given by the intrinsic torsion $\tilde{D}^{(T)}$ 
 \begin{align}
(D_{\mathbb{X}}^{(T)}\,\mathbb{V})^{E} & =
X^A\,(E^N{}_A\,\de_N)\,V^E + V^E\,(E^N{}_A\,\de_N)\,X^A \,+\nn\\
& - f^{AE}{}_B\,V^B\,S_A + f_A{}^D{}_E\,(E^N{}_D\,\de_N)\,X^E - T_A{}^E{}_B\,X^A\,V^B = \nn\\
& = (D_{\mathbb{X}}\,\mathbb{V})'^E-T_A{}^E{}_B\,X^A\,V^B,\\
(D_{\mathbb{X}}^{(T)}\,\mathbb{V})_{E}  &= (D_{\mathbb{X}}\,\mathbb{V})'_{E} - (E^N{}_E\,\de_N)\,(\tfrac{1}{60}\,T_B{}^F{}_G\,f_{CF}{}^G)\,X^B\,V^C \,+\nn\\
& - (\tfrac{1}{60}\,T_B{}^F{}_G\,f_{CF}{}^G)\,(E^N{}_E\,\de_N)\,X^B\,V^C+ Z_E.
\end{align}
\item[2.] Assuming the twisting to preserve the section choice and replacing the intrinsic torsion with an arbitrary flux $F$, we get the \emph{flux-deformed $E_{8(8)}$ Dorfman derivative} $D^{(F)}$ is
\begin{align}
(D_{\mathbb{X}}^{(F)}\,\mathbb{V})^M & = (D_{\mathbb{X}}\,\mathbb{V})^M - F_{PQ}{}^M\,\xi^P\,V^Q,\\
(D_{\mathbb{X}}^{(F)}\,\mathbb{V})_M & =
 (D_{\mathbb{X}}\,\mathbb{V})_M - \de_M\,(\tfrac{1}{60}\,T_P{}^S{}_R\,f_{QS}{}^R)\,X^P\,V^Q \,+\nn\\
& - (\tfrac{1}{60}\,T_P{}^S{}_R\,f_{QS}{}^R)\,\de_M\,X^P\,V^Q - \tfrac{1}{248}\,T_P{}^Q{}_Q\,U_M\,X^P.
\end{align}
\end{itemize}
Now, we can perform the two computations. Let us start with the first one. The next section is reserved to the second one. Consider the Leibniz identity for $\tilde{D}^{(T)}$:
\begin{equation}
[\tilde{D}^{(T)}_{\mathbb{X}},\tilde{D}^{(T)}_{\mathbb{V}}]\,\mathbb{W} - \tilde{D}^{(T)}_{\tilde{D}^{(T)}_{\mathbb{X}}\mathbb{V}}\,\mathbb{W} = 0.
\end{equation}
This is automatically satisfied, because the deformation given by the intrinsic torsion can be absorbed by twisting the parameters, as showed before. Now, set
\begin{equation}
\mathbb{X} = (X,0), \quad 
\mathbb{V} = (V,0), \quad
\mathbb{W} = (W,0),
\end{equation}
with $X, V, W$ constant
\begin{equation}
\de\,X = \de\,V = \de\,W = 0.
\end{equation}
Expanding, one finds the same integrability condition as in the $n\leqslant 7$ case with an additional term, which is the last line in the following expression:
\begin{align}
& ([T_F,T_E]-T_E{}^A{}_F\,T_A)^C{}_B + (E_{[F}{}^N\,\de_N)\,T_{E]}{}^C{}_B \,+\nn\\
& + (E^N{}_A\,\de_N)\,T_E{}^A{}_F\,\delta_B^C + f_B{}^{CE}\,f^A{}_{DE}\,(E^N{}_A\,\de_N)\,T_E{}^D{}_F \,+\nn\\
& - \tfrac{1}{60}\,f_B{}^{CA}\,(E^N{}_A\,\de_N)\,T_E{}^G{}_D\,f_F{}^D{}_G = 0.
\end{align}

\subsection{$E_{8(8)}$ flux constraints: first component}

In this section we analyse the constraints the flux has to satisfy in order for the Leibniz identity to be satisfied by the first component of the flux-deformed $E_{8(8)}$ Dorfman derivative. Compute the antisymmetric and the symmetric part of the Leibniz identity:
\begin{equations}
& [D^{(F)}_{\mathbb{X}},D^{(F)}_{\mathbb{Y}}] = D^{(F)}_{D^{(F)}_{\mathbb{X}}\mathbb{Y}}, \label{LeibnizAnti}\\
& D^{(F)}_{\frac{1}{2}(D^{(F)}_{\mathbb{X}}\mathbb{Y} + D^{(F)}_{\mathbb{Y}}\mathbb{X})} = 0, \label{LeibnizSym}
\end{equations}
with
\begin{equation}
\mathbb{X} = (X,S), \quad
\mathbb{Z} = (Z,T), \quad
\mathbb{V} = (V,U).
\end{equation}
Consider the first components of both the two parts. Select the terms proportional to $F$ and $\de\,F$ (the others sum to zero because the undeformed Dorfman derivative satisfies the Leibniz rule). The terms are of the following possible type, the second (first) sign for the (anti)symmetric part,
\begin{equations}
& F\,\de\,V, \\
& F\,V\,(Z\,\de\,X \mp \zeta\,\de\,X),\\
& F^2\,V\,X\,Z, \;\; \de\,F\,V\,X\,Z, \\
& F\,V\,(S\,Z \mp T\,X).
\end{equations}
In order to eliminate the $\de\,V$ terms, one has to impose the \emph{section constraint}:
\begin{equation}\label{E8firstsectionconstfluxtmp}
F_{MN}{}^P\,\mathcal{E}_P{}^p = 0.
\end{equation}
The terms $V\,(Z\,\de\,X \mp Z\,\de\,X)$ reproduce the expected antisymmetric/symmetric \emph{linear constraint} (which are equivalent, as in the case $n\leqslant 7$) with an additional term -- the same in both the cases -- proportional to $\frac{1}{60}$. Explicitly, the constraint is
\begin{align}
\mathcal{E}_P{}^p\,(&\delta_{(R}^P\,F_{S)M}{}^N -f^M{}_N{}^B\,f^P{}_{AB}\,F_{(RS)}{}^A \,+\nn\\
& + f^P{}_{RB}\,f_S{}^{AB}\,F_{AN}{}^M -\tfrac{1}{60}\,f^M{}_N{}^P\,f_{SA}{}^B\,F_{RB}{}^A) = 0.\label{E8linearconstfluxtmp}
\end{align}
Similarly, the terms $V\,X\,Z$ reproduce the antisymmetric/symmetric part of the \emph{Bianchi constraint} with an additional term. The complete constraint, writing $(F_A)_N{}^M = F_{AN}{}^M$, is 
\begin{align}
& ([F_R,F_S]+F_{RS}{}^A\,F_A)_N{}^M + \de_{[R}\,F_{S]N}{}^M \,+\nn\\
& - f^M{}_{NC}\,f_A{}^{BC}\,\de_B\,F_{RS}{}^A - \tfrac{1}{60}\,f^M{}_N{}^C\,\de_C\,(F_{RP}{}^Q\,f_R{}^P{}_Q) = 0.
\end{align}

The new piece of information is the coefficient of $V\,(S\,Z \pm T\,X)$:
\begin{equation}
(\tfrac{1}{496}\,f^M{}_N{}^Q\,F_{PA}{}^A - f^{MQ}{}_A\,F_{PN}{}^A - \tfrac{1}{2}\,f_P{}^{QA}\,F_{AN}{}^M - f_N{}^{QA}\,F_{PA}{}^M)\,V^N\,Z^P\,S_Q.
\end{equation}
The invariance of $f^{MQ}{}_N$ under $F_P$ implies
\begin{equation}
F_{PA}{}^M\,f^{AQ}{}_N + F_{PA}{}^Q\,f^{MA}{}_N - F_{PN}{}^A\,f^{MQ}{}_A = \tfrac{1}{248}\,F_{PA}{}^A\,f^{MQ}{}_N.
\end{equation}
Saturating with $S_Q$, which is on section, the second term drops out by means of the section constraint:
\begin{equation}
(-f^{MQ}{}_A\,F_{PN}{}^A-f_N{}^{QA}\,F_{PA}{}^M)\,S_Q = -\tfrac{1}{248}\,f^M{}_N{}^Q\,F_{PA}{}^A\,S_Q.
\end{equation} 
Replacing in the coefficient,
\begin{equation}
-\tfrac{1}{2}\,(-\tfrac{1}{248}\,f^{MQ}{}_N\,F_{PA}{}^A + f^{QA}{}_P\,F_{AN}{}^M)\,V^N\,Z^P\,S_Q.
\end{equation}
Therefore, we obtain the following \emph{new constraint}
\begin{equation}\label{E8NewConstr}
(\tfrac{1}{248}\,f^{QM}{}_N\,F_{PA}{}^A + f^{QA}{}_P\,F_{AN}{}^M)\,\mathcal{E}_Q{}^q = 0.
\end{equation}

Consider the trace of the linear constraint \eqref{E8linearconstfluxtmp} in the indices $M,N$:
\begin{equation}\label{E8TraceLinConstr}
\mathcal{E}_{(R}{}^p\,F_{S)M}{}^M = f^P{}_R{}^B\,f^A{}_{SB}\,\mathcal{E}_P{}^p\,F_{AM}{}^M.
\end{equation}
In particular, separating the symmetric, the antisymmetric and taking the trace, 
\begin{equations}
& f^{RSN}\,\mathcal{E}_R{}^p\,F_{SM}{}^M = 0,\label{TraceLinear1}\\
& \mathcal{E}_{(R}{}^p\,F_{S)M}{}^M = \tfrac{1}{2}\,f^P{}_{(R}{}^B\,f^A{}_{S)B}\,\mathcal{E}_P{}^p\,F_{AM}{}^M,\label{TraceLinear2}\\
& \mathcal{E}^{Rp}\,F_{RM}{}^M = 0.
\end{equations}
But the previous expressions are the section condition on the product $\text{Tr}\,F \otimes x$, for any $x$ on section.
\begin{equation}
Y^{MP}{}_{QN}\,\mathcal{E}_M{}^m\,F_{PR}{}^R = 0.
\end{equation}
We will call the previous condition \emph{second section constraint} for the flux in order to distinguish it from the (first) section constraint \eqref{E8firstsectionconstfluxtmp}.

Moreover, notice that the antisymmetric part can be obtained by tracing the new constraint. This shows that the new constraint and the linear one are not completely independent. We realised by a numerical analysis (building explicitly the structure constants) that \emph{the linear constraint and the extra constraint are together completely equivalent to the second section constraint}. Summarising, the constraints the flux has to satisfy in order for the first component of the $E_{8(8)}$ Dorfman derivative to be consistent are the following:
\begin{itemize}
\item[1.] \emph{First section constraint} 
\begin{equation}\label{E8ISectionConstr}
F_{MN}{}^P\,\mathcal{E}_P{}^p=0.
\end{equation}
\item[2.] \emph{Second section constraint} 
\begin{equation}\label{E8IISectionConstr}
Y^{MP}{}_{QN}\,\mathcal{E}_M{}^m\,F_{PR}{}^R = 0.
\end{equation}\label{E8BianchiConstr}
\item[3.] \emph{Bianchi constraint}
\begin{align}
& ([F_R,F_S]+F_{RS}{}^A\,F_A)_N{}^M + \de_{[R}\,F_{S]N}{}^M \,+\nn\\
& - f^M{}_{NC}\,f_A{}^{BC}\,\de_B\,F_{RS}{}^A - \tfrac{1}{60}\,f^M{}_N{}^C\,\de_C\,(F_{RP}{}^Q\,f_R{}^P{}_Q) = 0.
\end{align}
\end{itemize}

\subsection{$E_{8(8)}$ flux constraints: second component}

It remains to consider second component of the Leibniz identity \eqref{LeibnizAnti}--\eqref{LeibnizSym}. As before, select the terms proportional to $F$ or $\de\,F$. There are various possible terms. Most of them vanishes, using the constraints on the flux we derived in the analysis of the first component. It is useful to parametrise the flux in analogy with the intrinsic torsion in terms of $\vtheta_A$ and $\vtheta_{AB}$ in \eqref{E8Tcomponents}
\begin{equation}\label{Fdecomposition}
F_{MN}{}^P = \vphi_M\,\delta_N^P + \tfrac{1}{2}\,f^{RP}{}_N\,f_{RM}{}^S\,\vphi_S - f^{RP}{}_N\,\vphi_{MR}.
\end{equation}
Tracing and inverting, one can extract the two components 
\begin{equation} \label{DualExtraConstraint}
\vphi_M = \tfrac{1}{248}\,F_{MN}{}^N, \quad
\vphi_{MN} = -\tfrac{1}{60}\,F_{MR}{}^S\,f_N{}^R{}_S - \tfrac{1}{2}\,\tfrac{1}{248}\,f_{MN}{}^R\,F_{RS}{}^S.
\end{equation}
Since the parametrisation is the same as that of the intrinsic torsion, we know that the flux lies in the representations $\mathbf{1}\oplus\mathbf{248}\oplus\mathbf{3875}$ as the embedding tensor if and only if the rank-two component is symmetric:
\begin{equation}\label{SymmetryPhi}
\vphi_{MN} = \vphi_{NM}.
\end{equation}

The possible terms in the Leibniz identity are the following:
\begin{itemize}
\item[--] $\de\,U$, which is zero on the section constraint \eqref{E8IISectionConstr};
\item[--] $U\,Z\,\de\,X$ and $U\,X\,\de\,Z$, which are zero on the trace of the linear constraint \eqref{E8TraceLinConstr};
\item[--] $U\,X\,Z$, which is zero on the trace of the Bianchi constraint \eqref{E8BianchiConstr};
\item[--] $U\,Z\,S$ and $U\,X\,T$, which are zero on the trace of the new constraint \eqref{E8NewConstr};
\item[--] $V\,Z\,\de\,\de\,X$ and $V\,X\,\de\,\de\,Z$, which are zero using the linear constraint \eqref{E8linearconstfluxtmp}, the Jacobi identity \eqref{E8Jacobi}, the $f$ invariance and the $F$ decomposition \eqref{Fdecomposition};
\item[--] $V\,\de\,Z\,\de\,X$, which is zero using \eqref{E8linearconstfluxtmp} and \eqref{E8Jacobi};
\item[--] $V\,Z\,\de\,S$ and $V\,X\,\de\,T$, which are zero using \eqref{E8NewConstr};
\item[--] $V\,Z\,\de\,X$ and $V\,X\,\de\,Z$, which are zero using \eqref{E8linearconstfluxtmp}, the \eqref{E8BianchiConstr}, \eqref{E8Jacobi} and \eqref{Fdecomposition};
\item[--] $V\,Z\,X$, which is zero using the \eqref{E8BianchiConstr} and \eqref{Fdecomposition};
\item[--] $V\,S\,\de\,Z$ and $V\,T\,\de\,X$, which are zero using \eqref{E8linearconstfluxtmp}, \eqref{E8NewConstr} and \eqref{SymmetryPhi};
\item[--] $V\,S\,Z$ and $V\,T\,X$, which are the derivative of the previous coefficient, so they are zero.
\end{itemize}
Therefore, we conclude that \emph{the second component of the $E_{8(8)}$ flux-deformed Dorfmann derivative satisfies the Leibniz identity if and only if the flux satisfies the same constraints it has to satisfy in order for the first component to satisfy the Leibniz identity \eqref{E8ISectionConstr}, \eqref{E8IISectionConstr}, and \eqref{E8BianchiConstr}, and the flux sits in the same representation as the embedding tensor.}

As an example, let us show in details that the terms $V\,S\,\de\,Z$ are equal to zero, using the already known constraints. The terms we are interested in are 
\begin{equation}\label{CoefficientToVanish}
-\tfrac{1}{2}\,(S_q\,\mathcal{E}_Q{}^q)\,V^N\,\de_N\,Z^R\,[\tfrac{1}{248}\,\delta_M^Q\,F_{RB}{}^B + \tfrac{1}{60}\,f^Q{}_R{}^A\,(F_{AB}{}^C\,f_M{}^B{}_C) + \tfrac{1}{30}\,f^Q{}_M{}^A\,(F_{RB}{}^C\,f_A{}^B{}_C)].
\end{equation}
Consider the new constraint \eqref{E8NewConstr}, multiply it by $f_M{}^B{}_C$, and use the Cartan-Killing metric:
\begin{equation}\label{DualNewConstraint}
\tfrac{1}{248}\,\mathcal{E}_Q{}^q\,\delta_M^Q\,F_{RA}{}^A = \tfrac{1}{60}\,\mathcal{E}_Q{}^q\,f^Q{}_R{}^A\,(F_{AB}{}^C\,f_M{}^B{}_C).
\end{equation}
Replacing the last term in\eqref{CoefficientToVanish} with the previous expression,
\begin{equation}
-\tfrac{1}{2}\,S_q\,\mathcal{E}_Q{}^q\,V^N\,\de_N\,Z^R\,\tfrac{1}{30}\,[f^Q{}_R{}^A\,(F_{AB}{}^C\,f_M{}^B{}_C) + f^Q{}_M{}^A\,(F_{RB}{}^C\,f_A{}^B{}_C)].
\end{equation}
Now, replace the components of the flux
\begin{equation}
F_{MA}{}^A = 248\,\vphi_M, \quad
F_{MB}{}^C\,f_N{}^B{}_C = -60\,(\vphi_{MN} -\tfrac{1}{2}\,f_{MN}{}^A\,\vphi_A),
\end{equation}
so that the coefficient becomes
\begin{equation}
-\tfrac{1}{2}\,S_q\,\mathcal{E}_Q{}^q\,V^N\,\de_N\,Z^R\,\tfrac{1}{30}\,(\tfrac{30}{248}\,f^Q{}_{[R}{}^A\,f_{M]}{}^C{}_A\,\vphi_C + 60\,f^Q{}_{R}{}^A\,\vphi_{AM} + 60\,f^Q{}_{M}{}^A\,\vphi_{RA}).
\end{equation}
Now, using the Jacobi identity, the first term can be rewritten as 
\begin{equation}
\mathcal{E}_Q{}^q\,f^Q{}_{[R}{}^A\,f_{M]}{}^C{}_A\,\vphi_C = -\mathcal{E}_Q{}^q\,f_{RM}{}^A\,f^{QB}{}_A\,\vphi_B = 0,
\end{equation}
which vanishes, thanks to the expression \eqref{TraceLinear1}. In order to prove the last two terms to vanish, consider the dual of the new constraint \eqref{DualNewConstraint}, replacing the flux component:
\begin{equation}
\vphi_R\,\mathcal{E}_M{}^q = \mathcal{E}_Q{}^q\,f^Q{}_R{}^A\,(\vphi_{AM} - \tfrac{1}{2}\,f_{AM}{}^B\,\vphi_B)
\end{equation}
and take the symmetric part in $R,M$. The result is
\begin{equation}
\vphi_{(R}\,\mathcal{E}_{M)}^q = \tfrac{1}{2}\,\mathcal{E}_Q{}^q\,f^Q{}_{(R}{}^A\,f_{M)}{}^B{}_A\,\vphi_B -\mathcal{E}_Q{}^q\,f^Q{}_{(R}{}^A\,\vphi_{M)A}.
\end{equation}
But the left-hand side is precisely the expression \eqref{TraceLinear2}, if one replaces the flux components. Thus, we arrive to 
\begin{equation}
\mathcal{E}_Q{}^q\,f^Q{}_{(R}{}^A\,\vphi_{M)A} = 0,
\end{equation}
which is what is needed the last two terms in the coefficient to vanish, \emph{if} we assume the component $\vphi_{MN}$ to be symmetric.

\subsection{Consistency of the definition of the flux on the second component} 

This section is devoted to check that the second component of the parallelisability condition \eqref{E8ParallCond} is consistent with the definition of the flux, which is fixed only using the first component, as in the $n\leqslant 7$ case. Before starting, let us recall for convenience some definitions and expressions useful for the following:
\begin{equations}
& \hat{\mathbb{E}}_A = (E^M{}_A,h'_{MA})\;\;\text{(double frame)}, \\
& h'_{MA} = h'_{MA}(E,r) = \tfrac{1}{60}\,r^{-1}\,f_{AD}{}^C\,E_N{}^D\,\de_M\,E^N{}_C\;\;\text{(second component)},\\
& D_{\hat{\mathbb{E}}_A}\,\hat{\mathbb{E}}_B = - X_{AB}{}^C\,\hat{\mathbb{E}}_C \;\;\text{(parallelisability condition)},\label{E8ParallCondHere}\\
& (C^M{}_N, k_{MN})\;\;\text{(twisting matrix)},\\
& k_{MN} = h'_{MN}(C,\vrho),\\
& \hat{\mathbb{E}}_A = (C^M{}_N\,E^N{}_A, \vrho^{-1}\,h'_{MA} + k_{MN}\,E^N{}_A)\;\;\text{(dressing of double frame)},\\
& D_{\mathbb{E}_A}\,\mathbb{E}_B = - T_{AB}{}^C\,\mathbb{E}_C + (0,-\tfrac{1}{60}\,r^{-1}\,\de_M\,T_A{}^D{}_C\,f_{BD}{}^C)\;\;\text{(intrinsic torsion)}.
\end{equations}
Observe that the following is an alternative equivalent form of the second component of the double frame
\begin{align}
\phantom{h'_{AM}} &\phantom{=} - E_M{}^B\,r^{-1}\,W_{BA} = \nn\\
& = 
-E_M{}^B\,r^{-1}\,(-\tfrac{1}{60}\,f_A{}^C{}_D\,W_B{}^D{}_C) = \nn\\
& = E_M{}^B\,r^{-1}\,\tfrac{1}{60}\,f_A{}^C{}_D\,(E^P{}_B\,E^N{}_C\,\de_P\,E_N{}^D) = \nn\\
& = \tfrac{1}{60}\,r^{-1}\,f_A{}^C{}_D\,E^N{}_C\,\de_M\,E_N{}^D = \nn\\
&= - \tfrac{1}{60}\,r^{-1}\,f_A{}^C{}_D\,E_N{}^D\,\de_M\,E^N{}_C = \nn\\
& = \tfrac{1}{60}\,r^{-1}\,f_{AD}{}^C\,E_N{}^D\,\de_M\,E^N{}_C = h'_{MA}.
\end{align}

Consider the first component of the parallelisation condition \eqref{E8ParallCondHere}. On a side,
\begin{equation}
(D_{\hat{\mathbb{E}}_A}\,\hat{\mathbb{E}}_B)^M = - X_{AB}{}^C\,\hat{E}^M{}_C = - X_{AB}{}^C\,E^N{}_C\,C^M{}_N.
\end{equation}
On the other,
\begin{align}
(D_{\hat{\mathbb{E}}_A}\,\hat{\mathbb{E}}_B)^M &= C^M{}_N\,((D_{\mathbb{E}_A}\,\mathbb{E}_B)^N - F_{PQ}{}^N\,E^P{}_A\,E^Q{}_B) = \nn\\
& = C^M{}_N\,(- T_A{}^C{}_B\,E^N{}_C- F_{PQ}{}^N\,E^P{}_A\,E^Q{}_B).
\end{align}
Therefore, equating and inverting, one finds 
\begin{equation}\label{E8DefFlux}
F_{PQ}{}^M = (X_{AB}{}^C-T_A{}^C{}_B)\,E^M{}_C\,E_P{}^A\,E_Q{}^B,
\end{equation}
which is assumed as a \emph{definition} for the flux  in the uplift procedure. It will be useful to define also the flattened version of the flux:
\begin{equation}
\tilde{F}_{AB}{}^C = X_{AB}{}^C - T_{AB}{}^C.
\end{equation}

Now, consider the second component. On a side,
\begin{equation}
(D_{\hat{\mathbb{E}}_A}\,\hat{\mathbb{E}}_B)_M = - X_{AB}{}^C\,h'_{MC}\,\vrho^{-1} - X_{AB}{}^C\,k_{MN}\,E^N{}_C.
\end{equation}
On the other,
\begin{align}
(D_{\hat{\mathbb{E}}_A}\,\hat{\mathbb{E}}_B)_M &=
\vrho^{-1}\,[(D_{\mathbb{E}_A}\,\mathbb{E}_B)_M - \tfrac{1}{248}\,F_{PQ}{}^Q\,E^P{}_A\,h'_{MB} \,+\nn\\
& - \tfrac{1}{60}\,\de_M\,(E^P{}_A\,F_{PR}{}^S)\,E^Q{}_B\,f_{QS}{}^R] \,+\nn\\
& + k_{MN}\,[(D_{\mathbb{E}_A}\,\mathbb{E}_B)^N - F_{PQ}{}^N\,E^P{}_A\,E^Q{}_B].
\end{align}
By equating, we recognise in the term proportional to $k_{NM}$ the same expression as in the first component, so that we can drop it. By replacing the second component of the intrinsic torsion, it remains
\begin{align}
-(X_{AB}{}^C-& T_A{}^C{}_B)\,h'_{MC} = - \tfrac{1}{60}\,r^{-1}\,\de_M\,T_A{}^D{}_C\,f_{BD}{}^C - \tfrac{1}{248}\,(X_{AC}{}^C-T_A{}^C{}_C)\,h'_{MB} \,+\nn\\
& - \tfrac{1}{60}\,\de_M\,(E_R{}^E\,E^S{}_F\,(X_{AE}{}^F-T_A{}^F{}_E))\,f_{BC}{}^D\,E^S{}_C\,E^R{}_D\,r^{-1}.
\end{align}
Replacing $h_{MA}$ and recognising the components of the flattened flux, we obtain
\begin{align}
0 =&\, \tfrac{1}{60}\,E_N{}^E\,\de_M\,E^N{}_F\,r^{-1}\,[- \tfrac{1}{248}\,\tilde{F}_{AC}{}^C\,f_{BE}{}^F \,+\nn\\
& + \tilde{F}_{AB}{}^C\,f_{CE}{}^F + \tilde{F}_{AE}{}^C\,f_{BC}{}^F - \tilde{F}_{AC}{}^F\,f_{BE}{}^C].
\end{align}
But the term in the bracket is an identity, thanks to the invariance of the structure constants under the adjoint action of the flux (see Appendix \ref{fFinvariance}):
\begin{equation}
{(n_+-n_-)\,\tfrac{1}{248}\,\tilde{F}_{AC}{}^C\,f_{BE}{}^F}_{\big{|}_{\substack{n_+ = 1, \\ n_-=2}}} =  - \tilde{F}_{AB}{}^C\,f_{CE}{}^F - \tilde{F}_{AE}{}^C\,f_{BC}{}^F + \tilde{F}_{AC}{}^F\,f_{BE}{}^C,
\end{equation}
$n_{\pm}$ being the number of upper/lower indices of the $f$ on which $F$ acts. Therefore, we conclude that the second component is consistent with the definition of the flux fixed by the first one.\footnote{This result can be extended to the case in which the flux has a non-integrable component, as  in the case of uplift to massive type IIA supergravity \cite{Ciceri:2016dmd}.}

\subsection{$E_{8(8)}$ first and second section constraint}

This and the following Sections are devoted to show that the flux defined by \eqref{E8DefFlux} satisfies the consistency constraints, similarly to the $n\leqslant 7$ case. In order to verify if the flux satisfies the first section constraint \eqref{E8ISectionConstr}, we need to update the expressions \eqref{SectionFirst1} and \eqref{SectionFirst2}. Consider a twisting matrix 
\begin{equations}
C^M{}_N &= Z^M{}_N\,\vrho^{-1}, \\
k_{MN} &= -C_M{}^P\,\vrho^{-1}\,(-\tfrac{1}{60}\,f_N{}^R{}_S\,C^{M'}{}_P\,C^{N'}{}_R\,\de_{M'}\,C_{N'}{}^S = \nn\\
& = \tfrac{1}{60}\,\vrho^{-2}\,f_N{}^Q{}_S\,\de_M\,C_Q{}^S,
\end{equations}
assuming it to satisfy the section choice $C^M{}_N\,\mathcal{E}_M{}^m = 0$. $Z$ and $\vrho$ are the analogue of $U$ and $r$ with respect to $E$. The twisted parameters are
\begin{equation}
V'^M = C^M{}_N\,V^N, \quad
X^M = C^M{}_N\,X^N,
\end{equation}
\begin{equation}
S'_M = C_M{}^N\,\vrho^{-1}\,S_N + k_{MN}\,X^N = S_M\,\vrho^{-1} + k_{MN}\,X^N.
\end{equation}
The dressed formula for the Lie derivative is
\begin{equation}
L_{X',S'}\,V'^M = C^M{}_N\,(L_{X,S}\,V^N - F_{PQ}{}^M\,X^P\,V^Q) = C^M{}_N\,L_{X,S}^{(F)}\,V^N,
\end{equation}
the flux being the torsion associated to $C^M{}_N$ and $k_{MN}$:
\begin{equation}
-C^M{}_N\,F_{PQ}{}^M = L_{C_P,k_P}\,C^M{}_Q.
\end{equation}
Let $(\hat{E}^M{}_A,\hat{h}_{MA})$ be double frame satisfying the parallelisability condition
\begin{equation}\label{E8ParallCond1}
L_{\hat{E}_A,\hat{h}_A}\,\hat{E}^M{}_B = -X_{AB}{}^C\,\hat{E}^M{}_C.
\end{equation}
Suppose $(\hat{E}^M{}_A,\hat{h}_{MA})$ to be the double frame 
$(E^M{}_A,h'_{MA})$, twisted by $(C^M{}_N,k_{MN})$:
\begin{equation}
\hat{E}^M{}_A = C^M{}_N\,E^N{}_A, \quad
\hat{h}_{MA} = h'_{MA}\,\vrho^{-1} + k_{MN}\,E^N{}_A.
\end{equation}
The untwisted double frame satisfies the parallelisability condition with the flux-deformed derivative: 
\begin{equation}\label{E8ParallCond2}
L_{E_A,h'_A}^{(F)}\,E^M{}_B = -X_{AB}{}^C\,E^M{}_C.
\end{equation}
Define the Killing vectors
\begin{equation}
K_A{}^m = \hat{E}^M{}_A\,\mathcal{E}_M{}^m = E^M{}_A\,\mathcal{E}_M{}^m,
\end{equation}
defined equivalently in terms of $E$ or $\hat{E}$ because $C$ preserves the section choice. Then,
\begin{equation}
L_{\hat{E}_A,\hat{h}_A}\,\hat{E}^M{}_B\,\mathcal{E}_M{}^m = \mathcal{L}_{K_A}\,K_B{}^m = L_{E_A,h'_A}\,E^M{}_B\,\mathcal{E}_M{}^m
\end{equation}
Indeed, not only the $Y$-tensor term drops out, thanks to the section constraint, but also the ancillary term:
\begin{equation}
\delta_{\hat{h}_A}\,E^M{}_B\,\mathcal{E}_M{}^m = - f^{NM}{}_P\,\hat{h}_{NA}\,E^P{}_B\,\mathcal{E}_M{}^m = 
- (f^{NM}{}_P\,\hat{h}_{NA}\,\mathcal{E}_M{}^m)\,E^P{}_B = 0,
\end{equation}
where the fact that the first index in $\hat{h}_{NA}$ and in $h'_{NA}$ are on section is used. Therefore, if we saturate \eqref{E8ParallCond1} with the section, we get
\begin{equation}
-X_{AB}{}^C\,K_C{}^m = -X_{AB}{}^C\,\hat{E}^M{}_C\,\mathcal{E}_M{}^m = L_{\hat{E}_A,\hat{h}_A}\,\hat{E}^M{}_B\,\mathcal{E}_M{}^m = \mathcal{L}_{K_A}\,K_B{}^m.
\end{equation}
Instead, if we do the same with \eqref{E8ParallCond2},
\begin{align}
-X_{AB}{}^C\,K_C{}^m &= -X_{AB}{}^C\,E^M{}_C\,\mathcal{E}_M{}^m = L_{E_A,h_A}^{(F)}\,E^M{}_B\,\mathcal{E}_M{}^m =\nn\\
& =  L_{E_A,h'_A}\,E^{}_B{}\,\mathcal{E}_M{}^m - F_{NP}{}^M\,E^N{}_A{}\,E^P{}_B{}\,\mathcal{E}_M{}^m = \nn\\
& = \mathcal{L}_{K_A}\,K_B{}^m - (F_{NP}{}^M\,\mathcal{E}_M{}^m)\,E^N{}_A\,E^P{}_B.
\end{align}
Therefore, the previous result implies that
\begin{equation}
F_{NP}{}^M\,\mathcal{E}_M{}^m = 0,
\end{equation}
which is the first section constraint \eqref{E8ISectionConstr}.

It is very simple to extend the proof of the second section constraint in the $n=8$ case. Indeed, the new term in the torsion
\begin{equation}
T_A{}^C{}_B = \dots + f_B{}^{CD}\,(-\tfrac{1}{60}\,f_A{}^E{}_F\,W_D{}^F{}_E)
\end{equation}
is antisymmetric in the last two indices, so that it drops out when we take the trace, and we remain with the same expression as in the case $n\leqslant 7$. In the remaining of the proof there is no distinction between $n=8$ and $n\leqslant 7$.

\subsection{$E_{8(8)}$ Bianchi constraint}

The linear constraint \eqref{FluxLinearConstraint} has a new term in the right-hand-side:
\begin{align}
& F_{MS}{}^P\,\mathcal{E}_N{}^r - Y^{QR}{}_{SM}\,F_{QN}{}^P\,\mathcal{E}_R{}^r \,+\nn\\
& - Y^{QR}{}_{SN}\,F_{MQ}{}^R\,\mathcal{E}_R{}^r + Y^{PR}{}_{SQ}\,F_{MN}{}^Q\,\mathcal{E}_R{}^r = \nn\\
& = - F_{SM}{}^P\,\mathcal{E}_N{}^r
+ Y^{PR}{}_{QN}\,F_{SM}{}^Q\,\mathcal{E}_R{}^r
-\tfrac{1}{60}\,f_M{}^T{}_Q\,f_N{}^{PR}\,F_{ST}{}^Q\,\mathcal{E}_R{}^r.
\end{align}
Therefore, following the same strategy as in the case $n\leqslant 7$, the  flux-deformed Lie derivative on the flux itself reads
\begin{align}
L_{X,S}^{(F)}\,F_{MN}{}^P &= X^Q\,\de_Q\,F_{MN}{}^P + F_{QN}{}^P\,\de_M\,X^Q \,+\nn\\
& - F_{QM}{}^P\,\de_N\,X^Q + Y^{PR}{}_{SN}\,F_{QM}{}^S\,\de_R\,X^Q \,+\nn\\
& -\tfrac{1}{60}\,f_M{}^T{}_S\,f_N{}^{PR}\,F_{QT}{}^S\,\de_R\,X^Q + X^Q\,\delta_{F_Q}\,F_{MN}{}^P + \delta_S\,F_{MN}{}^P.
\end{align}
Also the Bianchi constraint \eqref{FluxBianchi} has a new term in the right-hand-side:
\begin{align}
\delta_{F_Q}\,F_{MN}{}^P &= - \de_Q\,F_{MN}{}^P + \de_M\,F_{QN}{}^P - \de_N\,F_{QM}{}^P \,+\nn\\
& +  Y^{PR}{}_{SN}\,\de_R\,F_{QM}{}^S + \tfrac{1}{60}\,f_M{}^T{}_S\,f_N{}^{PR}\,\de_R\,F_{QT}{}^S.
\end{align}
Therefore, we can write
\begin{align}
& \text{The Bianchi constraint holds} \Leftrightarrow \nn\\
\Leftrightarrow &\, L_{X,S}^{(F)}\,F_{MN}{}^P = X^Q\,\de_Q\,F_{MN}{}^P + F_{QN}{}^P\,\de_M\,X^Q \,+\nn\\
& - F_{QM}{}^P\,\de_N\,X^Q + Y^{PR}{}_{SN}\,F_{QM}{}^S\,\de_R\,X^Q \,+\nn\\
& + X^Q\,(- \de_Q\,F_{MN}{}^P + \de_M\,F_{QN}{}^P - \de_N\,F_{QM}{}^P \,+\nn\\
& +  Y^{PR}{}_{SN}\,\de_R\,F_{QM}{}^S
+ \tfrac{1}{60}\,f_M{}^T{}_S\,f_N{}^{PR}\,\de_R\,F_{QT}{}^S) + \delta_S\,F_{MN}{}^P \nn\\
\Leftrightarrow &\, L_{X,S}^{(F)}\,F_{MN}{}^P =
\de_M\,(X^Q\,F_{QN}{}^P) - \de_N\,(X^Q\,F_{QM}{}^P) + Y^{PR}{}_{SN}\,\de_R\,(X^Q\,F_{QM}{}^S) \,+\nn\\
& -\tfrac{1}{60}\,f_M{}^T{}_S\,f_N{}^{PR}\,\de_R\,(X^Q\,F_{TQ}{}^S) + \delta_S\,F_{MN}{}^P \nn\\
\Leftrightarrow &\, L_X^{(F)}\,F_{MN}{}^P =
T_M{}^P{}_N\{\de_M\,(E_A{}^Q\,F_{QN}{}^T)\} + \delta_S\,F_{MN}{}^P.
\end{align}
Also in this case the new term fits precisely the intrinsic torsion out of the connection $\de_M\,(X^Q\,F_{QN}{}^P)$, in such a way that we recover the same expression as in the case $n\leqslant 7$, up to the ancillary term. We find \eqref{Bianchi_iff1} by replacing $(X^M,S_M) \rightarrow (E_F{}^M,0)$.

Since the expression \eqref{Bianchi_iff2} continues to hold, it remains to verify if the relation \eqref{Bianchi_iff3} holds too. This is the case because, although both the left-hand side and the right-hand side receive the new contribution, by definition of $E_{8(8)}$ intrinsic torsion, they are precisely the same, thanks to this identity:
\begin{align}
& \tfrac{1}{60}\,E^M{}_A\,E^N{}_B\,E_P{}^C\,f^{TP}{}_N\,f_M{}^R{}_S\,(\de_T\,E^Q{}_F\,F_{QR}{}^S + E^Q_F\,\de_T\,F_{QR}{}^S) = \nn\\
=&\, -\tfrac{1}{60}\,f_A{}^H{}_G\,E^M{}_D\,E^N{}_H\,\de_M\,E_N{}^E\,\tilde{F}_{FE}{}^G \,+\nn\\
& + \tfrac{1}{60}\,f_A{}^H{}_G\,\tilde{F}_{FH}{}^E\,E^M{}_D\,E^N{}_E\,\de_M\,E_N{}^G \,+\nn\\
& - \tfrac{1}{60}\,f_A{}^H{}_G\,E^M{}_D\,\de_M\,\tilde{F}_{FH}{}^G.
\end{align}
Now, the left-hand side is the new contribution in $E^M{}_A\,E^N{}_B\,E_P{}^C\,T_M{}^P{}_N\{\de_M\,(E^Q{}_F\,F_{QN}{}^P)\}$; the right-hand-side is the new contribution in $T_A{}^C{}_B\{[W_A,\tilde{F}_F]_B{}^C + E^M{}_A\,\de_M\,\tilde{F}_{FB}{}^C\}$. To verify it, observe that the structure constants are invariant under the action of the zero-weighted twisting matrix $U^M{}_A$. 

It remains to reconsider the Bianchi identity \eqref{BianchiT} for the torsion out of the Weitzenb\"ock connection. It can be computed starting from the Leibniz identity on the same footing as in the case $n\leqslant 7$, the result being 
\begin{align}
& E^{}_F\,\de_M\,T_A{}^C{}_B + T_F{}^D{}_B\,T_A{}^C{}_D - T_A{}^D{}_B\,T_F{}^C{}_D + T_F{}^D{}_A\,T_D{}^C{}_B \,=\nn\\
& = E^M{}_A\,\de_M\,T_F{}^C{}_B - E^M{}_B\,\de_M\,T_F{}^C{}_A + Y^{CD}{}_{EB}\,E^M{}_D\,\de_M\,T_F{}^E{}_A \,+\nn\\
& - \tfrac{1}{60}\,f_E{}^{FD}\,E_D{}^M\,\de_M\,T_A{}^G{}_H\,f_B{}^H{}_G.
\end{align}
A new term appears, the last one, and this is precisely what is needed to reconstruct the torsion in the right-hand side of the \eqref{BianchiT}. 

In this way, we have completed the proof that the flux satisfies the Bianchi constraint in the case $n=8$ too.

\section{Towards non-maximal uplifts}\label{6}

Until now we have only considered consistent truncations preserving maximal supersymmetry, which is, say, $\mathcal{N}=8$ supercharges in four dimensions, and $\mathcal{N}=16$ supercharges in three dimensions. 

In the following Section, without exhausting the topic at all, we briefly review a simple example of reduction of a maximal theory to a theory with less amount of supersymmetry charges. Then, in next Section, we give a sketch of how one has to modify the setup of the uplift problem in order to include non-maximal theories.\footnote{The results about the uplift problem applied to half-maximal truncation in four dimensions using exceptional field theory appeared in \cite{Rovere:2025jks} after this thesis was completed. See also the references within.}

\subsection{Example of non-maximal theory}

Consider the truncation of $\mathcal{N}=8$ maximal supergravity theory in four dimensions to the theory with $\mathcal{N}=2$ supercharges (and correspondingly the dual one, with $\mathcal{N}=6$) \cite{Andrianopoli:2008ea}. The fields in maximal supergravity in four dimensions can be expressed in terms of irreducible representations of the maximal compact subgroup $\text{SU}(8)$ of the duality group $E_{7(7)}$:
\begin{itemize}
\item[--] The metric tensor $g_{\mu\nu}$ (spin $s=2$) is a scalar with respect to $\text{SU}(8)$;
\item[--] The eight gravitini $\psi_\mu^I$ (spin $s=3/2$),  $I=1,\dots,8$, are in the fundamental representation;
\item[--] The vectors $A_\mu^{IJ}$ (spin $s=1$) are in the antisymmetric rank-two representation, whose dimension is indeed ${{8}\choose{2}}=28$;
\item[--] The spin $s=\frac{1}{2}$ fields $\chi^{IJK}$ in the antisymmetric rank-three  representation, whose dimension is indeed ${{8}\choose{3}}=56$;
\item[--] The scalars $\vphi^{IJKL}$ ($s=0$) are in the antisymmetric rank-four representation, whose dimension is indeed ${{8}\choose{4}}=70$.
\end{itemize}
The rank $r$ of these $\text{SU}(8)$ representations increases as the spin  $s$ decreases, according to $s = 2-\frac{r}{2}$. $\text{SU}(8)$ is also the $R$-symmetry, rotating the supercharges. 

The fields in $\mathcal{N}=2$ truncation are obtained by selecting which fields are left invariant by
\begin{equation}
Z=\begin{pmatrix} I_2 & 0 \\ 0 & -I_6 \end{pmatrix}\in\text{SU}(8),
\end{equation}
which, acting on the fundamental representation,  generates a $\mathbb{Z}_2$ subgroup of $\text{SU}(8)$, together with the $8\times 8$ identity matrix $I_8$. For the gravitini,
\begin{equation}
\psi_\mu^I \rightarrow Z^I{}_J\,\psi_\mu^J
\end{equation}
$Z$ leaves two gravitini invariant out of the eight of the full theory. Indeed, splitting $I=(a,i)$, where $a=1,2$ and $i=1,\dots,6$, 
\begin{equation}
\psi_\mu^a \rightarrow \psi_\mu^a, \quad
\psi_\mu^i \rightarrow -\psi_\mu^i.
\end{equation} 
The number of degrees of freedom of the invariant gravitini is $2\times 2^{4/2-1}\,(4-3)=4$. Similarly for the vectors:
\begin{equation}
A_\mu^{IJ} \rightarrow Z^I{}_K\,Z^J{}_L\,A_\mu^{KL},
\end{equation}
whose splitting is
\begin{equation}
A_\mu^{ab} \rightarrow A_\mu^{ab},\quad
A_\mu^{ai} \rightarrow -A_\mu^{ai},\quad
A_\mu^{ij} \rightarrow A_\mu^{ij}.
\end{equation}
So, there are $1+15$ invariant vectors, with $32$ degrees of freedom. For spin $\frac{1}{2}$,
\begin{equation}
\chi^{IJK} \rightarrow Z^I{}_L\,Z^J{}_M\,Z^K{}_N\,\chi^{LMN},
\end{equation}
whose splitting is
\begin{equation}
\chi^{abi} \rightarrow -\chi^{abi},\quad
\chi^{aij} \rightarrow \chi^{aij},\quad
\chi^{ijk} \rightarrow -\chi^{ijk}.
\end{equation}
So, there are $2\,\frac{6(6-1)}{2}=30$ invariant $\frac{1}{2}$ spin, with $60$ degrees of freedom. For scalars, 
\begin{equation}
\vphi^{IJKL} \rightarrow Z^I{}_M\,Z^J{}_N\,Z^K{}_N\,Z^L{}_Q\,\vphi^{MNPQ},
\end{equation}
whose splitting is
\begin{equation}
\vphi^{abij} \rightarrow \vphi^{abij},\quad
\vphi^{aijk} \rightarrow -\vphi^{aijk},\quad
\vphi^{ijkl} \rightarrow \vphi^{ijkl}.
\end{equation}
So, there are $\frac{6(6-1)}{2}+{{6}\choose{4}}=30$ scalars. The graviton is invariant, since is a $\text{SU}(8)$ scalar. Summarising, the selected fields are
\begin{equation}
1\;\text{graviton}, \quad
2\;\text{gravitini}, \quad
16\;\text{vectors}, \quad
30\;\text{spin}\;\tfrac{1}{2},\quad
30\;\text{scalars}.
\end{equation}
Consistently, bosonic and fermionic degrees of freedom match
\begin{equation}
\#\,\text{bosons} = 2 + 32 + 30 = 4 + 60 = \#\,\text{fermions}.
\end{equation}

Alternatively, the selected fields are obtained by studying branching of the $\text{SU}(8)$ irreducible representations  in terms of representations of $\text{SU}(6) \otimes \text{SU}(2) \otimes \text{U}(1)$:
\begin{equations}
&\mathbf{8} \rightarrow (\mathbf{6},\mathbf{1})_{1/2} \oplus (\mathbf{1},\mathbf{2})_{3/2},\\
& \mathbf{28} \rightarrow (\mathbf{15},\mathbf{1})_{1} \oplus (\mathbf{1},\mathbf{1})_{-3} \oplus (\mathbf{6},\mathbf{2})_{-1},\\
&\mathbf{56} \rightarrow (\mathbf{20},\mathbf{1})_{3/2} \oplus (\mathbf{6},\mathbf{1})_{-5/2} \oplus (\mathbf{15},\mathbf{2})_{-1/2},\\
&\mathbf{70} \rightarrow (\mathbf{15},\mathbf{1})_{-2} \oplus (\overline{\mathbf{15}},\mathbf{1})_{2} \oplus (\mathbf{20},\mathbf{2})_0,
\end{equations}
where the subscript is the $\text{U}(1)$-charge. The $\mathcal{N}=6$ truncation is obtained by selecting the $\text{SU}(2)$ singlets. So, we read off 6 gravitini, 15+1 vectors, 20+6 spin $\frac{1}{2}$ and 15+15 scalars. The $\mathcal{N}=2$ truncation is obtained by selecting the complement fields for the fermionic fields and the same bosonic fields, and this matches with the direct previous computation.

\subsection{Towards uplifts of non-maximal theories}\label{nonMax}

For each supersymmetric solution of supergravity in ten or eleven dimensions of the type $\text{AdS}_d \times \mathscr{M}$, there exists a consistent reduction to a pure gauged $d$-dimensional supergravity which admits that solution and which enjoys the same (maximal) supersymmetry \cite{Cvetic:1999au, Gauntlett:2007ma, Louis:2014gxa,Malek:2017njj}. Given the internal manifold $\mathscr{M}$, the fields of the higher-dimensional theory can be decomposed in irreducible representations of the \emph{structure group} $G_S$ of $\mathscr{M}$ \cite{Cassani:2019vcl}.\footnote{If $\mathscr{M}$ is parallelisable, $G_S$ is trivial, and this is the case of Scherk-Schwarz reduction: the algebra generators are the global frame, the structure constants are the components of the intrinsic torsion, the commutators between the generators are the Lie derivative of the components of the global frame along themselves.} In the reduction, one selects the singlets with respect to $G_S$, that is, one considers the invariants of $G_S$ and expresses the fields of the reduced theory in terms of such invariants.

Since the metric tensor describes a coset space $G/H$, the commutant between $G_S$ and the group defining the coset has to be taken into account in order to select the singlets coming from the metric. The \emph{commutant} (or \emph{centraliser}) $\text{Comm}_G(H)$ of a subgroup $H$ of a group $G$ is the set of all the elements in $G$ which commute with every element in $H$
\begin{equation}
\text{Comm}_G(H)=\{g\in G : g\,h=h\,g,\;\forall \;h\in H\}.
\end{equation}
Therefore, the scalars coming from the reduction of the metric are described by the coset
\begin{equation}\label{CosetNonMax}
\text{Comm}_G(G_S)/\text{Comm}_H(G_S).
\end{equation}
Using the invariants of $G_S$, one can construct a basis of vectors $K_A$ (and the corresponding dual one-forms $A^A$), which are the vectors coming from the reduction of the metric. The index $A$ runs over the singlets of $G_S$. Starting from $K_A$ one defines the intrinsic torsion $-L_{K_A}\,K_B$, where $L_{K_A}$ is the generalised Lie derivative, imposing it to be a constant singlet of $G_S$ and to be equal to the component $X_{AB}{}^C$ of the embedding tensor
\begin{equation}\label{NonMaxPar}
L_{K_A}\,K_B = -X_{AB}{}^C\,K^C.
\end{equation}
One has to decompose the fundamental representation of the duality group $E_{n(n)}$ of the higher-dimensional theory -- say eleven-dimensional supergravity reduced to $d$ dimensions with $n=11-d$ internal dimensions -- with respect to $G_S$ in order to find the values which the index $A$ takes in $K_A$. The duality group $G_d$ of the reduced theory is the commutant of $G_S$ with respect to $E_{n(n)}$
\begin{equation}
G_d = \text{Comm}_{E_{n(n)}}(G_S),
\end{equation}
which indeed is the numerator in \eqref{CosetNonMax}. The \emph{gauge group} $G_g$ of the reduced theory is contained in $\text{Comm}_{E_{n(n)}}(G_S)$, since the gauge vectors are taken among the vectors coming from the metric:
\begin{equation}
G_g \subseteq \text{Comm}_{E_{n(n)}}(G_S).
\end{equation}

Let $H_n$ being the maximal compact subgroup of $E_{n(n)}$. Then, the coset of the scalars of the maximal reduction reads $E_{n(n)}/H_n$, and the $R$-symmetry group $G_{\mathcal{N}}$ with $\mathcal{N}$ supercharges is the largest subgroup of $H_n$ such that the generalised spinor bundle (where spinors and supercharges live) contains $\mathcal{N}$ singlets of $G_S$, since the gravitini of the reduced theory must be singlets of $G_S$, as the bosonic fields. 

In a maximal reduction, the internal manifold $\mathscr{M}$ is described by a coset $G/H$, and the vectors $K_A$ generate all the tangent bundle $T\mathscr{M}$. The numerator is the gauge group. The structure group is trivial $G_S = \mathbbm{1}$. Indeed, the maximal reductions are those in which  $\mathscr{M}$ is parallelisable (also in the generalised sense). 

In the non-maximal case,  $\mathscr{M}$ is not a coset, but one can locally write  \cite{Meinrenken2003}
\begin{equation}
\mathscr{M} \sim G/H \times \mathbb{R}^{n-\dim\,G+\dim\,H}
\end{equation}
and the vectors $K_A$ generate only the coset part $G/H$. The analogue to the (generalised) parallelisability condition in the non-maximal case \eqref{NonMaxPar} is less stringent, since there are a fewer $K_A$ than in the maximal case, since $A$ runs over the dimension of $\text{Comm}_{E_{n(n)}}(G_S)$, which is the whole duality group $E_{n(n)}$ in the maximal case only.

The \emph{embedding tensor} $\vtheta$ links the vectors/dual one-forms coming from the metric $K_A$ or $A^A$ to the gauge fields $K_\alpha = K_\alpha{}^m\,\de_m$ or $A^\alpha = \diff y^m\,A_m{}^\alpha$, where $\alpha$ is in the adjoint representation of $G_g$, and $y^m$ are coordinates in $\mathscr{M}$:
\begin{equation}
K_A = \vtheta_A{}^\alpha\,K_\alpha, \quad
A^\alpha = A^A\,\vtheta_A{}^\alpha.
\end{equation}

As an example, consider the eleven-dimensional supergravity as higher-dimensional theory, and its solution $\text{AdS}_4 \times S^3 \times S^3 \times \Sigma$ as a four-dimensional reduction, where $\Sigma$ is an interval (locally equivalent to $\mathbb{R}$). Since $S^3 \sim \text{SO}(3)$, and $\mathfrak{so}(3) \oplus \mathfrak{so}(3) \sim \mathfrak{so}(4)$, this reduction is an $\text{SO}(4)$ four-dimensional gauged supergravity in AdS. This reduction is \emph{half-maximal}, preserving half of the supersymmetry charges of the higher-dimensional theory: since the eleven-dimensional spinor corresponds to eight four-dimensional spinors, this reduction preserves $\mathcal{N}=4$ supercharges and $\text{SO}(4)_R$ rotates the four gravitini. The internal space is the seven-dimensional manifold 
\begin{equation}
S^3 \times S^3 \times \Sigma \sim \text{SO}(3) \times \text{SO}(3) \times \mathbb{R} \sim \text{SO}(4) \times \mathbb{R},
\end{equation}
where the isomorphisms hold \emph{locally}.\footnote{$S^d$ can be described as a coset space $\text{SO}(d+1)/\text{SO}(d)$. Nevertheless, locally $S^3$ is also equivalent to $\text{SO}(3)$, since $\text{SO}(3)$ is locally the same as $\text{SU}(2)$ (they have the same algebra); the most general matrix in $\text{SU}(2)$ can be parametrised by $U= \begin{pmatrix} \alpha & \beta \\ -\beta^* & \alpha^* \end{pmatrix}$, where $\alpha, \beta \in \mathbb{C}$, with $|\alpha|^2 + |\beta|^2 = 1$; but the last one is the equation on $S^3$ embedded in $\mathbb{R}^4$ with coordinates $\text{Re}\,\alpha$, $\text{Im}\,\alpha$, $\text{Re}\,\beta$, $\text{Im}\,\beta$.}
The structure group is\footnote{The choice of the structure group is constrained by the number of supersymmetry charges one wants to preserve. Consider the $R$-symmetry group of the maximal theory $\text{SU}(8)$. Breaking it with respect to $G_S$ one has to obtain $\mathcal{N}$ singlets of $G_S$ in the spinor bundle. In the half-maximal case $\mathcal{N}=4$, the largest choice is $\text{SU}(4)$. So, $G_S$ should be $\text{SU}(4)$ or a subgroup of it (in the maximal case is trivially $\mathbbm{1}$).}
\begin{equation}
G_S = \text{SU}(4)_S.
\end{equation}
The duality group of the maximal theory is $E_{7(7)}$. The duality group of the half-maximal theory is 
\begin{equation}
G_d = \text{Comm}_{E_{7(7)}}(\text{SU}(4)_S) = \text{SU}(4)_R \times \text{SL}(2,\mathbb{R}).
\end{equation}
The gauge group is
\begin{equation}
G_g = \text{SO}(4) \subseteq \text{Comm}_{E_{7(7)}}(\text{SU}(4)_S)
\end{equation}
and the reduction is half-maximal 
\begin{equation}
\mathcal{N}=4.
\end{equation}
The maximal compact subgroup of $E_{7(7)}$ is $H_d = \text{SU}(8)$. The $R$-symmetry group is 
\begin{equation}
G_{\mathcal{N}=4} = \text{SU}(4)_R.
\end{equation}
Since $\text{Comm}_{\text{SU}(8)}(\text{SU}(4)_S)=\text{SU}(4)_R \times \text{U}(1)$, the coset of scalars is 
\begin{equation}
\frac{\text{Comm}_{E_{7(7)}}(\text{SU}(4)_S)}{\text{Comm}_{\text{SU}(8)}(\text{SU}(4)_S)} = \frac{\text{SL}(2,\mathbb{R})}{\text{U}(1)}.
\end{equation}
The fundamental representation of $E_{7(7)}$ is $\textbf{56}$, whose branching with respect to $\text{SU}(4)_S \times \text{SU}(4)_R \times \text{SL}(2,\mathbb{R})$ is 
\begin{equation}
\textbf{56} \rightarrow (\textbf{6},\textbf{1},\textbf{2}) \oplus (\textbf{1},\textbf{6},\textbf{2}) \oplus \dots
\end{equation}
So, there are twelve vectors $K_A$, sitting in $(\textbf{1},\textbf{6},\textbf{2})$ and the index $A$ splits into an index in the antisymmetric rank two of $\text{SU}(4)$, which is the same as the fundamental of $\text{SO}(6)$, and one in the $1/2$ spin of $\text{SL}(2,\mathbb{R})$.

\newpage
\quad
\thispagestyle{empty}

\newpage 
\section{Covariant fracton gauge theory}\label{5}

\subsection{Introduction}

Covariant fracton gauge theory, introduced for the first time in \cite{Blasi:2022mbl}, is a family of gauge theories of a rank-two symmetric tensor $h_{\mu\nu}$, with mass dimension $[h_{\mu\nu}] = 2$ and with gauge symmetry given by the double derivative of a scalar parameter $\lambda$
\begin{equation}\label{GaugeIntro}
\delta_\lambda\,h_{\mu\nu} = \de_\mu\,\de_\mu\,\lambda.
\end{equation} 
One can treat these theories as a sort of ``higher rank electrodynamics", where the gauge transformation involves a second derivative instead of a first derivative, or as linearised gravity, where the gauge field $h_{\mu\nu}$ is viewed as the perturbation of the metric $g_{\mu\nu} = \eta_{\mu\nu} + \kappa^{4/(d-2)}\,h_{\mu\nu}$, and the gauge invariance is restricted to the longitudinal diffeomorphism only: the vector field generating the diffeomorphism is taken as the partial derivative of a scalar parameter 
\begin{equation}
\delta_\xi\,h_{\mu\nu} = \de_\mu\,\xi_\nu + \de_\nu\,\xi_\mu \rightarrow \de_\mu\,\de_\nu\,\lambda, \;\text{if}\;\xi_\mu \rightarrow \tfrac{1}{2}\,\de_\mu\,\lambda.
\end{equation}
One can study the most general gauge invariant action quadratic in the derivatives of $h_{\mu\nu}$, obtaining a generalisation of the action of linearised Einstein gravity (massless Fierz-Pauli theory \cite{Fierz:1939ix}). The latter reads
\begin{equation}\label{FP}
S^{(\textsc{fp})} = \tfrac{1}{2}\int\diff^d x\,(\de_\vrho\,h_{\mu\nu}\,\de^\vrho\,h^{\mu\nu} - 2\,\de^\nu\,h_{\mu\nu}\,\de_\vrho\,h^{\mu\vrho} + 2\,\de_\mu\,h_\nu{}^\nu\,\de_\vrho\,h^{\mu\vrho} - \de_\mu\,h_\nu{}^\nu\,\de^\mu\,h_\vrho{}^\vrho),
\end{equation}  
which is invariant under the whole group of linearised diffeomorphisms. It is convenient to introduce the analogue of the gauge-invariant Maxwell field strength:\footnote{The gauge invariant field strength $f_{\mu\nu\vrho}$ was used introduced in \cite{Rovere:2024nwc} in studying the \textsc{brst} cohomology of covariant fracton theory and it was used in \cite{Rovere:2025nfj} in writing the action and comparing with the linearisation of the torsion formulation of General Relativity and its extensions. As pointed out in \cite{Hinterbichler:2025ost}, a formally equivalent tensor appeared previously in \cite{Deser:2006zx}, where the theory of partially massless spin-two field in de Sitter space is investigated.}
\begin{equation}\label{ActionIntro}
f_{\mu\nu\vrho} = \de_\mu\,h_{\nu\vrho} - \de_\nu\,h_{\mu\vrho},
\end{equation}
such that
\begin{equation}
\delta_\lambda\,f_{\mu\nu\vrho} = 0.
\end{equation}
Notice that $f_{\mu\nu\vrho}$ satisfies a cyclic identity and a Bianchi identity
\begin{equation}
f_{\mu\nu\vrho}+f_{\nu\vrho\mu}+f_{\vrho\mu\nu} = 0, \quad
\de_\sigma\,f_{\mu\nu\vrho} + \de_\mu\,f_{\nu\sigma\vrho} + \de_\nu\,f_{\sigma\mu\vrho} = 0.
\end{equation}
The cyclic identity implies that there are only two independent quadratic scalar contractions of $f_{\mu\nu\vrho}$, which can be chosen to be
\begin{equation}
f^{\mu\nu\vrho}\,f_{\mu\nu\vrho}, \quad
f^{\mu\nu}{}_\nu\,f_{\mu\vrho}{}^\vrho.
\end{equation} 
Therefore, the most general quadratic action, left invariant by the gauge transformation, is an arbitrary combination of the two independent contractions of two field strengths:
\begin{equation}\label{FractonActionWithF}
S^{\text{(fr)}} = \int\diff^d x\,(\alpha\,f^{\mu\nu\vrho}\,f_{\mu\nu\vrho} + \beta\,f^{\mu\nu}{}_\nu\,f_{\mu\vrho}{}^\vrho),
\end{equation}
where $\alpha$ and $\beta$ are free constant parameters.\footnote{Note that covariant fracton theory actually depend only on one free parameter, as either $\alpha$ or $\beta$ can be absorbed by a field redefinition. We will keep both the constants for future convenience.} Making the field strength explicit,
\begin{equation}
S^{\text{(fr)}} = \int\diff^d x\,(-2\,\alpha\,h^{\mu\nu}\,\de^2\,h_{\mu\nu} -\beta\,h\,\de^2\,h- (2\,\alpha-\beta)\,\de_\mu\,h^{\mu\nu}\,\de^\vrho\,h_{\vrho\nu} + 2\,\alpha\,h\,\de_\mu\,\de_\nu\,h^{\mu\nu}),
\end{equation}
where $h=h_\mu{}^\mu$. The action \eqref{FP} is recovered when 
\begin{equation}\label{FractonToEinstein}
\tfrac{\beta}{\alpha} = -2,
\end{equation}
with the normalisation $\alpha = \frac{1}{4}$.

The study of this model was stimulated by the relation with \emph{fractons}. ``Fractons" are called excitations with restricted mobility \cite{Pretko:2017xar, Burnell:2021reh}, which appear in  lattice spin models \cite{Haah:2011drr, Vijay:2015mka}, whose low-energy continuous limit \cite{Affleck:1986} can be captured by effective field theories \cite{Seiberg:2020bhn}. In these models not only the charge is conserved, but also the dipole moment. These features can be described by a pair of gauge fields, $\vphi$ and $A_{ij}$, where $i,j = 1,\dots d$ are space indices, $A_{ij}$ being a symmetric rank-two tensor, with scalar gauge transformations 
\begin{equation}
\delta \vphi = \de_t\,\lambda, \quad
\delta A_{ij} = \de_i\,\de_j\,\lambda.
\end{equation}
Indeed, a coupling of the form 
\begin{equation}
-\vrho\,\vphi + J^{ij}\,A_{ij}
\end{equation}
for some fractonic matter current $\vrho$ and $J^{ij}$, leads to a higher-derivative continuity equation 
\begin{equation}\label{ContinuityEq}
\de_t\,\vrho + \de_i\,\de_j\,J^{ij} = 0.
\end{equation}
It implies that both the charge and the dipole moment 
\begin{equation}
Q = \int \diff^{d-1} x\,\vrho,\quad
\vec{P}=\int\diff^{d-1} x\,\vec{x}\,\vrho
\end{equation}
are conserved
\begin{equation}
\frac{\diff Q}{\diff t} = \frac{\diff \vec{P}}{\diff t} = 0.
\end{equation}
Observed that the dipole-moment conservation implies for a single charged particle to be fixed in space, whereas dipoles are free to move. Indeed, for a single particle with charge $q$ in the point $\vec{x}_0$ the charge density is 
\begin{equation}
\vrho(\vec{x}) = q\,\delta^{d-1}(\vec{x}-\vec{x}_0),
\end{equation}
so that $\vec{P} = q\,\vec{x}_0$. Since $\vec{P}$ is conserved, the position $\vec{x}_0$ cannot change. Instead, for a pair of charges $q_1 = -q_2 = q$ at positions $\vec{x}_1$ and $\vec{x}_2$ forming a dipole, only the relative position is fixed: since 
\begin{equation}
\vrho(\vec{x})=q\,\delta^{d-1}(\vec{x}-\vec{x}_1)-q\,\delta^{d-1}(\vec{x}-\vec{x}_2),
\end{equation}
then $\vec{P} = q\,(\vec{x}_1-\vec{x}_2)$.

These models manifestly break the Lorentz invariance. Nevertheless, they can be recovered in the framework of the gauge theory of the field $h_{\mu\nu}$, with gauge transformation \eqref{GaugeIntro}, by setting \cite{Bertolini:2022ijb}
\begin{equation}
h_{0\mu} = \de_\mu\,\vphi, \quad 
h_{ij}=A_{ij}.
\end{equation}
The covariant minimal coupling 
\begin{equation}
j^{\mu\nu}\,h_{\mu\nu}
\end{equation}
for some covariant fractonic current $j^{\mu\nu}$, which, as a consequence of gauge invariance, satisfies the equation 
\begin{equation}
\de_\mu\,\de_\nu\,j^{\mu\nu} = 0,
\end{equation}
becomes the original continuity equation \eqref{ContinuityEq}, if 
\begin{equation}
\vrho = \de_t\,j^{00} + \de_i\,j^{0i}, \quad 
J^{ij} = j^{ij}.
\end{equation}

The gauge transformations \eqref{GaugeIntro} can be translated in \textsc{brst} formalism by promoting the gauge parameter to a scalar anticommuting ghost $\lambda$:
\begin{equation}\label{BRSfractonrules}
s\,h_{\mu\nu} = \de_\mu\,\de_\nu\,\lambda, \quad
s\,\lambda = 0.
\end{equation}
These transformations are trivially nilpotent.\footnote{In \cite{Fecit:2025eet} one can also find the Batalin-Vilkovisky (\textsc{bv}) formulation for covariant fracton theory and some first-quantised worldline models are introduced and investigated, reproducing the BV transformation at spacetime level.}

As an example of \emph{fracton matter} covariantly couple to the fracton gauge field, consider a real scalar field $\vphi$ with mass dimension $[\vphi]=\frac{d-2}{2}$, in such a way that the kinetic term $\de\,\vphi\,\de\,\vphi$ does not depend on any mass scale. Introduce a scaled fracton gauge field $h^{(\mu)}_{\mu\nu}$, where $\mu$ is a mass scale ($[\mu]=1$), such that the kinetic term $\sim \de\,h^{(\mu)}\,\de\,h^{(\mu)}$ does not explicitly depend on the mass scale. This implies $[h^{(\mu)}_{\mu\nu}] = \frac{d-2}{2}$, and so we can define $h^{(\mu)}_{\mu\nu} = \mu^{\frac{d}{2}-3}\,h_{\mu\nu}$. We require the coupling between $h^{(\mu)}_{\mu\nu}$ and the Noether current $j^{\mu\nu}$ to be minimal. We can argue that $j \sim \de^m\,\vphi^n$, with $m, n$ positive integers, and with the derivatives arranged in all the possible ways. The integers $m,n$ should satisfy $m\geqslant 2$, since we need two free indices, and obviously $n \geqslant 1$. Since $[h_{\mu\nu}^{(\mu)}\,j^{\mu\nu}]=d$, $j^{\mu\nu}$ should have dimension $[j^{\mu\nu}]=\frac{d+2}{2}$. The unique possibility for each $d$ is $m=2$, $n=1$, so that the current reads
\begin{equation}\label{CovConsLawFracton}
j_{\mu\nu} = \de_\mu\,\de_\nu\,\vphi.
\end{equation}
The simplest theory with this Noether current is defined by the following higher derivative action:
\begin{equation}
S = \frac{1}{2\,\mu^2}\int \diff^d x\,\de_\mu\,\de_\nu\,\vphi\,\de^\mu\,\de^\nu\,\vphi.
\end{equation}
Indeed, the action is invariant under the global shift
\begin{equation}
\delta\,\vphi = \mu^2\,\lambda^{(\mu)},
\end{equation}
$\lambda^{(\mu)}$ being a constant with dimension $[\lambda^{(\mu)}]=\frac{d}{2}-3$, and the associated Noether current is the desired one.

A more interesting example of fracton matter can be obtained by considering the covariant version of the fracton matter introduced in \cite{Pretko:2020cko}. Consider a complex scalar field $\vphi$ and the following $\text{U}(1)$ transformation
\begin{equation}
\delta\,\vphi = i\,\lambda\,\vphi, \quad
\delta\,\overline{\vphi} = -i\,\lambda\,\overline{\vphi},
\end{equation}
where $\overline{\vphi}$ is the complex conjugate of $\vphi$ and $\lambda$ is the fracton gauge parameter in \eqref{GaugeIntro}. The fundamental observation is that the combination
\begin{equation}
\vphi\,\de_\mu\,\de_\nu\,\vphi + \de_\mu\,\vphi\,\de_\nu\,\vphi - i\,h_{\mu\nu}\,\vphi^2
\end{equation}
is covariant. Indeed, 
\begin{equation}
\delta\,(\vphi\,\de_\mu\,\de_\nu\,\vphi + \de_\mu\,\vphi\,\de_\nu\,\vphi - i\,h_{\mu\nu}\,\vphi^2) = 2\,i\,\lambda\,(\vphi\,\de_\mu\,\de_\nu\,\vphi + \de_\mu\,\vphi\,\de_\nu\,\vphi - i\,h_{\mu\nu}\,\vphi^2).
\end{equation}
Similarly for  the complex conjugate:
\begin{equation}
\delta\,(\overline{\vphi}\,\de_\mu\,\de_\nu\,\overline{\vphi} + \de_\mu\,\overline{\vphi}\,\de_\nu\,\overline{\vphi} + i\,h_{\mu\nu}\,\overline{\vphi}^2) = -2\,i\,\lambda\,(\overline{\vphi}\,\de_\mu\,\de_\nu\,\overline{\vphi} + \de_\mu\,\overline{\vphi}\,\de_\nu\,\overline{\vphi} + i\,h_{\mu\nu}\,\overline{\vphi}^2).
\end{equation}
This means that the following action is invariant:
\begin{align}
S_{\text{matter}} &= \int\diff^d x\,
(\overline{\vphi}\,\de^\mu\,\de^\nu\,\overline{\vphi} + \de^\mu\,\overline{\vphi}\,\de^\nu\,\overline{\vphi} + i\,h^{\mu\nu}\overline{\vphi}^2)\,\times\nn\\
& \qquad\quad \times (\vphi\,\de_\mu\,\de_\nu\,\vphi + \de_\mu\,\vphi\,\de_\nu\,\vphi - i\,h_{\mu\nu}\,\vphi^2) = \nn\\
&= \int\diff^d x\,[\overline{\vphi}\,\de^\mu\,\de^\nu\,\overline{\vphi}\,\vphi\,\de_\mu\,\de_\nu\,\vphi + \de_\mu\,\vphi\,\de_\nu\,\vphi\,\de^\mu\,\overline{\vphi}\,\de^\nu\,\overline{\vphi}  \,+\nn\\
& \qquad\quad + (\overline{\vphi}\,\de^\mu\,\de^\nu\,\overline{\vphi}\,\de_\mu\,\vphi\,\de_\nu\,\vphi + \text{c.c.}) \,+\nn\\
& \qquad\quad+ (i\,\overline{\vphi}^2(\vphi\,\de_\mu\,\de_{\nu}\,\vphi + \de_\mu\,\vphi\,\de_\nu\,\vphi) + \text{c.c.})\,h^{\mu\nu} \,+\nn\\
& \qquad\quad + h_{\mu\nu}\,h^{\mu\nu}\,(\overline{\vphi}\,\vphi)^2].
\end{align}
The linear term in $h_{\mu\nu}$ individuates the Noether current
\begin{equation}
j_{\mu\nu} = {\tfrac{\delta\,S}{\delta\,h^{\mu\nu}}}{\big{|}_{h\rightarrow 0}} = i\,\overline{\vphi}^2(\vphi\,\de_\mu\,\de_\nu\,\vphi + \de_\mu\,\vphi\,\de_\nu\,\vphi) + \text{c.c.},
\end{equation}
which is on-shell weakly conserved, since
\begin{equation}
\de^\mu\,\de^\nu\,j_{\mu\nu} = i\,(\overline{\vphi}\,\tfrac{\delta\,S}{\delta\,\overline{\vphi}} - \vphi\,\tfrac{\delta\,S}{\delta\,\vphi})|_{h\rightarrow 0},
\end{equation}
where the equations of motion are
\begin{equations}
& \tfrac{\delta\,S}{\delta\,\vphi} = \tfrac{\de\,\mathcal{L}}{\de\,\vphi} - \de_\mu\,(\tfrac{\de\,\mathcal{L}}{\de\,\de_\mu\vphi}) + \de_\mu\de_\nu\,(\tfrac{\de\,\mathcal{L}}{\de\,\de_\mu\de_\nu\vphi}) = 0,\\
& \tfrac{\delta\,S}{\delta\,\overline{\vphi}} = \tfrac{\de\,\mathcal{L}}{\de\,\overline{\vphi}} - \de_\mu\,(\tfrac{\de\,\mathcal{L}}{\de\,\de_\mu\overline{\vphi}}) + \de_\mu\de_\nu\,(\tfrac{\de\,\mathcal{L}}{\de\,\de_\mu\de_\nu\overline{\vphi}}) = 0,
\end{equations}
$\mathcal{L}$ being the Lagrangian density $S_{\text{matter}} = \int \diff^d x\,\mathcal{L}$.

We briefly mention that in \cite{Fecit:2025eet} some worldline models of covariant fracton field theory are presented. In worldline approach, intensively employed in string field theory, one defines a phase space, and \textsc{brst}-quantises it, defining a nilpotent \textsc{brst} charge $Q$ in a suitable subspace $\mathcal{H}$ of the Hilbert space. Then, one considers the most general state $\psi \in \mathcal{H}$, parametrising its components as spacetime fields $f^i$, $\psi = f^i\,\psi_i$, where $\{\psi_i\}_i$ is a basis for $\mathcal{H}$. The \textsc{brst} charge on the state gives the \textsc{brst} transformations on the components
\begin{equation}
g^i\,\psi_i = Q\,\psi = s\,f^i\,\psi_i
\Leftrightarrow
s\,f^i = g^i 
\end{equation} 
and the components $\{ f_i \}_i$ can be identified with the \textsc{bv} spectrum of a field theory. The action of this field theory is the spacetime integral of the mean value of the \textsc{brst} charge on the state
\begin{equation}
S = \int \diff^d x\,\langle \psi,Q\,\psi\rangle.
\end{equation}
The inner product could be represented as a Berezin integral over the first-quantised ghost $c$, which can be treated as a fermion variable
\begin{equation}
\langle\psi,\vphi \rangle = i\int\diff c\,\overline{\psi}\,\vphi.
\end{equation}
As a possible outlook, one can study the Siegel gauge $b\,\psi = 0$ \cite{Siegel:1985tw}, in this setting, where $b$ is the first-quantised antighost associated to $c$, comparing with the \textsc{brst} covariant gauge fixing, where the possible covariant gauge fixing conditions are: 
\begin{equation}
\de^\mu\,\de_\nu\,h_{\mu\nu} = 0, \quad
h_\mu{}^\mu = 0.
\end{equation}

The \textsc{bv} formulation of covariant fracton theory, studied in \cite{Fecit:2025eet} for the first time, amounts in including antifields $h_{\mu\nu}^*$ and $\lambda^*$ besides the fracton gauge field $h_{\mu\nu}$ and the fracton ghost $\lambda$, with ghost number $-1$ and $-2$ respectively. The nilpotent \textsc{brst-bv} transformations, extending \eqref{BRSfractonrules}, read
\begin{equations}
s\,\lambda &= 0,\\
s\,h_{\mu\nu} &= \de_\mu\,\de_\nu\,\lambda,\\
s\,h^*_{\mu\nu} &= -4\,\alpha\,\partial^2\,h_{\mu\nu}+(2\,\alpha-\beta)\,(\partial_\mu\,\partial_\lambda\,h^\lambda{}_\nu + \partial_\nu\,\partial_\lambda\,h^\lambda{}_\mu) \nonumber \\
& \phantom{=} + 2\,\beta\,\partial_\mu\,\partial_\nu\,h-2\,\beta\,\eta_{\mu\nu}\,(\partial^2\,h-\partial_\lambda\,\partial_\sigma\,h^{\lambda\sigma}), \\
s\,\lambda^* &= \de_\mu\,\de_\mu\,h^{*\mu\nu},
\end{equations}
where, as usual in \textsc{bv} formulation, the transformation of $h_{\mu\nu}^*$ reproduces the equations of motion.

\bigskip

The rest of the discussion about covariant fracton theory is divided in two parts: in the first one we address the problem of \emph{anomalies}, using the \textsc{brst} approach \cite{Rovere:2024nwc}.  The topic of anomalies in \textsc{brst} formulation, and the so-called \emph{Stora-Zumino method} in computing anomalies, are also shortly reviewed.  The second part is reserved to the investigation of the relation \cite{Rovere:2025nfj} between covariant fracton gauge theory and an extension of General Relativity in torsion formulation (studied in Section \ref{GenRelTorsion}), which we will call M{\o}ller-Hayashi-Shirafuji theory \cite{Moller1961a, Moller1961b, Moller1978, Hayashi:1967se, Hayashi:1977jd, Hayashi:1979qx}.

\subsection{Anomalies in BRST formalism and Stora-Zumino method}\label{BRSAnomalies}

An anomaly is a breakdown of a classical symmetry in a field theory, at quantum level. If $\Gamma$ is the quantum effective action, and $\mathscr{A}$ is the anomaly functional, then \cite{Bertlmann:2000, Piguet:1995er, Weinberg:1996kr}
\begin{equation}
s\,\Gamma = \mathscr{A}.
\end{equation}
Applying $s$ at both sides of the previous expression, and using the nilpotency of $s$, one obtains a \emph{consistency condition} an anomaly has to fulfill (found for the first time by Wess and Zumino in a different formulation \cite{Wess:1971yu}):
\begin{equation}
s\,\mathscr{A} = 0.
\end{equation}
So, the problem in computing all the possible anomalies amounts in solving the above equation, which does not depend on the peculiar field theory one started with, but only on the enjoyed symmetries by the theory, encoded in the \textsc{brst} operator. The task is to\emph{cohomologically} solve the consistency condition, identifying solutions which differ in $s$-exact terms:
\begin{equation}
\mathscr{A} \sim \mathscr{A} + s\,\mathscr{B}.
\end{equation}
In other words, 
\begin{center}
\emph{Anomalies are elements in the \textsc{brst} cohomology at ghost number one.}
\end{center}
Indeed, if $\mathscr{A} = s\,\mathscr{B}$, $-\mathscr{B}$ is a counterterm which can be added to the action in order to cancel the anomaly out:
\begin{equation}
s\,\Gamma = s\,\mathscr{B} \Leftrightarrow s(\Gamma - \mathscr{B}) = 0.
\end{equation}
It is quite misleading to say that ``a symmetry is anomalous". Rather, if there is an anomaly, no regularisation scheme exists such that all of the classical symmetries are simultaneously preserved in the quantum theory. \textsc{brst} formalism clarifies this aspect, because, using a single operator $s$, we do not distinguish which symmetry is violated. 

An anomaly is a UV effect, since it regards the small scales. So, we can parametrise the anomaly $\mathscr{A}$ as an integral of a local expression:
\begin{equation}
\mathscr{A} = \int \omega^{(d)}_1,
\end{equation}
where $\omega^{(d)}_1$ is a $d$-form with ghost number one in $d$ dimensions. In general, $\omega^{(p)}_g$ denotes a $p$-form of ghost number $g$. The consistency condition becomes
\begin{equation}
s\,\omega^{(d)}_1 = -\diff \omega^{(d-1)}_2,
\end{equation}
for some $\omega^{(d-1)}_2$. Thus, we are interested in computing the local $s$-cohomology on fields and ghosts at ghost number one modulo the external differential $\diff$, which is denoted with $H_1(s|\diff)$:
\begin{equation}
\omega^{(d)}_1 \in H_1(s|\diff)
\end{equation}
Using many times the nilpotency of $s$ and the triviality of the $\diff$-cohomology, one gets a descent of equations, whose bottom equation defines the local $s$-cohomology on zero-forms of ghost number $d+1$, since there is no differential form of negative degree:
\begin{equations}
s\,\omega^{(d)}_1 &= -\diff \omega^{(d-1)}_2,\\
s\,\omega^{(d-1)}_2 &= -\diff \omega^{(d-2)}_3,\\
&\vdots\nn\\
s\,\omega^{(1)}_{d} &= -\diff \omega^{(0)}_{d+1},\\
s\,\omega^{(0)}_{d+1} &= 0.
\end{equations}
Solving the last equation, one goes back up the descent, finding the anomaly $\omega_1^{(d)}$.

Stora and Zumino \cite{Stora:1976kd, Stora:1976LM, Stora:1984, Zumino:1983ew, Zumino:1983rz, Manes:1985df} observed that, if one defines 
\begin{equation}
\delta = s + \diff,
\end{equation}
on the space of \emph{polyforms} of total degree given by the sum of the form degree and of the ghost number 
\begin{equation}
\text{total degree} = \text{form degree} + \text{ghost number}
\end{equation}
($\delta$ increases the total degree by one), the descent can be recasted in a single equation, called \emph{Stora-Zumino equation},
\begin{equation}\label{StoraZuminoEquation}
\delta\,\bm{\omega}_{d+1} =
(s+\diff)\,(\omega^{(d)}_1 + \omega^{(d-1)}_2 + \dots + \omega^{(1)}_d + \omega^{(0)}_{d+1}) = 0,
\end{equation}
where $\bm{\omega}_{d+1}$ is a polyform of total degree $d+1$. The equation \eqref{StoraZuminoEquation} restricts $\bm{\omega}_{d+1}$ to be an element in the $\delta$-cohomology on local polyforms of total degree $d+1$.
\begin{equation}
\bm{\omega}_{d+1} \in H_{d+1}(\delta).
\end{equation}
This means that the $s$-cohomology modulo $\diff$ of ghost number one is isomorphic to the $\delta$-cohomology on polyforms of total degree $d+1$:
\begin{equation}
H_1^{(d)}(s|\diff) \simeq H_{d+1}(\delta).
\end{equation}
Now, the task is finding solutions of the Stora-Zumino equation. It is important to observe that there is a crucial difference between ordinary differential forms and polyforms:
\begin{center}
\emph{ordinary differential forms with form degree} $> d$ \emph{do vanish in} $d$ \emph{dimensions}, \\ \emph{whereas polyforms of total degree} $>d$ \emph{could not vanish in} $d$ \emph{dimensions}.
\end{center}

The simplest example in which a solution of the Stora-Zumino equation \eqref{StoraZuminoEquation} can be exhibited is the Yang-Mills theory (see Section \ref{YMpolyform}). The analogue of the Pontrygin invariant and its relation with the Chern-Simons form using $\bm{F}$ and $\bm{A}$ is
\begin{equation}\label{YMChern}
\text{Tr}\,\bm{F}^{\frac{d+2}{2}} = \delta\,Q_{d+1}(\bm{A},\bm{F}).
\end{equation}
Using the horizontality condition \eqref{HorizontalityYM}, it becomes
\begin{equation}
\text{Tr}\,F^{\frac{d+2}{2}} = \delta\,Q_{d+1}(\bm{A},F).
\end{equation}
But the left-hand side vanishes in $d$ dimensions, being a $d+2$--form. Thus, 
\begin{equation}
\delta\,Q_{d+1}(\bm{A},F) = 0,
\end{equation}
which is the Stora-Zumino equation, $Q_{d+1}(\bm{A},F)$ being a polyform of total degree $d+1$, since the Chern form in $d+1$ is a $(d+1)$--form. Thus, the ghost number one component $Q_1^{(d)}$ of $Q_{d+1}(\bm{A},F)$ is an anomaly. In the abelian case, this is the famous \emph{ABJ anomaly} \cite{Bell:1969ts}.

Summarising, we used two main properties in finding the solution:
\begin{itemize}
\item[1.] The set of polyforms with $\delta$ enjoys the same algebraic relations as the set of ordinary differential forms with $\diff$;
\item[2.] The Yang-Mills polycurvature is horizontal.
\end{itemize}
This method can be also successfully used to compute anomalies in (super)gravity theories \cite{Bonora:1984pz, Bardeen:1984pm, Langouche:1984gn, Baulieu:1985md, Sorella:1992dr, Sorella:1993kq} and in covariant fracton gauge theory. In the gravitational case, the set up is analogous to the Yang-Mills case, both in curved formulation (with affine connection) and in flat formulation (with spin connection). Covariant fracton gauge theory are less known and we will provide the details of the computations 
in the next sections. It is also possible to capture more exotic theories using an extension of the Stora-Zumino method: in \cite{Imbimbo:2023sph} anomalies in four-dimensional $\mathcal{N}=1$ conformal supergravity are investigated; in \cite{Rovere:2024wtv} the case of Kodaira-Spencer gravity is discussed; and in \cite{Rovere:2024nwc} there is the extension to conformal gravity. 

\subsection{BRST fracton cohomology and anomalies}\label{BRSTFracton1}

We want to study the failure of the covariant conservation law \eqref{CovConsLawFracton} in covariant fracton gauge theory at quantum level, that is
\begin{equation}
\de_\mu\,\de_\nu\,\langle j^{\mu\nu}\rangle= \mathscr{A},
\end{equation}
where $\mathscr{A}$ is an anomaly. Using the \textsc{brst} formalism, we have to study the \textsc{brst} cohomology of the model and find solutions of the anomalous descent, which can be equivalently formulated using polyforms, according to the Stora-Zumino approach. 

Let us study the local \textsc{brst} cohomology on the space of $h_{\mu\nu}$, $\lambda$ and their derivatives. The \textsc{brst} transformations \eqref{BRSfractonrules} imply
\begin{equation}
s\,[\tfrac{1}{(p+2)!}\,\de_{(\alpha_1}\dots\de_{\alpha_p}\,h_{\mu\nu)}] = \de_{\alpha_1}\dots\de_{\alpha_p}\,\de_\mu\,\de_\nu\,\lambda,\quad
s\,\hat{h}_{\mu\nu,\alpha_1\dots\alpha_p} = 0,
\end{equation}
where $\hat{h}_{\mu\nu,\alpha_1\dots\alpha_p} := \de_{\alpha_1}\dots\de_{\alpha_p}\,h_{\mu\nu}-\tfrac{1}{(p+2)!}\,\de_{(\alpha_1}\,\dots\de_{\alpha_p}\,h_{\mu\nu)}$. Therefore 
\begin{equation}
\{\tfrac{1}{(p+2)!}\,\de_{(\alpha_1}\,\dots\de_{\alpha_p}\,h_{\mu\nu)},\de_{\alpha_1}\dots\de_{\alpha_p}\,\de_\mu\,\de_\nu\,\lambda\}
\end{equation}
is a doublet for each $p$, so it does not belong to the $s$-cohomology, which turns out to be generated by $\{\lambda, \de_\mu\,\lambda, \hat{h}_{\mu\nu,\alpha}, \hat{h}_{\mu\nu,\alpha\beta}, \dots \}$. Actually, $\hat{h}_{\mu\nu,\alpha\alpha_1\dots\alpha_p}$ can be replaced by the derivatives of $f_{\mu\nu\alpha}$, since $\hat{h}_{\mu\nu,\alpha\alpha_1\dots\alpha_p}$ can be expressed in terms of the derivatives of $f_{\mu\nu\vrho}$, according to the following formula:
\begin{align}
\hat{h}_{\mu\nu,\alpha\alpha_1\dots\alpha_p}& = 
\tfrac{N_p-1}{(p+1)!\,2\,N_p}\,\de_{(\alpha_1}\dots\de_{\alpha_p}\,f_{\alpha)(\mu\nu)}\,+\nn\\
&-\tfrac{1}{(p+1)!\,4\,N_p},\de_{(\alpha_1}\dots\de_{\alpha_{p-1}|}\,\de_{(\mu}\,f_{\nu)|\alpha_p\alpha)},
\end{align}
where $N_p = \frac{(p+2)(p+3)}{2}$, with $p=1,2,\dots$. So, the $s$-cohomology is generated by
\begin{equation}
\{\lambda, \lambda_\mu, f_{\mu\nu\alpha}, \de_\beta\,f_{\mu\nu\alpha}, \de_\beta\,\de_\gamma\,f_{\mu\nu\alpha},\dots \},
\end{equation}
where $\lambda_\mu := \de_\mu\,\lambda$. The following properties will be useful 
\begin{equation}
[\lambda]= 0,\quad
[\lambda_\mu]= 1,\quad
[f_{\mu\nu\vrho}]= 3,\quad
[\de_\sigma\,f_{\mu\nu\vrho}]= 4, \;\;\dots
\end{equation}
\begin{equation}
\lambda^2 = 0, \quad \lambda_\mu\,\lambda^\mu  = 0,
\end{equation}
\begin{equation}
\lambda_{\mu_1}\,\lambda_{\mu_2}\dots\,\lambda_{\mu_{d+1}} = 0 \;\;\text{in}\;d\,\text{dimensions}.
\end{equation}

Let us define the following differential forms:
\begin{equation}
h^{(1)}_\mu := h_{\alpha\mu}\,\diff x^\alpha, \quad
f^{(2)}_\mu := \tfrac{1}{2}\,f_{\alpha\beta\mu}\,\diff x^\alpha\,\diff x^\beta.
\end{equation}
$h^{(1)}_\mu$ and $f^{(2)}_\mu$ are respectively a  one-form and a two-form of ghost number zero, taking values in the cotangent bundle of the spacetime. Both have mass dimension one. $f^{(2)}_\mu$ is the external differential of $h^{(1)}_\mu$ by definition: $f^{(2)}_\mu = \diff h^{(1)}_\mu$. The differential Bianchi identity is a consequence of the nilpotency of the de Rham differential: $ \diff f^{(2)}_\mu = 0$. Notice that the following relations hold
\begin{equation}
\diff x^\mu\,h^{(1)}_\mu = 0, \quad 
\diff x^\mu\,f^{(2)}_\mu = 0.
\end{equation}
The \textsc{brst} transformations \eqref{BRSfractonrules} can be rewritten as 
\begin{equation}
s\,h^{(1)}_\mu = -\,\diff\lambda_\mu,\quad
s\,\lambda = 0,\quad
s\,f^{(2)}_\mu = 0.
\end{equation}

The following polyform of total degree one can be defined
\begin{equation}
H_\mu := h^{(1)}_\mu + \lambda_\mu,
\end{equation}
and $\lambda$ can be viewed as a single component polyform of total degree one. The action of $\delta = s + \diff$ on the space $\{ H_\mu, f^{(2)}_\mu, \lambda, \diff\lambda \}$ consists into two doublets
\begin{equation}
\delta\,H_\mu = f^{(2)}_\mu, \quad
\delta\,f^{(2)}_\mu = 0, \quad
\delta\,\lambda = \diff\lambda, \quad
\delta\,\diff\lambda= 0.
\end{equation}
(Notice that the first transformation could be viewed as an horizontality condition for the field strength $f^{(2)}_\mu$.)
So, the $\delta$-cohomology seems to be trivial. Nevertheless, if we specify the spacetime dimension $d$, there are some polyforms which vanish. So, we are actually working on a quotient space, on which the cohomology is not guaranteed to be trivial. Moreover, passing in components, the two doublets are not independent.

As already observed, a polyform of total degree $>d$ is not necessarily vanishing in $d$ dimensions. We can produce vanishing polyforms in $d$ dimensions by arranging $H_\mu$, $f^{(2)}_\mu$ and $\diff\lambda$ in a polyform whose minimum form degree is at least $d+1$. For example, each polyform involving at least $d + 1$ $\diff x^\mu\,H_\mu$ vanishes in $d$ dimensions, 
\begin{equation}
(\diff x^\mu\,H_\mu)^{d+1} = 0\quad\text{in}\;d\;\text{dimensions,}
\end{equation}
since $\diff x^\mu\,H_\mu = \diff \lambda$. Moreover,
\begin{equation}
H_{\mu_1}\,\dots H_{\mu_{d+1}} = 0 \quad\text{in}\;d\;\text{dimensions.}
\end{equation}

In solving the $d$-dimensional descent we are interested in the $s$-cohomology on the $p$-forms of ghost number $d+1-p$, where $0 \leqslant p \leqslant d$. Let us denote an arbitrary cohomological class in that sector with ${\omega^{\natural}}^{(p)}_{d+1-p}$. The most general expression reads
\begin{align}
{\omega^{\natural}}^{(p)}_{d+1-p} &= A^{\mu_1\dots\mu_{d-p}}{}_{\alpha_1\dots\alpha_p}\,\lambda\,\de_{\mu_1}\,\lambda\dots\de_{\mu_{d-p}}\,\lambda\,\diff x^{\alpha_1} \dots \diff x^{\alpha_p} \,+\nn\\
& + B^{\mu_1\dots\mu_{d+1-p}}{}_{\alpha_1\dots\alpha_p}\,\de_{\mu_1}\,\lambda\dots\de_{\mu_{d+1-p}}\,\lambda\,\diff x^{\alpha_1} \dots \diff x^{\alpha_p}, 
\end{align}
where $A^{\mu_1\dots\mu_{d-p}}{}_{\alpha_1\dots\alpha_p}$ and $B^{\mu_1\dots\mu_{d+1-p}}{}_{\alpha_1\dots\alpha_p}$ are totally antisymmetric in the $\mu$'s and $\alpha$'s, having dimension $2\,p-d$ and $2\,p-1-d$ respectively. Since we can only use $\eta_{\mu\nu}$, $\vepsilon_{\mu_1\dots\mu_d}$ and $f_{\alpha\beta\mu}$ and its derivatives to construct them, and $f_{\alpha\beta\mu}$ has dimension three, $A^{\mu_1\dots\mu_{d-p}}{}_{\alpha_1\dots\alpha_p}$ and $B^{\mu_1\dots\mu_{d+1-p}}{}_{\alpha_1\dots\alpha_p}$ have vanishing dimension or at least dimension three. $A$ [$B$] has vanishing mass dimension only in even [odd] spacetime dimension; the corresponding $B$ [$A$] vanishes, since it would have mass dimension $-1$ [$1$]. Therefore, we can build
\begin{equations}
& {\omega^{\natural}}^{(n)}_{n+1} = \lambda\,\diff\lambda\dots\diff\lambda, \quad
{\tilde{\omega}}^{\natural(n)}_{n+1} = \lambda\,{\star(\diff\lambda\dots\diff\lambda)}, \quad \text{if}\; d = 2\,n, \label{EvenCohoClass}\\
& {\omega^{\natural}}^{(n)}_{n} = \diff\lambda\dots\diff\lambda, \quad \text{if}\; d = 2\,n-1,\label{OddCohoClass}
\end{equations}
where $\star$ denotes the Hodge dual, and $\diff \lambda = \diff x^\mu\,\de_\mu\,\lambda$ and ${\star\diff\lambda} = \de_\mu\,\lambda\,\vepsilon^\mu{}_{\alpha_1\dots\alpha_{d-1}}\,\diff x^{\alpha_1}\dots\diff x^{\alpha_{d-1}}$. By dimensional analysis, there are no cocycles with higher ghost number. Always using dimensional analysis, one can see that the previous cocycle is the unique nontrivial sector for the cohomology in odd dimension. But it is a total derivative. So, the anomaly descent in odd dimension has no nontrivial solution. In conclusion, \emph{there is no fracton anomaly in odd spacetime dimension}. 

In even spacetime dimension, the first class $\lambda\,(\diff\lambda)^n$ does not provide a solution of the descent. Indeed, its external differential $(\diff\lambda)^{n+1}$ sits in the cohomology on ($n+1$)-forms of the same ghost number, so that it cannot be written as an $s$ variation. The opposite parity class $\lambda\,{\star(\diff\lambda)^n}$ provides a solution of the descent only in the two-dimensional case. If $d\geqslant 4$, the descent starts with $s\,\omega^{(n+2)}_{n-1} = 0$, $s\,\omega^{(n+3)}_{n-2} = -\,\diff \omega^{(n+2)}_{n-1}$. Indeed, the cohomology on the $n+1$ forms of ghost number $n$ is trivial. 

Let us find possible solutions of the Stora-Zumino equation \eqref{StoraZuminoEquation} in covariant fracton theory. Consider the case $d=2$. Observe that
\begin{equation}
\delta\,(H_\mu\,{\star\diff x^\mu}) = f_\mu^{(2)}\,{\star\diff x^\mu} = 0,
\end{equation}
since it is a three-form in two-dimensions. Therefore,
\begin{equation}
\delta\,(\lambda\,H_\mu\,{\star\diff x^\mu}) = 
H_\nu\,\diff x^\nu\,H_\mu\,{\star\diff x^\mu} = H_\mu\,H_\nu\,\vepsilon^\mu{}_\vrho\,\diff x^\nu\,\diff x^\vrho =0,
\end{equation}
where in the last step we notice that the dual is $H_\mu\,H^\mu = 0$, since $H_\mu$ is anticommuting. Thus, the component of ghost number one of $\lambda\,H_\mu\,{\star\diff x^\mu}$, which reads
\begin{equation}\label{Anomaly2d}
\omega^{(2)}_1 = \lambda\,\vepsilon^{\mu\nu}\,h^{(1)}_\mu\,\diff x_\nu,
\end{equation}
is an anomaly. Taking the dual, it is proportional to $\lambda\,h_\mu{}^\mu\,\diff^2 x$. Notice that the general discussion on the $s$-cohomology (the sector on zero and two-forms of ghost number three and one in two dimensions are trivial and the cohomology class $\lambda\,\diff\lambda$ does not provide a solution of the descent) shows that the previous anomaly is the \emph{unique} nontrivial  consistent anomaly in two dimensions.

In order to extend the previous result to arbitrary even dimensions $d= 2\,n$, one may consider $\lambda\,H_{\mu_1}\dots H_{\mu_n}\,{\star(\diff x^{\mu_1}\dots \diff x^{\mu_n})}$. But it is easy to see that it fails to be $\delta$ invariant. This allows to conclude that $\lambda\,{\star(\diff\lambda)^n}$ alone does not bring to solution of the descent if $d \geqslant 4$. Nevertheless, if $\lambda\,H_{\mu}$ is replaced by $f^{(2)}_\mu$, preserving the total degree and the mass dimension, one gets a $\delta$ cocycle for any $n \geqslant 2$. Indeed,
\begin{align}
& \delta\,(f^{(2)}_{\mu_1}\,H_{\mu_2}\dots H_{\mu_n}\,{\star(\diff x^{\mu_1}\,\diff x^{\mu_2}\dots\diff x^{\mu_n})}) = \nn\\
= &\,(n-1)\,f^{(2)}_{\mu_1}\,f^{(2)}_{\mu_2}\,H_{\mu_3}\dots H_{\mu_n}\,{\star(\diff x^{\mu_1}\,\diff x^{\mu_2}\,\diff x^{\mu_3}\dots\diff x^{\mu_n})} = 0,
\end{align}
which vanishes since $f^{(2)}_{\mu_1}\,f^{(2)}_{\mu_2}$ is symmetric and $\vepsilon^{\mu_1\mu_2\dots}$ antisymmetric. This is a one-line proof that the component of ghost number one
\begin{align}
\omega_1^{(2n)} &= (n-1)\,f^{(2)}_{\mu_1}\,h_{\mu_2}^{(1)}\dots h_{\mu_{n-1}}^{(1)}\,\de_{\mu_n}\,\lambda\,{\star(\diff x^{\mu_1}\,\diff x^{\mu_2}\dots\diff x^{\mu_{n-1}}\diff x^{\mu_n})},
\end{align}
is an anomaly in $2\,n \geqslant 4$ dimensions.

Many other anomalies are in principle allowed if we consider also the Hodge dual of $f_\mu^{(2)}$, or in general contractions involving the form indices of $f^{(2)}_\mu$. These contractions are unavoidably dimensional dependent. Nevertheless, the anomaly we found before is the unique one in four dimensions. Let us see why. As it follows from the general discussion of the fracton \textsc{brst} cohomology, in four dimensions the problem simply consists in writing down the most general class in the \textsc{brst} cohomology on four-forms of ghost number one. There are two possible structures $\de\,f\,\lambda\,\vepsilon\,(\diff x)^4$ and $f\,\de\,\lambda\,\,\vepsilon\,(\diff x)^4$, but they bring to the same integrated anomaly, as it follows by integrating by parts. Therefore, we consider without loss of generality the second structure. In writing down all the possible contractions, it is easier to consider the dual $f\,\de\,\lambda\,\diff^4 x$, where $\diff^4x$ is the measure of integration in four dimensions. It remains to contract the four indices of $f\,\de\,\lambda$ in all the possible ways allowed by the symmetries of $f$. The result is a unique possibility, $f_{\mu\nu}{}^\nu\,\de^\mu\,\lambda\,\diff^4 x$. But this is precisely the anomaly found before. 

\subsection{Fractons and gravity}\label{FractonTorsion}

In this section covariant fracton gauge theory is formulated in a way in which it will be natural its embedding in gravitational theories in torsion formulation. 

The equations of motion obtained from the action \eqref{FractonActionWithF} impose the following symmetric tensor to vanish
\begin{align}
E^{\text{(fr)}}_{\mu\nu} := -\frac{1}{2}\,\frac{\delta\,S_{\text{fr}}}{\delta\,{h^{(\text{fr})}}^{\mu\nu}} &= 
\alpha\,\de_\vrho\,{f^{\text{(fr)}}}^\vrho{}_{(\mu\nu)} 
-\tfrac{\beta}{2}\,\de_{(\mu}\,{f^{\text{(fr)}}}_{\nu)\vrho}{}^\vrho 
+\beta\,\eta_{\mu\nu}\,\de_\vrho\,{f^{\text{(fr)}}}^{\vrho\sigma}{}_\sigma = \label{EqFractonf}\\
&= 2\,\alpha\,\de^2\,h_{\mu\nu}^{\text{(fr)}} - \tfrac{2\,\alpha-\beta}{2}\,\de_{(\mu}\,\de^\vrho\,h^{\text{(fr)}}_{\nu)\vrho} -\beta\,\de_\mu\,\de_\nu\,h^{\text{(fr)}\,\vrho}{}_\vrho \,+\nn\\
&\quad + \beta\,\eta_{\mu\nu}\,(\de^2\,h^{\text{(fr)}\,\vrho}{}_\vrho-\de_\vrho\,\de_\sigma\,h^{\text{(fr)}\,\vrho\sigma}).
\end{align}
It is useful to define the following tensor
\begin{align}
R^{\text{(fr)}}_{\mu\nu\vrho\sigma} &= 
\tfrac{1}{2}\,\de_\sigma\,\de_\mu\,h^{(\text{fr})}_{\nu\vrho}
-\tfrac{1}{2}\,\de_\sigma\,\de_\nu\,h^{(\text{fr})}_{\mu\vrho}
-\tfrac{1}{2}\,\de_\vrho\,\de_\mu\,h^{(\text{fr})}_{\nu\sigma}
+\tfrac{1}{2}\,\de_\vrho\,\de_\nu\,h^{(\text{fr})}_{\mu\sigma} =\nn\\
&= \tfrac{1}{2}\,\de_\sigma\,f^{\text{(fr)}}_{\mu\nu\vrho} -\tfrac{1}{2}\,\de_\vrho\,f^{\text{(fr)}}_{\mu\nu\sigma}, 
\end{align}
which enjoys the same symmetries as the Riemann tensor, being antisymmetric in $\mu\nu$ and in $\vrho\sigma$ and symmetric in $\mu\nu\leftrightarrow\vrho\sigma$. The corresponding ``Ricci tensor" and ``Ricci scalar" respectively read
\begin{align}
R^{(\text{fr})}_{\mu\nu} &= {R^{\text{(fr)}}}^\vrho{}_{\mu\vrho\nu} = -\tfrac{1}{4}\,\de_\vrho\,{f^{\text{(fr)}}}^\vrho{}_{(\mu\nu)} - \tfrac{1}{4}\,\de_{(\mu}\,{f^{\text{(fr)}}}_{\nu)\vrho}{}^\vrho,\label{Ricci}\\
R^{(\text{fr})} &= {R^{(\text{fr})}}_\mu{}^\mu = -\de_\mu\,{f^{\text{(fr)}}}^{\mu\nu}{}_\nu. \label{RicciScalar}
\end{align}
Writing $\de_{(\mu}\,{f^{\text{(fr)}}}_{\nu)\vrho}{}^\vrho$  and $\de_\mu\,{f^{\text{(fr)}}}^{\mu\nu}{}_\nu$ in terms of the ``Ricci tensor" and of the ``Ricci scalar" in the equations of motion \eqref{EqFractonf}, one recognises the analogue of the linearised Einstein tensor: 
\begin{equation}\label{EqFractonf2}
E^{\text{(fr)}}_{\mu\nu} = 
2\,\beta\,(R^{\text{(fr)}}_{\mu\nu} - \tfrac{1}{2}\,R^{\text{(fr)}}\,\eta_{\mu\nu})
+(\alpha + \tfrac{\beta}{2})\,\de_\vrho\,{f^{\text{(fr)}}}^\vrho{}_{(\mu\nu)},
\end{equation}
where the last term vanishes in the case of linearised Einstein gravity.

\subsection{M{\o}ller-Hayashi-Shirafuji theory}\label{MHSTorsion}

The relation \eqref{HE} shows that the Einstein-Hilbert action consists in a linear combination of three quadratic scalar contractions of the Weitzenb\"ock torsion with precise relative coefficients. In \cite{Moller1961a, Moller1961b, Moller1978, Hayashi:1967se, Hayashi:1979qx} the theory defined by the same action with arbitrary coefficients is considered:
\begin{equation}\label{MollerAction}
S_{\textsc{mhs}} = \frac{1}{2}\int \diff^d x\,e\,(\alpha_1\,T^{\mu\nu\vrho}\,T_{\mu\nu\vrho} + \alpha_2\,T^{\mu\nu\vrho}\,T_{\mu\vrho\nu} + \alpha_3\,T^{\mu\nu}{}_\nu\,T_{\mu\vrho}{}^\vrho).
\end{equation}
We will call the theory described by this action \emph{M{\o}ller-Hayashi-Shirafuji theory} (\textsc{mhs}).\footnote{Hayashi and Shirafuji called it \emph{New General Relativity} \cite{Hayashi:1979qx}. See Appendix \ref{Conventions} for a comparison with definitions and conventions in this paper and in \cite{Moller1978} and in \cite{Hayashi:1979qx}.} 

The peculiar combination picking the case of General Relativity, which is 
\begin{equation}\label{GenRelCond2}
\alpha_1 = \tfrac{1}{4}, \quad 
\alpha_2 = \tfrac{1}{2}, \quad
\alpha_3 = -1,
\end{equation}
is captured by formally requiring the local Lorentz invariance, which indeed is the necessary requirement in the vielbein formulation of General Relativity, beside diffeomorphism invariance, in order to recover the right number of degrees of freedom \cite{Cho:1975dh}. 

The \textsc{mhs} action is the most general quadratic action in the Weitzenb\"ock torsion, since there are only three possible independent quadratic scalar contractions of the torsion. Indeed, the torsion has three irreducible components:\footnote{\label{Young} We denote with $(\lambda_1,\dots,\lambda_n)$ the Young tableau with $n$ rows and $\lambda_i$ boxes in the $i$th row, $i=1,\dots,n$. $(\lambda_1,\dots,\lambda_n)_t$ is the corresponding traceless part.}
\begin{equation}
T_{\mu\vrho\nu} \rightarrow (1,1) \otimes (1) = (1,1,1) \otimes (2,1)_t \otimes (1).
\end{equation}
$(1,1,1)$ is the totally antisymmetric part $T_{[\mu\nu\vrho]}$, and $(1)$ is the unique trace $T_{\mu\nu}{}^\nu$. Namely, the contraction out of two traces is $T_{\mu\nu}{}^\nu\,T^{\mu\vrho}{}_\vrho$; the contraction out of two totally antisymmetric tensors is $T^{[\mu\nu\vrho]}\,T_{[\mu\nu\vrho]}$, which is proportional to $T^{\mu\nu\vrho}\,T_{[\mu\nu\vrho]}$; the scalar contraction involving the remaining irreducible component can be replaced by $T^{\mu\nu\vrho}\,T_{\mu\nu\vrho}$; and the second contraction can be replaced by $T^{\mu\nu\vrho}\,T_{\mu\vrho\nu}$, since 
\begin{equation}
T^{\mu\nu\vrho}\,T_{\mu\vrho\nu} = \tfrac{1}{2}\,T^{\mu\nu\vrho}\,T_{\mu\nu\vrho} - \tfrac{1}{4}\,T^{\mu\nu\vrho}\,T_{[\mu\nu\vrho]}.
\end{equation}
Therefore, the three contractions in \eqref{MollerAction} can be chosen as an independent basis. 
 
Using the variations
\begin{equation}
\delta\,e = -e\,e_\mu{}^a\,\delta\,e^\mu{}_a, \quad
\delta\,T_\mu{}^\vrho{}_\nu = e_{[\mu}{}^a\,D_{\nu]}\,\delta^\vrho{}_a,
\end{equation}
one can compute the equations of motion of the tree pieces in \eqref{MollerAction} \cite{Moller1978}:
\begin{equations}
\frac{e_{\mu a}}{e}\,\frac{\delta}{\delta e^\nu{}_a}\,\int \diff^dx\,e\,T^{\alpha\beta\gamma}\,T_{\alpha\beta\gamma} =&\, - g_{\mu\nu}\,T^{\alpha\beta\gamma}\,T_{\alpha\beta\gamma} 
-2\,T_{\alpha\mu\beta}\,T^\alpha{}_\nu{}^\beta - 4\,T_{\mu\nu}{}^\alpha\,T_{\alpha\beta}{}^\beta \,+\nn\\
& + 4\,T_{\mu\alpha\beta}\,T_\nu{}^{\alpha\beta} - 4\,D_\alpha\,T_{\mu\nu}{}^\alpha,\\
\frac{e_{\mu a}}{e}\,\frac{\delta}{\delta e^\nu{}_a}\,\int \diff^dx\,e\,T^{\alpha\beta\gamma}\,T_{\alpha\gamma\beta} =&\, - g_{\mu\nu}\,T^{\alpha\beta\gamma}\,T_{\alpha\gamma\beta} 
-2\,T_{\mu\alpha\nu}\,T^{\alpha\beta}{}_\beta +2\,T_{\mu\alpha\beta}\,T_\nu{}^{\beta\alpha} \,+\nn\\
& - 2\,T_{\nu\mu\alpha}\,T^{\alpha\beta}{}_\beta -2\,D_\alpha\,T_\mu{}^\alpha{}_\nu - 2\,D_\alpha\,T_{\nu\mu}{}^\alpha,\\
\frac{e_{\mu a}}{e}\,\frac{\delta}{\delta e^\nu{}_a}\,\int \diff^dx\,e\,T^{\alpha\beta}{}_\beta\,T_{\alpha\gamma}{}^\gamma =&\, g_{\mu\nu}\,T^{\alpha\beta}{}_\beta\,T_{\alpha\gamma}{}^\gamma + 2\,g_{\mu\nu}\,D_\alpha\,T^{\alpha\beta}{}_\beta -2\,D_\nu\,T_{\mu\alpha}{}^\alpha.
\end{equations}
Therefore, the equations of motion $E_{\mu\nu} := \frac{e_{\mu a}}{e}\frac{\delta S_{\textsc{mhs}}}{\delta e^\nu{}_a} = 0$, decomposed in symmetric and antisymmetric part $E_{\mu\nu} = E_{\mu\nu}^{(\text{sym})} + E_{\mu\nu}^{\text{(anti)}}$, with $E_{\mu\nu}^{\text{(sym)}} = E_{\nu\mu}^{\text{(sym)}}$, and $E_{\mu\nu}^{\text{(anti)}} = -E_{\nu\mu}^{\text{(anti)}}$, respectively read
\begin{equations}
E_{\mu\nu}^{\text{(sym)}} =&\, 2\,\alpha_1\,T_{\mu\alpha\beta}\,T_{\nu}{}^{\alpha\beta} 
+ \alpha_2\,T_{\mu\alpha\beta}\,T_{\nu}{}^{\beta\alpha} -\alpha_1\,T_{\alpha\mu\beta}\,T^\alpha{}_{\nu}{}^\beta\,+\nn\\
& - (\alpha_1+\tfrac{\alpha_2}{2})(T_{(\mu\nu)}{}^\alpha\,T_{\alpha\beta}{}^\beta + D_\alpha\,T_{(\mu\nu)}{}^\alpha) -\tfrac{\alpha_3}{2}\,D_{(\mu}\,T_{\nu)\alpha}{}^\alpha \,+\nn\\
& - \tfrac{1}{2}\,g_{\mu\nu}\,(\alpha_1\,T^{\alpha\beta\gamma}\,T_{\alpha\beta\gamma} + \alpha_2\,T^{\alpha\beta\gamma}\,T_{\alpha\gamma\beta}-\alpha_3\,T^{\alpha\beta}{}_\beta\,T_{\alpha\gamma}{}^\gamma
-2\,\alpha_3\,D_\alpha\,T^{\alpha\beta}{}_\beta),\label{MollerEqSym}\\
E_{\mu\nu}^{\text{(anti)}} =&\,(\tfrac{\alpha_2}{2}- \alpha_1)\,(T_{[\mu\nu]}{}^\alpha\,T_{\alpha\beta}{}^\beta + D_\alpha\,T_{[\mu\nu]}{}^\alpha) 
-(\alpha_2+\tfrac{\alpha_3}{2})\,(T_{\mu\alpha\nu}\,T^{\alpha\beta}{}_\beta + D_\alpha\,T_{\mu}{}^\alpha{}_{\nu}).\label{MollerEqAnti}
\end{equations}
where in the antisymmetric part we use the trace of the Bianchi identity \eqref{TraceBianchiWeitzenbock}. Using the choice \eqref{GenRelCond2}, the antisymmetric equation is trivial (and this is the unique case) and the symmetric part, using \eqref{RicciWeitzenbock} and \eqref{RicciScalarWeitzenbock}, is recognised to be the Einstein equations. To make this manifest, use \eqref{RicciWeitzenbock} and \eqref{RicciScalarWeitzenbock} to replace $D_{(\mu}\,T_{\nu)\alpha}{}^\alpha$ and $D_\alpha\,T^{\alpha\beta}{}_\beta$:
\begin{align}
D_{(\mu}\,T_{\nu)\alpha}{}^\alpha=&\, -2\,R_{\mu\nu} - T_{\mu\alpha\beta}\,T_{\nu}{}^{\alpha\beta} - T_{\mu\alpha\beta}\,T_{\nu}{}^{\beta\alpha} \,+\nn\\
& + \tfrac{1}{2}\,T_{\alpha\mu\beta}\,T^\alpha{}_{\nu}{}^\beta + T_{(\mu\nu)}{}^\alpha\,T_{\alpha\beta}{}^\beta + D_\alpha\,T_{(\mu\nu)}{}^\alpha,\label{DTRicci}\\
D_\alpha\,T^{\alpha\beta}{}_\beta =&\, -\tfrac{1}{2}\,R-\tfrac{1}{8}\,T^{\alpha\beta\gamma}\,T_{\alpha\beta\gamma} - \tfrac{1}{4}\,T^{\alpha\beta\gamma}\,T_{\alpha\gamma\beta} - \tfrac{1}{2}\,T^{\alpha\beta}{}_\beta\,T_{\alpha\gamma}{}^\gamma.\label{DTRicciScalar}
\end{align}
Moreover, it will be useful for the following to decompose the torsion in the components in transforming in the Young tableaux $(2,1)$ and $(1,1,1)$, the first one including the trace:
\begin{equation}\label{Tdec}
T_{\mu\nu\vrho} = f_{\mu\vrho\nu} + t_{\mu\nu\vrho},
\end{equation}
where\footnote{The inverse relations read: $f_{\mu\nu\vrho} = \tfrac{1}{3}\,(T_{\mu\nu\vrho} + T_{\vrho\mu\nu} - 2\,T_{\nu\vrho\mu})$ and $t_{\mu\nu\vrho} = \tfrac{1}{3}\,(T_{\mu\nu\vrho} + T_{\vrho\mu\nu} + T_{\nu\vrho\mu})$.}
\begin{equation}
f_{\mu\nu\vrho} = -f_{\nu\mu\vrho}, \quad f_{[\mu\nu\vrho]} = 0, \quad
t_{\mu\nu\vrho} = \tfrac{1}{6}\,t_{[\mu\nu\vrho]}.
\end{equation}
These symmetries imply that the following identities are satisfied:
\begin{equations}
f_{\alpha\beta\gamma}\,f^{\alpha\gamma\beta} - \tfrac{1}{2}\,f_{\alpha\beta\gamma}\,f^{\alpha\beta\gamma} = 0,\label{ftids1}\\
f^{\alpha\beta}{}_\beta\,f_{\alpha\mu\nu} - f^{\alpha\beta}{}_\beta\,f_{\mu\nu\alpha} = 0,\\
f_{\alpha\beta\gamma}\,t^{\alpha\beta\gamma} = 0,\\
f_{(\mu}{}^{\alpha\beta}\,t_{\nu)\alpha\beta} + \tfrac{1}{2}\,f^{\alpha\beta}{}_{(\mu}\,t_{\nu)\alpha\beta} = 0,\\
-\tfrac{1}{2}\,f_{\alpha\beta\mu}\,f^{\alpha\beta}{}_\nu + f_{\mu\alpha\beta}\,f_\nu{}^{\alpha\beta} - f_{\mu\alpha\beta}\,f_\nu{}^{\beta\alpha} = 0,\\
f_{\alpha\beta\mu}\,f^{\alpha\beta}{}_\nu + f_{\alpha\beta(\mu}\,f_{\nu)}{}^{\alpha\beta} = 0\label{ftids6}.
\end{equations}
Using the replacement rules \eqref{DTRicci}--\eqref{DTRicciScalar}, the decomposition of the torsion \eqref{Tdec}, and the identities \eqref{ftids1}--\eqref{ftids6} in the equations of motion \eqref{MollerEqSym} and \eqref{MollerEqAnti}, one obtains
\begin{align}
E_{\mu\nu}^{\text{(sym)}} =&\,\alpha_3\,(R_{\mu\nu}-\tfrac{1}{2}\,R\,g_{\mu\nu}) - (4\,\alpha_1-\alpha_2+\tfrac{\alpha_3}{2})\,f_{(\mu}{}^{\alpha\beta}\,t_{\nu)\alpha\beta} \,+\nn\\
& + (\alpha_1 - \alpha_2 - \tfrac{\alpha_3}{4})\,(t_{\mu\alpha\beta}\,t_\nu{}^{\alpha\beta}-\tfrac{1}{2}\,g_{\mu\nu}\,t_{\alpha\beta\gamma}\,t^{\alpha\beta\gamma}) \,+\nn\\
& + (2\,\alpha_1+\alpha_2+\alpha_3)\,(\tfrac{1}{2}\,D_\alpha\,f^\alpha{}_{(\mu\nu)} + \tfrac{1}{2}\,f^{\alpha\beta}{}_\beta\,f_{\alpha(\mu\nu)} \,+\nn\\
& + f_{\mu\alpha\beta}\,f_\nu{}^{\beta\alpha} - \tfrac{1}{4}\,g_{\mu\nu}\,f_{\alpha\beta\gamma}\,f^{\alpha\beta\gamma}),\label{MollerEomSymFinal}\\
E_{\mu\nu}^{\text{(anti)}}  =&\,\tfrac{1}{2}\,(2\,\alpha_1 + \alpha_2 + \alpha_3)\,(D_\alpha\,f^\alpha{}_{[\mu\nu]} + f^{\alpha\beta}{}_\beta\,f_{\alpha[\mu\nu]} \,+\nn\\ 
& + (-2\,\alpha_1 + 2\,\alpha_2 + \tfrac{\alpha_3}{2})\,(D_\alpha\,t^\alpha{}_{\mu\nu} + f^{\alpha\beta}{}_\beta\,t_{\alpha\mu\nu}).\label{MollerEomAntiFinal}
\end{align}
Using the choice \eqref{GenRelCond2}, all the combinations of the constants $\alpha_1, \alpha_2, \alpha_3$ vanish, and the symmetric part gives the Einstein equations.

Let us now study the linearisation of the M{\o}ller-Hayashi-Shirafuji theory \cite{Moller1978, Hayashi:1979qx}. Define
\begin{equation}
e_\mu{}^a = \delta_\mu^a + \vepsilon\,\overline{h}_\mu{}^a, \quad
g_{\mu\nu} = \eta_{\mu\nu} + \vepsilon\,h_{\mu\nu},
\end{equation}
where $\vepsilon$ is infinitesimal. Moreover, defining
\begin{equation}
\overline{h}_\mu{}^\nu = \overline{h}_\mu{}^a\,\delta_a^\nu, \quad
\overline{h}_{\mu\nu} = \overline{h}_\mu{}^\vrho\,\eta_{\vrho\nu} = \overline{h}_\mu{}^a\,\delta_a{}^\vrho\,\eta_{\vrho\nu}
\end{equation}
and combining the linearisation of $g_{\mu\nu}$ and $e_\mu{}^a$, we get that $h_{\mu\nu}$ is the symmetric part $\overline{h}_{\mu\nu} + \overline{h}_{\nu\mu}$. So, denoting with $b_{\mu\nu} = \overline{h}_{\mu\nu} - \overline{h}_{\nu\mu}$ the antisymmetric part, we can decompose 
\begin{equation}\label{DecHBar}
\overline{h}_{\mu\nu} = \tfrac{1}{2}\,(h_{\mu\nu} + b_{\mu\nu}).
\end{equation}
Similarly for the inverse vielbein:
\begin{equation}
e^\mu{}_a = \delta^\mu_a - \vepsilon\,\overline{h}_a{}^\mu, \quad \overline{h}_a{}^\nu\,\delta^a_\mu = \overline{h}_\mu{}^a\,\delta_a^\nu, \quad \overline{h}_b{}^\mu\,\delta^a_\nu = \overline{h}_\nu{}^a\,\delta_b^\mu,
\end{equation}
which are fixed in order to recover $e_\mu{}^a\,e^\nu{}_a = \delta_\mu^\nu$ and $e_\mu{}^a\,e^\mu{}_b = \delta^a_b$. 

The linearised equations of motion are
\begin{align}
\0{E}_{\mu\nu}^{\text{(sym)}} =&\,\alpha_3\,(\0{R}_{\mu\nu}-\tfrac{1}{2}\,\0{R}\,\eta_{\mu\nu}) + \tfrac{1}{2}\,(2\,\alpha_1+\alpha_2+\alpha_3)\,\de_\alpha\,\0{f}^\alpha{}_{(\mu\nu)},\label{MollerEomSymLin}\\
\0{E}_{\mu\nu}^{\text{(anti)}} =&\,(- 2\,\alpha_1 + 2\,\alpha_2 + \tfrac{\alpha_3}{2})\,\de_\alpha\,\0{t}^\alpha{}_{\mu\nu} + \tfrac{1}{2}\,(2\,\alpha_1 + \alpha_2 + \alpha_3)\,\de_\alpha\,\0{f}^\alpha{}_{[\mu\nu]},\label{MollerEomAntiLin}
\end{align} 
where
\begin{align}
\0{T}_{\mu\nu\vrho} &= \0{f}_{\mu\vrho\nu} + \0{t}_{\mu\nu\vrho} = \de_{\mu}\,\overline{h}_{\vrho\nu}- \de_{\vrho}\,\overline{h}_{\vrho\nu} = \\
&= \tfrac{1}{2}\,(\de_\mu\,h_{\vrho\nu} - \de_\vrho\,h_{\mu\nu} + \de_\mu\,b_{\vrho\nu} - \de_\vrho\,b_{\mu\nu}),\\
\0{f}_{\mu\nu\vrho} &= \tfrac{1}{2}\,\de_{[\mu}\,h_{\nu]\vrho} + \tfrac{1}{6}\,(\de_{[\mu}\,b_{\nu]\vrho} + 2\,\de_\vrho\,b_{\nu\mu}),\\
\0{t}_{\mu\nu\vrho} &= \tfrac{1}{6}\,\de_{[\mu}\,b_{\vrho\nu]},\\
\0{R}_{\mu\nu} &= \tfrac{1}{2}\,\de_\vrho\,\0{T}_{(\mu\nu)}{}^\vrho - \tfrac{1}{2}\,\de_{[\mu}\,\0{T}_{\nu)\vrho}{}^\vrho = 
-\tfrac{1}{2}\,\de_\vrho\,\0{f}^\vrho{}_{(\mu\nu)} - \tfrac{1}{2}\,\de_{(\mu}\,\0{f}_{\nu)\vrho}{}^\vrho =\nn\\
&= -\tfrac{1}{2}\,\de^2\,h_{\mu\nu} + \tfrac{1}{2}\,\de^\alpha\,\de_{(\mu}\,h_{\nu)\alpha} - \tfrac{1}{2}\,\de_\mu\,\de_\nu\,h_\alpha{}^\alpha, \\
\0{R} &= -2\,\de_\mu\,\0{T}^{\mu\nu}{}_\nu = -2\,\de_\mu\,\0{f}^{\mu\nu}{}_\nu = \de_\mu\,\de_\nu\,h^{\mu\nu} - \de^2\,h_\mu{}^\mu.
\end{align}

Since the \textsc{mhs} theory is formulated in a quadratic way, unlike General Relativity in conventional formulation, the linearised equations of motion can be obtained directly by varying the action \eqref{MollerAction} with the torsion $T_{\mu\nu\vrho}$ replaced by the linearised one $\0{T}_{\mu\nu\vrho}$:
\begin{equation}\label{MollerActionLin}
\frac{1}{2}\int\diff^d\,x\,(\alpha_1\,\0{T}^{\mu\nu\vrho}\,\0{T}_{\mu\nu\vrho}+\alpha_2\,\0{T}^{\mu\nu\vrho}\,\0{T}_{\mu\vrho\nu} +\alpha_3\,\0{T}^{\mu\nu}{}_\nu\,\0{T}_{\mu\vrho}{}^\vrho).
\end{equation}
This action can be obtained directly working at the linear level in the following way. One has to look for the most general quadratic action in the derivatives of the perturbation $\overline{h}_{\mu\nu}$, invariant under linearised diffeomorphisms, which are encoded in the following \textsc{brst} rules:\footnote{They are the linearisation of the \textsc{brst} rules for the vielbein and the  diffeomorphism ghost, which are generated by the Lie derivative: $s\,e_\mu{}^a = \mathcal{L}_\xi\,e_\mu{}^a = \xi^\nu\,\de_\nu\,e_\mu{}^a + e_\nu{}^a\,\de_\mu\,\xi^\nu$, $s\,\xi^\mu = \frac{1}{2}\,\mathcal{L}_\xi\,\xi^\mu = \xi^\nu\,\de_\nu\,\xi^\mu$, setting $\xi^\mu = \vepsilon\,\0{\xi}^\mu$.}
\begin{equation}\label{BRSruleLinMHS1}
s\,\overline{h}_{\mu\nu} = \de_\mu\,\0{\xi}_\nu, \quad s\,\0{\xi}^\mu = 0.
\end{equation}
or equivalently, using \eqref{DecHBar},
\begin{equation}\label{BRSruleLinMHS2}
s\,h_{\mu\nu} = \de_\mu\,\0{\xi}_\nu + \de_\nu\,\0{\xi}_\mu,\quad
s\,b_{\mu\nu} = \de_\mu\,\0{\xi}_\nu - \de_\nu\,\0{\xi}_\mu,\quad \quad s\,\0{\xi}^\mu = 0.
\end{equation}
So, one has to study the local $s$-cohomology of ghost number zero on the jet space of $\overline{h}_{\mu\nu}$ and its derivatives. The result \cite{Rovere:2025nfj} is that the $s$-cohomology is generated by the derivatives of the antisymmetrised combination $\de_{[\mu}\,\overline{h}_{\vrho]\nu}$, which is the linearised torsion $\0{T}_{\mu\nu\vrho}$. Thus, one has to consider the most general action quadratic in $\0{T}_{\mu\nu\vrho}$, which is a combination of the three independent quadratic scalar contractions $\0{T}^{\mu\nu\vrho}\,\0{T}_{\mu\nu\vrho}$, $\0{T}^{\mu\nu\vrho}\,\0{T}_{\mu\vrho\nu}$, $\0{T}^{\mu\nu}{}_\nu\,\0{T}_{\mu\vrho}{}^\vrho$ with arbitrary coefficients, and this is the action \eqref{MollerActionLin}.

The case of General Relativity \eqref{GenRelCond2} can be obtained directly by varying the massless Fierz-Pauli action \eqref{FP}, which is the most general quadratic action in the derivatives of the symmetric part $h_{\mu\nu}$, invariant under linearised diffeomorphisms
\begin{equation}
s\,h_{\mu\nu} = \de_\mu\,\0{\xi}_\nu + \de_\nu\,\0{\xi}_\mu, \quad s\,\0{\xi}^\mu = 0.
\end{equation}
Indeed, the symmetric part is the only propagating part of $\overline{h}_{\mu\nu}$ in General Relativity. If one starts with $\overline{h}_{\mu\nu}$, the antisymmetric $b_{\mu\nu}$ can be eliminated by modifying the \textsc{brst} rules  \eqref{BRSruleLinMHS2}. One introduces a new \textsc{brst} differential operator $s_{\textsc{fp}}$, adding an antisymmetric shift in the $b_{\mu\nu}$ transformation
\begin{equation}
s_{\textsc{fp}}\,h_{\mu\nu} = \de_\mu\,\0{\xi}_\nu + \de_\nu\,\0{\xi}_\nu,\quad
s_{\textsc{fp}}\,b_{\mu\nu} = \de_\mu\,\0{\xi}_\nu - \de_\nu\,\0{\xi}_\nu + 2\,\0{\Omega}_{\nu\mu}, \quad s_{\textsc{fp}}\,\0{\xi}_\mu = 0, \quad s_{\textsc{fp}}\,\0{\Omega}_{\mu\nu} = 0,
\end{equation}
or equivalently, putting the first two transformations together,
\begin{equation}
s_{\textsc{fp}}\,\overline{h}_{\mu\nu} = \de_\mu\,\0{\xi}_\nu + \0{\Omega}_{\nu\mu}, \quad s_{\textsc{fp}}\,\0{\xi}_\mu = 0, \quad s_{\textsc{fp}}\,\0{\Omega}_{\mu\nu} = 0,
\end{equation}
where $\0{\Omega}_{\mu\nu} = -\0{\Omega}_{\nu\mu}$ is anticommuting.  Redefining $\tilde{\Omega}_{\nu\mu} =: \de_\mu\,\0{\xi}_\nu - \de_\nu\,\0{\xi}_\mu + 2\,\0{\Omega}_{\nu\mu}$, one realises that the transformation of the antisymmetric part becomes $s_{\textsc{fp}}\,b_{\mu\nu} = \tilde{\Omega}_{\mu\nu}$, $s_{\textsc{fp}}\,\tilde{\Omega}_{\mu\nu} = 0$, so that $(b_{\mu\nu},\tilde{\Omega}_{\nu\mu})$ is a trivial doublet, and the local $s_{\textsc{fp}}$-cohomology does not depend on it \cite{Brandt:1989rd, Piguet:1995er}. On the other hand, the transformation of the symmetric part is the same as before  $s_{\textsc{fp}}\,h_{\mu\nu} = s\,h_{\mu\nu}$. Therefore, the local $s_{\textsc{fp}}$-cohomology on the jet space of $\{h_{\mu\nu},\0{\xi}_\mu,b_{\mu\nu},\tilde{\Omega}_{\mu\nu}\}$ and their derivatives is equivalent to the local $s$-cohomology on the jet space of the starting field space $\{h_{\mu\nu},\0{\xi}_\mu\}$. 

The linearised torsion $\0{T}_{\mu\nu\vrho}$ is not invariant under the modified transformations, but it transforms with respect to $\0{\Omega}_{\mu\nu}$ as a gauge two-form with a spectator index 
\begin{equation}
s_{\textsc{fp}}\,\0{T}_{\mu\nu\vrho} = \de_\mu\,\0{\Omega}_{\nu\vrho}-\de_\vrho\,\0{\Omega}_{\nu\mu}.
\end{equation}
Requiring the arbitrary combination in \eqref{MollerActionLin} to be invariant with respect to the above transformations, one selects the peculiar combination \eqref{GenRelCond2}, which indeed gives the Einstein-Hilbert action in the non-linear case. Consistently, the action \eqref{MollerActionLin} with coefficients \eqref{GenRelCond2} does not depend on the antisymmetric part $b_{\mu\nu}$ by integrating by parts.

\subsection{Particle spectrum and fracton embedding}

The space of the solutions of covariant fracton gauge theory $\mathscr{M}_{\text{fr}}$ (\emph{moduli space}) is the submanifold in the space of fields $\{h^{\text{(fr)}}_{\mu\nu}\}$, identified by the equations \eqref{EqFractonf} or \eqref{EqFractonf2}, modulo the gauge invariance, which amounts to linearised longitudinal diffeomorphisms,\footnote{In this section $\lambda$ is a simple gauge parameter. We maintain the same notation of the corresponding ghost because this should not be cause confusion.}
\begin{equation}\label{FractonGauge}
\delta_{\text{fracton}}\,h_{\mu\nu}^{\text{(fr)}} = \de_\mu\,\de_\nu\,\lambda,\;\;\forall\;\alpha, \beta.
\end{equation}
This means that two solutions $h^{\text{(fr)}}_{\mu\nu}(x)$ and $h'^{\text{(fr)}}_{\mu\nu}(x)$ of the equations of motion, which differ by a fracton gauge transformation are identified in the moduli space. 

When $\frac{\beta}{\alpha} = -2$, as in \eqref{FractonToEinstein}, the theory reduced to linearised Einstein gravity, and the gauge invariance is extended to all linearised diffeomorphisms:
\begin{equation}\label{EinsteinGauge}
\delta_{\text{diff}}\,h_{\mu\nu}^{\text{(fr)}} = \de_\mu\,\xi_\nu + \de_\nu\,\xi_\mu,\;\;\text{if}\;\; \tfrac{\beta}{\alpha} = -2.
\end{equation}
There is also a case in which the fracton theory is traceless, that is, it does not depend on $h^{\text{(fr)}\vrho}{}_\vrho$, when $h_{\mu\nu}^{\text{(fr)}}$ is decomposed in its traceless part and in its trace. One can check that this happens if and only if
\begin{equation}
\tfrac{\beta}{\alpha} = -\tfrac{2}{d-1}.
\end{equation}
In that case, there is an additional gauge symmetry, whose effect is to shift the trace, corresponding to a linearised Weyl scaling of $h^{\text{(fr)}}_{\mu\nu}$:
\begin{equation}\label{WeylGauge}
\delta_{\text{Weyl}}\,h^{\text{(fr)}}_{\mu\nu} = 2\,\sigma\,\eta_{\mu\nu}, \;\;\text{if}\;\; \tfrac{\beta}{\alpha} = -\tfrac{2}{d-1}.
\end{equation}

Let us find the general solution of covariant fracton gauge theory. For definiteness, we consider the four dimensional case from now on. The problem was addressed in \cite{Afxonidis:2023pdq, Afxonidis:2024tph}, where the particle content of the theory was studied: except for some special cases, the theory describes five propagating degrees of freedom, with helicities $0, \pm 1, \pm 2$ respectively. Working in Fourier space, $p^\mu$ denotes the momentum and $\tilde{h}^{\text{(fr)}}_{\mu\nu}$ the Fourier transform of the $h^{\text{(fr)}}_{\mu\nu}$. The equations \eqref{EqFractonf2} in Fourier space are:
\begin{align}
K_{\mu\nu}{}^{\vrho\sigma}(p^2;\alpha,\beta)\,\tilde{h}^{\text{(fr)}}_{\vrho\sigma} &= 2\,\alpha\,p^2\,\tilde{h}^{\text{(fr)}}_{\mu\nu} - \tfrac{1}{2}\,(2\,\alpha - \beta)\,p_\alpha\,p_{(\mu}\,\tilde{h}^{\text{(fr)}}_{\nu)}{}^\alpha \,+\nn\\
& - \beta\,p_\mu\,p_\nu\,\tilde{h}^{\text{(fr)}} + \beta\,\eta_{\mu\nu}\,(p^2\,\tilde{h}^{\text{(fr)}} - p_\alpha\,p_\beta\,\tilde{h}^{\text{(fr)}\,\alpha\beta}) = 0,\label{FractonEomsFourier}
\end{align}
where $K(p^2;\alpha,\beta)$, viewed as a $10\times 10$ matrix, identifies a system of homogeneous linear equations in the $10$ variables $\{\tilde{h}^{\text{(fr)}}_{\mu\nu}\}$. Trivial or pure-gauge solutions are those which solve the system without any condition on the momentum, and they correspond to the gauge invariance of the theory. The number of independent trivial solutions is given by the dimension of the kernel of off-shell $K(p^2;\alpha,\beta)$ (null eigenspace, whose dimension is the number of variables decreased by the matrix rank). One can show that
\begin{equation}
\#\,\text{trivial solutions} = \dim\,\text{ker}\,K(p^2 \neq 0;\alpha,\beta)=
\begin{cases}
6 &\text{if}\;\alpha=0,\\
4 &\text{if}\;\tfrac{\beta}{\alpha}=-2,\\ 
2 &\text{if}\;\tfrac{\beta}{\alpha}=-\tfrac{2}{3},\\
1 &\text{otherwise}.
\end{cases}
\end{equation}
$p_\mu\,p_\nu\,\lambda$ is always a trivial solution; in the linearised Einstein case $\frac{\beta}{\alpha}=-2$, this solution is encompassed in $p_{(\mu}\,\xi_{\nu)}$, which are four independent trivial solutions; in the traceless case $\frac{\beta}{\alpha}=-\frac{2}{3}$, besides $p_\mu\,p_\nu\,\lambda$, $2\,\eta_{\mu\nu}\,\sigma$ is trivial too. There is one peculiar case more, $\alpha = 0$, in which there is no kinetic term for the fracton gauge field, so that the theory is expected to have no propagating degrees of freedom in this case \cite{Blasi:2022mbl}.

In the massless case, the momentum can be canonically parametrised as $p^\mu = (1,0,0,1)$, such that $p^2 = 0$. The dimension of the kernel in the massless case gives the number of independent solutions \emph{without} taking into account the gauge redundancy. The result is
\begin{equation}
\dim\,\text{ker}\,K(p^2=0,\alpha,\beta) = \begin{cases} 
8 \quad \text{if}\;\frac{\beta}{\alpha}=2,\\
6 \quad\text{otherwise}.
\end{cases}
\end{equation}
The case $\frac{\beta}{\alpha}=2$ is special, because the contraction $p^\alpha\,\tilde{h}^{(\text{fr})}_{\alpha\mu}$ disappears from the equations \eqref{FractonEomsFourier}. Subtracting the number of trivial solutions to the dimension of the kernel in the various cases, one gets the number of degrees of freedom of the theory:
\begin{equation}
\#_{\text{dof}} = 
\begin{cases}
6-6 = 0, &\text{if}\;\alpha=0,\\
6-4 = 2, &\text{if}\;\tfrac{\beta}{\alpha}=-2\;\text{(linearised gravity)},\\
6-2 = 4, &\text{if}\;\tfrac{\beta}{\alpha}=-\tfrac{2}{3}\;\text{(traceless limit)},\\
8-1 = 7, &\text{if}\;\tfrac{\beta}{\alpha}=2, \\
6-1 = 5, &\text{otherwise.}
\end{cases}
\end{equation}
where we see that, as expected,  the theory has no propagating degrees of freedom when $\alpha = 0$. In the case of linearised Einstein gravity, the two degrees of freedom must describe the graviton, with helicities $\pm 2$. In the third case, we expect that the tracelessness condition eliminates a scalar particle, so that in the general case, which has five degrees of freedom, at least a scalar particle is included in the spectrum. Since the general case encompasses also the second, the graviton is also included. Only two degrees of freedom remain to be identified: they turn out to describe a particle with helicities $\pm 1$, as confirmed by the explicit solutions of the equations of motion. The two degrees of freedom more in the special case $\tfrac{\beta}{\alpha}=2$ have again helicities $\pm 1$, as we will see in the end of this section. 

Solving the equations of motion means finding a basis for the kernel in the massless case, eliminating the trivial solutions. In the general case, apart from the trivial solutions $p_\mu\,p_\nu\,\lambda$, the other five independent elements in the basis can be chosen to be equal to
\begin{equation}\label{BasisFracton}
\text{diag}\,(-1,-\tfrac{2\,\alpha + \beta}{2\,\beta},-\tfrac{2\,\alpha+\beta}{2\,\beta},1), \;\; \lambda_{\mu\nu}^{\pm 1}, \;\; \lambda_{\mu\nu}^{\pm 2}, \;\;\text{if}\;\beta \neq 0, \alpha \neq 0, \tfrac{\beta}{\alpha} = \pm 2, -\tfrac{2}{3},
\end{equation}
where $\lambda_{\mu\nu}^{s}$ are polarisation matrices with helicity $s=\pm 1, \pm 2$ (the explicit expressions are in \eqref{lambda1}--\eqref{lambda2}), and the diagonal matrix has vanishing helicity. Therefore, in the general case, covariant fracton gauge theory describes particles with helicities $0,\pm 1, \pm 2$ \cite{Afxonidis:2023pdq}. If $\beta = 0$, the scalar solution in \eqref{BasisFracton} is replaced by
\begin{equation}
\text{diag}\,(0,1,1,0),
\end{equation}
which again has vanishing helicity. If $\frac{\beta}{\alpha} = -2$, the trivial solutions $p_{(\mu}\,\xi_{\nu)}$ are completed to a basis for the kernel by $\lambda_{\mu\nu}^{\pm 2}$, as expected in linearised gravity. If $\frac{\beta}{\alpha} = -\frac{2}{3}$, the scalar solution in \eqref{BasisFracton} becomes $\eta_{\mu\nu}$, since $-\frac{2\,\alpha + \beta}{2\,\beta}\rightarrow 1$, so it is trivial, and the theory describes only the modes $\pm 1, \pm 2$. Finally, in the special case $\frac{\beta}{\alpha} = 2$, the basis has eight independent elements: one of them is the trivial one $p_\mu\,p_\nu\,\lambda$; the remaining seven ones can be chosen as
\begin{equation}
\text{diag}\,(-1,-1,-1,1), \;\;
\begin{pmatrix}
0 & -1 & \mp i & 0 \\
-1 & 0 & 0 & 0 \\
\mp i & 0 & 0 & 0 \\
0 & 0 & 0 & 0
\end{pmatrix}, \;\;
\begin{pmatrix}
0 & 0 & 0 & 0 \\
0 & 0 & 0 & -1 \\
0 & 0 & 0 & \mp i \\
0 & -1 & \mp i & 0
\end{pmatrix}, \;\; \lambda_{\mu\nu}^{\pm 2},
\end{equation}
where the diagonal solution has vanishing helicity, $\lambda_{\mu\nu}^{\pm 2}$ has helicities $\pm 2$, and the remaining two pairs of solutions have both helicities $\pm 1$. So, compared to the general case, there is one $\pm 1$ helicity particle more.

As a last observation, consider that, the polarisation matrices $\lambda_{\mu\nu}^{s}$ \eqref{lambda0}--\eqref{lambda2}, in the general solution \eqref{BasisFracton}, are transversal in the massless case, meaning that
\begin{equation}
p^\mu\,\lambda_{\mu\nu}^s = 0, \;\; s=\pm 1, \pm 2,\;\; \text{if}\; p^\mu = (1,0,0,1).
\end{equation}
Instead, the scalar mode is longitudinal 
\begin{equation}
p^\mu\,\text{diag}\,(-1,-\tfrac{2\,\alpha + \beta}{2\,\beta},-\tfrac{2\,\alpha + \beta}{2\,\beta},1)_{\mu\nu} = p_\nu.
\end{equation}
Nonetheless, this implies that the arbitrary solution $\tilde{h}_{\mu\nu}^{(\text{fr})}$ on-shell satisfies the condition
\begin{equation}\label{FractonCond}
p^\alpha\,p_{[\mu}\,\tilde{h}^{\text{(fr)}}_{\nu]\alpha} = 0 \;\;\text{if}\; p^\mu (1,0,0,1),
\end{equation}
which will be useful in the following. 

Let us now find the particle spectrum for the \textsc{mhs} theory (see also \cite{Hayashi:1979qx}), following the same method as in the previous section. It will be convenient to redefine the constants $\alpha_1$ and $\alpha_2$ as
\begin{equation}\label{NewMHSConstants}
\tilde{\alpha}_1 = 2\,\alpha_1 + \alpha_2, \quad
\tilde{\alpha}_2 = 2\,\alpha_1 - \alpha_2.
\end{equation}
The moduli space $\mathscr{M}_{\textsc{mhs}}$ of the linearised \textsc{mhs} theory is the submanifold in the field space $\{h_{\mu\nu}$, $b_{\mu\nu}\}$ identified by the equations of motion \eqref{MollerEomSymLin}--\eqref{MollerEomAntiLin}, modulo the gauge invariance, which amounts to linearised diffeomorphisms acting both on $h_{\mu\nu}$ and $b_{\mu\nu}$,\footnote{In this section $\xi_\mu$, and $\Omega_{\mu\nu}$ are simple gauge parameters. We maintain the same notation of the corresponding ghosts (omitting the dot denoting the linearisation) because this should not be cause confusion.} 
\begin{equation}\label{MHSGauge}
\delta_{\text{diff}}\,h_{\mu\nu} = \de_\mu\,\xi_\nu + \de_\nu\,\xi_\mu, \;
\delta_{\text{diff}}\,b_{\mu\nu} = \de_\mu\,\xi_\nu - \de_\nu\,\xi_\mu,\;\;\forall\;\tilde\alpha_1,\tilde\alpha_2,\alpha_3,
\end{equation}
as in \eqref{BRSruleLinMHS2}. As in covariant fracton gauge theory, there are some particular cases in which the gauge symmetry is enlarged. Linearised Einstein gravity is reproduced when the parameters are as in \eqref{GenRelCond2}, or equivalently $\tilde\alpha_2 = 0, \frac{\alpha_3}{\tilde\alpha_1}=-1$, according to \eqref{NewMHSConstants}, up to an overall constant. In this case there is no dependence on $b_{\mu\nu}$, so it can be shifted away, and this is the effect of local Lorentz invariance, as previously discussed:
\begin{equation}
\delta_{\text{Lorentz}}\,h_{\mu\nu} = 0, \;
\delta_{\text{Lorentz}}\,b_{\mu\nu} = \Omega_{\mu\nu},\;\;\text{if}\;\;\tilde\alpha_2 = 0,\;\tfrac{\alpha_3}{\tilde\alpha_1}=-1.
\end{equation}
The traceless limit in \textsc{mhs} theory is reached if and only if
\begin{equation}
\tfrac{\alpha_3}{\tilde\alpha_1}=-\tfrac{1}{d-1}.
\end{equation}
in which case there is no dependence on the trace $h^\vrho{}_\vrho$ and the gauge invariance includes a Weyl scaling:
\begin{equation}
\delta_{\text{Weyl}}\,h_{\mu\nu} = 2\,\sigma\,\eta_{\mu\nu},\;\;
\delta_{\text{Weyl}}\,b_{\mu\nu} = 0,\;\;\text{if}\;\;\tfrac{\alpha_3}{\tilde\alpha_1}=-\tfrac{1}{d-1}.
\end{equation}
A last limit case is when the theory does not depend on $h_{\mu\nu}$ at all. This happens when $\tilde\alpha_1 = \alpha_3 = 0$, in which case the symmetric equation \eqref{MollerEomSymLin} trivialises, and \textsc{mhs} theory reduces to the theory for a free two-form gauge field (Ramond-Kalb field $\textsc{rk}$ \cite{Kalb:1974yc}):
\begin{equation}
\delta_{\textsc{rk}}\,h_{\mu\nu} = \Delta_{\mu\nu}, \;
\delta_{\textsc{rk}}\,b_{\mu\nu} = \de_\mu\,\xi_\nu - \de_\nu\,\xi_\mu,\;\;\text{if}\;\;\tilde\alpha_1 = \alpha_3 = 0,
\end{equation}
where $\Delta_{\mu\nu}$ is any symmetric rank-two tensor, shifting $h_{\mu\nu}$.

In order to find the general solution of \eqref{MollerEomSymLin}--\eqref{MollerEomAntiLin}, we write them in Fourier space.  Again, we consider from now on the four-dimensional case. The equations read
\begin{equations}
&\tfrac{1}{2}\,\tilde\alpha_1\,p^2\,\tilde{h}_{\mu\nu} - \tfrac{1}{4}\,(\tilde\alpha_1 - \alpha_3)\,p_\alpha\,p_{(\mu}\,\tilde{h}_{\nu)}{}^\alpha - \tfrac{1}{2}\,\alpha_3\,p_\mu\,p_\nu\,\tilde{h} \,+\nn\\
&\quad + \tfrac{1}{2}\,\alpha_3\,\eta_{\mu\nu}\,(p^2\,\tilde{h} - p_\alpha\,p_\beta\,\tilde{h}^{\alpha\beta}) + \tfrac{1}{4}\,(\tilde\alpha_1 + \alpha_3)\,p_\alpha\,p_{(\mu}\,\tilde{b}_{\nu)}{}^\alpha = 0,\label{MHSFourier1}\\
& \tfrac{1}{2}\,\tilde\alpha_2\,p^2\,\tilde{b}_{\mu\nu} - \tfrac{1}{4}\,(\tilde\alpha_1 - 2\,\tilde\alpha_2 + \tilde\alpha_3)\,p^\alpha\,p_{[\mu}\,\tilde{b}_{\nu]\alpha} - \tfrac{1}{4}\,(\tilde\alpha_1 + \alpha_3)\,p^\alpha\,p_{[\mu}\,\tilde{h}_{\nu]\alpha} = 0.\label{MHSFourier2}
\end{equations}
This is a system of homogeneous linear equations in the $10+6$ variables $(\tilde{h}_{\mu\nu},\tilde{b}_{\mu\nu})$.

Looking at the equations alone it is possible to deduce the embedding of the fracton moduli space in the \textsc{mhs} one. Indeed, one sees that, if\footnote{In direct space, this means that the linearised torsion $\0{T}_{\mu\nu\vrho}$, corresponding to the choice \eqref{FractonId}, $\0{T}_{\mu\nu\vrho} = \tfrac{1}{2}\,f_{\mu\vrho\nu}^{\text{(fr)}}$. Moreover, there is no dependence in the projected equations on $\tilde{\alpha}_2$, and that the old constants $\alpha_1$ and $\alpha_2$ appear only in the combination $2\,\alpha_1 + \alpha_2 = \tilde{\alpha}_1$, which means that, had we varied the action with, say, $\alpha_2= 0$, we would have obtained the same result. This is not surprising, since there are only two independent quadratic scalar contractions if the totally antisymmetric part of the torsion is vanishing.}
\begin{equation}\label{FractonId}
\tilde{h}_{\mu\nu} = \tilde{h}_{\mu\nu}^{\text{(fr)}}, \quad
\tilde{b}_{\mu\nu} = 0, \quad
\tilde{\alpha}_1 = 4\,\alpha, \quad
\alpha_3 = 2\,\beta,
\end{equation}
then the first equation \eqref{MHSFourier1} reduces to the fracton equation \eqref{FractonEomsFourier}, and the second equation \eqref{MHSFourier2} reduces to the condition \eqref{FractonCond}. This shows that the fracton moduli space, if $\frac{\beta}{\alpha} \neq 2$, is contained in the subsector with vanishing $b_{\mu\nu}$ of the linearised \textsc{mhs} moduli space, upon using the identification in \eqref{FractonId},
\begin{equation}\label{Embedding}
\mathscr{M}_{\text{fr}}(\tfrac{\beta}{\alpha}\neq 2) \subseteq  {\mathscr{M}_{\textsc{mhs}}^{\text{(lin)}}}_{\big{|}_{b=0}} \subset \mathscr{M}_{\textsc{mhs}}^{\text{(lin)}}.
\end{equation}
Observe that the condition $b_{\mu\nu}= 0$ in \eqref{FractonId} breaks the gauge invariance \eqref{MHSGauge}, but it leaves a residual gauge freedom when $\xi_\mu$ is exact, that is, when it generates longitudinal diffeomorphisms $\xi_\mu \propto \de_\mu\,\lambda$. This is consistently the fracton gauge invariance \eqref{FractonGauge}.

Actually, we can prove an even stronger statement: not only the fracton moduli space is included in the $b_{\mu\nu} = 0$ subsector, but it is also isomorphic to the same subsector. To see why, let us find the most general solution of the full equations \eqref{MHSFourier1}--\eqref{MHSFourier2}. At the very end, we will see that, upon restricting in the $b_{\mu\nu} = 0$ subsector, the most general solution coincides with the fracton one, if $\frac{\beta}{\alpha} \neq 2$. 

A $16 \times 16$ matrix $K(p;\tilde\alpha_1,\tilde\alpha_2,\alpha_3)$ can be introduced, describing the system \eqref{MHSFourier1}--\eqref{MHSFourier2} in the variables $(\tilde{h}_{\mu\nu},\tilde{b}_{\mu\nu})$, viewed as a vector with $10+6$ entries. Except for the special cases in which the gauge invariance is enlarged, the dimension of the kernel in the massless case is 10, and the number of trivial solutions is 4, encoded in $(\tilde{h}_{\mu\nu},\tilde{b}_{\mu\nu})=(p_{(\mu}\,\xi_{\nu)},p_{[\mu}\,\xi_{\nu]})$.
So, there are six degrees of freedom, which can be chosen to be equal to 
\begin{equation}
(\text{diag}\,(-1,-\tfrac{\tilde\alpha_1+\alpha_3}{2\,\alpha_3},-\tfrac{\tilde\alpha_1+\alpha_3}{2\,\alpha_3},1), O), \;\; (\lambda_{\mu\nu}^{\pm 1},O), \;\; (\lambda_{\mu\nu}^{\pm 2},O),\;\; (O,\lambda_{\mu\nu}^0),
\end{equation}
where $O$ is the $4\times 4$ null matrix, and $\lambda_{\mu\nu}^s$ are the polarisation matrices in \eqref{lambda0}--\eqref{lambda2}, with helicities $s=0,\pm 1,\pm 2$. Therefore, $\textsc{mhs}$ theory describes particles with helicities $0, \pm 1, \pm 2$, encoded in the symmetric part $\tilde{h}_{\mu\nu}$, and a scalar particle, encoded in the antisymmetric part $\tilde{b}_{\mu\nu}$ -- this is the expected result, since, as well-known, a two-form gauge field in four dimensions is dual to a scalar particle. Moreover, the subsector with vanishing antisymmetric part is the same as in covariant fracton gauge theory, when the special case $\frac{\beta}{\alpha} = 2$ is excluded, since, using the identification of the constants in \eqref{FractonId}, $\frac{\tilde\alpha_1+\alpha_3}{2\alpha_3} = \frac{2\alpha + \beta}{2\beta}$, so that the first solution is the same as the first solution in \eqref{BasisFracton}. Therefore, the statement in \eqref{Embedding} holds also in the stronger version
\begin{equation}
\mathscr{M}_{\text{fr}}(\tfrac{\beta}{\alpha} \neq 2) = {\mathscr{M}_{\textsc{mhs}}^{(\text{(lin)}}}_{\big{|}_{b=0}}.
\end{equation}

\subsection{Comment on BRST formulation}\label{BRSTorsion}

In this concluding section the \textsc{brst} rules \eqref{BRSfractonrules} defining covariant fracton gauge theory are formulated in an equivalent way, off-shell realising the embedding of covariant fractons in linearised \textsc{mhs} theory at the level of symmetries.

Consider two doublets $(B_{\mu\nu},\tilde{b}_{\mu\nu})$ and $(r_\mu,\tilde{\xi}_\mu)$, with the following transformations
\begin{equations}
& s\,B_{\mu\nu} = \tilde{b}_{\mu\nu}, \quad s\,\tilde{b}_{\mu\nu} = 0,\\
& s\,r_\mu = \tilde{\xi}_\mu, \quad s\,\tilde{\xi}_\mu = 0,
\end{equations}
requiring that both $\tilde{b}_{\mu\nu}$ and $r_\mu$ are commuting with ghost number zero, so that $B_{\mu\nu}$ is anticommuting with ghost number $-1$ and $\tilde{\xi}_\mu$ is anticommuting with ghost number one. Then, using the doublet theorem \cite{Brandt:1989rd, Piguet:1995er}, the local $s$-cohomology on the jet space $\{h^{(\text{fr})}_{\mu\nu}, \lambda \}$ is equivalent to the local $s$-cohomology on the enlarged space including the two doublets $\{h^{(\text{fr})}_{\mu\nu}, \lambda, B_{\mu\nu}, \tilde{b}_{\mu\nu}, r_\mu, \tilde{\xi}_{\mu} \}$. If we redefine $\tilde{b}_{\mu\nu}$ and $\tilde{\xi}_\mu$ in this way
\begin{equation}
\tilde{b}_{\mu\nu} =: b_{\mu\nu} - \de_\mu\,r_\nu + \de_\nu\,r_\mu, \quad 
\tilde{\xi}_\mu =: \0{\xi}_\mu - \tfrac{1}{2}\,\de_\mu\,\lambda,
\end{equation}
for some $b_{\mu\nu}$ and $\0{\xi}_\mu$, the last one being anticommuting, then the consistency of the \textsc{brst} transformations implies that 
\begin{equation}
s\,b_{\mu\nu} = \de_\mu\,\0{\xi}_\nu - \de_\nu\,\0{\xi}_\mu, \quad
s\,\0{\xi}_\mu = 0.
\end{equation}
Notice that
\begin{equation}
s\,(h^{(\text{fr})}_{\mu\nu} + \de_\mu\,r_\nu + \de_\nu\,r_\mu) = \de_\mu\,\0{\xi}_\nu + \de_\nu\,\0{\xi}_\mu
\end{equation}
is the same transformation of the linearised metric $h_{\mu\nu}$ under linearised diffeomorphisms, with ghost $\0{\xi}_\mu$. So, upon defining 
\begin{equation}
h_{\mu\nu} := h_{\mu\nu}^{\text{(fr)}} + \de_\mu\,r_\nu + \de_\nu\,r_\mu, 
\end{equation}
the \textsc{brst} rules \eqref{BRSfractonrules} are equivalent to the transformations of linearised \textsc{mhs} theory \eqref{BRSruleLinMHS2}, together with the transformation of a one-form field $r_\mu$, whose ghost is the fracton ghost $\lambda$, shifted by the diffeomorphism ghost $\0{\xi}_\mu$, and with the transformation of a two-form field $B_{\mu\nu}$, whose transformation is the antisymmetrised derivatives of $r_\mu$, shifted by $b_{\mu\nu}$:
\begin{equation}
\begin{cases}
s\,\lambda = 0 ,\\
s\,h_{\mu\nu}^{(\text{fr})} = \de_\mu\,\de_\nu\,\lambda
\end{cases} 
\Leftrightarrow
\begin{cases}
s\,\0{\xi}_\mu = 0, \quad s\,\lambda = 0,\\
s\,h_{\mu\nu} = \de_\mu\,\0{\xi}_\nu + \de_\nu\,\0{\xi}_\mu,\\
s\,b_{\mu\nu} = \de_\mu\,\0{\xi}_\nu - \de_\nu\,\0{\xi}_\mu,\\
s\,r_\mu = \0{\xi}_\mu - \tfrac{1}{2}\,\de_\mu\,\lambda, \\
s\,B_{\mu\nu} = b_{\mu\nu} - \de_{\mu}\,r_{\nu} + \de_{\nu}\,r_{\mu}.
\end{cases}
\end{equation}
The local $s$-cohomology on $\{h_{\mu\nu}^{\text{(fr)}},\lambda\}$ is equivalent to that on $\{B_{\mu\nu}, h_{\mu\nu}, b_{\mu\nu}, r_\mu, \0{\xi}_\mu, \lambda\}$. In other words, the doublet $(r_\mu,\tilde{\xi}_\mu)$ implements the condition ``$\0{\xi}_\mu = \tfrac{1}{2}\,\de_\mu\,\lambda$", which is implied by ``$r_\mu = 0$"; and the doublet $(B_{\mu\nu},\tilde{b}_{\mu\nu})$ implements the condition ``$b_{\mu\nu} = \de_\mu\,r_\nu - \de_\nu\,r_\mu$", which is implied by ``$B_{\mu\nu} = 0$". In this sense, the covariant fracton gauge transformations can be thought as the restriction of linearised diffeomorphisms to the longitudinal ones, as they were firstly introduced \cite{Blasi:2022mbl}, and covariant fracton gauge theory can be thought as the restriction of the linearised \textsc{mhs} theory in the case in which the antisymmetric part of the linearised vielbein is $\diff$-exact. As a consequence, the linearised torsion is cohomologically equivalent to the fracton field strength in this formulation. Indeed, the linearised vielbein reads
\begin{equation}
\overline{h}_{\mu\nu} := \tfrac{1}{2}\,(h_{\mu\nu} + b_{\mu\nu}) = \tfrac{1}{2}\,h_{\mu\nu}^{\text{(fr)}} + \de_\mu\,r_\nu + \tfrac{1}{2}\,s\,B_{\mu\nu},
\end{equation}
so that the linearised torsion becomes
\begin{equation}
\0{T}_{\mu\vrho\nu} = \de_{[\mu}\,\overline{h}_{\nu]\vrho} = \tfrac{1}{2}\,f_{\mu\nu\vrho}^{\text{(fr)}} + s\,(\tfrac{1}{2}\,\de_{[\mu}\,B_{\nu]\vrho}),
\end{equation}
which, in particular, implies that the totally antisymmetric part of the torsion is cohomologically trivial
\begin{equation}
\0{t}_{\mu\nu\vrho} = s\,(\tfrac{1}{6}\,\de_{[\mu}\,B_{\vrho\nu]}).
\end{equation}
This is the reason why the constants $\alpha_1$ and $\alpha_2$ appear only in the combination $2\,\alpha_1 + \alpha_2$ in embedding covariant fracton gauge theory in linearised \textsc{mhs} theory.

\newpage
\quad
\thispagestyle{empty}

\newpage

\section{Outlooks on covariant fractons}\label{6bis}

A list of possible outlooks and future directions regarding the research on covariant fracton field theory is provided in the following.
\begin{itemize}
\item[--] Among the applications of General Relativity in torsion formulation, there is the linearised dual gravity, whose action is found in \cite{West:2001as}, starting from the Hilbert-Einstein action in terms of Weitzenb\"ock torsion, as in \eqref{HE}. An alternative form is introduced, such that, at linear level, it is the action for the dual graviton $\tilde{h}_{\mu_1\dots\mu_{d-3},\alpha}$.\footnote{See Appendix \ref{DualGraviton} for details on dual gravity.}

A problem which would be interesting to address is studying the dual version of covariant fracton gauge theory on the same line, finding the corresponding dual fracton gauge symmetry. Since a link is established between covariant fractons and M{\o}ller-Hayashi-Shirafuji extension of teleparallel gravity, it is natural to consider the modified action  in \cite{West:2001as}, by inserting arbitrary relative coefficients, and then selecting the sector in the space of solutions with vanishing (dual) antisymmetric part. This should correspond to the particle spectrum of the dual formulation of covariant fracton gauge theory, which in turn is required to be the same as in the direct formulation. 
\item[--] It is well known that gravitational anomalies \cite{Alvarez-Gaume:1983ihn} can be computed in the same fashion as Yang-Mills anomalies, by treating the Levi-Civita connection as a gauge field for the $\text{GL}(d)$ rotation $\de_\nu\,\xi^\nu$, $\xi^\mu$ being the diffeomorphism vector parameter \cite{Bardeen:1984pm}. In \textsc{brst} formalism, there is a unique non-trivial cocycle both in Yang-Mills case and in gravitational one, given by the Chern-Simons polynomial. As discussed in \cite{Imbimbo:2023sph} and \cite{Imbimbo:2025ffw,}, the same paradigm holds and it is generalised in (conformal) $\mathcal{N}=1$ supergravity. Namely, in \cite{Imbimbo:2023sph} it is shown that $\mathcal{N}=1$ superconformal $a$-anomaly can be expressed as a super-Chern-Simons polynomial. In \cite{Imbimbo:2025ffw,} this result is extended to all four-dimensional superconformal anomalies, including the $c$-anomaly, and the fundamental mechanism is individuated, which allows to understand in a new way the Stora-Zumino method for computing anomalies. One considers the graded ring $P$ of polynomials in the polyconnections $\bm{A}$ and polycurvatures $\bm{F}$ of a field theory. The nilpotent operator $\delta = \diff + s$ acts as a coboundary operator on $P$, increasing by one the polynomial degree. The horizontality condition $\bm{F}=F$, which holds both in Yang-Mills case and in gravitational one, can be thought as a constraint on the polynomial ring $P_n$
of polynomials of degree $n$, so that an ideal $W_n$ of polynomials, which vanish thanks to the horizontality condition, is defined. So, the relevant space is the quotient $P_n/W_n$. Anomalies in $d$ dimensions are cocycles in the $\delta$-cohomology in $P_{d+1}/W_{d+1}$, which turns out to be isomorphic to the $\delta$-cohomology on $W_{d+2}$. According to this paradigm, the torsion constraint in first-order formulation of gravitational theories, which allows to write the spin connection in terms of the vielbein and its inverse, can be treated on the same footing as the horizontality condition for the Riemann polycurvature. In $d$ dimensions, the ideal $W_{d+2}$ is generated by the cubic monomials in $\bm{T}$ and $\bm{R}$, and by $\bm{T}$ itself. But we know that an alternative formulation of gravity is allowed, in which the torsion constraint is replaced by the flat condition on the Riemann curvature. It is natural to ask which the geometric structure of anomalies is in gravitational theories in this formulation.

\item[--] Galilean transformations can be obtained as a limit of Lorentz transformations by sending the speed of light to infinity (small speeds compared to the speed of light). In 1965, Lévi-Leblond studied the opposite limit (the speed of light tending to zero) and, inspired by \emph{Alice in Wonderland}, called the result ``Carroll transformations" \cite{Levy-Leblond:1965dsc}. As Lorentz transformation, Galilean or Carroll transformations form groups too. The algebra of these groups can be obtained performing the İnönü-Wigner contraction of the Poincaré algebra \cite{inonu1953contraction}.

In Carroll limit, the invariant interval $\diff s^2 = - c^2\,\diff t^2 + \diff x^2$, taking the limit as $c$ tends to zero, becomes $\diff s^2 = \diff x^2$. This means that there exists an absolute space, on which all observers agree. In Carrollian physics, temporal variations dominate over spatial gradients. In the zero-speed limit, light cones are shrunk to the time axis, meaning any causal connection between event points with distinct spatial parts is excluded. In other words, if the speed of light is zero and the only allowed transmissions are those with speed less than that of light, no transmission is allowed, and no event point can be the cause or effect of any event point. For this reason, the Carrollian limit is also called ``ultralocal”. 

In the Carrollian limit, as in the Galilean one, the metric becomes degenerate, because the eigenvalue corresponding to the temporal coordinate is zero. The zero-signature limit of the metric naturally appears in the Hamiltonian formulation of gravity (\emph{Arnowitt-Deser-Misner formulation} \cite{Arnowitt:1959ah}, intensively studied by DeWitt as a first step in canonical quantisation of gravity \cite{DeWitt:1967yk}). In this formulation, the Hamiltonian of gravitational theory is made up of two terms, a ``kinetic" one and a ``potential" one. The potential term, given by the analogue of integrand of the Hilbert-Einstein action for the three-dimensional spacial part $\gamma_{ij}$ of the metric, is proportional to the signature of the metric, so that in the zero-signature metric only the kinetic term remains, which reads $\tfrac{1}{2}\,G_{ih,mn}\,\pi^{ij}\,\pi^{mn}$, where $\pi^{ij}$ is the canonical momentum, and $G_{ij,mn} = |\gamma|^{-1/2}\,(\gamma_{im}\,\gamma_{jn}+\gamma_{in}\,\gamma_{jm}-\gamma_{ij}\,\gamma_{mn})$ is the DeWitt ``supermetric". Teitelboim \cite{Teitelboim:1978wv} and Isham \cite{Isham:1975ur} argued that the zero-signature theory could be the starting point for a perturbation theory of gravity. Henneaux studied a covariant formulation of such a theory, amounting to a non-Riemannian geometry with a degenerate metric \cite{Henneaux:1979vn}.\footnote{The author thanks A. Campoleoni for providing the text of this paper.} Remarkably, the Hamiltonian constraints obey the Carroll algebra in this limit.

In \cite{Henneaux:2021yzg}, the Carrollian contraction of Lorentz-invariant theories is studied in general. It is shown that there always exist two independent contractions of the same Lorentz-invariant theory, which produce two independent Carrollian theories. To do this, one considers the action of the Lorentz-invariant theory in Hamiltonian formalism, in which Lorentz invariance is not manifest. The two limits consist in turning on only the electric or only the magnetic part of the energy, and are distinguished by how the fields are rescaled with respect to the speed of light. In the magnetic formulation, the action necessarily depends on the kinetic momentum of the field, which acts as a Lagrange multiplier that cannot be eliminated. In the second part of the paper, various examples are discussed, in which the actions in the electric or magnetic limit are written in a manifestly Carroll-invariant way. In particular, the case of linearised gravity theory is briefly discussed. 

A remarkable application of Carrollian limit in High Energy Physics is the so-called \emph{Belinskij-Khalatnikov-Lifshitz limit} (\textsc{bkl}) \cite{Belinsky:1982pk}. The ultrarelativistic limit of gravity captures the dynamics of the gravitational field couple to $p$-form fields near a time singularity, where spacial gradients are set to zero and time derivatives dominate, because the zero-signature limit or the Carroll limit are equivalent to the limit of large Newton constant $G $.\footnote{The potential term in the gravitational Hamiltonian is $\text{sgn}\,|\gamma|^{1/2}\,R(\gamma)$, where $R(\gamma)$ is the spacial Ricci scalar and $\text{sgn}$ is the metric signature. Requiring that this term has the dimension of a energy density in $d$ dimensions, and using as dimensionful constant the gravitational one $2\,\kappa = \frac{16\pi G}{c^4}$, we found that potential term is rescaled as $(2\,\kappa)^{-2}\,\text{sgn}\,|\gamma|^{1/2}\,R(\gamma) = \frac{c^8}{(16 \pi G)^2}\,\text{sgn}\,|\gamma|^{1/2}\,R(\gamma)$, so that all the three limits $G \rightarrow \infty$, $c \rightarrow 0$, and $\text{sgn} \rightarrow 0$ send the potential term to zero.}  This the \textsc{bkl} limit. In particular, for bosonic eleven-dimensional supergravity, it is equivalent to the electric Carrollian limit for both gravity and the three-form potential \cite{Damour:2002et}.

At this point, it would be interesting to see what happens with covariant fraction gauge theory. It has already been observed that the ultralocal nature of Carrollian physics can be thought of as fractonic behaviour, since the latter is characterized by reduced or absent mobility due to the conservation of the dipole moment. Various Lorentz-breaking models have been studied \cite{Bidussi:2021nmp, Figueroa-OFarrill:2023vbj}, but the possibility of obtaining them from covariant formulation of fracton theory, as well as the electric/magnetic Carroll limit of covariant fracton gauge theories, and its differences compared to linearised gravity, has not been explored yet.
\end{itemize}

\newpage
\quad
\thispagestyle{empty}

\newpage

\appendix
\firstsectioneqnums  

\section{Differential forms and Hodge dual}\label{Conventions}

In the following we summarise some useful formulas for treating $p$--form gauge theories using the formalism of differential forms. All the formulas are displayed in arbitrary $d$ dimensions. A $p$--form is defined by
\begin{equation}
A^{(p)} = \frac{1}{p!}\,A_{\mu_1\dots\mu_p}\,\diff x^{\mu_1}\dots\diff x^{\mu_p},
\end{equation} 
where the components $A_{\mu_1\dots\mu_p}$ are completely antisymmetrised. They can be commuting or anticommuting. We denote the statistic degree by $(-)^{|A|}$ or $(-)^A$.

The external differential is defined by
\begin{equation}
\diff = \diff x^\mu\,\de_\mu.
\end{equation}
The internal product is defined by
\begin{equation}
\iota_\xi\,\vphi = 0, \quad \iota_\xi\,\diff x^\mu = \xi^\mu,
\end{equation}
for any scalar $\vphi$ and for any vector field $\xi = \xi^\mu\,\de_\mu$.

$\diff$ is an odd operator since it increases by one the form degree, that is $|\diff|=1$; $\iota_\xi$ is odd or even if $\xi$ is commuting or anticommuting, that is $|\iota_\xi|=|\xi|+1$. The \emph{Cartan formula} gives the Lie derivative when it acts on differential forms:
\begin{equation}
\mathcal{L}_\xi = \iota_\xi\,\diff + (-)^{|\xi|}\,\diff\,\iota_\xi = [\iota_\xi,\diff],
\end{equation}
where the graded commutator is defined by
\begin{equation}
[a,b]=a\,b -(-)^{|a||b|}\,b\,a.
\end{equation}
$\mathcal{L}_\xi$ is even or odd if $\xi$ is commuting or anticommuting, that is $|\mathcal{L}_\xi|=|\xi|$. One can check that
\begin{equation}
[\diff,\mathcal{L}_\xi]=0,\quad
[\iota_\eta,\mathcal{L}_\xi]=\iota_{\mathcal{L}_\xi\eta},\quad
[\mathcal{L}_\xi,\mathcal{L}_\eta]=\mathcal{L}_{\mathcal{L}_\xi\eta}.
\end{equation}

Denote with $g_{\mu\nu}$ the metric tensor, with $g$ its determinant, and with $\text{sgn}(g)$ its signature, so that $g = \text{sgn}(g)\,|g|$. $\vepsilon_{\mu_1\dots\mu_d}$ is Levi-Civita symbol. It has the same components in all the coordinate systems. So it cannot be a true tensor. One chooses a system of coordinates in a reference sequence. The expression
\begin{equation}
\frac{\de x'^{\mu_1}}{\de x^{\alpha_1}}\dots\frac{\de x'^{\mu_d}}{\de x^{\alpha_d}}\,\vepsilon^{\alpha_1\dots\alpha_d}
\end{equation}
is totally antisymmetric in $\alpha_1\dots\alpha_d$, so it is proportional to $\vepsilon^{\alpha_1\dots\alpha_d}$. If $\alpha_1\dots\alpha_d$ is the reference sequence, then the previous expression is equal to $\det\frac{\de x'}{\de x}$ by definition. Then, using the transformation of $g$, we deduce that $|g|^{-1/2}\,\vepsilon^{\alpha_1\dots\alpha_d}$ is a true tensor. Therefore, we can write
\begin{equation}
\vepsilon^{\mu_1\dots \mu_d} = g\,g^{\mu_1 \alpha_1}\dots g^{\mu_d \alpha_d}\,\vepsilon_{\alpha_1\dots \alpha_d}.
\end{equation}
The contraction formula is
\begin{equation}
\vepsilon^{\mu_1\dots\mu_p\mu_{p+1}\dots\mu_d}\,\vepsilon_{\nu_1\dots\nu_p\mu_{p+1}\dots\mu_d} = \text{sgn}(g)\,(d-p)!\,\delta^{\mu_1}_{[\nu_1}\dots\delta^{\mu_p}_{\nu_p]},
\end{equation}
where $[\dots]$ denotes the total antisymmetrisation (without any numerical factor). 

The invariant measure of integration is
\begin{equation}
\diff^d x\,\sqrt{|g|} = \diff x^0 \dots \diff x^{d-1}\,\sqrt{|g|} = \frac{\sqrt{|g|}}{d!}\,\vepsilon_{\alpha_1\dots\alpha_d}\,\diff x^{\alpha_1}\dots\diff x^{\alpha_d}.
\end{equation}
Vice versa,
\begin{equation}
\diff x^{\alpha_1}\dots \diff x^{\alpha_d} = \text{sgn}(g)\,\vepsilon^{\alpha_1 \dots \alpha_d}\,\diff^d x.
\end{equation}

Let us define the \emph{Hodge dual}, which depends on the metric:
\begin{equation}
{\star{(\diff x^{\alpha_1}\,\dots \diff^{\alpha_p})}} = \frac{\sqrt{|g|}}{(d-p)!}\,\vepsilon^{\alpha_1\dots\alpha_p}{}_{\beta_{p+1}\dots\beta_d}\,\diff x^{\beta_{p+1}}\,\dots\diff x^{\beta_d},
\end{equation}
so that
\begin{equation}
{\star{A^{(p)}}} = \frac{1}{p!}\,A_{\mu_1\dots\mu_p}\,{\star{(\diff x^{\mu_1}\dots\diff x^{\mu_p})}}.
\end{equation}
The inverse is
\begin{equation}
\diff x^{\alpha_1}\dots \diff x^{\alpha_p} =  \frac{\text{sgn}(g)\,(-)^{p(d-p)}}{(d-p)!\,\sqrt{|g|}}\,\vepsilon^{\alpha_1\dots \alpha_p}{}_{\beta_{p+1}\dots\beta_d}\,{\star{(\diff x^{\beta_{p+1}}\dots\diff x^{\beta_d})}}.
\end{equation}
In particular,
\begin{equation}
\diff^d x\,\sqrt{|g|} = \star{1}.
\end{equation}

Given a $p$--form $A^{(p)}$ and a $q$-form $B^{(q)}$, the following identities holds:
\begin{equations}
A^{(p)}\,B^{(q)} &= (-)^{AB}\,(-)^{Bp}\,(-)^{Aq}\,(-)^{pq}\,B^{(q)}\,A^{(p)},\\
A^{(p)}\,{\star{B^{(q)}}} &= (-)^{AB}\,(-)^{Bp}\,(-)^{A(d-q)}\,(-)^{p(d-q)}\,{\star{B^{(q)}}}\,A^{(p)},\\
A^{(p)}\,{\star B^{(p)}} &= (-)^{Ap}\,(-)^{AB}\,(-)^{Bp}\,B^{(p)}{\star{A^{(p)}}},\\
{\star{\star{A^{(p)}}}} &= (-)^{p(d-p)}\,\text{sgn}(g)\,A^{(p)}.
\end{equations}
Some consequences
\begin{equations}
A^{(p)} &= {\star{ B^{(d-p)}}} \Rightarrow
{\star{A^{(p)}}} = (-)^{p(d-p)}\,\text{sgn}(g)\,B^{(d-p)},\\
\Rightarrow A^{(p)}\,{\star{A^{(p)}}} &= \text{sgn}(g)\,B^{(d-p)}\,{\star{B^{(d-p)}}}.
\end{equations}
Explicitly,
\begin{equation}
A^{(p)}\,{\star{B^{(p)}}} = \frac{1}{p!}\,(-)^{Bp}\,A^{\mu_1\dots\mu_p}\,B_{\mu_1\dots\mu_p}\,\sqrt{|g|}\,\diff^d x.
\end{equation}

The nilpotent \emph{external differential} is
\begin{equation}
\diff A^{(p)} = \frac{1}{p!}\,\de_\mu\,A_{\mu_1\dots\mu_p}\,\diff x^{\mu}\diff x^{\mu_1}\dots \diff x^{\mu_p},
\end{equation}
or equivalently
\begin{equation}
\diff A^{(p)} = \frac{\text{sgn}(g)}{p!(d-p-1)!\,\sqrt{|g|}}\,\de_\mu\,A_{\mu_1\dots\mu_p}\,
\vepsilon_{\beta_{p+2}\dots\beta_d}{}^{\mu\mu_1\dots\mu_p}\,{\star{(\diff x^{\beta_{p+2}}\dots\beta_d)}}.
\end{equation}

The external differential of the dual is
\begin{equation}
\diff{\star{A^{(p)}}} = \frac{\text{sgn}(g)}{(p-1)!\,\sqrt{|g|}}\,\de_\mu\,(\sqrt{|g|}\,A_{\mu_1 \dots \mu_p}\,g^{\mu_1\beta_1}\dots g^{\mu_{p-1}\beta_{p-1}}\,g^{\mu_p \mu})\,{\star{(\diff x_{\beta_1}\dots \diff x_{\beta_{p-1}})}}.
\end{equation}
The dual of the external differential is
\begin{equation}
{\star{\diff A^{(p)}}} = \frac{(-)^{p(d-p)}}{p!(d-p-1)!\,\sqrt{|g|}}\,\de_\mu\,A_{\mu_1\dots\mu_p}\,\vepsilon_{\beta_{p+2}\dots\beta_d}{}^{\mu\mu_1\dots\mu_p}\,\diff x^{\beta_{p+2}}\dots\diff x^{\beta_d}.
\end{equation}
The dual of the external differential of the dual is
\begin{equation}
{\star{\diff{\star{A^{(p)}}}}} = \frac{(-)^{(p-1)(d-p+1)}}{(p-1)!\,\sqrt{|g|}}\,\de_\mu(\sqrt{|g|}\,A_{\mu_1\dots\mu_p}\,g^{\mu_1\beta_1}\,\dots g^{\mu_{p-1}\beta_{p-1}}\,g^{\mu_p\,\mu})\,\diff x_{\beta_1}\dots \diff x_{\beta_{p-1}}.
\end{equation}
If the metric is constant and $p=1$, the last formula becomes
\begin{equation}
{\star{\diff{\star{A^{(1)}}}}} = \de_\mu\,A^\mu = \text{div}\,A^{(1)}.
\end{equation}

\section{Dual graviton}\label{DualGraviton}

The are several works on the topic of dual graviton (see for example \cite{Thierry-Mieg:1980ihu, Curtright:1980yk, Aulakh:1986cb, Labastida:1986gy, Hull:2001iu, Bekaert:2002dt, Henneaux:2019zod,West:2001as, West:2002jj, Bekaert:2002uh, Bekaert:2004dz, Hohm:2018qhd}). Without presumption of exhaustiveness, we review here some aspects regarding the gauge symmetry and the number of degrees of freedom.

As a preliminary, let us introduce the ``double differential forms". A $p$-form $\xi^{(p)}$ has components given by a totally antisymmetric rank-$p$ tensor $\xi_{\mu_1\dots\mu_p}$,
\begin{equation}
\xi^{(p)} = \tfrac{1}{p!}\,\diff x^{\mu_1}\diff x^{\mu_p}\,\xi_{\mu_1\dots\mu_p}.
\end{equation}
$\xi^{(p,q)}$, where $p, q$ are non-negative integers such that $p \geqslant q$, has components given by a tensor represented by the Young tableau with two column, the first with $p$ boxes, and the second with $q$ boxes.\footnote{It could be useful to report the formulas for the dimensions of the representation of $\text{GL}(n)$ and $\text{SO}(n)$ associated to a Young tableau $(\lambda_1,\lambda_2,\dots,\lambda_k)$, with $k$ rows with length $\lambda_i$, $i=1,\dots,k$. Denote the lengths of the columns with $\vrho_j$, $j=1,\dots,\lambda_1$. The \emph{Hooke number} of the $(i,j)$ entry in the tableau is $h_{ij} = 1+\lambda_i+\vrho_j-i-j$. Defining $\text{den}=\prod_{(i,j)} h_{ij}$, where the product is extended to all the boxes in the diagram, the $\text{GL}(n)$ dimension is \cite{Labastida:1986gy, Murtaza:1973bp}
\begin{equation} 
\text{dum}_{\text{GL}(n)}(\lambda_1,\dots,\lambda_k) = \text{den}^{-1}\,\prod_{i,j} (n-i-j).
\end{equation}
Instead, the $\text{SO}(n)$ dimension is 
\begin{equation}
\text{dim}_{\text{SO}(n)}(\lambda_1,\dots,\lambda_k) = \text{den}^{-1}\,\prod_{i=1}^k \frac{(n-k-1+\lambda_i)!}{(n-2i)!}\prod_{i\leqslant j = 1}^k (n+\lambda_i + \lambda_j)!
\end{equation}
} In the following we will represent such two-columned Young tableaux by writing in square brackets the lengths of the two columns $[p,q]$. Therefore, $\xi_{\mu_1\dots\mu_p,\alpha_1\dots\alpha_q}$ is antisymmetric in $\mu_1\dots\mu_p$, and antisymmetric in $\alpha_1\dots\alpha_q$, with totally antisymmetric part vanishing $\xi_{[\mu_1\dots\mu_p,\alpha_1\dots\alpha_q]} = 0$. Formally, $\xi^{(p,q)}$ can be represented as a differential form with $p$ differentials on the left and $q$ on the right: the left and right differentials implement the antisymmetry in the first $p$ indices and in the last $q$ ones, 
\begin{equation}
\xi^{(p,q)} = \tfrac{1}{p!}\,\diff x^{\mu_1}\dots\diff x^{\mu_p}\,\xi_{\mu_1\dots\mu_p,\alpha_1\dots\alpha_q}\,\diff x^{\alpha_1}\dots\diff x^{\alpha_q}\,\tfrac{1}{q!}.
\end{equation}
An ordinary differential form can be thought as, say, a left-differential form, $\xi^{(p)} = \xi^{(p,0)}$. One can introduce a left-differential and a right-differential:
\begin{equations}
\overset{\rightarrow}{\diff}\,\xi^{(p,q)} &= \tfrac{1}{(p+1)!}\,\diff x^\beta\,\diff x^{\mu_1}\dots\diff x^{\mu_p}\,\de_{[\beta}\,\xi_{\mu_1\dots\mu_p],\alpha_1\dots\alpha_q}\,\diff x^{\alpha_1}\dots\diff x^{\alpha_q}\,\tfrac{1}{q!},\\
\xi^{(p,q)}\,\overset{\leftarrow}{\diff} &= \tfrac{1}{p!}\,\diff x^{\mu_1}\,\dots\diff x^{\mu_p}\,\,\de_{[\beta|}\,\xi_{\mu_1\dots\mu_p,|\alpha_1\dots\alpha_q]}\,\diff x^{\alpha_1}\dots\diff x^{\alpha_p}\,\diff x^\beta\,\tfrac{1}{(q+1)!},
\end{equations}
in such a way that their components are tensors described by the Young tableaux $[p+1,q]$ and $[p,q+1]$ respectively. Using this formalism, the linearised graviton $h_{\mu\alpha}$ is a $(1,1)$--differential form
\begin{equation}
h^{(1,1)} = \diff x^{\mu}\,h_{\mu,\alpha}\,\diff x^{\alpha}.
\end{equation}
which, by definition, has components with vanishing antisymmetric part $h_{[\mu,\alpha]}=0$, meaning that $h_{\mu,\alpha}=h_{\mu\alpha}$. The components of the linearised Riemann tensor
\begin{equation}
R_{\mu\nu,\alpha\beta} = \tfrac{1}{2}\,(\de_{\mu}\,\de_\beta\,h_{\nu\alpha} - \de_\mu\,\de_\alpha\,h_{\nu\beta} - \de_\nu\,\de_\beta\,h_{\mu\alpha} + \de_\nu\,\de_\alpha\,h_{\mu\beta}),
\end{equation}
whose symmetry are according to $[2,2]$, fits in the following $(2,2)$--form
\begin{equation}
R^{(2,2)} = \overset{\rightarrow}{\diff}\,h^{(1,1)}\,\overset{\leftarrow}{\diff}.
\end{equation}
The linearised diffeomorphism transformation generated by $\xi_\mu$ can be written in terms the right-differential of $\xi^{(1)} = \diff  x^\mu\,\xi_\mu$:\footnote{Indeed, $\xi^{(1)}\,\overset{\leftarrow}{\diff} = \diff x^\mu\,\de_\beta\,\xi_\mu\,\diff x^\beta$, but the components has vanishing totally antisymmetric part by definition $\de_{[\beta}\,\xi_{\mu]}=0$, so that $\xi^{(1)}\,\overset{\leftarrow}{\diff} = \diff x^\mu\,\de_{(\beta}\,\xi_{\mu)}\,\diff x^\beta$.}
\begin{equation}
\delta\,h_{\mu\alpha} = \de_{(\mu}\,\xi_{\alpha)} \Leftrightarrow 
\delta\,h^{(1,1)} = \xi^{(1)}\,\overset{\leftarrow}{\diff}.
\end{equation}
Since the right-differential is nilpotent as well as the left one, the Riemann $(2,2)$--form is manifestly gauge invariant. 

We can consider the Hodge dual of the Riemann $(2,2)$--form in order to introduce a dual theory of gravity, at linearised level, in the same way as the Cremmer-Julia duality for $p$-forms. In principle, it is possible to consider the left-dual, the right-dual, or the double dual, both from the left and from the right. The left-dual theory and the right-dual one are equivalent, since double differential forms are symmetric in left and right. Moreover, it has been shown that the theory obtained using a double dual is completely equivalent to linearised gravity (Fierz-Pauli) \cite{Henneaux:2019zod}. We will consider for definiteness the left-dual theory, defining:
\begin{equation}
\tilde{R}^{(d-2,2)} = {\star R^{(2,2)}},
\end{equation}
with components
\begin{equation}
\tilde{R}_{\alpha_1\dots\alpha_{d-2},\vrho\sigma} = \tfrac{1}{2}\,\vepsilon_{\alpha_1\dots\alpha_{d-2}}{}^{\mu\nu}\,R_{\mu\nu,\vrho\sigma}.
\end{equation}
Since $\star\star = (-)^{p(d-p)}\,\text{sgn}$, where $\text{sgn}$ is the sign of the metric, ${\star \tilde{R}^{(d-2,2)}} = -R^{(2,2)}$, choosing the convention $(-,+,\dots,+)$ for the signature of the metric. The Ricci tensor is the trace $R_{\mu\nu} = R_{\mu\lambda,\nu}{}^\lambda$. Similarly, one could define the Ricci tensor of the dual Riemann tensor $\tilde{R}_{\alpha_1\dots\alpha_{d-3},\mu} = \tilde{R}_{\alpha_1\dots\alpha_{d-3}\alpha}{}^\alpha{}_\mu$. The algebraic Bianchi identity $R_{[\mu\nu,\vrho]\sigma}=0$, corresponds to the vanishing of the dual Ricci tensor $\tilde{R}_{\alpha_1\dots\alpha_{d-3},\mu}=0$, which are the dual Einstein equations in vacuum. Similarly, the dual algebraic Bianchi identity $\tilde{R}_{[\alpha_1\dots\alpha_{d-2},\vrho]\sigma}=0$, or $\tilde{R}_{\alpha_1\dots\alpha_{d-3}[\alpha_{d-2},\vrho\sigma]} = 0$, corresponds to the Einstein equations $R_{\mu\nu} = 0$. The differential Bianchi identity, which can be written as $R^{(2,2)}\,\overset{\leftarrow}{\diff}=0$, corresponds to the analogous identity for the dual Riemann $\tilde{R}^{(d-2,2)}\,\overset{\leftarrow}{\diff}=0$.

The dual graviton is defined in such a way that the dual Riemann has the same definition as the Riemann $(2,2)$--form in terms of the linearised graviton $h^{(1,1)}$. This means that the dual graviton must be a $(d-3,1)$--form $\tilde{h}^{(d-3,1)}$, such that
\begin{equation}
\tilde{R}^{(d-2,2)} = \tfrac{1}{2}\,\overset{\rightarrow}{\diff}\,\tilde{h}^{(d-3,1)}\,\overset{\leftarrow}{\diff}.
\end{equation}
The components $\tilde{h}_{\mu_1\dots\mu_{d-3},\alpha}$ are a tensor described by the Young tableau $[d-3,1]$, which means that is totally antisymmetric in $\mu_1,\dots,\mu_{d-3}$, and it has vanishing totally antisymmetric part $\tilde{h}_{[\mu_1\dots\mu_{d-3},\alpha]}=0$.

The gauge transformations of the dual graviton are fixed by requiring the invariance of the dual Riemann form. Consider for definiteness the first non-trivial case, the five-dimensional one -- in three dimensions the dual graviton is a vector, in four dimensions gravity is self-dual. The five-dimension dual graviton is given by $\tilde{h}_{\alpha\beta,\gamma}$, with symmetry described by the Young tableau $[2,1]$, so that $\tilde{h}_{\alpha\beta,\gamma} = - \tilde{h}_{\beta\alpha,\gamma}$, and $\tilde{h}_{[\alpha\beta,\gamma]}=0$. The most natural gauge transformation we can be written as \cite{Hull:2001iu}
\begin{equation}\label{5dDualGravitonGauge1}
\delta\,\tilde{h}_{\alpha\beta,\gamma} = 
\de_{[\alpha}\,\xi_{\beta]\gamma} + \de_\gamma\,\hat{\xi}_{\alpha\beta} - \tfrac{1}{3!}\,\de_{[\gamma}\,\hat{\xi}_{\alpha\beta]},
\end{equation}
where $\xi_{\alpha,\beta}$ is symmetric, and $\hat{\xi}_{\alpha\beta}$ is antisymmetric, that is, they are the components of $\xi^{(1,1)}$ and $\xi^{(2)}$ respectively; the relative coefficients between the first and the last two terms is arbitrary, since it can be absorbed in the definition of the two parameters; the last term ensures the condition $\tilde{h}_{[\alpha\beta,\gamma]}=0$ to be fulfilled.\footnote{The transformation \eqref{5dDualGravitonGauge1} can be equivalently written in the following way \cite{Curtright:1980yk, Aulakh:1986cb}
\begin{equation}\label{5dDualGravitonGauge2}
\delta\,\tilde{h}_{\alpha\beta,\gamma} = \de_{[\alpha}\,\xi_{\beta],\gamma} + \tfrac{2}{3}\,\de_\gamma\,\hat{\xi}_{\alpha\beta} + \tfrac{1}{3}\,\de_{[\alpha|}\,\hat{\xi}_{\gamma|\beta]}.
\end{equation}} By explicit evaluation, one can see that the dual Riemann components are invariant $\delta\,\tilde{R}_{\mu\nu\vrho,\sigma\tau}=0$. The symmetry \eqref{5dDualGravitonGauge1} is reducible, that is, there is a gauge symmetry for the gauge symmetry itself:
\begin{equation}
\delta\,\xi_{\alpha,\beta} = \de_{(\alpha}\,\xi_{\beta)}, \quad
\delta\,\hat{\xi}_{\alpha\beta} = -\de_{[\alpha}\,\xi_{\beta]},
\end{equation}
for a parameter $\xi^{(1)}$ with components $\xi_\alpha$. Thus, there is a tower of gauge parameters: $\{\xi^{(1,1)},\xi^{(2)}\}$ is the first generation; $\{\xi^{(1)}\}$ is the second generation. The elements of each generation can be obtained by iteratively deleting a box in a Young tableau, starting from the tableau $[2,1]$ of the gauge field, in such a way that the resulting diagram is still a Young tableau. The sequence in the five-dimensional case is $[2,1]\rightarrow   \{[1,1], [2] \}\rightarrow [1]$, where $[p]$ denotes a single-columned Young tableau, with $p$ boxes. The gauge transformations are obtained by putting a derivative in correspondence with the removed box in the diagram. 

We can argue the generalisation to arbitrary spacetime dimensions \cite{Labastida:1986gy}. Starting from the dual graviton $\tilde{h}_{\mu_1\dots\mu_{d-3},\alpha}$, with Young tableau $[d-3,1]$, the tower of gauge parameters is $\{\xi^{(d-3-i,1)},\xi^{(d-2-i)}\}_{i=1,\dots,d-4}$, and the last generation, which has a single term, is $\{\xi^{(1)}\}$. The gauge transformation of the dual graviton can be written as
\begin{equation}
\delta\,\tilde{h}^{(d-3,1)} = \overset{\rightarrow}{\diff}\,\xi^{(d-4,1)} + \xi^{(d-3)}\,\overset{\leftarrow}{\diff},
\end{equation}
in such a  way that the dual Riemann form is manifestly invariant
\begin{equation}
\delta\,\tilde{R}^{(d-2,2)} = \tfrac{1}{2}\,\overset{\rightarrow}{\diff}\,(\overset{\rightarrow}{\diff}\,\xi^{(d-4,1)}+\xi^{(d-3)}\,\overset{\leftarrow}{\diff})\,\overset{\leftarrow}{\diff} = 0, 
\end{equation}
since both the left- and the right-differential are nilpotent.

Let us count the degrees of freedom in dual gravity. We expect to obtain the same number as in gravity, that is (compare with the table in the end of Section \ref{Counting})
\begin{equation}\label{OnShellDofGravity}
\#\,h_{\mu\nu} = \tfrac{d(d-3)}{2}.
\end{equation}
We define the dimension $C^{(i)}$ of the $i$th generation of gauge parameters as the sum of the number of components of $\xi^{(d-3-i,1)}$ and $\xi^{(d-2-i)}$, which together form an antisymmetric rank-$(d-3-i)$ tensor with a spectator index more,
\begin{equation}
C^{(i)} = \,\scriptstyle{{\binom{d}{d-3-i}}\,d}, \quad i = 1,\dots,d-4.
\end{equation}
Notice that the formula gives also the right dimension for the last generation, if $i=d-3$. The number of components of the dual graviton (without taking into account the gauge redundancy) is the $\text{GL}(d)$-dimension of the Young tableau $[d-3,1]$
\begin{equation}
C^{(0)} = \,\scriptstyle{{\binom{d}{d-3}}\,d - {\binom{d}{d-2}}}\, = \tfrac{(d+1)d(d-1)(d-3)}{6}, \quad d>3.
\end{equation}
Now, remember that the on-shell \emph{massive} graviton is described by a symmetric, transversal, traceless rank-2 tensor, whose number of components is
\begin{equation}
\#_{\text{massive}}\,h_{\mu\nu} = \tfrac{d(d+1)}{2}-d-1 = \tfrac{(d-2)(d+1)}{2},
\end{equation}
which indeed gives in four dimensions the expected five polarisations. The transversality condition can be supplied by decreasing by one the number of spacetime dimensions. Indeed, for a symmetric, traceless, rank-two tensor in $d-1$ dimensions the number of components is 
\begin{equation}
\#_{\text{massive}}\,h_{\mu\nu} = {\tfrac{d(d+1)}{2}-1}_{\big{|}_{d\rightarrow d-1}} = \tfrac{(d-2)(d+1)}{2},
\end{equation}
which is the same number as before. In the massless case one has to subtract the trivial or pure-gauge solutions, whose number is given by the number of independent components of the gauge parameter
\begin{equation}
\#_{\text{massless}}\,h_{\mu\nu} = \#_{\text{massive}}\,h_{\mu\nu} - d = \tfrac{d(d-3)}{2},
\end{equation}
which is \eqref{OnShellDofGravity}. The elimination of the pure-gauge solution can be supplied by further decreasing by one the number of spacetime dimension 
\begin{equation}
\#_{\text{massless}}\,h_{\mu\nu} = {\tfrac{d(d+1)}{2}-1}_{\big{|}_{d\rightarrow d-2}} = \tfrac{d(d-3)}{2}.
\end{equation}
In the case of dual graviton, we proceed similarly, by imposing the tracelessness and the divergencelessness on the dual graviton and on all the tower of gauge parameters. The condition for the dual graviton are
\begin{equation}
\de^{\alpha_1}\,\tilde{h}_{\alpha_1\alpha_2\dots\alpha_{d-3},\beta} = 0, \quad
\tilde{h}^{\alpha}{}_{\alpha_2\dots\alpha_{d-3},\alpha} = 0.
\end{equation}
Notice that the other possible divergence $\de^\beta\,\tilde{h}_{\alpha_1\dots\alpha_{d-3},\beta}$ is not independent, since $\tilde{h}_{[\alpha_1\dots\alpha_{d-3},\beta]}=0$. The two conditions above correspond to fix the components of a $(d-4,1)$--form $\overline{\xi}^{(d-4,1)}$ and of a $(d-3)$--form $\overline{\xi}^{(d-3)}$ respectively, as if the first generation of gauge parameters were doubled.\footnote{The new parameters play the r\^ole of antighost in \textsc{brst} formalism \cite{Thierry-Mieg:1980ihu}.} Now, the transversality condition is imposed on all the elements in the first generation (notice that they are all traceless):
\begin{equation}
\de^{\alpha_1}\,\xi_{\alpha_1\dots\alpha_{d-4},\beta} = 0, \quad
\de^{\alpha_1}\,\hat{\xi}_{\alpha_1\dots\alpha_{d-3}} = 0, \quad
\de^{\alpha_1}\,\overline{\xi}_{\alpha_1\dots\alpha_{d-4},\beta} = 0,\quad
\de^{\alpha_1}\,\hat{\overline{\xi}}_{\alpha_1\dots\alpha_{d-3}} = 0.
\end{equation}
This corresponds to add to the second generation two more $(d-5,1)$--forms and two more $(d-4)$--forms, so that the elements in the second generation are tripled. In general, dimension of $i$th generation becomes $(1+i)\,C^{(i)}$. In particular, the last-but-one generation becomes
\begin{equation}
\xi_{\alpha,\beta}, \quad \hat{\xi}_{\alpha\beta}, \quad \overline{\xi}_{(j)\alpha,\beta}, \quad \hat{\overline{\xi}}_{(j)\alpha\beta}, \quad j = 1,\dots, d-4.
\end{equation}
The conditions to be imposed are
\begin{equation}
\de^\alpha\,(\xi_{\alpha,\beta} + \hat{\xi}_{\alpha\beta}) = 0, \quad
\de^\alpha\,(\overline{\xi}_{(j)\alpha,\beta} + \hat{\overline{\xi}}_{(j)\alpha\beta} = 0,\quad j = 1,\dots, d-4,
\end{equation}
which fix $d-3$ one--forms $\overline{\xi}^{(1)}_{(j)}$, $j=1,\dots,d-3$. Therefore, the dimension of the last generation becomes $(d-2)\,d = (d-3+1)\,C^{(d-3)}$. Finally, the number of on-shell degrees of freedom of the dual graviton in $d$ dimensions is given by the alternating sum of the extended dimensions $(1+i)\,C^{(i)}$, $i=0,\dots,d-3$, \cite{Labastida:1986gy, Bekaert:2002dt}
\begin{equation}
\#\,\tilde{h}_{\alpha_1\dots\alpha_{d-3},\beta} = \sum_{i=0}^{d-3} (-)^i\,(1+i)\,C^{(i)} = \tfrac{d(d-3)}{2},
\end{equation}
whose result remarkably agrees with \eqref{OnShellDofGravity}.

\section{Algebras}\label{A2}

An \emph{algebra} $\mathscr{A}$ is a vector space on a field $\mathbb{K}$, equipped with a \emph{product} 
\begin{equation}
* : \mathscr{A} \times \mathscr{A} \rightarrow \mathscr{A},
\end{equation}
which is bilinear
\begin{equation}
x * (y + \alpha\,z) = x * y + \alpha\,x * z, \quad \forall\; x, y, z \in \mathscr{A}, \;\; \forall\;\alpha \in \mathbb{K}.
\end{equation}
The product is
\begin{itemize}
\item[--] \emph{commutative} if $x * y = y * x$,
\item[--] \emph{associative} if $x * (y * z) = (x * y) * z$, 
\item[--] \emph{alternative} if $x * (y * x) = (x * y) * x$,
\item[--] \emph{flexible} if $x * (y * y) = (x * y) * y$,  
\item[--] \emph{power-associative} if $x * (x * x) = (x * x) * x$.
\end{itemize}
Notice that
\begin{equation}
\text{associativity} \Rightarrow
\text{alternativity} \Rightarrow
\text{flexibility} \Rightarrow
\text{power-associativity}.
\end{equation}
Every algebra has a zero element $0$, which is such that
\begin{equation}
x * 0 = 0 * x = 0, \quad \forall\;x \in \mathscr{A}.
\end{equation}
$\mathscr{A}$ is a \emph{division algebra} if 
\begin{equation}
x * y = 0 \Leftrightarrow x = 0 \;\text{or}\; y = 0.
\end{equation}
$\mathscr{A}$ is a \emph{unital algebra} if it contains a unit element $1$, defined by
\begin{equation}
x * 1 = 1 * x = x, \quad \forall\;x \in\mathscr{A}.
\end{equation}
A \emph{conjugation} is a map $\bar{\cdot} : \mathscr{A} \rightarrow \mathscr{A}$, such that 
\begin{align}
\bar{\bar{x}} = x, \quad
\overline{x * y} = \bar{x} * \bar{y}, \quad
x * \bar{x} = \bar{x} * x \in \mathbb{K}, \quad \forall\;x, y\in\mathscr{A}.
\end{align}
The last property defines a square norm $|x|^2 = x * \bar{x}$, for any element $x$ in the algebra. An algebra equipped with a norm is a \emph{normed algebra}. 

An important theorem, due to Hurwitz, states that there are precisely four normed division algebras, up to isomorphisms. They are the \emph{real numbers}, the \emph{complex numbers}, the \emph{quaternions} and the \emph{octonions}:
\begin{equation}
\mathscr{A}\;\text{is a normed division algebra} 
\Leftrightarrow \mathscr{A} = \R, \C, \mathbb{H}, \mathbb{O}.
\end{equation}
They are also unital algebras. Moreover, $\R$ and $\C$ are commutitive and associative; $\mathbb{H}$ is associative, but not commutative; $\mathbb{O}$ is not associative nor commutitative, but it is alternative. $\R$ is the unique among normed division algebra which is totally ordered, that it, a ordering of its elements can be defined (the relation of $<$ or $>$ make sense only in the real numbers).

A \emph{Leibniz algebra} is an algebra whose product satisfies the \emph{Leibniz identity} or \emph{Leibniz rule}:
\begin{equation}
x * (y * z) = (x * y) * z + y * (x * z), \quad
\forall \;x, y, z \in \mathscr{A}.
\end{equation}

A \emph{Jordan algebra} is a commutative algebra whose product satisfies the \emph{Jordan identity}:
\begin{align}
& x * y = y * x, \quad \forall \; x,y \in \mathscr{A},\\
& x * ( (x*x) * y) = (x * x) * (x * y), \quad \forall \; x,y \;\in \mathscr{A}.
\end{align}

A \emph{Lie algebra} is an algebra whose product is anticommutative and satisfying the \emph{Jacobi identity}
\begin{equations}
& x * y = - y * x, \quad \forall \; x, y \in \mathscr{A},\\
& x * (y * z) + y * (z * x) + z * (x * y) = 0, \quad
\forall \; x, y, z \in \mathscr{A}.
\end{equations}
In this case the product is usually denoted with $[\cdot,\cdot]$ and it is called \emph{Lie bracket}. Using this notation the two properties defining a Lie algebra read
\begin{equations}
& [x,y] = -[x,y],\nn\\
& [x,[y,z]] + [y,[z,x]] + [z,[x,y]] = 0.
\end{equations}

Anticommutativity and the Jacobi identity implies the Leibniz identity. So, all the Lie algebras are Leibniz algebras:
\begin{equation}
\mathscr{A}\;\text{is a Lie algebra} \Rightarrow
\mathscr{A}\;\text{is a Leibniz algebra}.
\end{equation}
Instead, the converse is not true. Leibniz algebras are the most simple generalisation of Lie algebras. More general examples are provided by the so-called $L_\infty$ \emph{algebras}. 

The commutator between matrices 
\begin{equation}
[A,B] = A\,B - B\,A
\end{equation}
is a Lie bracket, because it is anticommutative by definition and it satisfies the Jacobi identity:
\begin{align}
[A,[B,C]] + [B,[C,A]] + [C,[A,B]] &= 
A\,B\,C - A\,C\,B - B\,C\,A + C\,B\,A \,+ \nn\\
& + B\,C\,A - C\,B\,A - C\,A\,B + B\,A\,C \,+ \nn\\
& + C\,A\,B - B\,A\,C - A\,B\,C + A\,C\,B = 0.
\end{align}

The \emph{derivative algebra} $\text{Der}\,\mathscr{A}$ of an algebra $\mathscr{A}$ is the set of possible derivatives on $\mathscr{A}$, that is bilinear maps $D : \mathscr{A} \rightarrow\mathscr{A}$, satisfying the \emph{Leibniz rule}
\begin{equation}
D(x * y) = D(x) * y + x * D(x), \quad
\forall\; x, y \in \mathscr{A}.
\end{equation}

If we use as a Lie bracket the commutator defined by means of the composition $\circ$ between linear maps, then the derivative algebra of any algebra is a Lie algebra. Indeed, by definition, this product is antisymmetric and it satisfies the Jacobi identity. It remains to show that, if $D, D'$ are derivative, then $[D,D']$ is a derivative too:
\begin{align}
[D,D']\,(x * y) &= (D \circ D' - D' \circ D)\,(x * y) =\nn \\
& = D\,(D'(x) * y + x * D'(y)) - D'\,(D(x) * y + x * D(y)) = \nn \\
& = (D \circ D')(x) * y + D'(x) * D(y) + D(x) * D'(y) + x * (D \circ D')(y) \,+\nn\\
& - (D' \circ D)(x) * y - D(x) \circ D'(y) - D'(x) \circ D(y) - x * (D \circ D')(y) = \nn\\
& = (D\circ D' - D' \circ D)(x) * y + x * (D\circ D' - D' \circ D)(y) = \nn\\
& = [D,D'](x) * y + x * [D,D'](y),
\end{align}
that is $[D,D']$ satisfies the Leibniz rule.

An \emph{automorphism} of an algebra $\mathscr{A}$ is a bijective map $\vphi : \mathscr{A} \rightarrow \mathscr{A}$, preserving the structure of the algebra:
\begin{equation}
\vphi(x * y) = \vphi(x) * \vphi(y).
\end{equation}
The set $\text{Aut}\,\mathscr{A}$ of the automorphisms of $\mathscr{A}$ is closed under the composition $\circ$ between map. Indeed,
\begin{equation}
(\vphi \circ \vphi')(x * y) = \vphi(\vphi'(x) * \vphi'(y)) = (\vphi \circ \vphi')(x) * (\vphi \circ \vphi')(y).
\end{equation}
If $\mathscr{A}$ is a Lie algebra, $\text{Aut}\,\mathscr{A}$ is a Lie group and its algebra is the derivative algebra of $\mathscr{A}$ itself:
\begin{equation}
\text{Lie}\,\text{Aut}\,\mathscr{A} = \text{Der}\,\mathscr{A}.
\end{equation}

A representation $R$ of an algebra $\mathscr{A}$ on a vector space $V$ is a linear map $R : \mathscr{A} \rightarrow \text{End}\,V$, which satisfies a compatibility condition between the commutator out of the compositions between linear maps and the product which defines the algebra:
\begin{equation}
[R(x),R(y)] = R(x * y).
\end{equation}

Suppose that $\mathscr{A}$ is a Leibniz algebra. Then 
\begin{equation}
\text{ad}_x(\cdot) := x * \cdot, \quad x \in \mathscr{A}
\end{equation}
defines a representation of $\mathscr{A}$. Indeed, the Leibniz rule
\begin{equation}
\text{ad}_x(\text{ad}_y(z)) = \text{ad}_{\text{ad}_x(y)}(z) - \text{ad}_y\,(\text{ad}_x(z)),
\end{equation}
where $x,y,z \in \mathscr{A}$, can be written as
\begin{equation}
(\text{ad}_x * \text{ad}_y)(z) = \text{ad}_{\text{ad}_x(y)}(z),
\end{equation}
which is the compatibility condition. This representation is called \emph{adjoint representation}. 

Define the exponential of $\text{ad}_x$ 
\begin{equation}
\text{Ad}_x := e^{\text{ad}_x} = \sum_{n=0}^\infty \frac{1}{n!}\,(\text{ad}_x)^n,
\end{equation}
which makes sense if there is an integer $m$ such that $(\text{ad}_x)^m = 0$. An automorphism $\vphi$ is \emph{internal} if there is $y \in \mathscr{A}$ such that
\begin{equation}
\vphi(x) = \text{Ad}_y\,x.
\end{equation}

If $\mathscr{A}$ is represented by matrices, the following relation holds:
\begin{equation}
e^{\text{ad}_x}\,y = e^x\,y\,e^{-x}, \quad x,t \in \mathscr{A}.
\end{equation}

More in general, if $D_x \in \text{Der}\,\mathscr{A}$, $x \in \mathscr{A}$, then the set of $D_x$ defines a representation of $\mathscr{A}$, thanks to the Leibniz identity. The following relation holds: 
\begin{equation}
e^{D_x} * e^{D_y} = e^{D_{x * y}}.
\label{expD}
\end{equation}

\emph{Proof.} As a preliminary step, we prove by induction that
\begin{equation}
D^n_{x * y} = \sum_{j=0}^n \frac{n!}{j!\,(n-j)!}\,D^j_x * D^{n-j}_y.
\end{equation}
If $n=0$, both sides of the expression give $x * y$. If $n=1$, the expression is equivalent to the Leibniz identity. Now consider the case $n+1$ and use the expression for $n$:
\begin{align}
D^{n+1}_{x * y} &= D_{x * y}\,D^n_{x * y} 
= D_{x*y}\,\sum_{j=0}^n\frac{n!}{j!\,(n-j)!}\,D^j_{x} * D^{n-j}_y = \nn\\
& = \sum_{j=0}^n \frac{n!}{j!\,(n-j)!}\,\left(D^{j+1}_x * D^{n-j}_y + D^j_x * D^{n-j+1}_y \right) = \nn\\
& = \sum_{j=1}^{n+1} \frac{n!}{(j-1)!\,(n+1-j)!}\,D^j_x * D^{n+1-j}_y \,+\nn\\
& + \sum_{j=0}^n \frac{n!}{j!\,(n-j)!}\,D^j_x * D^{n+1-j}_y = \nn \\
& = D^{n+1}_x * D^0_y + D^0_x * D^{n+1}_y \,+\nn\\
& + \sum_{j=1}^n \left[\frac{n!}{(j-1)!\,(n+1-j)!}+\frac{n!}{j!\,(n-j)!}\right]\,D^j_x * D^{n+1-j}_y = \nn\\
& = D^{n+1}_x * D^0_y + D^0_x * D^{n+1}_y \,+\nn\\
& + \sum_{j=1}^n \frac{(n+1)!}{j!\,(n+1-j)!}\,[D^j_x * D^{n+1-j}_y = \nn\\
& = \sum_{j=0}^{n+1} \frac{(n+1)!}{j!\,(n+1-j)!}\,D^j_x * D^{n+1-j}_y.
\end{align} 
Now, observe that
\begin{align}
e^{D_{x * y}} &= \sum_{n=0}^m\,\frac{1}{n!}\,D^n_{x * y} = \nn\\
& = \sum_{n=0}^m \sum_{j=0}^n \frac{1}{n!}\,\frac{n!}{j!\,(n-j)!}\,D^j_x * D^{n-j}_y = \nn\\
& = \sum_{n,m=0}^M \frac{1}{n!}\,D^n_x * \frac{1}{m!}\,D^m_x = e^{D_x} * e^{D_y},
\end{align}
where in the last-but-one step we replaced an horizontal sum  $(j,n-j)$ with a vertical one  $(n,m)$. This completes the proof.

\section{Leibniz algebras}\label{A3}

Consider a Leibniz algebra $V$ with the $\Ll \cdot,\cdot \Lr$ The Leibniz identity is 
\begin{equation}\label{LeibnizId}
\Ll x, \Ll y,z \Lr \Lr = 
\Ll \Ll x,y \Lr, z \Lr + \Ll y, \Ll x,z \Lr \Lr, \quad  \forall\;x,y,z \in V,
\end{equation}
The antisymmetruc part of the product, denoted with $[\cdot,\cdot]$, is a Lie product. Denote the symmetric part with $\lbrace \cdot, \cdot \rbrace$. It is a trivial extension of $\Ll \cdot,\cdot \Lr$, in the sense that it commutes with every element in the algebra: 
\begin{equation} \label{TrivialProp2}
\Ll \lbrace x,y \rbrace, z\Lr = 0, \quad\forall\;\;x,y\in V,
\end{equation}
This is a direct consequence of the Leibniz identity \eqref{LeibnizId}, if it is written in the following equivalent way
\begin{equation}
\Ll x, \Ll y,z \Lr \Lr - \Ll y, \Ll x,z \Lr \Lr  =
\Ll \Ll x,y \Lr, z \Lr.
\end{equation}
The left-hand side, which is antisymmetric in $x \leftrightarrow y$, selects the antisymmetric part of $\Ll x,y \Lr$ in the right-hand side.

The following proposition shows that in general the antisymmetric part of the Leibniz product does not satisfy the Jacobi identity, but the obstruction can be expressed in terms of symmetric products.

\bigskip

\emph{Prop.} For each $x,y,z\in V$,
\begin{equation}
[[x,y],z]+[[y,z],x]+[[z,x],y]] = \tfrac{1}{3}\,\left(\lbrace [x,y],z \rbrace + \lbrace [y,z],x \rbrace + \lbrace [z,x],y \rbrace \right).
\end{equation}

\bigskip

\emph{Proof.} Define the \emph{Jacobiator} as
\begin{equation}
J(x,y,z) = [[x,y],z] + [[y,z],x] + [[z,x],y].
\end{equation}
Notice that
\begin{align}
[[x,y],z] &= \Ll [x,y],z \Lr - \lbrace [x,y], z\rbrace = \nonumber \\
& = \Ll \Ll x,y \Lr, z \Lr - \Ll \lbrace x,y \rbrace, z \Lr - \lbrace [x,y],z \rbrace = \nonumber \\
& = \Ll \Ll x,y \Lr, z \Lr - \lbrace [x,y],z \rbrace.
\label{Prop1}
\end{align}
In the last step, \eqref{TrivialProp2} is used. Then,
\begin{align}
J(x,y,z) &= \tfrac{1}{4}\big{(} \Ll \Ll x,y \Lr,z\Lr
- \Ll \Ll y,x \Lr, z \Lr - \Ll z, \Ll x,y \Lr \Lr + \Ll z, \Ll y,x \Lr \Lr \big{)} + \text{c.p.} = \nonumber \\
& = \tfrac{1}{4}\big{(} \Ll \Ll x,y \Lr,z\Lr
- \Ll y, \Ll x,z \Lr \Lr + \Ll x, \Ll y,z \Lr \Lr 
- \Ll z, \Ll x,y \Lr \Lr + \Ll z, \Ll y,x \Lr \Lr \big{)} + \text{c.p.},
\end{align}
where c.p. means cyclic permutation. In the first step, we wrote $[\cdot,\cdot]$ in terms of $\Ll \cdot,\cdot \Lr$. In the second one, we used the Leibniz identity on the second term. Expanding the cyclic permutations of all the terms, except the first, all the terms drop out. Indeed, writing $xyz$ in place of $\Ll x, \Ll y,z \Lr \Lr$ for brevity,
\begin{align}
& \quad\; - \Ll y, \Ll x,z \Lr \Lr + \Ll x, \Ll y,z \Lr \Lr 
- \Ll z, \Ll x,y \Lr \Lr + \Ll z, \Ll y,x \Lr \Lr + \text{c.p} = 
\nonumber \\
& = -yxz-xzy-zyx+xyz+yzx+zxy \,+\nonumber \\
& \quad\;-zxy-xyz-yzx+zyx+yxz+xzy = 0.
\end{align}
Therefore, it remains
\begin{align}
J(x,y,z) & = \tfrac{1}{4} \Ll \Ll x,y \Lr,z\Lr
+ \text{c.p.} = \nonumber \\
& = \tfrac{1}{4}\big{(} [[x,y],z] + \lbrace [x,y],z\rbrace\big{)} + \text{c.p.} = \nonumber \\
& =\tfrac{1}{4}\,J(x,y,z) + \tfrac{1}{4}\,\big{(}\lbrace [x,y],z\rbrace + \text{c.p.}\big{)},
\end{align}
where \eqref{Prop1} and the definition of Jacobiator are used. In conclusion, 
\begin{equation}
J(x,y,z) = \tfrac{1}{3}\,\lbrace [x,y],z\rbrace + \text{c.p.},
\end{equation}
as we wanted.

\section{More details on the generalised Lie derivative}\label{A4}

Extending the definition of generalised Lie derivative on a scalar $\vphi$ in the obvious way
\begin{equation}
L_X\,\vphi = \xi^M\,\de_M\,\vphi,
\end{equation}
one can also define the generalised Lie derivative on a one-form with components $U_M$, with density weight $\lambda$, by requiring $V^M\,U_M$ to be a scalar:
\begin{equation}
L_X\,U_M = X^N\,\de_N\,U_M + U_N\,\de_M\,X^N + (\lambda+\omega)\,\de_N\,X^N\,U_M - Y^{NP}{}_{QM}\,\de_P\,X^Q\,U_N,
\end{equation}

Let us find which properties an arbitrary $Y$ tensor has to enjoy in order for the Leibniz identity \eqref{LeibnizIdentityAlgebra} to be satisfied by the generalised Lie derivative \eqref{GenLieDerY}. Assume both $V$ and $L_X\,V$ to have density weight $\lambda$, and $X$ to have vanishing density weight. Let us compute the antisymmetric and the symmetric part of the Leibniz identity,
\begin{equation}
([L_X,L_V]-L_{\frac{1}{2}(L_X V - L_V X)})\,W^M, \quad 
L_{\frac{1}{2}(L_X V + L_V X)}\,W^M 
\end{equation}
imposing them to vanish. The terms to be set to zero are either proportional to $W$ or proportional to its derivatives $\de\,W$. To eliminate the latter ones, one has to impose the section constraint \eqref{SectionConstraintY}. Using it, we remain with
\begin{align}
([L_X,L_V]-L_{\frac{1}{2}(L_X V - L_V X)})\,W^M &= [(A^M{}_N + B^M{}_N)^{PQ}{}_{RS}-(A^M{}_N + B^M{}_N)^{QP}{}_{SR}]\,\times\nn\\
& \times W^N\,\de_P\,X^R\,\de_Q\,V^S + \tfrac{1}{4}\,(C^M{}_N)^{PQ}{}_{RS}\,\times\nn\\
& \times W^N\,(X^R\,\de_P\,\de_Q\,V^S -V^R\,\de_P\,\de_Q\,X^S),
\end{align}
for the antisymmetric part, and
\begin{equation}
L_{\frac{1}{2}(L_X V + L_V X)}\,W^M  = [(A^M{}_N)^{PQ}{}_{RS}+(A^M{}_N)^{QP}{}_{SR}]\,W^N\,\de_P\,X^R\,\de_Q\,V^S
\end{equation}
for the symmetric one, where
\begin{equations}
& (A^M{}_N)^{PQ}{}_{RS} = \tfrac{1}{2}\,Y^{MQ}{}_{TN}\,T^{TP}{}_{RS} - \tfrac{1}{2}\,Y^{MQ}{}_{SR}\,\delta^P_N,\\
& (B^M{}_N)^{PQ}{}_{RS} = Y^{MQ}{}_{ST}\,Y^{TP}{}_{RN}-T^{MQ}{}_{RN}\,\delta^P_S,\\
& (C^M{}_N)^{PQ}{}_{RS} = Y^{M(P|}{}_{TN}\,Y^{T|Q)}{}_{SR}-Y^{M(P}{}_{SR}\,\delta^{Q)}_N.
\end{equations}
Therefore, aside the section constraint, we have to impose the following conditions on the $Y$ tensor
\begin{equations}
& [(A^M{}_N)^{PQ}{}_{RS}+(A^M{}_N)^{QP}{}_{SR}]\,\mathcal{E}_P{}^p\,\mathcal{E}_Q{}^q = 0,\\
& [(A^M{}_N+B^M{}_N)^{PQ}{}_{RS}+(A^M{}_N+B^M{}_N)^{QP}{}_{SR}]\,\mathcal{E}_P{}^p\,\mathcal{E}_Q{}^q = 0,\\
& (C^M{}_N)^{PQ}{}_{RS}\,\mathcal{E}_P{}^p\,\mathcal{E}_Q{}^q = 0.
\end{equations}
One can verify that the $Y$ tensor in \eqref{YTensor} satisfies all the previous constraints. Therefore, the section constraint is a necessary and sufficient condition for the Leibniz identity to be true for the generalised Lie derivative.

\section{Invariance of the structure constant} \label{fFinvariance}\label{A5}

The flux $F_{MA}{}^B$ has a nonvanishing trace component $F_{MA}{}^A$. This has an implication in the equation expressing the invariance the structure constants under the action of the flux, that is,
\begin{equation}
\delta_{F_M}\,f_{PQR} = 0.
\end{equation}
Acting on $f_{PQR}$ with the flux, we replace alternatively the indices of the structure constant in this way
\begin{equation}
F_{MP}{}^A\,f_{AQR} + F_{MQ}{}^A\,f_{PAR} + F_{MR}{}^A\,f_{PQA} =: (*)
\end{equation}
But we have also to include a term proportional to the trace, which should be $\alpha\,F_{MA}{}^A\,f_{PQR}$, for some constant $\alpha$. Indeed, if we saturated $(*)$ with $f^{PQR}$, we would obtain
\begin{align}
(*)\,f^{PQR} &= F_{MP}{}^A\,f_{AQR} + F_{MQ}{}^A\,f_{PAR}\,f^{PQR} + F_{MR}{}^A\,f_{PQA}\,f^{PQR} = \nn\\
& = F_{MP}{}^A\,(-60\,\delta_A^P) + F_{MQ}{}^A\,(-60\,\delta_A^P) + F_{MQ}{}^A\,(-60\,\delta_A^Q) + F_{MR}{}^A\,(-60\,\delta_A^R) = \nn\\
& = -60\,(F_{MA}{}^A + F_{MA}{}^A + F_{MA}{}^A) = -60\time 3\,F_{MA}{}^A.
\end{align}
Moreover, 
\begin{equation}
\alpha\,F_{MA}{}^A\,f_{PQR}\,f^{PQR} = \alpha\,F_{MA}{}^A\,(-60\,\delta^R_R) = -60\times 248\,\alpha\,F_{MA}{}^A.
\end{equation}
Therefore
\begin{equation}
-60\times 3\,F_{MA}{}^A = -60\times 248\,\alpha\,F_{MA}{}^A \Rightarrow \alpha = \tfrac{3}{248}.
\end{equation}
Finally,
\begin{equation}
F_{MP}{}^A\,f_{AQR} + F_{MQ}{}^A\,f_{PAR} + F_{MR}{}^A\,f_{PQA} = \tfrac{3}{248}\,F_{MA}{}^A\,f_{PQR}.
\end{equation}

Analogously, starting with $f^{PQ}{}_R$, we should write:
\begin{align}
& F_{MA}{}^P\,f^{AQ}{}_R + F_{MA}{}^Q\,f^{PA}{}_R - F_{MR}{}^A\,f^{PQ}{}_A = \alpha\,F_{MA}{}^A\,f^{PQ}{}_R,\nn\\
\Rightarrow & \, F_{MA}{}^P\,f^{AQ}{}_R\,f_{PQ}{}^R + F_{MA}{}^Q\,f^{PA}{}_R\,f_{PQ}{}^R - F_{MR}{}^A\,f^{PQ}{}_A\,f_{PQ}{}^R = \alpha\,F_{MA}{}^A\,f^{PQ}{}_R\,f_{PQ}{}^R,\nn\\
\Rightarrow & \,F_{MA}{}^P\,(-60\,\delta_P^A) + F_{MA}{}^Q\,(-60\,\delta_Q^A) - F_{MR}{}^A\,(-60\,\delta_A^R) = \alpha\,F_{MA}{}^A\,(-60\times 248),\nn\\
\Rightarrow &\, F_{MA}{}^A = 248\,\alpha\,F_{MA}{}^A 
\Rightarrow \alpha = \tfrac{1}{248},
\end{align}
so that, 
\begin{equation}
F_{MA}{}^P\,f^{AQ}{}_R + F_{MA}{}^Q\,f^{PA}{}_R - F_{MR}{}^A\,f^{PQ}{}_A = \tfrac{1}{248}\,F_{MA}{}^A\,f^{PQ}{}_R.
\end{equation}

In general
\begin{equation}
(n_+-n_-)\,F_{MA}{}^A = 248\,\alpha\,F_{MA}{}^A 
\Rightarrow \alpha = \tfrac{n_+-n_-}{248},
\end{equation}
where $n_\pm$ is the number of upper/lower indices of the structure constants on which the flux acts.

\section{Theorems for computing BRST cohomologies}\label{A6}

In this Appendix some useful results in computing \textsc{brst} cohomologies are summarised and proved \cite{Brandt:1989gy, Brandt:1989gv, Brandt:1989rd, Baulieu:1985md, Piguet:1995er}.

\emph{Lemma.} (\emph{Basic lemma}) Let $V$ the space of local polynomial in the fields and in the ghosts in a field theory, collectively denoted by $\vphi$, and in their derivatives; $s$ the \textsc{brst} operator; $\mathcal{N}$ a linear operator on $V$, such that $\mathcal{N} = [s,r]$, for some odd operator $r$. Suppose $V$ to be decomponible with respect to $\mathcal{N}$, 
\begin{equation}
f = \sum_\lambda\,f_\lambda, \quad\forall\;f \in V, \quad \mathcal{N}\,f_\lambda = \lambda\,f_\lambda.
\end{equation}
Then, the $s$-cohomology is generated by the zero modes of $\mathcal{N}$: every $f \in V$ such that $s\,f = 0$ can be written as $f = f_0 + s\,\tilde{f}$, for some $\tilde{f} \in V$.

\bigskip

\emph{Proof.} The Jacobi identity
\begin{equation}
[s,[s,r]] + [s,[r,s]] + [r,[s,s]] = 0
\end{equation}
implies that $s$ and $\mathcal{N}$ are compatible
\begin{equation}
[\mathcal{N},s]=0,
\end{equation}
since $s$ is nilpotent. This means that the eigenspaces of $\mathcal{N}$ are stable under the action of $s$:
\begin{equation}
\mathcal{N}\,(s\,f_\lambda) = s\,\mathcal{N}\,f_\lambda = \lambda\,(s\,f_\lambda).
\end{equation}
If $s\,f = 0$ and $f = \sum_\lambda f_\lambda$, then $s\,f_\lambda = 0$, for each $\lambda$, because elements in distinct eigenspaces are independent. Then,
\begin{equation}
f = f_0 + \sum_{\lambda \neq 0} f_\lambda = f_0 + \sum_{\lambda \neq 0} \tfrac{1}{\lambda}\,\mathcal{N}\,f_\lambda  = f_0 + \sum_{\lambda \neq 0} \tfrac{1}{\lambda}\,(r\,s+s\,r)\,f_\lambda = f_0 + s\,\left(\sum_{\lambda \neq 0} \tfrac{1}{\lambda}\,r\,f_\lambda \right),
\end{equation}
as we wanted to show. 

\bigskip

\emph{Definition.} A trivial doublet $(u,v)$ is a pair of fields such that
\begin{equation}
s\,u = v,\quad s\,v = 0,
\end{equation}

\emph{Theorem.} (\emph{Doublet theorem}) Let $(u,v)$ a trivial doublet. Suppose that $u$ and $v$ are not involved in the transformation of any other field or ghost $w$ in the field theory. Then, $s$-cohomology does not depend on $(u,v)$ and on their derivatives.

\bigskip

\emph{Proof.} Define the following even operator
\begin{equation}
\mathcal{N} = v\,\frac{\delta}{\delta v} + u\,\frac{\delta}{\delta u},
\end{equation}
which counts the number of $u$'s and $v$'s in a monomial:
\begin{equation}
\mathcal{N}\,u = u, \quad
\mathcal{N}\,v = v, \quad
\mathcal{N}\,w = 0,
\end{equation}
where $w$ is any other ghost or field other than $(u,v)$. Consider the following odd operator
\begin{equation}
r = u\,\frac{\delta}{\delta v},
\end{equation}
which is such that 
\begin{equation}
r\,u = 0, \quad
r\,v = u, \quad
r\,w = 0.
\end{equation}
Thus, on the space of $u,v,w$ and their derivatives,
\begin{equation}
\mathcal{N} = [s,r],
\end{equation}
since $s\,w$ does not depend on $u,v$ by hypothesis.
Then, we can use the basic lemma, which says that the $s$-cohomology is generated by the zero modes of $\mathcal{N}$. They are the monomials which does not depend on $(u,v)$. So, the $s$-cohomology does not involve either $u$ or $v$, as we wanted to show. 

\bigskip

\emph{Observation.} Suppose we have a set of ghosts and fields $\vphi_{1-p}^{(p)}$, in $d$ dimensions, with the following \textsc{brst} transformations,
\begin{equations}
& s\,\vphi_{1-p}^{(p)} = -\,\diff \vphi_{2-p}^{(p-1)}, \quad 1 \leqslant p \leqslant d,\\
& s\,\vphi_1^{(0)} = 0.
\end{equations}
which are trivially nilpotent. In particular,  
\begin{equation}
s\,\diff\,\vphi_{2-p}^{(p-1)} = 0, \quad 1 \leqslant p \leqslant d.
\end{equation}
Therefore $(\vphi_{1-p}^{(p)},\diff \vphi_{2-p}^{(p-1)})$ are trivial doublets. So, they does not belong to the cohomology, which turns out to be generated by $\vphi_1^{(0)}$ and its derivatives. In particular, the $s$-cohomology of differential forms of any ghost number, with form degree greater than one, is trivial.

\bigskip

\emph{Theorem.} (\emph{Filtering theorem}) Let $\mathcal{N}$ be an even operator with nonnegative integer eigenvalues. Suppose the space of fields and ghosts to be decomponible with respect to $\mathcal{N}$. Moreover, suppose that 
\begin{equation}
s = \sum_n s_n, \quad \text{with}\;\; [\mathcal{N},s_n]=n\,s_n.
\end{equation}
Then,
\begin{itemize}
\item[1.] $\sum_{m=0}^n s_m\,s_{n-m} = 0, \;\;\forall\;n\in\mathbb{N}$.
\item[2.] $H_g^{(p)}(s) \simeq H \subseteq H_g^{(p)}(s_0), \;\;\forall\;g\in\mathbb{Z}, \;\; \forall\;p\in\mathbb{N}$.
\item[3.] $H_g^{(p)}(s|\diff) \simeq H'' \subseteq H_g^{(p)}(s_0|\diff), \;\;\forall\;g\in\mathbb{Z}, \;\; \forall\;p\in\mathbb{N}$.
\end{itemize}

\bigskip 

\emph{Corollary.} (\emph{Algebraic Poincaré lemma.}) Using the previous results, we can produce an algebraic proof of the Poincaré lemma on the space of $A$ and $F$ and its external differential:
\begin{equations}
\diff\,A &= F - A^2,\\
\diff\,F &= -[A,F].
\end{equations}
We split $\diff = \diff_0 + \diff_1$, in such a way that
\begin{equations}
& \diff_0\,A = F, \quad \diff_1\,A = -A^2,\\
& \diff_0\,F = 0, \quad \diff_1\,F = -[A,F].
\end{equations}
$A$ and $F$ form a trivial doublet with respect to $\diff_0$. Therefore, the $\diff_0$-cohomology is trivial. The operator
\begin{equation}
\mathcal{N} = A\,\frac{\delta}{\delta A} + F\,\frac{\delta}{\delta F},
\end{equation} 
is such that 
\begin{equation}
[\mathcal{N},\diff_n]= n\,\diff_n, \quad n= 0,1.
\end{equation}
Then, the $\diff$-cohomology is trivial by means of the filtering theorem.

\section{Polarisation matrices}\label{polarisation}

See for example \cite{VanNieuwenhuizen:1981ae}. Choosing the generator of $z$-axis rotation as
\begin{equation}
J_z = \begin{pmatrix}
0 & 0 & 0 & 0 \\
0 & 0 & -i & 0 \\
0 & i & 0 & 0 \\
0 & 0 & 0 & 0
\end{pmatrix},
\end{equation}
the polarisation vectors with helicities $\pm 1$ are
\begin{equation}
\vepsilon_\mu^{\pm} = \tfrac{1}{\sqrt{2}}\,(0,1,\pm i,0),
\end{equation}
such that
\begin{equation}
J_z\,\vepsilon_\mu^{\pm} = \pm\,\vepsilon_\mu^{\pm}.
\end{equation}
Define the polarisation matrices as
\begin{equations}
\lambda_{\mu\nu}^0 = \vepsilon_{(\mu}^+\,\vepsilon_{\nu)}^- &= \begin{pmatrix}
0 & 0 & 0 & 0 \\
0 & 0 & -i & 0 \\
0 & i & 0 & 0 \\
0 & 0 & 0 & 0
\end{pmatrix}, \label{lambda0}\\
\lambda_{\mu\nu}^{\pm 1} = \tfrac{1}{\sqrt{2}}\,p_{(\mu}\,\vepsilon_{\nu)}^{\pm} &= \frac{1}{2}\begin{pmatrix}
0 & -1 & \mp i & 0 \\
-1 & 0 & 0 & 1 \\
\mp i & 0 & 0 & \pm i \\
0 & 1 & \pm i & 0 
\end{pmatrix}, \label{lambda1}\\
\lambda_{\mu\nu}^{\pm 2} = \vepsilon_{(\mu}^{\pm}\,\vepsilon_{\nu)}^{\pm} &= \frac{1}{2}\begin{pmatrix}
0 & 0 & 0 & 0 \\
0 & 1 & \pm i & 0 \\
0 & \pm i & -1 & 0 \\
0 & 0 &  0 & 0 
\end{pmatrix}.\label{lambda2}
\end{equations}
They are such that
\begin{equation}
[J_z,\lambda_{\mu\nu}^{s}] = s\,\lambda_{\mu\nu}^{s}.
\end{equation}
Two useful formulas.
\begin{align}
\begin{pmatrix}
0 & 0 & 0 & 0 \\
0 & a & b & 0 \\
0 & b & -a & 0 \\
0 & 0 & 0 & 0
\end{pmatrix} &= (a-i\,b)\,\lambda^{+2} + (a + i\,b)\,\lambda^{-2}, \\
\begin{pmatrix}
0 & -a & -b & 0 \\
-a & 0 & 0 & a \\
-b & 0 & 0 & b \\
0 & a & b & 0
\end{pmatrix} &= (a-i\,b)\,\lambda^{+1} + (a + i\,b)\,\lambda^{-1}.
\end{align}

\nocite{xact}
\nocite{xTras}
\nocite{Frob:2020gdh}

\newpage
\quad
\thispagestyle{empty}

\newpage

\printbibliography

@article{Affleck:1986,
author = "Affleck, I.",
title = "{Dyon Analogs in Antiferromagnetic Chains}",
journal        = "Phys. Rev. Lett.",
volume         = "57",
year           = "1986",
pages          = "1048",
}

@article{Alvarez-Gaume:1983ihn,
    author = "Alvarez-Gaume, Luis and Witten, Edward",
    editor = "Salam, A. and Sezgin, E.",
    title = "{Gravitational Anomalies}",
    reportNumber = "HUTP-83/A039",
    doi = "10.1016/0550-3213(84)90066-X",
    journal = "Nucl. Phys. B",
    volume = "234",
    pages = "269",
    year = "1984"
}

@article{Afxonidis:2023pdq,
    author = "Afxonidis, Evangelos and Caddeo, Alessio and Hoyos, Carlos and Musso, Daniele",
    title = "{Fracton gravity from spacetime dipole symmetry}",
    eprint = "2311.01818",
    archivePrefix = "arXiv",
    primaryClass = "hep-th",
    doi = "10.1103/PhysRevD.109.065013",
    journal = "Phys. Rev. D",
    volume = "109",
    number = "6",
    pages = "065013",
    year = "2024"
}

@article{Afxonidis:2024tph,
    author = "Afxonidis, Evangelos and Caddeo, Alessio and Hoyos, Carlos and Musso, Daniele",
    title = "{Analysis of fracton-gravity stability}",
    eprint = "2406.19268",
    archivePrefix = "arXiv",
    primaryClass = "hep-th",
    doi = "10.1103/PhysRevD.110.085016",
    journal = "Phys. Rev. D",
    volume = "110",
    number = "8",
    pages = "085016",
    year = "2024"
}

@book{Arnold:1989who,
    author = "Arnold, V. I.",
    title = "{Mathematical Methods of Classical Mechanics}",
    doi = "10.1007/978-1-4757-2063-1",
    publisher = "Springer",
    series = "Graduate Texts in Mathematics",
    year = "1989"
}

@article{Bae:2015eoa,
      author         = "Bae, Jinbeom and Imbimbo, C. and Rey, Soo-Jong and
                        Rosa, Dario",
      title          = "{New Supersymmetric Localizations from Topological
                        Gravity}",
      journal        = "JHEP",
      volume         = "03",
      year           = "2016",
      pages          = "169",
      doi            = "10.1007/JHEP03(2016)169",
      eprint         = "1510.00006",
      archivePrefix  = "arXiv",
      primaryClass   = "hep-th",
      SLACcitation   = "%%CITATION = ARXIV:1510.00006;%%"
}

@article{Bardeen:1984pm,
    author = "Bardeen, W. A. and Zumino, B.",
    title = "{Consistent and Covariant Anomalies in Gauge and Gravitational Theories}",
    reportNumber = "FERMILAB-PUB-84-038-T, LBL-17639, UCB-PTH-84-12, FERMILAB-PUB-84-038-T-REV, LBL-17639-REV, UCB-PTH-84-12-REV",
    doi = "10.1016/0550-3213(84)90322-5",
    journal = "Nucl. Phys. B",
    volume = "244",
    pages = "421--453",
    year = "1984"
}

@article{Baulieu:1985md,
      author         = "Baulieu, L. and Bellon, Marc P.",
      title          = "{$p$ Forms and Supergravity: Gauge Symmetries in Curved
                        Space}",
      journal        = "Nucl. Phys.",
      volume         = "B266",
      year           = "1986",
      pages          = "75-124",
      doi            = "10.1016/0550-3213(86)90178-1",
      reportNumber   = "LPTENS 85/7",
      SLACcitation   = "%%CITATION = NUPHA,B266,75;%%"
}

@article{Baulieu:1988xs,
    author = "Baulieu, L. and Singer, I. M.",
    title = "{Topological Yang-Mills Symmetry}",
    reportNumber = "PAR-LPTHE-88/18, LPTENS-88/15",
    doi = "10.1016/0920-5632(88)90366-0",
    journal = "Nucl. Phys. B Proc. Suppl.",
    volume = "5",
    pages = "12--19",
    year = "1988"
}

@article{Becchi:1974md,
    author = "Becchi, C. and Rouet, A. and Stora, R.",
    title = "{Renormalization of the Abelian Higgs-Kibble Model}",
    reportNumber = "CPT-74-P.634-MARSEILLE, CNRS-CPT-74-P634",
    doi = "10.1007/BF01614158",
    journal = "Commun. Math. Phys.",
    volume = "42",
    pages = "127--162",
    year = "1975"
}

@article{Becchi:1974xu,
    author = "Becchi, C. and Rouet, A. and Stora, R.",
    title = "{The Abelian Higgs-Kibble Model. Unitarity of the S Operator}",
    reportNumber = "CPT-74/P.625-MARSEILLE",
    doi = "10.1016/0370-2693(74)90058-6",
    journal = "Phys. Lett. B",
    volume = "52",
    pages = "344--346",
    year = "1974"
}

@article{Becchi:1975nq,
    author = "Becchi, C. and Rouet, A. and Stora, R.",
    title = "{Renormalization of Gauge Theories}",
    reportNumber = "CPT-75-P.723-MARSEILLE",
    doi = "10.1016/0003-4916(76)90156-1",
    journal = "Annals Phys.",
    volume = "98",
    pages = "287--321",
    year = "1976"
}

@article{Becchi:1997jg,
    author = "Becchi, C. and Giusto, S. and Imbimbo, C.",
    editor = "Derendinger, J. P. and Lucchesi, C.",
    title = "{The Holomorphic anomaly of topological strings}",
    eprint = "hep-th/9801100",
    archivePrefix = "arXiv",
    reportNumber = "GEF-TH-4-1998, GEF-TH-98-04",
    doi = "10.1002/(SICI)1521-3978(199901)47:1/3<195::AID-PROP195>3.0.CO;2-8",
    journal = "Fortsch. Phys.",
    volume = "47",
    pages = "195--200",
    year = "1999"
}

@article{Bell:1969ts,
    author = "Bell, J. S. and Jackiw, R.",
    title = "{A \textsc{pcac} puzzle: $\pi^0 \to \gamma \gamma$ in the $\sigma$ model}",
    doi = "10.1007/BF02823296",
    journal = "Nuovo Cim. A",
    volume = "60",
    pages = "47--61",
    year = "1969"
}

@article{Berman:2012vc,
    author = "Berman, David S. and Cederwall, Martin and Kleinschmidt, Axel and Thompson, Daniel C.",
    title = "{The gauge structure of generalised diffeomorphisms}",
    eprint = "1208.5884",
    archivePrefix = "arXiv",
    primaryClass = "hep-th",
    reportNumber = "QMUL-PH-12-14, AEI-2012-085",
    doi = "10.1007/JHEP01(2013)064",
    journal = "JHEP",
    volume = "01",
    pages = "064",
    year = "2013"
}

@article{Berman:2020tqn,
    author = "Berman, David S. and Blair, Chris D. A.",
    title = "{The Geometry, Branes and Applications of Exceptional Field Theory}",
    eprint = "2006.09777",
    archivePrefix = "arXiv",
    primaryClass = "hep-th",
    reportNumber = "QMUL-PH-20-13",
    doi = "10.1142/S0217751X20300148",
    journal = "Int. J. Mod. Phys. A",
    volume = "35",
    number = "30",
    pages = "2030014",
    year = "2020"
}

@book{Bertlmann:2000,
    author = {Bertlmann, R. A.},
    title = "{Anomalies in Quantum Field Theory}",
    publisher = {Oxford University Press},
    year = {2000},
    month = {11},
    isbn = {9780198507628},
    doi = {10.1093/acprof:oso/9780198507628.001.0001},
    url = {https://doi.org/10.1093/acprof:oso/9780198507628.001.0001},
}

@article{Bertolini:2022ijb,
    author = "Bertolini, Erica and Maggiore, Nicola",
    title = "{Maxwell theory of fractons}",
    eprint = "2209.01485",
    archivePrefix = "arXiv",
    primaryClass = "hep-th",
    doi = "10.1103/PhysRevD.106.125008",
    journal = "Phys. Rev. D",
    volume = "106",
    number = "12",
    pages = "125008",
    year = "2022"
}

@article{Bertolini:2023juh,
    author = "Bertolini, Erica and Blasi, Alberto and Damonte, Andrea and Maggiore, Nicola",
    title = "{Gauging Fractons and Linearized Gravity}",
    eprint = "2304.10789",
    archivePrefix = "arXiv",
    primaryClass = "hep-th",
    doi = "10.3390/sym15040945",
    journal = "Symmetry",
    volume = "15",
    number = "4",
    pages = "945",
    year = "2023"
}

@article{Bertolini:2024apg,
    author = "Bertolini, Erica and Kim, Hyungrok",
    title = "{Covariant interacting fractons}",
    eprint = "2410.01397",
    archivePrefix = "arXiv",
    primaryClass = "hep-th",
    reportNumber = "DIAS-STP-24-21",
    doi = "10.1103/PhysRevD.111.025006",
    journal = "Phys. Rev. D",
    volume = "111",
    number = "2",
    pages = "025006",
    year = "2025"
}

@article{Blasi:2022mbl,
    author = "Blasi, Alberto and Maggiore, Nicola",
    title = "{The theory of symmetric tensor field: From fractons to gravitons and back}",
    eprint = "2207.05956",
    archivePrefix = "arXiv",
    primaryClass = "hep-th",
    doi = "10.1016/j.physletb.2022.137304",
    journal = "Phys. Lett. B",
    volume = "833",
    pages = "137304",
    year = "2022"
}

@article{Bonora:1984pz,
    author = "Bonora, L. and Pasti, P. and Tonin, M.",
    title = "{Gravitational and Weyl Anomalies}",
    reportNumber = "DFPD-12-84",
    doi = "10.1016/0370-2693(84)90421-0",
    journal = "Phys. Lett. B",
    volume = "149",
    pages = "346--350",
    year = "1984"
}

@article{Brandt:1989gy,
    author = "Brandt, F. and Dragon, N. and Kreuzer, M.",
    title = "{Completeness and Nontriviality of the Solutions of the Consistency Conditions}",
    reportNumber = "DESY-89-089, ITP-UH-5-89",
    doi = "10.1016/0550-3213(90)90037-E",
    journal = "Nucl. Phys. B",
    volume = "332",
    pages = "224--249",
    year = "1990"
}

@article{Brandt:1989gv,
    author = "Brandt, F. and Dragon, N. and Kreuzer, M.",
    title = "{Lie algebra cohomology}",
    reportNumber = "DESY-89-088, ITP-UH-6-89",
    doi = "10.1016/0550-3213(90)90038-F",
    journal = "Nucl. Phys. B",
    volume = "332",
    pages = "250--260",
    year = "1990"
}

@article{Brandt:1989rd,
    author = "Brandt, F. and Dragon, N. and Kreuzer, M.",
    title = "{All consistent Yang-Mills anomalies}",
    reportNumber = "DESY-89-076",
    doi = "10.1016/0370-2693(89)90211-6",
    journal = "Phys. Lett. B",
    volume = "231",
    pages = "263--270",
    year = "1989"
}

@article{Brans:1961sx,
    author = "Brans, C. and Dicke, R. H.",
    editor = "Hsu, Jong-Ping and Fine, D.",
    title = "{Mach's principle and a relativistic theory of gravitation}",
    doi = "10.1103/PhysRev.124.925",
    journal = "Phys. Rev.",
    volume = "124",
    pages = "925--935",
    year = "1961"
}

@article{Burnell:2021reh,
    author = "Burnell, Fiona J. and Devakul, Trithep and Gorantla, Pranay and Lam, Ho Tat and Shao, Shu-Heng",
    title = "{Anomaly inflow for subsystem symmetries}",
    eprint = "2110.09529",
    archivePrefix = "arXiv",
    primaryClass = "cond-mat.str-el",
    reportNumber = "MIT-CTP/5336",
    doi = "10.1103/PhysRevB.106.085113",
    journal = "Phys. Rev. B",
    volume = "106",
    number = "8",
    pages = "085113",
    year = "2022"
}

@article{Buscher:1987sk,
    author = "Buscher, T. H.",
    title = "{A Symmetry of the String Background Field Equations}",
    reportNumber = "ITP-SB-87-21",
    doi = "10.1016/0370-2693(87)90769-6",
    journal = "Phys. Lett. B",
    volume = "194",
    pages = "59--62",
    year = "1987"
}

@article{Buscher:1987qj,
    author = "Buscher, T. H.",
    title = "{Path Integral Derivation of Quantum Duality in Nonlinear Sigma Models}",
    reportNumber = "ITP-SB-87-61",
    doi = "10.1016/0370-2693(88)90602-8",
    journal = "Phys. Lett. B",
    volume = "201",
    pages = "466--472",
    year = "1988"
}

@book{Castellani:1991et,
    author = "Castellani, L. and D'Auria, R. and Fre, P.",
    title = "{Supergravity and superstrings: A Geometric perspective. Vol. 1: Mathematical foundations}",
    year = "1991"
}

@article{Chern:1944, 
author = "Chern, S.-S.", 
title= "{A simple instrinsic proof of the Gauss Bonnet formula for closed Riemannian manifolds}",
journal = "Annals of Mathematics, Second Series",
pages = "747-752",
volume= "45", 
DOI = "10.2307/1969302", 
number = "4", 
year = "1944", 
month = "Oct"
}

@article{Cremmer:1997xj,
    author = "Cremmer, E. and Lu, Hong and Pope, C. N. and Stelle, K. S.",
    title = "{Spectrum generating symmetries for BPS solitons}",
    eprint = "hep-th/9707207",
    archivePrefix = "arXiv",
    reportNumber = "CTP-TAMU-29-97, IMPERIAL-TP-96-97-54, LPTENS-97-34, SISSA-92-97-EP",
    doi = "10.1016/S0550-3213(98)00057-1",
    journal = "Nucl. Phys. B",
    volume = "520",
    pages = "132--156",
    year = "1998"
}

@article{Cremmer:1997ct,
    author = "Cremmer, E. and Julia, B. and Lu, Hong and Pope, C. N.",
    title = "{Dualization of dualities. 1.}",
    eprint = "hep-th/9710119",
    archivePrefix = "arXiv",
    reportNumber = "CTP-TAMU-27-97, LPTENS-97-27, SISSA-132-97-EP",
    doi = "10.1016/S0550-3213(98)00136-9",
    journal = "Nucl. Phys. B",
    volume = "523",
    pages = "73--144",
    year = "1998"
}

@article{Cremmer:1978km,
    author = "Cremmer, E. and Julia, B. and Scherk, Joel",
    title = "{Supergravity Theory in 11 Dimensions}",
    reportNumber = "LPTENS-78-10",
    doi = "10.1016/0370-2693(78)90894-8",
    journal = "Phys. Lett. B",
    volume = "76",
    pages = "409--412",
    year = "1978"
}

@article{Cremmer:1979up,
    author = "Cremmer, E. and Julia, B.",
    title = "{The SO(8) Supergravity}",
    reportNumber = "LPTENS 79/6",
    doi = "10.1016/0550-3213(79)90331-6",
    journal = "Nucl. Phys. B",
    volume = "159",
    pages = "141--212",
    year = "1979"
}

@article{Deser:2006zx,
    author = "Deser, Stanley and Waldron, A.",
    title = "{Partially Massless Spin 2 Electrodynamics}",
    eprint = "hep-th/0609113",
    archivePrefix = "arXiv",
    doi = "10.1103/PhysRevD.74.084036",
    journal = "Phys. Rev. D",
    volume = "74",
    pages = "084036",
    year = "2006"
}

@article{Fierz:1939ix,
    author = "Fierz, M. and Pauli, W.",
    title = "{On relativistic wave equations for particles of arbitrary spin in an electromagnetic field}",
    doi = "10.1098/rspa.1939.0140",
    journal = "Proc. Roy. Soc. Lond. A",
    volume = "173",
    pages = "211--232",
    year = "1939"
}

@article{Frob:2020gdh,
    author = {Fr\"ob, M. B.},
    title = "{FieldsX -- An extension package for the xAct tensor computer algebra suite to include fermions, gauge fields and BRST cohomology}",
    eprint = "2008.12422",
    archivePrefix = "arXiv",
    primaryClass = "hep-th",
    month = "8",
    year = "2020"
}

@inproceedings{Gaillard:1997rt,
    author = "Gaillard, Mary K. and Zumino, Bruno",
    title = "{Nonlinear electromagnetic selfduality and Legendre transformations}",
    booktitle = "{A Newton Institute Euroconference on Duality and Supersymmetric Theories}",
    eprint = "hep-th/9712103",
    archivePrefix = "arXiv",
    reportNumber = "LBL-41110, LBNL-41110, UCB-PTH-97-58",
    pages = "33--48",
    month = "12",
    year = "1997"
}

@article{Gaillard:1981rj,
    author = "Gaillard, Mary K. and Zumino, Bruno",
    title = "{Duality Rotations for Interacting Fields}",
    reportNumber = "LAPP-TH-37, CERN-TH-3078",
    doi = "10.1016/0550-3213(81)90527-7",
    journal = "Nucl. Phys. B",
    volume = "193",
    pages = "221--244",
    year = "1981"
}

@article{Gliozzi:1976qd,
    author = "Gliozzi, F. and Scherk, Joel and Olive, David I.",
    title = "{Supersymmetry, Supergravity Theories and the Dual Spinor Model}",
    reportNumber = "CERN-TH-2253",
    doi = "10.1016/0550-3213(77)90206-1",
    journal = "Nucl. Phys. B",
    volume = "122",
    pages = "253--290",
    year = "1977"
}

@article{Green:1984sg,
    author = "Green, Michael B. and Schwarz, John H.",
    title = "{Anomaly Cancellation in Supersymmetric D=10 Gauge Theory and Superstring Theory}",
    reportNumber = "CALT-68-1182",
    doi = "10.1016/0370-2693(84)91565-X",
    journal = "Phys. Lett. B",
    volume = "149",
    pages = "117--122",
    year = "1984"
}

@book{Green:1987sp,
    author = "Green, Michael B. and Schwarz, J. H. and Witten, Edward",
    title = "{Superstring Theory. Vol. 1: Introduction}",
    isbn = "978-0-521-35752-4",
    series = "Cambridge Monographs on Mathematical Physics",
    month = "7",
    year = "1988"
}

@book{Green:1987mn,
    author = "Green, Michael B. and Schwarz, J. H. and Witten, Edward",
    title = "{Superstring Theory. Vol 2: Loop Amplitudes, Anomalies and Phenomenology}",
    isbn = "978-0-521-35753-1",
    month = "7",
    year = "1988"
}

@article{Gunaydin:1985cu,
    author = "Gunaydin, M. and Romans, L. J. and Warner, N. P.",
    title = "{Compact and Noncompact Gauged Supergravity Theories in Five-Dimensions}",
    reportNumber = "CALT-68-1270",
    doi = "10.1016/0550-3213(86)90237-3",
    journal = "Nucl. Phys. B",
    volume = "272",
    pages = "598--646",
    year = "1986"
}

@article{Haah:2011drr,
    author = "Haah, Jeongwan",
    title = "{Local stabilizer codes in three dimensions without string logical operators}",
    eprint = "1101.1962",
    archivePrefix = "arXiv",
    primaryClass = "quant-ph",
    reportNumber = "CALT-68-2816, CALT-68-2816",
    doi = "10.1103/physreva.83.042330",
    journal = "Phys. Rev. A",
    volume = "83",
    number = "4",
    pages = "042330",
    year = "2011"
}

@article{Hinterbichler:2025ost,
    author = "Hinterbichler, Kurt and Joyce, Austin",
    title = "{A Partially Massless Superconductor}",
    eprint = "2507.15932",
    archivePrefix = "arXiv",
    primaryClass = "hep-th",
    month = "7",
    year = "2025"
}

@article{Hohm:2013vpa,
    author = "Hohm, Olaf and Samtleben, Henning",
    title = "{Exceptional Field Theory I: $E_{6(6)}$ covariant Form of M-Theory and Type IIB}",
    eprint = "1312.0614",
    archivePrefix = "arXiv",
    primaryClass = "hep-th",
    reportNumber = "MIT-CTP-4519",
    doi = "10.1103/PhysRevD.89.066016",
    journal = "Phys. Rev. D",
    volume = "89",
    number = "6",
    pages = "066016",
    year = "2014"
}

@article{Hohm:2013uia,
    author = "Hohm, Olaf and Samtleben, Henning",
    title = "{Exceptional field theory. II. E$_{7(7)}$}",
    eprint = "1312.4542",
    archivePrefix = "arXiv",
    primaryClass = "hep-th",
    reportNumber = "MIT-CTP-4522",
    doi = "10.1103/PhysRevD.89.066017",
    journal = "Phys. Rev. D",
    volume = "89",
    pages = "066017",
    year = "2014"
}

@article{Hohm:2014fxa,
    author = "Hohm, Olaf and Samtleben, Henning",
    title = "{Exceptional field theory. III. E$_{8(8)}$}",
    eprint = "1406.3348",
    archivePrefix = "arXiv",
    primaryClass = "hep-th",
    reportNumber = "MIT-CTP-4557",
    doi = "10.1103/PhysRevD.90.066002",
    journal = "Phys. Rev. D",
    volume = "90",
    pages = "066002",
    year = "2014"
}

@article{Hohm:2017wtr,
    author = "Hohm, Olaf and Musaev, Edvard T. and Samtleben, Henning",
    title = "{O($d+1, d+1$) enhanced double field theory}",
    eprint = "1707.06693",
    archivePrefix = "arXiv",
    primaryClass = "hep-th",
    doi = "10.1007/JHEP10(2017)086",
    journal = "JHEP",
    volume = "10",
    pages = "086",
    year = "2017"
}

@article{Imbimbo:2009dy,
      author         = "Imbimbo, C.",
      title          = "{The Coupling of Chern-Simons Theory to Topological
                        Gravity}",
      journal        = "Nucl. Phys.",
      volume         = "B825",
      year           = "2010",
      pages          = "366-395",
      doi            = "10.1016/j.nuclphysb.2009.09.022",
      eprint         = "0905.4631",
      archivePrefix  = "arXiv",
      primaryClass   = "hep-th",
      reportNumber   = "GEF-TH-05-2009",
      SLACcitation   = "%%CITATION = ARXIV:0905.4631;%%"
}

@article{Imbimbo:2014pla,
      author         = "Imbimbo, C. and Rosa, D.",
      title          = "{Topological anomalies for Seifert 3\--ma\-ni\-folds}",
      journal        = "JHEP",
      volume         = "07",
      year           = "2015",
      pages          = "068",
      doi            = "10.1007/JHEP07(2015)068",
      eprint         = "1411.6635",
      archivePrefix  = "arXiv",
      primaryClass   = "hep-th",
      SLACcitation   = "%%CITATION = ARXIV:1411.6635;%%"
}

@article{Imbimbo:2018duh,
    author = "Imbimbo, C. and Rosa, D.",
    title = "{The topological structure of supergravity: an application to supersymmetric localization}",
    eprint = "1801.04940",
    archivePrefix = "arXiv",
    primaryClass = "hep-th",
    reportNumber = "KIAS-P18006",
    doi = "10.1007/JHEP05(2018)112",
    journal = "JHEP",
    volume = "05",
    pages = "112",
    year = "2018"
}

@article{Inverso:2017lrz,
    author = "Inverso, Gianluca",
    title = "{Generalised Scherk-Schwarz reductions from gauged supergravity}",
    eprint = "1708.02589",
    archivePrefix = "arXiv",
    primaryClass = "hep-th",
    doi = "10.1007/JHEP06(2021)148",
    journal = "JHEP",
    volume = "12",
    pages = "124",
    year = "2017",
    note = "[Erratum: JHEP 06, 148 (2021)]"
}

@book{Jordan:1955,
    author = "Jordan, Pascual",
    title = "{Schwerkraft und Weltall: Grundlagen der theoretischen Kosmologie}",
    publisher = "Vieweg",
    address = "Braunschweig",
    year = "1955"
}

@article{Julia:1980gr,
    author = "Julia, B.",
    title = "{Group disintegrations}",
    reportNumber = "LPTENS-80-16",
    journal = "Conf. Proc. C",
    volume = "8006162",
    pages = "331--350",
    year = "1980"
}

@article{Kalkman:1993zp,
    author = "Kalkman, J.",
    title = "{BRST model for equivariant cohomology and representatives for the equivariant Thom class}",
    doi = "10.1007/BF02096949",
    journal = "Commun. Math. Phys.",
    volume = "153",
    pages = "447--463",
    year = "1993"
}

@article{Kalb:1974yc,
    author = "Kalb, Michael and Ramond, Pierre",
    title = "{Classical direct interstring action}",
    doi = "10.1103/PhysRevD.9.2273",
    journal = "Phys. Rev. D",
    volume = "9",
    pages = "2273--2284",
    year = "1974"
}

@article{Kaluza:1921tu,
    author = "Kaluza, Th.",
    title = {{Zum Unit{\"a}tsproblem der Physik}},
    eprint = "1803.08616",
    archivePrefix = "arXiv",
    primaryClass = "physics.hist-ph",
    reportNumber = "HUPD-8401",
    doi = "10.1142/S0218271818700017",
    journal = "Sitzungsber. Preuss. Akad. Wiss. Berlin (Math. Phys. )",
    volume = "1921",
    pages = "966--972",
    year = "1921"
}

@article{Klein:1926tv,
    author = "Klein, Oskar",
    editor = "Taylor, J. C.",
    title = "{Quantum Theory and Five-Dimensional Theory of Relativity. (In German and English)}",
    doi = "10.1007/BF01397481",
    journal = "Z. Phys.",
    volume = "37",
    pages = "895--906",
    year = "1926"
}

@article{Koepsell:1999uj,
    author = "Koepsell, K. and Nicolai, H. and Samtleben, H.",
    title = "{On the Yangian [Y(e(8))] quantum symmetry of maximal supergravity in two-dimensions}",
    eprint = "hep-th/9903111",
    archivePrefix = "arXiv",
    reportNumber = "AEI-103, LPTENS-99-08",
    doi = "10.1088/1126-6708/1999/04/023",
    journal = "JHEP",
    volume = "04",
    pages = "023",
    year = "1999"
}

@article{Langouche:1984gn,
    author = "Langouche, F. and Schucker, T. and Stora, R.",
    title = "{Gravitational Anomalies of the Adler-bardeen Type}",
    reportNumber = "CERN-TH-3898, LAPP-TH-110",
    doi = "10.1016/0370-2693(84)90057-1",
    journal = "Phys. Lett. B",
    volume = "145",
    pages = "342--346",
    year = "1984"
}

@article{Lavau:2020pwa,
    author = "Lavau, Sylvain and Stasheff, Jim",
    title = "{From Lie algebra crossed modules to tensor hierarchies}",
    eprint = "2003.07838",
    archivePrefix = "arXiv",
    primaryClass = "math-ph",
    doi = "10.1016/j.jpaa.2022.107311",
    journal = "J. Pure Appl. Algebra",
    volume = "227",
    pages = "107311",
    year = "2023",
    note = "[Erratum: J.Pure Appl.Algebra 227, 107428 (2023)]"
}

@article{Lavau:2017tvi,
    author = "Lavau, Sylvain",
    title = "{Tensor hierarchies and Leibniz algebras}",
    eprint = "1708.07068",
    archivePrefix = "arXiv",
    primaryClass = "hep-th",
    doi = "10.1016/j.geomphys.2019.05.014",
    journal = "J. Geom. Phys.",
    volume = "144",
    pages = "147--189",
    year = "2019"
}

@article{Majorana:1932ga,
    author = "Majorana, Ettore",
    title = "{Oriented atoms in a variable magnetic field}",
    doi = "10.1007/BF02960953",
    journal = "Nuovo Cim.",
    volume = "9",
    pages = "43--50",
    year = "1932"
}

@article{Manes:1985df,
    author = "Manes, J. and Stora, R. and Zumino, B.",
    title = "{Algebraic Study of Chiral Anomalies}",
    reportNumber = "LBL-19138, UCB-PTH-85/7, LAPP-TH-137",
    doi = "10.1007/BF01208825",
    journal = "Commun. Math. Phys.",
    volume = "102",
    pages = "157",
    year = "1985"
}

@article{Murtaza:1973bp,
    author = "Murtaza, G. and Rashid, M. A.",
    title = "{Duality of a young diagram describing a representation and dimensionality formulas}",
    doi = "10.1063/1.1666463",
    journal = "J. Math. Phys.",
    volume = "14",
    pages = "1196--1198",
    year = "1973"
}

@book{Piguet:1995er,
    author = "Piguet, O. and Sorella, S. P.",
    title = "{Algebraic renormalization: Perturbative renormalization, symmetries and anomalies}",
    doi = "10.1007/978-3-540-49192-7",
    journal = "Lect.Notes Phys.Monogr.", 
    pages = "1-134",
    volume = "28",
    year = "1995",
    publisher="Springer Netherlands"
}

@article{Pernici:1984xx,
    author = "Pernici, M. and Pilch, K. and van Nieuwenhuizen, P.",
    title = "{Gauged Maximally Extended Supergravity in Seven-dimensions}",
    reportNumber = "ITP-SB-84-31",
    doi = "10.1016/0370-2693(84)90813-X",
    journal = "Phys. Lett. B",
    volume = "143",
    pages = "103--107",
    year = "1984"
}

@article{Pretko:2017xar,
    author = "Pretko, Michael",
    title = "{Higher-Spin Witten Effect and Two-Dimensional Fracton Phases}",
    eprint = "1707.03838",
    archivePrefix = "arXiv",
    primaryClass = "cond-mat.str-el",
    doi = "10.1103/PhysRevB.96.125151",
    journal = "Phys. Rev. B",
    volume = "96",
    number = "12",
    pages = "125151",
    year = "2017"
}

@article{Pretko:2020cko,
    author = "Pretko, Michael and Chen, Xie and You, Yizhi",
    title = "{Fracton Phases of Matter}",
    eprint = "2001.01722",
    archivePrefix = "arXiv",
    primaryClass = "cond-mat.str-el",
    doi = "10.1142/S0217751X20300033",
    journal = "Int. J. Mod. Phys. A",
    volume = "35",
    number = "06",
    pages = "2030003",
    year = "2020"
}

@article{Rarita:1941mf,
    author = "Rarita, William and Schwinger, Julian",
    title = "{On a theory of particles with half integral spin}",
    doi = "10.1103/PhysRev.60.61",
    journal = "Phys. Rev.",
    volume = "60",
    pages = "61",
    year = "1941"
}

@article{Seiberg:2020bhn,
    author = "Seiberg, Nathan and Shao, Shu-Heng",
    title = "{Exotic Symmetries, Duality, and Fractons in 2+1-Dimensional Quantum Field Theory}",
    eprint = "2003.10466",
    archivePrefix = "arXiv",
    primaryClass = "cond-mat.str-el",
    doi = "10.21468/SciPostPhys.10.2.027",
    journal = "SciPost Phys.",
    volume = "10",
    number = "2",
    pages = "027",
    year = "2021"
}

@article{Siegel:1985tw,
    author = "Siegel, Warren and Zwiebach, Barton",
    title = "{Gauge String Fields}",
    reportNumber = "UCB-PTH-85/30",
    doi = "10.1016/0550-3213(86)90030-1",
    journal = "Nucl. Phys. B",
    volume = "263",
    pages = "105--128",
    year = "1986"
}

@article{Sorella:1992dr,
    author = "Sorella, S. P.",
    title = "{Algebraic characterization of the Wess-Zumino consistency conditions in gauge theories}",
    eprint = "hep-th/9302136",
    archivePrefix = "arXiv",
    reportNumber = "UGVA-DPT-1992-08-781",
    doi = "10.1007/BF02099759",
    journal = "Commun. Math. Phys.",
    volume = "157",
    pages = "231--243",
    year = "1993"
}

@article{Sorella:1993kq,
    author = "Sorella, Silvio P. and Tataru, Liviu",
    title = "{A Closed form for consistent anomalies in gauge theories}",
    eprint = "hep-th/9307169",
    archivePrefix = "arXiv",
    reportNumber = "TUW-93-15",
    doi = "10.1016/0370-2693(94)90205-4",
    journal = "Phys. Lett. B",
    volume = "324",
    pages = "351--358",
    year = "1994"
}

@article{Stora:1976kd,
    author = "Stora, R.",
    title = "{Continuum Gauge Theories}",
    reportNumber = "CPT-76-P.854, CNRS-CPT-76-P-854",
    journal = "Conf. Proc. C",
    volume = "7607121",
    pages = "201",
    year = "1976"
}

@inbook{Stora:1976LM,
author = "Stora, R", 
editor= "M. Levy, P. Mitter",
title = "Continuum Gauge Theories",
bookTitle = "New developments in quantum field theory and statistical mechanics, Cargese 1976, Ed. NATO ASI Ser B vol. 26",
year= "1977",
Publisher= "Plenum Press",
pages= "201-224",
doi= "10.1007/978-1-4615-8918-1"
}

@inbook{Stora:1984,
author="Stora, R.",
editor="'t Hooft, G. and Jaffe, A. and Lehmann, H. and Mitter, P. K. and Singer, I. M. and Stora, R.",
title="Algebraic Structure and Topological Origin of Anomalies",
bookTitle="Progress in Gauge Field Theory",
year="1984",
publisher="Springer US",
address="Boston, MA",
pages="543--562",
doi="10.1007/978-1-4757-0280-4_19",
url="https://doi.org/10.1007/978-1-4757-0280-4_19"
}

@article{Strickland-Constable:2013xta,
    author = "Strickland-Constable, Charles",
    title = "{Subsectors, Dynkin Diagrams and New Generalised Geometries}",
    eprint = "1310.4196",
    archivePrefix = "arXiv",
    primaryClass = "hep-th",
    reportNumber = "ZMP-HH-13-19",
    doi = "10.1007/JHEP08(2017)144",
    journal = "JHEP",
    volume = "08",
    pages = "144",
    year = "2017"
}

@article{Trigiante:2016mnt,
    author = "Trigiante, Mario",
    title = "{Gauged Supergravities}",
    eprint = "1609.09745",
    archivePrefix = "arXiv",
    primaryClass = "hep-th",
    doi = "10.1016/j.physrep.2017.03.001",
    journal = "Phys. Rept.",
    volume = "680",
    pages = "1--175",
    year = "2017"
}

@article{Tyutin:1975qk,
    author = "Tyutin, I. V.",
    title = "{Gauge Invariance in Field Theory and Statistical Physics in Operator Formalism}",
    eprint = "0812.0580",
    archivePrefix = "arXiv",
    primaryClass = "hep-th",
    reportNumber = "LEBEDEV-75-39",
    year = "1975"
}

@article{VanNieuwenhuizen:1981ae,
    author = "Van Nieuwenhuizen, P.",
    title = "{Supergravity}",
    doi = "10.1016/0370-1573(81)90157-5",
    journal = "Phys. Rept.",
    volume = "68",
    pages = "189--398",
    year = "1981"
}

@article{Vijay:2015mka,
    author = "Vijay, Sagar and Haah, Jeongwan and Fu, Liang",
    title = "{A New Kind of Topological Quantum Order: A Dimensional Hierarchy of Quasiparticles Built from Stationary Excitations}",
    eprint = "1505.02576",
    archivePrefix = "arXiv",
    primaryClass = "cond-mat.str-el",
    doi = "10.1103/PhysRevB.92.235136",
    journal = "Phys. Rev. B",
    volume = "92",
    number = "23",
    pages = "235136",
    year = "2015"
}

@book{Weinberg:1996kr,
    author = "Weinberg, Steven",
    title = "{The quantum theory of fields. Vol. 2: Modern applications}",
    doi = "10.1017/CBO9781139644174",
    isbn = "978-1-139-63247-8, 978-0-521-67054-8, 978-0-521-55002-4",
    publisher = "Cambridge University Press",
    month = "8",
    year = "2013"
}

@article{Wess:1971yu,
    author = "Wess, J. and Zumino, B.",
    title = "{Consequences of anomalous Ward identities}",
    doi = "10.1016/0370-2693(71)90582-X",
    journal = "Phys. Lett. B",
    volume = "37",
    pages = "95--97",
    year = "1971"
}

@article{Witten:1988ze,
    author = "Witten, Edward",
    title = "{Topological Quantum Field Theory}",
    reportNumber = "IASSNS-HEP-87-72",
    doi = "10.1007/BF01223371",
    journal = "Commun. Math. Phys.",
    volume = "117",
    pages = "353",
    year = "1988"
}

@article{Witten:1988xj,
    author = "Witten, Edward",
    title = "{Topological Sigma Models}",
    reportNumber = "IASSNS-HEP-88/7",
    doi = "10.1007/BF01466725",
    journal = "Commun. Math. Phys.",
    volume = "118",
    pages = "411",
    year = "1988"
}

@article{Labastida:1988zb,
    author = "Labastida, J. M. F. and Pernici, M. and Witten, Edward",
    title = "{Topological Gravity in Two-Dimensions}",
    reportNumber = "IASSNS-HEP-88/29",
    doi = "10.1016/0550-3213(88)90094-6",
    journal = "Nucl. Phys. B",
    volume = "310",
    pages = "611--624",
    year = "1988"
}

@inproceedings{Witten:1986,
	author = "Witten, Edward",
	title = "{Physics and Geometry}", 
	reportNumber = "PRE30537",
	booktitle = "International Congress of Mathematicians", 		address = "Berkeley",
	month = "August",
	year = "1986"
}

@inproceedings{Zumino:1981pt,
    author = "Zumino, Bruno",
    title = "{Duality rotations}",
    booktitle = "{Nuffield Workshop on Quantum Structure of Space and Time}",
    reportNumber = "LBL-13564, C81-08-20-7",
    month = "11",
    year = "1981"
}

@inproceedings{Zumino:1983ew,
    author = "Zumino, B.",
    title = "{Chiral Anomalies and Differential Geometry: Lectures given at Les Houches, August 1983}",
    booktitle = "{Les Houches Summer School on Theoretical Physics: Relativity, Groups and Topology}",
    reportNumber = "UCB-PTH-83-16, LBL-16747",
    pages = "1291--1322",
    month = "10",
    year = "1983"
}

@article{Zumino:1983rz,
    author = "Zumino, B. and Wu, Y.-S. and Zee, A.",
    title = "{Chiral Anomalies, Higher Dimensions, and Differential Geometry}",
    reportNumber = "LBL-16443, DOE-ER-40048-18-P3",
    doi = "10.1016/0550-3213(84)90259-1",
    journal = "Nucl. Phys. B",
    volume = "239",
    pages = "477--507",
    year = "1984"
}

@misc{xact,
      author         = "Mart{\'\i}n-Garc{\'\i}a, Jos{\'e} M. and Garc{\'\i}a-Parrado, Alfonso and Stecchina, Alessandro and Wardell, Barry and Pitrou, Cyril and Brizuela, David and Yllanes, David and Faye, Guillaume and Stein, Leo and Portugal, Renato and Nutma, Teake and B{\"a}ckdahl, Thomas and others",
      title          = "{xAct: Efficient tensor computer algebra for the Wolfram Language}",
      howpublished   = "\\\url{http://www.xact.es}",
      year           = "2020",
      url            = "http://www.xact.es"
}

@article{xTras,
    author = "Nutma, Teake",
    title = "{xTras : A field-theory inspired xAct  package for mathematica}",
    eprint = "1308.3493",
    archivePrefix = "arXiv",
    primaryClass = "cs.SC",
    reportNumber = "AEI-2013-236",
    doi = "10.1016/j.cpc.2014.02.006",
    journal = "Comput. Phys. Commun.",
    volume = "185",
    pages = "1719--1738",
    year = "2014"
}

@article{Cartan1922,
    author = "Cartan, E.",
    title = "{Sur une généralisation de la notion de courbure de Riemann et les espaces à torsion}",
    journal = "C. R. Acad. Sci.",
    volume = "174",
    pages = "593",
    year = "1922"
}

@article{Cartan1923,
    author = "Cartan, E.",
    title = "{Sur une généralisation de la notion de courbure de Riemann et les espaces à torsion}",
    journal = "Ann. Ec. Norm. Sup.",
    volume = "40",
    pages = "325-412",
    year = "1923"
}

@article{Cartan1924,
    author = "Cartan, E.",
    title = "{Sur une généralisation de la notion de courbure de Riemann et les espaces à torsion}",
    journal = "Ann. Ec. Norm. Sup.",
    volume = "41",
    pages = "1-25",
    year = "1924"
}

@article{Cartan1925,
    author = "Cartan, E.",
    title = "{Sur une généralisation de la notion de courbure de Riemann et les espaces à torsion}",
    journal = "Ann. Ec. Norm. Sup.",
    volume = "40",
    pages = "17-88",
    year = "1925"
}

@article{Einstein1928,
    author = "Einstein, A.",
    title = "{Riemann-Geometrie mit Aufrechterhaltung des Begriffes des Fernparallelismus}",
    doi = "10.1002/3527608958.ch36",
    journal = "Sitzber. Preuss. Akad. Wiss.",
    volume = "17",
    pages = "217–221",
    year = "1928"
}

@article{Weitzenbock1928,
    author = "Weitzenb{\"o}ck, R.",
    title = "{Differential invariants in Einstein’s theory of teleparallelism}",
    doi = "10.1002/3527608958.ch36",
    journal = "Sitzber. Preuss. Akad. Wiss.",
    volume = "17",
    pages = "466-474",
    year = "1928"
}

@article{Moller1961a,
    author = "M{\o}ller, C.",
    title = "{Conservation laws and absolute parallelism in general relativity}",
    journal = "Mat. Fys. Skr. Dan. Vid. Selck.",
    volume = "1",
    pages = "3--50",
    year = "1961"
}

@article{Moller1961b,
    author = "M{\o}ller, C.",
    title = "{Further remarks on the localization of the energy in the general theory of relativity}",
    doi = "10.1016/0003-4916(61)90148-8",
    journal = "Ann. Phys.",
    volume = "12",
    pages = "118–133",
    year = "1961"
}

@article{Moller1978,
    author = "M{\o}ller, C.",
    title = "{On the crisis in the theory of gravitation and a possible solution}",
    journal = "Mat. Fys. Medd. Dan. Vid. Selsk.",
    volume = "39",
    pages = "3-31",
    year = "1978"
}

@article{Hayashi:1967se,
    author = "Hayashi, K. and Nakano, T.",
    editor = "Hsu, Jong-Ping and Fine, D.",
    title = "{Extended translation invariance and associated gauge fields}",
    doi = "10.1143/PTP.38.491",
    journal = "Prog. Theor. Phys.",
    volume = "38",
    pages = "491--507",
    year = "1967"
}

@article{Hayashi:1977jd,
    author = "Hayashi, Kenji",
    title = "{The Gauge Theory of the Translation Group and Underlying Geometry}",
    reportNumber = "Print-77-0143 (MPI,MUNICH)",
    doi = "10.1016/0370-2693(77)90840-1",
    journal = "Phys. Lett. B",
    volume = "69",
    pages = "441--444",
    year = "1977"
}

@article{Cho:1975dh,
    author = "Cho, Y. M.",
    editor = "Hsu, Jong-Ping and Fine, D.",
    title = "{Einstein Lagrangian as the Translational Yang-Mills Lagrangian}",
    reportNumber = "EFI 75-61-CHICAGO",
    doi = "10.1103/PhysRevD.14.2521",
    journal = "Phys. Rev. D",
    volume = "14",
    pages = "2521",
    year = "1976"
}

@article{Hayashi:1979qx,
    author = "Hayashi, K. and Shirafuji, T.",
    editor = "Hsu, Jong-Ping and Fine, D.",
    title = "{New general relativity.}",
    reportNumber = "PRINT-79-0197 (TOKYO-U.,KOMABA)",
    doi = "10.1103/PhysRevD.19.3524",
    journal = "Phys. Rev. D",
    volume = "19",
    pages = "3524--3553",
    year = "1979",
    note = "[Addendum: Phys.Rev.D 24, 3312--3314 (1982)]"
}

@article{Shirafuji:1995xc,
    author = "Shirafuji, Takeshi and Nashed, Gamal G. L. and Hayashi, Kenji",
    title = "{Energy of general spherically symmetric solution in tetrad theory of gravitation}",
    eprint = "gr-qc/9601044",
    archivePrefix = "arXiv",
    reportNumber = "STUPP-95-141, KITASATO-95-1",
    doi = "10.1143/PTP.95.665",
    journal = "Prog. Theor. Phys.",
    volume = "95",
    pages = "665--678",
    year = "1996"
}

@article{deAndrade:1997gka,
    author = "de Andrade, V. C. and Pereira, J. G.",
    title = "{Gravitational Lorentz force and the description of the gravitational interaction}",
    eprint = "gr-qc/9703059",
    archivePrefix = "arXiv",
    doi = "10.1103/PhysRevD.56.4689",
    journal = "Phys. Rev. D",
    volume = "56",
    pages = "4689--4695",
    year = "1997"
}

@ARTICLE{Bott1958-vt,
  title     = "{On the parallelizability of the spheres}",
  author    = "Bott, R and Milnor, J",
  doi = "http://dx.doi.org/10.1090/S0002-9904-1958-10166-4",
  journal   = "Bull. New Ser. Am. Math. Soc.",
  publisher = "American Mathematical Society (AMS)",
  volume    =  64,
  number    =  3,
  pages     = "87--89",
  year      =  1958,
  language  = "en"
}

@article{Milnor1958,
    author = "Milnor, J.",
    title = "{Some Consequences of a Theorem of Bott}",
    doi = "http://dx.doi.org/10.2307/1970255",
    journal = "Ann. Math.",
    volume = "68",
    pages = "444",
    year = "1958"
}

@article{Kervaire1958,
    author = "Kervaire, M.A.",
    title = "{Non-parallelizability of the n-sphere for $n > 7$}",
    doi = "http://dx.doi.org/10.1073/pnas.44.3.280",
    journal = "Proc. Natl. Acad.Sci.",
    volume = "44",
    pages = "280",
    year = "1958"
}

@article{Lee:2014mla,
    author = "Lee, Kanghoon and Strickland-Constable, Charles and Waldram, Daniel",
    title = "{Spheres, generalised parallelisability and consistent truncations}",
    eprint = "1401.3360",
    archivePrefix = "arXiv",
    primaryClass = "hep-th",
    reportNumber = "IMPERIAL-TP-14-DW-01, ZMP-HH-14-3",
    doi = "10.1002/prop.201700048",
    journal = "Fortsch. Phys.",
    volume = "65",
    number = "10-11",
    pages = "1700048",
    year = "2017"
}

@article{deWit:1986oxb,
    author = "de Wit, B. and Nicolai, H.",
    title = "{The Consistency of the S**7 Truncation in D=11 Supergravity}",
    reportNumber = "CERN-TH-4359/86",
    doi = "10.1016/0550-3213(87)90253-7",
    journal = "Nucl. Phys. B",
    volume = "281",
    pages = "211--240",
    year = "1987"
}

@article{deWit:1982bul,
    author = "de Wit, B. and Nicolai, H.",
    title = "{N=8 Supergravity}",
    reportNumber = "CERN-TH-3291",
    doi = "10.1016/0550-3213(82)90120-1",
    journal = "Nucl. Phys. B",
    volume = "208",
    pages = "323",
    year = "1982"
}

@article{Nastase:1999cb,
    author = "Nastase, Horatiu and Vaman, Diana and van Nieuwenhuizen, Peter",
    title = "{Consistent nonlinear K K reduction of 11-d supergravity on AdS(7) x S(4) and selfduality in odd dimensions}",
    eprint = "hep-th/9905075",
    archivePrefix = "arXiv",
    reportNumber = "ITP-SB-99-19",
    doi = "10.1016/S0370-2693(99)01266-6",
    journal = "Phys. Lett. B",
    volume = "469",
    pages = "96--102",
    year = "1999"
}

@article{Nastase:1999kf,
    author = "Nastase, Horatiu and Vaman, Diana and van Nieuwenhuizen, Peter",
    title = "{Consistency of the AdS(7) x S(4) reduction and the origin of selfduality in odd dimensions}",
    eprint = "hep-th/9911238",
    archivePrefix = "arXiv",
    reportNumber = "ITP-SB-99-56",
    doi = "10.1016/S0550-3213(00)00193-0",
    journal = "Nucl. Phys. B",
    volume = "581",
    pages = "179--239",
    year = "2000"
}

@article{Cvetic:2000nc,
    author = "Cvetic, Mirjam and Lu, Hong and Pope, C. N. and Sadrzadeh, A. and Tran, Tuan A.",
    title = "{Consistent SO(6) reduction of type IIB supergravity on S**5}",
    eprint = "hep-th/0003103",
    archivePrefix = "arXiv",
    reportNumber = "CTP-TAMU-08-00, UPR-879-T",
    doi = "10.1016/S0550-3213(00)00372-2",
    journal = "Nucl. Phys. B",
    volume = "586",
    pages = "275--286",
    year = "2000"
}

@article{Cvetic:2000ah,
    author = "Cvetic, Mirjam and Lu, Hong and Pope, C. N. and Sadrzadeh, A. and Tran, Tuan A.",
    title = "{S**3 and S**4 reductions of type IIA supergravity}",
    eprint = "hep-th/0005137",
    archivePrefix = "arXiv",
    reportNumber = "CTP-TAMU-13-00, UPR-888-T",
    doi = "10.1016/S0550-3213(00)00466-1",
    journal = "Nucl. Phys. B",
    volume = "590",
    pages = "233--251",
    year = "2000"
}

@article{Hitchin:2003cxu,
    author = "Hitchin, Nigel",
    title = "{Generalized Calabi-Yau manifolds}",
    eprint = "math/0209099",
    archivePrefix = "arXiv",
    doi = "10.1093/qjmath/54.3.281",
    journal = "Quart. J. Math. Oxford Ser.",
    volume = "54",
    pages = "281--308",
    year = "2003"
}

@phdthesis{Gualtieri:2003dx,
    author = "Gualtieri, Marco",
    title = "{Generalized complex geometry}",
    eprint = "math/0401221",
    archivePrefix = "arXiv",
    school = "Oxford U.",
    year = "2003"
}

@article{Gualtieri:2007ng,
    author = "Gualtieri, Marco",
    title = "{Generalized complex geometry}",
    eprint = "math/0703298",
    archivePrefix = "arXiv",
    reportNumber = "10-03-2007",
    month = "3",
    year = "2007"
}

@article{Hitchin:2010qz,
    author = "Hitchin, Nigel",
    title = "{Lectures on generalized geometry}",
    eprint = "1008.0973",
    archivePrefix = "arXiv",
    primaryClass = "math.DG",
    month = "8",
    year = "2010"
}

@article{Bursztyn:2005vwa,
    author = "Bursztyn, Henrique and Cavalcanti, Gil R. and Gualtieri, Marco",
    title = "{Reduction of Courant algebroids and generalized complex structures}",
    eprint = "math/0509640",
    archivePrefix = "arXiv",
    doi = "10.1016/j.aim.2006.09.008",
    journal = "Adv. Math.",
    volume = "211",
    pages = "726--765",
    year = "2007"
}

@article{Hitchin:2005in,
    author = "Hitchin, Nigel",
    title = "{Brackets, forms and invariant functionals}",
    eprint = "math/0508618",
    archivePrefix = "arXiv",
    month = "8",
    year = "2005"
}

@article{Courant1990,
    author = "Courant, Theodore James",
    title = "{Dirac manifolds}",
       reportNumber = "IMPERIAL-TP-14-DW-01, ZMP-HH-14-3",
    doi = "10.1090/S0002-9947-1990-0998124-1",
    journal = "Trans. Amer. Math. Soc.",
    volume = "319",
    pages = "631-661",
    year = "1990"
}

@article{Courant1988,
  title={Beyond poisson structures},
  author={Courant, Ted and Weinstein, Alan},
  journal={Action hamiltoniennes de groupes. Troisieme th{\'e}oreme de Lie (Lyon, 1986)},
  volume={27},
  pages={39--49},
  year={1988}
}

@article{Dorfman1987,
  title={Dirac structures of integrable evolution equations},
  author={Dorfman, Irene Ya},
  journal={Physics Letters A},
  volume={125},
  number={5},
  pages={240--246},
  year={1987},
  publisher={Elsevier}
}

@article{Coimbra:2011nw,
    author = "Coimbra, Andre and Strickland-Constable, Charles and Waldram, Daniel",
    title = "{Supergravity as Generalised Geometry I: Type II Theories}",
    eprint = "1107.1733",
    archivePrefix = "arXiv",
    primaryClass = "hep-th",
    reportNumber = "IMPERIAL-TP-11-DW-01",
    doi = "10.1007/JHEP11(2011)091",
    journal = "JHEP",
    volume = "11",
    pages = "091",
    year = "2011"
}

@article{Hull:2007zu,
    author = "Hull, C. M.",
    title = "{Generalised Geometry for M-Theory}",
    eprint = "hep-th/0701203",
    archivePrefix = "arXiv",
    reportNumber = "IMPERIAL-TP-2007-CMH-01",
    doi = "10.1088/1126-6708/2007/07/079",
    journal = "JHEP",
    volume = "07",
    pages = "079",
    year = "2007"
}

@article{PiresPacheco:2008qik,
    author = "Pires Pacheco, Paulo and Waldram, Daniel",
    title = "{M-theory, exceptional generalised geometry and superpotentials}",
    eprint = "0804.1362",
    archivePrefix = "arXiv",
    primaryClass = "hep-th",
    reportNumber = "IMPERIAL-TP-08-DW-01",
    doi = "10.1088/1126-6708/2008/09/123",
    journal = "JHEP",
    volume = "09",
    pages = "123",
    year = "2008"
}

@article{Coimbra:2011ky,
    author = "Coimbra, Andr{\'e} and Strickland-Constable, Charles and Waldram, Daniel",
    title = "{$E_{d(d)} \times \mathbb{R}^+$ generalised geometry, connections and M theory}",
    eprint = "1112.3989",
    archivePrefix = "arXiv",
    primaryClass = "hep-th",
    reportNumber = "IMPERIAL-TP-11-DW-02",
    doi = "10.1007/JHEP02(2014)054",
    journal = "JHEP",
    volume = "02",
    pages = "054",
    year = "2014"
}

@article{Coimbra:2012af,
    author = "Coimbra, Andre and Strickland-Constable, Charles and Waldram, Daniel",
    title = "{Supergravity as Generalised Geometry II: $E_{d(d)} \times \mathbb{R}^+$ and M theory}",
    eprint = "1212.1586",
    archivePrefix = "arXiv",
    primaryClass = "hep-th",
    reportNumber = "IMPERIAL-TP-12-DW-01",
    doi = "10.1007/JHEP03(2014)019",
    journal = "JHEP",
    volume = "03",
    pages = "019",
    year = "2014"
}

@article{Grana:2008yw,
    author = "Grana, Mariana and Minasian, Ruben and Petrini, Michela and Waldram, Daniel",
    title = "{T-duality, Generalized Geometry and Non-Geometric Backgrounds}",
    eprint = "0807.4527",
    archivePrefix = "arXiv",
    primaryClass = "hep-th",
    reportNumber = "SPHT-T08-118, IMPERIAL-TP-08-DW-02",
    doi = "10.1088/1126-6708/2009/04/075",
    journal = "JHEP",
    volume = "04",
    pages = "075",
    year = "2009"
}

@article{Scherk:1979zr,
    author = "Scherk, Joel and Schwarz, John H.",
    editor = "Salam, A. and Sezgin, E.",
    title = "{How to Get Masses from Extra Dimensions}",
    reportNumber = "LPTENS-79-2",
    doi = "10.1016/0550-3213(79)90592-3",
    journal = "Nucl. Phys. B",
    volume = "153",
    pages = "61--88",
    year = "1979"
}

@article{Scherk:1978ta,
    author = "Scherk, Joel and Schwarz, John H.",
    title = "{Spontaneous Breaking of Supersymmetry Through Dimensional Reduction}",
    reportNumber = "LPTENS 78/28",
    doi = "10.1016/0370-2693(79)90425-8",
    journal = "Phys. Lett. B",
    volume = "82",
    pages = "60--64",
    year = "1979"
}

@article{Samtleben:2008pe,
    author = "Samtleben, Henning",
    editor = "Derendinger, J. P. and Orlando, Domenico and Uranga, Angel",
    title = "{Lectures on Gauged Supergravity and Flux Compactifications}",
    eprint = "0808.4076",
    archivePrefix = "arXiv",
    primaryClass = "hep-th",
    reportNumber = "ENSL-00315624",
    doi = "10.1088/0264-9381/25/21/214002",
    journal = "Class. Quant. Grav.",
    volume = "25",
    pages = "214002",
    year = "2008"
}

@article{Ciceri:2016dmd,
    author = "Ciceri, Franz and Guarino, Adolfo and Inverso, Gianluca",
    title = "{The exceptional story of massive IIA supergravity}",
    eprint = "1604.08602",
    archivePrefix = "arXiv",
    primaryClass = "hep-th",
    doi = "10.1007/JHEP08(2016)154",
    journal = "JHEP",
    volume = "08",
    pages = "154",
    year = "2016"
}

@article{Marcus:1983hb,
    author = "Marcus, Neil and Schwarz, John H.",
    editor = "Salam, A. and Sezgin, E.",
    title = "{Three-Dimensional Supergravity Theories}",
    reportNumber = "CALT-68-1018",
    doi = "10.1016/0550-3213(83)90402-9",
    journal = "Nucl. Phys. B",
    volume = "228",
    pages = "145",
    year = "1983"
}

@article{Gell-Mann:1960mvl,
    author = "Gell-Mann, Murray and Levy, M",
    title = "{The axial vector current in beta decay}",
    doi = "10.1007/BF02859738",
    journal = "Nuovo Cim.",
    volume = "16",
    pages = "705",
    year = "1960"
}

@article{Giddings:1983es,
    author = "Giddings, Steven and Abbott, James and Kuchar, Karel",
    title = "{Einstein's theory in a three-dimensional space-time}",
    doi = "10.1007/BF00762914",
    journal = "Gen. Rel. Grav.",
    volume = "16",
    pages = "751--775",
    year = "1984"
}

@article{Achucarro:1986uwr,
    author = "Achucarro, A. and Townsend, P. K.",
    editor = "Salam, A. and Sezgin, E.",
    title = "{A Chern-Simons Action for Three-Dimensional anti-De Sitter Supergravity Theories}",
    reportNumber = "Print-87-0078 (CAMBRIDGE)",
    doi = "10.1016/0370-2693(86)90140-1",
    journal = "Phys. Lett. B",
    volume = "180",
    pages = "89",
    year = "1986"
}

@article{Achucarro:1989gm,
    author = "Achucarro, A. and Townsend, P. K.",
    title = "{Extended Supergravities in $d$ = (2+1) as {Chern-Simons} Theories}",
    reportNumber = "PRINT-89-0627 (CAMBRIDGE)",
    doi = "10.1016/0370-2693(89)90423-1",
    journal = "Phys. Lett. B",
    volume = "229",
    pages = "383--387",
    year = "1989"
}

@article{Banados:1992wn,
    author = "Banados, Maximo and Teitelboim, Claudio and Zanelli, Jorge",
    title = "{The Black hole in three-dimensional space-time}",
    eprint = "hep-th/9204099",
    archivePrefix = "arXiv",
    reportNumber = "PRINT-92-0151 (CHILE), IASSNS-HEP-92-29",
    doi = "10.1103/PhysRevLett.69.1849",
    journal = "Phys. Rev. Lett.",
    volume = "69",
    pages = "1849--1851",
    year = "1992"
}

@article{Deser:1983tn,
    author = "Deser, Stanley and Jackiw, R. and 't Hooft, Gerard",
    title = "{Three-Dimensional Einstein Gravity: Dynamics of Flat Space}",
    reportNumber = "BRX-TH-137, PRINT-83-0884 (BRANDEIS)",
    doi = "10.1016/0003-4916(84)90085-X",
    journal = "Annals Phys.",
    volume = "152",
    pages = "220",
    year = "1984"
}

@article{Gott:1982qg,
    author = "Gott, J. Richard and Alpert, Mark",
    title = "{General relativity in a (2+1)-dimensional space-time}",
    doi = "10.1007/BF00762539",
    journal = "Gen. Rel. Grav.",
    volume = "16",
    pages = "243--247",
    year = "1984"
}

@inproceedings{Jackiw:1988sd,
    author = "Jackiw, R.",
    title = "{Quantum Gravity in Flatland}",
    booktitle = "{17th International Colloquium on Group Theoretical Methods in Physics}",
    reportNumber = "MIT-CTP-1622",
    month = "7",
    year = "1988"
}

@article{Ouvry:1988mm,
    author = "Ouvry, Stephane and Stora, Raymond and van Baal, Pierre",
    title = "{On the Algebraic Characterization of Witten's Topological Yang-Mills Theory}",
    reportNumber = "CERN-TH-5224/88, LAPP-TH-233/88, IPNO-TH-88-59",
    doi = "10.1016/0370-2693(89)90029-4",
    journal = "Phys. Lett. B",
    volume = "220",
    pages = "159--163",
    year = "1989"
}

@article{Staruszkiewicz:1963zza,
    author = "Staruszkiewicz, Andrzej",
    title = "{Gravitation Theory in Three-Dimensional Space}",
    journal = "Acta Phys. Polon.",
    volume = "24",
    pages = "735--740",
    year = "1963"
}

@article{Witten:1988hc,
    author = "Witten, Edward",
    title = "{(2+1)-Dimensional Gravity as an Exactly Soluble System}",
    reportNumber = "IASSNS-HEP-88-32",
    doi = "10.1016/0550-3213(88)90143-5",
    journal = "Nucl. Phys. B",
    volume = "311",
    pages = "46",
    year = "1988"
}

@article{Witten:1989sx,
    author = "Witten, Edward",
    title = "{Topology Changing Amplitudes in (2+1)-Dimensional Gravity}",
    reportNumber = "IASSNS-HEP-89/1",
    doi = "10.1016/0550-3213(89)90591-9",
    journal = "Nucl. Phys. B",
    volume = "323",
    pages = "113--140",
    year = "1989"
}

@article{Yang:1954ek,
    author = "Yang, Chen-Ning and Mills, Robert L.",
    editor = "Hsu, Jong-Ping and Fine, D.",
    title = "{Conservation of Isotopic Spin and Isotopic Gauge Invariance}",
    doi = "10.1103/PhysRev.96.191",
    journal = "Phys. Rev.",
    volume = "96",
    pages = "191--195",
    year = "1954"
}

@article{Hull:2009mi,
    author = "Hull, Chris and Zwiebach, Barton",
    title = "{Double Field Theory}",
    eprint = "0904.4664",
    archivePrefix = "arXiv",
    primaryClass = "hep-th",
    reportNumber = "IMPERIAL-TP-2009-CH-02, MIT-CTP-4031",
    doi = "10.1088/1126-6708/2009/09/099",
    journal = "JHEP",
    volume = "09",
    pages = "099",
    year = "2009"
}

@article{Berman:2010is,
    author = "Berman, David S. and Perry, Malcolm J.",
    title = "{Generalized Geometry and M theory}",
    eprint = "1008.1763",
    archivePrefix = "arXiv",
    primaryClass = "hep-th",
    reportNumber = "QMUL-PH-10-10",
    doi = "10.1007/JHEP06(2011)074",
    journal = "JHEP",
    volume = "06",
    pages = "074",
    year = "2011"
}

@article{Hull:2006va,
    author = "Hull, C M",
    title = "{Doubled Geometry and T-Folds}",
    eprint = "hep-th/0605149",
    archivePrefix = "arXiv",
    reportNumber = "IMPERIAL-TP-06-CH-02",
    doi = "10.1088/1126-6708/2007/07/080",
    journal = "JHEP",
    volume = "07",
    pages = "080",
    year = "2007"
}

@article{Aldazabal:2013sca,
    author = "Aldazabal, Gerardo and Marques, Diego and Nunez, Carmen",
    title = "{Double Field Theory: A Pedagogical Review}",
    eprint = "1305.1907",
    archivePrefix = "arXiv",
    primaryClass = "hep-th",
    doi = "10.1088/0264-9381/30/16/163001",
    journal = "Class. Quant. Grav.",
    volume = "30",
    pages = "163001",
    year = "2013"
}

@article{Hohm:2013bwa,
    author = {Hohm, Olaf and L{\"u}st, Dieter and Zwiebach, Barton},
    title = "{The Spacetime of Double Field Theory: Review, Remarks, and Outlook}",
    eprint = "1309.2977",
    archivePrefix = "arXiv",
    primaryClass = "hep-th",
    reportNumber = "MIT-CTP-4494, LMU-ASC-59-13, MPP-2013-241",
    doi = "10.1002/prop.201300024",
    journal = "Fortsch. Phys.",
    volume = "61",
    pages = "926--966",
    year = "2013"
}

@article{Andrianopoli:2008ea,
    author = "Andrianopoli, L. and D'Auria, R. and Ferrara, S. and Grassi, P. A. and Trigiante, M.",
    title = "{Exceptional N=6 and N=2 AdS(4) Supergravity, and Zero-Center Modules}",
    eprint = "0810.1214",
    archivePrefix = "arXiv",
    primaryClass = "hep-th",
    reportNumber = "CERN-PH-TH-2008-206, UCB-PTH-08-69, DISTA-2008",
    doi = "10.1088/1126-6708/2009/04/074",
    journal = "JHEP",
    volume = "04",
    pages = "074",
    year = "2009"
}

@article{Cvetic:1999au,
    author = "Cvetic, Mirjam and Lu, Hong and Pope, C. N.",
    title = "{Four-dimensional N=4, SO(4) gauged supergravity from D = 11}",
    eprint = "hep-th/9910252",
    archivePrefix = "arXiv",
    reportNumber = "CTP-TAMU-42-99, UPR-864-T",
    doi = "10.1016/S0550-3213(99)00828-7",
    journal = "Nucl. Phys. B",
    volume = "574",
    pages = "761--781",
    year = "2000"
}

@article{Gauntlett:2007ma,
    author = "Gauntlett, Jerome P. and Varela, Oscar",
    title = "{Consistent Kaluza-Klein reductions for general supersymmetric AdS solutions}",
    eprint = "0707.2315",
    archivePrefix = "arXiv",
    primaryClass = "hep-th",
    doi = "10.1103/PhysRevD.76.126007",
    journal = "Phys. Rev. D",
    volume = "76",
    pages = "126007",
    year = "2007"
}

@article{Louis:2014gxa,
    author = "Louis, Jan and Triendl, Hagen",
    title = "{Maximally supersymmetric AdS$_{4}$ vacua in N = 4 supergravity}",
    eprint = "1406.3363",
    archivePrefix = "arXiv",
    primaryClass = "hep-th",
    reportNumber = "ZMP-HH-14-15, CERN-PH-TH-2014-103",
    doi = "10.1007/JHEP10(2014)007",
    journal = "JHEP",
    volume = "10",
    pages = "007",
    year = "2014"
}

@article{Malek:2017njj,
    author = "Malek, Emanuel",
    title = "{Half-Maximal Supersymmetry from Exceptional Field Theory}",
    eprint = "1707.00714",
    archivePrefix = "arXiv",
    primaryClass = "hep-th",
    reportNumber = "LMU-ASC-39-17",
    doi = "10.1002/prop.201700061",
    journal = "Fortsch. Phys.",
    volume = "65",
    number = "10-11",
    pages = "1700061",
    year = "2017"
}

@article{Cassani:2019vcl,
    author = "Cassani, Davide and Josse, Gr{\'e}goire and Petrini, Michela and Waldram, Daniel",
    title = "{Systematics of consistent truncations from generalised geometry}",
    eprint = "1907.06730",
    archivePrefix = "arXiv",
    primaryClass = "hep-th",
    doi = "10.1007/JHEP11(2019)017",
    journal = "JHEP",
    volume = "11",
    pages = "017",
    year = "2019"
}

@article{Meinrenken2003,
  title={Group actions on manifolds},
  author={Meinrenken, Eckhard},
  journal={Lecture Notes, University of Toronto, Spring},
  year={2003}
}

@article{Blair:2024ofc,
    author = "Blair, Chris D. A. and Pico, Martin and Varela, Oscar",
    title = "{Infinite and finite consistent truncations on deformed generalised parallelisations}",
    eprint = "2407.01298",
    archivePrefix = "arXiv",
    primaryClass = "hep-th",
    reportNumber = "IFT-UAM/CSIC-24-99",
    doi = "10.1007/JHEP09(2024)065",
    journal = "JHEP",
    volume = "09",
    pages = "065",
    year = "2024"
}

@article{Bekaert:2002uh,
    author = "Bekaert, Xavier and Boulanger, Nicolas and Henneaux, Marc",
    title = "{Consistent deformations of dual formulations of linearized gravity: A No go result}",
    eprint = "hep-th/0210278",
    archivePrefix = "arXiv",
    reportNumber = "DFPD-02-TH-25, ULB-TH-02-31",
    doi = "10.1103/PhysRevD.67.044010",
    journal = "Phys. Rev. D",
    volume = "67",
    pages = "044010",
    year = "2003"
}

@article{Bekaert:2004dz,
    author = "Bekaert, Xavier and Boulanger, Nicolas and Cnockaert, Sandrine",
    title = "{No self-interaction for two-column massless fields}",
    eprint = "hep-th/0407102",
    archivePrefix = "arXiv",
    reportNumber = "ULB-TH-04-21, DAMTP-2004-67, DFPD-04-TH-14",
    doi = "10.1063/1.1823032",
    journal = "J. Math. Phys.",
    volume = "46",
    pages = "012303",
    year = "2005"
}

@article{West:2001as,
    author = "West, Peter C.",
    title = "{E(11) and M theory}",
    eprint = "hep-th/0104081",
    archivePrefix = "arXiv",
    reportNumber = "KCL-MTH-01-08",
    doi = "10.1088/0264-9381/18/21/305",
    journal = "Class. Quant. Grav.",
    volume = "18",
    pages = "4443--4460",
    year = "2001"
}

@article{Hohm:2018qhd,
    author = "Hohm, Olaf and Samtleben, Henning",
    title = "{The dual graviton in duality covariant theories}",
    eprint = "1807.07150",
    archivePrefix = "arXiv",
    primaryClass = "hep-th",
    doi = "10.1002/prop.201900021",
    journal = "Fortsch. Phys.",
    volume = "67",
    number = "5",
    pages = "1900021",
    year = "2019"
}

@article{West:2002jj,
    author = "West, Peter C.",
    title = "{Very extended E(8) and A(8) at low levels, gravity and supergravity}",
    eprint = "hep-th/0212291",
    archivePrefix = "arXiv",
    reportNumber = "KCL-MTH-02-31",
    doi = "10.1088/0264-9381/20/11/328",
    journal = "Class. Quant. Grav.",
    volume = "20",
    pages = "2393--2406",
    year = "2003"
}

@article{Henneaux:2019zod,
    author = "Henneaux, Marc and Lekeu, Victor and Leonard, Amaury",
    title = "{A note on the double dual graviton}",
    eprint = "1909.12706",
    archivePrefix = "arXiv",
    primaryClass = "hep-th",
    doi = "10.1088/1751-8121/ab56ed",
    journal = "J. Phys. A",
    volume = "53",
    number = "1",
    pages = "014002",
    year = "2020"
}

@article{Hull:2001iu,
    author = "Hull, C. M.",
    title = "{Duality in gravity and higher spin gauge fields}",
    eprint = "hep-th/0107149",
    archivePrefix = "arXiv",
    reportNumber = "QMUL-PH-01-01",
    doi = "10.1088/1126-6708/2001/09/027",
    journal = "JHEP",
    volume = "09",
    pages = "027",
    year = "2001"
}

@article{Curtright:1980yk,
    author = "Curtright, Thomas",
    title = "{Generalize Gauge Fields}",
    reportNumber = "EFI-80-04",
    doi = "10.1016/0370-2693(85)91235-3",
    journal = "Phys. Lett. B",
    volume = "165",
    pages = "304--308",
    year = "1985"
}

@article{Aulakh:1986cb,
    author = "Aulakh, C. S. and Koh, I. G. and Ouvry, S.",
    title = "{Higher Spin Fields With Mixed Symmetry}",
    reportNumber = "Print-86-0213 (ORSAY), IPNO-86-17",
    doi = "10.1016/0370-2693(86)90518-6",
    journal = "Phys. Lett. B",
    volume = "173",
    pages = "284--288",
    year = "1986"
}

@article{Thierry-Mieg:1980ihu,
    author = "Thierry-Mieg, Jean",
    title = "{{BRS} Structure of the Antisymmetric Tensor Gauge Theories}",
    reportNumber = "HUTMP 79/B86",
    doi = "10.1016/0550-3213(90)90497-2",
    journal = "Nucl. Phys. B",
    volume = "335",
    pages = "334--346",
    year = "1990"
}

@article{Labastida:1986gy,
    author = "Labastida, J. M. F. and Morris, T. R.",
    title = "{Massless Mixed Symmetry Bosonic Free Fields}",
    reportNumber = "PRINT-86-1128 (IAS,PRINCETON)",
    doi = "10.1016/0370-2693(86)90143-7",
    journal = "Phys. Lett. B",
    volume = "180",
    pages = "101--106",
    year = "1986"
}

@article{Bekaert:2002dt,
    author = "Bekaert, Xavier and Boulanger, Nicolas",
    title = "{Tensor gauge fields in arbitrary representations of GL(D,R): Duality and Poincare lemma}",
    eprint = "hep-th/0208058",
    archivePrefix = "arXiv",
    reportNumber = "ULB-TH-02-23",
    doi = "10.1007/s00220-003-0995-1",
    journal = "Commun. Math. Phys.",
    volume = "245",
    pages = "27--67",
    year = "2004"
}

@article{Henneaux:2021yzg,
    author = "Henneaux, Marc and Salgado-Rebolledo, Patricio",
    title = "{Carroll contractions of Lorentz-invariant theories}",
    eprint = "2109.06708",
    archivePrefix = "arXiv",
    primaryClass = "hep-th",
    doi = "10.1007/JHEP11(2021)180",
    journal = "JHEP",
    volume = "11",
    pages = "180",
    year = "2021"
}

@article{Levy-Leblond:1965dsc,
    author = "L{\'e}vy-Leblond, Jean-Marc",
    title = "{Une nouvelle limite non-relativiste du groupe de Poincar{\'e}}",
    journal = "Ann. Inst. H. Poincare Phys. Theor. A",
    volume = "3",
    number = "1",
    pages = "1--12",
    year = "1965"
}

@article{inonu1953contraction,
  title={On the contraction of groups and their representations},
  author={Inonu, Erdal and Wigner, Eugene Paul},
  journal={Proceedings of the National Academy of Sciences},
  volume={39},
  number={6},
  pages={510--524},
  year={1953}
}

@article{DeWitt:1967yk,
    author = "DeWitt, Bryce S.",
    editor = "Fang, Li-Zhi and Ruffini, R.",
    title = "{Quantum Theory of Gravity. 1. The Canonical Theory}",
    doi = "10.1103/PhysRev.160.1113",
    journal = "Phys. Rev.",
    volume = "160",
    pages = "1113--1148",
    year = "1967"
}

@article{Arnowitt:1959ah,
    author = "Arnowitt, Richard L. and Deser, Stanley and Misner, Charles W.",
    title = "{Dynamical Structure and Definition of Energy in General Relativity}",
    doi = "10.1103/PhysRev.116.1322",
    journal = "Phys. Rev.",
    volume = "116",
    pages = "1322--1330",
    year = "1959"
}

@article{Henneaux:1979vn,
    author = "Henneaux, Marc",
    title = "{Geometry of Zero Signature Space-times}",
    reportNumber = "PRINT-79-0606 (PRINCETON)",
    journal = "Bull. Soc. Math. Belg.",
    volume = "31",
    pages = "47--63",
    year = "1979"
}

@article{Isham:1975ur,
    author = "Isham, C. J.",
    title = "{Some Quantum Field Theory Aspects of the Superspace Quantization of General Relativity}",
    reportNumber = "Print-76-0255 (KING S COLL.)",
    doi = "10.1098/rspa.1976.0138",
    journal = "Proc. Roy. Soc. Lond. A",
    volume = "351",
    pages = "209--232",
    year = "1976"
}

@inproceedings{Teitelboim:1978wv,
    author = "Teitelboim, Claudio",
    title = "{Surface deformations, their suqare root and the signature of spacetime}",
    booktitle = "{7th International Group Theory Colloquium: The Integrative Conference on Group Theory and Mathematical Physics}",
    reportNumber = "Print-78-1134 (PRINCETON)",
    month = "12",
    year = "1978"
}

@article{Damour:2002et,
    author = "Damour, T. and Henneaux, M. and Nicolai, H.",
    title = "{Cosmological billiards}",
    eprint = "hep-th/0212256",
    archivePrefix = "arXiv",
    reportNumber = "IHES-P-02-80, AEI-2002-092, ULB-TH-02-33, IHES-P-02-08",
    doi = "10.1088/0264-9381/20/9/201",
    journal = "Class. Quant. Grav.",
    volume = "20",
    pages = "R145--R200",
    year = "2003"
}

@article{Belinsky:1982pk,
    author = "Belinsky, V. and Khalatnikov, I. and Lifshitz, E.",
    title = "{A General Solution of the Einstein Equations with a Time Singularity}",
    doi = "10.1080/00018738200101428",
    journal = "Adv. Phys.",
    volume = "31",
    pages = "639--667",
    year = "1982"
}

@article{Bidussi:2021nmp,
    author = "Bidussi, Leo and Hartong, Jelle and Have, Emil and Musaeus, J{\o}rgen and Prohazka, Stefan",
    title = "{Fractons, dipole symmetries and curved spacetime}",
    eprint = "2111.03668",
    archivePrefix = "arXiv",
    primaryClass = "hep-th",
    doi = "10.21468/SciPostPhys.12.6.205",
    journal = "SciPost Phys.",
    volume = "12",
    number = "6",
    pages = "205",
    year = "2022"
}

@article{Figueroa-OFarrill:2023vbj,
    author = "Figueroa-O'Farrill, Jos{\'e} and P{\'e}rez, Alfredo and Prohazka, Stefan",
    title = "{Carroll/fracton particles and their correspondence}",
    eprint = "2305.06730",
    archivePrefix = "arXiv",
    primaryClass = "hep-th",
    reportNumber = "EMPG-23-09",
    doi = "10.1007/JHEP06(2023)207",
    journal = "JHEP",
    volume = "06",
    pages = "207",
    year = "2023"
}

@article{Imbimbo:2023sph,
    author = "Imbimbo, Camillo and Rovere, D. and Warman, A.",
    title = "{Superconformal anomalies from superconformal Chern-Simons polynomials}",
    eprint = "2311.05684",
    archivePrefix = "arXiv",
    primaryClass = "hep-th",
    month = "11",
    year = "2023"
}

@article{Rovere:2024wtv,
    author = "Rovere, Davide",
    title = "{Kodaira-Spencer anomalies with Stora-Zumino method}",
    eprint = "2403.17071",
    archivePrefix = "arXiv",
    primaryClass = "hep-th",
    doi = "10.1007/JHEP01(2025)073",
    journal = "JHEP",
    volume = "01",
    pages = "073",
    year = "2025"
}

@article{Rovere:2024nwc,
    author = "Rovere, Davide",
    title = "{Anomalies in covariant fracton theories}",
    eprint = "2406.06686",
    archivePrefix = "arXiv",
    primaryClass = "hep-th",
    doi = "10.1103/PhysRevD.110.085012",
    journal = "Phys. Rev. D",
    volume = "110",
    number = "8",
    pages = "085012",
    year = "2024"
}

@article{Inverso:2024xok,
    author = "Inverso, Gianluca and Rovere, Davide",
    title = "{How to uplift D = 3 maximal supergravities}",
    eprint = "2410.14520",
    archivePrefix = "arXiv",
    primaryClass = "hep-th",
    doi = "10.1007/JHEP02(2025)130",
    journal = "JHEP",
    volume = "02",
    pages = "130",
    year = "2025"
}

@article{Rovere:2025nfj,
    author = "Rovere, Davide",
    title = {{Covariant Fractons and Weitzenb{\"o}ck Torsion}},
    eprint = "2505.21022",
    archivePrefix = "arXiv",
    primaryClass = "hep-th",
    month = "5",
    year = "2025"
}

@article{Fecit:2025eet,
    author = "Fecit, Filippo and Rovere, Davide",
    title = "{Worldline Formulations of Covariant Fracton Theories}",
    eprint = "2508.14591",
    archivePrefix = "arXiv",
    primaryClass = "hep-th",
    month = "8",
    year = "2025"
}

@article{Imbimbo:2025ffw,
    author = "Imbimbo, Camillo and Porro, Ludovico",
    title = "{One Ring to Rule Them All: A Unified Topological Framework for 4D Superconformal Anomalies}",
    eprint = "2507.16505",
    archivePrefix = "arXiv",
    primaryClass = "hep-th",
    month = "7",
    year = "2025"
}

@article{Rovere:2025jks,
    author = "Rovere, Davide and Sterckx, Colin",
    title = "{How to uplift non-maximal gauged supergravities}",
    eprint = "2510.24850",
    archivePrefix = "arXiv",
    primaryClass = "hep-th",
    month = "10",
    year = "2025"
}

\newpage
\quad
\thispagestyle{empty}

\newpage

\end{document}